\documentclass[11pt]{article}
\usepackage{amsmath,amssymb}
\usepackage[T1]{fontenc}
\usepackage{lmodern}
\makeatletter
\def\dedicatory#1{\def\@dedicatory{#1}}
\let\@dedicatory=\@empty
\makeatother

\numberwithin{equation}{section}
\let\dsl\displaylines
\let\bs\boldsymbol
\def\ex#1{\exp(#1)}
\def\ba{{\bar a}}
\def\hi{\wh\imath}

\let\iy\infty
\let\op\oplus
\let\ot\otimes
\let\ov\overline
\let\os\overrightarrow
\let\pa\partial
\def\pp#1#2{\frac{\pa #1}{\pa #2}}
\let\q\quad
\def\qh#1{\quad\hbox{#1}\quad}
\let\td\tilde
\let\tm\times
\let\ts\textstyle
\let\ul\underline
\let\wh\widehat
\let\wt\widetilde
\let\a\alpha
\let\b\beta
\def\brr#1#2{\overset{#1}{#2}}
\def\br#1#2{\overset{\rm #1}{#2}}
\def\brgg#1{\br{gauge(G2)}{\nabla_{#1}}}
\def\brgn#1{\br{gauge}{\nabla_{#1}}}
\def\brg#1{\br{gauge}{\nabla^{#1}}}
\let\d\delta
\let\D\varDelta
\def\ddiv{\operatorname{div}\nolimits}
\let\ve\varepsilon
\let\g\gamma
\def\gauge{\operatorname{gauge}\nolimits}
\def\grad{\operatorname{grad}\nolimits}
\def\gtk#1{g^{[#1]}}
\def\gtn#1{{\t g}^{(#1)}}
\let\G\varGamma
\def\ink#1{_{[#1]}}
\let\ka\kappa
\def\kin{\operatorname{kin}\nolimits}
\let\la\lambda
\let\La\varLambda
\let\n\nabla
\def\Nabla{\mathop{\nabla}\limits}
\let\o\omega
\let\om\omega
\let\O\varOmega
\let\F\varPhi
\let\vf\varphi
\let\Ps\varPsi
\let\PS\varPsi
\def\scal{\operatorname{scal}\nolimits}
\let\si\sigma
\let\Si\varSigma
\let\stpr\overrightarrow
\let\t\theta

\def\tv#1{\wt{\ov#1}{}}
\let\z\zeta
\def\SU{\text{SU}}
\def\U{\text{U}}
\def\(#1){{(#1)}}
\def\[#1]{{[#1]}}
\def\Ft#1#2{\F^{#1}_{\td #2}}
\def\gv{\nad{gauge}}
\def\gw{\nad{gauge(G2)}}
\def\crt{_{\rm crt}}
\def\pl{_{\rm Pl}}
\def\m{{\mu\nu}}

\def\cL{\mathcal L}
\def\cLY{\mathcal L_{\rm YM}}

\def\C{\mathbb C}
\def\R{\mathbb R}
\def\fg{\mathfrak g}
\def\fh{\mathfrak h}
\def\fm{\mathfrak m}
\def\tA{{\tilde A}}
\def\ta{{\tilde a}}
\def\tB{{\tilde B}}
\def\tb{{\tilde b}}
\def\tC{{\tilde C}}
\def\dg #1,#2,#3,{{#1{}_{#2}}^{\!#3}}
\def\dgs#1,#2,{_{#1}{}^{#2}}
\def\gd #1,#2,#3,{{#1{}^{#2}}_{\!#3}}
\def\gds#1,#2,{^{#1}{}_{\!#2}}
\def\gdg #1,#2,#3,#4,{{{#1{}^{#2}}_{\!#3}}{}^{\!#4}}
\def\dgd #1,#2,#3,#4,{{{#1{}_{#2}}^{\!#3}}{}_{\!#4}}
\def\nad#1#2{\overset{\rm #1}{#2}}

\def\falg{{\mathrel{\lower5pt\hbox{${\scriptstyle\sim}$}\hskip-5pt g}}}
\def\fal#1{{\mathrel{\lower4pt\hbox{${\scriptstyle\sim}$}\hskip-7.5pt #1}}}
\def\tl{\mathopen{\hbox{$\textstyle[$}}}
\def\tp{\mathclose{\hbox{$\textstyle]$}}}

\def\dwa#1{\textstyle{\substack{#1}}}
\def\falkag{\mathrel{\lower6pt\hbox{${\scriptstyle\sim}$}\hskip-5pt g}}
\def\falkaL{\mathrel{\lower4pt\hbox{${\scriptstyle\sim}$}\hskip-6pt L}}
\newdimen\krwys
\def\dkr#1{\leavevmode\setbox0\hbox{#1}\krwys=\dp0\advance\krwys by2pt
\rlap{\lower\krwys\hbox to\wd0{\hfil.\hfil}}#1}
\newdimen\krop
\setbox0=\hbox{$\scriptstyle\a$}
\krop=\wd0
\def\cdt{\hbox to\krop{$\scriptstyle\hfil\cdot\hfil$}}
\def\lw#1 {\lower#1pt\hbox\bgroup$\scriptstyle}
\def\eg{$\egroup}
\def\eff{\operatorname{eff}\nolimits}
\def\hor{\operatorname{hor}}
\def\sgn{\operatorname{sgn}\nolimits}

\def\ver{\operatorname{ver}}

\def\Int{\operatorname{int}\nolimits}

\def\Ad{\operatorname{Ad}}

\def\SO{{\rm SO}}
\def\Sp{{\rm Sp}}

\def\pap#1#2{\frac{\pa #1}{\pa #2}}
\def\udots{\lower1pt\hbox to4pt{\hss$\cdot$\hss}\raise2pt\hbox to4pt{\hss$\cdot$\hss}\raise5pt\hbox to4pt{\hss$\cdot$\hss}}
\def\hb#1{\widehat{\overline{#1}}}
\def\mpl{m\pl}

\def\rF{\text{F}}
\def\rE{\text{E}}
\def\rA{\text{A}}
\def\rB{\text{B}}
\def\rC{\text{C}}
\def\rD{\text{D}}
\def\rE{\text{E}}

\def\rH{\text{H}}
\def\rM{\text{M}}
\def\rN{\text{N}}

\def\rMC{\text{MC}}
\def\rAB{\text{AB}}
\def\rBA{\text{BA}}
\def\rCB{\text{CB}}
\def\rBC{\text{BC}}
\def\rBN{\text{BN}}
\def\rDB{\text{DB}}
\def\rDN{\text{DN}}
\def\rDM{\text{DM}}
\def\rYM{\text{YM}}
\def\rDE{\text{DE}}
\def\rBD{\text{BD}}
\def\rDC{\text{DC}}
\def\rAC{\text{AC}}
\def\rCA{\text{CA}}
\def\rCM{\text{CM}}
\def\rAM{\text{AM}}

\def\rMA{\text{MA}}
\def\rMB{\text{MB}}
\def\rMN{\text{MN}}
\def\rNM{\text{NM}}
\def\rBM{\text{BM}}
\def\rAD{\text{AD}}

\let\ti\textit
\def\beq#1 #2\e{\begin{equation}\label{#1}#2\end{equation}}
\def\bea#1 #2\e{\begin{align}\label{#1}#2\end{align}}
\def\bml#1 #2\e{\begin{multline}\label{#1}#2\end{multline}}
\def\bmlg#1\e{\begin{multline*}#1\end{multline*}}
\def\bg#1 #2\e{\begin{gather}\label{#1}#2\end{gather}}
\let\bal\aligned \let\eal\endaligned
\let\bga\gathered \let\ega\endgathered
\def\bma{\left(\begin{array}{c|c}} \def\ema{\end{array}\right)}
\def\bca{\begin{cases}}
\def\eca{\end{cases}}
\def\bit{\begin{itemize}}
\def\eit{\end{itemize}}

\let\nn\nonumber
\def\lb#1 {\label{#1}}
\let\er\eqref
\def\ER(#1){\eqref{#1}}

\def\up#1{\uppercase{#1}}
\def\E{\expandafter\up}
\def\ap{approximat}

\def\cf{coefficient}
\def\cm{composition}
\def\cfn{confinement}
\def\cn{connection}

\def\ct{constant}
\def\cd{coordinate}
\def\co{cosmolog}
\def\ci{covariant}
\def\cvt{curvature}
\def\dc{dependen}
\def\dv{derivative}
\def\di{di\-men\-sion}
\def\ef{effective}
\def\elm{electromagneti}
\def\el{element}
\def\e{equation}
\def\exi{existence}

\def\fw{following}
\def\f{function}
\def\fn{fundamental}
\def\gn{generalization}
\def\gz{generalized}
\def\gr{gravitation}
\def\Hm{Higgs' mechanism}

\def\ia{interaction}
\def\iv{invarian}
\def\KK{Kaluza--Klein }
\def\KWK{Kerner--Wong--Kopczy\'nski }
\def\KC{Killing--Cartan }
\def\lg{lagrangian}

\def\LC{Levi-Civita }
\def\Li{Lie algebra}
\def\MR{Moffat--Ricci}
\def\mo{morphi}
\def\nA{non-Abelian}
\def\JT{Nonsymmetric Jordan--Thiry Theory}
\def\nos{nonsymmetric}
\def\NK{Nonsymmetric Kaluza--Klein Theory}
\def\pc{particle}

\def\pt{potential}
\def\qe{quintessence}
\def\rp{represent}
\def\sf{satisf}
\def\so{solution}
\def\spt{space-time}

\def\sn{spontaneous}
\def\ssb{spontaneous symmetry breaking}
\def\sc{structur}
\def\st{such that }
\def\sy{symmetr}
\def\s{symmetr}
\def\tr{transform}
\def\tf{transformation}
\def\un{unification}
\def\Un{Universe}
\def\v{variation}
\def\wrt{with respect to }
\def\YM{Yang--Mills}
\let\TM\texttrademark

\def\dah#1{{\textstyle{\hat {\scriptstyle #1}}}}
\def\mPsi{\varPsi}
\def\mPhi{\varPhi}
\def\eq#1 {\label{#1}}
\let\al\aligned
\def\ealn{\begin{align}}
\def\gH{\mathfrak h}
\def\gG{\mathfrak g}
\def\gM{\mathfrak m}
\def\hj{{\wh\jmath}}
\let\d\delta
\def\G{\varGamma}
\let\u\tilde
\def\Id{\text{Id}}
\def\rank{\operatorname{rank}}
\def\gag{\overset{\text{\rm gauge}}\nabla}
\def\dr#1{_{\rm #1}}
\def\dq#1{^{\rm #1}}
\def\dpl{\mathbin{\overset.+}}
\def\dt{dependent}
\def\sb{symmetry breaking}
\def\qs{quintessence}

\title{The Nonsymmetric
Kaluza--Klein Theory and Modern Physics.\\
A novel approach II}
\author{M. W. Kalinowski\\
Bioinformatics Laboratory, Medical Research Centre,
Polish Academy of Sciences,\\
e-mail: markwkal1@gmail.com}
\dedicatory{To the memory of my teacher Professor Adam Bielecki\\
\vskip12pt \rightline{{\bf Motto:} W fizyce teoretycznej potrzebujemy ka\.zdej matematyki,\hskip30pt}
\rightline{a g\l\'ownie takiej, kt\'orej jeszcze nie ma.\hskip30pt}
\rightline{$($In theoretical physics we need all the mathematics,\hskip30pt}
\rightline{especially such which does not exist up to now.$)$\hskip30pt}\rightline{\rm Adam Bielecki\hskip30pt}\vskip18pt}
\begin{document}
\maketitle

\makeatletter
\vtop{\centering{\footnotesize\itshape\@dedicatory\@@par}%
\global\dimen@i\prevdepth}\prevdepth\dimen@i
\makeatother

\begin{abstract}
\looseness-1
In the paper we consider the Nonsymmetric Kaluza--Klein (Jordan--Thiry) Theory
and hierarchy of symmetry breaking, within Grand Unified Theories. In this way
we try to construct Unified Field Theory. We consider also a quintessence and
skewon fields as possible Dark Matter particles. Both particles are massive
with zero and one spin. It means, with a scalar and a pseudovector particle.
They are interacting only gravitationally. They are
really a part of gravity. In this way they are geometrized.
We find a natural way to get a cosmological constant
(Dark Energy) and the fifth force. We consider also an effective gravitational
``constant'' $G_{\rm eff}$ and a test particle movement in the theory,
i.e.\ a generalized Kerner--Wong--Kopczy\'nski equation with a scalar field~$\rho$.
We consider an inflation in cosmological model with a scalar field as an
inflaton. We consider also a tower of scalar (massive) fields as an additional Dark
Matter derived in the paper. A partial unification of physical fundamental
interactions (a~bosonic part) has been developed, which is a unification of a
bosonic part of a Standard Model with nonsymmetric gravity (NGT) in Nonsymmetric
Kaluza--Klein (Jordan--Thiry) Theory scheme.
We find an exact solution of field equations of the unification with remarkable
properties, which describes ``charge without charge'' and ``mass without mass''
with cosmological constant. We consider masses of~$W^\pm$ and $Z^0$ bosons
getting an agreement with an experiment. We get also an agreement for an
experiment for a mass of Higgs' boson and a value of a Weinberg angle.
We prove that a cosmological constant is not zero and calculate it.
We consider a soliton-bag model of hadrons in the context of our unification.
We consider a nonlocal quantization of the theory.
We do as much as possible to present our formalism friendly for readers.
\end{abstract}

\section{Introduction}
The paper is a continuation of Ref.~\cite{11a}. For this it has number~II.
It is not a review paper. It is a path to a Unified Field Theory, which should be geometrical.
In this paper we consider a \NK\ and the \NK\ with
a \ssb\ and \Hm\ as a unification of NGT (\E\nos\ \E\gr al Theory) with \YM'
fields and Higgs' fields (see Refs \cite{1}--\cite{5}), for NGT see Ref.~\cite6.
This is NGT, not NewNGT (NNGT) (see Ref.~\cite{Febb}).
We consider a hierarchy of two stages of \sb\ in the \NK\ and the \NK\ with
a \ssb\ and \Hm\ within Grand Unified Theories.
We develop a hierarchy of two stages of the \sb\ in our theory.
This approach is a preliminary version of a Unified Field Theory.
For further development of the \E\nos\ \KK (Jordan--Thiry) Theory see
Refs. \cite{7}--\cite{8}. For Grand Unified models see Refs \cite{al}, \cite{be}, \cite{t}.
In Ref.~\cite{11a} all the formalism was developed under the assumption that
$\rho=1$ ($\Psi=0$). Thus our extension is significant ($\rho\ne1$ ($\Psi\ne0$)).

Let us remind to the reader that NGT (\E\nos\ \E\gr\ Theory) known also as
Moffat theory is a theory which uses mathematical apparatus of Einstein Unified
Theory as a pure \gr al theory. Moreover, A.~Einstein considers also his
unified field theory as a generalized theory of gravity (see Ref.~\cite{p}).

We construct the Nonsymmetric Jordan--Thiry Theory
unifying NGT, the Yang--Mills' field, the Higgs' fields and scalar
forces in a geometric manner. In this way we get masses from higher
dimensions. We discuss spontaneous symmetry breaking, the Higgs'
mechanism and a mass generation in the theory. The scalar field
${\varPsi}$ (as
in the classical Jordan--Thiry Theory) is connected to the effective
gravitational constant. This field is massive and has Yukawa-type
behaviour. The field~$\Ps$ can be considered as a quintessence field.
Moreover, to make the formalism friendly for readers and more understandable
we repeat some results from Ref.~\cite{11a}. In this way it is easy to see
where the assumption $\rho\ne1$ ($\Psi\ne0$) enters.

We consider a mass of a quintessence particle, various properties of a
quintessence field. In Ref.~\cite{xx} we consider some \co ical effects
of a quintessence field.

The \nos\ \KK\ (Jordan--Thiry) Theory unifies the gauge invariance
principle with the coordinate invariance principle but in more than
four-dimen\-sional space-time. In particular in the case of electromagnetic and
\gr al interactions in 5-dimensional theory.

A general nonabelian Yang--Mills fields have been unified with gravity in
$(n+4)$-\di al space-time ($n$---a \di\ of gauge group).
The theory uses a \nos\ metric defined on a metrized (in a \nos\
way) principal fibre bundle over a \spt\ with a structural group $\U(1)$ in an
\elm c case and in general case nonabelian semi-simple compact group~$G$. The
\cn\ on \spt\ and on a metrized principal fibre bundle is compatible with
this metric. This \cn\ is similar to a \cn\ from Einstein's Unified Field
Theory, however we use its higher \di al analogue. This \cn\ is
right-\iv t \wrt an action of the group~$G$ (a~gauge group).

In the \elm c case the metric and the \cn\ are bi-\iv t \wrt the group~$\U(1)$.
The theory has been developed to
include a scalar field leading to an effective \gr\ constant and \spt\
dependent cosmological terms. It is possible to extend the theory to include
Higgs' fields and spontaneous symmetry breaking of a gauge group to get
massive vector boson fields.

The theory is fully relativistic and unifies \elm c field, gauge fields,
Higgs' field and scalar forces with NGT (\E\nos\ Gravitation Theory)
in a nontrivial way.
By `in a nontrivial way' we mean that we get from the theory
something more than NGT, ordinary Kaluza--Klein (ordinary Jordan--Thiry) Theory, classical
electrodynamics, Yang--Mills' field theory with Higgs' field and spontaneous
symmetry breaking. These new features are some kind of ``interference
effects'' between all of them.
This theory unifies two important approaches in higher-dimensional
philosophy: Kaluza--Klein principle and a \di al reduction principle.
A~Dark Matter and a Dark Energy are ``interference effects'' in our \un s.
They are geometrized.

There is a controversy between an existence of Dark Matter and modification
of theory of gravity. The truth probably is in the middle. In our approach,
dark matter \pc s are coming from \JT\ which modifies General Relativity.
Dark Energy (\E\co ical \E\ct, \E\qs) is also coming from \JT. Dark Matter
and Dark Energy in our approach are strongly combined. In particular a
\E\co ical \E\ct\ is a stationary value of a selfinteraction \pt\ of
a~field~$\Psi$ (a~\E\qe). The \E\co ical \E\ct\ induces a mass of a skewon
field. A~scalar field~$\vf$ (a~fluctuation around a~stationary value of
$\Psi,\Psi_0$) is also massive. In this way a~Dark Matter and Dark Energy is
a~part of extended gravity coming from Unified Theory. Our scalar Dark Matter
is a Self-Interacting Dark Matter (SIDM). In Section~5 we consider many
consequences of the Dark Matter from the point of view of nonlinear phenomena.

For a Dark Matter Problem see Refs \cite{k,l1,m,r}. In Ref.~\cite{m} someone
finds a correlation coming to \fn\ scale of an acceleration in galaxies. This
supports a claim for modification of~GR. In Ref.~\cite{r} someone denied it,
i.e.\ there is an absence of such a scale. This supports existence of Dark
Matter. A Dark Energy (\E\co ical \E\ct) has been discovered and reported
in Refs \cite{n,p1}.

According to modern ideas our Universe is contemporary described by
Robertson--Walker spatially flat metric. A~model of the Universe is
Friedman--Lema\^\i tre model with a \co ical \ct\ (a~Dark Energy). The model
contains pressureless dust matter (see Ref.~\cite{yy,yy1}):
\begin{itemize}


\item ordinary matter (barionic matter) 4.9\%


\item Dark Matter (cold Dark Matter) 26.8\%


\item Dark Energy 68.3\%.
\end{itemize}
A Dark Energy is coming from a \co ical \ct\ which can have a dynamical origin
as in the \E\nos\ \KK (Jordan--Thiry) Theory.
A~Dark Matter is needed on the level of galaxies, clusters of
galaxies and on the level of whole Universe. This matter is considered as
pressureless dust. Moreover, we expect a Dark Matter to consist of some \el ary
\pc s to be discovered. However, up to now we did not detect such \pc s. They
must interact very weakly with our detectors (i.e.\ with ordinary matter).
There are several concepts of such \pc s. In the \nos\ \KK (Jordan--Thiry) Theory
they are: scalarons (massive) and massive skewons (see Section~4).

The model of the Universe in NKK(J-T)T is $\Lambda$CDM-model.
The recent Hubble \ct\ (Hubble parameter) crisis (see Ref.~\cite{m2d}) can be
solved (in principle) in NKK(J-T)T inspired \co y (see Appendix~C).
Moreover, a Dark
Matter problem (Cold Dark Matter) seems to be solved in our approach
because scalarons (spin zero)
\pc s and skewons (spin one)---pseudovector \pc s interact very weakly with an
ordinary matter (only \gr ally).

The beautiful theories such as Kaluza--Klein theory (a~Kaluza miracle) and
its descendents should pass the following test if they are treated as real
unified theories. They should incorporate chiral fermions. Since the
fundamental scale in the theory is a Planck's mass, fermions should be
massless up to the moment of spontaneous symmetry breaking. Thus they should
be zero modes. In our approach they can obtain masses on a \di al reduction
scale. Thus they are zero modes in $(4+n_1)$-\di al case. In this way
$(n_1+4)$-\di al fermions are not chiral (according to very well known
Witten's argument on an index of a Dirac operator). Moreover, they are
not zero modes after a \di al reduction, i.e., in 4-\di al case. It means we
can get chiral fermions under some assumptions.

Using a quite old dictionary \cite{ga} we paraphrase a notion of Unified Field
Theory: {\it Unified Field Theory---any theory which attempts to express \gr al
theory and fundamental interactions theories within a single unified framework
to generalize Einstein's general theory of relativity alone and classical
theories describing fundamental interactions}. In original statement it
considers \elm c \ia s. In our case this single unified framework is a
multidimensional analogue of a geometry from Einstein's Unified Field Theory
(treated as a \gz\ gravity) defined on principal fibre bundles with base
manifolds $E$ or $E\tm M$ and structural groups $G$ or~$H$ ($E$~means \spt,
$M=G/G_0$ is a manifold of vacuum states). Thus the definition from an old
dictionary (paraphrased by us) is still valid. Thus this is a unified
description of \gr al, strong, \elm c and weak \ia s. (In our approach it is
possible to get the fifth force, i.e.\ non\ct\ \gr al ``\ct''.)
For a historical development
of unified theories see Refs \cite{de},~\cite{eps}. Let us mention that our
multidimensional connection is right-\iv t \wrt an action of the group ($G$~or~$H$).
For some philosophical considerations see Refs~\cite{q},~\cite{a}.

We expect some nonrelativistic effects leading
to nonnewtonian gravity.

Let us give some general remarks on \un\ of \ia s. In the case of \gr al and
\elm c fields we write down Einstein and Maxwell \e s. If we switch off gravity
we get only Maxwell \e s. If we switch off \elm sm we get only vacuum Einstein
\e s. Such approach is very well known in physics. For example we can write
down \e s of continuous mechanics with elasticity and thermoelasticity (see
Ref.~\cite{m2A}). If we switch off a heat production with established
temperature we get only elasticity. Such approach is advocated by R.~P.~Feynman
in his lectures of physics (see Ref.~\cite{m2B}). Classical (5-\di al) \KK
Theory (\s ic) belongs to those approaches.

In our approach of \un, i.e.\ \E\nos\ \KK (Jordan--Thiry) Theory the situation
is different. For example if we switch off gravity in a \un\ of bosonic GSW
model in NKK(J-T)T we still have important effects coming from the fact that
the theory has a \nos\ metric, i.e.\ a possibility of obtaining a correct mass
for a Higgs' boson and a finite renormalization of a Weinberg angle. They are
``interference effects'' in \un\ theory (see Ref.~\cite{11a}). Those effects
are absent in previously mentioned approaches.

The real motivation of this work is to find a geometric unifications of
fundamental interactions of Nature and to find applications in Modern
Cosmology including inflation, Dark Matter and Dark Energy using a paradigm
of physics, which unifies two fundamental concepts of invariance in physics:
coordinate invariance principle and gauge invariance principle. The first is
known in General Relativity and in vaiable alternative theories of gravitation. The
second is known in Electrodynamics and Yang--Mills field theory. Yang--Mills
field theory governs Standard Model, i.e.: Glashow--Salam--Weiberg (GSW) model of
electro-weak interactions and the theory of strong interactions (QCD) using
$\text{SU}(2)_L\times \U(1)$ and SU(3)$_c$ groups as gauge groups. It governs
also any Grand Unified Theorems.

We give here a solution for Dark Matter and Dark Energy problem. Our Dark Matter
\pc s are a scalar \pc\ (field~$\Ps$ or~$\vf$ or~$q_0$) and a vector \pc\ massive skewon
field. Dark energy is a \co ical \ct\ as a stationary value of a potential
of self-\ia\ of $\Ps$~field.

Let us notice that in Ref.~\cite{11a} we develop a geometrical unification
within \NK\ of gravity (described by NGT) and Glashow--Salam--Weinberg model\break
(a~bosonic part of GSW).

In the mentioned paper we get correct values of masses of $W^{\pm}$ and $Z^0$
bosons and a mass of a Higgs' particle. We get a correct value of a Weinberg angle.
It means we explain the fact that $\theta_W$ is smaller than $\frac\pi6$ ($30^\circ$)
by a finite renormalization. In that paper we develop a dielectric model of a
\cfn\ in QCD (\cfn\ from higher \di s). In the paper we use a real version of
the \E\nos\ \KK (Jordan--Thiry) Theory. Moreover, all the formulae are correct
in complex and hypercomplex Hermitian version if we change
$$
g_\[\a\b] \to ig_\[\a\b] \qh{or }g_\[\a\b]\to Jg_\[\a\b]
$$
where $i^2=-1$, $J^2=1$.

We can do also:
\begin{align*}
\ell_{ab} &= h_{ab} + i\mu k_{ab}, \q k_{ab}=-k_{ba} \hbox{ or }\\
\ell_{ab} &= h_{ab}+J\mu k_{ab},\\
\ell_{\td a\td b} &= {h^0}_{\td a\td b}+ i\z {k^0}_{\td a\td b}, \q
{k^0}_{\td a\td b}=-{k^0}_{\td b\td a} \quad \hbox{or}\\
\ell_{\td a\td b} &= {h^0}_{\td a\td b} + J\z {k^0}_{\td a\td b}
\end{align*}
and we have $\ell^+_{ab}=\ell_{ab}$, $\ell^+_{\td a\td b}=\ell_{\td a\td b}$,
$\g^+_{AB} = \g_{AB}$, $\ka^+_{\td A\td B}=\ka_{\td A\td B}$.

Our claim is as follows: Unified Field
Theory = \E\nos\ \KK (Jordan--Thiry) Theory + Grand Unified Theory.
In some sense this is ToE (Theory of Everything).

We discuss cosmological models involving field $\varPsi$ which plays a
r\^ole of a quintessence field (see Refs \cite{5,xx}). We find inflationary models of the Universe.
We consider the
field~$\varPsi$ as a quintessence field building some cosmological models
with a quintessence.
We speculate on a future of the Universe based on our simple model with
a quintessence. In this paper we consider $\Lambda CDM$ \co ical model coming
from our theory. This model includes inflation.

We consider an infinite tower of scalar fields ${\varPsi} _{k}(x)$ coming
from the expansion of the field ${\varPsi} (x,y)$ on the manifold $M=G/G_{0}$
into harmonics of the Beltrami--Laplace operator.
Due to Friedrichs' theory we can diagonalize an infinite matrix
(with a~scale $\frac{hc}r$) of
masses for ${\varPsi} _{k}$ transforming them into new fields
${\varPsi}'_k$. The truncation procedure means here to take a zero mass mode
${\varPsi} _{0}$ and equal it to ${\varPsi} $ from
the preceding section. This is an extension of our Dark Matter.
Our Dark Matter consists of two \pc s: a~\qe\ \pc\ (spin zero) and a skewon
\pc\ (spin one---vector \pc). Moreover, we can extend a spectrum of a Dark
Matter as an expansion of $\Ps(x,y)$ field ($x\in E$, $y\in M=G/G_0$). We
choose in the paper the simplest possibility: $\Ps$~is a function of~$x$ only.
In Appendix B we give a more general approach.

Let us give some details of our \co ical models in the \E\nos\ \KK (Jordan--Thiry)
Theory. According to Refs \cite{5},~\cite{xx} we have several inflationary scenarios.
Our models are starting from nonzero vacuum energy (due to \co ical \ct\ present
in the theory) with zero remaining fields (including also $\Ps$~field). We can
have second order phase transition in an evolution of the Universe. It means,
we go from one de-Sitter phase to the second one with creation of energy from
vacuum energy. We can calculate an amount of an inflation.
Moreover, the most interesting scenario is a \qe\ inflationary scenario
with an inflation driven by the scalar field~$\Ps$ (or~$\vf$). In the case
of $\PS$-inflation field we start from zero value field~$\Ps$ coming to
$\Ps=\Ps_0$, which corresponds to minimum of the self-interacting \pt\ for the
field~$\Ps$ (or~$\vf$). This gives us a value of contemporary \co ical \ct. Due
to the \ct s $\xi,\z$ and also skew\s ic fields $k_{ab}$, $k^0_{\td a\td b}$
defined on~$H$ and $M=G/G_0$ we can turn the obtained  value of~$\wt\La$ to the
desired one from observational data. Small oscillations of the field~$\vf$
around zero value give us \qe\ \pc s---scalar \pc s---a component of Dark
Matter. In this way \qe\ inflationary scenario gives us inflationary model with
a calculable fluctuation spectrum and a \co ical \ct\ for our contemporary epoch.
Simultaneously we get candidates for Dark Matter \pc s which interact almost only
\gr ary. All important features of inflationary models can be preserved.

Dark Matter and Dark Energy are part of an extended \nos, scalar tensor
gravity. The field~$\Ps$ (introduced in a Jordan--Thiry manner) plays several
roles. It induces non\ct\ \gr al ``\ct'', plays a role of an inflaton, being
a source of an effective \co ical \ct\ and also a Dark Matter \pc. It is very
economous. In order to avoid some misunderstanding we recapitulate differences
between \E\nos\ Kaluza--Klein (Jordan--Thiry) Theory and ordinary \s ic version.

In the theory we consider a scalar \cvt\ on a many-\di al manifold obtained
from the metrized (\s ically or non\s ically) principal fibre bundles over
\spt~$E$ or over $E\tm M=E\tm G/G_0$. This scalar \cvt\ is projected on~$E$ or
$E\tm M$. Afterwards we obtain from Palatini \v al principle classical field
\e s. In our approach we do not consider many-\di al Einstein \e s. Our \e s are
four-\di al. We consider in our theory two cases: ordinary and with \ssb. Both
cases can be considered in \s ic and non\s ic case. In the case of the \E\nos\
\KK (Jordan--Thiry) Theory we are getting new features which are impossible to
obtain in the \s ic case. They are dielectric model of a charge \cfn\ extended
(see \cite{2,4,11a}) to a colour \cfn, e.g.\ in QCD ($G=\SU(3)$). In the case
with \ssb\ and \Hm, we get in the \nos\ case a correct pattern of masses for
$W^\pm,Z^0$ and Higgs' boson with a correct value of a Weinberg angle (this
parameter is not a phenomenological parameter) corrected by a finite
renormalization (see Ref.~\cite{11a}). This is impossible in the \s ic case.
The mass of a Higgs' boson is too low and a correct finite renormalization
of~$\t_W$ (Weinberg angle) does not work. In the \s ic case we get an enormous
\co ical \ct. This value can be taken under control in the \nos\ case. In the
\nos\ case we get a theory of a Dark Matter and Dark Energy, which we do not
get in the \s ic case. It means we get a massive pseudovector field and massive
scalar field as Dark Matter \pc s. We get also a \co ical \ct\ as Dark Energy.
In the \s ic case this is not possible. Scalar field plays several roles in
the theory. It is an inflaton field and a \qe\ field as a source of a Dark
Energy---a~\co ical \ct. According to E.~Witten argument (see~\cite{del}) on
chiral fermions we cannot get those fermions in \KK Theory. Moreover, we can
avoid those arguments considering higher dimensional spinors on $E\tm M=E\tm
G/G_0$ as zero-modes on $E\tm M\tm H$ (see Ref~\cite{x}). We get Yukawa terms in
the theory for fermions. In our theory fermions are coupled to the horizontal
part of \LC \cn\ on $E\tm M\tm H$. This \cn\ is \nos\ (non-zero torsion).
Moreover, the \cn\ is metric \wrt \s ic part of a metric. We consider as usual
test \pc\ motion as many-\di al geodesic in \LC \cn\ \wrt \s ic part of a
metric projected on $E\tm M$. We obtain new additional charges similar to electric
charge connected to a \gn\ of a Lorentz force term for Higgs' field. We get also
such Lorentz force term for \nA\ gauge field as in Ref.~\cite{alf}. Equation
of motion for a test \pc\ can be obtained from a \v al principle, simultaneously
it can be obtained from equation of fields (via conservation laws) as in General
Relativity.

Our theory
due to an existence of a scalar field~$\Ps$ (or~$\vf$) is going to the non\ct\
\gr al \ct\ $G_{\rm eff}$. Moreover, only in the \nos\ case these results can be
considered seriously as in Section~5. It means we get the so-called fifth
force. Let us notice also that in the \nos\
case we can have to do with two vacuum states (in general)---true and false
vacuum states. The theory is a classical field theory. Moreover, we want to
quantize the theory using nonlocal quantization methods (see Ref.~\cite{11a}
for details).

Let us give the following remark (coming from architecture) concerning
a construction of a Unified Field Theory (\un\ of all \fn\ physical \ia s). The
construction is similar to the construction of a gothic medieval cathedral
church. During the building works it is hard to distinguish between finished
parts of the cathedral church and something which is only helping to do the
construction. This is the case of Ref.~\cite5. It is an assembly. Moreover,
it is better to say a falsework or a scaffolding.

In the classical \KK (Jordan--Thiry) Theory an \elm c field curves the fifth
\di. In the \E\nos\ (5-\di al) \KK (Jordan--Thiry) Theory it curves and
twists the fifth \di. A~\YM' field curves additional \di s in the \nA\ \KK
Theory (\s ic). In the case of the \E\nos\ \KK (Jordan--Thiry) Theory a~\YM'
field curves and twists additional \di s. In the \E\nos\ \KK (Jordan--Thiry)
Theory with \ssb\ and \Hm\ \YM' field and Higgs' field curve and twist
additional \di s. In the \s ic case (does not exist as a mathematical construction)
they curve additional \di s. The mathematical construction of the previous one
can be obtained as a \s ic limit of the \E\nos\ Theory. In this way an \elm c
field, \YM' fields and Higgs' fields obtain a geometrical interpretation as a
\cvt\ and torsion in higher \di s.

Let us give the following comment. Grand Unified Field Theories (GUT) do not
give a conclusion on a proton decay. In some of them a proton decays, in other
not. Our approach can incorporate both cases (bosonic parts). An experiment does not conclude
``is proton stable or not''. A mathematical background for our considerations
can be found in Section~1 of Ref.~\cite{11a}.

The paper is organized as follows. In the second section we give \el s
of the \NK\ in general non-Abelian case and with \ssb\ and \Hm. In the third
section we give hierarchy of \sb. The fourth section is devoted to the
\JT\ over $V=E\tm G/G_0$. In the fifth section we consider some
consequences of an existence of a scalar field~$\Ps$. It means, some
non-Newtonian gravity in nonrelativistic limits. We give some application of
$G\dr{eff}$ in a galactic flat velocity curve problem without a Dark Matter.
In the sixth section
we consider geodetic equations as test \pc\ equations in the \JT.

\looseness-1
In Appendix A we give some elements of an evolution of a field~$\Ps$ (or~$\vf$)
in a ``quintessence inflation'' and some \co ical consequences of a \qe.
We describe a $\La CDM$ model in our approach.
In Appendix~B we give a theory of an additional Dark Matter (a~tower of scalar fields).
A~tower of scalar fields is coming from the assumption that $\rho=\rho(x,y)$,
$x\in E$, $y\in M=G/G_0$. This is a new approach to Dark Matter appearance.
It works as scalar massive \pc s---Dark Matter even if a \co ical \ct\ is zero.
This approach can be used only with a \ssb\ and \Hm.
In Appendix C we give a partial geometrization and unification of a bosonic
part of Standard Model. We find an exact \so\ of the model with remarkable
properties describing ``mass without mass'' and ``charge without charge''
with a \co ical \ct.
We prove that a cosmological constant is not zero and calculate it.

We obtain values of masses for $W^\pm$ bosons, $Z^0$~bosons and Higgs' bosons
with an agreement with an experiment. We get also value of a Weinberg angle
with an agreement with experiment (as in Ref.~\cite{11a}). Soliton-bag models
of hadrons are discussed in a context of our unification. We consider a nonlocal
quantization of the theory.

The paper contains also Conclusions and Prospects for Further Research.

\advance\abovedisplayskip by-1pt
\advance\belowdisplayskip by-1pt
\section{Elements of the \NK\ in general \nA\ case and with \sn\ \s y breaking and \Hm}\label{s:els}
This section gives us notations and general assumption of the theory in the
case of $\rho=1$ ($\Psi=0$). For the convenience of the reader we repeat
here some results from Ref.~\cite{11a}.
We find an exact solution of field equations of the unification with remarkable
properties describing ``charge without charge'' and ``mass without mass''
with cosmological constant.

Let $P$ be a principal fibre bundle over a \spt\ $E$ with a structural
group~$G$ which is a semisimple Lie group. On a \spt~$E$ we define a \nos\
tensor $g_\m=g_\(\m)+g_\[\m]$ \st
\beq2.1
\bal
g&=\det(g_\m)\ne0\\
\wt g&=\det(g_\(\m))\ne0.
\eal
\e
$g_\[\m]$ is called as usual a skewon field (e.g.\ in NGT, see Refs
\cite{6,14,15,16}). In Refs \cite{15,b} we have the so-called Einstein--Strauss
theory.
We define on $E$ a \nos\ \cn\ compatible with $g_\m$ \st
\beq2.2
\ov Dg_{\a\b}=g_{\a\d}\gd \ov Q,\d,\b\g,(\ov\G)\ov \t{}^\g
\e
where $\ov D$ is an exterior covariant \dv\ for a \cn\ $\gd\ov\o,\a,\b,=
\gd\ov \G,\a,\b\g,\ov \t{}^\g$ and $\gd\ov Q,\a,\b\d,$ is its torsion. We suppose also
\beq2.3
\gd\ov Q,\a,\b\a,(\ov \G)=0.
\e
Lower case Greek letters $\a,\b,\mu,\nu=1,2,3,4$. In this way we have a usual
convention for \spt\ indices.
We introduce on $E$ a second \cn
\beq2.4
\gd\ov W,\a,\b,=\gd\ov W,\a,\b\g,\ov \t{}^\g
\e
\st
\bg2.5
\gd \ov W,\a,\b,=\gd \ov\o,\a,\b,-\tfrac23\,\gd \d,\a,\b,\ov W\\
\ov W=\ov W_\g\ov \t{}^\g=\tfrac12(\gd \ov W,\si,\g\si,-\gd \ov W,\si,\si\g,)
\ov \t{}^\g. \lb2.6
\e

Now we turn to \nos\ metrization of a bundle $P$. We define a \nos\ tensor
$\g$ on a bundle manifold $P$ \st
\beq2.6a
\g=\pi^* g\op \ell_{ab}\t^a\ot \t^b
\e
where $\pi$ is a projection from $P$ to $E$. On $P$ we define a \cn~$\o$
(a~1-form with values in a Lie algebra $\fg$ of~$G$). In this way we can
introduce on~$P$ (a~bundle manifold) a frame  $\t^A=(\pi^*(\ov \t{}^\a),
\t^a)$ \st
$$
\t^a=\la\o^a,\q \o=\o^a X_a, \q a=5,6,\dots,n+4, \q
n=\dim G=\dim\fg, \q \la={\rm const.}
$$
Thus our \nos\ tensor looks like
\bg2.7
\g=\g_{AB} \t^A\ot\t^B, \q A,B=1,2,\dots,n+4,\\
\ell_{ab}=h_{ab}+\mu k_{ab}, \lb2.8
\e
where $h_{ab}$ is a bi\iv t Killing--Cartan tensor on~$G$ and $k_{ab}$ is a
right-\iv t skew-\s ic tensor on~$G$, $\mu={\rm const}$.

We have
\beq2.9
\bal
h_{ab}&=\gd C,c,ad,\gd C,d,bc,=h_{ba}\\
k_{ab}&=-k_{ba}.
\eal
\e
Thus we can write
\bea2.10
\ov \g(X,Y)=\ov g(\pi'X,\pi'Y)+\la^2h(\o(X),\o(Y))\\
\ul \g(X,Y)=\ul g(\pi'X,\pi'Y)+\la^2k(\o(X),\o(Y)) \lb2.11
\e
($\gd C,a,bc,$ are structural \ct s of the Lie algebra $\fg$) (see Eq.~\er{4.9} for comparison).

$\ov \g$ is the \s ic part of $\g$ and $\ul \g$ is the anti\s ic part of~$\g$.
We have as usual
\beq2.12
[X_a,X_b]=\gd C,c,ab,X_c
\e
and
\beq2.13
\O=\frac12 \gd H,a,\m,\t^\mu \land \t^\nu X_a
\e
is a curvature of the \cn\ $\o$,
\beq2.14
\O=d\o+\frac12[\o,\o].
\e
The frame $\t^A$ on $P$ is partially nonholonomic. We have
\beq2.15
d\t^a=\frac\la2\Bigl(\gd H,a,\m,\t^\mu\land \t^\nu - \frac1{\la^2}\,
\gd C,a,bc,\t^b\land\t^c\Bigr)\ne0
\e
even if the bundle $P$ is trivial, i.e.\ for $\O=0$. This is different than
in an \elm c case (see Ref.~\cite3). Our \nos\ metrization of a
principal fibre bundle gives us a right-\iv t structure on~$P$ \wrt an action
of a group~$G$ on~$P$ (see Ref.~\cite3 for more details). Having $P$ \nos
ally metrized one defines two \cn s on~$P$ right-\iv t \wrt an action of a
group $G$ on~$P$. We have
\beq2.16
\g_{AB}=\left(\begin{array}{c|c}
g_{\a\b}&0\\
\hline
0&\ell_{ab}\end{array}\right)
\e
in our lift horizontal frame $\t^A$.
\bg2.17
D\g_{AB}=\g_{AD}\gd Q,D,BC,(\G)\t^C\\
\gd Q,D,BD,(\G)=0 \label{2.18}
\e
where $D$ is an exterior covariant \dv\ \wrt a \cn\ $\gd \o,A,B,=\gd \G,A,BC,
\t^C$ on~$P$ and $\gd Q,A,BC,(\G)$ its torsion.

Let us notice the following fact. A~metric (\nos) tensor \er{2.16} is diagonal.
It does not mean that the tensor is trivial. For the frame \er{2.15} is
nonholonomic, the vector field (a~gauge field) is present and calculations
of a \cn\ and a \cvt\ of a \cn\ (a~little different formulae than in the
holonomic case) gives the results given here (see Ref.~\cite3). In this case
we can understand a power of differential forms calculations. They are much easier
than the ordinary ones.

In this case nondiagonal parts of a metric tensor are not coming from quantum
fluctuations. They are effects of a nonzero \cvt\ of a gauge \cn.

One can solve
Eqs~\er{2.17}--\er{2.18} getting the following results
\beq2.19
\gd \o,A,B,=\left(
\begin{array}{c|c}
\pi^*(\gd \ov\o,\a,\b,)-\ell_{db}g^{\mu\a}\gd L,d,\mu\b,\t^b&\gd L,a,\b\g,\t^\g\\
\hline
\ell_{bd}g^{\a\b}(2\gd H,d,\g\b,-\gd L,d,\g\b,)\t^\g & \gd \wt\o,a,b,
\end{array}\right)
\e
where $g^{\mu\a}$  is an inverse tensor of $g_{\a\b}$
\beq2.20
g_{\a\b}g^{\g\b}=g_{\b\a}g^{\b\g}=\d^\g_\a,
\e
$\gd L,d,\g\b,=-\gd L,a,\b\g,$ is an Ad-type tensor on~$P$ \st
\beq2.21
\ell_{dc}g_{\mu\b}g^{\g\mu}\gd L,d,\g\a,+\ell_{cd}g_{\a\mu}g^{\mu\g}
\gd L,d,\b\g,=2\ell_{cd}g_{\a\mu}g^{\mu\g}\gd H,d,\b\g,,
\e
$\gd\wt\o,a,b,=\gd \wt\G,a,bc,\t^c$ is a \cn\ on an internal space (typical
fibre) compatible with a metric $\ell_{ab}$ \st
\bg2.22
\ell_{db}\gd \wt\G,d,ac,+\ell_{ad}\gd \wt\G,d,cb,=-\ell_{db}\gd C,d,ac,\\
\gd \wt\G,a,ba,=0, \q \gd \wt\G,a,bc,=\gd -\wt\G,a,cb, \lb2.23
\e
and of course $\gd \wt Q,a,ba,(\wt\G)=0$ where $\gd \wt Q,a,bc,(\G)$ is a
torsion of the \cn~$\gd \wt\o,a,b,$.

We also introduce an inverse tensor of $g_\(\a\b)$
\beq2.24
g_\(\a\b)\wt g{}^\(\a\g)=\d^\g_\b.
\e
We introduce a second \cn\ on~$P$ defined as
\beq2.25
\gd W,A,B,=\gd \o,A,B,-\frac4{3(n+2)}\,\gd\d,A,B,\ov W.
\e
$\ov W$ is a horizontal one form
\bg2.26
\ov W=\hor\ov W\\
\ov W=\ov W_\nu\t^\nu =\tfrac12(\gd\ov W,\si,\nu\si, - \gd \ov W,\si,\si\nu,).
\lb2.27
\e

\advance\abovedisplayskip by1pt
\advance\belowdisplayskip by1pt
In this way we define on $P$ all analogues of four-\di al quantities from NGT
(see Refs \cite{6,14}). It means, $(n+4)$-\di al analogues from Moffat
theory of \gr, i.e.\ two \cn s and a \nos\ metric $\g_{AB}$. Those quantities
are right-\iv t \wrt an action of a group~$G$ on~$P$. One can calculate a
scalar curvature of a \cn\ $\gd W,A,B,$ getting the following result (see
Refs \cite{1,3}):
\beq2.28
R(W)=\ov R(\ov W)-\frac{\la^2}4 \bigl(2\ell_{cd}H^cH^d - \ell_{cd}L^{c\m}
\gd H,d,\m,\bigr) + \wt R(\wt \G)
\e
where
\beq2.29
R(W)=\g^{AB}\bigl(\gd R,C,ABC,(W)+\tfrac12 \, \gd R,C,CAB,(W)\bigr)
\e
is a Moffat--Ricci curvature scalar for the \cn~$\gd W,A,B,$,
$\ov R(\ov W)$ is a Moffat--Ricci curvature scalar for the \cn~$\gd \ov W,\a,\b,$,
and $\wt R(\wt \G)$ is a Moffat--Ricci curvature scalar for the \cn~$\gd \wt\o,a,b,$,
\bg2.30
H^a=g^\[\m]\gd H,a,\m,\\
L^{a\m}=g^{\a\mu}g^{\b\nu}\gd L,a,\a\b,. \lb2.31
\e
Usually in ordinary (\s ic) Kaluza--Klein Theory one has
$\la=2\frac{\sqrt{G_N}}{c^2}$, where $G_N$ is a Newtonian \gr al \ct\ and
$c$~is the speed of light. In our system of units $G_N=c=1$ and $\la=2$. This
is the same as in \NK\ in an \elm c case (see Refs \cite{3,4}). In the \nA\
Kaluza--Klein Theory which unifies GR and \YM\ field theory we have a \YM\
lagrangian and a \co ical term. Here we have
\beq2.32
\cLY=-\frac1{8\pi}\,\ell_{cd}\bigl(2H^cH^d-L^{c\m}\gd H,d,\m,\bigr)
\e
and $\wt R(\wt\G)$ plays a role of a \co ical term.

$\gd L,a,\a\b,$ is an induction tensor for \YM' field (defined on a gauge
bundle, not on a \spt~$E$) and is defined by Eq.~\er{2.21} which can be solved
getting Eq.~\er{2.125}. Due to a difference between $\gd H,a,\a\b,$ and
$\gd L,a,\a\b,$ in the \E\nos\ \KK (Jordan--Thiry) Theory we get a dielectric
model of a colour \cfn. In the case of the ordinary \nA\ \KK Theory one gets
$\gd L,a,\a\b, = \gd H,a,\a\b,$ (see Refs \cite{2, 8, 11a}).

Lagrangian \er{2.32} in the five-\di al case ($G=U(1)$) gives nonsingular
(finite energy) \so s which are absent in the ordinary (\s ic) \KK theory
(the \so\ is spherically \s ic and stationary with nonsingular electric and
\gr al fields, approaching asymptotically Reissner--Nordstr\"om \so, see~Ref.~\cite4).

If $G=U(1)$, i.e.\ \E\nos\ Kaluza--Klein (Jordan--Thiry) in an \elm c case
we have to do with dielectric model of a charge \cfn, see Ref.~\cite8.

Let us notice the \fw\ fact. In Refs \cite{alf,beta,gama} one uses \gn s of the
\KK theory to an arbitrary \nA\ group. In this case one is getting also an
enormous \co ical \ct. In the case of a \nos\ \nA\ \KK theory this problem
can be solved as described here. We can even nivel the \ct\ to zero.

Our theory is a classical field theory and a \co ical \ct\ is coming from the
geometry.

Let us notice the \fw\ fact. $\gd H,d,\mu\nu,$~is a \cvt\ of the \cn~$\o$
defined on a fibre bundle~$P$,
$$
H=\gd H,a,\mu\nu, \t^\mu \land \t^\nu X_a.
$$
If we take a local section $e:E\to P$ we get $\gd F,a,\mu\nu,\ov\t{}^\mu
\land \ov\t{}^\nu X_a = e^*(\gd H,a,\mu\nu, \t^\mu\land \t^\nu)X_a$. In terms
of differential forms one gets
$$
e^*H = F = dA + [A,A] = \gd F,a,\mu\nu, \ov\t{}^\mu\land\ov\t{}^\nu X_a,
$$
where $A=e^*\o$. This is the same as a conventional strength of \YM\ field.
The relations of both formalisms can be found in Appendix~A of Ref.~\cite8
and in Section~2, \ti{Elements of geometry}, of Ref.~\cite{11a}. The indices
$\mu,\nu$ are of the same nature as $\a,\b$. All kinds of indices are explained
above (and in more general case below). In this case Greek lower case letters
are connected to \spt~$E$.

In order to incorporate a \sn\ \s y breaking and \Hm\ in our geometrical \un\
of \gr\ and \YM' fields we consider a fibre bundle $P'$ over a base manifold
$E\times G/G_0$, where $E$ is a \spt, $G_0\subset G$, $G_0,G$ are semisimple Lie
groups. Thus we are going to combine a Kaluza--Klein theory with a \di al
reduction procedure.

Let $P'$ be a principal fibre bundle over $V=E\times M$ with a structural
group~$H$ and with a projection~$\pi$, where $M=G/G_0$ is a homogeneous
space, $G$~is a semisimple Lie group and $G_0$ its semisimple Lie subgroup.
Let us suppose that $(V,\g)$ is a manifold with a \nos\ metric tensor
\beq3.1
\g_{AB}=\g_\(AB)+\g_\[AB].
\e
The signature of the tensor $\g$ is ${(}+{-}-{-},
\underbrace{{}-{-}-\cdots-}_{n_1})$. Let us
introduce a natural frame on~$P$
\beq3.2
\t^{\tilde A}=(\pi^*(\t^A),\t^a=\la\o^a), \q \la={\rm const.}
\e
It is convenient to introduce the following notation. Capital Latin indices
with tilde $\wt A,\wt B,\wt C$ run $1,2,3,\dots,m+4$, $m=\dim H+\dim M
=n+\dim M=n+n_1$, $n_1=\dim M$, $n=\dim H$. Lower Greek indices $\a,\b,\g,\d
=1,2,3,4$ and lower Latin indices
$a,b,c,d=n_1+5,n_2+5,\dots,\break n_1+6,\dots,m+4$. Capital Latin indices
$A,B,C=1,2,\dots,n_1+4$. Lower Latin indices with tilde $\wt a,\wt b,\wt c$
run $5,6,\dots,n_1+4$. The symbol $\ov{\hbox to5pt{\hfill\vphantom{$\t$}}}$
over $\ov\t{}^A$ and other quantities indicates
that these quantities are defined on~$V$. We have of course
$$
n_1=\dim G-\dim G_0=n_2-(n_2-n_1),
$$
where $\dim G=n_2$, $\dim G_0=n_2-n_1$, $m=n_1+n$. In this case ($P'$)
$H$~plays the role of a group~$G$ for~$P$.

On the group $H$ we define a bi-\iv t (\s ic) Killing--Cartan tensor
\beq3.3
h(A,B)=h_{ab}A^aB^b.
\e
We suppose $H$ is semisimple, it means $\det(h_{ab})\ne0$. We define a skew-\s
ic right-\iv t tensor on~$H$
$$
k(A,B)=k_{bc}A^bB^c, \q k_{bc}=-k_{cb}.
$$

Let us turn to the \nos\ metrization of~$P'$.
\beq3.4
\ka(X,Y)=\g(X,Y)+\la^2\ell_{ab}\o^a(X)\o^b(Y)
\e
(see Eq.~\er{4.9} for comparison), where
\beq3.5
\ell_{ab}=h_{ab}+\xi k_{ab}
\e
is a \nos\ right-\iv t tensor on~$H$. One gets in a matrix form (in the
natural frame \er{3.2})
\beq3.6
\ka_{\tA\tB}=\left(\begin{array}{c|c}
\g_{AB}&0\\ \hline 0&\ell_{ab}
\end{array}\right),
\e
$\det(\ell_{ab})\ne0$, $\xi={\rm const}$ and real, then
\beq3.7
\ell_{ab}\ell^{ac}=\ell_{ba}\ell^{ca}=\gd\d,c,b,.
\e
The signature of the tensor $\ka$ is $(+,-{-}-,\underbrace{-\cdots-}_{n_1},
\underbrace{{}-{-}\cdots-}_n)$.
As usual, we have commutation relations for Lie algebra of~$H$, $\fh$
\beq3.8
[X_a,X_b]=\gd C,c,ab,X_c.
\e
This metrization of $P'$ is right-\iv t \wrt an action of $H$ on $P'$.

Let us take a local section $e:E\to P$ and attach to it a frame~$v^a$, $a=
5,6,\dots,n+4$, selecting $X^\mu={\rm const}$ on a fibre in such a way that
$e$~is given by the condition $e^*v^a=0$ and the \fn\ fields $\z_a$ \st
$v^a(\z_b)=\d^a_b$ satisfy
\beq ast
[\z_a,\z_b] = \frac1\la \,\gd C,c,bc,\z_c.
\e
Thus we have
$$
\o=\frac1\la \,v^aX_a + \pi^\ast(\gd A,a,\mu,\ov\t{}^\mu)X_a
$$
where $e^*\o = A = \gd A,a,\mu, \ov\t{}^\mu X_a$. In this frame the tensor
$\g$ takes the form
\beq plus
\g_{AB}= \left(\begin{array}{c|c}
g_{\a\b}+\la^2 \ell_{ab}\gd A,a,\a, \gd A,b,\b, & \la \ell_{cb}\gd A,c,\a,\\
\hline
\la \ell_{ac}\gd A,c,\b, &\ell_{ab}
\end{array}\right) \q\q (\ell_{ab}=h_{ab}+\mu k_{ab}).
\e
This frame is also unholonomic
\beq 2ast
dv^a = -\frac1{2\la}\,\gd C,a,bc, v^b\land v^c.
\e
On the same footing we can consider a tensor $\ka_{\td A\td B}$ getting the
formula
$$
\ka_{\td A\td B}=\left(\begin{array}{c|c}
\g_{AB}+\la^2\ell_{ab}\gd A,a,A, \gd A,b,B, & \la\ell_{cb}\gd A,c,A,\\
\hline
\la\ell_{ac}\gd A,c,B, & \ell_{ab}
\end{array}\right).
$$
Moreover, now $\o$ is defined on principal fibre bundle over $E\tm M$ and a local
section~$e:E\tm M\to P'$ gives us
$$
e^\ast \o = \gd A,a,A,\ov\t{}^A X_a = (\gd A,a,\a,\ov\t{}^\a +\gd A,a,\ta,\ov\t{}^\ta)X_a.
$$
Similarly as in the previous case we attach a frame $v^a$ selecting on a fibre
$X^A={\rm const}$ in such a way that $e$~is given by the condition $e^*v^a=0$.
A dual frame to $v^a$ is the same as before, i.e.\ the formula \er{ast}, and
the formula \er{2ast} is also satisfied.

It is easy to see that now
$$
\o= \frac1\la\,v^aX_a + \pi^*(\gd A,a,\mu,\ov\t{}^\mu)X_a+ \pi^*(\gd A,a,\tilde m,
\ov\t{}^{\td m})X_a
$$
and $\ka_{\tA\tB}$ looks like
$$
\ka_{\tA\tB}=\left(\begin{array}{c|c|c}
g_{\a\b}+\la^2\ell_{ab}\gd A,a,\a, \gd A,b,\b, &
\la\ell_{ac}\gd A,c,\a, & \la\ell_{ac}\gd A,c,\ta, \\
\hline
\la\ell_{ac}\gd A,c,\b, & r^2g_{\ta\tb}+\la^2\ell_{ab}\gd A,a,\ta,\gd A,b,\tb, & 0 \\
\hline
\la\ell_{ac}\gd A,c,\tb, & 0 & \ell_{ab}
\end{array}\right).
$$
$g_{\ta\tb}$ is a metric tensor defined on $M=G/G_0$ and $r$ is a real
parameter, the radius of~$M$. A~construction of the \nos\ tensor $g_{\ta\tb}$
will be considered later.

Let us rewrite the last formula in the \fw\ way:
\beq 3plus
\ka_{\tA\tB}=\left(\begin{array}{c|c|c}
g_{\a\b}+\la^2\ell_{ab}\gd A,a,\a, \gd A,b,\b, &
\la\ell_{ac}\gd A,c,\a, & \la\ell_{ac}\gd \F,c,\ta, \\
\hline
\la\ell_{ac}\gd A,c,\b, & r^2g_{\ta\tb}+\la^2\ell_{ab}\gd \F,a,\ta,\gd \F,b,\tb, & 0 \\
\hline
\la\ell_{ac}\gd \F,c,\tb, & 0 & \ell_{ab}
\end{array}\right).
\e
In this way a multi\di al field $\gd A,a,A,$ is a source of a gauge field
$\gd A,a,\a,$ and a scalar field $\gd\F,c,\tb,$. Now it is easy to see that
a vector (gauge) field and a scalar (Higgs) field are coming from tensor
$\g_{AB}$ or~$\ka_{\tA\tB}$.

For comparison see
\beq 2plus
\ka_{\tA\tB} = \left(\begin{array}{c|c}
\g_{AB} & 0 \\\hline 0 & \ell_{ab}
\end{array}\right)=
\left(\begin{array}{c|c|c}
\g_{\a\b} & 0 & 0 \\\hline
0 & r^2g_{\ta\tb} & 0 \\\hline
0 & 0 & \ell_{ab}
\end{array}\right)
\e
in a unholonomic frame used by us.

In order to connect two formalisms one writes
$$
\g_{AB} \t^A\otimes \t^B.
$$
Taking for
$$
\t^\a=\pi^* (\ov\t{}^\a)
$$
and for
$$
\t^a = v^a+\pi^\ast(\gd A,a,\mu,\ov\t{}^\mu),
$$
it is easy to see that we get formula \er{plus}.

In the more general case with a \ssb\ one writes
$$
\ka_{\tA\tB}\t^{\tA}\otimes\t^{\tB}.
$$
Taking for
$$
\t^\a = \pi^*(\ov\t{}^\a)
$$
and for
$$
\t^a = v^a+\pi^* (\gd A,a,\mu,\ov\t{}^\mu)+\pi^*(\gd \F,a,\td m,\ov\t{}^{\td m})
$$
it is easy to see that we get formula \er{3plus}.

We do not repeat consideration from the previous part of the paper because we
consider a \nos\ \KK Theory on principal bundle over $E\tm M=E\tm G/G_0$, \ti{not\/}
on~$E$. Due to this and some additional \s ies we get also Higgs' field. The
Higgs' field is a part of vector field  on $E\tm M$. A~scalar part is over~$M$.
Roughly speaking, the scalar field is a part of a vector field over~$M$ (it is
not a~surprise). Moreover, in order to get a desired \ssb\ and \Hm\ we should add
some additional assumptions concerning a \cn\ on a principal bundle $P'$ (see the
next part of the paper).

Some calculations using a tensor in a partially nonholonomic frame seem to be
very tedious (it is a power of differential forms calculus to avoid this).

Now we should \nos ally metrize $M=G/G_0$. $M$~is a homogeneous space for~$G$
(with left action of group~$G$). Let us suppose that the Lie algebra of~$G$,
$\fg$ has the following reductive decomposition
\beq3.9
\fg=\fg_0 \mathrel{\dot+} \fm
\e
where $\fg_0$ is a Lie algebra of $G_0$ (a~subalgebra of~$\fg$) and $\fm$
(the complement to the subalgebra~$\fg_0$) is $\Ad G_0$ \iv t, $\dot+$
means a direct sum. Such a decomposition might be not unique, but we assume
that one has been chosen. Sometimes one assumes a stronger condition
for~$\fm$, the so called \s y requirement,
\beq3.10
[\fm,\fm]\subset \fg_0.
\e
Let us introduce the following notation for generators of $\fg$:
\beq3.11
Y_i\in\fg, \q Y_{\tilde\imath}\in\fm, \q Y_{\hat a}\in \fg_0.
\e
This is a de\cm\ of a basis of $\fg$ according to \er{3.9}. We define a \s ic
metric on~$M$ using a \KC form on~$G$ in a classical way. We call this tensor
$h_0$.

Let us define a tensor field $h^0(x)$ on $G/G_0$, $x\in G/G_0$, using tensor
field $h$ on~$G$. Moreover, if we suppose that $h$ is a bi\iv t metric on~$G$
(a~\KC tensor) we have a simpler construction.

The complement $\fm$ is a tangent space to the point $\{\ve G_0\}$ of~$M$,
$\ve$~is a unit element of~$G$. We restrict $h$ to the space $\fm$ only. Thus
we have $h^0(\{\ve G_0\})$ at one point of~$M$. Now we propagate
$h^0(\{fG_0\})$ using a left action of the group $G$
$$
h^0(\{fG_0\})=(L_f^{-1})^\ast (h^0(\{\ve G_0\})).
$$
$h^0(\{\ve G_0\})$ is of course $\Ad G_0$ \iv t tensor defined on~$\fm$
and $L_f^\ast h^0=h^0$.

We define on $M$ a skew-\s ic 2-form $k^0$. Now we introduce a natural frame
on~$M$. Let $\gd f,i,jk,$ be structure \ct s of the Lie algebra $\fg$, i.e.
\beq3.12
[Y_j,Y_k]=\gd f,i,jk, Y_i.
\e
$Y_j$ are generators of the \Li~$\fg$. Let us take a local section $\si:V\to
G/G_0$ of a natural bundle $G\mapsto G/G_0$ where $V\subset M=G/G_0$. The
local section $\si$ can be considered as an introduction of a \cd\ system
on~$M$.

Let $\o_{MC}$ be a left-\iv t Maurer--Cartan form and let
\beq3.13
\gd \o,\si,MC,=\si^\ast \o_{MC}.
\e
Using de\cm\ \er{3.9} we have
\beq3.14
\gd \o,\si,MC,=\gd \o,\si,0,+\gd \o,\si,\fm,=
\t^{\hi}Y_{\hi}+\ov \t{}^{\td a}Y_{\td a}.
\e
It is easy to see that $\ov \t{}^{\td a}$ is the natural (left-\iv t) frame
on~$M$ and we have
\bea3.15
h^0&=\gd h,0,\td a\td b,\ov\t{}^{\td a}\otimes \ov\t{}^{\td b}\\
k^0&=\gd k,0,\td a\td b,\ov\t{}^{\td a}\land \ov\t{}^{\td b}. \lb3.16
\e
According to our notation $\wt a,\wt b=5,6,\dots,n_1+4$.

Thus we have a \nos\ metric on $M$
\beq3.17
\g_{\td a\td b}=r^2\bigl(\gd h,0,\td a\td b,+\z \gd k,0,\td a\td b,\bigr)
=r^2g_{\td a\td b}.
\e
Thus we are able to write down the \nos\ metric on $V=E\tm M=E\tm G/G_0$
\beq3.18
\g_{AB}=\left(\begin{array}{c|c}
g_{\a\b}&0\\
\hline
0 & r^2g_{\td a\td b}
\end{array}\right)
\e
where
\begin{align*}
g_{\a\b}&=g_\(\a\b)+g_\[\a\b]\\
g_{\td a\td b}&=\gd h,0,\td a\td b,+\z \gd k,0,\td a\td b,\\
\gd k,0,\td a\td b,&=-\gd k,0,\td b\td a,\\
\gd h,0,\td a\td b,&=\gd h,0,\td b\td a,,
\end{align*}
$\a,\b=1,2,3,4$, $\wt a,\wt b=5,6,\dots,n_1+4=\dim M+4=\dim G-\dim G_0+4$.
The frame $\ov\t{}^{\td a}$ is unholonomic:
\beq3.19
d\ov\t{}^{\td a}=\frac12\,\gd \ka,\td a,\td b\td c, \ov\t{}^{\td b}\land
\ov \t{}^{\td c}
\e
where $\gd \ka,\td a,\td b\td c,$ are \cf s of nonholonomicity and depend  on
the point of the manifold $M=G/G_0$ (they are not \ct\ in general). They
depend on the section~$\si$ and on the \ct s $\gd f,\td a,\td b\td c,$.

We have here three groups $H,G,G_0$. Let us suppose that there exists a
homomorphism $\mu$ between $G_0$ and~$H$,
\beq3.20
\mu:G_0 \to H
\e
\st a centralizer of $\mu(G_0)$ in $H$, $C^\mu$ is isomorphic to~$G$.
$C^\mu$, a centralizer of $\mu(G_0)$ in $H$, is a set of all \el s of~$H$
which commute with \el s of $\mu(G_0)$, which is a subgroup of~$H$. This
means that $H$~has the following structure, $C^\mu=G$.
\beq3.21
\mu(G_0)\otimes G\subset H.
\e
If $\mu$ is a iso\mo sm between $G_0$ and $\mu(G_0)$ one gets
\beq3.22
G_0\otimes G\subset H.
\e
Let us denote by $\mu'$ a tangent map to $\mu$ at a unit \el. Thus $\mu'$ is
a differential of~$\mu$ acting on the \Li\ \el s. Let us suppose that the
\cn~$\o$ on the fibre bundle $P'$ is \iv t under group action of~$G$ on the
manifold $V=E\tm G/G_0$. According to Refs \cite{13,18,19,20} this means the
following.

Let $e$ be a local section of $P'$, $e:V\subset U\to P'$ and $A=e^*\o$. Then
for every $g\in G$ there exists a gauge \tf\ $\rho_g$ \st
\beq3.23
f^*(g)A=\Ad_{\rho_g^{-1}}A+\rho_g^{-1}\,d\rho_g,
\e
$f^*$ means a pull-back of the action $f$ of the group $G$ on the
manifold~$V$. According to Refs \cite{16,13,18,19,20,21}
(see also Refs \cite{22,23,**,*1,*2,24,25}) we are able to write
a general form for such an~$\o$. Following Ref.~\cite{20} we have
\beq3.24
\o=\wt \o_E +\mu'\circ \gd \o,\si,0,+\F\circ \gd \o,\si,\fm,.
\e
(An action of a group $G$ on $V=E\tm G/G_0$ means left multiplication on a
homogeneous space $M=G/G_0$.)
where $\gd \o,\si,0,+\gd \o,\si,\fm,=\gd \o,\si,MC,$ are components of the
pull-back of the Maurer--Cartan form from the de\cm~\er{3.14}, $\wt\o_E$ is a
\cn\ defined on a fibre bundle $Q$ over a \spt~$E$ with structural
group~$C^\mu$ and a projection~$\pi_E$. Moreover, $C^\mu=G$ and  $\wt\o_E$ is
a 1-form with values in the \Li~$\fg$. This \cn\ describes an ordinary \YM'
field gauge group $G=C^\mu$ on the \spt~$E$. $\F$~is a \f\ on~$E$ with values
in the space $\wt S$ of linear maps
\beq3.25
\F:\fm \to \fh
\e
\sf ying
\beq3.26
\F([X_0,X])=[\mu'X_0,\F(X)], \q X_0\in\fg_0.
\e
Thus
\beq3.27
\bal
\wt\o_E=\gd\wt\o,i,E,Y_i, &\q Y_i\in\fg,\\
\gd\o,\si,0,=\t^{\hi}Y_{\hi}, &\q Y_{\hi}\in\fg_0,\\
\gd\o,\si,\fm,=\ov\t{}^{\td a}Y_{\td a}, &\q Y_{\td a}\in\fm.
\eal
\e

Let us write condition \er{3.24} in the base of left-\iv t form
$\t^{\hi},\ov\t{}^{\td a}$, which span respectively dual spaces
to~$\fg_0$ and~$\fm$ (see Refs \cite{24,25}). It is easy to see that
\beq3.28
\F\circ \gd\o,\si,\fm,=\gd \F,a,\td a,(x)\ov\t{}^{\td a}X_a, \q X_a\in\fh
\e
and
\beq3.29
\mu'=\gd\mu,a,\hi, \t^{\hi}X_a.
\e
From \er{3.26} one gets
\beq3.30
\F_{\td b}^c (x)f_{\hi\td a}^{\td b}=
\mu^a_{\hi}\F^b_{\td a}(x)\gd C,c,ab,
\e
where $f^{\td b}_{\hi\td a}$ are structure \ct s of the \Li~$\fg$ and
$\gd C,c,ab,$ are structure \ct s of the \Li~$\fh$. Eq.~\er{3.30} is a
constraint on the scalar field $\F^a_{\td a}(x)$. The $\SU(3)$ coloured
Higgs cannot appear if we do not want it. The structure of the multiplet of
scalar fields can be arranged in such a way that we get desired pattern of
\ssb. It means that we do not have any phenomenological problems.

For a curvature of~$\o$ one
gets
\bml3.30a
\O=\frac12\,\gd H,C,AB,\t^A\land \t^BX_C=
\frac12\,\gd\wt H,i,\m,\t^\mu\land\t^\nu \a^c_iX_c
+\gv{\n_\mu} \F^c_{\td a}\t^\mu\land \t^{\td a}X_c\\
{}+\frac12\,\gd C,c,ab,\F^a_{\td a}\F^b_{\td b}\t^{\td a}\land \t^{\td b}
X_c - \frac12\,\F^c_{\td d}f^{\td d}_{\td a\td b}\t^{\td a}\land\t^{\td b}X_c.
\e
Thus we have
\bg3.31
\gd H,c,\m,=\gd \a,c,i,\gd \wt H,i,\m,\\
\gd H,c,\mu\td a,=\gv{\n_\mu} \F^c_{\td a}=-\gd H,c,\td a\mu, \lb3.32 \\
\gd H,c,\td a\td b,=\gd C,c,ab,\cdot\F^a_{\td a}\F^b_{\td b} - \mu^c_{\hi}
f^{\hi}_{\td a\td b} - \F^c_{\td d}\gd f,\td d,\td a\td b, \lb3.33
\e
(see Section 4 for comparison), where $\gv{\n_\mu}$ means gauge \dv\ \wrt the \cn\ $\wt\o_E$ defined on a
bundle~$Q$ over a \spt~$E$ with a \sc al group~$G$
\beq3.34
Y_i=\a^c_i X_c.
\e
$\gd \wt H,i,\m,$ is the curvature of the \cn\ $\ov\o_E$ in the base
$\{Y_i\}$, generators of the \Li\ of the Lie group~$G$, $\fg$, $\a^c_i$ is
the matrix which connects $\{Y_i\}$ with $\{X_c\}$. Now we would like to
remind that indices $a,b,c$ refer to the \Li~$\fh$, $\wt a,\wt b,\wt c$ to
the space~$\fm$ (tangent space to~$M$), $\wh \imath,\wh \jmath,\wh k$ to the
\Li~$\fg_0$ and $i,j,k$ to the \Li\ of the group $G$, $\fg$. The matrix
$\a^c_i$ establishes a direct relation between generators of the \Li\ of the
subgroup of the group~$H$ iso\mo c to the group~$G$.

The \cn\ on principal fibre bundle over $E\tm G/G_0$ consists of two parts:
vector fields (gauge fields) over~$E$ and scalar fields (vector fields over
$M=G/G_0$). Due to this an ordinary \lg\ of multi\di al vector fields is
divided  into some parts involving ordinary gauge fields, Higgs' fields
(scalar fields) and interacting terms. We called the scalar field Higgs' field,
because we get \ssb\ and \Hm.

Let us come back to a construction of the \NK\ on a manifold~$P'$. We should
define \cn s. First of all, we should define a \cn\ compatible with a \nos\
tensor $\g_{AB}$, Eq.~\er{3.18},
\bg3.35
\gd \ov\o,A,B,=\gd \ov\G,A,BC,\t^C\\
\ov D\g_{AB}=\g_{AD}\gd\ov Q,D,BC,(\ov\G)\t^C \lb3.36 \\
\gd \ov Q,D,BD,(\ov \G)=0 \nn
\e
where $\ov D$ is the exterior covariant \dv\ \wrt $\gd \ov\o,A,B,$ and $\gd
\ov Q,D,BC,(\ov\G)$ its torsion.

Using \er{3.18} one easily finds that the \cn\ \er{3.35} has the following
shape
\beq3.37
\gd \ov\o,A,B,=\left(\begin{array}{c|c}
\pi^*_E(\gd \ov\o,\a,\b,) & 0 \\ \hline
0 & \gd \wh{\bar\o},\td a,\td b,
\end{array}\right)
\e
where $\gd \ov\o,\a,\b,=\gd \ov \G,\a,\b\g,\ov\t{}^\g$ is a \cn\ on the
\spt~$E$ and $\gd\wh{\ov \o},\td a,\td b,=\gd \wh{\ov\G},\td a,\td b\td c,
\ov \t{}^{\td c}$ on the manifold $M=G/G_0$ with the following properties
\bg3.38
\ov Dg_{\a\b}=g_{\a\d}\gd \ov Q,\d,\b\g,(\ov \G)\ov \t{}^\g=0\\
\gd \ov Q,\a,\b\a,(\ov\G)=0 \lb3.39 \\
\wh{\ov D}g_{\td a\td b}=g_{\td a\td d}\gd\wh{\ov Q},\td d,\td b\td c,(\wh{\ov\G}).
\lb3.40 \\
\gd\wh{\ov Q},\td d,\td b\td d,(\wh{\ov\G})=0 \nonumber
\e

$\ov D$ is an exterior covariant \dv\ \wrt a \cn~$\gd \ov\o,\a,\b,$. $\gd \ov
Q,\a,\b\g,$ is a tensor of torsion of a \cn~$\gd \ov\o,\a,\b,$. $\wh{\ov
D}$~is an exterior covariant \dv\ of a \cn~$\gd \wh{\ov\o},\td a,\td b,$ and
$\gd \wh{\ov Q},\td a,\td b\td c,(\wh{\ov\G})$ its torsion.

On a \spt\ $E$ we also define the second affine \cn\ $\gd \ov W,\a,\b,$ \st
\beq3.41
\gd \ov W,\a,\b,= \gd \ov\o,\a,\b, - \frac23\,\gd \d,\a,\b,\ov W,
\e
where
$$
\ov W=\ov W_\g \ov\t{}^\g = \tfrac12(\gd \ov W,\si,\g\si,-\gd \ov W,\si,\g\si,).
$$
We proceed a \nos\ metrization of a principal fibre bundle $P'$ according
to \er{3.18}. Thus we define a right-\iv t \cn\ \wrt an action of the
group~$H$ compatible with a tensor $\ka_{\tA\tB}$
\bg3.42
D\ka_{\tA\tB}=\ka_{\tA\td D}\gd Q,\td D,\tB\tC,(\G)\t^{\tC}\\
\gd Q,\td D,\td B\td D,(\G)=0 \nn
\e
where $\gd \o,\tA,\tB,=\gd \G,\tA,\tB\tC,\wt \t{}^\tC$. $D$ is an exterior
covariant \dv\ \wrt the \cn\ $\gd \o,\tA,\tB,$ and $\gd Q,\tA,\tB\tC,$ its
torsion. After some calculations one finds
\beq3.43
\gd \o,\tA,\tB,= \bma
\pi^*(\gd \ov\o,A,B,)-\ell_{db}\g^{MA}\gd L,d,MB,\t^b & \gd L,a,BC,\t^C \\
\hline
\ell_{bd}\g^{AB}(2\gd H,d,CB,-\gd L,d,CB,)\t^C & \gd \wt\o,a,b,
\ema
\e
(comp. to Eq. \er{4.11}), where
\bg3.44
\gd L,d,MB,=-\gd L,d,BM,\\
\ell_{dc}\g_{MB}\g^{CM}\gd L,d,CA, + \ell_{cd}\g_{AM}\g^{MC}\gd L,d,BC,
=2\ell_{cd}\g_{AM}\g^{MC}\gd H,d,BC,, \lb3.45
\e
$\gd L,d,CA,$ is Ad-type tensor \wrt $H$ (Ad-\ci\ on~$\ul P'$)
\bg3.46
\gd \wt\o,a,b,=\gd \wt\G,a,bc,\t^c\\
\ell_{db}\gd \wt\G,d,ac,+\ell_{ad}\gd \wt\G,d,cb,=-\ell_{db}\gd C,d,ac, \lb3.47 \\
\gd \wt\G,d,ac,=-\gd \wt\G,d,ca,, \q \gd \wt\G,d,ad,=0. \lb3.48
\e

The \cn\ \er{3.43} is a \cn\ defined on a \nos ally metrized principal fibre
bundle over $E\tm M$, i.e.\ it is $(n+n_1+4)$-\di al. It contains vector field
(gauge field) over~$E$ and also Higgs' field (scalar field) unified with \nos\
gravity.

We define on $P'$ a second \cn
\beq3.49
\gd W,\tA,\tB, = \gd \o,\tA,\tB, - \frac4{3(m+2)}\,\gd \d,\tA,\tB,\ov W.
\e
Thus we have on $P'$ all $(m+4)$-\di al analogues of geometrical quantities from
NGT, i.e.
$$
\gd W,\tA,\tB,, \q \gd \o,\tA,\tB, \qh{and} \ka_{\tA\tB}.
$$

Let us calculate a Moffat--Ricci curvature scalar for the \cn\ $\gd W,\tA,\tB,$
\beq3.50
R(W)=\ka^{\tA\tB}\bigr(\gd R,\tC,\tA\tB\tC,(W)+\tfrac12 \gd R,\tC,\tC\tA\tB,
(W)\bigr)
\e
(see Eq.~\er{4.16} for comparison), where $\gd R,\tC,\tC\tA\tB,(W)$ is a curvature tensor for a \cn\ $\gd
W,\tA,\tB,$ and $\ka^{\tA\tB}$ is an inverse tensor for $\ka_{\tA\tB}$
\beq3.51
\ka^{\tA\tC}\ka_{\tA\tB} = \ka^{\tC\tA}\ka_{\tB\tA}=\gd \d,\tC,\tB,.
\e
Using results from Ref.~\cite1 one gets (having in mind some analogies from a
theory with a base space $E$ to the theory with the base space
$V=E\tm M=E\tm G/G_0$)
\beq3.52
R(W)=\ov R(\ov W)+\frac1{r^2}\,R(\wh{\ov \G})+\frac1{\la^2}\,
\wt R(\wt\G) - \frac{\la^2}4\,\ell_{ab}\bigl(2H^aH^b - L^{aMN}\gd H,b,MN,\bigr)
\e
(see Eq.~\er{4.17} for comparison)
where $\ov R(\ov W)$ is a \MR\ \cvt\ scalar on the \spt~$E$ for a \cn~$\gd\ov
W,\a,\b,$, $R(\wh{\ov \G})$ is a \MR\ \cvt\ scalar for a \cn~$\gd
\wh{\ov\o},\td a,\td b,$ on a homogeneous space $M=G/G_0$, $\wt R(\wt\G)$
is a \MR\ \cvt\ scalar for a \cn~$\gd\wt\o,a,b,$,
\beq3.53
H^a=\g^\[AB]\gd H,a,\[AB],=g^\[\a\b]\gd H,a,\a\b,+\frac1{r^2}\,
g^\[\td a\td b]\gd H,a,\td a\td b,
\e
\bml3.54
L^{aMN}=\g^{AM}\g^{BN}\gd L,a,AB,=\gd \d,M,\mu,\gd \d,N,\g,g^{\a\mu}
g^{\b\g}\gd L,a,\a\b,\\
{}+ \frac1{r^2}\bigl(g^{\a\mu}g^{\td b\td n}\gd L,a,\a\td b,
+g^{\td a\td n}g^{\b\g}\gd L,a,\td a\b,\bigr)\gd \d,M,\mu,\gd \d,N,\td n,
+\frac1{r^4}\,g^{\td a\td m}g^{\td b\td n}\gd L,a,\td a\td b,\gd \d,M,\td m,
\gd \d,N,\td n,.
\e
One finds that
\bml3.55
-\ell_{ab}L^{aMN}\gd H,b,MN,=-\ell_{ab}\Bigr(g^{\a\mu}g^{\b\nu}\gd L,a,\a\b,
\gd H,b,\m,+\frac2{r^2}\,g^{\a\mu}g^{\td b\td n}\gd L,a,\a\td b,\gd
H,b,\mu\td n,+\frac1{r^4}\,g^{\td a\td m}g^{\td b\td n}\gd L,a,\td a\td b,
\gd H,b,\td m\td n,\Bigr)\\
{}=-\ell_{ab}\Bigl(L^{a\m}\gd H,b,\m,+\frac2{r^2}\,g^{\td b\td n}
\gd L,\a\mu,\td b,\gd H,b,\mu\td n,+\frac1{r^4}\,g^{\td a\td m}g^{\td b\td n}
\gd L,a,\td a\td b,\gd H,b,\td m\td n,\Bigr).
\e
We get conditions from Eq.~\er{3.45}
\bea3.56
\ell_{dc}g_{\mu\b}g^{\g\mu}\gd L,d,\g\a,
+ \ell_{cd}g_{\a\mu}g^{\mu\g}\gd L,d,\b\g,
&= 2\ell_{cd}g_{\a\mu}g^{\mu\g}\gd H,d,\b\g, \\
\ell_{dc}g_{\td m\td b}g^{\td c\td m}\gd L,d,\td c\td a,
+ \ell_{cd}g_{\td a\td m}g^{\td m\td c}\gd L,d,\td b\td c,
&= 2\ell_{cd}g_{\td a\td m}g^{\td m\td c}\gd H,d,\td b\td c, \lb3.57 \\
\ell_{dc}g_{\mu\b}g^{\g\mu}\gd L,d,\g\td a,
+ \ell_{cd}g_{\td a\td m}g^{\td m\td c}\gd L,d,\b\td c,
&= 2\ell_{cd}g_{\td a\td m}g^{\td m\td c}\gd H,d,\b\td c, \lb3.58 \\
L^{a\m}&=g^{\a\mu}g^{\b\nu}\gd L,a,\a\b, \lb3.59 \\
\gd L,a\mu,\td b,&=g^{\a\mu}\gd L,a,\mu\td b, \lb3.60
\e
(see Eqs~\er{4.22}--\er{4.24} for comparison).

For $\ell_{ab}H^aH^b=h_{ab}H^aH^b$ we have the following:
\beq3.61
h_{ab}H^aH^b = h_{ab}\gd H,a,0,\gd H,b,0, + \frac2{r^2}\,h_{ab}
\gd H,a,0,\gd H,b,1, + \frac1{r^4}\,h_{ab}\gd H,a,1,\gd H,b,1,
\e
where
\beq3.62
\gd H,a,0,=g^{\a\b}\gd H,a,\a\b,, \q \gd H,a,1,=g^\[\td a\td b]\gd H,a,\td
a\td b,.
\e

Finally, we have for a density of $R(W)$, i.e.
\bml3.63
\sqrt{|\ka|}\,R(W) = \sqrt{-g}\,r^{n_1} \sqrt{|\wt g|}\,\sqrt{|\ell|}\,R(W)\\
{}=\sqrt{-g}\,r^{n_1} \sqrt{|\wt g|}\,\sqrt{|\ell|}\,
\biggl(\ov R(\ov W)+\frac{\wt R(\wt \G)}{\la^2} + \frac1{r^2}\,R(\wh{\ov\G})
+\frac{\la^2}4 \, \ell_{ab}\bigl(2\gd H,a,0,\gd H,b,0, - L^{a\m}
\gd H,b,\m,\bigr)\\ {}+ \frac{\la^2}{4r^2} \, \ell_{ab}\bigl(4\gd H,(a,0,
\gd H,b),1, - 2g^{\td b\td n}\gd L,a\mu,\td b,\gd H,b,\mu\td n,\bigr)
+\frac{\la^2}{4r^2}\,\ell_{ab}\bigl(2\gd H,a,1,\gd H,b,1, - g^{\td a\td m}
g^{\td b\td n}\gd L,a,\td a\td b,\gd H,b,\td m\td n,\bigr)\biggr).
\e

We define an integral of action
\beq3.64
S \sim \int_U \sqrt{|\ka|}\,R(W)\,d^{m+4}x,
\e
where
$$
U=M \tm G \tm \ov V, \q \ov V\subset E, \q d^{m+4}x=d^4x\,d\mu_H(h)\,dm(y),
$$
$d\mu_H(h)$ is a bi\iv t measure on a group $H$ and $dm(y)$ is a measure
on~$M$ induced by a bi\iv t measure on~$G$. $\ov R(\ov W)$ is a \MR\ \cvt\
scalar for a \cn~$\gd\ov W,\a,\b,$ on~$E$.

Let us consider Eqs \er{3.56}--\er{3.58} modulo \e s \er{3.31}--\er{3.33}.
One gets
\beq3.65
\ell_{ij}g_{\mu\b}g^{\g\mu}\gd \wt L,i,\g\a, + \ell_{ji}g_{\a\mu}g^{\mu\g}
\gd \wt L,i,\b\g, = 2\ell_{ji}g_{\a\mu}g^{\mu\g}\gd \wt H,i,\b\g,
\e
where $\ell_{ij}=\ell_{cd}\gd\a,c,i,\gd \a,d,j,$ is a right-\iv t \nos\
metric on the group~$G$ and
\beq3.66
\gd L,c,\m,=\gd \a,c,i, \gd \wt L,i,\m,.
\e
$\gd \wt L,i,\m,$ plays a role of an induction tensor for the \YM' field with
the gauge group~$G$. $\gd \wt H,i,\m,$~is of course the tensor of strength of
this field. The polarization tensor is defined as usual
\beq3.67
\gd \wt L,i,\m,=\gd \wt H,i,\m, - 4\pi \gd \ov{\wt M},i,\m,.
\e
We introduce two $\Ad_G$-type 2-forms with values in the \Li~$\fg$ (of~$G$)
\bea3.68
\wt L&=\tfrac12\,\gd \wt L,i,\m,\t^\mu \land \t^\nu Y_i\\
\ov{\wt M}&=\tfrac12\,\gd \ov{\wt M},i,\m,\t^\mu \land \t^\nu Y_i\lb3.69
\e
and we easily write
\beq3.70
\wt L=\wt\O_E - 4\pi \ov{\wt M}=\wt\O_E - \tfrac12\,\wt Q
\e
where $\wt Q=\frac12\,\gd \wt Q,i,\m,\t^\mu\land \t^\nu Y_i$, $\gd \wt Q,i,\m,
=\gd \a,i,c, \gd Q,c,\m,$.
$\wt\O_E$ is a 2-form of a \cvt\ of a \cn\ $\wt\o_E$ (Eq.~\er{3.27}) in
Eq.~\er{3.30a} (the first term of this \e).

In this way we get a geometrical interpretation of a \YM' induction tensor in
terms of the \cvt\ tensor and torsion in additional \di s. Afterwards we get
\bg3.71
\ell_{cd}g_{\td m\td b}g^{\td c\td m}\gd L,d,\td c\td a,
+\ell_{cd}g_{\td a\td m}g^{\td m \td c}\gd L,d,\td b\td c,
=2\ell_{cd}g_{\td a\td m}g^{\td m\td c}\bigl(\gd C,d,ab,\F^a_{\td b}
\F^b_{\td c} - \gd \mu,d,\hi, \gd f,\hi,\td b\td c, - \F^d_{\td d}
\gd f,\td d,\td b\td c,\bigr),\\
\ell_{cd}g_{\mu\b}g^{\g\mu}\gd L,d,\g\td a,+\ell_{cd}g_{\td a\td m}
g^{\td m\td c}\gd L,d,\b\td c,
=2\ell_{cd}g_{\td a\td m}g^{\td m\td c}\gv{\n_\b}\F^d_{\td c}. \lb3.72
\e

The formula \er{3.71} is coming from formula \er{3.57} by using formula
\er{3.33}. Formulae \er{3.56}--\er{3.58} are coming from the formula \er{3.45}.

Let us rewrite an action integral
\bg3.73
S=-\frac1{V_1V_2r^{n_1}} \int_U \bigl(R(W)\,d^nx\bigr)\,d^{n_1}x\,d^4x,
\q U=V\tm M\tm H, \q V\subset E,\\
V_1=\int_H \sqrt{|\ell|}\,d^nx \lb3.74 \\
V_2=\int_M \sqrt{|\wt g|}\,d^{n_1}x. \lb3.75
\e
Thus we get
\beq3.76
S=-\int_V \sqrt{-g}\,d^4x \,\cL(\ov W,g,\wt A,\F)
\e
where
\bg aa
\cL(\ov W,g,\wt A,\F) \hskip320pt \nn \\
\hskip20pt{}= \ov R(\ov W) + \frac{\la^2}4 \Bigr(
8\pi \cLY(\wt A)+\frac2{r^2}\,\cL_{\rm kin}(\gv\n \F) + \frac1{r^4}\,\wh V(\F)
-\frac4{r^2}\,\cL_{\rm int}(\F,\wt A)\Bigr)+\la_c \lb3.77 \\
\cLY(\wt A)=-\frac1{8\pi}\,\ell_{ij}\bigl(2\wt H{}^i \wt H{}^j - \wt L^{i\m}
\gd \wt H,j,\m,\bigr) \lb3.78
\e
(see Eqs~\er{4.34}, \er{4.48}, \er{4.69}, \er{4.73}, \er{4.79} for comparison,
this makes our presentation friendly for the reader)
is the lagrangian for the \YM' field with the gauge group~$G$
(see Eq.~\er{2.32}),
\bml3.79
\cL_{\rm kin}(\gv\n \F)=\frac1{V_2}\int_M \sqrt{|\wt g|}\,d^{n_1}x
\bigl(\ell_{ab}g^{\td b\td n}\gd L,a\mu,\td b,\gv{\n_\mu} \F^b_{\td n}\bigr)\\
{}=\ell_{ab}g^{\a\mu}\,\frac1{V_2}\int_M \sqrt{|\wt g|}\,d^{n_1}x
\bigl(g^{\td b\td n}\gd L,a,\a\td b,\gv{\n_\mu} \F^b_{\td n}\bigr)
\e
is a kinetic part of a lagrangian for a scalar field $\F^a_{\td a}$. It is
quadratic in gauge \dv\ of $\F^a_{\td a}$ and is \iv t \wrt the action of
groups $H$ and~$G$.
\bml3.80
\wh V(\F)=\frac{\ell_{ab}}{V_2} \int_M \sqrt{|\wt g|}\,d^{n_1}x
\Bigl[ 2g^\[\td m\td n]\bigl(\gd C,a,cd,\F^c_{\td m}\F^d_{\td n} -
\gd\mu,a,\hi,\gd f,\hi,\td m\td n,- \F^a_{\td e}\gd f,\td e,\td m\td n,\bigr)\\
{}g^\[\td a\td b]\bigl(\gd C,b,ef,\F^e_{\td a}\F^f_{\td b}
-\gd \mu,b,\wh\jmath,\gd f,\wh\jmath,\td a\td b,
-\Ft ba\gd f,\td d,\td a\td b,\bigr)
-g^{\td a\td m}g^{\td b\td n}\gd L,a,\td a\td b,
\bigl(\gd C,b,cd,\F^c_{\td m}\F^d_{\td n}-\gd \mu,b,\hi,\gd f,\hi,\td m\td n,
-\F^b_{\td e}\gd f,\td e,\td m\td n,\bigr)\Bigr]
\e
is a self-interacting term for a field $\F$. It is \iv t \wrt the action of
the groups $H$ and~$G$. This term is a polynomial of fourth order in $\F$'s
(a~Higgs' field \pt\ term)
\beq3.81
\cL_{\rm int}(\F,\wt A)=h_{ab}\gd \mu,a,i,\wt H{}^i \ul g^\[\td a\td b]
\bigl(\gd C,b,cd,\F^c_{\wt a} \F^d_{\td b} - \gd \mu,b,\hi,\gd f,\hi,\td a\td
b,-\F^b_{\td d}\gd f,\td d,\td a\td b,\bigr)
\e
where
\beq3.82
\ul g^\[\td a\td b]=\frac1{V_2}\int_M \sqrt{|\wt g|}\,d^{n_1}x \,g^\[\td a\td
b]
\e
is the term describing non-minimal coupling between the scalar field $\F$ and
the \YM' field. This term is also \iv t \wrt the action of the groups $H$
and~$G$.
\beq3.83
\la_c = \frac1{\la^2}\,\wt R(\wt \G)+\frac1{r^2V_2} \int_M \sqrt{|\wt g|}\,
\wh{\ov R}(\wh{\ov \G})\,d^{n_1}x = \frac1{\la^2}\,\wt R(\wt \G)+ \frac1{r^2}\,\ul{\wt P}.
\e

Let us pass to \sn\ \s  y breaking and \Hm\ in our theory. In order to do
this we look for the critical points (the minima) of the \pt\ $V(\F)$.
However, our field \sf ies the constraints
\beq3.97
\Ft cb\gd f,\td b,\hi\td a, - \gd \mu,a,\hi,\Ft ba \gd C,c,ab,=0.
\e
Thus we must look for the critical points of
\beq3.98
V'=\wh V+\gd \psi,\hi\td d,c,\bigl(\Ft cb\gd f,\td b,\hi\td a,
-\gd \mu,a,\hi, \Ft ba \gd C,c,ab,\bigr)
\e
where $\gd \psi,\hi\td d,c,$ is a Lagrange multiplier.
It is easy to see that, if
\beq3.101
\gd H,a,\td m\td n,=0
\e
then
\beq3.102
\frac{\d V'}{\d \F}=0
\e
if \er{3.97} is \sf ied.

$\gd H,a,\td n\td m,$ is a part of the curvature of $\o$ over
a~manifold~$M$. Thus it means that $\F\crt$ \sf ying Eq.~\er{3.101} is a
``pure gauge''. If the \pt\ $\wh V(\F)$ is positively defined, then we have the
absolute minimum of~$\wh V$
\beq3.103
\wh V(\F\crt^0)=0.
\e
But apart from this \so\ there are some others due to an influence of \nos\
metric on~$H$ and~$M$. The details strongly depend on \ct s $\xi$,~$\z$ and on
groups $G,G_0,H$. There are also some critical which are minima. Moreover, we
expect the second critical point $\F\crt^1\ne \F\crt^0$ \st
$\wh V(\F\crt^1)\ne0$ and
\bg3.104
\gd H,a,\td m\td n,(\F\crt^1)\ne0 \\
\frac{\d V'}{\d\F}(\F\crt^i)=0, \q i=0,1. \lb3.105
\e
This means that $\F\crt^1$ is not a ``pure gauge'' and a gauge configuration
connected to $\F\crt^1$ is not trivial. This indicates that the local minimum
is not a vacuum state. It is a ``false vacuum'' in contradiction to ``true
vacuum'' for the absolute minimum $\F\crt^0$.

Now we answer the question of what is a \s y breaking if we choose one of the
critical values of $\F\crt^0$ (we choose one of the degenerated vacuum states
and the \sn\ breaking of the \s y takes place). It was
shown that if $\gd H,a,\td m\td n,=0$ and Eq.~\er{3.97} is \sf ied then the
\s y is reduced to~$G_0$. In the case of the second minimum (local
minimum---false vacuum) the unbroken \s y will be in general different.

Let us call it $G_0'$ and its \Li~$\fg_0'$. This will be the \s y which
preserves $\F\crt^1$ and the constraint \er{3.97}. It is easy to see that the
\Li\ of this unbroken group preserves $\F\crt^1$ under Ad-action. For
the \s y group of $\wh V$ is larger than $G$ (it is~$H$) we expect some scalars
which remain massless after the \s y breaking in both cases (i.e., $i=0,1$,
``true'' and ``false'' vacuum case). They became massive only through radiative
corrections. They are often referred as the pseudo-Goldstone bosons.

We can solve explicitly Eq.~\er{2.21} (or \er{3.56} and \er{3.65}), see
Refs \cite{16}, \cite{ka}, getting
\beq2.125
\aligned
L^n{}_{\om\mu}&=H^n{}_{\om\mu}+\mu h^{na}k_{ad}H^d{}_{\om\mu}
+\bigl(H^n{}_{\a\om}\gtn{\a\d}g\ink{\d\mu}-H^n{}_{\a\mu}\gtn{\a\d}g\ink{\d\om}
\bigr)\\
&-2\mu h^{na}k_{ad}\gtn{\d\tau}\gtn{\a\b}H^d{}_{\d\a}g\ink{\tau\om}g\ink{\b\mu}
-2\mu h^{na}k_{ad}\gtn{\d\b}\gtn{\a\tau}H^d{}_{\b[\om}g_{\mu]\tau}g\ink{\d\a}\\
&+2\mu^2h^{na}h^{bc}k_{ac}k_{bd}\gtn{\a\b}
H\gds d,\a\lower2pt\hbox{$\ts[$}\om,g_{[\mu\lower2pt\hbox{$\ts]$}\b]}
\endaligned
\e
(see also Refs \cite{8,11a}).

The condition \er{3.58} can be explicitly solved.
One gets
\beq2.126
\aligned
L^n{}_{\om\u m} &= \brgn\om \F\gds n,\u m,+\xi k\gds n,d,\brgn\om \F\gds d,\u m,
-\Bigl(\z\brgn\om \F\gds n,\u a,h^{o\u a\u d}k_{o\u d\u m}+\gtn{\a\mu}
\brgn\a\F\gds n,\u m,g\ink{\mu\om}\Bigr)\\
&-2\xi\z k\gds n,d,\brgn\om \F\gds d,\u d,\gtn{\d\a}g\ink{\a\om}h^{o\u d\u a}
k\gds o,\u a\u m,\\
&+\xi k\gds n,d,\Bigl(\z^2 h^{\u d\u a}\brgn\om \F\gds d,\u a,
k\gds o,\u d\u b,k\gds o,\u m\u c,h^{o\u c\u b}
+\brgn\b \F\gds d,\u m,\gtn{\d\b}g\ink{\d\a}g\ink{\om\mu}\gtn{\a\mu}\Bigr)\\
&-\xi^2k^{nb}k_{bd}\Bigl(\z\brgn\om \F\gds d,\u a,h^{o\u a\u b}k\gds o,\u m\u b,
+\gtn{\a\b}\brgn\a \F\gds d,\u m,g\ink{\om\b}\Bigr),
\endaligned
\e
where
\beq2.127
k^{nb}=h^{na}h^{bp}k_{ap}.
\e
The condition \er{3.57} can be explicitly solved too. One gets
\beq2.128
\aligned
L\gds n,\u w\u m,&=H\gds n,\u w\u m,+\mu k\gds n,d,H\gds d,\u w\u m,
+\z\bigl(h^{o\u a\u d}H\gds n,\u a\u w, k\gds o,\u d\u m,-h^{o\u a\u d}
H\gds n,\u a\u m,k\gds o,\u a\u w,\bigr)\\
&-2\mu \z^2h^{o\u d\u c}h^{o\u a\u b}H\gds d,\u d\u a,k\gds o,\u c\u w,
k\gds o,\u b\u m,
-2\mu\z k\gds n,d, h^{o\u a\u p}h^{o\u d\u b}H\gds d,\u b[\u w,k\gds o,\u m]\u p,
k_{\u d\u a}\\
&+2\mu^2\z k^{nb}k_{bd}H\gds d,\u a[\u w,k\gds o,\u m]\u p,h^{o\u p\u a}.
\endaligned
\e

Using the above formulae we can express the Yang--Mills \lg\ in the \E\nos\
Nonabelian \KK\ Theory and the \lg\ for Higgs' field.

The Yang--Mills \lg\ reads
\beq2.129
\bal
\hskip-20pt
\cL\dr{YM}&=\frac1{8\pi}\,\biggl(
h_{nk}H^{k\om\mu}H\gds n,\om\mu,-2h_{cd}H^cH^d+2h_{nk}H^{k\om\mu}
H\gds n,\d\om,g\ink{\a\mu}\gtn{\a\d}\\
&+\mu\Bigl[2k_{nk}H^{k\om\mu}H\gds n,\d\om,\gtn{\d\a}g\ink{\a\mu}
-2k_{kd}H^{k\om\mu}H\gds d,\d\a,\gtn{\d\b}\gtn{\a\rho}g\ink{\b\om}g\ink{\rho\mu}\\
&-k_{kd}H^{k\om\mu}H\gds d,\eta\om,\gtn{\eta\b}\gtn{\a\rho}
g\ink{\mu\a}g\ink{\b\rho}
+k_{kd}H^{k\om\mu}H\gds d,\eta\mu,\gtn{\eta\d}\gtn{\a\rho}g\ink{\d\rho}
g\ink{\om\d}\Bigr]\\
&+\mu^2\Bigl[k_{nk}k\gds n,d,H^{k\om\mu}H\gds d,\eta\mu,\gtn{\rho\b}\gtn{\eta\a}
g\ink{\om\b}g\ink{\a\rho}\\
&-2k_{nk}k\gds n,d,H^{k\om\mu}H\gds d,\d\a,\gtn{\d\eta}\gtn{\a\rho}
g\ink{\eta\om}g\ink{\rho\mu}
-k_{nk}k\gds n,d,H^{k\om\mu}H\gds d,\eta\om,\gtn{\rho\a}\gtn{\eta\b}
g\ink{\mu\a}g\ink{\b\rho}\\
&+k\dgs k,b,k_{bd}H^{k\om\mu}H\gds d,\a\om,\gtn{\a\b}g\ink{\mu\a}
-k\dgs k,b,k_{bd}H^{k\om\mu}H\gds d,\a\mu,\gtn{\a\b}g\ink{\om\b}
+k\gds p,n,k_{pk}H^{k\om\mu}H\gds n,\om\mu,\Bigr]\\
&+\mu^3 \Bigl[k_{nk}k^{nb}k_{bd}H^{k\om\mu}H\gds d,\a\om,\gtn{\a\b}
g\ink{\mu\b}
-k_{nk}k^{nb}k_{bd}H^{k\om\mu}H\gds d,\a\mu,H^{k\om\mu}
\gtn{\a\b}g\ink{\om\b}\Bigr]\biggr)
\eal
\e

The kinetic term for Higgs' field is given by
\beq2.130
\bal
\cL\dr{kin}(\nabla\F)&=\frac{1}{V_2}\int_M {\sqrt{|\wt g|}}\,d^{n_1}x\,
\Bigl[l_{nk}g^{\om\mu}g^{\u m\u p}\brgn \mu \F\gds k,\u p,
\Bigl\{\brgn\om \F\gds n,\u m,+\xi k\gds n,d,\brgn\om \F\gds d,\u m,\\
&-\z\brgn\om \F\gds d,\u a,h^{o\u a\u q}k\gds o,\u q\u m,
-\brgn\a \F\gds a,\u m,\gtn{\a\eta}g\ink{\eta\om}\\
&-2\xi\z\brgn\d \F\gds d,\u a,k\gds n,d,\gtn{\d\a}g\ink{\a\om}
h^{o\u d\u q}k\gds o,\u q\u m,\\
&-\xi\bigl(\z^2 k\gds n,d,\brgn\om \F\gds d,\u a,h^{o\u b\u q}h^{o\u a\u w}
k\gds o,\u q\u m,k\gds o,\u w\u b,
+k\gds n,d,\brgn\b \F\gds d,\u m,\gtn{\a\nu}\gtn{\b\rho}g\ink{\nu\om}
g\ink{\rho\a}\bigr)\\
&+\xi^2\bigl(\z k^{nb}k_{bd}\brgn\om \F\gds d,\u a,h^{o\u a\u q}
k\gds o,\u q\u m,+\brgn\a \F\gds d,\u m,\gtn{\a\b}g\ink{\b\om}\bigr)\Bigr\}
\Bigr]
\eal
\e

In the case of $g_{\mu\nu}=\eta_{\mu\nu}$ (a Minkowski tensor)
one gets
\beq2.131
\bal
\cL\dr{kin}(\nabla\F)&=\frac{1}{V_2}\int_M {\sqrt{|\wt g|}}\,d^{n_1}x\,
\Bigl[l_{nk}g^{\u m\u p}\brg\om \F\gds k,\u p,
\Bigl\{\brgn\om \F\gds n,\u m,
+\xi k\gds n,d,\brgn\om \F\gds d,\u m,\\
&-\z\brgn\om \F\gds d,\u a,k\gds o\u a,\u m,
-\xi\z^2 k\gds n,d,k\gds o\u b,\u m, k\gds o\u a,\u b,\brgn\om\F\gds d,\u a,\\
&+\xi^2\z k^{nb}k_{bd}k\gds o\u a,\u m,\brgn\om \F\gds d,\u a,\Bigr\}\Bigr],
\eal
\e
where $\brg\om \F\gds k,\u p,=\eta^{\om\mu}\brgn\mu \F\gds k,\u p,$,
$k\gds o\u a,\u b,=h^{o\u a\u c}k_{o\u c\u b}$.

A mass matrix for broken gauge bosons can be calculated.
\beq2.132
\bal
{}M_{ij}^2(\F\dr{crt}^k)&=\frac{\a_s^2}{\hbar c}\,\frac1{V_2}
\int_M \sqrt{|\wt g|}\,d^{n_1}x\\
&\Bigl\{l_{np}g^{\u m\u p}\brr{(k)}B{}\gds p,\u pi,
\Bigl(\brr{(k)}B{}\gds d,\u mj,
+\xi k\gds n,d,\brr{(k)}B{}\gds d,\u mj,
-\z \brr{(k)}B{}\gds d,\u aj,k\gds o\u a,\u m,\\
&-\xi\z^2k\gds n,d,k\gds o\u b,\u m,k\gds o\u a,\u b,\brr{(k)}B{}\gds d,\u aj,
+\xi^2\z k^{nb}k_{bd}k\gds o\u a,\u m,\brr{(k)}B{}\gds d,\u aj,\Bigr)\Bigr\}
\eal
\e
$k=0,1$, where
\beq2.133
\brr{(k)}B{}\gds b,\u ni,=\bigl[\d\gds\u m,\u n, C\gds b,ms,\a\gds s,i,
+\d\gds b,m,f\gds \u m,\u ni,\bigr][\F^k\dr{crt}]\gds m,\u m,.
\e

In the case of symmetric theory ($l_{ab}=h_{ab}$, $g_{\u a\u b}=h\gds
o,\u ab,$) one gets
\beq2.134
M^2_{ij}=\frac{\a^2_s}{\hbar c}\,\frac1{V_2} \int_M\sqrt{|\wt g|}\,d^{n_1}x
\,\bigl\{h_{bn}h^{o\u m\u p} B\gds b,\u pi,B\gds n,\u mj,\bigr\}.
\e

The Higgs' potential is given by
\beq2.135
\bal
\wh V(\F)&=\frac1{V_2}\int_M\sqrt{|\wt g|}\,d^{n_1}x\biggl\{g^{\u w\u p}g^{\u m\u q}
\Bigl[h_{nk}H\gds n,\u w\u m,+2\z h_{nk}H\gds n,\u d\u w,k\gds o\u d,\u m,\\
&+\mu\z\Bigl(2k_{nk}H\gds n,\u d\u w,k\gds o\u d,\u m,
+\z\bigl(-2k_{kd}H\gds d,\u d\u a,k\gds o\u d,\u w,k\gds o\u a,\u m,\\
&-k_{kd}H\gds d,\u l\u w,k\gds o,\u m\u a,k^{o\u l\u a}+k_{kd}H\gds d,\u l\u m,
k^{o\u l\u a}k\gds o,\u w\u a,\bigr)\Bigr)\\
&+\mu^2\z\Bigl(k\dgs n,b,k_{bd}H\gds d,\u a\u w,k\gds o,\u m,{}^{\u a}
-k\dgs k,b,k_{bd}H\gds d,\u a\u m,k\gds o,\u w,{}^{\u a}
+\z\bigl(k_{nk}k\gds n,d,H\gds d,\u l\u m,k\gds o,\u w,{}^{\u r}k\gds \u l,\u r,\\
&-2k_{nk}k\gds n,d,H\gds d,\u d\u a,k\gds o\u d,\u w,k\gds o\u a,\u m,
-k_{nk}k\gds n,d,H\gds d,\u l\u w,k\gds o,\u m,{}^{\u r}k\gds \u d,\u r,
\bigr)\Bigr)\\
&+\mu^3\z \bigl(k_{nk}k^{nb}k_{bd}H\gds d,\u a\u w,k\gds o,\u m,{}^{\u a}
-k_{nk}k^{nb}k_{bd}H\gds d,\u a\u m,k\gds o,\u w,{}^{\u a}\bigr)\Bigr]
\cdot H\gds k,\u p\u q,\\
&-2h_{cd}\bigl(H\gds c,\u p\u q,\gtk{\u p\u q}\bigr)
\bigl(H\gds d,\u a\u b,\gtk{\u a\u b}\bigr)\biggr\}
\eal
\e
or
\beq2.136
\bal
\wh V(\F)&=\frac1{V_2}\int_M\sqrt{|\wt g|}\,d^{n_1}x
\Bigl(P\dgs kl,[\u p\u q][\u a\u b],H\gds k,\u p\u q,H\gds l,\u a\u b,
-2h_{kl}\bigl(H\gds k,\u p\u q,\gtk{\u p\u q}\bigr)\bigl(H\gds l,\u a\u b,
\gtk{\u a\u b}\bigr)\Bigr)\\
&=\frac1{V_2}\int_M\sqrt{|\wt g|}\,d^{n_1}x \, Q\dgs sk,[\u c\u d][\u p\u q],
H\gds s,\u c\u d,H\gds k,\u p\u q, \\
&Q\dgs sk,[\u c\u d][\u p\u q],=Q\dgs ks,[\u p\u q][\u c\u d],=
-Q\dgs sk,[\u d\u c][\u p\u q],=-Q\dgs sk,[\u c\u d][\u q\u p],=
Q\dgs sk,[\u d\u c][\u q\u p],
\eal
\e
\beq2.137
\bal
P\dgs sk,[\u c\u d][\u p\u q],
&=g^{[\u c\ts[\u p}g^{\u d]\u q\ts]}h_{sk}-2\z h_{sk}k\gds o[\u d,|\u e|,g^{\u c
][\u p}g^{|\u e|\u q]}\\
&+\mu\z\Bigl(-2k_{sk}k\gds o[\u d,|\u e|,g^{\u c][\u p}g^{|\u e|\u q]}
+\z\bigl(2k_{sk}k\gds o[\u c,|\u e|,k\gds o\u d],\u f,g^{\u e[\u p}
g^{|\u f|\u q]}\\
&-k_{sk}k\gds o,\u e\u a,k^{o[\u d|\u a|}g^{\u c][\u p}g^{|\u e|\u q]}
-k_{sk}k^{o[\u c|\u a|}g^{\u d][\u q}g^{|\u e|\u p]}k\gds o,\u e\u a,\bigr)\Bigr)\\
&+\mu^2\z\Bigl(-k_{bs}k\dgs k,b,k\gds o[\u d,|\u a|,g^{\u c][\u p}
g^{|\u a|\u q]}
-k_{bs}k\dgs k,b,k\gds o,\u a,{}^{[\u c}g^{|\u a|\ts[\u p}g^{\u d]\u q\ts]}\\
&+\z\bigl(k\gds n,s,k_{nk}k\gds o,\u a,{}^{\u r}k\gds o[\u c,|\u r|,
g^{\u a\ts[\u p}g^{\u d]\u q\ts]}
-2k\gds n,s,k_{nk}k\gds o[\u c,|\u e|,k\gds o\u d],\u f,g^{\u e[\u p}g^{|\u f|
\u q]}\bigr)\Bigr)\\
&+\mu^3 \z\bigl(-k_{bs}k^{nb}k_{nk}k\gds o,\u e,{}^{[\u d}g^{\u c][\u p}
g^{|\u e|\u q]}
-k_{bs}k^{nb}k_{ns}k\gds o,\u e,{}^{[\u c}g^{|\u e|\ts[\u p}g^{\u d]\u q\ts]}\bigr)\\
&+\mu^2g^{[\u c\ts[\u p}g^{\u d]\u q\ts]}k\gds p,s,k_{pk}\\
Q\dgs sk,[\u c\u d][\u p\u q],&=P\dgs sk,[\u c\u d][\u p\u q],
-2h_{sk}\gtk{\u c\u d}\gtk{\u p\u q}
\eal
\e

One can decompose the Higgs field into independent components in the following way
\beq2.138
\F^c_{\u b}=\wt\F{}^c_{\u b}=\sum_{(\ul n{}_i,\ul n{}'_j)}
\Bigl(\F_{\ul m{}_j}^{(\ul n{}_i,\ul n{}'_j)}\Bigr)_{\u b}^c
\e
where
\bea2.139
\Ad_G\bigr|_{G_0}&=\sum_i\oplus \ul n{}_i\oplus\Ad_{G_0} \\
\Ad_H\bigr|_{G_0\oplus G}&=\sum_j\oplus\bigl(\ul n{}'_j\otimes \ul m{}_j
\bigr). \label{2.140}
\e
$\ul n{}_i$ are irreducible representations of $G_0$ and $\ul m{}_j$ are
irreducible representations of~$G$. In the sum \er{2.138} $\ul n{}_i$ and $\ul
n{}'_j$ are identical representations of~$G_0$ from \er{2.139} and \er{2.140}
decomposition. In this way a constraint is satisfied identically and we
can put $\wt\F{}_{\u b}^c$ given by \er{2.138} into \er{2.127}--\er{2.128}, \er{2.129}--\er{2.136}. Thus the
analysis of a mass spectrum in the theory can be simplified (a~little).

For $k=0$, $\F^0\dr{crt}$, $H\gds k,\u p\u q,=0$ one gets the following
matrix for Higgs' bosons
\beq2.141
\bal
m\gds 2\u h,f,{}\gds \u e,a,&=\frac{-1}{V_2}\int_M\bigl\{
8Q\dgs sk,[\u e\u a][\u h\u q],C\gds s,ac,C\gds k,ef,
(\F^0\dr{crt})\gds c,\u a,(\F^0\dr{crt})\gds e,\u q,\\
&-2Q\dgs as,[\u p\u q][\u h\u a],f\gds \u e,\u p\u q,
C\gds s,ef,(\F^0\dr{crt})\gds e,\u a,
+4Q\dgs sf,[\u e\u a][\u p\u q],f\gds \u h,\u p\u q,
C\gds s,ea,(\F^0\dr{crt})\gds a,\u a,\\
&+Q\dgs af,[\u c\u d][\u p\u q],f\gds e,\u c\u d,f\gds \u h,\u p\u q,
\bigr\}\sqrt{|\u g|}\,d^{n_1}x
\eal
\e

For $k=1$, $\F^1\dr{crt}$, $H\gds k,\u p\u q,\ne0$ and $\F^1\dr{crt}$ (if
exists) satisfies the following equation:
\beq2.142
2Q\dgs sk,[\u e\u a][\u p\u q],C\gds s,ac,
(\F^1\dr{crt})\gds c,\u a,=Q\dgs ak,[\u c\u d][\u p\u q],f\gds \u e,\u c\u d,
\e
and a supplementary condition
\beq2.143
\F_{\u b}^c f_{\hi\u d}^{\u b}-\mu_{\hi}^a \F_{\u a}^b C_{ab}^c=0.
\e

A mass matrix for Higgs' bosons looks like
\bea2.144
m\gds 2\u h,f,{}\gds \u e,a,&=\frac{-1}{V_2}\int_M
\bigl(4Q\dgs sk,[\u e\u h][\u p\u q],
H\gds k,\u p\u q,(\F^1\dr{crt})C\gds s,af,\bigr)\sqrt{|\u g|}\,d^{n_1}x.\\
H_{\u m\u n}^b(\F\dr{crt}^k)&=C_{cd}^b
(\F\dr{crt}^k)_{\u n}^c(\F\dr{crt}^k)_{\u m}^d
-\mu_{\hi}^b f_{\u n\u m}^{\hi}
-(\F\dr{crt}^k)_{\u c}^b f_{\u n\u m}^{\u c}. \label{2.145}\\
H\gds b,\u m\u n,(\F\dr{crt}^0)&=0 \label{2.146}
\e

Finally, let us give a formula for a \lg\ of an electromagnetic field in the
NKK.
\beq2.147
\cL\dr{em}=\frac1{8\pi}\Bigl(2\bigl(\gtk{\mu\nu}F_{\mu\nu}\bigr)^2-
\bigl(g^{\mu\a}g^{\nu\b}-g^{\nu\b}\gtn{\mu\a}+g^{\nu\b}g^{\mu\om}
\gtn{\tau\a}g_{\om\tau}\bigr)F_{\a\b}F_{\mu\nu}\Bigr)
\e

In order to calculate \co ical terms in the theory it is necessary to know
Einstein--Kaufmann \cn s on a group~$G$ and on a homogeneous manifold
$M=G/G_0$. Let us give those \cn s. On a group~$G$ a right invariant
Einstein--Kaufmann \cn\ reads:
\bml2.148
\G_{wm}^n=-\tfrac12\,C_{wm}^n +\tfrac12\bigl(K\dgs wm,n,-2\mu^2k\dgs [m,a,
K_{w]ab}k^{nb}\bigr)\\
{}+h^{me}\Bigl\{\mu K\dgs e(w,a,k_{m)a}+\mu^2k\dgs c,b,\bigl[k\dgs (m,c, k\dgs w,c, K_{w)ab}
k\dgs e,a,-K_{eab}k\dgs(w,a,k\dgs m),c,\bigr]\Bigr\}
\e
where
\beq2.149
K_{abc}=-\mu \bigl(\wt\nabla_a k_{bc}-\wt\nabla_b k_{ca}+\wt\nabla_c k_{ab}
\bigr).
\e
$\wt\nabla_a$ means a Riemannian covariant derivative on a Lie-semisimple
group $G$ \wrt a biinvariant Killing tensor $h_{ab}$.

One gets
\beq2.150
\wt\nabla k_{bc}=-\tfrac12\bigr(C_{bc}^f k_{fc}+C_{ca}^f k_{bf}\bigr)
\e
and
\beq2.151
K_{abc}=\mu\bigl(C_{ba}^f k_{fc}+C_{ac}^fk_{fb}-C_{bc}^f k_{fa}\bigr).
\e
Let us remind that $k_{ab}$ is a right-invariant anti\sy ic tensor on~$G$.

If we write a \cn\ $\G$ on~$G$ in the form
\beq2.152
\G_{wm}^n = -\tfrac12\, C_{wm}^n +u\gds n,wn,,
\e
one gets
\beq2.153
R_{bd}=\wt R_{bd}+\wt\nabla_a u\gds a,bd,-\wt\nabla_du\gds a,ba,
+\tfrac12\bigl(\wt\nabla_b u\gds a,ad,-\wt\nabla_d u\gds a,ab,\bigr)
\e
where
\bg2.154
\wt\nabla_a u\gds c,ed,=-\tfrac12 \bigl(C_{ea}^fu\gds c,fd,
+C_{da}^f u\gds c,ef,-C_{fa}^c u\gds f,ed,\bigr)\\
\bal
u\gds n,wm,&=\tfrac12\,\mu\bigl(L\dgs wm,n,-2\mu k\dgs [m,a,L_{w]ab}
k^{nb}\bigr)\\
&-\mu^2 h^{ne}\bigl(L\dgs e(w,a, k_{m)a}+\mu^2 k\dgs c,b,
\bigl[k\dgs (m,c,L_{w)ab}k\dgs e,a,-L_{eb}k\dgs(w,a,
k\dgs m), c,\bigr]\bigr)
\eal \label{2.155}
\e
where
\beq2.156
L_{abc}=C_{ba}^f k_{fc}+C_{ac}^fk_{fb}-C_{bc}^fk_{fa}.
\e

Finally
\beq2.157
\bga
R_{bd}=\wt R_{bd}-\tfrac12\,C_{db}^f(u\gds a,af,+u\gds a,fa,)
-\tfrac14\,C_{fa}^a(2u\gds f,ba,+u\gds f,ab,)+\tfrac14\,(C_{fb}^au\gds f,ad,
-C_{bd}^fu\gds a,fd,), \\
\wt R_{bd}=-\tfrac14\,h_{bd}. \ega
\e
For example, for $G=\text{SO}(3)$,
\beq2.158
\wt R_{bd}(\text{SO}(3))=\tfrac12\,\d_{bd}.
\e

If
\bea2.159
{}&k_{ab}=C_{ab}^f V_f  \\
&\wh \nabla_k V_f=0 \label{2.160}
\e
one gets
\beq2.161
\bal
u\gds n,wm,&=\frac\mu2\,C\gds sn,w,C\gds p,ms,V_p-\mu^3C\gds p,as,V_p
C^{rnb}V_rC^q{}\dgs[m,a,V_qC\gds s,bw],\\
&{}+\mu^4C^f{}\dgs c,b,V_f \bigl[C^p{}\dgs(m,c,V_pC\gds s,bw),
C\gds q,as,V_qC^{rna}V_r \\
&\quad{}- C^s{}\dgs b,n,C\gds p,as,V_p C^q{}\dgs (w,a,
V_q C\gds r,m),V_r\bigr].
\eal
\e

Let us remind to the reader that
\beq2.162
\wt\nabla_kV_f=-\tfrac12\,C\gds f,fk,V_e.
\e
$u\gds n,wm,$ can be calculated explicitly in a general form. One gets
\bg2.163
\bal
u\gds n,wm,&=\tfrac12\,\mu\bigl(C\gds f,mw,k\dgs f,n,+C^f{}\dgs w,n,
k_{fm}-C^f{}\dgs m,n,k_{fw}\bigr)\\
&-\tfrac12\,\mu^2\Bigl[C\gds f,bw,k_{fa}(k^{na}k\dgs m,b,-k^{nb}k\dgs
m,a,)+C\gds f,mb,k_{fa}(k^{na}k\dgs w,b,-k^{nb}k\dgs w,a,)\\
&-2k^{nb}k\dgs m,a,C\gds f,ab,k_{fw}-\bigl(k\dgs f,a,k_{mc}C^f{}\dgs
w,n,+2k_{fm}C^{anf}k_{wa}\\
& - C^f{}\dgs w, a, k\dgs f,a,k_{ma}
+C^f{}\dgs m,n,k\dgs f, a,k_{wa}-C^f{}\dgs m,a,k\dgs f,a,
k_{wa}\bigr)\Bigr]\\
&+\tfrac12\,\mu^4\Bigl[3k\dgs c,b,C\gds f,ab,k\dgs f,n,k\dgs w,a,
k\dgs m,a,+C\gds f,bw,k_{fa}k\dgs m,c,(k\dgs c,a,k^{nb}
-k\dgs c,b,k^{na})\\
&+C\gds f,bm,k_{fa}k\dgs w,a,(k\dgs c,a,k^{nb}-k\dgs c,b,k^{na})
+C\gds f,ab,k\dgs c,b,k\dgs w,c,k\gds a, m,k\gds n, f,\\
&+C\gds fn,b,k_{fa}k\dgs m,c,(k\dgs c,a,k\dgs w,b,-k\dgs c,b,
k\dgs w,a,)\\
&+C\gds an,f,k\dgs b,f,(k\dgs c,b,k_{ma}
k\dgs w,c,-k\dgs w,b,k\dgs m,c,k_{ac})\Bigr]
\eal \\
R_G=l^{ab}R_{ab}. \label{2.164}
\e
In the case of the Einstein--Kaufmann \cn\ on a $M=G/G_0$ manifold one gets
\bml2.165
\wh{\G}{}\gds \u n,\u w\u m,=\left\{
\begin{matrix} \wt n\\\wt w\wt m\end{matrix}\right\}
+\frac12\bigl(K\dgs\u w\u m,\u n,-2\wt g\dgs[\u m,\u a],K_{\u w]\u a\u b}
\wt g_[{}\gds \u n\u b,],\bigr)\\
{}+h^{0\u n\u e}\left\{K_{\u e}{}\gds \u a,(\u w, \wt g_{|\u m|\u a)}
+\wt g{}_{[\u c}{}\gds \u b,], \bigl[\wt g\dgs [(|\u m|,\u c,{}_]K_{\u w)\u a\u b}
\wt g_{[\u e}{}\gds \u a,],-K_{\u c\u a\u b}\wt g_{[(\u w}{}\gds \u a,],
\wt g_{[\u m)}{}\gds\u c,],\bigr]\right\}
\e
where
\beq2.166
K_{\u a\u b\u c}=-\wt\nabla_{\u a}\wt g_{[\u b\u c]}-\wt\nabla_{\u b}
\wt g\ink{\u c\u a}+\wt\nabla_{\u c}\wt g\ink{\u a\u b}=
\z\bigl(-\wt\nabla_{\u a}k_{\u b\u c}^0-\wt\nabla_{\u b}k_{\u c\u a}^0
+\wt\nabla_{\u c}k_{\u a\u b}^0\bigr)
\e
where $\wt\nabla$ means a covariant derivative \wrt the Riemannian \cn\ on a
manifold $M=G/G$ with a left-invariant metric tensor $h_{\u a\u b}^0$,
$\left\{\begin{matrix} \u a\\\u b\u c\end{matrix}\right\}$ mean Christoffel symbols
built from $h_{\u a\u b}^0$. In this way a \co ical term reads
\beq2.167
\ul P=\frac1{V_1}\int_M \sqrt{|\wt g|}\, \wh{R}(\wh{\G})\,d^{n_1}x
\e
where
\bg2.168
\wh{R}(\wh{\G})=g^{\u a\u b}\wh{R}_{\u a\u b}(\wh{\G}) \\
\wh{R}_{\u b\u d}=\wt R_{\u b\u d}+\wt\nabla_{\u a}u\gds \u a,\u b\u d,
-\wt\nabla_{\u d}u\gds\u a,\u b\u a,+\tfrac12\,\bigl(\wt\nabla_{\u b}
u\gds\u a,\u a\u d,-\wt\nabla_{\u d}u\gds\u a,\u a\u b,\bigr)\label{2.169}
\e
where $\wt R_{\u b\u d}$ is a Ricci tensor for a Riemannian \cn\ on~$M$ formed
from $g_{(\u a\u b)}=h_{\u a\u b}^0$.

Let us calculate $\wt R\SU(3)$. Using results from Ref.~\cite3 one gets
\beq 2.75
\wt R\SU(3) = 2\ell^{\SU(3)bk} \gd C,e,ka,\gd\wt\G,a,be, + \ell^{\SU(3)bk}
\gd C,e,ba,\gd\wt\G,a,ek, + \ell^{\SU(3)bk}\gd\wt\G,a,fk, \gd\wt\G,f,ba,
\e
where $\ell^{\SU(3)bk}$ is an inverse tensor of a \nos\ tensor on $\SU(3)$
\beq 2.76
\gathered
\ell^{\SU(3)bk}\gd\ell,\SU(3),bi, =\ell^{\SU(3)kb} \gd\ell,\SU(3),ib, =\d^k_i,\\
\gd\ell,\SU(3),ij, = \dg h,ij,\SU(3), + \mu \gd k,\SU(3),ij,,
\endgathered
\e
$\dg h,ij,\SU(3),$ is a \KC tensor on $\SU(3)$ and $\gd k,\SU(3),ij,$ is a
right-invariant tensor on the group $\SU(3)$. One gets
\beq 2.77
\gathered
\gd\ell,\SU(3),db, \gd\wt \G,d,ac, + \gd\ell,\SU(3),ad,\gd\wt\G,d,bc,
= -\gd\ell,\SU(3),db, \gd C,d,bc,\\
\gd\wt\G ,b,ac, = -\gd\wt\G,b,ca, \q \gd\wt\G,d,ad, =0.
\endgathered
\e
$\gd\wt\G,a,bc,$ is a \nos\ Einstein--Kaufmann \cn\ compatible with
$\gd\ell,\SU(3),ij,$ on the group $\SU(3)$, right-invariant \wrt $\SU(3)$
action. ($U(g)$ is an adjoint \rp ation of $\SU(3)$.)

It is easy to see that
\beq 2.78
\gd\wt\G,a,bc, (g) = \gd\wt \G,a',b'c',(e) U^a_{a'}(g^{-1})U^{b'}_b(g) U^c_{c'}(g),
\e
$g\in\SU(3)$, $e$~is a unit \el\ of $\SU(3)$. We also have $\gd\ell,\SU(3),ab,
(g) = U^{a'}_aU^{b'}_b(g)\gd\ell,\SU(3),ab,(e)$. Let us notice that $\SU(2)
\subset \SU(3)$ and $\SU(2)$ is locally isomorphic to $\SO(3)$ (their Lie
algebras are iso\mo c). On $\SO(3)$ we
have a skew\nos\ tensor $\gd k,\SO(3),ab, = {\ve_{abc}\cdot V_c}$ right-invariant
\wrt an action of a group $\SO(3)$ where
\beq 2.79
\aligned
V_1(\a,\b,\g) &= \sin\a \sin\b \\
V_2(\a,\b,\g) &= - \sin\a \cos\b \\
V_3(\a,\b,\g) &= \cos\a
\endaligned
\e
where $0\le\a\le\pi$, $0\le\b\le2\pi$, $0\le\g\le 2\pi$ are usual Euler angles
\beq 2.80
\ov V_1=V_1(0,0,0)=0, \q \ov V_2 = V_2(0,0,0)=0, \q \ov V_3 = V_3(0,0,0) =1.
\e

In terms of Euler angles we have the \fw\ \rp ation of $\SO(3)$ (locally isomorphic to $\SU(2)$) Lie algebra:
\beq 2.86
\aligned
& \wh L_x = -i\biggl(\sin\a \,\pp{}\b + \cot\b \cos\a \,\pp{}\a + \frac{\cos\a}{\cos\b}\,\pp{}\g\biggr) \\
& \wh L_y = -i\biggl(\cos\a \,\pp{}\b - \cot\b \sin\a \,\pp{}\a - \frac{\sin\a}{\sin\b}\,\pp{}\g\biggr) \\
& \wh L_z = -i\,\pp{}\a \\
& [\wh L_x,\wh L_y] = i\wh L_z, \q\ [\wh L_x,\wh L_z] = [\wh L_y,\wh L_z] =0.
\endaligned
\e
Let us define skew\nos\ tensor (right-invariant \wrt $\SU(3)$ group action)
in the \fw\ way:
\beq 2.81
\aligned
\gd k,\SU(3),ij, &= \ve_{ijk}V_k \qh{for} i,j,k=1,2,3,\\
\gd k,\SU(3),ij, &= 0 \q\hbox{in the remaining cases.}
\endaligned
\e
$\ve_{abc}$ is a usual anti\sy ic tensor \st $\ve_{123}=1$.

Let us consider the \fw\ realization of a Lie algebra of $\SU(3)$ ($A_2$):
\beq 2.82
[\la_i,\la_j] = 2i f_{ijk}\la_k = \gd C,k,ij, \la_k, \q i,j,k= 1,2,3,\dots,8
\e
\st (see Ref.~\cite{mcm})
$$
[\la_1,\la_2] = 2i \ve_{123}\la_3 = \gd C,3,12,\la_3
$$
is a subalgebra of Lie algebra of $\SU(3)$ ($A_1$) (a Lie algebra of $\SU(2)$).

In this way
\beq 2.83
\gd\ell,\SU(3),ij, = \gd h,\SU(3),ij, + \mu \gd k,\SU(3),ij,.
\e
One gets
\beq 2.84
\wt R\SU(3)(g) = \wt R\SU(3)(e) = \wt R\SU(3)= -\frac54
+\frac{2(2\mu^3+7\mu^2+25\mu+20)}{(\mu^2+4)^2}
\e
where
\beq2.85
\wt R\SU(2) = \frac{2(2\mu^3+7\mu^2+25\mu+20)}{(\mu^2+4)^2}
\e
(see Ref.~\cite3).

\allowdisplaybreaks
\section{Two stages of a \sb\ and its hierarchy with some \gn s}

Let us incorporate in our scheme two stages of a symmetry breaking and its hierarchy. In order
to do this let us consider a case of the manifold
\begin{equation}
M\cong M_0\times M_1 \eq1
\end{equation}
where $\cong$ is a diffeomorphic equivalence,
\ealn
\dim M_0& = \ov n_0, \quad \dim M_1= \ov n_1,\\
\dim M&= \ov n_0 + \ov n_1, \eq2 \\
M_0&=G_{1}/G_0\,, \quad M_1=G_2/G_1\,. \eq3
\end{align}
Every manifold $M_0,M_1$ is a manifold of vacuum states if the \s y is breaking from
$G_{1}$ to~$G_0$, or $G_2$ to $G_1$.

Thus
\begin{equation}
G_0\subset G_1\subset G_2=G. \eq4
\end{equation}
We will consider the situation when
\begin{equation}
M= G/G_0. \eq5
\end{equation}
This is a constraint in the theory. From \ER(4) one gets
\begin{equation}
\gG_0\subset \gG_1\subset \gG_2 =\gG \eq6
\end{equation}
and
\begin{equation}
\gG_{1}=\gG_0 \dpl \gM_0, \quad \gG_{2}=\gG_1 \dpl \gM_1. \eq7
\end{equation}
The relations \ER(1) and \ER(5) mean that there is a diffeomorphism $g$ onto $G/G_0$
such that
\begin{equation}
g: M_0\times M_1 = G_{1}/G_0 \times G/G_1 \to G/G_0=M. \eq8
\end{equation}
This diffeomorphism is a deformation of a product \ER(1) in $G/G_0$. The
theory has been constructed (in Section~2) for the case considered before with $G_0$
and~$G$. The multiplet of Higgs' fields $\mPhi$ breaks the \s y from $G$
to~$G_0$ (equivalently from $G$ to~$G_0'$ in the false vacuum case).
$\gG_0,\gG_1$ mean Lie algebras for groups $G_0,G_1$ and $\gM_0,\gM_1$ complements in a
decomposition~\ER(7). On every manifold~$M_0,M_1$ we introduce radii $r_0,r_1$
(a~``size'' of a manifold) in such a
way that $r_0>r_{1}$. On the manifold $G/G_0$ we define the radius~$r$ as
before. The diffeomorphism $g$ induces a contragradient transformation for a
Higgs field $\mPhi$ in such a way that
\begin{equation}
g^\ast \mPhi=(\mPhi_0,\mPhi_1). \eq9
\end{equation}

In this way we get the following decomposition for a kinetic part of the
field~$\mPhi$ and for a potential of this field:
\ealn
\cL_{\textrm {kin}}(\gag\mPhi)&= \cL_{\text{\rm kin}}^0(\gag\mPhi_0) + \cL_{\text{\rm kin}}^1(\gag\mPhi_1) \eq10 \\
\wh V(\mPhi)&=V^0(\mPhi_0) + V^1(\mPhi_1)\eq11
\end{align}
where
\begin{equation}
\bal
V^i(\mPhi_i)&=\frac{l_{ab}}{V_i} \intop_{M_i} \sqrt{|\wt g_i|}\,d^{\bar n_i}x
\biggl[2g^{[\u m_i\u n_i]} \bigl(C^a_{cd}\mPhi^c_{i\u m_i}\mPhi^d_{i\u n_i}
-\mu^a_{\hi_i}f^{\hi_i}_{\u m_i\u n_i}-\mPhi^a_{i\u e_i}
f^{\u e_i}_{\u m_i\u n_i}\bigr)\\
&\times g^{[\u a_i\u b_i]}
\bigl(C^b_{ef}\mPhi^e_{i\u a_i}\mPhi^f_{i\u b_i}
-\mu^b_{\hj_i} f^{\hj_i}_{\u a_i\u b_i}-\mPhi^b_{i\u a_i}
f^{\u d_i}_{\u a_i\u b_i}\bigr)\\
&-g^{\u a_i\u m_i}_i g^{\u b_i\u n_i}_i L^a_{\u a_i\u b_i}
\bigl(C^b_{cd}\mPhi^c_{i\u m_i}\mPhi^d_{\u n_i}
-\mu^b_{\hi_i}f^{\hi_i}_{\u m_i\u n_i}-\mPhi^b_{\u e_i}f^{\u e_i}
_{\u m_i\u n_i}\bigr)\biggr],
\eal \eq15
\end{equation}
$f^{\hj_i}_{\u a_i\u b_i}$ are structure \ct s of the Lie algebra
$\gG_i$, $i=0,1$,
\ealn
V_i&=\intop_{M_i}\sqrt{|\wt g_i|}\,d^{\bar n_i}x, \quad i=0,1, \eq13 \\
\wt g_i&=\det(g_{i\u b_i\u a_i}), \quad i=0,1. \eq14
\end{align}
$g_{i\u b_i\u a_i}$ is a non\s ic tensor on a manifold~$M_i$, $i=0,1$.

The scheme of the \s y breaking acts as follows from the group $G_{2}$
to~$G_1$ ($G_1'$) (if the \s y has been broken up to $G_{2}$). The potential
$V^0(\mPhi_0)$ has a minimum (global or local) for $\mPhi^k_{0\rm crt}$,
$k=0,1$. The value of the remaining part of the sum \ER(11) for the field
$\mPhi_1$, is small for the scale of energy is much lower ($r_0>r_1$).
Thus the minimum of $V^0(\mPhi_0)$ is an approximate minimum of the
remaining part of the sum \ER(11). In this way we have a
truncation of the Higgs potential. This gives in principle a pattern of a
\s y breaking. However, this is only an approximate \s y breaking. The real
\s y breaking is from $G$ to $G_0$ (or to $G_0'$ in a false vacuum case). The
important point here is the diffeomorphism~$g$.
\ealn
g^\ast \mPhi^b&=\bigl(\mPhi^b_0,\mPhi^b_1\bigr) \eq16 \\
\mPhi^b_i&=\mPhi^b_{i\u a_i}\wt \theta^{\u a_i},
\quad i=0,1. \eq 17
\end{align}

The shape of $g$ is a true indicator of a reality of the \s y breaking
pattern. If
\begin{equation}
g=\Id+\d g \eq18
\end{equation}
where $\d g$ is in some sense small and $\Id$ is an identity, the sums
\ER(10)--\ER(11) are close to Eqs \eqref{3.79}--\eqref{3.80}. The smallness of $\d
g$ is a criterion of a practical application of the \s y breaking pattern
\ER(4). For example we can define a norm in a space of~$g$ and $\|\d g\|\ll 1$.
It seems that there are a lot of possibilities for the condition
\ER(8). Moreover, a smallness of $\d g$ plus some natural conditions for
groups~$G_1, G_2=G$ can narrow looking for grand unified models. Let us notice that
the decomposition of~$M$ results in decomposition of \co ical terms (see Eq.~\eqref{3.83})
\begin{equation}
\ul{\wt P}=\ul{\wt P}_0 + \ul{\wt P}_1\eq19
\end{equation}
where
\begin{equation}
\ul{\wt P}_i=\frac1{r^2_iV_i}\intop_{M_i}\sqrt{|\wt g_i|}\,
\wh{\ov R}_i(\wh{\ov\G}_i)\,d^{n_i}x, \quad i=0,1, \eq20
\end{equation}
where $\wh{\ov \G}_i$, $i=0,1$, is a non\s ic connection on $M_i$, $i=0,1$, compatible with the
non\s ic tensor $g_{i\u a_i\u b_i}$, $i=0,1$, and $\wh{\ov R}_i(\wh{\ov \G}_i)$, $i=0,1$, its
curvature scalar.
The truncation procedure can be proceeded in several ways. Finally let us
notice that the energy scale of broken gauge bosons is fixed by
radii~$r_0,r_1$ at any stage of the \s y breaking in our scheme.

Let us consider Eq.\ \ER(9) in more details. One gets
\begin{equation}
A^{\u a}_{i\u a_i}(y)\mPhi^b_{\u a}(y)=
\mPhi^b_{\u a_i}(y_i), \quad y\in M,\ y_i\in M_i,\quad i=0,1, \eq21
\end{equation}
where
\begin{equation}
g^\ast(y)=\biggl(A_0\biggm|A_1\biggr), \eq22
\end{equation}
\begin{equation}
A_i=\bigl(A^{\u a}_{i\u a_i}\bigr)_{\u a=1,2,\dots,n_1,\,\u a_i=1,2,\dots,\bar
n_i}\,,\ i=0,1, \eq23
\end{equation}
is a matrix of Higgs' fields transformation.

According to our assumptions one gets also:
\begin{equation}
\Bigl(\frac{r_i}r\Bigr)^2 g_{i\u a_i\u b_i}(y_i)=
A^{\u a}_{\u a_i}(y)A^{\u b}_{\u b_i}(y)g_{\u a\u b}(y),\quad i=0,1. \eq24
\end{equation}
For $g$ is an invertible map we have $\det g^\ast(y)\ne0$.

We have also
\begin{equation}
n_1= \ov n_0 + \ov n_1 \eq25
\end{equation}
and
\begin{equation}
\mPhi^b_{\u a}(y)= \wt A_{0\u a}^{\u a_0}(y)\mPhi^b_{\u a_0}(y_0) +
\wt A_{1\u a}^{\u a_1}(y)\mPhi^b_{\u a_1}(y_1) \eq26
\end{equation}
or
\begin{gather}
{g^\ast}^{-1}(y)=\left(\vcenter{\offinterlineskip
\hbox to40pt{\hfil$\wt A_0$\hfil \vrule height10pt depth4pt width0pt}
\hrule width40pt
\hbox to40pt{\hfil$\wt A_1$\hfil \vrule height10pt depth4pt width0pt}
}\right) \eq27 \\
\wt A_i=\bigl(\wt A^{\u a_i}_{i\u a}\bigr)_{\u a_i=1,2,\dots,\bar n_i,\,\u a=1,2,\dots,
n_1}, \quad i=0,1, \eq28
\end{gather}
such that
\ealn
&g(y_0,y_{1})=y \eq29 \\
&(y_0,y_1)=g^{-1}(y) \eq29a
\end{align}

For an inverse tensor $g^{\u a\u b}$ one easily gets
\begin{equation}
\Bigl(\frac{1}{r^2}\Bigr)g^{\u a\u b}=
\Bigl(\frac{1}{r_0^2}\Bigr) \wt A^{\u a}_{0\u a_0} g^{\u a_0\u b_0}_0
A^{\u b}_{0\u b_0} +
\Bigl(\frac{1}{r_1^2}\Bigr) \wt A^{\u a}_{1\u a_1} g^{\u a_1\u b_1}_1
A^{\u b}_{1\u b_1}\,. \eq30
\end{equation}
We have
\begin{equation}
r^{2n_1}\det(g_{\u a\u b})=r^{2\bar n_0}_0\cdot
r^{2\bar n_1}_1\cdot \det(g_{0\u a_0\u b_0})\cdot \det(g_{1\u a_1\u b_1}). \eq31
\end{equation}
In this way we have for the measure
\begin{equation}
d\mu(y)=d\mu_0(y_0)\, d\mu_1(y_1) \eq32
\end{equation}
where
\ealn
d\mu(y)&=\sqrt{\det g}\,r^{n_1}\,d^{n_1}y \eq33 \\
d\mu_i(y_i)&=\sqrt{\det g_i}\,r^{\bar n_i}_i\,d^{\bar n_i}y_i\,, \quad i=0,1. \eq34
\end{align}
In the case of $\cL\dr{int}(\mPhi,\wt A)$ one gets (see Eq.~\eqref{3.81})
\begin{equation}
\cL\dr{int}(\mPhi,\wt A)= \cL\dr{int}(\mPhi_0,\wt A) + \cL\dr{int}(\mPhi_1,\wt A)\eq35
\end{equation}
where
\begin{equation}
\cL\dr{int}(\mPhi_i,\wt A)=h_{ab}\mu^a_{\ul i}\wt H^{\ul i}\ul g^{[\u a_i\u b_i]}
_i \bigl(C^b_{cd}\mPhi^c_{i\u a_i}\mPhi^d_{i\u b_i}
-\mu^b_{\hi}f^{\hi}_{\u a_i\u b_i}-\mPhi^b_{\u d_i}f^{\u d_i}_{\u a_i\u b_i}\bigr),
\quad i=0,1,
\eq36
\end{equation}
where
\begin{equation}
\ul g^{[\u a_i\u b_i]}_i = \frac 1{V_i}\intop_{M_i} \sqrt{|\wt g_i|}\,d^{\bar n_i}x\,
g^{[\u a_i\u b_i]}_i, \quad i=0,1. \eq37
\end{equation}
Moreover, to be in line in the full theory we should consider
groups $H_0,H_1$ in such a way that
\begin{equation}
H_0\subset H_1=H. \eq38
\end{equation}
We have
\begin{equation}
G_0\subset H_0, \quad G_1\subset H_1. \eq39
\end{equation}
We should have
\begin{equation}
G_0 \otimes G_{1}\subset H_0, \eq40
\end{equation}
$G_1$ is a centralizer of $G_0$ in $H_0$.

In a Standard Model we have only one stage of the \sb.
Thus we have
\begin{equation}\label{41}
\aligned
G_0&=U(1)\dr{em} \otimes SU(3)\dr{c},\\
G_1&=SU(2)\dr{L} \otimes U(1)\dr{Y} \otimes SU(3)\dr{c}.
\endaligned
\end{equation}
This is known from the Standard Model.
$G2$ is a group which plays the role of $H$ in the case of a \s y
breaking from $SU(2)\dr{L}\otimes U(1)\dr{Y}$ to $U(1)\dr{em}$. However, in this
case because of a factor $U(1)$, $M=S^2$. Thus $M_0=S^2$ and $G2\subset H_0$.

In this approach we try to consider additional dimensions connecting to the
manifold~$M$ more seriously, i.e.\ as physical dimensions, additional space-like
dimensions. We remind to the reader that gauge-dimensions connecting to the
group~$H$ have different meaning. They are dimensions connected to local
gauge \s ies and they cannot be directly observed. Simply saying
we cannot travel along them. In the case of a manifold~$M$ this possibility still
exists. However, the manifold~$M$ is diffeomorphically equivalent to the product of
some manifolds~$M_0,M_1$, sizes~$r_0,r_1$. The
possibility of this ``travel'' is considered in Ref.~\cite{S2}.

The radii $r_0,r_1$ represent energy scales of \s y breaking. The lowest energy
scale is a scale of weak interactions (Weinberg--Glashow--Salam model)
$r_0\simeq 10^{-16}$~cm. In this case this is a radius of a sphere~$S^2$.

It is interesting to ask on a stability of a \s y breaking pattern \wrt
quantum fluctuations. This hard problem strongly depends on the details
of the model. Especially on the Higgs sector of the practical model. In order
to preserve this stability on both stages of the \s y breaking we should
consider remaining Higgs' fields (after \sb) with zero mass.
They can stabilize the \sb\ in the range of energy
\beq gwiaz
\frac1{r_0}\Bigl(\frac{\hbar}c\Bigr)<E <\frac1{r_{1}}\Bigl(\frac{\hbar}c\Bigr),
\e
i.e.\ for a \sb\ from $G_{1}$ to $G_0$.

In our approach there are some constraints of Higgs' fields (see Section~2).
Solving these constraints we obtain some of Higgs' fields as functions of
in\dt\ components. This could result in some cross terms in the potential
\ER(11) between $\mPhi$'s with different~$i$. For example a term
$$
V(\mPhi'_0,\mPhi'_1),
$$
where $\mPhi'$ means in\dt\ fields. This can cause some problems in a
stability of \sb\ pattern against radiative corrections.

Moreover, $G2$ has been considered as a
group $H$ in the \E\nos\ \KK\ (Jordan--Thiry) Theory. $G2$ is important
only for Glashow--Weinberg--Salam model of unification of electroweak
interactions. If we want to unify all fundamental interactions we need a
bigger group~$H$, \st $G2\subset H$. We need of course a group $G$ \st
\beq{Dn263}
\SU(2)_L \otimes \U(1)_{\rm em} \otimes \SU(3)_c \subset G.
\e
There are a lot of possibilities. One of the most promising is $G=\SO(10)$.
Moreover, we need also a group $G_0$ \st $M=G/G_0$.

In our world $G_0=\U(1)\dr{em} \otimes \SU(3)_c$. The group $H$ for $G=\SO(10)$
and $G_0=\U(1)_{\rm em}\otimes \SU(3)_c$ should be \st
\beq{Dn264}
\SO(10)\otimes \bigl(\U(1)_{\rm em}\otimes \SU(3)_c\bigr) \subset H.
\e
The simplest choice is $H=\SO(16)$. Why?

First of all $G2\subset \SO(16)$ and $\SO(10)\otimes\SO(6)\subset\SO(16)$.
Moreover, $\SO(6)\simeq\SU(4)$ and $\U(1)\times \SU(3) \subset\SU(4)$. Thus if we
identify $\U(1)$ with $\U(1)_{\rm em}$ and $\SU(3)$ with $\SU(3)_c$ we get
what we want. In this way
\begin{multline*}
M=\raise6pt\hbox{SO(10)}\!\bigg/\! \lower6pt\hbox{$\U(1)\otimes\SU(3)$}, \q
S^2\subset M,\\ \q \dim\SO(16)=120, \ \dim\SO(10)=45, \ n_1=\dim M=36.
\end{multline*}

Let us also notice that it would be interesting to consider
as~$G_2$ ($G_2$ is not $G2$!)
$$
G_2=\SU(2)\dr{L} \otimes \SU(2)\dr{R} \otimes \SU(4)
$$
suggested by Salam and Pati, where $\SU(4)$ unifies $\SU(3)\dr{c} \otimes
\U(1)\dr{Y}$.

The bosonic part of a standard model (QCD+GSW), i.e.\ strong and electro-weak
\ia s can be geometrized and
unified with \nos\ gravity (NGT) in the following way:
$H=\SU(3)_c \otimes G2$, $M=S^2$ with $\SO(3)$ invariance of $G2$ part of
a connection defined on a principal bundle over $E\tm S^2$ and $\SU(3)$ part
of a connection coming from connection defined on a principal fibre bundle
over~$E$ only (see Appendix~C for details). One can consider more stages of
\s y breaking (see Ref.~\cite{xy}).

It seems that in a \un\ of \fn\ \ia s we need two or maybe also three stages
of \s y breaking. In the last case we have
\bg3.44a
M = G/G_0= G_3/G_0, \\
M_0=G/G_2,\q M_1=G_2/G_1, \q M_2=G_3/G_2, \lb3.45a
\e
where
\beq3.46a
G_0\subset G_1\subset G_2\subset G_3 = G.
\e
The \sb\ is from $G=G_3$ to~$G_2$ and from $G_2$ to~$G_1$ and from $G_1$
to~$G_0$. We should consider a \tf\ (a~diffeomorphism):
\beq3.47a
g: M_0\tm M_1\tm M_2 = G/G_2\tm G_2/G_1 \tm G_3/G_2 \to
G/G_0=M
\e
and afterwards proceed similarly as in the case of two stages. A~typical
example of such a pattern of \sb\ is:
\beq3.48a
\SO(10) \to \SU(5) \to \SU(3)_c \otimes \SU(2)_L\otimes \U(1)\dr{Y} \to \SU(3)_c
\otimes \U(1)\dr{em}\q\ \hbox{(see Refs \cite{al}--\cite{t})}.
\e

In our case we have
\bea3.49a
G_0&=\SU(3)_c \ot U(1)\dr{em}\\
G&=G_3 = \SO(10) \lb 3.50a \\
G_2&=\SU(5) \lb 3.51a \\
G_1&=\SU(3)_c \ot \SU(2)_L \ot \U(1)\dr{Y}. \lb 3.52a
\e
Thus we get
\bea3.53a
M_0 &= \raise6pt\hbox{\SO(10)}\!\bigg/\! \lower6pt\hbox{$\SU(5)$}, \q \dim(M_0)=21, \\
M_1 &= \raise6pt\hbox{$\SU(5)$}\!\bigg/\! \lower6pt\hbox{$\SU(3)_c \ot \SU(2)_L \ot \U(1)\dr{Y}$}, \q \dim(M_1)=12, \lb 3.54a \\
M_2 &= \raise6pt\hbox{$\SU(3)_c \ot \SU(2)_L \ot \U(1)\dr{Y}$}\!\bigg/\! \lower6pt\hbox{$\SU(3)_c \ot \U(1)\dr{em}$}, \q \dim(M_2)=3, \lb 3.55a \\
M   &= \raise6pt\hbox{$\SO(10)$}\!\bigg/\! \lower6pt\hbox{$\SU(3)_c \ot \U(1)\dr{em}$}, \q \dim(M)=36. \lb 3.56a
\e
Moreover, we need also a chain of $H$ groups. We have as before from
Eq.~\er{Dn264} $H_1=\SO(16)$ and $M$ is also as before
\beq3.57a
M = \raise6pt\hbox{\SO(10)}\!\bigg/\! \lower6pt\hbox{$\SU(3)\ot \U(1)$}.
\e
Moreover
\bg3.58a
\SU(5) \ot \SO(10) \subset H_0 = \SO(17), \q \dim(H_0)=136 \\
\SU(3) \ot \SU(2) \ot \U(1)\ot \SU(5) \subset H_1 = \SO(16), \q \dim(H_1)=120 \lb3.59a \\
\SU(3) \ot \SU(2) \ot \U(1) \subset H_2 = \SU(3) \ot G2, \q \dim(H_2)=22 \lb3.60a
\e
and $S^2\subset M$ as before.

In the case where $G_2$ equals Salam--Pati group one gets
\beq3.63b
\aligned
G_3 &= \SO(10) = G, \q \dim(G) = 45 \\
G_2 &= \SU(2)_L \ot \SU(2)_R  \ot \SU(4), \q \dim(G_2)=21 \\
G_1 &= \SU(2)_L \ot \U(1)\dr{Y} \ot \SU(3)_c,  \q \dim(G_1)=12 \\
G_0 &= \SU(3)_c \ot \U(1)\dr{em}, \q \dim(G_0)=9,
\endaligned
\e
\beq3.64b
\aligned
M_0 &= \raise6pt\hbox{$G$}\!\bigg/\! \lower6pt\hbox{$G_2$} =
\raise6pt\hbox{$\SO(10)$}\!\bigg/\! \lower6pt\hbox{$\SU(2)_L \ot \SU(2)_R \ot \SU(4)$}, \q \dim(M_0)=24 \\
M_1 &= \raise6pt\hbox{$G_2$}\!\bigg/\! \lower6pt\hbox{$G_1$} =
\raise6pt\hbox{$\SU(2)_L \ot \SU(2)_R \ot \SU(4)$}\!\bigg/\! \lower6pt\hbox{$\SU(2)_L \ot \U(1)\dr{Y} \ot \SU(3)_c$}, \q \dim(M_1)=9 \\
M_2 &= \raise6pt\hbox{$G_1$}\!\bigg/\! \lower6pt\hbox{$G_0$} =
\raise6pt\hbox{$\SU(2)_L \ot \U(1)\dr{Y} \ot \SU(3)_c$}\!\bigg/\! \lower6pt\hbox{$\U(1)\dr{em} \ot \SU(3)_c$}, \q \dim(M_2) =3 \\
M &= \raise6pt\hbox{$\SO(10)$}\!\bigg/\! \lower6pt\hbox{$\U(1)\dr{em} \ot \SU(3)_c$}, \q \dim(M) = 36.
\endaligned
\e
We need also a chain of $H$ groups such that
\beq3.65b
\aligned
& \SU(2)_L \ot \SU(2)_R \ot \SU(4) \ot \SO(10) \subset H_0, \\
& \SU(2)_L \ot \U(1)\dr{Y} \ot \SU(3)_c \ot \SU(2)_L \ot \SU(2)_R \ot \SU(4) \subset H_1, \\
& \SU(3)_c \ot \U(1)\dr{em} \ot \SU(3)_c \ot \SU(2)_L \ot \U(1)_Y \subset H_2, \\
& \dim(H_0) \ge 66,\q \dim(H_1) \ge 33,\q \dim(H_2) \ge 21, \\
& {\rm Rank}(H_0) \ge 10, \q {\rm Rank}(H_1) \ge 9, \q {\rm Rank}(H_2) \ge 6.
\endaligned
\e

Moreover, a group from Salam--Pati model gives a possibility to have an
additional stage of the \sb, i.e.
\bml3.66b
\SU(2)_L \ot \SU(2)_R \ot \SU(4) \to \SU(2)_L \ot \SU(2)_R \ot \SU(3)_c
\ot \U(1)\dr{Y}\\ {}\to \SU(2)_L \ot \U(1)\dr{Y} \ot \SU(3)_c
\e
which gives us additional possibilities. In this way one gets
\beq3.67b
\aligned
G_4 &= \SO(10) = G, \q \dim(G) = 45 \\
G_3 &= \SU(2)_L \ot \SU(2)_R \ot \SU(4), \q \dim(G_3) = 21 \\
G_2 &= \SU(2)_L \ot \SU(2)_R \ot \SU(3)_c \ot \U(1)\dr{Y}, \q \dim(G_2) = 15 \\
G_1 &= \SU(2)_L \ot \SU(3)_c \ot \U(1)\dr{Y}, \q \dim(G_1) = 12 \\
G_0 &= \SU(3)_c \ot \U(1)\dr{em}, \q \dim(G_0) = 9
\endaligned
\e
\beq3.68b
\aligned
M_0 &= \raise6pt\hbox{$G$}\!\bigg/\! \lower6pt\hbox{$G_3$} =
\raise6pt\hbox{$\SO(10)$}\!\bigg/\! \lower6pt\hbox{$\SU(2)_L \ot \SU(2)_R \ot \SU(4)$}, \q \dim(M_0) = 24 \\
M_1 &= \raise6pt\hbox{$G_3$}\!\bigg/\! \lower6pt\hbox{$G_2$} =
\raise6pt\hbox{$\SU(2)_L \ot \SU(2)_R \ot \SU(4)$}\!\bigg/\! \lower6pt\hbox{$\SU(2)_L \ot \SU(2)_R \ot \SU(3)_c \ot \U(1)\dr{Y}$}, \q \dim(M_1) = 6 \\
M_2 &= \raise6pt\hbox{$G_2$}\!\bigg/\! \lower6pt\hbox{$G_1$} =
\raise6pt\hbox{$\SU(2)_L \ot \SU(2)_R \ot \SU(3)_c \ot \U(1)\dr{Y}$}\!\bigg/\! \lower6pt\hbox{$\SU(2)_L \ot \SU(3)_c \ot \U(1)\dr{Y}$}, \q \dim(M_2) =3\\
M_3 &= \raise6pt\hbox{$\SU(2)_L \ot \SU(3)_c \ot \U(1)\dr{Y}$}\!\bigg/\! \lower6pt\hbox{$\SU(3)_c \ot \U(1)\dr{em}$}, \q \dim(M_3) = 3.
\endaligned
\e
In this case we need a chain of groups $H$ such that
\beq3.69b
\aligned
& \SU(2)_L \ot \SU(2)_R \ot \SU(4) \ot \SU(10) \subset H_0, \\
& \SU(2)_L \ot \SU(2)_R \ot \SU(3)_c \ot \U(1)\dr{Y} \ot \SU(2)_L \ot \SU(2)_R \ot \SU(4) \subset H_1, \\
& \SU(2)_L \ot \SU(3)_c \ot \U(1)\dr{Y} \ot \SU(2)_L \ot \SU(2)_R \ot \SU(3)_c \ot \U(1)\dr{Y} \subset H_2, \\
& \SU(3)_c \ot \U(1)\dr{em} \ot \SU(2)_L \ot \SU(3)_c \ot \U(1)\dr{Y} \subset H_3, \\
& \dim(H_0) \ge 66, \q \dim(H_1) \ge 36, \q \dim(H_2) \ge 27, \q \dim(H_3) \ge 21, \\
& {\rm Rank}(H_0) \ge 10,\q {\rm Rank}(H_1) \ge 10,\q {\rm Rank}(H_2) \ge 9,\q {\rm Rank}(H_3) \ge 7.
\endaligned
\e

There are some more general and interesting \s y breaking patterns, i.e.,
\bg 3.61a
E6 \to \SU(3)_c\ot \SU(3)_L \ot \SU(3)_R \to
\SU(3)_c \ot \SU(2)_L \ot \U(1)\dr{Y} \to \SU(3)_c \ot \U(1)\dr{em}\\
\intertext{or}
E8 \to E7 \to E6 \to \SO(10) \to \SU(5) \to \SU(3)_c\ot \SU(2)_L
\ot \U(1)\dr{Y} \to \SU(3)_c\ot \U(1)\dr{em}. \lb3.62a
\e
In the second case we have to do with more than three stages of a \s y
breaking. Such a possibility is considered in Ref.~\cite{xy}.

The above patterns are inspired by string theory (see Ref.~\cite{m3k}).
However, an experiment does not give any trace of some new physical states
coming from those higher \s ies. According to our approach we have in the
first case a manifold of vacuum states
\beq 3.63a
M = \raise6pt\hbox{$E6$}\!\bigg/\! \lower6pt\hbox{$\SU(3)_c\ot \U(1)\dr{em}$} =
\raise6pt\hbox{$G$}\!\bigg/\! \lower6pt\hbox{$G_0$} =
\raise6pt\hbox{$G_3$}\!\bigg/\! \lower6pt\hbox{$G_0$}
\e

\goodbreak
\noindent and
\bea 3.64a
M_0 &= \raise6pt\hbox{$E_6$}\!\bigg/\! \lower6pt\hbox{$\SU(3)_c\ot \SU(3)_L \ot
\SU(3)_R$} = \raise6pt\hbox{$G$}\!\bigg/\! \lower6pt\hbox{$G_2$}\\
M_1 &= \raise6pt\hbox{$\SU(3)_c \ot \SU(3)_L \ot \SU(3)_R$}\!\bigg/\!
\lower6pt\hbox{$\SU(3)_c \ot \SU(2)_L \ot \U(1)\dr{Y}$} =
\raise6pt\hbox{$G_2$}\!\bigg/\! \lower6pt\hbox{$G_1$} \lb3.65a \\
M_2 &= \raise6pt\hbox{$\SU(3)_c \ot \SU(2)_L \ot \U(1)\dr{Y} $}\!\bigg/\!
\lower6pt\hbox{$\SU(3)_c \ot \U(1)\dr{em}$}=
\raise6pt\hbox{$G_1$}\!\bigg/\! \lower6pt\hbox{$G_0$} \lb3.66a
\e

In this case we need a diffeomorphism $g_1$ such that
\beq 3.67a
g_1 : M_0\tm M_1\tm M_2 = G/G_2 \tm G_2/G_1 \tm G_1/G_0 \to G/G_0 = M,
\e
$G=G_3=E6$, $G_2 = \SU(3)_c \ot \SU(3)_L \ot \SU(3)_R$, $G_1 = \SU(3)_c \ot
\SU(2)_L \ot \U(1)\dr{Y}$, $G_0 = \SU(3)_c\ot \U(1)\dr{em}$.

We need also some groups $H$ such that
\bg 3.68a
E6 \ot \SU(3)_c \ot \SU(3)_L \ot \SU(3)_R \subset H_0, \\
\SU(3)_c \ot \SU(3)_L \ot \SU(3)_R \ot \SU(3)_c \ot \SU(2)_L \ot \U(1)\dr{Y}
\subset H_1, \lb3.69a \\
\SU(3)_c \ot \SU(2)_L \ot \U(1)\dr{Y} \ot \SU(3)_c \ot \U(1)\dr{em} \subset \SU(3)_c
\ot G2 \subset H_2, \lb3.70a
\e
and of course
\beq 3.71a
S^2 \subset M.
\e
In the case \er{3.62a} we have
\bea 3.72a
M &= \raise6pt\hbox{$E8$}\!\bigg/\! \lower6pt\hbox{$\SU(3)_c \ot \U(1)\dr{em}$} \\
M_0 &=\raise6pt\hbox{$E8$}\!\bigg/\! \lower6pt\hbox{$E7$} \lb3.73a \\
M_1 &=\raise6pt\hbox{$E7$}\!\bigg/\! \lower6pt\hbox{$E6$} \lb3.74a \\
M_2 &=\raise6pt\hbox{$E6$}\!\bigg/\! \lower6pt\hbox{$\SO(10)$} \lb3.75a \\
M_3 &=\raise6pt\hbox{$\SO(10)$}\!\bigg/\! \lower6pt\hbox{$\SU(5)$} \lb3.76a \\
M_4 &=\raise6pt\hbox{$\SU(5)$}\!\bigg/\! \lower6pt\hbox{$\SU(3)_c \ot \SU(2)_L
\ot \U(1)\dr{Y}$} \lb3.77a \\
M_5 &=\raise6pt\hbox{$\SU(3)_c \ot \SU(2)_L \ot \U(1)\dr{Y}$}\!\bigg/\!
\lower6pt\hbox{$\SU(3)_c \ot U(1)\dr{em}$} \lb3.78a
\e
and we need a diffeomorphism
\beq 3.79a
g_2 : M_ 0\tm M_1 \tm M_2 \tm M_3 \tm M_4 \tm M_5 \to M.
\e
We have of course
\beq 3.79b
\bal
G &= E8 = G_6, \\
G_5 &= E7, \q G_4 = E6, \\
G_3 &= \SO(10), \q G_2 = \SU(5), \\
G_1 &= \SU(3)_c \ot \SU(2)_L \ot \U(1)\dr{Y}, \\
G_0 &= \SU(3)_c \ot \U(1)\dr{em}
\eal
\e
and
\beq 3.80a
\bal
G_6 \ot G_5 &\subset H_0 \\
G_5 \ot G_4 &\subset H_1 \\
G_4 \ot G_3 &\subset H_2 \\
G_3 \ot G_2 &\subset H_3 \\
G_2 \ot G_1 &\subset H_4 \\
G_1 \ot G_0 &\subset H_5 = \SU(3)_c\ot G2 \\
\hbox{and}\qquad S^2 &\subset M.
\eal
\e
The shape of those manifolds can be very complicated. One can try to construct
chains of groups~$H$ in both considered cases.

In the case of \er{3.61a} one gets

\noindent ${\rm Rank}(H_0) \ge 12$.
Taking the lowest value of the rank we get several possibilities for $H_0=
\SU(13)$, \SO(25), Sp(12), \SO(24).

\noindent ${\rm Rank}(H_1) \ge 10$ and similarly for lowest \di s
$H_1 = \SU(9)$, \SO(20), Sp(10).

\noindent ${\rm Rank}(H_2)\ge7$ and $H_2 = \SU(6)$, Sp(7), \SO(14).

\noindent Moreover
$$
\SU(3) \ot G2 \subset H_2.
$$
In the second case \er{3.62a} one gets
\beq3.91b
\aligned
{}&{\rm Rank}(H_0)\ge15 \hbox{ and in the lowest \di s }H_0= \SU(14),\
\SO(31),\ \Sp(15),\ \SO(30).\\
&{\rm Rank}(H_1)\ge13 \hbox{ and }H_1=\SU(13),\ \SO(27),\ \Sp(13),\ \SO(26).\\
&{\rm Rank}(H_2)\ge11 \hbox{ and }H_2= \SU(11),\ \SO(23),\ \Sp(11),\ \SO(22).\\
&{\rm Rank}(H_3)\ge9 \hbox{ and }H_3=\SU(8),\ \SO(19),\ \Sp(9),\ \SO(18).\\
&{\rm Rank}(H_4)\ge8 \hbox{ and }H_4 = \SU(7),\ \SO(15),\ \Sp(7),\ \SO(14).\\
&{\rm Rank}(H_5)\ge5 \hbox{ and } H_5 = \SU(6),\ \SO(11),\ \Sp(5),\ \SO(10).
\endaligned
\e

Moreover we should have
\beq3.92b
\SU(3) \ot G2 \subset H_5.
\e

We need additional investigations to choose those groups, possibly with higher
\di s.

In order to consider schemes \er{3.61a}, \er{3.62a} and also \er{3.48a}, we
generalize two stages of \sb\ to the case of $k$~stages of \sb\ for arbitrary~$k$.
In particular, $k=k_1=3$ or $k=k_2=6$.

On every manifold $M_i$ in both cases we construct a \ssb\ scheme and a gauge
field with Higgs' field as in Section~2. Afterwards we use $g_1$ or $g_2$
diffeomorphism to connect Higgs' field from every $M_i$ to a Higgs' field
on~$M$, i.e.\ for a \s y breaking from~$G$ to $\SU(3)_c\ot\U(1)\dr{em}$
(see Ref.~\cite{xy}).

\def\eq#1 {\label{9.#1}}
\def\ER(#1){\eqref{9.#1}}

Let us incorporate in our scheme a hierarchy of a symmetry breaking case \er{3.61a} and \er{3.62a}.
This scheme covers also three stages of \sb\ \er{3.48a}. In order
to do this let us consider a case of the manifold
\begin{equation}
M \cong M_0\times M_1\times \ldots \times M_{k-1},\q k=k_1=3 \hbox{ or }k=k_2=6 \eq1
\end{equation}
where
\ealn
\dim M_i& = \ov n_i, \quad i=0,1,2,\dots,k-1,\q k=k_1=3 \hbox{ or }k=k_2=6 \\
\dim M&= \sum_{i=0}^{k-1} \ov n_i, \eq2 \\
M_i&=G_{i+1}/G_i\,. \eq3
\end{align}
Every manifold $M_i$ is a manifold of vacuum states if the \s y is breaking from
$G_{i+1}$ to~$G_i$, $G_k=G$.

Thus
\begin{equation}
G_0\subset G_1\subset G_2\subset \ldots \subset G_k=G, \q k=k_1 \hbox { or }k=k_2. \eq4
\end{equation}
We will consider the situation when
\begin{equation}
M= G/G_0. \eq5
\end{equation}
This is a constraint in the theory. From the chain \ER(4) one gets
\begin{equation}
\gG_0\subset \gG_1\subset \ldots \subset \gG_k =\gG,\q k=k_1=3 \hbox{ or }k=k_2=6 \eq6
\end{equation}
and
\begin{equation}
\gG_{i+1}=\gG_i \dpl \gM_i, \quad i=0,1,\dots,k-1,\q k=k_1=3 \hbox{ or }k=k_2=6. \eq7
\end{equation}
The relation \ER(5) means that there is a diffeomorphism $g$ (in our case two
different diffeomorphisms for \er{3.61a} and \er{3.62a}) onto $G/G_0$ such that
\begin{equation}
g: \prod_{i=0}^{k-1}(G_{i+1}/G_i)\to G/G_0\,,\q k=k_1=3 \hbox{ or }k=k_2=6. \eq8
\end{equation}
This diffeomorphism is a deformation of a product \ER(1) in $G/G_0$ for two cases. The
theory has been constructed for the case considered before with $G_0$
and~$G$. The multiplet of Higgs' fields $\mPhi$ breaks the \s y from $G$
to~$G_0$ (equivalently from $G$ to~$G_0'$ in the false vacuum case).
$\gG_i$~mean Lie algebras for groups $G_i$ and $\gM_i$ a complement in a
decomposition~\ER(7). On every manifold~$M_i$ we introduce a radius $r_i$
(a~``size'' of a manifold) in such a
way that $r_i>r_{i+1}$. On the manifold $G/G_0$ we define the radius~$r$ as
before. The diffeomorphism $g$ induces a contragradient transformation for a
Higgs field $\mPhi$ in such a way that
\begin{equation}
g^\ast \mPhi=(\mPhi_0,\mPhi_1,\dots,\mPhi_{k-1}),\q k=k_1=3 \hbox{ or }k=k_2=6. \eq9
\end{equation}
The fields $\mPhi_i$, $i=0,\dots,k-1$, $k=k_1=3$ or $k=k_2=6$.

Eq.\ \er{9.10} generalizes Eqs \er{16}--\er{17}.
In this way we get the following decomposition for a kinetic part of the
field~$\mPhi$ and for a potential of this field:
\ealn
\cL_{\textrm {kin}}(\gag\mPhi)&=\sum_{i=0}^{k-1} \cL_{\text{\rm kin}}^i(\gag\mPhi_i) \eq10 \\
V(\mPhi)&=\sum_{i=0}^{k-1}V^i(\mPhi_i),\q k=k_1=3 \hbox{ or }k=k_2=6 \eq11
\end{align}
where
\ealn
V_i&=\intop_{M_i}\sqrt{|\wt g_i|}\,d^{\bar n_i}x \eq13 \\
\wt g_i&=\det(g_{i\u b_i\u a_i}). \eq14
\end{align}
$g_{i\u b_i\u a_i}$ is a non\s ic tensor on a manifold~$M_i$.
\begin{equation}
\bal
V^i(\mPhi_i)&=\frac{l_{ab}}{V_i} \intop_{M_i} \sqrt{|\wt g_i|}\,d^{\bar n_i}x
\biggl[2g^{[\u m_i\u n_i]} \bigl(C^a_{cd}\mPhi^c_{i\u m_i}\mPhi^d_{i\u n_i}
-\mu^a_{\hi_i}f^{\hi_i}_{\u m_i\u n_i}-\mPhi^a_{i\u e_i}
f^{\u e_i}_{\u m_i\u n_i}\bigr)\\
&\times g^{[\u a_i\u b_i]}
\bigl(C^b_{ef}\mPhi^e_{i\u a_i}\mPhi^f_{i\u b_i}
-\mu^b_{\hj_i} f^{\hj_i}_{\u a_i\u b_i}-\mPhi^b_{i\u a_i}
f^{\u d_i}_{\u a_i\u b_i}\bigr)\\
&-g^{\u a_i\u m_i}_i g^{\u b_i\u n_i}_i L^a_{\u a_i\u b_i}
\bigl(C^b_{cd}\mPhi^c_{i\u m_i}\mPhi^d_{\u n_i}
-\mu^b_{\hi_i}f^{\hi_i}_{\u m_i\u n_i}-\mPhi^b_{\u e_i}f^{\u e_i}
_{\u m_i\u n_i}\bigr)\biggr],
\eal \eq15
\end{equation}
$f^{\hj_i}_{\u a_i\u b_i}$ are structure \ct s of the Lie algebra
$\gG_i$. Eqs \er{9.10}--\er{9.15} generalize Eqs \er{10}--\er{14}.

The scheme of the \s y breaking acts as follows from the group $G_{i+1}$
to~$G_i$ ($G_i'$) (if the \s y has been broken up to $G_{i+1}$). The potential
$V^i(\mPhi_i)$ has a minimum (global or local) for $\mPhi^k_{i\rm crt}$,
$k=0,1$. The value of the remaining part of the sum \ER(11) for fields
$\mPhi_j$, $j<i$, is small for the scale of energy is much lower ($r_j>r_i$,
$j<i$). Thus the minimum of $V^i(\mPhi_i)$ is an approximate minimum of the
remaining part of the sum \ER(11). In this way we have a descending chain of
truncations of the Higgs potential. This gives in principle a pattern of a
\s y breaking. However, this is only an approximate \s y breaking. The real
\s y breaking is from $G$ to $G_0$ (or to $G_0'$ in a false vacuum case). The
important point here is the diffeomorphism~$g$.
\ealn
g^\ast \mPhi^b&=\bigl(\mPhi^b_0,\mPhi^b_1,\dots,\mPhi^b_{k-1}\bigr) \eq16 \\
\mPhi^b_i&=\mPhi^b_{i\u a_i}\wt \theta^{\u a_i},
\quad i=0,\dots,k-1,\q k=k_1 \hbox{ or }k=k_2. \eq 17
\end{align}

The shape of $g$ is a true indicator of a reality of the \s y breaking
pattern. If
\begin{equation}
g=\Id+\d g \eq18
\end{equation}
where $\d g$ is in some sense small and $\Id$ is an identity. The smallness of $\d
g$ is a criterion of a practical application of the \s y breaking pattern
\ER(4) (comp.\ \er{16} in two stages of a \sb).
It seems that there are a lot of possibilities for the condition
\ER(8). Let us notice that
the decomposition of~$M$ results in decomposition of \co ical terms
\begin{equation}
\ul{\wt P}=\sum_{i=0}^{k-1} \ul{\wt P}_i \eq19
\end{equation}
where
\begin{equation}
\ul{\wt P}_i=\frac1{r^2_iV_i}\intop_{M_i}\sqrt{|\wt g_i|}\,
\wh{\ov R}_i(\wh{\ov\G}_i)\,d^{n_i}x \eq20
\end{equation}
where $\wh{\ov \G}_i$ is a non\s ic connection on $M_i$ compatible with the
non\s ic tensor $g_{i\u a_i\u b_i}$ and $\wh{\ov R}_i(\wh{\ov \G}_i)$ its
curvature scalar. Eqs \er{9.19}--\er{9.20} generalize Eqs \er{19}--\er{20}.

The scalar field $\mPsi$ from Section~4 is now a function on a product
of~$M_i$,
$$
\mPsi=\mPsi(x,y_0,\dots,y_{k-1}),\quad
y_i\in M_i,\ i=0,1,\dots,k-1,\q k=k_1 \hbox{ or }k=k_2.
$$
The truncation procedure can be proceeded in several ways. Finally let us
notice that the energy scale of broken gauge bosons is fixed by a
radius~$r_i$ at any stage of the \s y breaking in our scheme.

Let us consider Eq.\ \ER(9) in more details. One gets
\begin{equation}
A^{\u a}_{i\u a_i}(y)\mPhi^b_{\u a}(y)=
\mPhi^b_{\u a_i}(y_i), \quad y\in M,\ y_i\in M_i \eq21
\end{equation}
where
\begin{equation}
g^\ast(y)=\Biggl(A_0\Biggm|A_1\Biggm|A_2\Biggm|\ldots\Biggm| A_{k-1}\Biggr), \eq22
\end{equation}
\begin{equation}
A_i=\bigl(A^{\u a}_{i\u a_i}\bigr)_{\u a=1,2,\dots,n_1,\,\u a_i=1,2,\dots,\bar
n_i}\,,\ i=0,1,2,\dots,k,\q k=k_1 \hbox{ or }k=k_2, \eq23
\end{equation}
is a matrix of Higgs' fields transformation.

According to our assumptions one gets also:
\begin{equation}
\Bigl(\frac{r_i}r\Bigr)^2 g_{i\u a_i\u b_i}(y_i)=
A^{\u a}_{\u a_i}(y)A^{\u b}_{\u b_i}(y)g_{\u a\u b}(y). \eq24
\end{equation}
For $g$ is an invertible map we have $\det g^\ast(y)\ne0$.

We have also
\begin{equation}
n_1=\sum_{i=0}^{k-1} \ov n_i,\q k=k_1 \hbox{ or }k=k_2 \eq25
\end{equation}
and
\begin{equation}
\mPhi^b_{\u a}(y)=\sum_{i=0}^{k-1} \wt A_{i\u a}^{\u a_i}(y)
\mPhi^b_{\u a_i}(y_i) \eq26
\end{equation}
or
\begin{gather}
{g^\ast}^{-1}(y)=\left(\vcenter{\offinterlineskip
\hbox to40pt{\hfil$\wt A_0$\hfil \vrule height10pt depth4pt width0pt}
\hrule width40pt
\hbox to40pt{\hfil$\wt A_1$\hfil \vrule height10pt depth4pt width0pt}
\hrule width40pt
\hbox to40pt{\hfil$\vdots$\hfil \vrule height10pt depth4pt width0pt}
\hrule width40pt
\hbox to40pt{\hfil$\wt A_{k-1}$\hfil \vrule height10pt depth4pt width0pt}
}\right) \eq27 \\
\wt A_i=\bigl(\wt A^{\u a_i}_{i\u a}\bigr)_{\u a_i=1,2,\dots,\bar n_i,\,\u a=1,2,\dots,
n_1} \eq28
\end{gather}
such that
\ealn
&g(y_0,\dots,y_{k-1})=y \eq29 \\
&(y_0,y_1,\dots,y_{k-1})=g^{-1}(y) \eq29a
\end{align}

For an inverse tensor $g^{\u a\u b}$ one easily gets
\begin{equation}
\Bigl(\frac{r_i^2}{r^2}\Bigr)g^{\u a\u b}=
\sum_{i=0}^{k-1} \wt A^{\u a}_{i\u a_i} g^{\u a_i\u b_i}_i
A^{\u b}_{i\u b_i}\,,\q k=k_1 \hbox{ or }k=k_2. \eq30
\end{equation}
We have
\begin{equation}
r^{2n_1}\det(g_{\u a\u b})=\prod_{i=0}^{k-1}
r^{2\bar n_i}_i\det(g_{i\u a_i\u b_i}). \eq31
\end{equation}
In this way we have for the measure
\begin{equation}
d\mu(y)=\prod_{i=0}^{k-1}d\mu_i(y_i) \eq32
\end{equation}
where
\ealn
d\mu(y)&=\sqrt{\det g}\,r^{n_1}\,d^{n_1}y \eq33 \\
d\mu_i(y_i)&=\sqrt{\det g_i}\,r^{\bar n_i}_i\,d^{\bar n_i}y_i\,. \eq34
\end{align}
Eqs \er{9.21}--\er{9.34} are \gn\ of Eqs \er{24}--\er{34}.

In the case of $\cL\dr{int}(\mPhi,\wt A)$ one gets
\begin{equation}
\cL\dr{int}(\mPhi,\wt A)=\sum_{i=0}^{k-1} \cL\dr{int}(\mPhi_i,\wt A) \eq35
\end{equation}
where
\begin{equation}
\cL\dr{int}(\mPhi_i,\wt A)=h_{ab}\mu^a_{\ul i}\wt H^{\ul i}\ul g^{[\u a_i\u b_i]}
_i \bigl(C^b_{cd}\mPhi^c_{i\u a_i}\mPhi^d_{i\u b_i}
-\mu^b_{\hi}f^{\hi}_{\u a_i\u b_i}-\mPhi^b_{\u d_i}f^{\u d_i}_{\u a_i\u b_i}\bigr)
\eq36
\end{equation}
where
\begin{equation}
\ul g^{[\u a_i\u b_i]}_i = \frac 1{V_i}\intop_{M_i} \sqrt{|\wt g_i|}\,d^{\bar n_i}x\,
g^{[\u a_i\u b_i]}_i, \quad i=0,1,2,\dots,k-1,\ k=k_1 \hbox{ or }k=k_2. \eq37
\end{equation}
Eqs \er{9.35}--\er{9.37} generalize Eqs \er{35}--\er{37}.

Moreover, to be in line in the full theory we should consider a chain of
groups $H_i$,\break $i=0,1,2,\dots,k-1$, in such a way that
\begin{equation}
H_0\subset H_1\subset H_2\subset \ldots \subset H_{k-1}=H. \eq38
\end{equation}
For every group $H_i$ we have the following assumptions
\begin{equation}
G_i\subset H_i \eq39
\end{equation}
and $G_{i+1}$ is a centralizer of $G_i$ in $H_i$. Thus we should have
\begin{equation}
G_i \otimes G_{i+1}\subset H_i, \quad i=0,1,2,\dots,k-1,\q k=k_1 \hbox{ or }k=k_2. \eq40
\end{equation}
Proposals of such chains of $H$ groups in both cases are given above.

In this approach we try to consider additional dimensions connecting to the
manifold~$M$ more seriously, i.e.\ as physical dimensions, additional space-like
dimensions. We remind to the reader that gauge-dimensions connecting to the
group~$H$ have different meaning. They are dimensions connected to local
gauge \s ies  and they cannot be directly observed. Simply saying
we cannot travel along them. In the case of a manifold~$M$ this possibility still
exists. This possibility was considered in the case of two stages of a \sb.
Now it is more general.
However, the manifold~$M$ is diffeomorphically equivalent to the product of
some manifolds~$M_i$, $i=0,1,2,\dots,k-1$, with some characteristic sizes~$r_i$,
$k=k_1$ or $k=k_2$.

The radii $r_i$ represent energy scales of \s y breaking. The lowest energy
scale is a scale of weak interactions (Weinberg--Glashow--Salam model)
$r_0\simeq 10^{-16}$~cm. In this case this is a radius of a sphere~$S^2$,
which we mentioned before.

It is interesting to ask on a stability of a \s y breaking pattern \wrt
quantum fluctuations. This difficult problem strongly depends on the details
of the model. Especially on the Higgs sector of the practical model. In order
to preserve this stability on every stage of the \s y breaking we should
consider remaining Higgs' fields (after \sb) with zero mass. According to
S.~Weinberg, they can stabilize the \sb\ in the range of energy
\beq 2gw
\frac1{r_i}\Bigl(\frac{\hbar}c\Bigr)<E <\frac1{r_{i+1}}\Bigl(\frac{\hbar}c\Bigr),
\quad i=0,1,2,\dots,k-1,\q k=k_1 \hbox{ or }k=k_2,
\e
i.e.\ for a \sb\ from $G_{i+1}$ to $G_i$.

\er{2gw} is a \gn\ to $k=k_1$ or $k=k_2$ of \er{gwiaz} for
two stages of \s y breaking.

It seems that in order to create a realistic grand unified model based on
non\s ic Kaluza--Klein (Jordan--Thiry) theory it is necessary to nivel \co ical
terms (see Section~4). This could be achieved in some models due to choosing \ct s~$\xi$ and
$\zeta$ and~$\mu$. After this we can control the value of those terms, which
are considered as a selfinteraction potential of a scalar field~$\mPsi$. The
scalar field~$\mPsi$ can play in this context a role of a \qs.

Let us notice that using the equation
\begin{equation}
{\mPhi} ^{c}_{\tilde b}(x) f^{\tilde b}_{\dah{i}\tilde a}= \mu
^{a}_{\dah{i}} {\mPhi} ^{b}_{\tilde a}(x) C^{c}_{ab},\eq gw
\end{equation}
and \ER(26) one gets
\begin{equation}
\sum_{i=0}^{k-1} \wt A^{\u a_i}_{i\u b} \mPhi^c_{\u a_i} f^{\u b}_{\hi\u a}
=C^c_{ab}\mu^a_{\hi}\sum_{i=0}^{k-1} \wt A^{\u a_i}_{\wt{ia}}\mPhi^b_{\u a_i}\,.
\eq47
\end{equation}

In this way we get constraints for Higgs' fields $\mPhi_0$, $\mPhi_1$, \dots,
$\mPhi_{k-1}$,
$$
\mPhi_i=(\mPhi^b_{\u a_i}), \quad i=0,1,\dots,k-1.
$$
Solving these constraints we obtain some of Higgs' fields as functions of
in\dt\ components. This could result in some cross terms in the potential
\ER(11) between $\mPhi$'s with different~$i$. For example a term
$$
V(\mPhi'_i,\mPhi'_j),
$$
where $\mPhi'$ means in\dt\ fields. This can cause some problems in a
stability of \sb\ pattern against radiative corrections. This can be easily
seen from Eq.~\eqref{9.gw} solved by in\dt~$\mPhi'$,
\ealn
\mPhi&=B\mPhi' \eq48 \\
\mPhi^c_{\u b}&=B^{c\u{\bar b}}_{\u b\bar c}{\mPhi'}^{\bar c}_{\u{\bar b}}
\eq49
\end{align}
where $B$ is a linear operator transforming in\dt~$\mPhi'$ into $\mPhi$.

We can suppose for a trial a condition similar to~\eqref{9.gw} for every
$i=0,\dots, k-1$, $k=k_1$ or $k=k_2$,
\begin{equation}
\mPhi^{c_i}_{\u b_i}f^{\u b_i}_{\hi_i\u a_i}=
\mu^{a_i}_{\hi_i} \mPhi^{b_i}_{\u a_i} C^{c_i}_{a_ib_i} \eq50
\end{equation}
where $C^{c_i}_{a_ib_i}$ are structure \ct s for the Lie algebra $\gH_i$ of
the group~$H_i$. $f^{\u b_i}_{\hi_i\u a_i}$ are structure \ct s of the Lie
algebra $\gG_{i+1}$, $\hi_i$~are indices belonging to Lie algebra $\gG_i$ and
$\wt a_i$ to the complement~$\gM_i$.

In this way
\begin{equation}
\mPhi^c_{\u b_i} = \mPhi^{c_i}_{\u b_i} \d^c_{c_i}\,. \eq51
\end{equation}
In this case we should have a consistency between \ER(50) and \ER(47) which
impose constraints on $C,f,\mu$ and $C^i,f^i,\mu^i$ where $C^i,f^i,\mu^i$
refer to $H_i$,~$G_{i+1}$. Solving \ER(50) via introducing in\dt\ fields
$\mPhi'_i$ one gets
\begin{equation}
\mPhi^{c_i}_{i\u b_i} = B^{c_i\u{\bar b}_i}_{i\bar c_i\u b_i}
{\mPhi'_i}^{\bar c_i}_{\u{\bar b}_i}. \eq52
\end{equation}

Combining \ER(49), \ER(51), \ER(52) one gets
\begin{equation}
B^{c\u{\bar b}}_{\u b\bar c}{\mPhi'}^{\bar c}_{\u{\bar b}}
=\sum_{i=0}^{k-1} \wt A^{\u a_i}_{i\u b} \d^c_{c_i} B^{c_i\u{\bar b}_i}_{i\bar
c_i\u b_i}{\mPhi'_i}^{\bar c_i}_{\u{\bar b}_i}\,.\eq53
\end{equation}
Eq.\ \ER(53) gives a relation between in\dt\ Higgs' fields $\mPhi'$ and
$\mPhi'_i$. Simultaneously it is a consistency condition between Eq.~\eqref{9.gw}
and Eq.~\ER(50). However, the condition \ER(50) seems to be too strong and
probably it is necessary to solve a weaker condition \ER(47) which goes
to the mentioned terms $V(\mPhi'_i,\mPhi'_j)$. The conditions \ER(50) plus a
consistency \ER(53) avoid those terms in the Higgs potential. This problem
demands more investigations. The above calculations can be easily applied
for the case of two stages \sb. For this we do not give them before.

It seems that the condition \ER(8) could be too strong. In order to find a
more general condition we consider a simple example of~\ER(4). Let
$G_0=\{e\}$ and $k=2$ (two stages of \s y breaking). In this case we have
\ealn
&\{e\}\subset G_1\subset G_2=G \eq54 \\
&M_0=G_1, \quad M_1=G/G_1 \eq55 \\
&g:G_1\times G/G_1 \to G. \eq56
\end{align}
In this way $G_1\times G/G_1$ is diffeomorphically equivalent to~$G$.

Moreover, we can consider a fibre bundle with base space $G/G_1$ and a
structural group~$G_1$ with a bundle manifold~$G$. This construction is known in
the theory of induced group representation (see~\cite{S3}). The
projection $\vf:G\to G/G_1$ is defined by $\vf(g)=\{gG_1\}$. The natural
extension of \ER(56) is to consider a fibre bundle $(G,G/G_1,G_1,\vf)$. In this
way we have in a place of~\ER(56) a local condition
\begin{equation}
g_U:G_1\times U\underset{\text{in}}\longrightarrow G \eq57
\end{equation}
where $U\subset G/G_1$ is an open set. Thus in a place of~\ER(8) we consider
a local diffeomorphism
\begin{equation}
g_U : M_0\times M_1\times \ldots \times
M_{k-1}\underset{\text{in}}\longrightarrow G/G_0\,, \q k=k_1 \hbox{ or }k=k_2. \eq58
\end{equation}
where
$$
U=U_0\times U_1\times \ldots \times U_{k-1},
$$
$U_i\subset M_i$, $i=0,1,2,\dots,k-1$, are open sets. Moreover we should
define projectors $\vf_i$, $i=0,1,2,\dots,k-1$, $k=k_1$ or $k=k_2$,
\begin{equation}
\vf_i: G/G_0 \to G_{i+1}/G_i, \eq59
\end{equation}
i.e.
\begin{gather}
\vf_i(\{\ov gG_0\})=\{\ov g_{i+1}G_i\}, \eq60 \\
\ov g\in G,\quad \ov g_{i+1}\in G_{i+1}, \quad
G_0 \subset G_i \subset G_{i+1} \subset G \nn
\end{gather}
in a unique way. This could give us a fibration of $G/G_0$ in $\prod\limits
_{i=0}^{k-1} (G_{i+1}/G_i)$.

Let us give some details of our two strings inspired chains of \s y
breaking in a hierarchy of \s y breaking. In the case \er{3.61a} one gets
\beq3.150b
\aligned
\dim(E6) &= \dim(G) =\dim(G_3)=78\\
\dim(G_2) &= \dim(\SU(3)_c \ot \SU(3)_L \ot \SU(3)_R) = 24\\
\dim(G_1) &= \dim(\SU(3)_c \ot \SU(2)_L \ot \U(1)\dr{em}) = 12\\
\dim(G_0) &= \dim(\SU(3)_c \ot \U(1)\dr{em}) = 9
\endaligned
\e
\beq3.151b
\aligned
\dim(M_0) &= \dim(E_0) - \dim(\SU(3)_c \ot \SU(3)_L \ot \SU(3)_R) = 78-24 = 54 =\ov n_0\\
\dim(M_1) &= \dim(G_2) - \dim(G_1) = 24 - 12 = 12 = \ov n_1\\
\dim(M_2) &= \dim(G_1) - \dim(G_0) = 12 - 9 =3 = \ov n_2\\
\dim(M) &= \dim(G) - \dim(G_0) = 78 - 9 = 69 = n_1
\endaligned
\e
Simultaneously we have \di s of Lie algebras $\fg_i$ and complements $\fm_i$
for $\dim(\fg_i)=\dim(G_i)$ and $\dim(M_i)=\dim(\fm_i)=\ov n_i$.

In the second case \er{3.62a} one gets
\beq3.152b
\aligned
\dim(E8) &= \dim(G) = \dim(G_6) = 248\\
\dim(G_5) &= \dim(E_7) = 133\\
\dim(G_4) &= \dim(E_6) = 78\\
\dim(G_3) &= \dim(\SO(10)) =45 \\
\dim(G_2) &= \dim(\SU(5)) = 24\\
\dim(G_1) &= \dim(\SU(3)_c \ot \SU(2)_L \ot \SU(1)\dr Y) = 12\\
\dim(G_0) &= \dim(\SU(3)_c \ot \U(1)\dr{em}) = 9
\endaligned
\e
\beq3.153b
\aligned
\dim(M) &= \dim(E8) - \dim(\SU(3)_c \ot \U(1)\dr{em}) = 248 - 9 =239 =n_1 \\
\dim(M_0) &= \dim(E8) - \dim(E7)= 248-133 = 115 = \ov n_0\\
\dim(M_1) &= \dim(E7) - \dim(E6) = 133 - 78 = 55 = \ov n_1\\
\dim(M_2) &= \dim(E6) - \dim(\SO(10)) = 78 - 45 = 33 = \ov n_2\\
\dim(M_3) &= \dim(\SO(10)) - \dim(\SU(5)) = 45 - 24= 21 = \ov n_3\\
\dim(M_4) &= \dim(\SU(5)) - \dim(\SU(3)_c \ot \SU(2)_L \ot \U(1)\dr Y) = 24 - 12 = 12 =\ov n_4 \\
\dim(M_5) &= \dim(\SU(3)_c \ot \SU(2)_L \ot \U(1)\dr Y) - \dim(\SU(3)_c\ot \U(1)\dr{em}) = 12 - 9 = 3=\ov n_5\hskip-30pt
\endaligned
\e
We have of course the same remarks for \di s of algebras and complements as in the case of
\er{3.61a}, i.e.\ $\dim(M_i)=\dim(\fm_i)=\ov n_i$, $\dim(G_i)=\dim(\fg_i)$.

\section{The Nonsymmetric Jordan--Thiry Theory
over $V=E\times G/G_{0}$. A~\co ical \ct\ (a~Dark Energy).
A~Dark Matter}

Let $P'$ be the principal fibre bundle with the structural group $H$,
over $V = E \times G/G_{0}$ with a projection $\pi $ and let us define on this bundle
a connection $\omega$. Let us suppose that $H$ is semisimple. On the
base $V = E \times G/G_{0}$ we define a nonsymmetric metric tensor such that:
\beq4.1
\bal
\gamma _{\rAB} &= \gamma _{(\rAB)} + \gamma _{[\rAB]},\cr
\gamma _{\rAB} \gamma ^{\rCB} &= \gamma _{\rBA} \gamma ^{\rBC} =
\delta^{\rC}_{\rA},\eal
\e
(see Section 2), where the order of the indices is important. We also define on $V$ the
connection $\overline{\omega}^{\rA}{}_{\rB}$,
\beq4.2
\overline{\omega }^{\rA}{}_{\rB} =
\overline{\varGamma}^{\rA}{}_{\rBC}\overline{\theta }^{\rC}
\e
such that:
\beq4.3
\ov{D}\gamma _{\text{\rm A+B}-} = \ov{D} \gamma _{\rAB} - \gamma
_{\rAD}\ov{Q}^{\rD}{}_{\rBC}(\overline{\varGamma}) \theta ^{\rC} = 0,
\e
where $\ov{D}$ is the exterior covariant derivative with respect to
$\overline{\omega }^{\rA}{}_{\rB}$ and
$\ov{Q}^{\rA}{}_{\rBC}(\overline{\varGamma})$ is the torsion of
$\overline{\omega }^{\rA}{}_{\rB}$. One easily finds that the
connection \er{4.2} has the following shape:
\beq4.4
\overline{\omega }^{\rA}{}_{\rB} = \left(\vcenter{\offinterlineskip\tabskip0pt
\halign{\strut\tabskip5pt plus10pt minus10pt
\hfill#\hfill&#&\hfill#\hfill\tabskip0pt\cr
$\vrule height0pt depth6pt width0pt \widetilde{\overline{\omega }}^\alpha{}_\beta $ &\vrule height10pt& $0$\cr
\noalign{\hrule}
$\vrule height12pt depth4pt width0pt 0$ &\vrule height14pt& $\widehat{\overline{\omega
}}^{\tilde a}{}_{\tilde b}$\cr}}\right),
\e
where $\widetilde{\overline{\omega }}^\alpha {}_\beta$ is the connection on the space-time $E$ and
$\widehat{\overline{\omega }}^{\tilde a}{}_{\tilde b}$ is the
connection on the manifold $M = G/G_{0}$ with the following properties:
\begin{gather}
\widetilde{{\ov{D}}} g_{\alpha +\beta -} = \widetilde{{\ov{D}}}
g_{\alpha \beta } - g_{\alpha \delta } \widetilde{\ov{Q}}^\delta
{}_{\beta \gamma }(\widetilde{{\overline{\varGamma}}}) \theta ^{\gamma } =0,\nonumber\\
\widetilde{\ov{Q}}^{\alpha }{}_{\beta \alpha
}(\widetilde{\overline{\varGamma}}) = 0,\label{4.5}\\
\widehat{{\ov{D}}} g_{\tilde a+\tilde b-} = \widehat{{\ov{D}}}
g_{\tilde a\tilde b}- g_{\tilde a\tilde d}\widehat{{\ov{Q}}}^{\tilde
d}{}_{\tilde b\tilde c}(\widehat{{\overline{\varGamma}}}) \overline{\theta
}^{\tilde c} = 0,\label{4.6}
\end{gather}
$\ov{D}$ is the exterior covariant derivative with respect to the connection
$\overline{\omega }^\alpha {}_\beta $ on $E$, $\ov{Q}^{\alpha }{}_{\beta
\gamma }(\overline{\omega})$ is the tensor of torsion for
$\overline{\omega }^{\alpha }{}_{\beta }$. $\widehat{\ov{D}}$ means the
exterior covariant derivative with respect to the connection
$\widehat{\overline{\omega }}^{\tilde a}{}_{\tilde b}$ and
$\widehat{\ov{Q}}^{\tilde a}{}_{\tilde b\tilde
c}(\widehat{\overline{\varGamma}})$ is the tensor of torsion for
$\widehat{\overline{\omega }}^{\tilde a}{}_{\tilde b}$. The second condition of
Eq.~\er{4.5} is not necessary to define Einstein connection on $E$. Moreover
we suppose it to be in line with some classical results. On the
space-time $E$ we also define the second affine connection
$\ov{W}^{\alpha }{}_{\beta }$ such that:
\beq4.7
\ov{W}^{\alpha }{}_{\beta } = \overline{\omega }^{\alpha }{}_{\beta } -
\frac{2}{3} \delta ^{\alpha }{}_{\beta }\ov{W},
\e
where
$$
\ov{W}= \ov{W}_{\gamma } \overline{\theta }^\gamma =
\frac{1}{2}(\ov{W}^{\sigma }{}_{\gamma \sigma } - \ov{W}^{\sigma
}{}_{\sigma \gamma })\overline{\theta }^\gamma .
$$
Thus on the space-time $E$ we have all the geometrical quantities from
N.G.T.: two connections $\overline{\omega }^{\alpha }{}_{\beta }$ and $\ov{W}^{\alpha }{}_{\beta }$ and the nonsymmetric metric
$g_{\alpha \beta}$. Now let us turn to the nonsymmetric
metrization of the bundle $P$. We have:
\beq4.8
\varkappa _{\tilde{\rA}\tilde{\rB}} =
\left(\vcenter{\offinterlineskip\tabskip0pt
\halign{\strut\tabskip5pt plus10pt minus10pt
\hfill#\hfill&\vrule height10pt#&\hfill#\hfill\tabskip0pt\cr
$\vrule height0pt depth6pt width0pt \gamma _{\rAB}$ && $0$\cr
\noalign{\hrule}
$\vrule height9pt depth4pt width0pt 0$ &&
$\rho^2\ell_{ab}$\cr}}\right)
\e
(see Section 2 for comparison), where $\rho = \rho (x)$ is a scalar field on $E$, $x\in E$ and
$$\displaylines{
\gamma _{\rAB} =
\left(\vcenter{\offinterlineskip\tabskip0pt
\halign{\strut\tabskip5pt plus10pt minus10pt
\hfill#\hfill&\vrule height10pt#&\hfill#\hfill\tabskip0pt\cr
$\vrule height0pt depth6pt width0pt g_{\alpha \beta }$ && $0$\cr
\noalign{\hrule}
$\vrule height9pt depth4pt width0pt 0$ &&
$r^2g_{\tilde{a}\tilde{b}}$\cr}}\right)\quad\
\hbox{and}\cr
\ell _{ab} = h_{ab} + \xi k_{ab} ,\quad\ \det (\ell _{ab}) \neq
0\cr}
$$
or
\beq4.9
\bal
\varkappa _{(\tilde{\rA}\tilde{\rB})} \theta ^{\tilde{\rA}} \otimes
\theta ^{\tilde{\rB}} &= \gamma _{(\rAB)} \theta ^{\rA} \otimes \theta
^{\rB} + \rho ^{2} h_{ab} \theta ^{a} \otimes \theta ^{b},\cr
\varkappa _{[\tilde{\rA}\tilde{\rB}]}\theta ^{\tilde{\rA}}\wedge \theta
^{\tilde{\rB}} &= \gamma _{[\rAB]} \theta ^{\rA}\wedge \theta ^{\rB} +
\rho ^{2} \xi k_{ab} \theta ^{a}\wedge \theta ^{b},\eal
\e
(see Section 2 where $\rho=1$),
where $\theta ^{a} = \lambda \omega ^{a}$. From the Kaluza--Klein Theory and
Jordan--Thiry we know
that $\lambda $ is proportional to $\sqrt{G_{N}}$, $\lambda \sim \sqrt{G_{N}}$.
We work with such a system of units that $\lambda = 2$.

\vskip4pt plus2pt
Now we define on $P'$, a connection $\omega^{\tilde{\rA}}{}_{\tilde{\rB}}$ right-invariant
(Ad-covariant) with respect to the action of group $H$ on $P'$ such that:
\beq4.10
D\gamma _{\tilde{\rA}+\tilde{\rB}-} = D\gamma _{\tilde{\rA}\tilde{\rB}} -
\gamma _{\tilde{\rA}\tilde{\rD}}
Q^{\tilde{\rD}}{}_{\tilde{\rB}\tilde{\rC}}
({\varGamma} ) \theta ^{\tilde{\rC}} = 0,
\e
where $\omega^{\tilde{\rA}}{}_{\tilde{\rB}} =
{\varGamma}^{\tilde{\rA}}{}_{\tilde{\rB}\tilde{\rC}} \theta
^{\tilde{\rC}}$. $D$ is the exterior covariant derivative with
respect to the connection $\omega^{\tilde{\rA}}{}_{\tilde{\rB}}$
and $Q^{\tilde{\rA}}{}_{\tilde{\rB}\tilde{\rC}}({\varGamma})$ is the tensor of torsion for
the connection $\omega^{\tilde{\rA}}{}_{\tilde{\rB}}$. After some
calculations one finds:
\bml4.11
\omega^{\tilde{\rA}}{}_{\tilde{\rB}}\\
=\left(\vcenter{\offinterlineskip\tabskip0pt
\halign{\tabskip5pt plus10pt minus10pt
\hfill#\hfill&\vrule height10pt#&\hfill#\hfill\tabskip0pt\cr
$\vrule height0pt depth9pt width0pt \pi^*(\overline{\omega
}^{\rA}{}_{\rB})-\rho^2\ell_{db}\gamma ^{\rMA}L^d{}_{\rMB}\theta ^b$ &&
$L^a{}_{\rBC}\theta ^C-{\textstyle{\frac 1\rho}}\gamma
_{\rBD}\widetilde{\gamma}^{(\rDC)}\rho_{,C}\theta ^a $\cr
\noalign{\hrule}
$\vrule height11pt depth4pt width0pt \rho^2\ell_{bd}\gamma
^{\rAB}(2H^d{}_{\rCB}-L^d{}_{\rCB})\theta ^C-\rho\widetilde{\gamma
}^{(\rAB)}\rho_{,\rB}\ell_{bc}\theta ^C$ &&
${\textstyle{\frac1\rho}}\gamma
_{\rBD}\widetilde{\gamma}^{(\rDC)}\rho_{,C}\delta ^a_b\theta
^{\rB}+\widetilde{\omega }^a{}_b$\cr}}\right),
\e
where $\widetilde{\gamma }^{(\rAB)}$ is the inverse tensor for $\gamma
_{(\rAB)}$
\beq4.12
\bal
\gamma _{(\rAB)} \widetilde{\gamma }^{(\rAC)} &= \delta
^{\rC}{}_{\rB},\cr
L^{d}{}_{\rMB} &= -L^{d}{}_{\rBM},\eal
\e
is an Ad-type tensor (Ad-covariant) on $P'$ such that
\bg4.13
\ell _{dc} \gamma _{\rMB} \gamma ^{\rCM} L^{d}{}_{\rCA} +
\ell _{cd} \gamma _{\rAM} \gamma ^{\rMC} L^{d}{}_{\rBC} =
2\ell _{cd} \gamma _{\rAM} \gamma ^{\rMC}
H^{d}{}_{\rBC},\\
\widetilde{\omega }^{a}{}_{b} = \widetilde{{\varGamma} }^{a}{}_{bc}
\theta ^{c},\nonumber\\
\ell _{db} \widetilde{{\varGamma} }^{d}{}_{ac} +
\ell _{ad}\widetilde{{\varGamma} }^{d}{}_{cb} = -\ell
_{db}C^{d}{}_{ac},\label{4.14}\\
\widetilde{{\varGamma} }^{d}{}_{ac} = -\widetilde{{\varGamma}
}^{d}{}_{ca} ,\quad\ \widetilde{{\varGamma} }^{d}{}_{ad} = 0.\nonumber
\e
We define on $P'$ a second connection:
\beq4.15
W^{\tilde{\rA}}{}_{\tilde{\rB}} = \omega ^{\tilde{\rA}}{}_{\tilde{\rB}} -
{\frac4{3(m+2)}} \delta ^{\tilde{\rA}}{}_{\tilde{\rB}}\ov{W}.
\e
Thus we have on $P'$ all $(m+4)$-dimensional analogues of geometrical
quantities from N.G.T., i.e.:
$$
W^{\tilde{\rA}}{}_{\tilde{\rB}} , \omega
^{\tilde{\rA}}{}_{\tilde{\rB}}\quad\hbox{and}\quad \varkappa
_{\tilde{\rA}\tilde{\rB}} .
$$
The connection \er{4.15} is analogous to the connection $W$
from NGT. The form $\ov{W}$ is horizontal one,
hor $\ov{W} = \ov{W}$ (in the sense of the connection $\omega $ on the bundle~$P'$).

Let us calculate the Moffat--Ricci curvature scalar for the
connection $W^{\tilde{\rA}}{}_{\tilde{\rB}}$.
\beq4.16
R(W) = \varkappa ^{\tilde{\rA}\tilde{\rB}}
\biggl(R^{\tilde{\rC}}{}_{\tilde{\rA}\tilde{\rB}\tilde{\rC}}(W)+
{\frac12}R^{\tilde{\rC}}{}_{\tilde{\rC}\tilde{\rA}\tilde{\rB}}(W)\biggr),
\e
where $R^{\tilde{\rA}}{}_{\tilde{\rB}\tilde{\rC}\tilde{\rD}}(W)$ is the curvature tensor for the connection
$W^{\tilde{\rA}}{}_{\tilde{\rB}}$.
One gets:
\begin{multline}\label{4.17}
R(W) ={}\ov{R}(\ov{W}) + {\frac1{r^{2}}}
R(\widehat{\overline{\varGamma}}) + {\frac1{\lambda ^{2} \rho ^{2}}}
\widetilde{R}(\widetilde{{\varGamma} })\\
{}- \frac{\lambda ^{2} \rho ^{2}}{ 4} \ell _{ab}
(2H^{a}H^{b}-L^{a\rMN}H^{b}{}_{MN})
- \frac{2\lambda ^{2}\widetilde{\rM}}{ \rho ^{2}} \widetilde{\gamma
}^{(\rBN)}\rho _{,\rB} \rho _{,\rN} + \ul P(\rho ),
\end{multline}
where:
\bg4.18
\bal
\ul P(\rho ) &= \frac{\lambda ^{2}n}{ 8\rho } \gamma _{\rDB}
\widetilde{\gamma }^{(\rDC)} \rho _{,\rC} \gamma
^{\rMN}\ov{Q}^{\rB}{}_{\rMN}(\overline{\omega }) \\
\noalign{\vskip4pt}
&+ \frac{\lambda ^{2}n}{ 4\rho ^{2}}\overline{\nabla}_{\rA} (\rho
\widetilde{\gamma}^{(\rAB)}\rho _{,\rB}) + \frac{\lambda ^{2}}{ 4}
n\gamma ^{\rBC}\overline{\nabla}_{\rC} \biggl(\frac{1}{ \rho }\gamma _{\rBD}
\widetilde{\gamma }^{(\rDE)}\rho _{,\rE}\biggr) \\
\noalign{\vskip4pt}
&{}+ \frac{\lambda ^{2}\rho }{ 8} \gamma ^{\rMN} \biggl\{\overline{\nabla}_{\rM}
\biggl(\frac{1}{ \rho }\gamma _{\rDN}\widetilde{\gamma }^{(\rDC)}\rho
_{,\rC}\biggr)
- \overline{\nabla}_{\rN} \biggl(\frac{1}{ \rho }\gamma _{\rDM}\widetilde{\gamma
}^{(\rDC)}\rho _{,\rC}\biggr)\biggr\},
\eal \tag{4.18}\\
\widetilde{\rM} = \ell ^{[dc]} \ell _{[dc]} - n(n-1),\tag{4.19}
\e
$\ov{Q}^{\rB}{}_{\rMN}(\overline{\omega })$ means the torsion of the connection
$\overline{\omega }^{\rA}{}_{\rB}$ on $V = E \times M = E \times G/G_{0}$
and $\overline{\nabla}_{\rA}$ is the covariant derivative with respect to this connection.
$\ov R(\ov W)$ is the Moffat--Ricci curvature scalar on the space-time $E$ for the
connection $\ov{W}^{\alpha }{}_{\beta }$.
$R(\widehat{\overline{\varGamma}})$ is the Moffat--Ricci curvature scalar for the
connection $\widehat{\overline{\omega }}^{\alpha }{}_{\beta }$ on the homogeneous space $M =
G/G_{0}$, $\widetilde{R}(\widetilde{{\varGamma} })$ is the
Moffat--Ricci curvature scalar for the connection $\widetilde{\omega
}^{a}{}_{b}$. Notice that in
the curvature scalar we pass from $\lambda =2$ to the arbitrary value of this
constant.
\setcounter{equation}{19}
\bg4.20
H^{a} = \gamma ^{[\rAB]} H^{a}{}_{[\rAB]} = g^{[\alpha \beta ]}
H^{a}{}_{\alpha \beta } + \frac{1}{ r^{2}} g^{[\tilde a\tilde b]}
H^{a}{}_{\tilde a\tilde b}, \\
\bal
L^{aMN} &= \gamma ^{\rAM} \gamma ^{\rBN} L^{a}{}_{\rAB} =
\delta ^{\rM}{}_{\mu } \delta ^{\rN}{}_{\gamma } g^{\alpha \mu }
g^{\beta \gamma } L^{a}{}_{\alpha \beta } \\
&+ \frac{1}{ r^{2}} (g^{\alpha \mu } g^{\tilde b\tilde n}
L^{a}{}_{\alpha \tilde{b}} + g^{\tilde a\tilde n} g^{\beta \gamma }
L^{a}{}_{\tilde a\beta }) \delta ^{\rM}{}_{\mu } \delta
^{\rN}{}_{\tilde n} \\
&+ \frac{1}{ r^{4}} g^{\tilde a\tilde m} g^{\tilde b\tilde n}
L^{a}{}_{\tilde a\tilde b} \delta ^{\rM}{}_{\tilde m} \delta
^{\rN}{}_{\tilde n}.\eal \label{4.21}
\e
Let us consider the condition \eqref{4.13}. One gets:
\bea4.22
\ell _{dc} g_{\mu \beta } g^{\gamma \mu } L^{d}{}_{\gamma \alpha } +
\ell _{cd} g_{\alpha \mu } g^{\mu \gamma } L^{d}{}_{\beta \gamma }
&=2\ell _{cd} g_{\alpha \mu } g^{\mu \gamma } H^{d}{}_{\beta \gamma},\\
\ell _{dc} g_{\tilde m\tilde b} g^{\tilde c\tilde m} L^{d}{}_{\tilde
c\tilde a} + \ell _{cd} g_{\tilde a\tilde m} g^{\tilde m\tilde c}
L^{d}{}_{\tilde b\tilde c} &= 2\ell _{cd} g_{\tilde a\tilde m}
g^{\tilde m\tilde c} H^{d}{}_{\tilde b\tilde c},\label{4.23}\\
\ell _{dc} g_{\mu \beta } g^{\gamma \mu } L^{d}{}_{\gamma \tilde a} +
\ell _{cd} g_{\tilde a\tilde m} g^{\tilde m\tilde c} L^{d}{}_{\beta \tilde
c} &= 2\ell _{cd} g_{\tilde a\tilde m} g^{\tilde m\tilde c}
H^{d}{}_{\beta \tilde c}.\label{4.24}
\e
One finds that:
\beq4.25
\bal
-\ell _{ab} L^{a\rMN} H^{b}{}_{\rMN} &= -\ell _{ab}
\biggl(g^{\alpha \mu } g^{\beta \nu } L^{a}{}_{\alpha \beta }
H^{b}{}_{\mu \nu }\\
&\qquad{} + \frac{2}{ r^{2}} g^{\alpha \mu } g^{\tilde b\tilde n}
L^{a}{}_{\alpha \tilde b} H^{b}{}_{\mu \tilde n}
+ \frac{1}{ r^{4}} g^{\tilde a\tilde m} g^{\tilde b\tilde n} L^{a}{}_{\tilde
a\tilde b} H^{b}{}_{\tilde m\tilde n}\biggr)\\
&=-\ell _{ab} \biggl(L^{a\mu \nu }H^{b}{}_{\mu \nu }+\frac{2}{ r^{2}}
g^{\tilde b\tilde n}L^{\alpha \mu }{}_{\tilde b}H^{b}{}_{\mu \tilde n}\\
&\qquad{}+ \frac{1}{ r^{4}}g^{\tilde a\tilde m}g^{\tilde b\tilde n}L^{a}{}_{\tilde
a\tilde b}H^{b}{}_{\tilde m\tilde n}\biggr),
\eal
\e
where:
$$
\aligned
L^{a \mu \nu } &= g^{\alpha \mu } g^{\beta \nu } L^{a}{}_{\alpha
\beta },\\
L^{a \mu }{}_{\tilde b} &= g^{\alpha \mu } L^{a}{}_{\alpha \tilde b}.
\endaligned
$$
For $\ell _{ab} H^{a} H^{b} = h_{ab} H^{a} H^{b}$, we have the following:
\beq4.26
h_{ab} H^{a} H^{b} = h_{ab} H^{a}{}_{0} H^{b}{}_{0} + \frac{2}{ r^{2}}
h_{ab} H^{a}{}_{0} H^{b}{}_{1} + \frac{1}{ r^{4}} h_{ab} H^{a}{}_{1}
H^{b}{}_{1},
\e
where
\beq4.27
H^{a}{}_{0} = g^{[\alpha \beta ]} H^{a}{}_{\alpha \beta }
\e
and
\beq4.28
H^{a}{}_{1} = g^{[\tilde a\tilde b]} H^{a}{}_{\tilde a\tilde b}.
\e
And finally we get for $R(W)$:
\beq4.29
\aligned
R(W) &= \ov{R} (\ov{W}) + \frac{1}{r^{2}}
R(\widehat{\overline{\varGamma}}) + \frac{4}{\lambda ^{2} r^{2}}
\widetilde{R}(\widetilde{{\varGamma} })
- \frac{\lambda ^{2} \rho ^{2}}{4} \ell _{ab} (2H^{a}{}_{0}H^{b}{}_{0}-
L^{a\mu \nu }H^{b}{}_{\mu \nu }) \\
&- \frac{\lambda ^{2} \rho ^{2}}{4r^{2}} \ell _{ab}
(4H^{(a}_{0}H^{b)}_{1}-2g^{\tilde b\tilde n}L^{a\mu }_{\tilde b}
H^{b}{}_{\mu \tilde n}) \\
&- \frac{\lambda ^{2} \rho ^{2}}{4r^{4}} \ell _{ab}
(2H^{a}{}_{1}H^{b}{}_{1}-g^{\tilde a\tilde m}g^{\tilde b\tilde
n}L^{a}{}_{\tilde a\tilde b}H^{b}{}_{\tilde m\tilde n}) \\
&- \frac{\widetilde{M}}{\rho ^{2}} g_{\gamma \delta }
\widetilde{g}^{(\delta \nu )} \rho _{,\nu } \widetilde{g}^{(\gamma
\beta )} \rho _{,\beta } + \ul{P}(\rho ),
\eal
\e
where $\widetilde{g}^{(\delta \nu )}$ is the inverse tensor of
$g_{(\alpha \beta )}$ such that:
\beq4.30
\widetilde{g}^{(\delta \nu )} g_{(\delta \mu )} = \delta ^{\nu
}{}_{\mu }.
\e
Let us calculate the density for the Moffat--Ricci curvature scalar for
the connection $W^{\tilde{\rA}}{}_{\tilde{\rB}}$. We have:
\beq4.31
\sqrt{|\varkappa|} R(W) = \sqrt{-g} r^{n_{1}} \sqrt{|\widetilde{g}|}
\sqrt{|\ell|} \rho ^{n} R(W),
\e
where
\beq4.32
g = \det (g_{\alpha \beta }) ,\quad \widetilde{g} = \det (g_{\tilde a\tilde
b}) ,\quad \ell = \det (\ell _{ab})
\e
and
\beq4.33
\varkappa = \det (\varkappa _{\tilde{\rA}\tilde{\rB}}) = g \cdot
r^{n_{1}} \cdot \widetilde{g} \cdot \rho ^{n} \ell .
\e
After some calculation one gets: 
\beq4.34
\bal
\sqrt{|\varkappa|} R(W) &= \sqrt{-g}r^{n_{1}}\sqrt{|\widetilde{g}|}
\ell \biggl\{\rho ^{n}\ov{R}(\ov{W}) + \frac{\widetilde{R}(\widetilde{{\varGamma} })
}{ \lambda ^{2}\rho ^{2-n}} + \frac{\rho ^{n}}{ r^{2}}
R(\widehat{\overline{\varGamma}}) \\
&+ \frac{\lambda ^{2}}{ 4} \rho ^{n+2} \ell _{ab} (2H^{a}{}_{0}H^{b}{}_{0}-
L^{a\mu \nu }H^{b}{}_{\mu \nu }) \\
&+ \frac{\lambda ^{2}}{ 4r^{2}} \rho ^{n+2} \ell _{ab}
(4H^{(a}_{0}H^{b)}_{1}-2g^{\tilde b\tilde n}L^{a\mu }{}_{\tilde b}
H^{b}{}_{\mu \tilde n}) \\
&+ \frac{\lambda ^{2}}{4r^{4}} \rho ^{n+2} \ell _{ab}
(2H^{a}_{1}H^{b}_{1}-g^{\tilde a\tilde m}g^{\tilde b\tilde n}
L^{a}{}_{\tilde a\tilde b}H^{b}{}_{\tilde m\tilde n}) \\
&+ \frac{\lambda ^{2}}{ 4} \rho ^{n-2} \bigl(\ov{M}\widetilde{g}^{(\gamma
\mu )} \rho _{,\gamma } \rho _{,\mu } \\
&+ n^{2}g^{[\mu \nu ]} g_{\delta \mu } \widetilde{g}^{(\delta \gamma )}
\rho _{,\nu } \rho _{,\gamma }\bigr)\biggr\}+ \partial
_{\rM}K^{\rM},
\eal
\e
where
\beq4.35
\ov{\rM}= (\ell ^{[dc]}\ell _{[dc]} -3n(n-1))
\e
and
$$
K^{\rM} = \frac{n}{2} \rho ^{n-1} \sqrt{|\varkappa|}
(5\widetilde{\gamma }^{(\rMC)}-\gamma ^{\rNM}\gamma _{\rDN}
\widetilde{\gamma }^{(\rDC)})\cdot \rho _{,\rC}.\eqno(4.35')
$$
If we define an integral of action:
$$
S \sim \int_U \sqrt{|\varkappa|} R(W) d^{m+4}x,\eqno(4.36)
$$
(where $d^{m+4}x = d^{4}x\, d\mu _{\rH}(h)\cdot dm(y)$, $d\mu _{\rH}(h)$ is a biinvariant measure on a
group $H$ and $dm(y)$ is a measure on $M$ induced by a biinvariant measure on
a group $G$, $x\in E$, $h\in H$, $y\in M)$, and the variation principle for $R(W)$ then the
full divergence $\partial _{\rM}K^{\rM}$ does not play any role. It could play a certain
role in topological problems.

One can write
$$
K^{\mu } = \frac{n}{ 2} \rho ^{n-1}\sqrt{g\widetilde{g}}
(5\tilde{g}^{(\mu \gamma )}-g^{\nu \mu }g_{\delta \nu
}\widetilde{g}^{(\delta \gamma )}) \rho _{,\gamma }
$$
and
$$
K^{\tilde m} = \frac{n}{ 2r^{2}} \rho ^{n-1}\sqrt{g\widetilde{g}}
(5g^{(\tilde m\tilde c)} - g^{\tilde n\tilde m} g_{\tilde d\tilde n}
\widetilde{g}^{(\tilde d\tilde c)} ) \rho _{,\tilde c}.
$$
Thus we really only have to deal with $B(W)$:
\beq4.37
B(W) =\sqrt{\varkappa} R(W) - \partial _{\rM}K^{\rM}.
\e
We have:
\beq4.38
\rho _{,\tilde a} = 0,
\e
for every $\widetilde a= 5, 6, 7,\dots ,n_{1}+4$.

It is worth to notice that the formulae \er{4.21}--\er{4.23} do not change
under the general conformal transformation---the redefinition of the
tensor $g_{\mu \nu }$. This is a general property of these formulae i.e.
\beq4.39
g_{\mu \nu } \rightarrow C\cdot g_{\mu \nu },
\e
where $C = C(x)$ is the conformal factor.

$R(W)$ is invariant with respect to the right action
of the group $H$ on ${P}$. Thus an integration over the group $H$ is trivial.
Let us consider the following two $\Ad_{H}$-type 2-forms with values in the
Lie algebra of $H,(\fh)$.
\beq4.40
\ov{L}=\frac{1}{2}L^{d}{}_{\rMB}\theta ^{\rM}\wedge\theta
^{\rB}X_{d}
\e
and
\beq4.41
\ov{Q}=\frac{1}{2}Q^{d}{}_{\rMB}\theta ^{\rM}\wedge\theta
^{\rB}X_{d}.
\e
One gets
\beq4.42
\ov{L}= {\varOmega} - \frac{1}{2}\ov{Q},
\e
where ${\varOmega} $ is the curvature of the connection $\omega $ on
$P'$ (over $E \times G/G_{0})$ and
$Q^{d}{}_{\rMB}$ is torsion in additional gauge dimensions. This generalizes our
considerations in Nonabelian Kaluza--Klein Theory to higher dimensional
space-time.

Let us repeat considerations from Section~2. One finds:
\beq4.43
\ell _{ij} g_{\mu \beta } g^{\gamma \mu } \widetilde{L}^{i}{}_{\gamma
\alpha } + \ell _{ji} g_{\alpha \mu } g^{\mu \gamma }
\widetilde{L}^{i}{}_{\beta \gamma } = 2 \ell _{ji} g_{\alpha \mu }
g^{\mu \gamma } \widetilde{H}^{i}{}_{\beta \gamma },
\e
where $\ell _{ij} = \ell _{cd} \alpha ^{c}{}_{i} \alpha ^{d}{}_{j}$ is a
right-invariant nonsymmetric metric on the group $G$ and
\beq4.44
L^{c}{}_{\mu \nu } = \alpha ^{c}{}_{i} \widetilde{L}^{i}{}_{\mu \nu},
\e
$\widetilde{L}^{i}{}_{\mu \nu }$ plays the role of an induction
tensor for the Yang--Mills' field with the gauge group $G$.
$\widetilde{H}^{i}{}_{\mu \nu }$ is of
course the tensor of strength of this field. The polarization
tensor is defined as usual
\beq4.45
\widetilde{L}^{i}{}_{\mu \nu } = \widetilde{H}^{i}{}_{\mu \nu } - 4\pi
\widetilde{M}^{i}{}_{\mu \nu }.
\e
We introduce two $\Ad_{G}$-type 2-forms with values in the Lie algebra
$\fg$ (of $G)$
\begin{align*}
\widetilde{L} &=\frac{1}{2}\widetilde{L}^{i}{}_{\mu \nu }\theta ^{\mu }\wedge
\theta ^{\nu }Y_{i},\\
\ov{\widetilde{M}} &=\frac{1}{2}\widetilde{M}^{i}{}_{\mu \nu }\theta ^{\nu
}\wedge\theta ^{\nu }Y_{i}
\end{align*}
and one easily writes
$$
\widetilde{L} = \widetilde{{\varOmega} }_{\rE} - 4\pi \ov{\widetilde{M}} =
\widetilde{{\varOmega} }_{\rE} -\frac{1}{2}\,\wt Q,
$$
where $\widetilde{Q} =\frac{1}{2}\widetilde{Q}^{i}{}_{\mu \nu }
\theta ^{\mu }\wedge\theta ^{\nu }Y_{i}$.

In this way we get a geometrical interpretation of a Yang--Mills'
induction tensor in terms of the curvature and torsion in additional
dimensions.

One gets:
\beq4.46
\ell _{dc} g_{\mu \beta } g^{\gamma \mu } L^{d}{}_{\gamma \tilde a} +
\ell _{cd} g_{\tilde a\tilde m} g^{\tilde m\tilde c} L^{d}{}_{\beta \tilde
c} = 2\ell _{cd} g_{\tilde a\tilde m} g^{\tilde m\tilde c}
\Nabla^{\gauge}{}_{\beta }({\varPhi} ^{d}{}_{\tilde c}).
\e
One finds:
\beq4.47
\ell _{dc} g_{\tilde m\tilde b}g^{\tilde c\tilde m} L^{d}{}_{\tilde c\tilde a}
+ \ell _{cd} g_{\tilde a\tilde m} g^{\tilde m\tilde c} L^{d}{}_{\tilde
b\tilde c}
= 2 \ell _{cd} g_{\tilde a\tilde m} g^{\tilde m\tilde c}
(C^{d}_{ab}{\varPhi} ^{a}_{\tilde b}{\varPhi} ^{b}_{\tilde b}-
\mu ^{d}_{\dah{i}}f^{\dah{i}}_{\tilde b\tilde c}-{\varPhi} ^{d}_{\tilde
d}f^{\tilde d}_{\tilde b\tilde c}),
\e
In this way we have natural communication to the \NK\ (see Section 2)
\beq4.48
\bal
B(W) &= \sqrt{-g|\widetilde{g}||\ell|}
\biggl\{r^{n_{1}}\rho ^{n}\ov{R}(\ov{W}) + \frac{\rho ^{n-2}}{\lambda ^{2}}
r^{n_{1}} \widetilde{R}(\widetilde{{\varGamma} })\\
\noalign{\vskip2pt}
&+ r^{n_{1}-2}\rho^n \widehat{\ov{R}}(\widehat{\overline{\varGamma}})
- r^{n_{1}}\rho ^{n+2} \frac{\lambda ^{2}}{4} \widetilde{\ell }_{ij}
(2\widetilde{H}^{i}\widetilde{H}^{j}-\widetilde{L}^{i\mu\nu
}\widetilde{H}^{j}{}_{\mu\nu }) \\
\noalign{\vskip2pt}
&+ \frac{\lambda ^{2}\rho ^{n+2}r^{n_{1}-2}}{2} \ell _{ab} g^{\tilde b
\tilde n} L^{a\mu}{}_{\tilde b } \Nabla^{\gauge}{}_\mu {\varPhi}
^{b}{}_{\tilde n} \\
\noalign{\vskip2pt}
&- \frac{\lambda ^{2}\rho ^{n+2}r^{n_{1}-4}}{4} \ell _{ab}
[2g^{[\tilde a \tilde b ]} (C^{a}_{cd}{\varPhi} ^{c}{}_{\tilde a }
{\varPhi} ^{d}{}_{\tilde b }-\mu ^{a}{}_{\dah{i}}f^{\dah{i}}{}_{\tilde
a \tilde b }-{\varPhi} ^{a}{}_{\tilde d }f^{\tilde d }{}_{\tilde a
\tilde b} ) \\
\noalign{\vskip2pt}
&\times g^{[\tilde n\tilde m ]} (C^{b}_{ef}{\varPhi} ^{e}{}_{\tilde
n}{\varPhi} ^{f}{}_{\tilde m }-\mu ^{b}{}_{\dah{i}}f^{\dah{j}}_{\tilde
n\tilde m }-{\varPhi} ^{b}{}_{\tilde e}f^{\tilde e}{}_{\tilde
n\tilde m }) \\
&- g^{\tilde a \tilde m } g^{\tilde b \tilde n} L^{a}{}_{\tilde a
\tilde b }\cdot (C^{b}_{cd}{\varPhi} ^{c}{}_{\tilde m }{\varPhi}
^{d}{}_{\tilde n}-\mu ^{b}{}_{\dah{i}}f^{\dah{i}}{}_{\tilde m \tilde n}-
{\varPhi} ^{b}{}_{\tilde e}f^{\tilde e}{}_{\tilde m \tilde n})]\\
\noalign{\vskip2pt}
&- \frac{\lambda ^{2}\rho ^{n+2}r^{n_{1}-2}}{2} h_{ab} \alpha ^{a}{}_{i}
\widetilde{H}^{i} g^{[\tilde a \tilde b ]} (C^{b}_{cd}{\varPhi}
^{c}{}_{\tilde a }{\varPhi} ^{d}{}_{\tilde b }-\mu ^{b}{}_{\dah{i}}
f^{\dah{i}}{}_{\tilde a \tilde b }-{\varPhi} ^{b}{}_{\tilde d }
f^{\tilde d }{}_{\tilde a \tilde b })\\
\noalign{\vskip2pt}
&-\phantom{\{} r^{n_{1}} \rho ^{n-2} (\ov{M} \widetilde{g}^{(\gamma \mu)}
\rho _{,\gamma } \rho _{,\mu} + n^{2}g^{[\mu\nu ]} g_{\delta \mu}
\widetilde{g}^{(\delta \gamma )} \rho _{,\nu } \rho _{,\gamma }
)\biggr\},
\eal
\e
where
\beq4.49
\widetilde{L}^{i\mu\nu } = g^{\alpha \mu} g^{\beta \nu }
\widetilde{L}^{i}{}_{\alpha \beta }
\e
and $\widetilde{L}^{i}{}_{\alpha \beta }$ obeys \er{4.43}.

One can define the following two $\Ad_{H}$-type 2-forms with value in the Lie
algebra of $H,(\fh)$
$$
\Check{L} = \frac{1}{2} L^{d}{}_{\tilde a \tilde b }\theta ^{\tilde
a }\wedge \theta ^{\tilde b }X_{d},\quad\
\Check{Q} = \frac{1}{2}Q^{d}{}_{\tilde a \tilde b }\theta ^{\tilde a
}\wedge\theta ^{\tilde b }X_{d},
$$
in such a way that
$$
\Check{L}=\Check{\Omega}-\frac{1}{2}\Check{Q},
$$
where $\Check{\Omega} = \frac{1}{2}H^{d}{}_{\tilde a \tilde b }\theta
^{\tilde a }\wedge \theta ^{\tilde b }X_{d}$. These forms are defined on
$P^\circ$ (over $G/G_{0})$. In
this way we get similar formulae for the Higgs' field and for
Yang--Mills' field.

Let us define the following quantity $M^{d}{}_{\tilde a \tilde b}$ such that
\beq4.50
L^{d}{}_{\tilde a \tilde b } = H^{d}{}_{\tilde a \tilde b } - 4\pi
M^{d}{}_{\tilde a \tilde b }.
\e
This quantity is an analogue of the polarization tensor for the Higgs'
field.

In this way we have
\beq4.50a
\Check{Q}= 8\pi \Check{M}= 4\pi M^{d}{}_{\tilde a \tilde b }
\theta ^{\tilde a }\wedge \theta ^{\tilde b }X_{d}.
\e
The form $\Check{M}$ is an $\Ad_{\rH}$-type 2-form.
\beq4.51
\widetilde{H}^{i} = g^{[\alpha \beta ]} \widetilde{H}^{i}{}_{\alpha
\beta },
\e
$L^{a}{}_{\alpha \beta }$ obeys \er{4.47} and scalar field
${\varPhi} ^{a}{}_{\tilde b}$ satisfies the following
constraints:
\beq4.52
{\varPhi} ^{c}{}_{\tilde b } f^{\tilde b }{}_{\dah{i}\tilde a } =
\mu ^{a}{}_{\dah{i}} {\varPhi} ^{b}{}_{\tilde a } C^{c}_{ab}
\e
(see Section 2).
Let us define the integral of action for $B(W)$
\beq4.53
S = - \frac{1}{ V_{1} V_{2}
r^{n_{1}}}\int_{{\mathbf U}}((B(W)\,d^{n}x)d^{n_{1}}x) d^{4}x,
\e
where ${\mathbf U} = \ov V \times M \times H$, $\ov V \subset E$.
\beq4.53a
V_{1} = \int_{H} \sqrt{-\ell}\,d^{n}x,
\e
where $d^{n}x = d\mu _{\rH}(h)$ is a biinvariant measure on a group $H$ and
\beq4.54
V_{2} = \int_{M} \sqrt{|\widetilde{g}|}\,d^{n_{1}}x,
\e
$d^{n_{1}}x = dm(y)$ is a measure on $M$ which is quasi-invariant with respect to
the action of $G$ on $M$ and $d^{4}x$ means integration over space-time
coordinates. After some calculations one gets:
\beq4.55
S = - \int_{\ov V}\sqrt{-g} B(\ov{W}, g, \widetilde{A}, \rho ,
{\varPhi})\, d^{4}x,\quad\ \ov V \subset E,
\e
where
\bml4.56
B(\ov{W}, g, \widetilde{A}, \rho , {\varPhi}) = \rho ^{n}\cdot
\ov{R}(\ov{W}) + \frac{\lambda^2}{4}\biggl[8\pi
\rho ^{n+2} {\cal L}_{\rYM}(\widetilde{A})
+ \rho ^{n+2} \frac{2}{ r^{2}} {\cal L}_{\kin} (\Nabla^{\gauge} {\varPhi}
)\\
{}+ \rho ^{n+2} \frac{1}{ r^{4}} \wh V({\varPhi} )
- \rho ^{n+2} \frac{4}{ r^{2}} {\cal L}_{\Int}({\varPhi}
,\widetilde{A}) + \rho ^{n-2} {\cal L}_{\scal}(\rho )\biggr]+ \lambda_{c},
\e
$\ov{R}(\ov{W})$ is the Moffat--Ricci curvature scalar on the space-time $E$ and plays
the role of the gravitational lagrangian
\beq4.57
{\cal L}_{\rYM}(\widetilde{A}) = - \frac{1}{8\pi } \ell _{ij}
(2\widetilde{H}^{i}\widetilde{H}^{j}-\widetilde{L}^{i\mu\nu }
\widetilde{H}^{j}{}_{\mu\nu })
\e
is the lagrangian for the Yang--Mills' field with the gauge group $G$ in
the Nonsymmetric-Nonabelian Kaluza--Klein Theory. $\widetilde{H}^{i}$ is defined in
\er{4.51} and $\widetilde{L}^{i\mu\nu }$ by \er{4.43}.
\beq4.58
\bal
{\cal L}_{\kin}(\Nabla^{\gauge} {\varPhi} ) &= \frac{1}{
V_{2}}\int_{M}\sqrt{|\widetilde{g}|} d^{n_{1}}x (\ell
_{ab}g^{\tilde b \tilde n }L^{a\mu}{}_{\tilde b
}\Nabla^{\gauge}{}_\mu {\varPhi} ^{b}{}_{\tilde n }) \\
& = \ell _{ab} g^{\alpha \mu} \frac{1}{V_{2}}\int_{M}\sqrt{|\widetilde{g}|}
d^{n_{1}}x (g^{\tilde b \tilde n }L^{a}{}_{\alpha \tilde b
}\Nabla^{\gauge}{}_\mu {\varPhi} ^{b}{}_{\tilde n })
\eal
\e
is the kinetic part of the lagrangian for the scalar field ${\varPhi}
^{b}{}_{\tilde n }$. It is
easy to see that ${\cal L}_{\kin}$ is a quadratic form with respect to
$\Nabla^{\gauge}{\varPhi}$ (gauge
derivative with respect to the connection $\widetilde{\omega }_{\rE})$ and is invariant with
respect to the action of the groups $H$ and $G$.
\beq4.59
\bal
\wh V({\varPhi} ) &= \frac{\ell _{ab}}{ V_{2}}\int_{M}\sqrt{|\widetilde{g}|}
d^{n_{1}}x \bigl[2g^{[\tilde m \tilde n ]}(C^{a}_{cd} {\varPhi}
^{c}{}_{\tilde m } {\varPhi} ^{d}{}_{\tilde n } - \mu ^{a}_{\dah{i}}
f^{\dah{i}}{}_{\tilde m \tilde n }
- {\varPhi} ^{a}{}_{\tilde e} f^{\tilde e}{}_{\tilde m \tilde n })
g^{[\tilde a \tilde b ]}\\
&\cdot (C^{b}_{ef}{\varPhi} ^{e}{}_{\tilde
a }{\varPhi} ^{f}{}_{\tilde b }-\mu ^{b}{}_{\dah{j}}f^{\dah{j}}{}_{\tilde a \tilde b }-{\varPhi} ^{b}{}_{\tilde a }f^{\tilde d
}{}_{\tilde a \tilde b }) \\
&- g^{\tilde a \tilde m } g^{\tilde b \tilde n } L^{a}{}_{\tilde a
\tilde b } (C^{b}_{cd}{\varPhi} ^{c}{}_{\tilde m }{\varPhi}
^{d}{}_{\tilde n }-\mu ^{b}_{\dah{i}}f^{\dah{i}}{}_{\tilde m \tilde n }-
{\varPhi} ^{b}{}_{\tilde e }f^{\tilde e }{}_{\tilde m \tilde n})\bigr]
\eal
\e
is the self-interacting term for the field ${\varPhi} $. It is invariant with
respect to the action of the groups $H$ and $G$. This term is a polynomial
of 4th order in ${\varPhi}$'s (a Higgs' field potential term).
\beq4.60
{\cal L}_{\Int}({\varPhi} ,\widetilde{A}) = h_{ab} \mu ^{a}_{i}
\widetilde{H}^{i}\ul{g}^{[\tilde a \tilde b ]} \cdot (C^{b}_{cd}
{\varPhi} ^{c}{}_{\tilde a }{\varPhi} ^{d}{}_{\tilde b }-
\mu ^{b}{}_{\dah{i}}f^{\dah{i}}{}_{\tilde a \tilde b }-{\varPhi}
^{b}{}_{\tilde d }f^{\tilde d }{}_{\tilde a \tilde b })
\e
is the term describing nonminimal coupling between the scalar field ${\varPhi} $
and the Yang--Mills' field where
\bg4.61
\ul{g}^{[\tilde a \tilde b ]} = \frac{1}{ V_{2}}\int_{M}\sqrt{|\widetilde{g}|}
d^{n_{1}}x g^{[\tilde a \tilde b ]},\\
\lambda _{c} = \frac{\rho ^{n}}{ \lambda ^{2}} \widetilde{R}
(\widetilde{{\varGamma} }) + \frac{\rho ^{n}}{ r^{2}V_{2}}\int_{M}\sqrt{|\widetilde{g}|}
\widehat{\ov{R}}(\widehat{\overline{\varGamma}})\, d^{n_{1}}x
= \frac{\rho ^{n}}{ \lambda ^{2}} \widetilde{R}(\widetilde{{\varGamma} }) +
\frac{\rho ^{n}}{ r^{2}} \widetilde{\ul P}.\label{4.62}
\e
This term is also invariant with respect to the action of groups $H$ and
$G$. One can write $\cL_{\rm kin}$ and~$V$ in different forms:
\begin{align}
{\cal L}_{\kin}(\Nabla^{\gauge}{\varPhi} ) ={}& \ell _{ab}
(g^{\beta \nu }L^{a\mu}{}_{\tilde b }\Nabla^{\gauge}{}_{\mu}{\varPhi}
^{b}{}_{\tilde n })_{av}\nonumber\\
&= \ell _{ab} g^{\alpha \mu} (g^{\tilde b \tilde n }L^{a}{}_{\alpha
\tilde b }\Nabla^{\gauge}{}_\mu {\varPhi} ^{b}{}_{\tilde n
})_{av},\label{4.63} \\
V({\varPhi} ) &= 2 \ell _{ab}([g^{([\tilde m \tilde n ][\tilde a
\tilde b ])} (C^{a}_{cd}{\varPhi} ^{c}{}_{\tilde m }{\varPhi}
^{d}{}_{\tilde n }-\mu ^{a}{}_{\dah{i}}f^{\dah{i}}{}_{\tilde m \tilde n }-
{\varPhi} ^{a}{}_{\tilde e }f^{\tilde e }{}_{\tilde m \tilde n })\nonumber \\
&\quad{}\times (C^{b}_{ef}{\varPhi} ^{e}{}_{\tilde a }{\varPhi}
^{f}{}_{\tilde b }-\mu ^{b}_{\dah{j}}f^{\dah{j}}{}_{\tilde a \tilde b }-
{\varPhi} ^{b}{}_{\tilde d }f^{\tilde d }{}_{\tilde a \tilde b })\nonumber\\
&\quad{}+ \ul{g}^{\tilde a \tilde m \tilde b \tilde n } L^{a}{}_{\tilde a
\tilde b } (C^{b}_{cd}{\varPhi} ^{c}{}_{\tilde m }{\varPhi}
^{d}{}_{\tilde n }-\mu ^{b}{}_{\dah{i}}f^{\dah{i}}{}_{\tilde m \tilde n }-
{\varPhi} ^{b}{}_{\tilde e }f^{\tilde e }{}_{\tilde m \tilde n
})])_{av},\label{4.64}
\end{align}
where
\beq4.65
\bga
(\ldots )_{av} = \frac{1}{ V_{2}} \int_{M}\sqrt{|\widetilde{g}|}\, dx^{n_{1}}
(\ldots ),\\
\ul{g}^{\tilde a \tilde m \tilde b \tilde n } = g^{\tilde a \tilde m }
g^{\tilde b \tilde n },\\
\ul{g}^{([\tilde m \tilde n ][\tilde a \tilde b ])} = g^{[\tilde m
\tilde n ]} \cdot g^{[\tilde a \tilde b ]}.
\ega
\e
The connection $\widehat{{\varGamma}}^{\tilde a }{}_{\tilde b
\tilde c }$ on $M$ satisfies the following compatibility
conditions:
\beq4.66
g_{\tilde d \tilde b }\widehat{\overline{\varGamma}}^{\tilde d
}{}_{\tilde a \tilde c } + g_{\tilde a \tilde d }
\widehat{\overline{\varGamma}}^{\tilde d }{}_{\tilde c \tilde b } =
g_{\tilde a \tilde d } \varkappa ^{\tilde d }{}_{\tilde b \tilde c }
+ g _{\tilde a \tilde b ,\tilde c },
\e
where $\varkappa ^{\tilde d }{}_{\tilde b \tilde c }$ are nonholonomicity coefficients,
``$,$'' means the action of a
vector field defined on $M$ and induced by a left invariant vector field
on $G$. For $g_{\tilde a \tilde b ,\tilde c }$ one easily finds
\beq4.67
g_{\tilde a \tilde b ,\tilde c }(x) = \zeta k^{0}{}_{\tilde a \tilde
b ,\tilde c } + h^{0}{}_{\tilde a \tilde b ,\tilde c }.
\e

For ${\cal L}_{\scal}(\rho )$ we have the following:
\beq4.67a
{\cal L}_{\scal}(\rho ) = -(\ov{M}\widetilde{g}^{(\gamma \mu)}
\rho _{,\gamma } \rho _{,\mu} + n^{2}g^{[\mu\nu ]} g_{\delta\mu}
\widetilde{g}^{(\delta \gamma )} \rho _{,\nu } \rho _{,\gamma }),
\e
where
\beq4.68
\ov{M} = \ell _{[dc]}\ell ^{[dc]}-3n(n-1).
\e
$B(W,\widetilde{A},{\varPhi} ,{\varPsi} )$ is invariant with respect to the right
action of the group $G$ on the bundle $Q(E, G)$. Thus we do not see any
additional dimensions. They can be easily dropped due to the
integration over the group $G$.

We obtained the lagrangian density
$B(\ov{W}, g, \widetilde{A}, \rho , {\varPhi})$ such that:
\beq4.69
\bal
B(\ov{W}, g, \widetilde{A}, \rho , {\varPhi})
&= \sqrt{-g}\cdot \biggl\{\rho ^{n} \cdot \ov{R}(\ov{W})
+ \frac{\lambda ^{2}}{ 4}\biggl[8\pi \rho ^{n+2} {\cal L}_{\rYM}
(\widetilde{A}) \\
\noalign{\vskip5pt}
&+ \rho ^{n+2} \frac{2}{ r^{2}} {\cal L}_{\kin}
(\Nabla^{\gauge} {\varPhi} )
- \rho ^{n+2} \frac{1}{ r^{4}} \wh V({\varPhi} ) - \rho ^{n+2} \frac{4}{ r^{2}}
{\cal L}_{\Int}({\varPhi} , \widetilde{A})\biggr]\\
\noalign{\vskip5pt}
&+ \frac{\lambda ^{2}}{ 4} \rho ^{n-2} {\cal L}_{\scal}(\rho ) +
\frac{4}{ \lambda ^{2}\rho ^{2-n}} \widetilde{R}(\widetilde{{\varGamma} })
+ \rho ^{n} \frac{1}{ r^{2}} \widetilde{\ul{P}}\biggr\},
\eal
\e
where ${\cal L}_{\rYM}$, ${\cal L}_{\kin}$, $\wh V$, ${\cal L}_{\Int}$,
${\cal L}_{\scal}$ and $\widetilde{\ul{P}}$ are defined earlier.
This lagrangian
density is a generalization of the lagrangian from Bergmann's paper
(see Ref.~\cite{27}). P.~G.~Bergmann considers the lagrangian for the
tensor-scalar theory of gravitation including Jordan--Thiry theory and
Brans--Dicke theory. Our lagrangian is more general for two reasons.
We have here nonsymmetric metric tensor $g_{\mu\nu }$ as a metric
$(\ov{R}(\ov{W})$ is the
lagrangian of gravitational field from the Nonsymmetric Theory of
Gravitation). The lagrangian \er{4.69} possesses also apart from lagrangian
for gauge field (in the Bergmann's paper it is an electromagnetic
field) lagrangian for Higgs' field coupled to Yang--Mills' field. We
also get two terms which play the role of the cosmological terms. In
the Bergmann's paper there are four arbitrary functions of scalar field
$\rho $, $f_{1}$, $f_{2}$, $f_{3}$, $f_{4}$. Here we have some functions
in front of the mentioned lagrangians:
\beq4.70
\bal
f_{1}(\rho ) &= \rho ^{n} ,\quad\ &f_{2}(\rho ) &= 8\pi \rho
^{n+2},\cr
f_{2'}(\rho ) &= - \frac{2}{ r^{2}} \rho ^{n+2} ,\quad\ &f_{2''}(\rho ) &= \frac{1}{ r^{4}} \rho
^{n+2},\cr
f_{2'''}(\rho ) &= \frac{4}{ r^{2}} \rho ^{n+2} ,\quad\ &f_{3}(\rho ) &= \rho
^{n+2},\cr
f_{4'}(\rho ) &= \frac{4}{ \lambda ^{2}\rho ^{2-n}} ,\quad\ &f_{4''}(\rho ) &= \frac{1}{ r^{2}} \rho
^{n-2}.
\eal
\e
Now we proceed with the conformal transformation for the metric
$g_{\mu\nu }$ and
the transformation of the scalar field $\rho $. This is only the
redefinition of $g_{\mu\nu }$ and $\rho $.
\bg4.71
\rho = e^{-{\varPsi}},\\
g_{\mu\nu } \rightarrow e^{n{\varPsi} }\cdot g_{\mu\nu } =
\frac{1}{ \rho ^{n}} g_{\mu\nu }.\label{4.72}
\e
This procedure comes of course from Ref.~\cite{27}. The only difference is
that $g_{\mu\nu }$ is now nonsymmetric. After transformations \er{4.71} and \er{4.72} we
get the following:
\bml4.73
B(\ov{W}, g, \widetilde{A}, {\varPsi} , {\varPhi})=
\sqrt{-g}\biggl\{\ov{R}(\ov{W}) + \frac{\lambda ^{2}}{ 4}\biggl(8\pi
e^{-(n+2){\varPsi} } {\cal L}_{\rYM}(\widetilde{A})
+ \frac{2e^{-2{\varPsi} }}{ r^{2}} {\cal L}_{\kin}
(\Nabla^{\gauge}{\varPhi})\\
\noalign{\vskip8pt}
{}- \frac{e^{(n-2){\varPsi} }}{ r^{4}} \wh V({\varPhi} )
- \frac{4e^{(n-2){\varPsi} }}{ r^{2}} {\cal L}_{\Int}({\varPhi} ,
\widetilde{A})\biggr)+ \frac{\lambda ^{2}}{ 4} {\cal L}_{\scal}({\varPsi} )
+ \frac{1}{ \lambda ^{2}} e^{(n+2){\varPsi} } \widetilde{R}(
\widetilde{{\varGamma} }) + \frac{e^{n{\varPsi} }}{ r^{2}}
\widetilde{\ul{P}}\biggr\},
\e
where
\beq4.74
\bga
{\cal L}_{\scal}({\varPsi} ) = -(\ov{M}\widetilde{g}^{(\gamma
\nu )} + n^{2}g^{[\mu \nu ]} g_{\delta \mu } \widetilde{g}^{(\delta
\gamma )}){\varPsi} _{,\nu } {\varPsi} _{,\gamma},\\
\ov{M}= \ell ^{[dc]}\ell _{[dc]}-3n(n-1).
\ega
\e
It is easy to see that the scalar field ${\varPsi} $ is chargeless (it has no
colour charges). However, it couples the gauge (Yang--Mills') field and
the Higgs' field due to the terms:
\begin{gather}
8\pi e^{-(n+2){\varPsi} } {\cal L}_{\rYM}(\widetilde{A}), \tag{\theequation a}\\
\noalign{\vskip4pt}
{}+ \frac{2e^{-2{\varPsi} }}{ r^{2}} {\cal L}_{\kin}
(\Nabla^{\gauge} {\varPhi} ),\tag{\theequation b}\\
{} - \frac{e^{(n-2){\varPsi} }}{ r^{4}} \wh V({\varPhi}),\tag{\theequation c}\\
\noalign{\vskip4pt}
{}- \frac{4e^{(n-2){\varPsi} }}{ r^{2}} {\cal L}_{\Int}({\varPhi} ,
\widetilde{A}),\tag{\theequation d}
\end{gather}
\refstepcounter{equation} \label{4.75}
It also couples the cosmological terms:
\begin{gather}
\frac{4}{ \lambda ^{2}} e^{(n+2){\varPsi} }
\widetilde{R}(\widetilde{{\varGamma}}),\tag{\theequation a}\\
\frac{e^{n{\varPsi} }}{ r^{2}} \widetilde{\ul{P}}.\tag{\theequation b}
\end{gather}
These six terms (\ref{4.74}a--d) and (\ref{4.75}a--b) suggest that the scalar field is
massive. This is different than in Brans--Dicke theory, where the
scalar field couples the trace of the energy-momentum tensor for
matter.

We can consider a more general case for the lagrangian in the
theory, i.e.
\beq4.76
(R(W)+\beta )\sqrt{|\varkappa|}.
\e
In this case we get an additional cosmological term $\beta \rho ^{n}$ or in terms of
the field ${\varPsi} $ and after a redefinition of the nonsymmetric metric
$g_{\mu \nu }$
\beq4.77
\Bigl(\beta \int dm(y)\sqrt{|\widetilde{g}|}\Bigr)e^{n{\varPsi}}.
\e
This term can be added to remaining cosmological terms in the theory.
Moreover we do not consider this term anymore because it is not in the
real spirit of the Kaluza--Klein (Jordan--Thiry) Theory.

Before passing to symmetry breaking in our theory we do some
cosmetic manipulations, connecting constants. The connection $\omega $ on the
fibre bundle $P$ has no correct physical dimensions. It is necessary to
pass in all formulae from $\omega $ to
$\alpha _{s}\frac{1}{\sqrt{\hbar c}} \omega $
\beq4.78
\omega \rightarrow \alpha _{s} \frac{1}{\sqrt{\hbar c}} \omega ,
\e
where $\hbar$ is the Planck's constant, $c$ is the velocity of light in the
vacuum and $\alpha _{s}$ is dimensionless coupling constant for the
Yang--Mills'
field if this field couples matter. For example in the electromagnetic
case $\alpha _{s} = \frac{1}{\sqrt{137}}$. We use $\alpha _{g} = \alpha ^{2}_{s}
= \frac{g^{2}}{\hbar c}$ where $g$ is a coupling constant
for a gauge field. The redefinition of $\omega$ is equivalent to a usual
treatment in a local
section $e:V\supset U\to P$, $e^*\omega=\frac{g}{\hbar c}A$.
Now our quantities have correct physical dimensions.

Using \er{4.78} one easily writes the integral of action \er{4.53}:
\bml4.79
S = - \frac{1}{ r^{2}}\int_{{\mathbf U}}\sqrt{-g}\,d^{4}x\biggl[
\ov{R}(\ov{W}) + \frac{8\pi \lambda ^{2}\alpha ^{2}_{s}}{ 4c\hbar} \cdot
\biggl(e^{-(n+2){\varPsi} } \cdot {\cal L}_{\rYM} \\
{}+ \frac{e^{-2{\varPsi}}}{ 4\pi r^2} {\cal L}_{\kin}-
\frac{e^{(n-2){\varPsi}}}{ 8\pi r^4}\wh V({\varPhi} ) -
\frac{e^{(n-2){\varPsi}}}{ 2\pi r^2} {\cal L}_{\Int} ({\varPhi}
,\widetilde{A})
+{\cal L}_{\scal}({\varPsi})\biggr)+\lambda _c\biggr].
\e
Thus we get the integral of action for the matter described by the
Yang--Mills' field and scalar field coupled to gravity. If we want to
be in line with the ordinary coupling between gravity and matter we
should put:
\beq4.80
\frac{8\pi \lambda ^{2} \alpha ^{2}_{s}}{ 4c\hbar} = \frac{8\pi G_{\rN}}{c^{4}}.
\e
One gets:
\beq4.81
\lambda = \frac{2}{ \alpha _{s}} \ell\pl = \frac{2}{ \sqrt{\alpha
_g}} \ell\pl,
\e
where $\ell\pl$ is the Planck's length $\ell\pl =
\sqrt{\dfrac{G_{\rN}\hbar}{ c^{3}}} \simeq 10^{-33}$~cm. In this case
we have:
\beq4.82
\lambda _{c} = \biggl(\frac{e^{(n+2){\varPsi} } \alpha ^{2}_{s}}{ \ell ^{2}\pl}
\widetilde{R}(\widetilde{{\varGamma} })+
\frac{e^{n{\varPsi} }}{ r^{2}}\widetilde{\ul{P}}\biggr)=\ov\la_{c0}(\Ps).
\e
For $\wh V({\varPhi} )$ one gets:
\beq4.83
\bal
\wh V({\varPhi} ) &= 2 \ell _{ab} \biggl(\ul{g}^{([\tilde m \tilde n
],[\tilde a \tilde b ])}
\times \biggl(\alpha _{s}\frac{1.}{\sqrt{\hbar c}}C^{a}_{cd}{\varPhi}
^{c}{}_{\tilde m }{\varPhi} ^{d}{}_{\tilde n }-\frac{1}{ \alpha _{s}}
\sqrt{\hbox{\dkr{$\hbar$}} c}\mu ^{a}{}_{\dah{i}}f^{\dah{i}}{}_{\tilde m \tilde n }
-{\varPhi} ^{a}{}_{\tilde e }f^{\tilde e }{}_{\tilde m \tilde n }\biggr) \\
\noalign{\vskip4pt}
&\times \biggl(\alpha _{s}\frac{1.}{\sqrt{\hbar c}}C^{b}_{ef}{\varPhi}
^{e}{}_{\tilde a }{\varPhi} ^{f}{}_{\tilde b }-\frac{1}{ \alpha _{s}}
\sqrt{\hbox{\dkr{$\hbar$}} c}\mu ^{b}{}_{\dah{j}}f^{\dah{j}}{}_{\tilde a \tilde b }
-{\varPhi} ^{b}{}_{\tilde d }f^{\tilde d }{}_{\tilde a \tilde
b}\biggr)\biggr)_{av}\\
\noalign{\vskip4pt}
&- \frac{\ell _{ab}}{ V_{2}} \int_{\rM}\sqrt{|\widetilde{g}|}\,d^{n_{1}}x
\biggl[g^{\tilde a \tilde n } g^{\tilde b \tilde m } L^{a}{}_{\tilde a
\tilde b }\\
\noalign{\vskip4pt}
&\times \biggl(\alpha _{s}\frac{1.}{ \sqrt{\hbar c}}C^{b}_{cd}
{\varPhi} ^{c}{}_{\tilde m }{\varPhi} ^{d}{}_{\tilde n }-
\frac{1}{ \alpha _{s}}\sqrt{\hbox{\dkr{$\hbar$}} c}\mu ^{b}{}_{\dah{i}}f^{hat
i}{}_{\tilde m \tilde n }-{\varPhi} ^{b}{}_{\tilde e }f^{\tilde e
}{}_{\tilde m \tilde n }\biggr)\biggr],
\eal
\e
where
\beq4.84
\ul{g}^{([\tilde m \tilde n ],[\tilde a \tilde b ])} =
g^{[\tilde m \tilde n ]} g^{[\tilde a \tilde b ]}.
\e
Let us pass to spontaneous symmetry breaking and the Higgs' mechanism
in our theory. In order to do this we look for the critical points
(the minima) of the potential $\wh V({\varPhi} )$. The scalar factor before $\wh V({\varPhi} )$ is
positive and has no influence on these considerations. However, our
field ${\varPhi} $ satisfies the constraints
\beq4.85
{\varPhi} ^{c}{}_{\tilde b } f^{\tilde b }{}_{\dah{i}\tilde a } -
\mu ^{a}{}_{\dah{i}} {\varPhi} ^{b}{}_{\tilde a } C^{c}_{ab} = 0.
\e
Thus we must look for the critical points of
\beq4.86
V' = \wh V + \psi ^{\dah{i}\tilde d }{}_{c} ({\varPhi} ^{c}{}_{\tilde b }
f^{\tilde b }{}_{\dah{i}\tilde d }-\mu ^{a}{}_{\dah{i}}
{\varPhi} ^{b}{}_{\tilde d }C^{c}_{ab}),
\e
where $\psi ^{\dah{i}\tilde d }{}_{c}$ is a Lagrange multiplier. Let us calculate
$\frac{\delta V'}{ \delta {\varPhi} }$. One finds
\beq4.87
\bal
\frac{\delta V' }{ \delta {\varPhi} ^{w}{}_{\tilde v }} &= \ell
_{ab}\biggl(\biggl\{\biggl[ 4\ul{g}^{([\tilde m \tilde n ],[\tilde a
\tilde b ])}
\biggl(\alpha _{s}\frac{1}{ \sqrt{\hbar c}}C^{a}_{cd}
{\varPhi} ^{c}{}_{\tilde m }{\varPhi} ^{d}{}_{\tilde n }-
\frac{1}{ \alpha _{s}}\sqrt{\hbar c}\mu ^{a}{}_{\dah{i}}
f^{\dah{i}}{}_{\tilde m \tilde n }-{\varPhi} ^{a}{}_{\tilde c }
f^{\tilde c }_{\tilde m \tilde n }\biggr)\biggr] \\
\noalign{\vskip4pt}
&-\ul{g}^{\tilde n \tilde a ,\tilde m \tilde b } \cdot
L^{a}{}_{\tilde n \tilde m } -\ul{g}^{\tilde m \tilde n ,\tilde r
\tilde p } \cdot \biggl( \frac{\delta L^{a}{}_{\tilde m \tilde n }}{
\delta H^{b}{}_{\tilde a \tilde b }}\biggr)\\
\noalign{\vskip4pt}
&\times \biggl(\alpha _{s}\frac{1}{ \sqrt{\hbar c}}C^{d}_{ce}
{\varPhi} ^{c}{}_{\tilde r }{\varPhi} ^{e}{}_{\tilde p }-
\frac{1}{ \alpha _{s}}\sqrt{\hbar c}\mu ^{a}{}_{\dah{i}}
f^{\dah{i}}{}_{\tilde r \tilde p }-{\varPhi} ^{d}{}_{\tilde c }
f^{\tilde c }_{\tilde r \tilde p }\biggr)\\
\noalign{\vskip4pt}
&\times \biggl(\alpha _{s}\frac{1}{ \sqrt{\hbar c}}C^{d}_{ce}
{\varPhi} ^{c}{}_{\tilde r }\delta ^{f}{}_{w}\delta ^{\tilde v
}{}_{\tilde b }-\delta ^{b}{}_{w}\delta ^{\tilde d }{}_{\tilde v }
f^{\tilde d }{}_{\tilde a \tilde b }\biggr)\biggr\}\biggr)_{av}\\
\noalign{\vskip4pt}
&+ \psi ^{\dah{i}\tilde d }{}_{c} (\delta ^{c}{}_{w}
f^{\tilde v }{}_{\dah{i}\tilde d }-\mu ^{a}{}_{\dah{i}}C^{c}{}_{aw}
\delta ^{\tilde v }{}_{\tilde d }),
\eal
\e
where
\beq4.88
H^{b}{}_{\tilde a \tilde b} = \biggl(\alpha _{s}\frac{1}{ \sqrt{\hbar
c}}C^{d}_{cd} {\varPhi} ^{c}{}_{\tilde a }{\varPhi} ^{d}_{\tilde b } -
\frac{1}{ \alpha _{s}}\sqrt{\hbar c}\mu ^{b}{}_{\dah{i}}
f^{\dah{i}}{}_{\tilde a \tilde b } - {\varPhi} ^{b}{}_{\tilde c }
f^{\tilde c }{}_{\tilde a \tilde b }\biggr)
\e
and $\dfrac{\delta L^{a}{}_{\tilde m \tilde n }}{
\delta H^{b}{}_{\tilde a \tilde b }}$ satisfies the following equation:
\beq4.89
\ell _{dc} g_{\tilde m \tilde b } g^{\tilde c \tilde m }
\frac{\delta L^{d}{}_{\tilde e \tilde a }}{ \delta H^{w}{}_{\tilde p
\tilde q }} + \ell _{cd} g_{\tilde a \tilde m } g^{\tilde m \tilde c }
\frac{\delta L^{d}{}_{\tilde b \tilde e }}{ \delta H^{w}{}_{\tilde p
\tilde q }}
= 2\ell _{cd} g_{\tilde a \tilde m } g^{\tilde m \tilde c }
\delta ^{d}{}_{w} \delta ^{\tilde p}{}_{\tilde b } \delta ^{\tilde
q}_{\tilde c }.
\e
It is easy to see that, if
\beq4.90
H^{a}{}_{\tilde m \tilde n } = 0,
\e
then
\beq4.91
\frac{\delta V' }{ \delta {\varPhi} } = 0,
\e
if \er{4.85} is satisfied. In this way we get results from Section 2, i.e.\
a mass matrix for broken gauge bosons can be calculated.
\beq A.21
\bal
{}M_{ij}^2(\F\dr{crt}^k)&=\frac{\a_s^2}{\hbar c}\,\frac1{V_2}
\int_M \sqrt{|\wt g|}\,d^{n_1}x\\
&\Bigl\{l_{np}g^{\u m\u p}\gd\br{(k)}B, p,\u pi,
\Bigl(\gd\br{(k)}B, d,\u mj,
+\gd\xi k, n,d,\gd\br{(k)}B,d,\u mj,
-\z \gd\br{(k)}B, d,\u aj,\gd k, o\u a,\u m,\\
&-\xi\z^2\gd k, n,d,\gd k, o\u b,\u m,\gd k, o\u a,\u b,\gd\br{(k)}B, d,\u aj,
+\xi^2\z k^{nb}k_{bd}\gd k, o\u a,\u m,\gd\br{(k)}B, d,\u aj,\Bigr)\Bigr\}
\eal
\e
$k=0,1$, where
\beq A.22
\gd\br{(k)}B, b,\u ni,=\bigl[\gd\d,\u m,\u n, \gd C, b,ms,\gd\a, s,i,
+\gd\d, b,m,\gd f, \u m,\u ni,\bigr]\gd[\F^k\dr{crt}], m,\u m,.
\e

In the case of symmetric theory ($l_{ab}=h_{ab}$, $g_{\u a\u b}=\gd h,
o,\u ab,$) one gets
\beq A.23
M^2_{ij}=\frac{\a^2_s}{\hbar c}\,\frac1{V_2} \int_M\sqrt{|\wt g|}\,d^{n_1}x
\,\bigl\{h_{bn}h^{o\u m\u p} \gd B, b,\u pi,\gd B, n,\u mj,\bigr\}.
\e

For $k=0$, $\F^0\dr{crt}$, $\gd H, k,\u p\u q,=0$ one gets the following
matrix for Higgs' bosons
\beq A.30
\bal
\gd m, 2\u h,f,\gd, \u e,a,&=\frac{-1}{V_2}\int_M\biggl\{
\frac{8\a^2_s}{\hbar c}\,\dg Q, sk,[\u e\u a][\u h\u q],\gd C, s,ac,\gd C, k,ef,
\gd(\F^0\dr{crt}), c,\u a,\gd(\F^0\dr{crt}), e,\u q,\\
&-\frac{2\a_s}{\sqrt{\hbar c}}\,\dg Q, as,[\u p\u q][\u h\u a],\gd f, \u e,\u p\u q,
\gd C, s,ef,\gd(\F^0\dr{crt}), e,\u a,
+\frac{4\a_s}{\sqrt{\hbar c}}\,\dg Q, sf,[\u e\u a][\u p\u q],\gd f, \u h,\u p\u q,
\gd C, s,ea,\gd(\F^0\dr{crt}), a,\u a,\\
&+\dg Q, af,[\u c\u d][\u p\u q],\gd f, e,\u c\u d,\gd f, \u h,\u p\u q,
\biggr\}\sqrt{|\u g|}\,d^{n_1}x
\eal
\e

For $k=1$, $\F^1\dr{crt}$, $\gd H, k,\u p\u q,\ne0$ and $\F^1\dr{crt}$ (if
exists) satisfies the following equation:
$$
\frac{2\a_s}{\sqrt{\hbar c}}\,\dg Q, sk,[\u e\u a][\u p\u q],\gd C, s,ac,
\gd(\F^1\dr{crt}), c,\u a,=\dg Q, ak,[\u c\u d][\u p\u q],\gd f, \u e,\u c\u d,
$$
and a supplementary condition
$$
\F_{\u b}^c f_{\hi\u d}^{\u b}-\mu_{\hi}^a \F_{\u a}^b C_{ab}^c=0.
$$

A mass matrix for Higgs' bosons looks like
\beq A.34
\bal
\gd m, 2\u h,f,\gd, \u e,a,&=\frac{-1}{V_2}\int_M
\biggl(\frac{4\a_s}{\sqrt{\hbar c}}\,\dg Q, sk,[\u e\u h][\u p\u q],
\gd H, k,\u p\u q,(\F^1\dr{crt})\gd C, s,af,\biggr)\sqrt{|\u g|}\,d^{n_1}x.\\
H_{\u m\u n}^b(\F\dr{crt}^k)&=\a_s\frac1{\sqrt{\hbar c}}\,C_{cd}^b
(\F\dr{crt}^k)_{\u n}^c(\F\dr{crt}^k)_{\u m}^d
-\frac1{\a_s}\sqrt{\hbar c}\,\mu_{\hi}^b f_{\u n\u m}^{\hi}
-(\F\dr{crt}^k)_{\u c}^b f_{\u n\u m}^{\u c}. \\
\gd H, b,\u m\u n,(\F\dr{crt}^0)&=0
\eal
\e

Let us pass to the cosmological terms in both cases of a symmetry breaking.
\beq4.92
\lambda _{ck} = e^{(n+2){\varPsi} } \frac{\alpha ^{2}_{s}}{ \ell ^{2}\pl}
\widetilde{R}(\widetilde{{\varGamma} }) + \frac{e^{n{\varPsi} } \widetilde{\ul{P}}
}{ r^{2}}
+ e^{(n-2){\varPsi} } \frac{4}{ \hbar cr^{2}}\biggl(\frac{\ell ^{2}\pl}{
r^{2}}\biggr)\wh V({\varPhi} ^{k}\dr{crt}),\quad\ k = 0, 1.
\e
These two terms are different and both
depend on the scalar field ${\varPsi} $. One gets:
\beq4.93
\lambda _{ck} = e^{(n+2){\varPsi} } \frac{\alpha ^{2}_{s}}{ \ell ^{2}\pl}
\widetilde{R}(\widetilde{{\varGamma} }) + e^{n{\varPsi} }
\frac{m^{2}_{\tilde A }}{ \alpha ^{2}_{s}} \biggl(\frac{c}{ \hbar}\biggr)^{2}
\widetilde{\ul{P}}
+ 4e^{(n-2){\varPsi} } \frac{m^{4}_{\tilde A }}{ \alpha ^{4}_{s}}
\biggl(\frac{\hbar^3}{ c^{5}}\biggr) \ell ^{2}\pl \wh V({\varPhi}
^{k}\crt).
\e
We get:
\beq4.94
\widetilde{R}(\widetilde{{\varGamma} }) = \frac{Q_{s}(\xi )}{ P_{s+1}(\xi )},
\quad \hbox{or}\quad \frac{Q_{s}(\xi )}{ P_{s}(\xi )},
\e
where $Q_{s}$ and $P_{s}$ are polynomials of the $s^{\text{\rm th}}$ order and $P_{s+1}$ is the
polynomial of $(s+1)^{\text{\rm th}}$ order with respect to $\xi $.
$Q_{s}$ and $P_{s+1}(P_{s})$ have no common divisors.
\beq4.95
\ell _{ab} = h_{ab} + \xi k_{ab}.
\e
In a similar way we can prove that
\beq4.96
\widehat{\ov{R}}(\widehat{\overline{\varGamma}}) =
\frac{W_{k}(x,\zeta )}{ V_{k+1}(x,\zeta )} ,\quad \hbox{or}\quad
\frac{W_{k}(x,\zeta )}{ V_{k}(x,\zeta )},\quad\ x \in G/G_{0},
\e
where $W_{k}(x,\zeta )$, $V_{k}(x,\zeta )$ are polynomials of the
$k^{\text{\rm th}}$ order with respect to
$\zeta $ with coefficients depending on $x\in G/G_{0}$ and $V_{k+1}(x,\zeta )$ is the polynomial
of the $(k+1)$ order with respect to $\zeta $ with coefficients depending on
$x\in G/G_{0}$. $W_{k}$ and $V_{k+1}(V_{k})$ have no common divisors.
\beq4.97
\widetilde{g}_{\tilde a \tilde b } = h^{0}_{\tilde a \tilde b } + \zeta
k^{0}_{\tilde a \tilde b }.
\e
However
\beq4.98
\bal
\widetilde{\ul P} &= \frac{\int\limits_{M}
\sqrt{|\widetilde{g}|}d^{n_1}x\widehat{\ov{R}}(\widehat{\overline{\varGamma}})
}{ \int\limits_{M}\sqrt{|\widetilde{g}|}d^{n_1}x}
= \frac{1}{ V_{1}} \int_{M}\sqrt{|\widetilde{g}|}d^{n_1}x \frac{W_{k}(x,\zeta )
}{ V_{k+1}(x,\zeta )} \\
&= \frac{R_{r}(\zeta )}{ S_{r+1}(\zeta )} \phi (\zeta ) ,\quad
\hbox{or}\quad \frac{R_{r}(\zeta )}{ S_{r}(\zeta )} \phi (\zeta)
\eal
\e
and $\phi (\zeta )$ is a function of $\zeta $ where $R_{r}$, $S_{r}$ are polynomials of the
$r^{\text{\rm th}}$
order with respect to~$\zeta $ and $S_{r+1}$ is the polynomial of
$(r+1)^{\text{\rm st}}$ order
with respect to~$\zeta $. In both cases we have similar asymptotic behaviour
with respect to $\xi $ and $\zeta $ if the function $\phi $ is bounded.
$R_{r}$ and $S_{r+1}(S_{r})$ have no common divisors.
\bea4.99
\widetilde{R}(\widetilde{{\varGamma} })&\sim \frac{C_{1}}{ \xi }\quad
\hbox{or}\quad \sim C_{1},\\
\widetilde{\ul{P}} &\sim \frac{C_{2}}{ \zeta }\quad \hbox{or}\quad \sim C_{2},\label{4.100}
\e
where $C_{1}$ and $C_{2}$ are constants. If the polynomials $Q_{s}$ and $R_{r}$ have real
roots $\xi _{0}$ and $\zeta _{0}$ such that
\beq4.101
Q_{s}(\xi _{0}) = 0
\e
and
\beq4.102
R_{r}(\zeta _{0}) = 0
\e
we get
\beq4.103
\lambda _{ck}(\xi _{0},\zeta _{0}) = 4e^{(n-2){\varPsi} }
\frac{m^{4}_{\tilde A }}{ \alpha ^{4}_{s}} \biggl(\frac{\hbar^{3}}{
c^{5}}\biggr) \ell ^{2}\pl \wh V({\varPhi} ^{k}\crt),\quad\ k = 0,1.
\e
In the case for sufficiently large $\xi $, $\zeta $ one gets:
$$
\lambda _{ck}(\xi ,\zeta ) = \biggl(e^{(n+2){\varPsi} }
\frac{C'_1}{ \xi }+e^{n{\varPsi} }\frac{C'_2}{ \zeta }\biggr) +
4\frac{m^{4}_{\tilde A }}{ \alpha ^{4}_{s}} \biggl(\frac{\hbar^{3}}{
c^{5}}\biggr) \cdot e^{(n-2){\varPsi} } \ell ^{2}\pl
\wh V({\varPhi} ^{k}\crt)
$$
or
\beq4.104
\lambda _{ck}(\xi ,\zeta ) = (e^{(n+2){\varPsi} }
C'_1+e^{n{\varPsi} }C'_2) +
4\frac{m^{4}_{\tilde A }}{ \alpha ^{4}_{s}} \biggl(\frac{\hbar^{3}}{
c^{5}}\biggr) \cdot e^{(n-2){\varPsi} } \ell ^{2}\pl
\wh V({\varPhi} ^{k}\crt),
\e
$k = 0, 1$; $C'_1, C'_2$ are constants ($m_{\tilde A} = \frac{\hbar c}{r}$ is
a scale of a mass of broken gauge bosons).

Thus in some cases we are able to make the first part of $\lambda _{ck}$ as small
as we want. $\wh V(\Phi^k_{\rm crt})$ is usually supposed to be zero for $k=0$.
From the observational data point of view we know that the
cosmological constant is small. Thus it occurs in the first or
second case (real roots of polynomials $Q_{s}$ and $R_{r}$ or in the limit of
large $\xi $ and $\zeta )$.
One gets:
\bg4.105
\lambda _{c0} \to 0,\\
\lambda _{c1} \to 4\frac{m^{4}_{\tilde A }}{ \alpha ^{4}_{s}} \biggl(\frac{\hbar^{3}}{
c^{5}}\biggr)\ell ^{2}\pl \wh V({\varPhi} ^{1}\crt).\label{4.106}
\e
This is this possibility to nivel a \co ical term.

Moreover, we have \co ical terms depending on a field~$\Ps$, and a value of
a \co ical \ct\ can be calculated later.

\def\lw#1 {\lower#1pt\hbox\bgroup$\scriptstyle}
\let\hp\hphantom
\def\eg{$\egroup}
\let\na\nabla

Let us consider Palatini variational principle for the action $S$.
\beq4.107
\delta S = 0.
\e
It is easy to see that \er{4.107} is equivalent to
\beq4.108
\delta \int_{{\mathbf U}} L(g,\ov{W}, \widetilde{A}, {\varPsi} ,
{\varPhi})\sqrt{-g}\,d^4 x= 0 ,\quad\ {\mathbf U} \subset E.
\e
We have the following independent quantities $g_{\mu\nu }$,
$\ov{W}^{\lambda }{}_{\mu\nu }$, $\widetilde{\omega }_{\rE}$, ${\varPsi} $ and ${\varPhi} $.
We vary with respect to the independent quantities. After some
calculations one gets:
\bg4.109
\ov{R}_{\mu\nu }(\ov{W}) - \frac{1}{ 2} g_{\mu\nu}\ov{R}(\ov{W})
= \frac{8\pi K}{ c^{4}} (\mathop{T_{\mu\nu }}\limits^{\gauge} +T_{\mu\nu }({\varPhi} )
+ \mathop{T_{\mu \nu }}\limits^{\scal} ({\varPsi} )+
\mathop{T_{\mu \nu }}\limits^{\Int} +g_{\mu \nu }\wt{\varLambda })
= 8\pi \mathop{T_{\mu\nu}}\limits^{\rm eff},\\
\falkag^{[\mu \nu ]}{}_{,\nu } = 0,\label{4.110}
\e
or
\bg4.110a
\overline{\nabla}_{\nu } g^{[\mu \nu ]} =0,\\
g_{\mu \nu ,\sigma } - g_{\xi\nu }\overline{\varGamma}^{\xi}{}_{\mu
\sigma } - g _{\mu \xi}\overline{\varGamma}^{\xi}{}_{\sigma \nu }=0,\label{4.111}
\e
We get a solution of Eq.\ \er{4.111}:
\beq D.2
\gd \ov\G{},\la,\m,=\gd\tv\G,\la,\m,+\gd\ov Q{},\la,\m,+\gd\D,\la,\m,
\e
where $\gd\tv\G,\la,\m,$ is a Levi-Civita \cn\ induced by $g_\(\a\b)$ on~$E$
and
\beq{D.3}
\gd\ov Q{},\nu,\g\mu,=\frac12 \Bigl(\dg K,\g\mu,\nu,
-2g_{\tl[\mu\cdt]}^{\hp{\tl[\mu}\a}K_{\lw2.5 \g\tp \a\b\eg}g^\[\nu\b]\Bigr)
\e
is a torsion of the \cn\ $\gd\ov\G,\mu,\la\nu,$,
\bg{D.4}
\gd\D,\nu,\g\mu,=\wt g{}^\(\nu\d)\Bigl\{K^{\hp{\d(\g}\a}_{\d(\g\cdt}
g_{\lw2.5 \[\mu)\a]\eg} + g^{\hp{[\rho}\b}_{[\rho\cdt]}\Bigl[g^{\hp{(\mu}\rho}
_{([\mu\cdt]}K_{\lw2.5 \g)\a\b\eg} g^{\hp{[\d}\a}_{[\d\cdt]}
- K_{\d\a\b} g^{\hp{([\g}\a}_{([\g\cdt]} g^{\hp{[\mu)}\rho}_{[\mu)\cdt]}
\Bigr]\Bigr\} \\
K_{\a\b\g}=-\tv\na_\a g_\[\b\g] - \tv\na_\b g_\[\g\a]
+ \tv\na_\g g_\[\a\b] \label{D.6}
\e
(see Ref. \cite{16}).
\beq4.112
\bal
\lefteqn{((n^{2}+2\ov M)\widetilde{g}^{(\alpha \mu )}-n^{2}g^{\nu \mu }
g_{\delta \nu }\widetilde{g}^{(\alpha \delta )})
\frac{\partial ^{2}{\varPsi} }{ \partial x^{\alpha } \partial x^{\mu }}}\\
\noalign{\vskip2pt}
&+ \frac{1}{ \sqrt{-g}} \partial _{\mu }\biggl\{\sqrt{-g}\biggl[n^{2}
\widetilde{g}^{(\alpha \mu )} - \frac{n^{2}}{ 2} g_{\delta \nu }(
g^{\nu \alpha } \widetilde{g}^{(\mu \delta )} + g^{\nu \mu }
\widetilde{g}^{(\mu \alpha )}) - 2M \widetilde{g}^{(\mu \alpha
)}\biggr]\biggr\}\\
\noalign{\vskip2pt}
&\times \frac{\partial {\varPsi} }{ \partial x^{\alpha }} - 8\pi (n+2)
e^{-(n+2){\varPsi} } {\cal L}_{\rYM}(\widetilde{A}) -
\frac{4e^{-2{\varPsi} }}{ r^{2}} {\cal L}_{\kin}({\varPhi} ,\widetilde{A})\\
\noalign{\vskip2pt}
&+ \frac{(n-2)}{ r^{4}} e^{(n-2){\varPsi} } \wh V({\varPhi} ) +
\frac{4(n-2)}{ r^{2}} e^{(n-2){\varPsi} } {\cal L}_{\Int}({\varPhi} ,
\widetilde{A}) \\
\noalign{\vskip2pt}
&- \frac{n}{ r^{2}} e^{n{\varPsi} } \widetilde{\ul{P}} -
\frac{(n+2)\alpha ^{2}_{s}}{ \ell ^{2}\pl} e^{(n+2){\varPsi} }
\widetilde{R}(\widetilde{{\varGamma} }) = 0,
\eal
\e
\bg4.113
\bal
\mathop{\nabla_{\mu}}\limits^{\gauge} (\widetilde{\ell
}_{ij}\widetilde{\falkaL}^{i\alpha \mu }) &= 2\falkag^{[\alpha
\beta ]} \mathop{\nabla_{\beta}}\limits^{\gauge}(\widetilde{h}_{ij}g^{[\mu \nu ]}
\widetilde{H}^{i}{}_{\mu \nu }) \\
\noalign{\vskip4pt}
&+ 2\sqrt{-g}\alpha _{s} \frac{1}{ \sqrt{\hbar c}}
\frac{e^{n{\varPsi} }}{ r^{2}} \biggl[ \ell _{ab}g^{\tilde b
\tilde n } g^{\mu \alpha } L^{a}{}_{\mu \tilde b }
({\varPhi} ^{d}{}_{\tilde c }C^{b}_{dc}\alpha ^{c}_{j}+{\varPhi}
^{b}{}_{\tilde a }f^{\tilde a }_{\tilde n j}) \\
\noalign{\vskip4pt}
&+ \biggl( \frac{\delta L^{a}{}_{\beta \tilde b }}{ \delta
\mathop{\nabla_{\alpha}}\limits^{\gauge}{\varPhi} ^{w}_{\tilde v }}\biggr)
\ell _{ab}g^{\tilde b \tilde n } g^{\beta \mu }
(\mathop{\nabla_{\mu}}\limits^{\gauge}{\varPhi} ^{b}{}_{\tilde n })
({\varPhi} ^{d}{}_{\tilde w }C^{w}_{dc}\alpha ^{c}_{j}+{\varPhi}
^{w}_{\tilde a }f^{\tilde a }{}_{\tilde n j}) \biggr]_{av}\\
\noalign{\vskip4pt}
&+ 4\sqrt{-g} \frac{e^{2n{\varPsi} }}{ r^{2}} h_{ab} \mu ^{a}_{k}
\widetilde{\ell }_{ij} \ell ^{ki} \widetilde{g}^{[\tilde a \tilde b ]}
\mathop{\nabla_{\mu}}\limits^{\gauge}\biggl\{g^{[\mu \alpha ]}\\
\noalign{\vskip4pt}
&\times \biggl[\frac{1}{ \alpha _{s}}\sqrt{\hbar c}C^{b}_{cd}
{\varPhi} ^{c}{}_{\tilde a }{\varPhi} ^{d}{}_{\tilde b }-
\alpha _{s}\biggl(\frac{\hbar }{ c}\mu ^{b}_{\dah{i}}f^{\dah{i}}{}_{\tilde
a \tilde b }-\alpha _{s}\frac{1}{ \sqrt{\hbar c}}{\varPhi}
^{b}{}_{\tilde d }f^{\tilde d }{}_{\tilde a \tilde b
}\biggr)\biggr]\biggr\} \\
\noalign{\vskip4pt}
&+ (n+2) \partial _{\beta }{\varPsi}
[\widetilde{\ell }_{ij}\widetilde{\falkaL}^{i\beta \alpha }-
2\falkag^{[\b \a ]}(\widetilde{h}_{ij}g^{[\mu \nu
]}\widetilde{H}^{i}{}_{\mu \nu })],
\eal
\e
\beq 4.114
\bal
\mathop{\nabla_{\mu}}\limits^{\gauge} (\ell _{ab}\falkaL^{a\mu }{}_{\tilde b })_{av}
&= - \sqrt{-g} \frac{e^{n{\varPsi} }}{ 2r^{2}}\biggl\{\biggl(\frac{\delta \wh V' }{ \delta
{\varPhi} ^{b}{}_{\tilde n}}\biggr) g_{\tilde b \tilde n}\\
\noalign{\vskip4pt}
&-2 \sqrt{-g} e^{n{\varPsi} } \mu ^{e}_{i} (\widetilde{H}^{i}{}_{\mu
\nu }g^{[\mu \nu ]}) h_{ed} \biggl( \frac{2}{ \alpha _{s}}\sqrt{\hbar
c}\ul{g}^{[\tilde a \tilde n]} C^{d}_{cb} {\varPhi}
^{c}{}_{\tilde a } g{}_{\tilde b \tilde n} \\
\noalign{\vskip4pt}
& - \alpha _{s} \frac{1}{ \sqrt{\hbar
c}}g^{[\tilde c \tilde d ]} f^{\tilde n}{}_{\tilde c
\tilde d } g_{\tilde b \tilde n}\biggr) + 2
\partial _{\mu }{\varPsi} \ell _{ab}\falkaL^{a\mu }{}_{\tilde b
}\biggr\}_{av},
\eal 
\e
where
\beq4.115
\bal
\mathop{T_{\alpha\beta}}\limits^{\gauge} &= - \frac{\widetilde{\ell }_{ij}}{ 4\pi }
\biggl\{g_{\gamma \beta } g^{\tau\rho} g^{\varepsilon \gamma }
\widetilde{L}^{i}{}_{\rho\alpha }
\widetilde{L}^{j}{}_{\tau\varepsilon} - 2 g^{[\mu \nu ]}
\widetilde{H}^{(i}_{\mu \nu } \widetilde{H}^{j)}_{\alpha \beta }\\
\noalign{\vskip2pt}
&- \frac{1}{ 4} g_{\alpha \beta } [\widetilde{L}^{i\mu
\nu }\widetilde{H}^{j}{}_{\mu \nu }-2(g^{[\mu \nu
]}\widetilde{H}^{i}{}_{\mu \nu })(g^{[\gamma \sigma ]}
\widetilde{H}^{j}{}_{\gamma \sigma })]\biggr\}
\eal
\e
is the energy-momentum tensor for the gauge (Yang--Mills') field with
the zero trace.
\bea4.116
\mathop{T_{\alpha \beta }}\limits^{\gauge} g^{\alpha \beta } &=0,\\
\noalign{\vskip2pt}
\mathop{T_{\alpha \beta }}\limits^{\scal} ({\varPsi} )
&= - \frac{e^{(n+2){\varPsi} }}{ 16\pi}
\biggl\{(g_{\varkappa \alpha }g_{\omega \beta }+g_{\omega \alpha }
g_{\varkappa \beta }) \nonumber\\
\noalign{\vskip2pt}
&\quad{}\times \widetilde{g}^{(\gamma \varkappa )} \widetilde{g}^{(\nu
\omega )} \cdot \biggl[\frac{n^{2}}{ 2}(g^{\xi\mu }g_{\nu \xi}-
\delta ^{\mu }_{\nu }){\varPsi} _{,\mu }+\ov{M}{\varPsi} _{,\nu
}\biggr] {\varPsi} _{,\gamma } \nonumber\\
\noalign{\vskip2pt}
&\quad{}- g_{\alpha \beta } \Bigl[\bigl(\ov{M}\widetilde{g}^{(
\nu\g )}+n^{2}g^{[\mu \nu ]}g_{\delta \mu }\widetilde{g}^{(\gamma
\delta )}\bigr){\varPsi} _{,\nu }{\varPsi} _{,\gamma }\Bigr]\biggr\}\label{4.117}
\e
is the energy-momentum tensor for the scalar field ${\varPsi} $ with nonzero
trace
\bg4.118
\mathop{T_{\alpha \beta }}\limits^{\scal} ({\varPsi} )
g^{\alpha \beta } \neq 0\\
K = G_{\rN} e^{-(n+2){\varPsi} } = G_{\eff}({\varPsi}).\label{4.119}
\e
It plays the role of an effective gravitational constant.
\bml4.120
T_{\mu \nu }({\varPhi} ) = \frac{e^{(n-4){\varPsi} }}{ 4\pi r^{2}}
\ell _{ab}\Bigl(g^{\tilde b \tilde n} L^{a}{}_{\mu \tilde
b } \mathop{\nabla_{\nu}}\limits^{\gauge} {\varPhi} ^{b}{}_{\tilde n}\Bigr)_{av} \\
{}- \frac{1}{ 2} g_{\mu \nu } \biggl(-\frac{e^{2(n-2){\varPsi} }}{ 8\pi r^{4}}
\wh V({\varPhi} )+ \frac{e^{(n-4){\varPsi}}}{ 4\pi r^{2}}\ell _{ab}
(g^{\tilde b \tilde n}g^{\alpha \beta }L^{a}{}_{\alpha
\tilde b }\mathop{\nabla_{\beta}}\limits^{\gauge} {\varPhi} ^{b}{}_{\tilde
n})_{av}\biggr).
\e
It is an energy-momentum tensor for the Higgs' field.
\beq4.121
\bal
\mathop{T_{\mu\nu }}\limits^{\Int} &= -\frac{e^{2(n-2){\varPsi} }}{ 2\pi r^{2}} h_{ab}
\mu ^{a}_{i} \widetilde{H}^{i}{}_{\mu \nu
}\widetilde{\ul{g}}^{[\tilde a \tilde b ]}\biggl(\frac{1}{ \alpha _{s}}
\sqrt{\hbar c} C^{b}_{cd} {\varPhi} ^{c}{}_{\tilde a } {\varPhi}
^{d}{}_{\tilde b } \\
&- \alpha _{s} \frac{1}{ \sqrt{\hbar c}} \mu ^{b}_{\dah{i}} f^{\dah{i}}{}_{\tilde a \tilde b } - \alpha _{s} \frac{1}{ \sqrt{\hbar c}}
{\varPhi} ^{b}{}_{\tilde d } f^{\tilde d }{}_{\tilde a \tilde b}\biggr)\\
&+ \frac{e^{2(n-2){\varPsi} }}{ 4\pi r^{2}} g_{\mu \nu }\biggl[h_{ab}
\mu ^{a}_{i} (\widetilde{H}^{i}_{\alpha \beta }g^{[\alpha \beta
]})\widetilde{\ul{g}}^{[\tilde a \tilde b ]} \\
&\qquad{}\times \biggl(\frac{1}{ \alpha _{s}}\sqrt{\hbar c}C^{b}_{cd}
{\varPhi} ^{c}_{\tilde a }{\varPhi} ^{d}_{\tilde b }-\alpha _{s}
\frac{\hbar}{ c}\mu ^{b}_{\dah{i}}f^{\dah{i}}{}_{\tilde a \tilde b }-
\alpha _{s}\frac{1}{ \sqrt{\hbar c}}{\varPhi} ^{b}{}_{\tilde d
}f^{\tilde d }{}_{\tilde a \tilde b }\biggr)\biggr].
\eal
\e
It is an energy-momentum tensor corresponding to the nonminimal
interaction term ${\cal L}_{\Int}(\widetilde{A},{\varPhi} )$.
\beq4.122
\wt{\varLambda} = \frac{1}{ 16\pi G_{\rN}} \biggl(\frac{e^{(2n+4){\varPsi} }
\alpha ^{2}_{s}}{ \ell ^{2}\pl} \widetilde{R}(\widetilde{{\varGamma} })
+\frac{e^{(n+2){\varPsi} }}{ r^{2}}\widetilde{\ul{P}}\biggr)
= 16\pi G_{\rN} e^{-(n+2){\varPsi}} \wt{\overline{\lambda }}_{c0}.
\e
It plays the role of the ``cosmological constant'', which now depends on
the scalar field~${\varPsi} $. The quantity
$\dfrac{\delta L^{a}_{\beta \tilde b }}{ \delta
\mathop{\nabla_{\alpha}}\limits^{\gauge} {\varPhi} ^{w}_{\tilde v }}$ satisfies the following
equation:
\bg4.123
\ell _{dc} g_{\mu \beta } g^{\gamma \mu } \frac{\delta L^{d}_{\gamma
\tilde a }}{ \delta \mathop{\nabla_{\alpha}}\limits^{\gauge}{\varPhi}
^{w}_{\tilde v }} + \ell _{cd} g_{\tilde a \tilde m} g^{\tilde
m\tilde c } \frac{\delta L^{d}_{\beta \tilde c }}{ \delta
\mathop{\nabla_{\alpha}}\limits^{\gauge} {\varPhi} ^{w}_{\tilde v}}
= 2\ell _{cd} g_{\tilde a \tilde m} g^{\tilde m\tilde c }
\delta ^{\alpha }{}_{\beta } \delta ^{\tilde v}{}_{\tilde c } \delta^{d}_{w},\\
\widetilde{\falkaL}^{i\mu \nu } = \sqrt{-g} g^{\beta \mu }
g^{\gamma \nu } \widetilde{L}^{i}_{\beta \gamma },\label{4.124}\\
\falkag^{[\mu \nu ]} = \sqrt{-g} g^{[\mu \nu ]},\label{4.124a}
\e
Equations \er{4.109}, \er{4.110} and \er{4.111} are gravitational
equations from N.G.T. with the following matter sources: Yang--Mills'
field (in the nonsymmetric version), Higgs' field, scalar field ${\varPsi} $ with
a presence of the cosmological term depending on scalar field ${\varPsi} $.
Equation \er{4.112} is the equation for the scalar field ${\varPsi} $. This field is
of course chargeless, but it interacts with Yang--Mills' field and
Higgs' field due to some terms in \er{4.112}. It interacts also with
cosmological terms, which effectively depend on ${\varPsi} $. This field due to
equation \er{4.119} has an interpretation as an effective gravitational
constant. Equation \er{4.113} is the equation for Yang--Mills' field. Now as
in the Nonsymmetric Kaluza--Klein Theory we have for this field two
tensors of strength $\widetilde{H}_{\mu \nu }$ and
$\widetilde{L}^{i}{}_{\mu \nu }$ (ordinary and an induction one) and
the nonsymmetric parts of metrics on $E$ and on $G$ induce the polarization
$\widetilde{M}^{i}{}_{\mu \nu }$. In the equation \er{4.113} we have sources connected to a
skewsymmetric part of metric $g_{\mu \nu }$ and to Higgs' field. Due to the
existence of a skewsymmetric part of metric $\ell _{ab}$ and
$g_{\tilde a \tilde b }$ the current
connected to Higgs' field is more complicated. Equation \er{4.114} is an
equation for Higgs' field. We write this equation in terms of tensor
\beq4.125
L^{a\mu }{}_{\tilde b } = g^{\alpha \mu } L^{a}{}_{\alpha \tilde b }.
\e
This tensor plays a similar role for $\mathop{\nabla_{\alpha}}\limits^{\gauge}
{\varPhi} ^{a}_{\tilde b }$ as $\widetilde{L}^{i\mu }{}_{\alpha }$ for
$\widetilde{H}^{i}{}_{\mu \alpha }$.Thus we
have in the theory an analogue of the polarization tensor
$M^{a}{}_{\alpha \tilde b }$ for Higgs' field
\beq4.126
L^{a}{}_{\alpha \tilde b } = \mathop{\nabla_{\alpha}}\limits^{\gauge}{\varPhi}
^{a}_{\beta } - \frac{4\pi }{ c} M^{a}{}_{\alpha \tilde b }.
\e
Let us consider the following two $\Ad_{H^-}$ type two-forms
$\widehat{L} = L^{a}{}_{\alpha \tilde b }\theta ^{\alpha }\wedge
\theta ^{\tilde b }X_{a}$
and $\widehat{M} = M^{a}{}_{\alpha \tilde b }\theta ^{\alpha }\wedge
\theta ^{\tilde b }X_{a}$. One gets $\widehat{L}=
\widehat{{\varOmega}} - \frac{4\pi}{ c} \widehat{M} =
\widehat{{\varOmega}} - \frac{1}{ 2} \widehat{Q}$, where
$$
\widehat{Q} = \mathop{\nabla_{\alpha}}\limits^{\gauge} {\varPhi} ^{a}{}_{\tilde b
}\theta ^{\alpha }\wedge \theta ^{\alpha }X_{a},\quad\ \widehat{Q} =
Q^{a}{}_{\alpha \tilde b }\theta ^{\alpha }\wedge \theta ^{b}X_{a}.
$$
In this way we get a geometrical interpretation of the polarization
2-form of the Higgs' field as a part of a torsion.

The field ${\varPsi} $ due to \er{4.119} is connected to the effective
gravitational
constant. However, it also enters the definition of energy-momentum
tensors $T_{\mu \nu }({\varPhi} )$ and $\mathop{T_{\mu \nu }}\limits^{\Int}$.
Thus it plays the role of the universal
factor. In the next section we deal with this field in details.

Let us notice that the left-hand side of Eq.~\er{4.113} can be rewritten in
the following way
\beq4.127
\mathop{\nabla_{\mu}}\limits^{\gauge} (\widetilde{l}_{ij} \widetilde{\falkaL}^{i\alpha
\mu } ) = \sqrt{-g} \mathop{\widetilde{\overline{\nabla}}_{\mu}}\limits^{\gauge}
(\widetilde{l}_{ij}\widetilde{L}^{i\alpha \mu }),
\e
where $\mathop{\widetilde{\overline{\nabla}}_{\mu}}\limits^{\gauge}$ means a covariant derivative with respect to the connection
$\widetilde{\overline{\omega }}^{\alpha }{}_{\beta }$ on $E$ and $\omega
_{\rE}$ on $Q(E,G)$ at once. One gets from Eq.~\er{4.109}:
\bg{D.15}
\tv R_{\b\g}=8\pi (\nad{eff}T_\(\b\g)-\tfrac12 g^{\mu\nu}\nad{eff}T_{\mu\nu}g_\(\b\g))
+\tfrac34 \tv\na_\d \gd\D,\d,\b\g, -\tfrac14 \tv\na_{(\g} \gd\D,\a,\b)\a, \\
-\tfrac12 \tv\na_\d \gd\ov Q{},\d,\b\g,+\tfrac14 \tv\na_{[\g}\gd\D,\a,\b]\a,
+ \tfrac23 \ov W_\[\b,\g]=8\pi (\nad{eff}T_\[\b\g]
-\tfrac12 g^{\mu\nu}\nad{eff}T_{\mu\nu}g_\[\b\g]). \label{D.16}
\e
One can eliminate $\ov W_\mu$ from the theory using \eqref{D.16} and getting
\beq{D.17}
\tfrac14 \tv\na_{\tl[\g}\gd\D,\a,{\b]|\a|,\mu\tp},
-\frac12\tv\na_\d \ov Q_\[\b\g,\mu] = 8\pi \bigl(\nad{eff}T_{\tl[\b\g],\mu\tp}
-\frac12 (g^{\a\nu}\nad{eff}T_{\a\nu}g_{\tl[\b\g]})_{,\mu\tp}\bigr),
\e
$8\pi\nad{eff}T_{\a\b}$ is the right-hand side of Eq.~\er{4.109}.

Now we can proceed considerations of two stages of symmetry breaking
and its hierarchy
from Section~3. This involves a scalar field $\Psi$ and \co ical terms.

In Ref.~\cite5 we developed a similar formalism with a scalar field $\rho(\Psi)$.
Moreover, we consider the formalism presented here as more profound.
Ref.~\cite5 contains more possibilities which we now neglect.

Let us recapitulate Section~4 (up to now).

The \E\nos\ \KK (Jordan--Thiry) Theory  unifies the Nonsymmetric
Gravitational Theory (NGT) and gauge fields (Yang--Mills' fields) including
spontaneous symmetry breaking and the Higgs' mechanism with scalar
forces connected to the gravitational constant and cosmological terms
appearing as the so-called quintessence. The theory is
geometric and unifies tensor-scalar gravity with massive gauge theory
using a multidimensional manifold in a Jordan--Thiry manner.
We use a nonsymmetric version of this theory.
The general scheme is the following. We introduce
the principal fibre bundle over the base $V = E \times G/G_{0}$ with the
structural group $H$, where $E$ is a space-time, $G$ is a compact semisimple
Lie group, $G_{0}$ is its compact subgroup and $H$ is a semisimple compact
group. The manifold $M = G/G_{0}$ has an interpretation as a ``vacuum states
manifold'' if $G$ is broken to $G_{0}$ (classical vacuum states). We define on
the space-time $E$, the nonsymmetric tensor $g_{\alpha\beta}$ from NGT. Simultaneously we introduce
on $E$ two connections from NGT $\ov{W}{}_{\beta \gamma}^\alpha $ and
$\overline{\varGamma}{}_{\beta \gamma}^\alpha $. On the homogeneous space $M$
we define the nonsymmetric metric tensor
\beq22.3
g_{\tilde{a}\tilde{b}} = h^{0}_{\tilde{a}\tilde{b}}
+ \zeta k^{0}_{\tilde{a}\tilde{b}}
\e
where $\zeta $ is the dimensionless constant. Now on the principal bundle $P'$ we define the connection~$\omega $, which
is the 1-form with values in the Lie algebra of $H$.

After this we introduce the nonsymmetric metric on $P'$
right-invariant with respect to the action of the group $H$, introducing
scalar field $\rho $ in a Jordan--Thiry manner.
The only difference is that
now our base space has more dimensions than four. It is
$(n_{1}+4)$-dimensional, where $n_{1} = \dim (M) = \dim (G)-\dim (G_{0})$. In other
words, we combine the nonsymmetric tensor $\gamma _{\rAB}$ on $V$ with the
right-invariant nonsymmetric tensor on the group $H$ using the connection
$\omega $ and the scalar field $\rho $. We suppose that the factor $\rho $ depends on a
space-time point only.  This is really the Jordan--Thiry
theory in the nonsymmetric version but with $(n_{1}+4)$-dimensional
``space-time''. After this we act in the classical manner. We introduce the linear connection
which is compatible with this nonsymmetric metric. This connection is
the multidimensional analogue of the connection
$\widetilde{\overline{\varGamma}}{}^{\alpha}_{\beta \gamma }$ on the space-time
$E$. Simultaneously we introduce the second connection~$W$. The
connection $W$ is the multidimensional analogue of the $\ov{W}$-connection from
NGT and Einstein's Unified Field Theory. Now we calculate the Moffat--Ricci
curvature scalar $R(W)$ for the connection $W$ and we get the following
result. $R(W)$ is equal to the sum of the Moffat--Ricci curvature on the
space-time $E$ (the gravitational lagrangian in Moffat's theory of
gravitation), plus $(n_{1}+4)$-dimensional lagrangian for the Yang--Mills'
field from the Nonsymmetric Kaluza--Klein Theory plus the Moffat--Ricci
curvature scalar on the homogeneous space $G/G_{0}$ and the Moffat--Ricci
curvature scalar on the group $H$ plus the lagrangian for the scalar
field~$\rho $. The only difference is that our Yang--Mills' field is defined
on $(n_{1}+4)$-dimensional ``space-time'' and the existence of the
Moffat--Ricci
curvature scalar of the connection on the homogeneous space
$G/G_{0}$.
All of these terms (including $R(\ov{W})$) are multiplied by some
factors depending on the scalar field~$\rho $.

This lagrangian depends on the point of $V = E \times G/G_{0}$ i.e.\ on the
point of the space-time $E$ and on the point
of $M = G/G_{0}$. The curvature
scalar on $G/G_{0}$ also depends on the point of~$M$.

We now go to the group structure of our theory. We assume $G$ invariance of
the connection $\omega $ on the principal fibre bundle $P'$, the so
called Wang condition. According to the Wang theorem the connection $\omega $
decomposes into the connection $\widetilde{\omega }_{\rE}$ on the principal
bundle $Q$ over space-time $E$ with structural group $G$ and the multiplet of
scalar fields ${\varPhi} $. Due to this decomposition the multidimensional
Yang--Mills' lagrangian decomposes into: a 4-dimensional Yang--Mills'
lagrangian with the gauge group $G$ from the Nonsymmetric Kaluza--Klein
Theory, plus a polynomial of 4th order with respect to the fields ${\varPhi}
$, plus a term which is quadratic with respect to the gauge derivative of
${\varPhi}$ (the gauge derivative with respect to the connection
$\widetilde{\omega }_{\rE}$ on a space-time $E$) plus a new term which is of
2nd order in the ${\varPhi} $, and is linear with respect to the Yang--Mills'
field strength. After this we perform the dimensional reduction procedure for
the Moffat--Ricci scalar curvature on the manifold $P'$. We average
$R(W)$ with respect to the homogeneous space $M = G/G_{0}$.  In this way we
get the lagrangian of our theory. It is the sum of the Moffat--Ricci
curvature scalar on $E$ (gravitational lagrangian) plus a Yang--Mills'
lagrangian with gauge group~$G$ from the Nonsymmetric Kaluza--Klein Theory
(see \cite7), plus a kinetic term for the scalar field~${\varPhi} $, plus a
potential $V({\varPhi} )$ which is of 4th order with respect to ${\varPhi} $,
plus ${\cal L}_{\Int}$ which describes a nonminimal interaction between the
scalar field ${\varPhi} $ and the Yang--Mills' field, plus cosmological
terms, plus lagrangian for scalar field $\rho $. All of these terms
(including $\ov{R}(\ov{W})$) are multiplied of course by some factors
depending on the scalar field $\rho $. We redefine tensor $g_{\mu \nu }$ and
$\rho$ and pass from scalar field $\rho $ to ${\varPsi} $.
After this we get lagrangian which is the sum of gravitational lagrangian,
Yang--Mills' lagrangian, Higgs' field lagrangian, interaction term ${\cal
L}_{\Int}$ and lagrangian for scalar field ${\varPsi} $ plus cosmological
terms. These terms depend now on the scalar field ${\varPsi} $. In this way
we have in our theory a multiplet of scalar fields $({\varPsi} ,{\varPhi} )$.
As in the Nonsymmetric-Nonabelian Kaluza--Klein Theory we get a polarization
tensor of the Yang--Mills' field induced by the skewsymmetric part of the
metric on the space-time and on the group $G$.  We get an additional term in
the Yang--Mills' lagrangian induced by the skewsymmetric part of the metric
$g_{\alpha \beta }$. We get also ${\cal L}_{\Int}$, which is absent in the
dimensional reduction procedure known up to now.  Simultaneously, our
potential for the scalar---Higgs' field has more complicated structure, due
to the skewsymmetric part of the metric on $G/G_{0}$ and on $H$.
This structure offers two kinds of critical points for
the minimum of this potential: ${\varPhi} ^{0}\crt$ and ${\varPhi}
^{1}\crt$. The first is known in the classical, symmetric dimensional
reduction procedure and corresponds to the trivial Higgs' field (``pure
gauge''). This is the ``true'' vacuum state of the theory.  The second,
${\varPhi} ^{1}\crt$, corresponds to a more complex configuration.  This
is only a local (no absolute) minimum of~$\wh V$. It is a ``false'' vacuum. The
Higgs' field is not a ``pure'' gauge here. In the first case the unbroken
group is always $G_{0}$. In the second case, it is in general different and
strongly depends on the details of the theory: groups $G_{0}$, $G$, $H$,
tensors $\ell _{ab}$, $g_{\tilde{a}\tilde{b}}$ and the constants $\zeta$,
$\xi $. It results in a different spectrum of mass for intermediate bosons.
However, the scale of the mass is the same and it is fixed by a constant $r$
(``radius'' of the manifold $M = G/G_{0})$. In the first case $\wh V({\varPhi}
^{0}\crt) =0$, in the second case it is, in general, not zero
$\wh V({\varPhi} ^{1}\crt) \neq 0$. Thus, in the first case, the cosmological
constant is a sum of the scalar curvature on $H$ and $G/G_{0}$, and in the
second case, we should add the value $\wh V({\varPhi} ^{1}\crt)$. We proved
that using the constant $\xi$ we are able in some cases to make the
cosmological constant as small as we want (it can change the sign).  Here we
can perform the same procedure for the second term in the cosmological
constant using the constant $\zeta $. It can change the sign too.

Let us notice the following fact. In Ref.~\cite{11a} we consider the GSW
(Glashow--Salam--Weinberg) model in a framework of the NKKT. In this case we
have to do only with one critical point (because of a simplicity of $M=S^2$)
and $\wh V(\Phi_{\rm crt})\ne0$.

The interesting point is that there exists an effective scale of masses,
which depends on the scalar field~${\varPsi} $.

Using Palatini variational principle we get an equation for fields
in our theory. We find a gravitational equation from N.G.T. with
Yang--Mills', Higgs' and scalar sources (for scalar field~${\varPsi} )$ with
cosmological terms. This gives us an interpretation of the scalar
field ${\varPsi} $ as an effective gravitational constant.

We get an equation for this scalar field ${\varPsi} $. Simultaneously we get
equations for Yang--Mills' and Higgs' field. We can discuss the change
of the effective scale of mass, $m_{\eff}=m_{\td A} e^{\frac n2 \Psi_0}$ with a relation to the change of
the gravitational constant~$G_{\eff}$.

In the ``true'' vacuum case we get that the scalar field ${\varPsi} $ is
massive and has Yukawa-type behaviour. In this way the weak
equivalence principle is satisfied. In the ``false'' vacuum case the
situation is more complex. It seems that there are possible some
scalar forces with infinite range. Thus the two worlds constructed
over the ``true'' vacuum and the ``false'' vacuum seem to be completely
different: with different unbroken groups, different mass spectrum for
the broken gauge and Higgs' bosons, different cosmological constants
and with different behaviour for the scalar field ${\varPsi} $. The last point
means that in the ``false'' vacuum case the weak equivalence principle
could be violated and the gravitational constant (Newton's constant)
would increase in distance between bodies. Here we work in a true vacuum case.

We are interested in properties of the scalar field $\Ps$ which
is a source of an inconstancy of an effective \gr al \ct.
Thus we consider a \lg\ of this field, neglecting \YM' field and Higgs' fields
from the full theory. A kinetic part of a \lg\ of the field $\Ps$ looks:
\beq22.7
\bal
\cL_{\rm scal}^{\rm kin}(\Ps)&=-\bigl(\ov M g^\(\g\nu)+n^2g^\[\m]
g_{\d\mu}\wt g^\(\d\g)\bigr)\Ps_{,\nu}\Ps_{,\g} \\
\ov M&=l^\[dc]l_\[dc]-3n(n-1) .
\eal
\e
This field couples to \co ical \ct s in the theory: $\wt R(\wt\G)$---a scalar
curvature of a \cn\ $\wt\G$ on group manifold~$H$, and to
$$
\wt{\ul P}=\frac 1{V_2}\int_{G/G_0} d^{n_1}x\, \hb R(\hb \G),
$$
where $V_2$ is a volume of a manifold $M=G/G_0$ and $\hb R(\hb\G)$ is a
scalar curvature of a \cn\ $\hb\G$ defined on this manifold,
and a full \lg\ for a field $\Ps$ looks
\beq22.12
L=\cL_{\rm scal}^{\rm kin}(\Ps)-\ov\la _{c0}(\Ps),
\e
where
\beq22.13
\ov\la_{c0}(\Ps)=\a_s^2\,\frac{\ex{(n+2)\Ps}}{\ell\pl^2}\,\wt R(\wt\G)+\frac{\ex{n\Ps}}{r^2}\,\wt{\ul P},
\e
$\ell\pl=\sqrt{\frac{G_N\hbar}{c^3}}\simeq 10^{-33}$\,cm is a Planck's length, and
$\a_s$ is a \di less coupling \ct. Due to \nos ity of a \cn\ $\wt\G$ and
$\hb\G$, $\wt R(\wt \G)$ and $\wt{\ul P}$ are \f\ of \ct s $\mu$ and $\z$ and can change
the signs. Explicit examples are for $G=\SU(2)$ and $M=S^2$. $\ov\la_{c0}$ can be
written in a different way
\beq22.14
\ov\la_{c0}=\ex{(n+2)\Ps}\,\frac{\a_s^2}{\ell\pl^2}\,\wt R(\wt\G)
+\ex{n\Ps}\,\frac{m_{\tilde A}^2}{\a_s^2}\Bigl(\frac c\hbar\Bigr)^2 \wt{\ul P},
\e
where $m_{\tilde A}$ is a scale of mass of broken gauge bosons. Moreover, we
get an equation for a scalar field~$\Ps$
\bml22.15
2\biggl[\Bigl((n^2+2\ov M)\wt g^\(\a\mu)-n^2g^{\m}g_{\d\nu}
\wt g^\(\a\d)\Bigr)\pp{^2\Ps}{x^\a \pa x^\mu}\\
{}+\frac1{\sqrt{-g}}\,\pa_\mu\biggl\{\sqrt{-g}\Bigl[n^2\wt g^\(\a\mu)
-\frac{n^2}2\,g_{\d\nu}\bigl(g^{\nu\a}\wt g^\(\mu\d)+g^{\nu\mu}\wt g^\(\mu\a)\bigr)
-2\ov M\wt g^\(\mu\a)\Bigr]\biggr\}\pp\Ps{x^\a}\biggr]\\
{}-\frac n{r^2}\,\ex{n\Ps}\wt{\ul P}
-\frac{(n+2)\a_s^2}{\ell\pl^2}\,\ex{(n+2)\Ps}\wt R(\wt\G)=0.
\e
We neglect in Eq.\ \er{22.15} terms involving \YM' fields and Higgs' fields.

We are interested in a propagation of this field in Riemannian
geometry. Thus Eq.~\er{22.15} simplifies. The \co ical terms in the full \lg\
of the theory can be considered as selfinteraction potential of a scalar
field~$\Ps$
\beq22.16
U(\Ps)=\ex{(n+2)\Ps}\,\frac{\a_s^2\wt R(\wt\G)}{\ell\pl^2}+
\frac{\wt {\ul P}}{r^2}\,\ex{n\Ps}=\g \ex{n\Ps}+\b \ex{(n+2)\Ps}.
\e
The \co ical ``\ct'' (\co ical term) is equal to
\beq22.17
\ov\la_{\rm co}(\Ps)=-\frac\g2\,\ex{n\Ps}-\frac\b2 \ex{(n+2)\Ps}=-\frac12 U(\Ps).
\e
The simplified \e\ for $\Ps$ looks:
\beq22.18
2\ov M \wt\nabla_\a(g^{\a\b}\pa_\b \Ps)-(n+2)\ex{(n+2)\Ps}\b
-n\ex{n\Ps}\g=0.
\e

In order to find a \co ical \ct\ in the theory we should minimize a
selfinteraction potential \wrt field~$\Ps$. One gets that for the value
\beq22.19
\ex{\Ps_0}=x_0=\sqrt{\frac{n|\g|}{(n+2)\b}}
\e
we get a minimum for $U$. In this way
\beq22.20
\ov\la_{\rm co}(\Ps_0)=\frac{x_0^n|\g|}{(n+2)}
\e
($\g<0$, $\b>0$). $\ov\la_{c0}(\Ps_0)$ can be considered as a \co ical \ct\
(a~true \ct).

Let us notice the following fact. If we write $\Ps=\Ps_0+\vf$ where we have
redefined field $\Ps$ we get
\beq22.21
2\wt\nabla_\a \bigl(g^{\a\b} \pa_\b \vf)-\frac{x_0^n|\g|n}{\ov M}\,
\ex{n\vf}(\ex{2\vf}-1)=0.
\e
Using Eq.\ \er{22.20} we get
\beq22.22
\wt\nabla_\a \bigl(g^{\a\b}\pa_\b \vf)-\frac{n(n+2)}{2\ov M}\,
\ov\la_{c0}\ex{n\vf}(\ex{2\vf}-1)=0.
\e
Taking a known value of a contemporary \co ical \ct\ as $\ov\la_{\rm co}$
($\wt\La=10^{-52}\frac1{\rm m^2}$ or $\wt\La=0.53\tm 10^{-26}\frac{\rm kg}{\rm m^3}$) we
come to the equation
\beq22.22a
\wt\nabla_\a \bigl(g^{\a\b}\pa_\b \vf\bigr)+\wt\ve \ex{n\vf}(\ex{2\vf}-1)=0,
\e
where $\wt\ve=\sgn\ov M$. In Eq.~\er{22.20} we use natural scales of space and
time \cd s.
\beq22.23
L=\sqrt{\frac{n(n+2)\ov\la_{c0}(\Ps_0)}{2|\ov M|}}
\simeq 10\,{\rm Mpc}, \q T=\frac Lc\simeq32\times10^6{\rm yr.}
\e
$T$ is of order of a geological time or a propagation time of temperature
perturbations from a centre of the Sun to its surface.

In linear \ap ion in Minkowski space one gets
\beq4.144
\eta^{\a\b}\pa_{\a\b}\vf + 2\wt\ve \vf=0
\e
or in ordinary \cd s
\beq4.145
\eta^{\a\b}\pa_{\a\b}\vf + \frac{\wt\ve n(n+2)\wt\La}{|\ov M|}\vf =0.
\e
Thus in  linear \ap ion field $\vf$ is massive with a mass
\beq4.146
m_0=\sqrt{\frac{n(n+2)\wt\La}{|\ov M|}}
\e
and a \co ical \ct\ is equal to
\beq4.147
\wt\La=\ov\la_{\rm co}=\frac{x_0^n|\g|}{n+2}\,.
\e

Let us consider Eq.~\er{4.112}, i.e.\ the equation for field~$\Ps$. Let us take
such a solution of~$\Ps$ that $\Ps=\Ps_0$ in such a way that $\Ps_0$ extremizes
(minimalizes) a self\ia\ term (a~\co ical term) for the field~$\Ps$.

One can easily calculate a \co ical ``\ct'' (a~\co ical term) from Eq.~\er{22.20}.
One gets
\beq4.147a
\wt\La=\ov\la_{c_0}(\Ps_0)=-\frac1{2(n+2)}\biggl(\frac n{n+2}\biggr)^{n/2}
\biggl(\frac{\wt{\ul P}}{r^2}\biggr)\cdot\biggl|\frac{\ell\pl^2}{\a_s^2r^2}\,
\frac{\wt{\ul P}}{\wt R(\wt\G)}\biggr|^{n/2}.
\e
Due to previous considerations from Section 3 we get the \ct\ of desired value
using $k^0_{\td a\td b}$, $k_{ab}$, $\gd h,0,\ta\tb,$  tensors and also \ct s $\xi$ and~$\z$.

In this way an effective \gr al \ct\ is equal to $G_{\rm eff}=K(\Ps)=
G_Ne^{-(n+2)\Ps}$. Moreover, if we consider our world (our Universe) as
a~world with a minimum (extremum) of a \co ical term we can rescale $G_{\rm eff}$
and $m_{\rm eff}$
in such a way that $G_N=G_{\rm eff}(\Psi_0)$ and $G_N=G_0 e^{-(n+2)\Ps_0}$, where
$G_0=G_{\rm eff}(0)$, $m_{\rm eff}=(m_{\td A}e^{\frac n2 \Psi_0})e^{\frac n2\vf}$,
$m^0_{\td A}=m_{\td A}e^{\frac n2 \Psi_0}$.

Thus effective \gr al $G_{\rm eff}$ and $m_{\rm eff}$ ``\ct s'':
$$
G_{\rm eff}= G_Ne^{-(n+2)\vf}, \quad m_{\rm eff}=m^0_{\td A}e^{\frac n2\vf},
$$
where $\vf$ is a field defined by us above. From Eq.~\er{4.112} it is easy
to see that this field is massive in a linear approximation with a mass
calculated by us. It is convenient to introduce a \qs\ field~$q_0$ with the
same mass. In all field equations we can write $\Ps=\Ps_0+\vf$ (introducing
$q_0$ if necessary, for a definition of $q_0$ see the second \e\ in Eq.~\er{5.63b}).

Let us consider a full field equation for a \gr al field, i.e.\ Eqs \er{D.15},
\er{D.16} and \er{D.17} after elimination of the field~$W_\mu$. In this case
we put in all the formulae $\Ps_0=\Ps+\vf$ with a rescaling of $G_{\rm eff}$
we mentioned before. The most important for us is Eq.~\er{D.17}.

Let us give the following remark. The \co ical \ct\ in the theory is of
a~dynamical origin. It is the minimum (extremum) of a self\ia\ \pt\ of
a~field~$\Ps$, $\la_{c0}(\Ps_0)=U(\Ps_0)$.

In linearized field \e s, especially in Eq.~\er{D.17}, we put in the place
of $U(\Ps)$ a value $U(\Ps_0)$ introducing the \co ical \ct.

Let us consider a linearization procedure for the full field equations \wrt
$h_{\mu\nu}$, i.e.\ (see also Ref.~\cite{g}, for comparison)
\beq4.148
g_{\mu\nu} = \eta_{\mu\nu}+h_\(\mu\nu)+h_\[\mu\nu], \q |h_{\mu\nu}|\ll 1.
\e
(weak field approximation). One gets
$$
g^{\mu\nu} \simeq \eta^{\mu\nu}-\eta^{\mu\a}\eta^{\nu\b}h_{\a\b},
$$
$\eta_{\a\b}$ is a Minkowski tensor (weak field approximation).
\bg4.149
\Box F_{\mu\nu\la} = 2\wt\La F_{\mu\nu\la}\\
F_{\mu\nu\la} = h_{[[\mu\nu],\la]} \label{4.150} \\
\dg h,[\mu\nu],{,\nu},=0. \label{4.151}
\e

Equation \eqref{4.149} can be rewritten in a more general way using
\beq4.168
\frac14 \,\tv\n_{[\g} \gd\D,\a,{\tl\b]|\a|,\mu\tp}, - \frac12\tv\n_\d
\tv Q_{[\b\g,\mu]} = 8\pi \nad{eff}T_{\tl[\b\g],\mu\tp},
\e
$8\pi \nad{eff}T_{\tl[\b\g],\mu\tp}$ is the right-hand side of Eq.~\er{D.17}.
One gets
\beq4.169
\Box F_{\mu\nu\la} = 2\wt\La F_{\mu\nu\la} + 16\pi\nad{eff\,lin}T_{\tl[\mu\nu],\la
\tp}
\e
where $\nad{eff\,lin}T_{\tl[\mu\nu],\la\tp}$ is a linearized version of the right-hand
side of \eqref{4.168}.

We give below formulae which we use for derivation of Eq.~\er{4.169}.
\begin{equation}\label{4.170}
\gathered
\gd \ov Q,\nu,\g\mu,= \frac12\Bigl(\dg K,\g\mu,\nu, - 2g^{\hp{\tl[\mu}\a}
_{\tl[\mu\cdt]}K_{\g\tp\a\b}g^\[\nu\b]\Bigr)
=\frac12 \Bigl(-\tv\n_\g g^{\hp{[\mu}\nu}_{[\mu\cdt]} - \tv\n_\mu
g^{\hp{[}\nu}_{[\cdt\g]}\\ {} + \wt g{}^\(\nu\rho) \tv\n_\rho g_\[\g\mu]
-2g^{\hp{\tl[\mu}\a}_{\tl[\mu\cdt]} \Bigl(-\tv\n_{\g\tp} g_\[\a\b] -\tv\n_\a
g_{[\b\g]\tp} + \tv\n_\b g_{[\g\tp \a]}\Bigr)g^\[\nu\b]\Bigr),
\endgathered
\end{equation}
$\tv\n$ is a covariant derivative \wrt \LC \cn\ generated by $g_\(\a\b)$ on~$E$.
\begin{gather}\label{4.171}
\gathered
\gd\D,\nu,\g\mu, = \wt g^\(\nu\d) \biggl\{\Bigl(-\tv\n_\d g^{\hp{[(\g}\a}
_{[(\g\cdt]} - \tv\n_{(\g} g^{\hp{[}\a}_{[\cdt\d]} + \wt g{}^\(\a\rho)
\tv\n_\rho g_\[\d(\g] \Bigr)\cdot g_\[\mu)\a]\\
{}+g^{\hp{[\rho}\b}_{[\rho\cdt]} \Bigr[g^{\hp{([\mu}\rho}_{([\mu\cdt]}
\Bigr(-\tv\n_{\g)} g_\[\a\b] - \tv\n_\a g_\[\b\g)] + \tv\n_\b g_\[\g)\a]\Bigr)
g^{\hp{[\d}\a}_{[\d\cdt]} \\
{}- \Bigl(-\tv\n_\d g_\[\a\b] - \tv\n_\a g_\[\b\d]
+\tv\n_\b g_\[\d\a]\Bigr) g^{\hp{([\g}\a}_{([\g\cdt]} g^{\hp{[\mu)}\rho}
_{[\mu)\cdt]}\Bigr]\biggr\},
\endgathered  \\
\gathered
\frac14\tv\n _{\tl\o} \wt g{}^\(\nu\d) \biggl\{\Bigl(-\tv\n_\d g^{\hp{[(\b\tp}\a}
_{[(\b\tp\cdt]} - \tv\n_{(\b\tp} g^{\hp{[}\a}_\[\cdt \d] + \wt g{}^\(\a\rho)
\tv\n_\rho g_{[\d(\b]\tp}\Bigr) g_\[|\nu|)\a] \\
{}+g^{\hp{[\rho}\b}_\[\rho\cdt] \Bigl[ g^{\hp{(|\nu|}\rho}_{(|\nu|\cdt]} \Bigl(
-\tv\n_{\d)}g_\[\a\b\tp] - \tv\n_\a g_\[\b\tp\d)] +\tv\n_{\b\tp}
g_\[|\d|)\a] \Bigr) g^{\hp{[\d}\a} _\[\d\cdt] \Bigr]\biggr\}_{,\mu\tp}\\
{}-\frac14 \tv\n_\nu\biggl(-\tv\n_{\tl\b} g^{\hp{[\o}\nu}_\[\o\cdt]
-\tv\n_{\tl \o} g^{\hp{[}\nu}_\[\cdt\tl\b] + \wt g{}^\(\nu\rho)
\tv\n_\rho g_\[\tl \b\o]\\
{}+ 2g^{\hp{\tl[\o}\a}_{\tl[\o\cdt]} \Bigl(-\tv\n_{[\b\tp}g_\[\a\rho]
-\tv\n_\a g_\[\rho\tp\b] +\tv\n_\rho g_\[[\b\tp\a]\Bigr) g^\[\nu\rho]\biggr)
_{,\mu\tp} = 8\pi \nad{eff}T_{\tl[\b\o],\mu\tp}
\endgathered \label{4.172}\\
\gathered
\tv R_{\b\g} = 8\pi \nad{eff}T_\(\b\g) + \frac34 \,\tv\n_\nu \biggl(\wt g{}^\(\nu\d)
\Bigl\{\Bigl(-\tv\n_\d g^{\hp{[(\b}\a}_\[(\b\cdt]
-\tv\n_{(\b} g^{\hp{[}\a}_\[\cdt\d] + \wt g{}^\(\a\rho)\tv\n_\rho
g_\[\d(\b]\Bigr) \cdot g_\[\g)\a] \\
{}+ g^{\hp{[\rho}\si}_\[\rho\cdt]\Bigl[ g^{\hp{([\g}\rho}_{([\g\cdt]}
\Bigl(-\tv\n_{(\b} g_\[\a\si] - \tv\n_\a g_\[\si(\b] + \tv\n_\si g_{(\b\a]}\Bigr)
g^{\hp{[\d}\a}_\[\d\cdt] \\
{}-\Bigl( -\tv\n_\d g_\[\a\si] - \tv\n_\a g_\[\si\d] + \tv\n_\si g_\[\d\a]\Bigr)
\cdot g^{\hp{[\g}\a}_\[\g\cdt] g^{\hp{[\g)}\rho}_\[\g)\cdt] \Bigr]\Bigr\}\biggr)\\
{}-\frac14\, \tv\n_{(\g} \biggl\{\wt g{}^\(\nu\d) \biggl\{\Bigl(-\tv\n_\d g^{\hp{[(\b}\a}
_\[(\b\cdt] -\tv\n_\b g^{\hp{[}\a}_\[\cdt\d] + \wt g{}^\(\a\rho) \tv\n_\rho
g_\[\d(\b] \Bigr)\cdot g_\[\nu)\a] \\
{}+ g^{\hp{[\rho}\si}_\[\rho\cdt] \Bigl( g^{\hp{([\nu}\rho} _{([\nu\cdt]}
\Bigl(-\tv\n_{\b)}g_\[\a\si] - \tv\n_\a g_\[\si\b)] + \tv\n_\si g_\[\b)\a]
\Bigr) g^{\hp{[\d} \a}_\[\d\cdt]\\
{}- \Bigl(-\tv\n_{\b)} g_\[\a\si] -\tv\n_\a g_\[\si\b)]
+\tv\n_\si g_\[\b)\a] \Bigr) g^{\hp{[\d}\a}_\[\d\cdt] \\
{}-\Bigl( -\tv\n_{\d} g_\[\a\si] - \tv\n_\a g_\[\si\d] + \tv\n_\si g_\[\d\a]
\Bigr)\cdot g^{\hp{[\g}\a}_\[\g\cdt] g^{\hp{[\nu)}\rho}_\[\nu)\cdt]
\Bigr]\biggr\}\biggr\}
\endgathered \label{4.173}
\end{gather}
$\tv R_{\b\g}$ is a Ricci tensor formed for $g_{(\a\b)}$.

Let us come back to our linearisation.
For $\wt {\ov R}_{\mu\nu}$ we have
\bg4.152
\wt{\ov R}_{\mu\nu} = -\frac12 \Box \ov h_\(\mu\nu) \\
\ov h_\(\mu\nu) = h_\(\mu\nu) - \frac12 \eta^{\a\b} h_\(\a\b) \eta_{\mu\nu} \label{4.153} \\
\gd h,(\mu\nu),{,\nu},=0 \label{4.154}\\
\Box \ov h_\(\mu\nu)=-16\pi (\nad{eff.lin}T_\(\mu\nu)), \nonumber
\e
where $\nad{eff.lin}T_\(\mu\nu)$ is a linearization of $(\nad{eff}T_\(\mu\nu)
- \frac12 g_\(\mu\nu) g^{\a\b}\nad{eff}T_{\a\b})$ in first order.

The field $h_\[{\mu\nu}]$ has a spin one (see Ref.~\cite{l}). In some sense $h_\[{\mu\nu}]$
is a different form of a vector field (Proca field with nonzero mass). In the case
of zero mass ($\wt\La=0$) it has spin zero.

In this way a skewon field $h_\[\mu\nu]$ is massive in a linear \ap ion due to
\co ical term.

Let us consider a new field $a^\rho$ such that
\beq4.177
\gd\ve,\mu\nu\la,\rho, F_{\mu\nu\la} = a_\rho, \quad a^\rho=\eta^{\a\rho}a_\a.
\e
One gets
\beq4.178
\pp{a_\rho}{x^\rho} = \eta^{\a\b} a_{\a,\b}=0
\e
and
\beq4.179
\gathered
\Box a_\rho=m^2 a_\rho + 16\pi \gd\ve,\mu\nu\la,\rho, \nad{lin}T
_{\tl[\mu\nu],\la\tp}\\
\pp{a_\rho}{x^\rho}=0, \quad m^2=2\wt\La.
\endgathered
\e
Thus $a_\rho$ is a Proca field with the same mass as a skewon field.
$\wt\La$~is a measured \co ical \ct.

Let us introduce a tensor of strength of $a_\rho$
\bg4.180
f_{\mu\nu} = \partial_\mu a_\nu - \partial_\nu a_\mu\\
h_{\tl[\mu\nu],\la\tp} = F_{\mu\nu\la} = \frac16\,\ve_{\mu\nu\la\rho}a^\rho
\label{4.181}
\e
Using $f_{\mu\nu}$ we get
\beq4.181a
\gd f,\mu,{\nu,\mu}, = \Box a_\nu
\e
and eventually
\beq4.182
\gathered
\gd f,\mu,{\nu,\mu}, = m^2a_\rho + 16\pi \gd\ve,\mu\nu\la,\rho, \cdot
\nad{eff\,lin}T_{\tl[\mu\nu],\la\tp}\\
m=m_{\rm skewon}
\endgathered
\e
$\ve_{\mu\nu\la\rho}$ is an anti\s ic symbol \st $\ve_{1234}=1$.

Let us consider Eq.~\er{4.182} with an \elm c field in first order of \ap ion.
\beq4.183
\Box a_\nu = m^2a_\nu + \frac{16\pi G_N}{c^4}\, \nad{em}T _{\tl[\mu\a],\la\tp}
\gd\ve,\mu\a\la,\nu,
\e
where
\beq4.184
\nad{em}T_\[\mu\nu] \cong -\frac1{16\pi} \,F^{\a\b}F_{\a\b} h_\[\mu\nu].
\e
One gets
\beq4.185
\Box a_\nu = m^2a_\nu - \frac{G_N}{c^4} \,F^{\a\b}F_{\a\b}\cdot
\gd\ve,\mu\g\la,\nu, h_{\tl[\mu\g],\la\tp}
\e
or
\beq4.186
\Box a_\nu = m^2a_\nu - \frac{G_N}{c^4} \,F^{\a\b}F_{\a\b}a_\nu.
\e
Let us notice the following fact: an \ia\ term
\beq4.187
\wt V_{\rm int}= \frac{G_N}{c^4}\,F^{\a\b}F_{\a\b}a_\nu
\e
is very weak because of the \ct\ in front of $\frac{G_N}{c^4}$.

One can derive Eq.~\er{4.186} from the following \lg\ in a Minkowski space
(remember we are in the first order of \ap ion of a weak field):
\beq4.188
\cL = -\frac14\, f^{\mu\nu}f_{\mu\nu} - \frac{m^2}2 \,a_\nu a^\nu
-\frac{G_N}{2c^4} (F^{\a\b}F_{\a\b}) a_\nu a^\nu.
\e
We neglect \e s for an \elm c field $F_{\a\b}$. Thus the \ia\ of our
``skewon'' field $a_\nu$ is very weak with an \elm c field (with any field).
It means that ``skewon'' Dark Matter is extremely hard to be detected. It
interacts really only \gr ally. We neglect all terms connecting to gauge
fields, Higgs' fields, except \elm c field. Those neglected terms are similar
in a form and can easily be derived. $a_\mu$ is really a pseudovector because
of $\ve_{\mu\nu\a\la}$ anti\s ic symbol in the definition. We have
$\gd P,\la,\mu, a_\la(Px)=-a_\mu(x)$, where $\gd P,\la,\mu,$ is a~space inverse
operation
$$
\gd P,\la,\mu,=\left(\begin{matrix}
-1 &\ &0 &\ & 0 &\ & 0\\
0 && -1 && 0 && 0\\
0 && 0 && -1 && 0\\
0 && 0 && 0 && 1 \\
\end{matrix}\right), \quad Px= P(\bs x,\bs y, \bs z, t) =
(-\bs x,-\bs y,-\bs z,t)=(-\bs{\vec r},t).
$$

Similar situation we have to do with the field~$\vf$ (small oscillation
around the minimum of a self-\ia\ \pt\ of the field~$\Ps$).
Using Eq.~\er{4.112} and the decomposition $\Ps=\Ps_0+\vf$ one gets in
Minkowski \spt\ for small~$\vf$:
\begin{gather}
(2\ov M\Box - m_0^2)\vf - 8\pi(n+2)e^{-(n+2)\Ps_0}\cLY(A)e^{(n+2)\vf}
-4\,\frac{e^{-2\Ps_0}}{r^2} \,\cL\dr{kin}(\Phi,A)e^{-2\vf} \hskip30pt\nonumber\\
\hskip50pt {}+\frac{(n-2)}{r^4}\,e^{(n-2)\Ps_0}\wh V(\Phi)e^{(n-2)\vf}
+\frac{4(n-2)}{r^2}\, e^{(n-2)\Ps_0}\cL\dr{int}(\Phi,A)e^{(n-2)\vf}=0 \lb 4.189 \\
\cL = \ov M\pa^\mu \vf \pa_\mu \vf - \frac12\,m_0^2\vf^2
+8\pi e^{-(n+2)\Ps_0}\cLY e^{-(n+2)\vf} + 2\,\frac{e^{-2\Psi_0}}{r^2}
\cL\dr{kin}(\Phi,A)e^{-2\vf} \hskip30pt \nonumber\\
\hskip50pt {}- \frac{e^{(n-2)\Ps_0}}{r^4}\wh V(\Phi)
e^{(n-2)\vf} - \frac{4e^{(n-2)\Ps_0}}{r^2} \cL\dr{int}(\Phi,A) e^{(n-2)\vf}=0.
\label{4.190}
\end{gather}
$m_0$ is the mass of a scalaron. In Ref.~\cite{xx} we consider a different
scheme of linearization of field \e s getting similar results.

The exponential factor due to smallness of the field~$\vf$ ($|\vf|
\ll1$) does not make strong \ia s between photons, gauge bosons,
Higgs' bosons and scalarons.

What is a mass of a skewon field? It is easy to see that (where we use a value
for a \co ical \ct\ known from observation)
\beq4.155
m_{\rm skewon} = \sqrt{\ov\la_{c0}(\Ps_0)} \simeq 10^{-5}\ {\rm eV}.
\e
Let us come back to the scalar field $\vf$ (or~$\Ps$). In theoretical considerations
it is better to work with the field~$\vf$. Moreover, in \co y it is better to use
$\Ps$~field. This field is known as a so called {\it quintessence field\/}. This field
is a source of a \co ical \ct. Moreover we can proceed an inflationary scenario using
quintessence field. Due to quintessence field we get also non-Newtonian limits
in \gr al physics. This results in very interesting behaviour ``of an effective
\gr al \ct'' $G_{\rm eff}$ (see Ref.~\cite5). This can explain anomalous acceleration
of Pioneer 10/11 spacecrafts (see Refs \cite7, \cite{7a}).
Quintessence field from the \E\nos\ \KK (Jordan--Thiry) Theory is a basic tool to get
inflationary models in \co y (see Refs \cite5, \cite{xx}). The mass of a quintessence
particle is easy to calculate. Quintessence \pc\ is a scalar \pc, a~scalaron.

One gets
\beq4.156
m_0=m_{\rm quintessence} = m_{\rm scalaron}= \frac12 \frac{\hbar}c \,
\sqrt{\frac{n(n+2)}{2|\ov M|}}\,\sqrt{\ov\la_{c0}(\Ps_0)}\,.
\e
Using the value of $\ov\la_{c0}$ one finally gets
\beq4.157
m_0\simeq \sqrt{n(n+2)}\cdot 0.17\cdot 10^{-39}\,\text{g}\quad (2\ov M=1)
\e
or
\beq4.157a
m_0\simeq \sqrt{n(n+2)}\cdot 0.95\cdot 10^{-5}\,\text{eV}\quad (2\ov M=1).
\e
For example, if we take $n=14(=\dim G2)$, one finally gets
\beq4.158
m_0\simeq 14.2\cdot 10^{-5}\,\text{eV}.
\e

This value is bigger than that considered by different authors. Moreover,
still sufficiently small. The \pc\ interacts only gravitationally and because
of this it is undetectable by using known experimental methods.
Taking a density of Dark Energy as $0.7$ of a critical density,
\beq4.159
\rho_c=1.88 h^2 \cdot 10^{-29}\frac{\text{g}}{\text{cm}^3}\,,
\e
one gets a number of quintessence \pc s per unit volume
\beq4.160
\ov n=\frac{h^2}{\sqrt{n(n+2)}}\cdot1.31 \cdot10^{10}\, \frac1{\text{cm}^3}
\e
where $h$ is a dimensionless Hubble \ct\ $0.7<h<1$. Taking $n=14$ and $h=0.7$
one finally gets
\beq4.161
\ov n=4\cdot10^8\,\frac1{\text{cm}^3}
\e
which is many orders of magnitude smaller than Loschmidt number. Thus a gas
of quintessence \pc s is not so dense from the point of view of our earth conditions.
However, if this number of \pc s per unit volume is considered in a container of
size 200\,Mpc, the gas can be considered as extremely dense.

In order to settle---is this gas dense or not---we should calculate a mean
scattering length. The scattering cross-section for a quintessence \pc
\beq4.162
\sigma=\frac1{\ov\la_{c0}(\Ps_0)}=10^{52}\,\text{m}^2.
\e
A mean scattering length
\beq4.163
l=\frac1{\sigma \ov n}
\e
where $\ov n$ is a number of quintessence \pc s per unit volume (Eq.~\er{4.161}).

One gets
\beq4.164
l=10^{-60}\,\text{m}.
\e
It means that a gas of quintessence \pc s is extremely dense (if we apply the Knudsen
criterion---a gas is dense if $l\ll L$, where $L$ is the size of the
container) even in the Solar System.

One can find a relation between a skewon mass and a quintessence particle mass.
It is
\beq4.165
m_{\rm quintessence} = \frac12 \sqrt{n(n+2)}\, m_{\rm skewon}.
\e

Let us notice the following fact. In our theory we have two natural
candidates for a Dark Matter. They are a skewon and a quintessence particle. Both
particles are massive with a mass of the same order ($10^{-5}$\,eV, see
\er{4.165}). They are weakly interacting with an ordinary matter, they
are a part of gravity. Skewon is a quant of a quantized (in a linear
approximation) skew-\sy ic part of the metric. In contradiction to a graviton
it obtains a mass due to a \co ical \ct. A~quintessence \pc\ is a quant of a quantized
scalar field $\varPsi$ in a linear approximation. The scalar field $\varPsi$ plays many
roles in the theory. It influences an effective \gr al \ct\ and \co ical terms.

In this way we get Self-Interacting Dark Matter, scalar Dark Matter and
pseudovector Dark Matter. The Self-\E\ia\ of scalar Dark Matter is quite
strong (see next section). The pseudovector Dark Matter self-interact weaker.

Simultaneously we explain \co ical \ct\ via our quintessence scenario.

Both \pc s interact \gr ally---they are part of gravity. A~Dark Matter problem
appears on many levels in the Universe. On the level of galaxies, clusters of
galaxies and on the level of \co ical models. It seems, it is quite universal.
This universality is similar to the universality of \gr al interactions. Some
researchers claim that because of this it is natural to change a \gr al theory
in order to give a pure \gr al explanation of the effect of a Dark Matter.
Moreover, in our proposal we have to do with a deformation of General
Relativity (due to \nos\ metric and a scalar field from the \JT).
The theory satisfies a Bohr correspondence principle. The additional
\gr al degrees of freedom are universal as a gravity itself. Thus both massive
\pc s corresponding to these degrees of freedom can be considered seriously
as a Dark Matter (gravitons cannot be ``a~Dark Matter'', they are massless, they
are a part of a radiation). \E\co ical models with such a Dark Matter are
examined in Appendix~A (preliminary).

Let us give the following remark. We suppose that due to \co ical evolution
of the field~$\Ps$ (or $\vf$ or~$q_0$) we get such a configuration that
$U(\Ps_0)=\min$ and this value corresponds to the measured \co ical \ct. We
put this value into field \e s for remaining fields and proceed a linearization
procedure. We can fit (in principle) the value of a \co ical \ct\ using
$\gd h,0,\ta\tb,, \z, \gd k,0,\ta\tb,,k_{ab},\xi$ to the measured value~$\wt\La$
(Eqs~\er{22.17}, \er{4.147a}), $2\wt\La=-U(\Psi_0)$. We can also tune $2\ov M$
to~$1$.

\def\eq#1 {\label{5.#1}}
\def\rf#1 {\label{5.#1}}
\def\bq{\begin{equation}}
\def\tin#1{\hbox to\parindent{\hfil #1)\enspace}}

\section{Effective gravitational ``\ct'' $G_{\rm eff}$ in various
solutions of equation for a field~$\Ps$ (or~$\vf$ or~$q_0$).
The fifth force problem. A~Dark Matter or the fifth force ($G\dr{eff}$) or
both?}

Let us give the following remark. In our Universe we have to do with the
following kinds of matter:
\begin{enumerate}
\def\labelenumi{{\rm\arabic{enumi})}}
\item radiation,
\item barionic matter (ordinary),
\item cold Dark Matter,
\item hot Dark Matter (e.g.\ massive neutrinos),
\item \co ical \ct.
\end{enumerate}

In several approaches to Dark Matter and \co ical \ct\ a notion of a \qe\
appears. This name is coming from aristotelian philosophy of four elements: earth, water, air,
fire, and the fifth element---\ti{la quinta essentia} (a ``matter'' beyond
the Moon sphere) = \qe. For in our theory we
get a dynamical origin of a \co ical \ct, we call a scalar field in a
particular normalization---$q_0$---a \qe. (This is only an analogy to aristotelian
philosophy.) It is not a different degree of
freedom than $\Ps$ or $\vf$. Moreover, it is better to use~$q_0$ from pure
practical point of view and in this section we use also~$q_0$, calling it
a~\qe. We can use also~$Q$---a~\qe\ connected directly to~$\Ps$,
i.e.\ $Q=\frac{\Ps}{\,\ov\b\,}$, $\ov \b=\frac{8\pi |\ov M|}{\mpl^2}$,
$\mpl=\bigl(\frac{\hbar c^5}{G_N}\bigr)^{1/2}$ is a Planck's mass.
A scalar field can be connected to the so-called fifth force (see Ref.~\cite{Novc}).
In this approach a \qe\ (the fifth \el) is a source of the fifth force.

Let us write an equation for the scalar field $\Ps$ in terms of the scalar
field $\vf$. One easily gets
\begin{equation}
g^{\a\b}({\wt\nabla}_\a \pa_\b \vf)-\frac{n(n+2)}{16\pi M}\,\ov\la_{c0}(\Ps_0)
m\pl^2e^{n\vf}(e^{2\vf}-1)=0 \eq63a
\end{equation}
or in terms of $q_0$
\begin{equation}
g^{\a\b}({\wt\nabla}_\a \pa_\b q_0)+\wt\ve\ov \a(\exp(2\ov\b q_0)-1)=0,
\eq63b
\end{equation}
where
$$
\Ps = \Ps_0+\frac{m\pl}{2\sqrt{2\pi|\ov M|}}\,q_0= \Ps_0+\ov\b q_0
=\Ps_0+\vf.
$$

A scalar field $\vf$ (or $q_0$) is a Self-Interacting Dark Matter (SIDM). Its
\ia\ is highly nonlinear and is going to nonlinear phenomena (see Eqs
\er{5.63a}--\er{5.63b}).

In order to find some influence of $q_0$ (\qe) field on the value of the \ef\
\gr al \ct\ we consider a field equation for the scalar $q_0$ field in empty
space. One gets
\begin{equation}
\biggl(\frac{\pa^2q_0}{\pa t^2}-{\stpr\nabla}{}^2 q_0\biggr)+
\wt \ve \ov \a \exp(n\ov\b q_0)\bigl(\exp(2\ov\b q_0)-1\bigr)=0 \eq63c
\end{equation}
where
\begin{align}
\ov\a&=\frac{\ov\la_{c0}(\Ps_0)n(n+2)\mpl^2}{16\pi\cdot\ov M} \rf64 \\
\ov\b&=\frac{\mpl}{2\sqrt{2\pi|\ov M|}} \rf65 \\
\wt\ve&=\sgn\ov M, \ \wt\ve{}^2=1.\rf66
\end{align}

Let us consider a static, spherically \sy ic case. In the spherical
coordinates one gets
\begin{equation}
0=\frac1{r^2} \frac{d}{dr}\biggl(r^2\frac{dq_0}{dr}\biggr) - \wt\ve\ov\a
\exp(n\ov\b q_0)\bigl(\exp(2\ov\b q_0)-1\bigr) \eq67
\end{equation}
where $q_0=q_0(r)$ is a function of $r$ only. In order to treat this equation
it is easier to come back to the old variable $\vf=\ov\b q_0$. One gets
\begin{equation}
\frac1{r^2} \frac{d}{dr}\biggl(r^2\frac{d\vf}{dr}\biggr) - \frac14\wt\ve\ov\la_{c0}(\Ps_0)
\exp(n\vf)\bigl(\exp(2\vf)-1\bigr)=0. \eq68
\end{equation}
We change the in\dc t variable $r$ into $\tau$
\begin{equation}
\frac1{\tau^2} \frac{d}{d\tau}\biggl(\tau^2\frac{d\vf}{d\tau}\biggr) - \wt\ve
\exp(n\vf)\bigl(\exp(2\vf)-1\bigr)=0 \eq69
\end{equation}
where
\begin{equation}
r=\frac2{\sqrt{\ov\la_{c0}(\Ps_0)}}\,\tau \eq70
\end{equation}
and
\begin{equation}
\ov\la_{c0}=\frac{n(n+2)\ov\la_{c0}(\Ps_0)}{8\pi\ov M}\,\mpl^2\,. \eq71
\end{equation}

We consider Eq.\ \er{5.69} in two regions:
\begin{itemize}
\item[1)] for small fields $\vf$,
\item[2)] for large fields $\vf$.
\end{itemize}

In the first region we get
\begin{equation}
\frac1{\tau^2} \frac{d}{d\tau}\biggl(\tau^2\frac{d\vf}{d\tau}\biggr) + \wt\ve
e^{n\vf}=0. \eq72
\end{equation}
In the second region we get
\begin{equation}
\frac1{\tau^2} \frac{d}{d\tau}\biggl(\tau^2\frac{d\vf}{d\tau}\biggr) - \wt\ve
e^{(n+2)\vf}=0. \eq73
\end{equation}
Let us notice that both equations have similar nature and can be reduced to
the equation
\begin{equation}
\frac1{x^2} \frac{d}{dx}\biggl(x^2\frac{dy}{dx}\biggr) +\ve \wt\ve e^y=0 \eq74
\end{equation}
where in the first region $\ve=1$,
\begin{align}
y&=n\vf, \rf75 \\
x&=\sqrt n\,\tau, \rf76
\end{align}
and in the second region $\ve=-1$,
\begin{align}
y&=(n+2)\vf, \rf77 \\
x&=\sqrt {n+2}\,\tau, \rf78
\end{align}
We can transform \er{5.74} into
\begin{equation}
x\,\frac{d^2y}{dx^2}+2\,\frac{dy}{dx}+\ve\wt\ve xe^y=0 \eq79
\end{equation}
which is the celebrated Emden--Fowler equation known in the theory of gaseous
spheres (see~\cite{131}).
Let us notice that the first region (small fields) means large distances and
the second region (large fields) means small distances.

In this way we should consider Eq.~\er{5.79} in the region of small and
large~$x$. In the case of $\ve\wt\ve=1$ the equation \er{5.74} has an exact
solution
\begin{equation}
y=\ln\biggl(\frac2{x^2}\biggr). \eq80
\end{equation}
Let us apply this to both regions (remembering that $\wt\ve$ in both cases has
a different sign).

One gets in the first region
\begin{equation}
q_0=-\frac{2\sqrt{2\pi|\ov M|}}{\mpl n} \ln\Biggl(\frac{r}{{2\sqrt2}/
{\sqrt{n\ov\la_{c0}}}}\Biggr) \eq81
\end{equation}
and
\begin{equation}
G\dr{eff}=G_N\Biggl(\frac{r}{{2\sqrt2}/
{\sqrt{n\ov\la_{c0}}}}\Biggr)^{(n+2)/n}. \eq82
\end{equation}
In the second region
\begin{equation}
q_0=-\frac{2\sqrt{2\pi|\ov M|}}{\mpl (n+2)} \ln\Biggl(\frac{r}{{2\sqrt2}/
{\sqrt{(n+2)\ov\la_{c0}}}}\Biggr) \eq83
\end{equation}
and
\begin{equation}
G\dr{eff}=G_N\biggl(\frac{r}{{2\sqrt2}/
{\sqrt{(n+2)\ov\la_{c0}}}}\biggr). \eq84
\end{equation}
In this way we get an interesting prediction for the behaviour of the
strength of \gr al interactions.
In this very special solution $G\dr{eff}$ is going to zero if $r\to0$ and to
infinity if $r\to\infty$.

Let us come to Eq.~\er{5.79} supposing $\ve\wt\ve=1$. Thus we get
\begin{equation}
x\,\frac{d^2y}{dx^2}+2\,\frac{dy}{dx}+ xe^y=0. \eq85
\end{equation}
Using an exact solution \er{5.80} we write
\begin{equation}
y=y_1+\wt y \eq86
\end{equation}
and consider Eq.~\er{5.85} for large $x$.

In this way we get an approximate solution (given by Chandrasekhar~\cite{132})
\begin{equation}
y=\ln\biggl(\frac2{\eta^2}\biggr)+\frac A{\sqrt\eta}\cos\biggl(\frac{\sqrt7}2 \ln\eta\biggr)
-2\ln\ov\d, \quad |A|\ll 1, \eq87
\end{equation}
where
$\eta=\frac x{\,\ov\d\,}$, $A$ and $\ov\d$ are integration \ct s, $\ov\d>0$. In this
way we get in the first region
\begin{equation}
G\dr{eff}=\ov G\dr{eff}\cdot\exp\biggl(-\frac{\ov A}{\sqrt \eta}\cos\biggl(\frac{\sqrt7}
2 \ln\eta\biggr)\biggr) \eq88a
\end{equation}
where $\ov A$ is a \ct\ ($|\ov A|\ll 1$) and
\begin{equation}
\eta=\frac{\sqrt{n\ov\la_{c0}}\,\ov\d}{2\sqrt2}\,r, \eq88b
\end{equation}
$\ov G\dr{eff}$ is given by the formula \er{5.82}.

In the second region
\begin{equation}
G\dr{eff}=\ov{\ov G}\dr{eff}\cdot\exp\biggl(-\frac{\ov{\ov A}}{\sqrt \eta}\cos\biggl(\frac{\sqrt7}
2 \ln\eta\biggr)\biggr)\eq88c
\end{equation}
where $\ov{\ov A}$ is a \ct\ ($|\ov{\ov A}|\ll1$), $\ov{\ov G}\dr{eff}$ is given by
the formula \er{5.84} and
\begin{equation}
\eta=\frac{\sqrt{(n+2)\ov\la_{c0}}\,\ov\d}{2\sqrt2}\,r. \eq88d
\end{equation}
In this way we have very interesting non-Newtonian behaviour of $G\dr{eff}$
for large distances. Let us notice that the length scale is completely
arbitrary, because it is given by an integration \ct~$\ov\d$.

Let us consider Eq.\ \er{5.63b} in Cartesian coordinates supposing flat \sy y
for a \qe\ field $q_0=q_0(z,t)$ (nonstatic). One gets
\begin{equation}
\biggl(\frac{\pa^2q_0}{\pa t^2}-\frac{\pa^2q_0}{\pa z^2}\biggr)
-\wt \ve \ov \a \exp(n\ov\b q_0)\bigl(\exp(2\ov\b q_0)-1\bigr)=0. \eq89
\end{equation}
Let us change \dc t and in\dc t variables to $\xi,\eta,\vf$:
\begin{align}
z&=\frac2{\sqrt{\ov\la_{c0}}}\,\xi \rf90 \\
t&=\frac2{\sqrt{\ov\la_{c0}}}\,\eta \rf91 \\
\vf&=\ov\b q_0. \rf92
\end{align}
One gets
\begin{equation}
\biggl(\frac{\pa^2\vf}{\pa \eta^2}-\frac{\pa^2\vf}{\pa \xi^2}\biggr)
-\wt\ve e^{n\vf}\bigl(e^{2\vf}-1\bigr)=0. \eq93
\end{equation}
Eq.\ \er{5.93} is an equation for flat scalar (\qe) waves in our theory.
Let us consider it for large and small field~$\vf$ (as before).

In this way one gets the equation
\begin{equation}
\biggl(\frac{\pa^2 y}{\pa T^2}-\frac{\pa^2y}{\pa x^2}\biggr)
+\ve\wt\ve e^y=0, \eq94
\end{equation}
where in the region of small field $\vf$ we have $\ve=1$ and
\begin{align}
y&=n\vf \rf95 \\
x&=\sqrt n\, \xi \rf96 \\
T&=\sqrt n\, \eta \rf97
\end{align}
and in the region of large field $\vf$, $\ve=-1$ and
\begin{align}
y&=(n+2)\vf \rf98 \\
x&=\sqrt {n+2}\, \xi \rf99 \\
T&=\sqrt {n+2}\, \eta. \rf100
\end{align}

Eq.\ \er{5.94} is the famous Liouville equation which can be transformed via a
B\"ack\-lund \tr ation into a two-dimensional wave equation and afterwards
solved exactly. The general solution depends on two arbitrary functions $f$
and~$g$ of one variable, sufficiently regular. It is possible to consider
several problems for this equation: Cauchy initial problem, Darboux problem
and Goursat problem.

The general solution of \er{5.94} looks like ($\ve\wt\ve=-1$)
\begin{equation}
y(T,x)=\ln\biggl[\frac{2g'(x-T)f'(x+T)}{\bigl(g(x-T)+f(x+T)\bigr)^2}\biggr]
\eq101
\end{equation}
where $g'$ and $f'$ are derivatives of $g$ and $f$.

Thus one gets in the first region (small field)
\begin{gather}
q_0(t,z)=\frac{2\sqrt{2\pi \ov M}}{\mpl n}\cdot
\ln\Biggl[\frac{2g'\bigl(\frac{z-t}a\bigr)f'\bigl(\frac{z+t}a\bigr)}
{\bigl(g\bigl(\frac{z-t}a\bigr)+f\bigl(\frac{z+t}a\bigr)\bigr)^2}\Biggr], \eq102 \\
a=\frac{2\sqrt2}{\sqrt{n\ov\la_{c0}}}\,, \eq103
\end{gather}
and
\begin{equation}
G\dr{eff}=G_N\Biggl(\frac{\Bigl(g\bigl(\frac{z-t}a\bigr)+f\bigl(\frac{z+t}a\bigr)\Bigr)^2}
{2g'\bigl(\frac{z-t}a\bigr)f'\bigl(\frac{z+t}a\bigr)}\Biggr)^{(n+2)/n}. \eq104
\end{equation}
In the second region (large field)
\begin{gather}
q_0(t,z)=\frac{2\sqrt{2\pi \ov M}}{\mpl (n+2)}\cdot
\ln\Biggl[\frac{2g'\bigl(\frac{z-t}b\bigr)f'\bigl(\frac{z+t}b\bigr)}
{\Bigl(g\bigl(\frac{z-t}b\bigr)+f\bigl(\frac{z+t}b\bigr)\Bigr)^2}\Biggr], \eq105 \\
b=\frac{2\sqrt2}{\sqrt{(n+2)\ov\la_{c0}}}\,, \eq106
\end{gather}
and
\begin{equation}
G\dr{eff}=G_N\cdot\frac{\Bigl(g\bigl(\frac{z-t}b\bigr)+f\bigl(\frac{z+t}b\bigr)\Bigr)^2}
{2g'\bigl(\frac{z-t}b\bigr)f'\bigl(\frac{z+t}b\bigr)}. \eq107
\end{equation}

In this way we get a spatio-temporal pattern of changing the \ef\ \gr al \ct\
for small and large field regions. In both cases we have
$\ve\wt\ve=-1$. However, in the small field region we have $\ve=1$ and because of
this $\wt\ve=-1$. In the case of large field region $\ve=-1$ and $\wt\ve=1$. In
order to be in line with our assumptions we should consider in the first case
such functions $f$ and~$g$ that the expression in~\er{5.105} is small and for
the second case vice versa.

Let us consider Eq.\ \er{5.63b} in cylindrical coordinates supposing cylindrical
\sy y for the field~$q_0$, $q_0=q_0(\rho)$. One gets
\begin{equation}
\frac1\rho\biggl(\rho\,\frac{dq_0}{d\rho}\biggr)-\ve \ov\a\exp(n\ov\b q_0)\bigl(\exp(2\ov\b
q_0)-1\bigr)=0. \eq108
\end{equation}
This equation can be \tr ed into
\begin{gather}
\frac1\tau\,\frac{d}{d\tau}\biggl(\tau\,\frac{d\vf}{d\tau}\biggr)-\wt\ve \exp(n\vf)\bigl(e^{2\vf}
-1\bigr)=0, \eq109 \\
\rho=\frac2{\sqrt{\ov\la_{c0}}}\,\tau. \eq110
\end{gather}
As usual we consider Eq.\ \er{5.109} in two regions for small and large fields.

In the first region
\begin{equation}
\frac1\tau\,\frac{d}{d\tau}\biggl(\tau\,\frac{d\vf}{d\tau}\biggr)+\wt\ve e^{n\vf}=0. \eq111
\end{equation}
In the second region we get
\begin{equation}
\frac1\tau\,\frac{d}{d\tau}\biggl(\tau\,\frac{d\vf}{d\tau}\biggr)-\wt\ve e^{(n+2)\vf}=0.
\eq112
\end{equation}
Both equations can be reduced to the equation
\begin{equation}
\frac1x\,\frac{d}{dx}\biggl(x\,\frac{dy}{dx}\biggr)+\ve\wt\ve e^y=0, \eq113
\end{equation}
where in the first region $\ve=1$ and
\begin{align}
y&=n\vf \rf114 \\
x&=\sqrt n\,\tau \rf115
\end{align}
and in the second region $\ve=-1$ and
\begin{align}
y&=(n+2)\vf \rf116 \\
x&=\sqrt{n+2}\,\tau. \rf117
\end{align}
We can \tr\ \er{5.113} into
\begin{equation}
x\,\frac{d^2y}{dx^2}+\frac{dy}{dx}+\ve\wt\ve xe^y=0 \eq118
\end{equation}
which is the equation considered in~\cite{133} for $\ve\wt\ve=1$.

Following H. Lemke we write down a solution to Eq.\ \er{5.118} in a compact
form in three cases (concerning an integration \ct\ introduced by H.~Lemke).
We adopt his solutions to our problem.

\smallskip
\noindent 1) $C=\ka^2>0$, $\ka$---arbitrary positive number:
\begin{equation}
y(x)=-2\ln\biggl(\frac x{\,\ov a\,}\biggl(\biggl(\frac{\ov a}{x}\biggr)^\ka +\biggl(\frac{x}{\,\ov a\,}\biggr)^\ka \biggr)\biggr)
+\ln(4\ka^2\ov a{}^2) \eq119
\end{equation}
\indent where $\ov a>0$ is an arbitrary \ct.

\smallskip
\noindent 2) $C=-\om^2<0$, $\om$ is an arbitrary positive number:
\begin{equation}
y(x)=-2\ln\bigl(2x\sin(\om \ln x + \d)\bigr)+\ln(4\om^2) \eq120
\end{equation}
\indent where $\d$ is an arbitrary \ct.

\smallskip
\noindent 3) $C=0$:
\begin{equation}
y(x)=-2\ln\biggl(\frac x{\,\ov a\,}\ln\biggl(\frac x{\,\ov a\,}\biggr)\biggr)-2\ln\ov a \eq121
\end{equation}
\indent where $\ov a>0$ is an arbitrary \ct.

\smallskip
Using these solutions we write down a spatial \dc ce of $G\dr{eff}$ in the
case of small and large fields.
In the case of small fields one gets
\refstepcounter{equation}\rf122
$$
\dsl{
\tin1 \hfill G\dr{eff}=G_N(4\ka^2\ov a{}^2)^{(n+2)/n}
\frac{(1+r^{2\ka})^{2(n+2)/n}}{r^{2(\ka-1)(n+2)/n}} \hfill (\theequation)
}
$$
where
\bq
r=\frac1{2\ov a}\sqrt{n\ov\la_{c0}}\,\rho. \eq123
\end{equation}
\refstepcounter{equation}\rf124
$$
\dsl{
\tin2 \hfill G\dr{eff}=\frac{G_N(\ov\d)^{2(n+2)/n}}{(\om^2)^{(n+2)/n}}
\, r^{2(n+2)/n}\bigl(\sin(\om\ln r)\bigr)^{2(n+2)/n} \hfill (\theequation)
}
$$
where
\begin{align}
&r=\frac1{2\ov\d}\sqrt{n\ov\la_{c0}}\,\rho \rf125 \\
&\om\ln\ov\d=-\d \rf126
\end{align}
\refstepcounter{equation}\rf127
$$
\dsl{
\tin 3\hfill G\dr{eff}=G_N\ov a^{2(n+2)/n}r^{2(n+2)/n}(\ln r)^{2(n+2)/n} \hfill (\theequation)
}
$$
and $r$ is given by Eq.\ \er{5.123}.

In the case of large field one gets:
\refstepcounter{equation}\rf128
$$
\dsl{
\tin1 \hfill G\dr{eff}=G_N(4\ka^2\ov a^2)\frac{(1+r^{2\ka})^2}{r^{2(\ka-1)}} \hfill (\theequation)
}
$$
where
\begin{equation}
r=\frac1{2\ov a}\sqrt{(n+2)\ov\la_{c0}}\,\rho. \eq129
\end{equation}
\refstepcounter{equation}\rf130
$$
\dsl{
\tin2 \hfill G\dr{eff}=G_N\biggl(\frac{\ov\d}{\om}\biggr)^2 r^2 \bigl(\sin(\om\ln r)\bigr)^2
\hfill  (\theequation)
}
$$
where
\begin{equation}
r=\frac1{2\ov\d}\sqrt{(n+2)\ov\la_{c0}}\,\rho , \quad \ln \ov\d=-\frac\d\om. \eq131
\end{equation}
\refstepcounter{equation}\rf132
$$
\dsl{
\tin3 \hfill G\dr{eff}=G_N\ov a^2r^2(\ln r)^2 \hfill  (\theequation)
}
$$
and $r$ is given by Eq.\ \er{5.129}.

Let us notice that in that spatial \dc ce
for large and small field we have $\ka$ and $\ov a$ ($\ov\d$, $\om$) as
integration \ct s. In this way integration \ct s induce a power law of this
\dc ce and also a scale. For sufficiently big~$n$ ($n>14$) there is no
significant difference between both cases. It means the \qe\ field behaves
everywhere as for large field case (in these solutions of course). It is
evident that the spatial \dc ce (in cylindrical \sy y case) of~$G\dr{eff}$
goes to some kind of the fifth force. However, we have to do not with a
universal law of Nature but rather with some kind of initial conditions.

Let us consider three our cases \er{5.119}, \er{5.120} and \er{5.121} for small and
large fields cases.

The small field case is such that
\bq
e^\vf<1, \qquad\text{i.e. }\vf<0. \eq137
\end{equation}
The large field case is if
\bq
e^\vf>1, \qquad\text{i.e. }\vf>0. \eq138
\end{equation}
For \er{5.119} we have for the small case
$$
f(r)>1,
$$
where
\bq
f(r)=r^{1-\ka}+r^{\ka+1}, \qquad r=\biggl(\frac x{\,\ov a\,}\biggr). \eq139
\end{equation}

Let us consider two cases
$$
0<\ka<1 \qquad\text{and}\qquad \ka>1.
$$
In the first case $f(r)\nearrow$ in $[0,+\infty)$ and we have simply
$r>r_0$ where $r_0$ satisfies the equation
\bq
r_0^{2\ka+1}+r_0-1=0. \eq140
\end{equation}
In the second case $\ka>1$,
$$
\lim_{r\to0} f(r)=+\infty \qquad\text{and}\qquad \lim_{r\to\infty}
f(r)=+\infty  .
$$
The function $f(r)$ has a minimum at
\bq
r_1=\biggl(\frac{\ka-1}{\ka+1}\biggr)^{1/\ka}. \eq141
\end{equation}
Let us calculate $f(r_1)$.
\bq
f(r_1)=\biggl(\frac{\ka+1}{\ka-1}\biggr)^{\ka-1/\ka}+\biggl(\frac{\ka-1}{\ka+1}\biggr)^{\ka+1/\ka}. \eq142
\end{equation}
It is easy to see that if $\ka>1$, then $\ka-1/\ka>0$. This means that
\bq
f(r_1)>1 \eq143
\end{equation}
and therefore $f(r)>1$ for all $r>0$. Thus simultaneously we get a solution
for large field only if $\ka<1$, i.e.
\bq
r<r_0. \eq144
\end{equation}
If $\ka=1$, we always have $f(r)\ge1$ (i.e.\ only a small field).

Let us consider \er{5.120}. In this case the small field condition reads
\bq
h(r)=r\sin(\om\ln r)>1, \quad \text{where }r=\frac x {\,\ov \d\,}\,. \eq145
\end{equation}
First of all we need $h(r)>0$. Let us observe that $|h(r)|\le r$ for
every~$r>0$. Next we see that the roots of $h(r)$ are the numbers
\bq
r_{0,k}=e^{k\pi/\om}, \quad k=0,\pm1,\pm2,\ldots \eq146a
\end{equation}
and
\bq
h(r)>0 \quad\text{if}\quad
r_{0,2k}<r<r_{0,2k+1},\quad k=0,\pm1,\pm2,\ldots \eq146b
\end{equation}
The maximum of the function $h(r)$ in the interval
$(r_{0,2k},r_{0,2k+1})$ is greater than
\bq
h(r_{0,2k+1/2})=e^{2k\pi/\om}\cdot e^{\pi/(2\om)}, \eq147
\end{equation}
but smaller than $r_{0,2k+1}$. Thus the maxima are smaller than~1 for
negative $k$ and greater than~1 if $k=0,1,2,\ldots$. It
means that in each interval $(r_{0,2k},r_{0,2k+1})$, $k=0,1,2,\ldots$, there
exist two numbers $r_{3,2k}$ and $r_{2,2k}$ such that $r_{3,2k}<r_{2,2k}$,
\bq
h(r_{3,2k})=h(r_{2,2k})=1 \eq148
\end{equation}
and
\bq
h(r)>1 \quad\text{if}\quad r_{3,2k}<r<r_{2,2k}\,. \eq149
\end{equation}

The condition for large fields, $0<h(r)<1$, is satisfied if
\bq
r_{0,2k}<r<r_{0,2k+1}, \quad k=-1,-2,\ldots \eq150
\end{equation}
or
\bq
r_{0,2k}<r<r_{3,2k}, \quad \text{or}\quad r_{2,2k}<r<r_{0,2k+1},
\quad k=0,1,2,\ldots \eq151
\end{equation}

In the third case, i.e.\ Eq.\ \er{5.121}, one has for the small field case
\bq
r \ln r>1 \eq152
\end{equation}
where $r=\frac x{\,\ov a\,}$. Let $r_4\ln r_4=1$, $r_4>1$. We have
$r>r_4=1.7632\ldots$. In the large field case
\bq
0<r<r_4=1.7632\ldots. \eq153
\end{equation}

Thus we have in general large fields on large distances.

In this way we have solutions for large and small distances. One can try to
connect them to get a solution for all distances. However in this case it is
necessary to be very careful, for our solutions depend on some integration
\ct s which can be different for both asymptotic regions.

Let us consider Eq.\ \er{5.63b} in two special cases:
\begin{itemize}
\item[I] $q_0=q_0(z)$ --- static and depending only on~$z$;
\item[II] $q_0=q_0(t)$ --- non-static and spatially \ct.
\end{itemize}

Let us consider also these cases for small and large fields~$\vf$. In all of
these cases we come to the following equation
\bq
\frac{d^2y}{dx^2} + \ve_1\ve\wt\ve e^y=0 \eq154
\end{equation}
where $\ve_1=1$ for case I and $\ve_1=-1$ for case II.

Eq.\ \er{5.154} can easily be reduced to the integral
\bq
x-x_0=\frac{\ve_3}{\sqrt2} \int \frac{dy}{\sqrt{\ve_2\om^2-\eta e^y}} \eq155
\end{equation}
where $\eta=\ve_1\ve\wt\ve$, $\eta^2=1$, $C=2\ve_2\om^2$, $\om\ge0$, $\ve_2^2=1$ is
an integration \ct, $\ve_3^2=1$ and $x_0$ also is an integration \ct.

After some calculation we get the following solutions:
\refstepcounter{equation}\rf156
$$
\dsl{
\text{A.}\hfill y(x)=2\biggl[\ln \om-\ln\left|\sinh\biggl(
\frac{\sqrt2\,(x-x_0)\om}2 \biggr)\right|\biggr], \quad \eta=-1,\ \ve_2=1\hfill (\theequation) \cr
\refstepcounter{equation}\rf157
\text{B.}\hfill y(x)=2\biggl[\ln \om-\ln\left|\sinh\biggl(
\frac{\sqrt2\,(x-x_0)\om}2 \biggr)\right|\biggr], \quad \eta=1,\ \ve_2=1\hfill (\theequation)
}
$$
where
\bq
\frac{\sqrt2\,(x-x_0)\om}2 >\ln(1+\sqrt2)\eq158
\end{equation}
or
\bq
\frac{\sqrt2\,(x-x_0)\om}2 <\ln(\sqrt2-1).\eq159
\end{equation}
\refstepcounter{equation}\rf160
$$
\dsl{
\text{C.}\hfill y(x)=2\biggl[\ln\om-\ln\left|
\cos\biggl(\frac{\sqrt2\,(x-x_0)\om}2\biggr)\right|\biggr],
\quad \eta=-1,\ \ve_2=-1. \hfill (\theequation)
}
$$

Let us apply these solutions to our problems. First of all let us consider a
static configuration with $z$ \dc ce only. In this case $\ve_1=1$ and $\eta=\ve
\wt\ve$. For the small field case one gets, $\ve=1$, $\eta=\wt\ve$,
\bq
G\dr{eff}=\frac{G_N}{\om^{2(n+2)/n}}\left|\sinh p\right|^{(n+2)/n} \eq161
\end{equation}
where
\bq
p=\frac{\om(z-z_0)}4 \,\sqrt{n\ov\la_{c0}}\,. \eq162
\end{equation}
For the large field case we get, $\ve=-1$, $\eta=-\wt\ve$,
\bq
G\dr{eff}=\frac{G_N}{\om^2}\left|\cos p\right| \eq163
\end{equation}
where
\bq
p=\frac{\om(z-z_0)}4 \,\sqrt{(n+2)\ov\la_{c0}}\,. \eq164
\end{equation}
In this case $\wt\ve=1$ for $\eta=-1$.

In a nonstatic configuration $\ve_1=-1$ and $\eta=-\ve\wt\ve$. For the small
field case ($\ve=1$), $\eta=-\wt\ve$,
\bq
G\dr{eff}=\frac{G_N}{\om^{2(n+2)/n}}\left|\sinh q\right|^{(n+2)/n} \eq165
\end{equation}
where
\bq
q=\frac{\ve_3\om(t-t_0)}4 \sqrt{n\ov\la_{c0}} \eq166
\end{equation}
and analogically for the large field case ($\ve=-1$), $\eta=\wt\ve$,
\bq
G\dr{eff}=\frac{G_N}{\om^2}\left|\cos q\right|, \eq167
\end{equation}
\bq
q=\frac{\om(t-t_0)}4 \,\sqrt{(n+2)\ov\la_{c0}}\,. \eq168
\end{equation}
In this case $\wt\ve=-1$ for $\eta=-1$.

Let us notice that in a static configuration for small field we have two
possibilities for $\eta=-1$ (no condition on~$p$) and $\eta=1$ (conditions
\er{5.158}--\er{5.159}). Thus without conditions we have $\wt\ve=-1$ and with conditions
$\wt\ve=1$. In a non-static configuration for small field we have vice versa
$\wt\ve=1$ without conditions and $\wt\ve=-1$ with conditions \er{5.158}--\er{5.159}.

Let us come back to the Eq.\ \er{5.89} and consider it in a travelling wave
scheme. In this way we have
\bq
q_0(z,t)=\wt q(z-vt) \eq169
\end{equation}
where $v$ is a velocity of the travelling wave (a soliton), $|v|<1$. Let us
consider this equation in both regimes (for small and large fields). In this
way we come to the expression
\bq
(1-v^2)\,\frac{d^2\chi}{d\xi^2}-\ve\wt\ve e^\chi=0 \eq170
\end{equation}
where $\chi$ is a shape function of a soliton. Changing an independent
variable from $\xi$ to~$\la$ one gets
\bq
\frac{d^2\chi}{d\la^2}-\ve_1\ve\wt\ve e^\chi=0 \eq171
\end{equation}
where $\ve_1=-1$, i.e.\ we get Eq.\ \er{5.154} with $\eta=-\ve\wt\ve$,
\bq
\la=\frac \xi {\sqrt{1-v^2}},\quad \xi=\sqrt{1-v^2}\,\la. \eq172
\end{equation}

In this way we adopt our solutions in both regimes: small and large
field (changing $\chi$ into~$\la$). For small field we get ($\ve=1$,
$\eta=-\wt\ve$)
\bq
q_0(z,t)=\frac2{\ov \b n}\bigl[\ln\om - \ln\left|\sinh p\right|\bigr] \eq173
\end{equation}
where
\bq
p=\frac{\om\sqrt{n\ov\la_{c0}}}{4\sqrt{1-v^2}}\,(z-vt). \eq174
\end{equation}
For large field we get ($\ve=-1$, $\eta=\wt\ve$)
\bq
q_0(z,t)=\frac2{\ov \b (n+2)}\bigl[\ln\om - \ln\left|\cos p\right|\bigr] \eq175
\end{equation}
where
\bq
p=\frac{\om\sqrt{(n+2)\ov\la_{c0}}}{4\sqrt{1-v^2}}\,(z-vt). \eq176
\end{equation}
For \er{5.173} we have $\eta=-\wt\ve$ and because of this $\wt\ve=1$ without any
conditions and if $\wt\ve=-1$ we have conditions \er{5.158}--\er{5.159}. In the case of
the formula \er{5.174} $\eta=-1$ and $\wt\ve=-1$.

We can write down formulas for $G\dr{eff}$ in the soliton case
\bq
G\dr{eff}=\frac{G_N}{\om^{2(n+2)/n}}\left|\sinh p\right|^{(n+2)/n} \eq177
\end{equation}
and $p$ is given by the formula \er{5.174} (with or without conditions
\er{5.158}--\er{5.159}). This is of course a small field case.

In the large field case
\bq
G\dr{eff}=\frac{G_N}{\om^2}\left|\cos p\right|, \eq178
\end{equation}
and $p$ is given by the formula \er{5.176}. In this case $\wt\ve=-1$. (Let us
notice that this is a case of SO(3) group in our theory.)

Let us notice that conditions \er{5.158}--\er{5.159} can be considered as conditions
for small field in $z$ or $t$ domains. Let us notice that in our solutions
concerning a behaviour of an \ef\ \gr\ \ct\ we get completely arbitrary
length or time scale (given by integration \ct s). In this way a spatial or
time \dc ce of $G\dr{eff}$ can be (except the solution \er{5.80} and
simultaneously the approximate solution in the case of spherical \sy y) such
that $G\dr{eff}$ can be really \ct\ on distances (or times) accessible in
experiments.

In order to connect our results to the ordinary \gr\ physics we consider
again Eq.~\er{5.67} in small field regime for initial conditions $\vf(0)=0$ and
$\frac{d\vf}{dx}(0)=0$. The first condition means that we want to have
$G\dr{eff}(0)=G_N$ and the second that the \qe\ field does not grow quickly.
The problem cannot be solved analytically. Moreover R.~Emden in
Ref.~\cite{131} did it for us numerically. We quote here his results adopted to our
notation, $\ve=+1$ and $\wt\ve=-1$ (see Table~1).

\def\nh{\cr\noalign{\hrule}}
$$
\vbox{\offinterlineskip
\halign{\strut\vrule\quad \hfil$#$\quad &\vrule\quad \hfil$#$\quad
&\vrule\quad $#$\hfil \quad&\vrule\quad \hfil$#$\quad \hfil\vrule\cr
\noalign{\hrule}
\quad x\hfil&\quad -y\hfil&\hfil e^{y}\quad &\quad G\dr{eff}/G_N\nh
0.00&0.00000&1.00000&1.00000\nh
0.25&0.01037&0.98969&0.98823\nh
0.50&0.04113&0.95971&0.95409\nh
0.75&0.09113&0.91290&0.90109\nh
1.00&0.15903&0.85296&0.83380\nh
1.25&0.24225&0.78486&0.75816\nh
1.50&0.33847&0.71285&0.67920\nh
1.75&0.44488&0.64090&0.60143\nh
2.00&0.55967&0.57140&0.52749\nh
2.50&0.80584&0.44671&0.39813\nh
3.00&1.06226&0.34537&0.29670\nh
3.50&1.31937&0.26730&0.22138\nh
4.00&1.57071&0.20790&0.16611\nh
4.50&1.81246&0.16325&0.12601\nh
5.00&2.04264&0.12968&0.09686\nh
6.00&2.46598&0.08493&0.05971\nh
7.00&2.84160&0.05833&0.03887\nh
8.00&3.17489&0.04180&0.02656\nh
9.00&3.47128&0.03108&0.01893\nh
10.00&3.73646&0.02384&0.01398\nh
100&8.59506&0.000175&5.0854\cdot10^{-5}\nh
1000&13.09847&0.000002&3.0683\cdot10^{-7}\nh
}}
$$
\centerline{\small Table 1. $G\dr{eff}$ (explanations in a text below)}
\smallskip \noindent
where
\ealn
&x=\frac12 \sqrt{n\ov\la_{c0}}\,r \cong \biggl(\frac r{10\,\text{Mpc}}\biggr), \rf179 \\
&\frac{G\dr{eff}}{G_N}=(e^{y})^{(n+2)/n}=(e^{y})^{8/7}. \rf180
\end{align}
We take $n=14$ (this is equal to $\dim G2$, $G2=H$ group for GSW model).

It is easy to see that for large $n$
\bq
\frac{G\dr{eff}}{G_N}=e^{y}. \eq181
\end{equation}

It is easy to see that on a distance of $1\,$Mpc $G\dr{eff}$ does not differ
from~$G_N$. Even on a distance of $10\,$Mpc it is about 10\% smaller. Thus in
the Solar System Newtonian \gr\ physics does not change. Even on the level of
a galaxy this change is minimal and cannot be observed. Moreover, there is an
important conclusion: on distances about $200\,$Mpc the strength of \gr\
interactions is about $10^{-5}$ times this on short distances measured in the
Solar System (and for $10^3$\,Mpc of $10^{-7}$). It is hard to tell how it influences a mass of a cluster of
galaxies if we realize that from any observational data only a product $GM$
has been obtained (not~$M$).

From the other side on distances of $100\,$Mpc the strength of \gr\
interactions is very weak (not only because of the distance). Thus if we
consider clusters of galaxies as substrat \pc s in cosmology then they do not
interact.

Let us consider Eq.\ \er{5.63b} in Cartesian coordinates for two-dimensional
static case (i.e. $\frac{\pa}{\pa z}=0$, $\frac{\pa}{\pa t}=0$). One gets
\bq
\biggl(\frac{\pa^2q_0}{\pa x^2}+\frac{\pa^2q_0}{\pa y^2}\biggr)
-\wt \ve \ov \a \exp(n\ov\b q_0)\bigl(\exp(2\ov\b q_0)-1\bigr)=0 \eq182
\end{equation}
(where $\ov\a$, $\ov\b$ are given by formulas \er{5.64} and \er{5.65}).

As usual, we come to the formula
\bq
\biggl(\frac{\pa^2\vf}{\pa x_1^2}+\frac{\pa^2\vf}{\pa x_2^2}\biggr) - \wt\ve e^{n\vf}
(e^{2\vf}-1)=0 \eq183
\end{equation}
where
\bq
\left.\begin{matrix} x\\y\end{matrix}\right\} = \frac2{\sqrt{\ov \la_{c0}}}\,x_i,
\qquad i=1,2. \eq183a
\end{equation}
We consider Eq.\ \er{5.183} for small and large fields and we get
\bq
\biggl(\frac{\pa^2\chi}{\pa z_1^2}+\frac{\pa^2\chi}{\pa z_2^2}\biggr) - \ve\wt\ve e^{\chi}
=0 \eq184
\end{equation}
where as usual for a small field $\ve=1$ and
\ealn
\chi&=n\vf,  \rf185 \\
z_i&=\sqrt n\,x_i \rf186
\end{align}
and for a large field $\ve=-1$ and
\ealn
\chi&=(n+2)\vf, \rf187 \\
z_i&=\sqrt {n+2}\,x_i\,. \rf188
\end{align}
Thus we come to the equation known as Liouville equation
\bq
\D \chi=e^\chi \eq184a
\end{equation}
if $\ve\wt\ve=1$.

This equation can be explicitly solved. First of all we change in\dc t
variables into
\ealn
Z&=\frac1{\sqrt2}(z_1+iz_2) \rf189 \\
\chi(Z)&=-\ln\biggl(\tfrac12(1-|g|^2)\biggr)+\tfrac12 \ln\left|\frac{dg}{dZ}\right|
\rf190
\end{align}
where $g$ is an arbitrary analytic function on a complex plane $Z$.

In this way we get for the small field case
\bq
G\dr{eff}=G_N\biggl(\frac{1-|g(Z)|^2}2\biggr)^{(n+2)/n}
\biggl(\left|\frac{dg}{dZ}(Z)\right|\biggr)^{-(n+2)/(2n)} \eq191
\end{equation}
where
\bq
Z=\sqrt{\frac{\ov\la_{c0}}{2n}}\cdot(x+iy). \eq192
\end{equation}

In the large field case
\bq
G\dr{eff}=G_N\biggl(\frac{1-|g(Z)|^2}2\biggr)
\biggl(\left|\frac{dg}{dZ}(Z)\right|\biggr)^{-1/2} \eq193
\end{equation}
where
\bq
Z=\sqrt{\frac{\ov\la_{c0}}{2(n+2)}}\cdot(x+iy). \eq194
\end{equation}
Eqs \er{5.191} and \er{5.193} can have very interesting behaviour for
$\frac{dg}{dZ}$ could have some singularities. The physical interpretation of
these singularities can be very interesting.

All solutions for $G\dr{eff}$ given here, depending on space or time \cd s
can be a source of non-Newtonian \gr al physics, e.g.\ a theory with a \gr al \pt
$$
V=-G\dr{eff}\cdot \frac Mr\,.
$$
In this way we can fit anomalous acceleration of Pioneer 10/11 spacecrafts
(see Refs \cite{7,7a}). Thus in our approach there is something similar to
MOND (see Ref.~\cite{c} or \cite{xX}). MOND = MOdified Newtonian Dynamics
introduced by M.~Milgrom (see Ref.~\cite{xxq}) in order to explain flat curve of galactic rotation.
In Ref.~\cite{c} we have to do with
bimetric \gr al theory. The difference between two \LC \cn s interacts
quadratically with a strength $a_0$ (a~universal \ct). In Ref.~\cite{xX} we
have to do with a tensor-vector-scalar theory.

Let us consider a problem of a flat velocity curve with a cold Dark Matter.
In the \E\nos\ \KK (Jordan--Thiry) Theory we have a lot of such Dark Matter.
Let us consider Dark Matter and ordinary barionic matter as collisionless dust
\pc s. It means, pressureless dust. What is a density distribution of such a
matter? One considers (see Refs~\cite{zdel,ro}) the so-called NFW
(Navarro--Frank--White) profile
\beq5.137a
\rho(r) = \frac{\rho_0}{\frac r{R_s}(1+\frac r{R_s})^2}
\e
where $\rho_0$ and $R_s$ are parameters. One considers also a density
distribution
$$
\rho(r)=\rho_0\biggl(1+\biggl(\frac r{r_c}\biggr)^2\biggr)^{-1},
$$
$r_c$---a core radius, $r$ is a distance from galaxy centre. A mass inside a sphere of the radius~$r$ for \er{5.137a} is
\beq5.138a
M_{\rm DM} (r)= \int_0^r 4\pi (r')^2 \rho(r')\,dr' =
4\pi \rho_0 R_s^3\biggl[\ln\biggl(\frac{R_s+r}{R_s}\biggr) -
\frac r{R_s+r}\biggr].
\e
A \gr al \pt\ of such a density distribution is (from the Poisson \e)
\beq5.139a
\gathered
V_{\rm DM}(r) = -\frac{4\pi G_N\rho_0R_s^3}r \ln\biggl(1+\frac r{R_s}\biggr),\\
\lim_{r\to\infty}V_{\rm DM}=0, \qquad \lim_{r\to0}V_{\rm DM}(r)=-4\pi G_N\rho_0
R_s^3.
\endgathered
\e

The orbital velocity in $V_{\rm DM}(r)$ is
$$
v(r)=\biggl(r\,\frac{dV_{\rm DM}}{dr}\biggr)^{1/2}
$$
and is going to a flat velocity curve.

Let us suppose that ordinary (barionic) matter has the same distribution but
with different parameters.

One gets
\bg5.140a
\rho_b(r) = \frac{\rho_{0b}}{\frac r{R_{sb}}(1+\frac r{R_{sb}})^2} \\
V_b(r) = -\frac{4\pi G_N\rho_{0b}R_{s0}^3}r \ln\biggl(1+\frac r{R_{sb}}\biggr)
\label{5.141a}
\e
If we want to model an existence of a cold Dark Matter by an exotic physics,
i.e.\ in our case with fifth force ($G\dr{eff}$) we should have
\beq5.142a
\wt V_b(r) = -\frac{4\pi G\dr\eff \rho_{0b}R_{s0}^3}r \ln\biggl(1+\frac r
{R_{sb}}\biggr).
\e
Eventually one gets from
\beq5.143a
\wt V_b(r) = V\dr{DM}(r)
\e
the equality
\beq5.144a
\frac{G\dr\eff}{G_N} = \biggl(\frac{\rho_0}{\rho_{0b}}\biggr)\biggl(\frac{R_s}
{R_{sb}}\biggr)^3 \frac{\ln(1+\frac r{R_s})}{\ln(1+\frac r{R_{sb}})}.
\e
In terms of the field $\vf$ we have
\beq5.145a
\vf(r) = -\frac1{(n+2)}\biggl[\ln\biggl(\frac{\rho_0}{\rho_{0b}}\biggr)
+3\ln\biggl(\frac{R_s}{R_{sb}}\biggr) + \ln\biggl(\frac{\ln(1+\frac r{R_s})}
{\ln(1+\frac r{R_{sb}})}\biggr)\biggr].
\e

It means, a field $\vf(r)$ should be a solution of a field \e\ \er{5.68} in order
to mimic a Dark Matter existence. We do not expect an exact solution. It is
enough to get a numerical solution of Eqs~\er{5.63a}, \er{5.63c}, \er{5.68},
\ap ing the \f~\er{5.145a}. The NFW profile is consistent with Milky Way and
M31 galaxies data. Moreover, we can also use a different profile, the so-called
Einasto profile (see Refs \cite{zom,zmu}), where
\beq5.147a
\rho(r) \sim \exp(-A r^\a).
\e
$A$ and $\a$ are parameters. In our considerations we suppose that
\beq5.148a
\rho_b(r) = \rho\dr{luminous matter (barionic)}+\rho\dr{dust (barionic)}.
\e
We can also add a part of a Dark Matter to $\rho_b(r)$, i.e.
\beq5.149a
\rho_{b'}(r) = (1-\b)\rho_b(r) + \b \rho\dr{DM}(r),
\e
where $0<\b<1$, and treat the remaining part of DM as an exotic physics matter
(using both NFW and Einasto profiles). This will be done in future papers.

Let us mimic MOND by $G\dr{eff}$. One gets
\beq5.149b
\frac{MG_N}{r^2} = a\mu\biggl(\frac a{a_0}\biggr)
\e
where $a$ is an acceleration and $a_0$ is a fundamental \ct\ from MOND theory
$$
a_0\simeq 2\cdot 10^{-10}{\rm\frac m{s^2}}\,,
$$
$\mu$ is a \f\ \st
\beq5.150a
\aligned
\mu(x\gg1)&=1\\
\mu(x\ll 1)&=x.
\endaligned
\e
One can use the following \f:
\beq5.151a
\aligned
\mu(x)&=x(1+x^2)^{-1/2}\\
\mu^{-1}(y)&=y(1-y^2)^{-1/2}.
\endaligned
\e

Let
$$
\frac{MG\dr{eff}}{r^2} = a
$$
be equivalent to \er{5.149b}.

One easily gets
\beq5.153a
a=a_0\,\frac{e^{(n+2)\vf}}{(1-e^{2(n+2)\vf})^{1/2}}\,.
\e
In this way a scalar field $\vf=\vf(r)$ rescales an acceleration to a
Newtonian value from a \ct~$a_0$ and gives us all achievements of MOND theory
(nonrelativistic) connected to the galactic rotation velocity curve (galaxy
rotation curve) (see Refs \cite{xxq,pe}).

One gets from $a=\frac{v^2(r)}r$:
\beq5.154a
v(r) = e^{(n+2)\vf/2} \cdot \frac{\sqrt{a_0r}}{\root4\of{1-e^{2(n+2)\vf}}}\,.
\e
This is our galactic rotation velocity curve in terms of the field~$\vf$.

Let us do some manipulations using Eq.~\er{5.154a}. One gets
\beq5.154b
\vf(r) = \frac1{(n+2)}\ln\Biggl(\frac{v^2(r)/r}{\sqrt{a_0^2+(v^2(r)/r)^2}}
\Biggr)
\e
If a rotation velocity curve is flat, i.e.\ $v(r)={\rm const.}=v_0$ one gets
\beq5.155b
\vf(r) = \frac1{(n+2)}\ln\Biggl(\frac{v_0^2/r}{\sqrt{a_0^2+(v_0^2/r)^2}}
\Biggr).
\e
Moreover, we have the following relation from Newtonian dynamics
$$
v^2(r)=r\,\frac{dV}{dr}
$$
and we get
\beq5.156b
\vf(r) = \frac1{(n+2)}\ln\Biggl(\frac{dV/dr}{\sqrt{a_0^2+(dV/dr)^2}}\Biggr).
\e
Eq.~\er{5.156b} can be used to get $\frac{dV}{dr}$:
\beq5.157b
\frac{dV}{dr} = \frac{a_0^2 e^{(n+2)\vf}}{\sqrt{1-e^{2(n+2)\vf}}}\,.
\e
The above \e s give us a taste of various possibilities among mentioned
approaches. For the latest measurement of matter distribution and rotation
curve in the Milky Way see Refs~\cite{Nova,Novb}.

Thus we get a very interesting situation. In one theory it is possible to
explain the same problem with quite different means (a~Dark Matter or the fifth
force, or both). Simultaneously we have also different theories describing MOND
as non-Newtonian limits (see Refs \cite{c,xX}).

It is also worth to mention a work of J.~W.~Moffat and E.~Rahvar (see
Ref.~\cite{zkap} which avoids a Dark Matter problem introducing an alternative
theory of gravity MOG (MOdified Gravity).

From the point of view of Philosophy of Science (Philosophy of Physics) this
is very interesting.

Let us notice that theories from Refs \cite{c,xX,zkap} are relativistic.
However they are not geometrical. The \E\nos\ \KK (Jordan--Thiry) Theory is
both relativistic and geometrical. After Einstein's General Relativity which
is both relativistic and geometrical we should find its extension which is also
both relativistic and geometrical as my \E\nos\ \KK (Jordan-Thiry) Theory.
Geometrical is considered here in a sense given in \ti{Conclusions and
Prospects}.

If we consider skewon and \qe\ \pc s as a Dark Matter in the Solar System, they
are interacting only \gr ally and they can have an influence on planetary orbits
in the Solar System. According to Ref.~\cite{Pit} we get the following
estimations:
\begin{align*}
\rho\dr{DM} &< 1.1 \tm 10^{-20} {\rm\frac g{cm^3}}\hbox{ on the Saturn orbit},\\
\rho\dr{DM} &< 1.4 \tm 10^{-20} {\rm\frac g{cm^3}}\hbox{ on the Mars orbit},\\
\rho\dr{DM} &< 14 \tm 10^{-19} {\rm\frac g{cm^3}}\hbox{ on the Earth orbit}.
\end{align*}
An amount of a Dark Matter inside the Saturn orbit is
$$
M\dr{DM} < 1.7\cdot 10^{-10}\,M_\odot
$$
($M_\odot$ is the mass of the Sun). This is in agreement with Eq.~\er{4.159}.

\advance\abovedisplayskip by-1pt
\advance\belowdisplayskip by-1pt
\section{Geodetic equations. A test particle movement}

Let us consider a geodetic equation on a manifold~$P'$ \wrt a \LC part of a
\cn\ $\gd \o,\tA,\tB,$, i.e.\ $\gd\wt\o,\tA,\tB,$. One gets from
\beq6.1
u^\tA \wt\n_\tA u^\tB=0.
\e
(Such an \e\ has usual interpretation as a test \pc\ \e\ of motion. In our
theory we consider such an \e\ for a $\gd\wt\o,\tA,\tB,$ \cn. The \e\
\er{6.1} is defined for a curve $\G\subset P'$, $\wt\n_u u=0$, where $u$ is
tangent to~$\G$.)
\bg6.2
\frac{\tv D u^\a}{d\tau} + \biggl(\frac{q^c}{m_0}\biggr) h_{cd}
\wt g^\(\a\d)\gd H,d,\b\d, u^\b + \biggl(\frac{q^c}{m_0}\biggr)h_{cd}
g^\(\a\d)\brgn \d \gd\F,d,\tb, u^\tb - \frac{\|q\|^2}{8m_0^2}\,\wt g
^\(\a\d)\biggl(\frac1{\rho^2}\biggr)_{,\d}=0 \\
\frac{\wt Du^\ta}{d\tau} + \frac1{r^2}\biggl(\frac{q^c}{m_0}\biggr)
h_{cd}h^{0\ta\tilde d}\cdot \frac{\br{gauge}D \gd\F,d,\tilde d,}{d\tau}
+\frac1{r^2}\biggl(\frac{q^c}{m_0}\biggr)h_{cd} h^{0\ta\tilde d}\bigl(
\gd c,d,ab,\gd\F,a,\tilde d, \gd\F,b,\tb, -\gd\mu,d,\hat{\imath},\cdot
\gd f,
\hat{\imath},d\tb, - \gd\F,d,\tilde e, \gd f,\tilde e,\tilde d\tb,\bigr)u^\tb =0,
\label{6.3}\\
2\rho^2 u^a = \frac{q^a}{m_0}\,,\nonumber\\
\|q\|^2= -h_{ab}q^aq^b,\nonumber
\e
where
\bg6.4
\frac{\br{gauge}D}{d\tau} \gd\F,d,\tilde d, = \brgn\b \gd \F,d,\tilde d,u^\b \\
\frac d{d\tau}\biggl(\frac{q^b}{m_0}\biggr)=0 \label{6.5}
\e
($\frac{q^b}{m_0}$ is an integral of motion).

$\wt{\ov D}$ means a covariant derivative along a line \wrt the \cn\ $\gd\wt
{\ov \o},\a,\b,$ on~$E$, $\wh{\wt D}$~means a covariant derivative along
a line \wrt the \cn\ $\gd\wh{\wt\o},\ta,\tb,$ (a~\LC \cn\ on $M=G/G_0$ \wrt
a~metric $h^{0\ta\tb}$), $r=\rm const$,
\beq6.6
u^\tA = (u^\a,u^\ta,u^a) = (\hor(u),\ver(u)).
\e
$q^a$ is a gauge (colour) charge of a test \pc, $m_0$ its mass.

Equation \er{6.1} has the following first integral of motion,
\beq6.7
g_\(\a\b)u^\a u^\b +r^2\gd h,0,\ta\tb,u^\ta u^\tb + \rho^2 h_{ab}u^a u^b
=\rm const.
\e
It is easy to see that a scalar field $\rho$ or $\Ps$ has an influence on
a~test \pc\ motion. Some possible applications of those \e s in the case
$\rho=1$ can be found in Refs \cite{5,11a}. More details concerning geodetic
\e s on~$P$ and also geodetic \e\ deviations can be found in Ref.~\cite{131}.
$u^\tb$ is a charge coupled to Higgs' field. For an application in GSW model
see \cite{11a} and \cite{131}.

Let us project the \e s \er{6.2}--\er{6.5} on $V=E\tm G/G_0$, i.e.\ let us take
a section $e$, ${e:V \to P}$. One gets
\bg6.7a
\frac{\wt{\ov D}u^\a}{d\tau} + \biggl(\frac{Q^c}{m_0}\biggr) u^\b \wt g{}
^\(\a\d) \gd F,d,\b\d, h_{cd} +\biggl(\frac{Q^c}{m_0}\biggr) u^b
h_{cd}\wt g{}^\(\a\d) e' \bigl(\brgn\d \gd\F,d,\tb,\bigr)
-\frac{\|Q\|^2}{8m_0} \,\wt g{}^\(\a\d) \biggl(\frac1{\rho^2}\biggr)_{,\d}=0\\
\frac{\wh{\wt D}u^\ta}{d\tau} + \frac1{r^2}\biggl(\frac{Q^c}{m_0}\biggr)
u^\b h_{cd}h^{0\ta\tilde d}e'\bigl(\brgn\b \gd\F,d,\tilde d,\bigr)
+\frac1{r^2}\biggl(\frac{Q^c}{m_0}\biggr)u^\tb h_{cd}h^{0\ta\tilde d}
e' (\gd H,d,\tilde d\tb,)=0 \label{6.8a}\\
e^\ast\o =\gd A,a,\mu, \ov\t{}^\mu X_a +\gd\F,a,\tb,\t^\tb X_a \label{6.9}\\
\frac{dQ^a}{d\tau} - \gd C,a,cb, Q^c \gd A,b,N, u^N =0\label{6.10}
\e
or
\beq6.11
\frac{dQ^a}{d\tau} - \gd C,a,cb,Q^c \gd A,b,\nu, u^\nu - \gd C,a,cb,Q^c
\gd\F,b,\ta, u^\ta=0
\e
where
\beq6.12
e'(q^cX_c) = Q^cX_c.
\e
We have
\begin{gather*}
e' \biggl(\frac{\br{gauge}D\gd \F,d,\tilde d,}{d\tau}\biggr)
=e'\bigl(\brgn\b \gd\F,d,\tilde d,\bigr)u^\b\\
\gd H,d,\ta\tb, = \gd C,d,ab, \gd \F,a,\ta, \gd\F,b,\tb, -\mu^d_{\hat\imath}
\gd f,\hat\imath,\tilde a\tb, -\gd\F,d,\tilde e,\gd f,e,\ta\tb,\\
\gd F,d,\b\d, \ov\t{}^\mu \wedge\ov \t{}^\nu X_d = e^\ast(\gd H,d,\b\d,
\t^\b \wedge \t^\d X_a)\\
\|Q\|^2 = -h_{ab}Q^a Q^b.
\end{gather*}
Equations \er{6.2}--\er{6.5} or \er{6.7a}--\er{6.8a} and \er{6.10} or \er{6.11}
are \KWK \e s extended to Jordan--Thiry Theory (see Refs \cite{1,5,11a} for
details and further references), i.e.\ an existence of scalar forces. In our
applications
$$
\rho = e^{-\Ps} = e^{-(\Ps_0+\vf)}
$$
and an additional term in \er{6.3} or \er{6.7a} has the form
\beq6.13
-\frac{\|q\|}{4m_0^2}\, e^{2(\Ps_0+\vf)}\wt g{}^\(\a\d)\vf_{,\d}.
\e
Let us notice that
\beq6.13a
\|q\|=\|Q\|.
\e
Moreover, in the stationary case (for $\Ps$) $\vf=0$ and the term disappears.

Let us consider Eqs \er{6.2}--\er{6.3} and Eq.\ \er{6.5} or Eqs \er{6.7a}--\er{6.8a}
and Eq. \er{6.11}. Let us suppose that $\gd h,0,\ta\tb, u^\ta u^\tb=1$.
In this way we write
\beq6.14
u^\ta = \frac{dx^\ta}{d\tau}.
\e
Let us consider an integral of motion \er{6.7}, i.e.
\beq6.15
g_\(\a\b)u^\a u^\b + \frac{\|q\|^2}{4\rho^2 m_0^2}
={\rm const}- r^2={\rm const}'
\e
and let us rewrite it in the following form
\beq6.16
m_0g_\(\a\b)u^\a u^\b + \frac{\|q\|^2}{4\rho^2 m_0}={\rm const}''.
\e
Usually
$$
E_p=m_0g_\(\a\b)u^\a u^\b
$$
is considered as an energy of a test \pc. Thus
\beq6.17
E_p+\frac{\|q\|^2}{4\rho^2m_0} = {\rm const}''
\e
is a bilans energy of test \pc\ and an external field~$\rho$ (or~$\Ps$).

In a differential form we can write \er{6.17} in the following way:
\bea6.18
\frac{dE_p}{d\tau} &= -\frac{\|q\|^2}{4m_0} \,\frac d{d\tau}\biggl(\frac1
{\rho^2}\biggr)\\
\frac d{d\tau}\biggl(\frac1 {\rho^2}\biggr)&=\biggl(\frac1
{\rho^2}\biggr)_{,\b}u^\b. \label{6.19}
\e
Consequently we get after projecting on $E\tm G/G_0$:
\begin{gather}
E_p+\frac{\|Q\|^2}{4\rho^2m_0} = {\rm const}'' \tag{\ref{6.17}$'$}\\
\frac{dE_p}{d\tau} = -\frac{\|Q\|^2}{4m_0} \,\frac d{d\tau}\biggl(\frac1
{\rho^2}\biggr)\tag{\ref{6.18}$'$}
\end{gather}
In this way we can write
\bg6.20
\bal
\frac{d^2x^\a}{d\tau^2} &+ {\a\brace\b\ \g}\frac{dx^\b}{d\tau}\,
\frac{dx^\g}{d\tau}+\biggl(\frac{Q^c}{m_0}\biggr)h_{cd} \wt g^\(\a\d)
\gd F,d,\b\d, \frac{dx^\b}{d\tau}\\
&+\biggl(\frac{Q^c}{m_0}\biggr)
h_{cd}\wt g^\(\a\d) e' \bigl(\brgn\d \gd \F,d,\tb,\bigr)
\frac{dx^\b}{d\tau} - \frac{\|Q\|^2}{8m_0^2}\,\wt g^\(\a\d)
\biggl(\frac1{\rho^2}\biggr)_{,\d}=0
\eal\\
\bal
\frac{d^2x^\ta}{d\tau^2} &+ {\wt a\brace \wt b\ \wt c}\frac{dx^\tb}{d\tau}
\cdot\frac{dx^{\tilde c}}{d\tau}\\
&+\frac1{r^2}
\biggl(\frac{Q^c}{m_0}\biggr)h_{cd} h^{0\ta\tilde d}\cdot e'\bigl(
\brgn\b \gd\F,d,\tilde d,\bigr)\cdot \frac{dx^\b}{d\tau}+\frac1{r^2}
\biggl(\frac{Q^c}{m_0}\biggr)h_{cd}h^{0\ta\tilde d}e'(\gd H,d,\td d\tb,)=0
\eal \label{6.21}
\e
where ${\a\brace\b\ \g}$ are Christoffel symbols constructed from $g_\(\a\b)$
on~$E$ and ${\ta\brace\tb\ \td c}$ are Christoffel symbols constructed from
$\gd h,0,\ta\tb,$ on $M=G/G_0$.

In this way we have a curve $\G'\subset V=E\tm G/G_0$, $(x^\a(\tau),x^\ta(\tau))$.
Simultaneously we write Eq.~\er{6.11}
\beq6.22
\frac{dQ^a}{d\tau} - \gd C,a,cb, Q^c \gd A,b,\nu, \frac{dx^\nu}{d\tau}
-\gd C,a,cb, Q^c e^\ast (\gd\F,b,\ta,)\frac{dx^\ta}{d\tau}=0.
\e
This \e\ describes a precession of a charge $Q^a$. A~charge $q^b$ is a \ct\
during a motion $(\frac{q^b}{m_0},m_0={\rm const})$ and a norm of~$q$, $\|q\|$
is also a \ct\ of motion. $Q^a$ is not a \ct, moreover $\|Q\|$ is \ct.
After projecting a curve $\G'$ on~$E$ we get a trajectory of a test \pc\
motion $\G''\subset E$, $x^\a(\tau)$.

Let us come back to the \e\ (\ref{6.18}$'$). This \e\ describes changing of an
energy $E_p$ during a motion due to field~$\rho$. It is a friction or an
amplification. This is an influence of the fifth force on a test \pc\
movement. Moreover
\beq6.23
G\dr{eff} = G_N\rho^{(n+2)} = G_Ne^{-(n+2)\Ps}.
\e
In this way we can connect this with a measured change of a \gr al \ct.
According to recent measurement we have (see Ref.~\cite{e}):
\beq6.23a
-4.2\cdot 10^{-14} < \frac{\dot G}G < 7.5\cdot 10^{-14} \,\frac1{y_r}
\e
(statistically non-zero). $G=G_{\rm astr}$ (see Appendix C).

Let us consider Eqs \eqref{6.1}--\eqref{6.5} in our partial \un\ in Appendix~C.
In this way we take under consideration the existence of a scalar field~$\rho$
(equivalently $\Psi$ or~$\vf$). Compare those \e s with Eqs (3.176)--(3.178),
(3.180) from Ref.~\cite{11a}. We have here an additional term
$$
-\frac{\|q\|^2}{8m_0^2}\,{\wt g}^{(\a\d)}\biggl(\frac1{\rho^2}\biggr)_{,\d}
$$
in Eq.~\eqref{6.2} and a different definition of~$q^a$ (including the field~$\rho$).

Let us apply those results to \KWK \e\ found in Ref.~\cite{11a} for GSW model
unified with gravity. Eq.~(C.49) will be changed by adding the term
$$
-\frac{\|Q\|^2}{8m_0^2}\,{\wt g}^{(\a\d)}\biggl(\frac1{\rho^2}\biggr)_{,\d},
\q \|Q\|={\rm const}
$$
on the left-hand side of the \e. Simultaneously let us notice the \fw\ fact. In
Appendix~C of this paper (not of Ref.~\cite{11a}) we consider a partial \un\
of \gr al, electro-weak and strong \ia s. We have a \cfn\ condition for strong
\ia s described by $\SU(3)_c$ gauge field.  In this way all colour charges of
a test \pc\ are zero and they are absent in the definitions of~$\|q\|$
and~$\|Q\|$. We have here only charges from GSW model. Thus the remaining \e s
of \KWK \e s, i.e. (C.50)--(C.51) and (C.53)--(C.58) from Ref.~\cite{11a} are
unchanged. For a full treatment see Appendix~C of this paper.

We called the above \e s \KWK \e s. Why? R.~Kerner found in Ref.~\cite{alf} an
\e\ of motion for a test \pc\ in coupled \gr al and nonabelian \YM' fields
starting from nonabelian \KK theory. This generalizes an \e\ with a Lorentz
force term from \elm sm. S.~K.~Wong (see Ref.~\cite{m6b}) found such an \e\
for special case of $G=\SU(2)$ in Minkowski space. After that this \e\ has been
called Wong \e. The Kerner result is more general for a gauge group in his case
is an arbitrary semi-simple Lie group. W.~Kopczy\'nski considers fibre bundle
approach for coupled \gr al and gauge fields in Ref.~\cite{m6a}. He derived
the \e\ on a principal fibre bundle and afterwards projected it on a base~$E$
(a~\spt). Our
generalization consists in extending a test \pc\ motion also in Higgs' field
and in a scalar field from Jordan--Thiry Theory.

\section{Conclusions and prospects for further research}

The \nos\ \KK\ (Jordan--Thiry) Theory has been developed within GUT (Grand Unified Theories).
It unifies the gauge invariance
principle with the coordinate invariance principle but in more than
four-dimensional space-time. In the paper we consider \nos\ Jordan--Thiry
Theory, i.e.\ $\rho\ne1$ ($\Psi\ne0$) in full details. The case with $\rho=1$
($\Psi=0$) is partially repeated from Ref.~\cite{11a} in order to make the
paper more readable and friendly for readers.

A general nonabelian Yang--Mills fields have been unified with gravity in
$(n+4)$-\di al space-time ($n$---a \di\ of gauge group).
The theory uses a \nos\ metric defined on a metrized (in a \nos\
way) principal fibre bundle over a \spt\ with a structural group $\U(1)$ in an
\elm c case and in general case nonabelian semi-simple compact group~$G$. The
\cn\ on \spt\ and on a metrized principal fibre bundle is compatible with
this metric. This \cn\ is similar to a \cn\ from Einstein's Unified Field
Theory, however we use its higher \di al analogue (see also \cite{d}, \cite{f})---the
so-called Einstein--Kaufman Theory. This \cn\ is
right-\iv t \wrt an action of the group~$G$ (a~gauge group).

The theory has been developed to
include a scalar field leading to an effective \gr al constant and \spt\
dependent cosmological terms. It is possible to extend the theory to include
Higgs' fields and spontaneous symmetry breaking of a gauge group to get
massive vector boson fields.

In Ref.~\cite5 we consider more possibilities to use scalar fields in NKK(JT)T.
We have decided to use the one described here as more profound. Simultaneously
we make the presentation friendly for readers as much as possible. To continue
an analogy to medieval architecture we are removing some assemblies (assembly
devices) from Ref.~\cite5 in order to get valiable \un\ theory. It is better
to say removing falseworks and scaffoldings.

The theory is fully relativistic and unifies \elm c field, gauge fields,
Higgs' field and scalar forces with NGT (\E\nos\ Gravitation Theory)
in a nontrivial way.
By `in a nontrivial way' we mean that we get from the theory
something more than NGT, ordinary Kaluza--Klein (Jordan--Thiry) Theory, classical
electrodynamics, Yang--Mills' field theory with Higgs' field and spontaneous
symmetry breaking. These new features are some kind of ``interference
effects'' between all of them.
This theory unifies two important approaches in higher-dimensional
philosophy: Kaluza--Klein principle and a \di al reduction principle.

Due to a concept of a hierarchy of a \s y breaking we can include GUT models
in the \E\nos\ \KK (Jordan--Thiry) Theory. We consider several possibilities.

The beautiful theories such as Kaluza--Klein theory (a~Kaluza miracle) and
its descendents should pass the following test if they are treated as real
unified theories. They should incorporate chiral fermions. Since the
fundamental scale in the theory is a Planck's mass, fermions should be
massless up to the moment of spontaneous symmetry breaking. Thus they should
be zero modes. In our approach they can obtain masses on a \di al reduction
scale. Thus they are zero modes in $(4+n_1)$-\di al case. In this way
$(n_1+4)$-\di al fermions are not chiral (according to very well known
Witten's argument on an index of a Dirac operator). Moreover, they are
not zero modes after a \di al reduction, i.e., in 4-\di al case. It means we
can get chiral fermions under some assumptions.

This has been obtained in Ref.~\cite{x}. We obtain Yukawa terms in the \lg\
for $\frac12$-spin fermion fields using higher dimensional spinor fields.
Our candidates for Dark Matter \pc s are a quintessence \pc\ and a skewon.
A~skewon is a pseudovector massive \pc\ (spin~1), a \qe\ \pc\ is a massive
scalaron (spin~0).

We expect some nonrelativistic effects leading to non-Newtonian gravity.

We consider the \E\nos\ \KK\ (Jordan--Thiry) Theory with spontaneous \sy
y breaking and Higgs' mechanism with a scalar field~$\varPsi$. We get inflation from this theory with a quintessence and several
testable \co ical predictions (see Refs \cite{xx},~\cite5).
We find a dynamical model for a \co ical \ct.
In the models of the Universe we get several phase
transitions of the second and of the first order. The real source of a quintessence is
a scalar field~$\varPsi$. Due to its very unusual selfinteraction potential
(coming from higher dimensions) it can proceed to the very unexpectable
features. Of course this is not the end of the story. This is really a
beginning.

In future papers we develop an interaction of fermion fields within our
hierarchy of \s y breaking. It means a fermion part of GUT's in the \E\nos\
\KK (Jordan--Thiry) Theory. We will also develop an idea of Dark Matter in our theory.

The future development should contain several examples of GUT's. It means, we
should find exact forms of a diffeomorphism~$g$
$$
g:M \to M_0\times M_1\tm \ldots \tm M_{k-1}, \q k=2,3,
$$
(see Section 3). The quantization procedure (our theory up
to now is a classical field theory) will be developed by using nonlocal
quantization according some ideas from Ref.~\cite{11a}.

Let us give the following comment. In the paper \cite{S3} one considers \s ic
\KK Theory (also non-Abelian) (see also Ref.~\cite{G} for some explanations
and Translation of Terminology in Table~1). Our construction (\nos) is a generalization
of this approach. In Ref.~\cite{y} one considers a possibility to travel in
higher dimensions a little different in spirit than in Ref.~\cite{S2}.
One can find the full treatment of hierarchy of the \s y breaking
in Refs~\cite{xy} and~\cite5.

For a modern treatment of geometry see Appendix A of Ref.~\cite8 and Section~2 {\it Elements
of geometry} of Ref.~\cite{11a}.

Let us give the following historical remark. In 1915 in G\"ottingen (Germany)
three people: A.~Einstein, D.~Hilbert and O.~Klein were discussing the
following problem: {\it What is a lagrangian for a \gr al field?} Eventually
they decided: {\it it is a scalar curvature $R_4$ for a \LC \cn\ on a
4-\di al manifold~$E$---induced by a metric tensor $g_{\mu\nu}=
g_{\nu\mu}$ on~$E$}. In this way an action for a \gr al field is $\int R_4\sqrt{-g}
\,d^4x$, known now as a Hilbert action. In~1921 (see Ref.~\cite{aa}) T.~Kaluza
obtained classical electrodynamics in a vacuum coupled to General Relativity
considered as a \lg\ of both fields $R_5$---a~scalar curvature on a 5-\di al
manifold equipped with a metric tensor and a 5-\di al \LC \cn. An action is of
course $\int R_5\sqrt{|\g|}\,d^5x$. The theory has been very much developed
to include \nA\ \YM' fields (for full bibliography see Ref.~\cite4).
Simultaneously General Relativity has been extended to \nos\ metric tensor
and \nos\ affine \cn\ in order to get a unified field theory (see Ref.~\cite1
for a bibliography). A~\lg\ of a unified field theory was $R_4$ but for
a generalized linear (affine) \cn\ and an action was as before $\int R_4
\sqrt{-g}\,d^4x$.

In the case of \nA\ gauge fields a \lg\ was $R_{n+4}$ (a~scalar curvature on\break
$(n+4)$-\di al manifold), an action being $R_{n+4}\sqrt{|\g|}\,d^{n+4}x$. The
so-called \di al reduction procedure has been added to the Kaluza--Klein
Theory (see Ref.~\cite{bb}), resulting in appearing of Higgs' fields, \ssb\ and \Hm,
with a~\lg\ $\frac1{V_2}\int_M R_{n+n_1+4}\sqrt{|\wt g|}\,d^nx$ (an average of
$(n+n_1+4)$-\di al curvature scalar on ${(n+n_1+4)}$-\di al manifold) and with
an action $\int R_{n+n_1+4} \sqrt{|\ov g|}\,d^{n+n_1+4}x$.

On every stage of the Kaluza--Klein theory it is possible to add a scalar
field~$\Ps$ (in a Jordan--Thiry manner). Our preliminary version of a unified
field theory uses these ideas, using
$$
\frac1{V_2} \int_M R_{n+n_1+4}\sqrt{|\wt g|}\,d^{n_1}x
$$
(an average of $(n+n_1+4)$-\di al curvature scalar of an
affine \cn\ on ${(n+n_1+4)}$-\di al manifold equipped with a \nos\ metric tensor)
as a~\lg, and\break $\int R_{n+n_1+4}\sqrt{|\ov \g|}\,d^{n+n_1+4}x$ as an action.
In all mentioned theories geodetic \e s \wrt \LC part of a \cn\ considered
is supposed to be a test \pc\ \e\ of motion.

Let us give the following remark. A unified field theory describes a local
physics (e.g.\ in the Solar System). Field \e s of this theory are proposed in
this paper (except fermions which are described in Ref.~\cite{x}). Moreover,
there is a problem of mass generating for some fields. In the paper
this is a \Hm. Moreover, due to the need for the existence of a Dark Matter
and Dark Energy some nonlocal physics should be engaged. It means, we consider
also \co y. We need massive \pc s as Dark Matter \pc s (massless are useless,
they are radiation). Skewon obtained a mass due to a \co ical \ct. \E\co ical
\ct\ is not zero due to \co ical evolution of a field~$\vf$ (an inflaton).
Thus a ToE (Theory of Everything) should contain also a Modern \E\co y.

Let us give the \fw\ remark. In our theory of gravity we have three \gr al
\pc s: graviton (massless) and two massive---scalaron and skewon. Gravitons
cannot form Dark Matter (they are massless). Scalarons and skewons are Dark
Matter. In this way Dark Matter has \gr al origin.

Moreover a full \un\ theory contains an alternative (\nos) theory of gravity
with a scalar field. Thus we have an interesting duality: effective alternative
theory of gravity without Dark Matter or with Dark Matter. Some intermediate
stages are possible.

Let us consider a problem of \gr al radiation. In the \E\nos\ \KK (Jordan--Thiry)
Theory we have to do with three \gr al \pc s: graviton, skewon and scalaron
(\qe\ \pc). The first one is massless, the remaining are massive. Moreover,
if the \co ical \ct\ is zero, they are massless. According to the analysis of
a secular motion of the binary system BPSR 1913+16 we have the \fw\ conclusion.
The energy loss of a binary system is consistent with a quadruple radiation
formula of GR modulo Kepler law and general relativistic effects: periastron
movement, Doppler effect etc.\ (see Refs \cite{xxa,xxb}).

It means, we have to do with only one kind of radiation-\gr al waves (gravitons).
There is not any trace of scalar (isotropic) radiation. Such radiation would
be possible if a scalaron and a skewon were massless. Thus there are not any
long range fields except ordinary gravity (symmetric part of a metric tensor)
(see Refs \cite{xxa,xxb}). The analysis of the BPSR 1913+16 has been done by
C.~Will in Ref.~\cite{xxg} by PPN (Parametrized PostNewtonian) formalism.
After the discovery of BPSR 1913+16 we found several binary systems with
relativistic effect which support the above claim. Recent observation with
discovery of \gr al waves supports existence of two polarizations of the waves
($\pm2$). There is no scalar (isotropic) polarization (see Refs \cite{xxc,
xxd,xxe,xxf}).

Let us notice the following fact. NGT (\E\nos\ \E\gr\ Theory) does not
constrain BPSR 1913+16 data. There is here only quadrupole radiation as in~GR.
This is important because the \E\nos\ \KK (Jordan--Thiry) Theory has NGT as
a limit. J.~W.~Moffat and his coworkers have developed PPN formalism in~NGT
(see Refs \cite{zk, zp, zalf, zbet, zgam}). They have applied the formalism
successfully for BPSR 1913+16 (see Refs \cite{zl, zm}). We do not have any
scalar radiation.

This strongly supports that a scalaron and a skewon are massive with
Yukawa-type long range behaviour. They are not of an infinite range type
fields. There is a research devoted to finding a scalar radiation.

Let us give some more general and philosophical considerations on \un\ and geometrization of
physical \ia s.

We describe here a scheme of \un\ of \fn\ physical \ia s. Simultaneously this
\un\ is geometrical. Thus this is in some sense a Unified Field Theory (see
Introduction). Moreover, contemporary ToE (Theory of Everything) should also
answer some additional astrophysical and \co ical problems (see Refs
\cite{k,l1,alfa}). In particular we face a problem of a Dark Matter and a Dark
Energy (\co ical \ct, \qe, vacuum energy). Our approach answer those questions
giving as Dark Matter \pc s: pseudovector massive bosons (skewons) and massive
scalarons. Simultaneously it is possible to extend the theory (still
geometrical) to include a tower of massive scalar fields (masses have a scale equal
to~$\frac 1r$, $r$~is a radius of a vacuum states manifold $M=G/G_0$). The
tower can be also used as a regulator field in quantized theory. Let us remind
to the reader that a quantization procedure for our theory will be a nonlocal
quantization (Yukawa, Efimov, Moffat, see Refs \cite{na,nb,ng,nd,no}).

In our theory we have the fifth force which manifests as a $G\dr{eff}$ (or $\wt G\dr{eff}$)
(effective \gr al ``\ct''), depending on a scalar field~$\vf$ (or~$\Ps$), the
same field which is a source of scalarons, our Dark Matter. The ``fifth force''
$\wt G\dr{eff}$ depends also on skewons (see Appendix~C). The field~$\vf$
(or~$\Ps$) is coming from higher dimensions. This field plays the role of an
inflaton field. In this way we have geometrized inflationary geometry giving
the \ti{reason d'\^etre} for an inflaton. Our Dark Matter and Dark Energy have
been geometrized. Dark Matter \pc s interact very weakly with ordinary
(barionic) matter. Really they are interacting almost only \gr ally, being
a part of a gravity. A~Dark Matter and the fifth force can explain a flat
rotation velocity curve for galaxies and a necessary non-barionic part of a
matter in our \Un. A~Dark Energy in our approach gives us a \co ical \ct\
important for an observed accelerated expansion of the \Un.

In the five-\di al theory we get nonsingular (finite energy) \so\ of field \e s
in the case of spherically \s ic and stationary case without and with a
\co ical \ct\ (see Refs \cite{4, 8}. These \so s have nonsingular electric and
Newtonian-like \gr al fields. We get also gravito-\elm c waves (generalized
plane-waves) \so s (see~Ref.~\cite8).

In this geometrical \un\ of GSW model we get correct masses for $W^\pm$, $Z^0$
and Higgs' bosons together with a value of Weinberg angle $\theta_W$ in
comparison with an experiment. We give a theory of a dielectric model of
confinement of a charge and a colour. Moreover, a problem of a flat rotation velocity
curves for galaxies can be explained using our Dark Matter, the fifth force or
both. In the approach it is possible to make a \co ical \ct\ equal to zero,
making place for inhomogeneous models of the \Un\ (Lema\^\i tre--Tolman model,
or even Szekers). In this way scalarons and skewons cannot be used as a Dark
Matter. They are a part of a radiation. Only an additional Dark Matter
(mentioned above a tower of scalar fields) can explain a missing mass etc. Some
recent constrains from observation do not settle a controversy: a Dark Matter
or modification of gravity (e.g.\ a~fifth force) or both. In our approach
(\E\nos\ Kaluza--Klein (Jordan--Thiry) Theory) we have Dark Matter, Dark Energy
strongly combined from deeper level of the theory and also the fifth force.

According to A.~Mukhanov (a private communication) \co ical and astrophysical
data cannot be the only data to choose an appropriate \un\ scheme of \fn\
\ia s. Our approach can explain and repeat all successes of Milgrom's MOND
(MOdified Newtonian Dynamics). Simultaneously two relativistic approaches,
different in their nature, have excerpted Newtonian, non-Newtonian (MOND)
nonrelativistic limits. A~different approach to gravity, the so-called MOG
(MOdified Gravity) can explain the mentioned data without a Dark Matter too.

In the end of Section 5 we consider several theories and also several
approaches coming to Dark Matter theories. Those theories are able to explain
an existence of a Dark Matter or to explain a problem of missing mass in galaxies
without a Dark Matter. Thus it will be interesting to consider a problem from
philosophical point of view in a wider sense.

What does it mean from the philosophical point of view? It means that there
are several theories different in nature (giving completely different picture of the
world) that can give an explanation of mentioned data. This is very interesting
because it touches our definition of truth in philosophy of science (in
scientific, physical-like theories).

Let us remind to the reader that we have several theories of the truth:
classical, pragmatical, conventional, coherent etc. All of these theories have
been applied in the philosophy of science. The classical truth is an
Aristotelian theory. It says: \ti{Veritas est adequatio rei et intelecto}
(the truth is equivalence between a think and a reality, someone tells a truth
if she (or~he) is saying as it is). Moreover, in the case of several physical
theories the situation is more complex. Every of the mentioned theories tells
something completely different (different worlds). Moreover, reality is the same. This
concerns also our approach. What kind of the theory should we use? Classical
theory is not enough. Let us mention that the classical theory of truth has
been formalized by A.~Tarski (see Ref.~\cite{nq}) by a relation of
satisfiability in a model of the theory (i.e.\ in model theory in foundation of
mathematics). The sentence is true if it is satisfiable in all models.
In some sense this is in the spirit of W.~Leibniz monadology. In his approach
a sentence is true if it is true in all possible worlds (monads).

This theory of truth is going to the G\"odel theorem and L\"owenheim--Skolem
theorem. Thus this is very important for deductive sciences (mathematics and
logic). If we formalize some physical theories, this will also be important
for empirical sciences (e.g.\ Hilbert axiomatization of elastic mechanics, general
relativity etc.). In some approaches to a true theory there is a graduation
of worlds in such a way that some sentences do not exist in worlds of lower order.
In this way we can consider graduate theory of true. We do not consider such
approaches in empirical sciences.

We should mention an incompleteness theorem by K.~G\"odel (see Ref.~\cite{nnr})
based really on the theory of truth by A.~Tarski. Moreover, we will not
consider the theorem for physical theories because this theorem is applicable
first of all for formal sciences such as logic and mathematics (see Refs
\cite{nnz,nnw,Novx}). It is worth to mention that recently someone applied some
G\"odel's ideas in quantum mechanics (see Ref.~\cite{nny}). However this
approach has nothing to do in our coherent theory of the truth, which we
advocate and explain below. Some
methodological problems in empirical sciences can be found in Refs \cite
{nnt,nnv} which are also beyond a coherent theory of the truth.

However in the case of empirical sciences the connection
with the reality is more important. The question is as follows. What kind of
the theory of truth should we use? The answer is not so easy. A~classical
theory of truth is not enough. Pragmatical theory of truth is obsolete for natural
(physical) sciences. Conventional theory of truth (a~truth is a convention) is
not enough, e.g.\ it is useful if we accept a convention that all real numbers
are accessible in an experiment. However not more. Thus we have only coherent
theory of truth. The theory is true if it ``works'' on more experimental facts
(wider reality). It is easy to see that the \E\nos\ \KK (Jordan--Thiry) Theory
satisfies this condition. The reality covered by the theory is wider than
MOND, bimetric MOND, scalar, tensor, vector MOND theory, MOG.

In order to explain the problem to the reader we give some examples from
literature and movies. In the book \ti{The Futurological Congress} (see
Ref.~\cite{np}) the author S.~Lem (a science-fiction writer) gives us
a world governed by hallucinogenic
manipulations on several levels. The world in one level is a hallucination on
the higher level. The reality (the truth) is on the highest level (if we know
this level). The second example is in the movie \ti{Matrix}, where we have to
do with a ``reality'' which is a simulation in a different reality. In this
movie and its subsequent parts we have to do with such simulations. All of
these examples can be summarized as a~dream inside a~dream. This dream is real
(a~reality) which is coherent (consistent) on all facts. This is ToE (Theory
of Everything) in the meaning of physical theories.

Roughly speaking, the book describes adventures of Ijon Tichy, a~hero of Lem's
books (see lemology---a~science devoted to S.~Lem's books), during the Eight
Futurological Congress in Costaricana (sent by Professor Tarantoga to the
Congress). Do not mix it with
real Costa Rica. He loses a consciousness during a revolution in Costaricana
and destruction of the Hotel Hilton where the congress took place. During that
time he has several visions concerning the world where chemical hallucinogens
influence human brain causing effects expected by manipulators.
There are some cures for an influence of hallucinogens (also some
chemical substances) which act on several levels. In some sense the levels are
``local truths''. In the end there is a ``real truth''. The world is going to
the climate catastrophe (a new Pleistocen).

The truth on several levels of manipulations gives us a picture of a reality
needed by manipulators (by chemical substances) to be considered as a real
truth.

In \ti{Matrix} we have similar ideas of different realities caused by computer
simulations. All the simulations are delivered directly to human brains
connected among themselves and to central computer Matrix. It is possible to
interrupt a simulation (a~program) getting a different reality. The analogy to
Lem's book is evident (even by different ``physical means''). In the case of
dreams (the so-called REM-dreams) this is also evident. During a dream we
consider it as a reality. After a walkup we know it was a dream. A~virtual
reality which is now better and better can realize such ideas giving us a
taste of a notion of coherent theory of the truth.

In scientific theories partial truth (partial realities) are inconsistent
among themselves as dreams, levels of hallucinogens, manipulations, computer
programs (in \ti{Matrix}). There is a coherent reality, e.g.\ Dark Matter with
the fifth force gravity in the \E\nos\ \KK (Jordan--Thiry) Theory.
Explanations of the flat rotation curve of a galaxy given by Refs \cite{c,xX}
are inconsistent. They are consistent only in a nonrelativistic (non-Newtonian)
limits (see Ref.~\cite{xxq}). They are inconsistent also in the case of MOG.
Coherent theory of the truth works here very well.

We have also some additional problems to be solved. It is a problem of a
chaotic motion of a test \pc\ in our partial \un\ (see Appendix~C) due to a
scalar field~$\rho$ ($\Psi$~or~$\vf$) and a possibility to travel in higher
\di s. In Appendix~C we consider a \so\ of a Hubble \ct\ crisis using $\wt G\dr{eff}$
and a classical (Newtonian-like) \co y. This approach will be developed after
obtaining new observation data.

The tower for scalar fields from Appendix B as a Dark Matter, i.e., $\rho=
\rho(x,y)$, $x\in E$, $y\in M=G/G_0$ (which is important in the case of zero
\co ical \ct) is applied if we consider \ssb\ and \Hm. The notion of
Self-Interacting Dark Matter is very fruitful in applications to dynamics of
galaxies with their satellites and cluster of galaxies. This is a scalar field
Dark Matter. In our approach a scalar field appears very naturally in \E\nos\
\KK (Jordan--Thiry) Theory coming from higher \di s. This supports our
approach very strongly (see Refs \cite{m7a, m6c, m6d, m6e, m6f} for example).
Let us notice the following unusual application of Dark Matter. Lisa Randall,
a theoretical physicist from Harvard University, claims that Dark Matter
caused death and extinction of dinosaurs about 66~Myrs
(see Ref.~\cite{m7k} with~\cite{m7l} and especially
the book \cite{m7aa}). Dark Matter can deform orbits of comets, meteors and
small planets resulting in their impacts on the Earth being a cause of the
death of dinosaurs. According to Ref.~\cite{m7aa} the Sun going through the
plane of the Milky Way (our Galaxy) intersects a cloud of Dark Matter. It
results in a deformation of comets orbits.

One can use some results from Ref.~\cite5 in developing \co ical models. In
particular a phase transition from a false vacuum to a true vacuum. In the
paper we abandon such possibilities concentrating on a slow roll inflationary
scenario in Appendix~A. The aim of our work is to construct some kind of
unified field theory, and those additional possibilities should be treated as
falseworks and scaffoldings in our construction. For this they are removed.

There is a very interesting connection between a Hubble \ct\ crisis and comet
motion in the Solar System via $\wt G\dr{eff}$. This can be connected to
extinction of dinosaurs (see Ref.~\cite{m7p}). In our approach such
possibility appears in Appendix~C, i.e.\ $\wt G\dr{eff}$ and a Hubble \ct\
crisis.

Let us comment Refs \cite{m7k, m7l, m7p}. Dark Matter influences the Oort cloud
where in the end of the Solar System comets are created. Initially circular
orbits were deformed. The periodic motion
of those comets is disturbed and such comets can fall on the Earth. In
particular cause the extinction of dinosaurs (see \cite{m7k, m7l, m7aa}). In the
case of a Hubble \ct\ crisis caused by changing of $G\dr{eff}$ (the increase of $G\dr{eff}$) we have also
an influence of this change on comets' orbits and afterwards an impact of
a comet on the Earth. This impact can be a cause of the extinction of dinosaurs.
Moreover, in our approach $G\dr{eff}$ (or~$\wt G\dr{eff}$) and Dark Matter
are strongly combined in the \E\nos\ \KK (Jordan--Thiry) Theory. Thus we can
have the same conclusion on the extinction of dinosaurs.

Present day the most stringent limits of $G\dr{eff}$ variation come from lunar
ranging data
$$
\frac{\dot G\dr{eff}}{G\dr{eff}} = (7.1 \pm 7.6)\cdot 10^{-14}\frac1{\rm yr}
$$
(see Ref.\ \cite{m8a}).

\looseness-1
In Appendix C where we develop a partial \un\ of NGT and bosonic part of a Standard
Model we derive an exact \so\ of field \e s, which describes ``charge without charge''
and ``mass without mass'' with a cosmological constant.
We prove that a cosmological constant is not zero and calculate it.
We get masses of $W^\pm$ bosons, $Z^0$~boson,
Higgs' boson and a value of Weinberg angle with good agreement with an experiment
as in Ref.~\cite{11a}. We discuss also a soliton-bag model of hadrons in a context
of our \un. We consider a nonlocal quantization of the theory.

\advance\abovedisplayskip by1pt
\advance\belowdisplayskip by1pt
\appendix
\section*{Appendix A}
\def\theequation{A.\arabic{equation}}

Let us consider a model of the \Un\ with Friedman--Robertson--Walker metric
(a~big-bang \co y)
\beq a1
ds^2 = dt^2 - a^2(t)\biggl(\frac{dr^2}{1-kr^2}+r^2(d\t^2+\sin^2\t\,d\vf^2)\biggr),\q
k=0,\pm1,
\e
where $a(t)$ is a scale factor. The \Un\ with scalar field~$\Ps$ only (see \eqref{4.74}
for a form of a \lg).

We get the following \e\ from Einstein's \e:
\beq a4
H^2 = \frac{8\pi}{3\mpl^2} - \frac k{a^2}\,.
\e
Moreover, in order to make additional considerations we consider also a
hydrodynamic energy-momentum tensor as a source of Einstein \e s.
\beq a2
T_{\mu\nu} = (p+\rho)u_\mu u_\mu - pg_{\mu\nu}.
\e
$p$ is the pressure and $\rho$ is the density of a matter, $u_\mu$~is the
four-velocity of a fluid.

From Einstein's \e
\bg a3
R_{\mu\nu} - \frac12 \,Rg_{\mu\nu} = 8\pi G_N T_{\mu\nu}\\
\dot \rho + 3H(p+\rho)=0, \label{a5}
\e
where $H=\frac{\dot a}a$ is a Hubble \ct\ (our evolution is de Sitter
exponential evolution, thus $H$ is really a \ct).

We get also
\beq a7
\frac{\ddot a}{a} = -\frac{4\pi}{3\mpl^2}(p+3\rho),
\e
$\mpl=(\frac{\hbar c^5}{G_N})^{1/2}$
is the Planck's mass. (We consider such a system of units that $c=1$,
$\hbar=1$.) Eq. \er{a4} can be rewritten as
\bg a8
\O-1 = \frac{k}{aH^2}\\
\O=\frac\rho{\rho_c}, \quad \rho_c=\frac{3H^2\mpl^2}{8\pi}\,,\label{a9}
\e
where $\rho_c$ is a critical density and $\O$ is the ratio of energy density
to the critical density.

Let us consider a dynamics of our field~$\Ps$ in the following way:
\beq a10
\aligned
\rho &= \frac{\ov M}2 \dot\Ps{}^2 +U(\Ps)\\
p &= \frac{\ov M}2 \dot\Ps{}^2 -U(\Ps).
\endaligned
\e
One gets
\bg a11
H^2 = \frac{8\pi}{3\mpl^2} \biggl(\frac{\ov M}2\,\dot \Ps{}^2+U(\Ps)\biggr)\\
{\ov M}\ddot\Ps +3{\ov M}H\dot\Ps +\frac{dU}{d\Ps}=0 \label{a12}\\
k=0.\nonumber
\e

Our field $\Ps$ is an inflation field (it plays several r\^oles). During
inflation, i.e.\ for $\ddot a>0$, we have
\beq a13
{\ov M}\dot\Ps{}^2 < U(\Ps).
\e
Usually one supposes also a slow-roll relations (see Ref.~\cite{alfa}) $\frac12\dot\Ps{}^2\ov M\ll
U(\Ps)$ and $\ddot\Ps \ov M\ll H\dot \Ps$. In this way one gets
\bg a14
H^2 = \frac{8\pi}{3\mpl^2}\,U(\Ps)\\
3H\ov M\dot \Ps \cong -U'(\Ps) \label{a15}
\e
According to Ref.~\cite{alfa} we introduce slow-roll parameters
\beq a16
\bga
\ve = \frac{\mpl^2}{16\pi}\biggl(\frac{U'}U\biggr)^2\\
\eta = \frac{\mpl^2}{8\pi}\biggl(\frac{U''}U\biggr).
\ega
\e
Our \ap ion is valid only if
\beq a17
\ve\ll 1, \q |\eta|\ll 1.
\e
An amount of inflation is supposed to be
\beq a18
N=\ln \frac{a_f}{a_i} = \int_{t_i}^{t_f} H\,dt,
\e
where $a_f$ is the final scale radius of the \Un, and $a_i$ is the initial
radius of the \Un. $H$~is nearly \ct\ during the inflation.
If the number $N\simeq 70$ the problem of flatness is solved. Moreover, in our
model we use $\Ps$ for several purposes.

We have $\Psi_i\cong0$, $\Ps_f=\Ps_0$ or in terms of the field~$\vf$ ($\Ps=
\Ps_0+\vf$)
\beq a19
\bga
\vf_i=-\Ps_0\\
\vf_f=0.
\ega
\e
In this moment our field $\vf$ is the so-called quintessential inflaton (it is
a source of a \co ical \ct---vacuum energy---a Dark Energy, a \qe) and we get
\beq a20
\la_{c0}(\Ps_0)=-\frac12\,U(\Ps_0)
\e
(see Section 4 for details). This is really a dynamical origin of a \co ical
\ct\ in slow-roll \ap ion.

Let us calculate $\ve$ and $\eta$ parameters in our theory (\E\nos\ \KK
(Jordan--Thiry) Theory). One gets
$$
\bga
\ve= \frac{\mpl^2}{16\pi}\biggl(\frac{U'}U\biggr)^2 = \frac{\mpl^2}{16\pi}
\frac{\Bigl((n+2)e^{2\Ps}\a_s^2 \frac{\wt R(\wt\G)}{\ell\pl^2}
+n\frac {\wt{\ul P}}{r^2}\Bigr)^2}{\Bigl(e^{2\Ps}\a_s^2 \frac{\wt R(\wt\G)}{\ell\pl^2}
+\frac {\wt{\ul P}}{r^2}\Bigr)^2}\\
\eta= \frac{\mpl^2}{8\pi}\biggl(\frac{U''}U\biggr) = \frac{\mpl^2}{8\pi}
\frac{\Bigl((n+2)^2 e^{2\Ps}\a_s^2 \frac{\wt R(\wt\G)}{\ell\pl^2}
+n^2\frac {\wt{\ul P}}{r^2}\Bigr)}{\Bigl(e^{2\Ps}\a_s^2 \frac{\wt R(\wt\G)}{\ell\pl^2}
+\frac {\wt{\ul P}}{r^2}\Bigr)^2}
\ega
$$
One gets $\ve(\Ps_0)=0$. Moreover, we can substitute $\Ps=\Ps_0+\vf$ and \ap e
a \pt\ $U(\Ps_0+\vf)$ as $U(\Ps_0)+\frac12\ov M m_0^2\vf^2$ getting typical slow-roll inflation.

In this Appendix we consider an application of a scalar field~$\vf$ (or~$\Psi$).
Moreover, in Ref.~\cite5 we give some general formalism of scalar fields in
NKK(JT)T. Some results from Ref.~\cite5 are not applicable here. Moreover, we
consider the formalism presented here as really legal. In Ref.~\cite5 we
consider several possibilities of scalar fields in NKK(JT)T which are not
developed longer.

\def\pd#1#2{\frac{d#1}{d#2}}
Let us consider an evolution of a field $\Ps$ (or $\vf$ or~$q_0$) in a \co ical
background in a more general case (without a slow-roll \ap ion). One gets
\beq a21
\pd{^2\Ps}{t^2}+3H\,\pd\Ps t+ \frac1{\ov M}\,U'(\Ps)=0,
\e
i.e.
$$
\pd{^2\Ps}{t^2} + 3H\,\pd\Ps t+\frac1{\ov M}\bigl(\g ne^{n\Ps}+\b(n+2)
e^{(n+2)\Ps}\bigr)=0
$$
or
\beq a22
\pd{^2\vf}{t^2} + 3H\,\pd\vf t+\frac1{\ov M}\biggl(\biggl(\g n
\sqrt{\frac{|\g|n}{\b(n+2)}}\biggr)^n \cdot e^{n\vf}(1-e^{2\vf})\biggr)=0
\e
under the following conditions
\beq a23
\bga
\vf_i=\vf(0)=-\Ps_0\\
\vf_f(t_f)=0
\ega
\e
or
\beq a24
\bga
\Ps_i(0)=0\\
\Ps_f(t_f)=\Ps_0={\rm arg\,min}\,U(\Ps).
\ega
\e
The evolution described here might not give enough inflation. Thus we need
additional inflationary schemes, i.e.\ a~tunnel effect from a false vacuum
state to a true vacuum state. We get
\beq a25
\la_{c1}(0)=\frac{\a_s^2}{\ell\pl^2}\,\wt R(\wt\G) +\frac {\wt{\ul P}}{r^2}+
\frac4{r^2}\biggl(\frac{\ell\pl^2}{r^2}\biggr) V(\Phi^1\dr{crt})
\e
(false vacuum, see Eq.\ \er{4.92}) and
\beq a26
\bga
\la_{c0}(0)= \frac{\a_s^2}{\ell\pl^2}\,\wt R(\wt \G) + \frac {\wt{\ul P}}{r^2}
\quad \hbox{(true vacuum)}\\
\la_{c1} > \la_{c0},
\ega
\e
after that we have an evolution of the field $\vf$ according to Eq.~\er{a22}.
A~time of a tunnel effect to proceed we suppose to be zero. The evolution of
the \Un\ is still exponential-de Sitter space-time.

In the end of the inflation and a further evolution of the field~$\Ps$
we are getting a \co ical \ct\
\beq a20a
\la_{c0}(\Ps_0) = -\frac12\,U(\Ps_0)
\e
which we can try to tune to the measured value of the \co ical \ct. From that moment we
can put in the full field \e\ $U(\Ps)=U(\Ps_0)=-2\la_{c0}(\Ps_0)$ and
proceed some calculations known in Section~4.

In Refs \cite5, \cite{xx} we consider several models of inflation. In all
approaches the field $\Ps$ or~$\vf$ (or~$q_i$) plays the r\^ole of an
inflaton field with an evolution to a value~$\Ps_0$ (a~minimum of the \pt).
Around $\Ps_0$ (or $0$ for~$\vf$) we can apply quadratic \ap ion for a \pt\
$U(\Ps_0+\vf)\cong -\frac{\ov M}2m_0^2\vf^2 + U(\Ps_0)$.

We get
\begin{gather}
U(\Ps_0)=\frac{2\g}{n+2}\biggl(\sqrt{\frac{n(-\g)}{(n+2)\b}}\biggr)^n \nonumber \\
\pd U\Ps (\Ps_0+\vf)= \g n\biggl(\sqrt{\frac{(-\g)n}{\b(n+2)}}\biggr)^n
\cdot e^{n\vf}(1-e^{2\vf}) \label{a28} \\
\pd U\Ps (\Ps_0)=0 \label{a29} \\
\pd{^2U}{\Ps^2}(\Ps_0) = n\g \biggl(\sqrt{\frac{(-\g)n}{\b(n+2)}}\biggr)^n
\cdot e^{n\vf}(n-(n+2)e^{2\vf}) \label{a30} \\
U(\Ps_0+\vf)=U(\Ps_0)+\frac12\,\pd{^2U}{\Ps^2}(\Ps_0)\vf^2+\ldots
= - 2\wt\La + \frac12\,\ov Mm_0^2\vf^2+\ldots \label{a31} \\
\ov Mm_0^2 = -2n\g \biggl(\sqrt{\frac{(-\g)n}{\b(n+2)}}\biggr)^n =
2n|\g|\biggl(\sqrt{\frac{|\g|n}{\b(n+2)}}\biggr)^n, \q
\g<0,\ \b>0. \label{a32}
\end{gather}

Let us give the following remark. The \co ical \ct\ in the theory is of a
dynamical origin. It is the minimum (extremum) of a self-\ia\ \pt\ of the
field~$\Psi$, $\la_{c0}(\Ps_0)=U(\Ps_0)$. We can introduce to the full field
\e\ $\Ps=\Ps_0+\vf$ and $\la_{c0}(\Ps_0)$.

In terms of the field $\vf$ we get:
\bg a39
\ve = \frac{\mpl^2}{16\pi} \,\frac{n^2(n+2)^2 (1-e^{2\vf})^2}
{((n+2)-ne^{2\vf})^2} \\
\eta = \frac{\mpl^2}{8\pi} \,\frac{n(n+2) (n-(n+2)e^{2\vf})}
{((n+2)-ne^{2\vf})}  = \frac{\mpl^2}{8\pi} \cdot \frac{n(n+2)(n(1-e^{2\vf})-
2e^2\vf)}{(n(1-e^{2\vf})+2e^{2\vf})} \label{e40} \\
\ve(\Ps_0) = \ve(\vf=0)=0 \label{e41} \\
\ve(\Psi=0)\cong \frac{\mpl^2}{16\pi}\biggl(\frac{\g n+\b(n+2)}{\g+\b}\biggr)^2
\cong\frac{\mpl^2}{16\pi}(n+2)^2 \label{e42} \\
|\eta(\Ps_0)| = |\eta(\vf=0)| =\frac{\mpl^2}{8\pi}\,n(n+2) \label{e43} \\
U(\Ps_0)=-2\la(\Ps_0)=-2\wt\La = 2\biggl(\frac{m_{\td A}}{\a_s^2 \mpl}\biggr)^n
\frac{nm_{\td A}|\wt{\fal P}|}{\a_s^2 }\biggl(\sqrt{\frac{|\wt{\fal P}|n}{(n+2)
\wt R(\wt\G)}}\biggr)^n = |\ov M|m_0^2. \label{e44}
\e

Let us consider an inflationary era in our model in quadratic \ap ion for
$U(\Ps_0+\vf)$.
\bg a52a
H^2 = \frac{4\pi m_0^2\vf^2}{3\mpl^2} + \frac{8\pi m_0^2}{3\mpl^2} \\
\aligned
3H\dot\vf |\ov M|+ m_0^2\vf &=0\\
3\biggl(\frac {\dot a}{a}\biggr)\dot\vf \,\ov M+m_0^2\vf &=0.
\endaligned \lb{a53a}
\e
One gets
\begin{gather*}
\biggl(\frac{\dot a}a\biggr)^2 = \frac{4\pi m_0^2\vf^2}{3\mpl^2} + \frac{8\pi}{3\mpl^2}\,m_0^2\\
\frac{\dot a}a = \sqrt{\vf^2+2} \cdot \biggl(\frac{2m_0\sqrt\pi}{\sqrt3\,\mpl}\biggr) \\
\dot\vf = -\frac{m_0\vf}{\ov M\sqrt{\vf^2+2}} \cdot \frac{\mpl}{2\sqrt{3\pi}}
\end{gather*}
and consequently
$$
\frac{\sqrt{\vf^2+2}}{\vf}\,d\vf = -\frac{m_0\mpl}{2\sqrt{3\pi}\,\ov M}\,dt
$$
One easily gets
\begin{gather*}
-\frac{m_0\mpl}{2\sqrt{3\pi}\,\ov M}(t-t_0) = \sqrt{\vf(t)^2+2} +
\frac1{\sqrt2}\ln\frac{\sqrt2 - \sqrt{\vf(t)^2+2}}{\sqrt2 + \sqrt{\vf(t)^2+2}}\\
a(t) = \exp\Biggl(\frac{2m_0\sqrt\pi}{\sqrt3\,\mpl} \int_{t_0}^t \sqrt{\vf(t)^2+2}\,dt\Biggr)
\end{gather*}
or
\begin{gather*}
\sqrt{\vf(t)^2+2} +
\frac1{\sqrt2}\ln\frac{\sqrt2 - \sqrt{\vf(t)^2+2}}{\sqrt2 + \sqrt{\vf(t)^2+2}}=
\frac1{\,\ov M\,}\vf_i - \frac{m_0\mpl}{2\ov M\sqrt{3\pi}}\,t \\
\ve=\frac{\mpl^2}{4\pi}\,\frac{\vf^2}{(2+\vf^2)^2}, \q
\vf^2\gg2, \ \ve=\frac{\mpl^2}{4\pi}\cdot\frac1{\vf^2}\\
\eta=\frac{\mpl^2}{4\pi}\,\frac{1}{2+\vf^2}, \q
\vf^2\gg2, \ \eta=\frac{\mpl^2}{4\pi}\cdot\frac1{\vf^2}.
\end{gather*}
And eventually
\begin{gather*}
\vf\simeq \vf_i - \frac{m_0\mpl}{2\sqrt{3\pi}}\,\ov M t\\
a(t)\simeq \exp\biggl(\frac{2m_0\sqrt\pi}{\sqrt3\,\mpl}\biggl(\vf_i t
- \frac{m_0\mpl \ov M}{4\sqrt{3\pi}}\,t^2\biggr)\biggr).
\end{gather*}

The slow-roll parameters are as follows
\beq a49
\ve=\eta=\frac{\mpl}{4\pi} \biggl(\frac1\vf\biggr)^2.
\e

To get $N\simeq70$ we should have
\beq a51
\vf_i>3\mpl.
\e
The inflation is successful if for the scalar field $\vf$
\beq a52
m_0\leq 10^{-6} \,\mpl,
\e
$\mpl\simeq 24\tm 10^{18}$\,GeV and in our case $m_0\simeq14.2 \tm 10^{-5}$\,eV.
The inflation presented here could be called a dark inflation because a scalar
field $\vf$ is a part of our Dark Matter.
In this way the condition is satisfied.
Moreover, an evolution of a field~$\vf$ does not end after a de Sitter phase,
exponential evolution phase. It ends in the moment
$\vf=0$, i.e.\ where $U(\Ps)=U(\Ps_0+\vf)$ reaches a minimum.

In further development we can consider an idea of \co ical models with a
\co ical \ct\ (vacuum energy) and cold Dark Matter and ordinary matter. Our
Dark Matter is of course a cold Dark Matter.

Let us remind to the reader that
our unification contains GR as a limit for $g_\[\mu\nu]=0$ and $\Ps=0$. Thus
we can consider ordinary Einstein \e s with a \co ical \ct\ (or without) equal
to $\wt\La=-\frac12 U(\Ps_0)$ which has been proceeded.
\bg a37
R_{\mu\nu} - \frac12\,Rg_{\mu\nu} + \wt\La g_{\mu\nu} = \frac{8\pi}{\mpl^2}
\br{matter}T_{\mu\nu} \qh{or}\\
R_{\mu\nu} - \frac12\,Rg_{\mu\nu} = \frac{8\pi}{\mpl^2}
\bigl(\br{matter}T_{\mu\nu} - \rho\dr{vac} g_{\mu\nu}\bigr), \label{a38}
\e
where $\rho\dr{vac}$ is a vacuum energy.

In this case one gets
\bea a53
H^2 &= H_0 \sum \frac{\rho_{0i}}{\rho_c} = H_0\sum_i \O_i a^{-3(1+W_i)}
 && \biggl(H^2=\frac{8\pi G_N}3 \sum\rho_i\biggr)\\
\rho_c&= \frac{3H_0^2}{8\pi G_N} &-& \hbox{ critical density} \nonumber\\
\O_i &= \frac{\rho_{0i}}{\rho_c}&-& \ \rho_{0i} - \hbox{initial density of a type
of a matter ``$i$''} \nonumber\\
\rho\dr{vac}&=\frac{\wt\La}{8\pi G_N}\,.\nonumber
\e
In our case we have
\begin{align*}
i=0, \q & \hbox{ordinary matter dust, $W=0$}\\
i=1, \q & \hbox{radiation, $W=\frac13$}\\
i=2, \q & \hbox{cold Dark Matter dust, $W=0$}\\
i=3, \q & \hbox{vacuum energy, $W=-1$}
\end{align*}

Our \co ical model is the so called $\La$CDM model and one gets
\bg a54
H(a) = \frac{\dot a}a = H_0 \sqrt{(\O_0+\O_2)a^{-3} + \O_1a^{-4}+\O_3}\\
\O_1 =10^{-4}. \label{a54a}
\e
Thus one finds ($\O_0+\O_2=\O_m$)
\bg a55
H(a)=H_0 \sqrt{\O_m a^{-3}+\O_3}\\
a(t)=\biggl(\frac{\O_m}{\O_3}\biggr)^{1/3}\sinh\biggl(\frac t{t_{\wt\La}}\biggr) \label{a56} \\
t_{\wt\La} = \frac2{(3H_0\sqrt{\O_3})} \label{a57}
\e
where
\beq a58
\O_i=\frac{\rho_{ic}(t-t_{\wt\La})}{\rho_c} = \frac{8\pi G_N\rho_{ic}(t-t_{\wt\La})}
{3H_0^2}
\e

We suppose that our model is spatially flat $k=0$. In further
considerations we will consider some predictions of the model considered with
perturbations of density of matter (both barionic and cold Dark Matter).
This gives us a fluctuation of spectrum of background radiation to compare
with observations. The value of~$W_i$ can be considered as a parameter of the
theory. The comparison of $\La$CDM with observation is really perfect.
Moreover, if we abandon the Robertson--Walker--Friedman and consider
inhomogeneous Lema\^\i tre--Tolman models, we can avoid an accelerating expansion
and introduction of a \co ical \ct.

Let us consider a density perturbation in our theory where the field~$\Ps$
plays the role of an inflaton and also a role of a \qe\ field. It means, the
energy of a field~$\Ps$ has been released and \pc s have been created. Before
an era of slow-roll inflation an additional energy has been released after
a tunnel effect from a false vacuum to a true vacuum state. Simultaneously
the field~$\Ps$ (alternatively the field~$\vf$) evolves to the minimum of
an energy of a self\ia\ energy $U(\Ps)=U(\Ps_0)$. According to Ref.~\cite{xxx}
we consider Einstein equation without a \co ical \ct\ (it is included in~$U(\Ps)$)
with sources of a field~$\Ps$. We divide $\Ps$ into two parts, $\wt\vf$ and
$\d\wt\vf$. $\wt\vf$~is a homogeneous part of the field~$\Ps$ (unperturbed) and $\d\wt\vf$
is its fluctuating part. We can repeat all calculations from Ref.~\cite{xxx},
getting the \fw\ results for a spectrum of curvature fluctuations
\beq a55a
P_{\cal R} = A^2_{\cal R}\biggl(\frac k{aH}\biggr)^{n_{\cal R}-1},
\e
$k$ is a wave vector number (do not mix this~$k$ with $k$~which is a curvature parameter).
(Let us remind to the reader that we are using linearized Einstein \e s around
homogeneous solution for metric and field~$\Ps$ or $\vf$.)

In particular we consider a perturbed metric:
\beq a52b
ds^2 = -(1+2\F)\,dt^2 + a^2(t)(1-2\Xi)\d_{ij}\,dx^i\,dx^j
= a^2(\tau)\bigl(-(1+2\F)d\tau^2 + (1-2\Xi)\d_{ij}\,dx^i\,dx^j\bigr).
\e
The source of Einstein \e s (with zero \co ical \ct) is an energy momentum
tensor for the field~$\vf$. For we consider a linear theory of perturbations
we get
\beq a53b
\gd dG,\nu,\mu, = 8\pi G_N\gd\d T,\nu,\mu,.
\e
$\tau$ is a conformal time
\beq a54b
\tau = \int a^{-1}\,dt,
\e
$\F$ and $\Xi$ are \pt s of \gr al perturbations.

$\d \gd G,\mu,\nu,$ is a linearized perturbed part of Einstein \e s around
a \co ical \so, $\gd\d T,\mu,\nu,$ is linearized perturbed energy momentum
tensor around~$\wt\vf$ (unperturbed) \wrt $\d\wt\vf$ (perturbation). Thus we
have
\beq a55b
\gd G,\nu,\mu, = 8\pi G_N \gd T,\nu,\mu,
\e
where
\bg a56b
\gd G,\nu,\mu, = \gd G_0{},\nu,\mu, + \gd \d G,\nu,\mu, \\
\gd T,\mu,\nu, = \gd T_0{},\mu,\nu, + \gd \d T,\mu,\nu, \lb a57b \\
\gd G_0{},\nu,\mu, = \gd T_0{},\mu,\nu, . \lb a58b
\e
The last \e\ gives us a \co ical \so\ with scalar sources, i.e.\ $\wt\vf$,
$\vf=\wt\vf+\d\wt\vf$,
\bg a59b
\gd T,\nu,\mu, = \gd T,\nu,\mu,(\vf), \q
\gd T_0{},\nu,\mu, = \gd T_0{},\nu,\mu,(\vf), \\
\gd G,\nu,\mu, = \gd G,\nu,\mu,(ds^2),\q
ds^2 \hbox{ from Eq.\ \er{a52a}}, \lb a60b \\
\gd G_0{},\nu,\mu,(ds^2,k=0),\q ds^2 \hbox{ from \er{a1}}.\lb a61b
\e

After some tedious calculations (see Ref.\ \cite{xxx}) one gets a spectrum
of the comoving curvature perturbations
\beq a62b
P_{\cal R}= \frac{4\pi}{\mpl^2\ve} \biggl(\frac H{2\pi}\biggr)^2
\biggl(\frac k{aH}\biggr)^{n_{\cal R}-1}
\e
where $k$ is a wave number of a scalar $\d\wt\vf$ in a Fourier decomposition,
$H=\frac{\dot a}a$ is a Hubble \ct, $\ve$ is a slow-roll
parameter \er{a16}, ``$\,\dot{\phantom H}\,$'' means a derivative \wrt a time~$t$.

Let us consider a primordial spectrum of \gr al waves in our approach. This
spectrum is flat for our Hubble \ct\ is really \ct
$$
P\dr{grav}=\frac2{m^2\pl}\biggl(\frac{H}{2\pi}\biggr)^2
$$
Thus
$$
n\dr{gr}=1.
$$

\beq a56a
n_{\cal R}-1 = \frac{d\ln R_{\cal R}}{d\ln k} = -6\ve+2\eta.
\e

Thus we get that a spectrum of a curvature is scale invariant. In this way we
reached classical Modern \E\co y from the \E\nos\ \KK (Jordan--Thiry) Theory
where the inflaton field (a~scalar field) has its origin from geometry, being
simultaneously a Dark Matter.

\def\eq#1 {\label{a#1}}
\def\rf#1 {\eqref{a#1}}
\def\({\left(}
\def\){\right)}
\let\ul\underline
\let\t\widetilde
\def\uP{\ul{\t P}}
\def\RG{\t R(\t\G)}
\def\gi#1{^{\text{\rm#1}}}
\def\il{inflation}
\def\arctg{\operatorname{arctg}}
\def\sh{\operatorname{sinh}}
\def\ars{\operatorname{arsinh}}
\def\Sh{\operatorname{sinh}}
\def\Ch{\operatorname{cosh}}

We follow Ref.~\cite{alfa} to perturb a spatial part of a metric
\begin{equation}
ds^2=a^2(\tau)\bigl[-d\tau^2+\bigl(\d_{dc}+2\t E{}^T_{dc}\bigr)dx^b\,dx^c\bigr], \eq237
\end{equation}
$\tau$ is a conformal time, $d,c=1,2,3$.

\advance\abovedisplayskip by-1pt
\advance\belowdisplayskip by-1pt
If we expand Einstein equations for \rf237 \ up to linear terms, we get
\begin{equation}
\frac{d^2E^T_{dc\vec k}}{d\tau^2}+2aH\frac{dE^T_{dc\vec k}}{d\tau}
+k^2E^T_{dc\vec k}=0, \eq238
\end{equation}
where $\t E{}^T_{dc}$ is a spatial perturbation of the metric, $a(\tau)$ is a
scale factor depending on a conformal time, $E^T_{dc\vec k}$ is a $\vec k$
Fourier component $\t E{}^T_{dc}$
\begin{equation}
\al
&\t E{}^T_{dc}=\frac1{(2\pi)^{3/2}}\int E^T_{dc\vec k}e^{-i\vec k\vec r}\,d^3\vec
k\\
&\vec k=(k_1,k_2,k_3), \quad |\vec k|^2=k^2, \quad \vec r=(x,y,z), \quad r^2=x^2+y^2+z^2.
\eal
\eq239
\end{equation}
The important quantity is an amplitude of \gr\ wave corresponding to
$E^T_{dc}$, i.e.
\begin{equation}
h_{dc\vec k}=aE^T_{dc\vec k}\,. \eq240
\end{equation}
$H$ is a Hubble \ct. The relation between a conformal time $\tau$ and
ordinary time~$t$ is given by
\begin{align}
a(\tau)&=-\frac1{H\tau}, \quad -\infty<\tau<0, \eq241 \\
a(t)&=a_0 e^{Ht}, \eq242
\end{align}
see Eq.\eqref{a54b} for comparison. Now $H$ is really \ct.
For $h_{dc\vec k}$ one gets
\begin{equation}
\frac{d^2h_{dc\vec k}}{d\tau^2}-\frac2{\tau^2}\,\frac{dh_{dc\vec k}}{d\tau}
+k^2h_{dc\vec k}=0. \eq243
\end{equation}
This equation can be easily solved. 
\begin{equation}
h_{dc\vec k}=h_{dck}=A_{dc}\bigl(\tau|k| \cos(\tau|k|)-\sin(\tau|k|)\bigr)+
B_{dc}\bigl(\tau|k| \sin(\tau|k|)+\cos(\tau|k|)\bigr) \eq245
\end{equation}
where $A_{dc}$ and $B_{dc}$ are \ct s, $d,c=1,2,3$.

Using Eq.\ \er{4.112} in a background \er{a237} with a solution \er{a245}
one gets for $\Psi(r,\tau)$
\begin{equation}
\pp{}r\biggl(a^4D\pp{\Psi(r,\tau)}r \biggr)
-\pp{}\tau \biggl(a^4D\pp{\Psi(r,\tau)}\tau \biggr)=0
\eq245a
\end{equation}
where
\begin{multline}
D=
\biggl(\det\Bigl(\Bigl[\d_{dc}+ \frac2{(2\pi)^{1/2}a}
\int_{-\infty}^{+\infty} dk\,k\cos(kr)\bigl[A_{dc}(k\cos(\tau k)-\sin(\tau k))\\
{}+B_{dc}(\tau k\sin(\tau k)+\cos(\tau k))\bigr]\Bigr]\Bigr)_{d,c=1,2,3}\biggr)^{1/2}
\eq245b
\end{multline}
In Eq.\ \er{4.112} we neglect all terms except a kinetic term,
especially self-interacting terms.

Let us consider Eq.~\er{a245b}. One gets
\beq aa75
D = \biggl(\det\biggl(\d_{dc} + \frac2{(2\pi)^{1/2}a} [A_{dc}J_1 + B_{dc}J_2]
\biggr)_{d,c=1,2,3}\biggr)^{1/2}
\e
where
\bea aa76
J_1 &= \int_{-\iy}^{+\iy} dk\,k\cos(kr)(k\cos(\tau k)-\sin(\tau k))\\
J_2 &= \int_{-\iy}^{+\iy} dk\,k\cos(kr)(\tau k\sin(\tau k)+\cos(\tau k))\lb aa77
\e
or
\bea aa78
J_1 &= \frac1{\tau^3} \lim_{x\to\iy} \int_{-x}^x dy\, y\cos(by)(y\cos y
-\tau \sin y) \\
J_2 &= \frac1{\tau^2} \lim_{x\to\iy} \int_{-x}^x dy\, y\cos(by)(y\sin y
- \cos y) \lb aa79
\e
where
\beq aa80
b=\frac r\tau.
\e
One gets
\bea aa81
J_1 &= \lim_{x\to\iy} \frac1{\tau^3} \int_{-x}^x dy\, y\cos(by)(y\cos y
- \tau\sin y)=0 \\
J_2 &= \lim_{x\to \iy} \frac2{\tau^2} \lim_{x\to\iy} \int_0^x dy\, y\cos(by)(y\sin y
- \cos y). \lb aa82
\e
\setbox9=\hbox{$\displaystyle{-\frac{x^2}2 \biggl(\frac{\cos(b+1)x}{b+1}
+ \frac{\cos(1-b)x}{1-b}\biggr)}$}
\setbox8=\hbox{\begin{picture}(156,17)(0,0)
\unitlength1pt
\put(0,0){\copy9}
\put(0,17){\line(6,-1){156}}
\end{picture}}
\setbox9=\hbox{$\displaystyle{x\biggl(\frac{\sin(1+b)x}{2(1+b^2)}(1-b)
+ \frac{\sin(1-b)x}{2(1-b)^2}(b+1)\biggr)}$}
\setbox7=\hbox{\begin{picture}(215,17)(0,0)
\unitlength1pt
\put(0,0){\copy9}
\put(20,20){\line(6,-1){175}}
\end{picture}}
\setbox9=\hbox{$\displaystyle{\biggl(\frac{\cos(b+1)x}{2(b+1)^3}(1-b)
+ \frac{\cos(1-b)x}{2(1-b)^3}(b+1)\biggr)}$}
\setbox6=\hbox{\begin{picture}(200,17)(0,0)
\unitlength1pt
\put(0,0){\copy9}
\put(19,19){\line(6,-1){175}}
\end{picture}}

\noindent
After some calculations one gets
\beq aa83
J_2 = \frac2{\tau^2}\Bigl[ \lim_{x\to\iy}K(x) - K(0)\Bigr]
\e
where
\bml aa84
K(x) = -\frac{x^2}2 \biggl(\frac{\cos(b+1)x}{b+1} + \frac{\cos(1-b)x}{1-b}\biggr)
+ x \biggl(\frac{\sin(1+b)x}{2(1+b)^2}(1-b) + \frac{\sin(1-b)x}{2(1-b)^2}(b+1)\biggr)\\
{}+ \frac{\cos(b+1)x}{2(b+1)^3}(1-b) + \frac{\cos(1-b)x}{2(1+b)^3}(b+1)
\e
and
\beq aa85
K(0) = \frac{2(1+3b^2)}{(1-b^2)^3}
\e
and eventually
\beq aa86
\bga
D = \biggl(\det\biggl(\d_{dc} + \frac{4B_{dc}}{(2\pi)^{1/2}a\tau^2} \biggl(
\lim_{x\to\iy} \int_0^x dy\,y\cos(by) (y\sin y-\cos y)\biggr)_{d,c=1,2,3}\biggr)^{1/2}\\
= \Biggl(\det\biggl(\d_{dc} + \frac{4B_{dc}}{(2\pi)^{1/2}}\biggl[
\lim_{x\to\iy}\copy8 \\
{} + \copy7 \\
{}+ \copy6
{}- \frac{2(1+3b^2)}{(1-b^2)^3}\biggr]\biggr)_{d,c=1,2,3}\Biggr)^{1/2}
\ega
\e

By removing the divergent part we get a regularized expression
\bml aa87
D\dr{reg} = \biggl(\det\biggl(\d_{dc} - \frac{8B_{dc} (1+3b^2)}{(2\pi)^{1/2}
a\tau^2(1-b^2)^3}\biggr)_{d,c=1,2,3}\biggr)^{1/2}\\
{}= \biggl(\det\biggl(\d_{bc} + \frac{8B_{dc} H(\tau^2+3r^2)\tau^3}
{(2\pi)^{1/2}(\tau^2-r^2)^3} \biggr)_{d,c=1,2,3}\biggr)^{1/2}.
\e
$H$ is the Hubble \ct\ for Early \E\Un\ (CMB or BAO).

One gets eventually
\bml aa88
D\dr{reg} = \biggl(1+\frac{4\sqrt2\, H(\tau^2+3r^2)\tau^3}{\pi^{1/2}(\tau^2
-r^2)^3} (B_{11}+B_{22}+B_{33}) \\
{}+ \frac{32\tau^6 H^2(\tau^2+3r^2)^2}{\pi
(\tau^2-r^2)^6}(B_{11}B_{22}+B_{33}B_{22}+B_{33}B_{11} -B_{13}B_{31}
-B_{12}B_{21}-B_{23}B_{32})\\
{}+ \frac{128\sqrt2\,H^3(\tau^2+3r^2)^3\tau^9}{\pi^{3/2}(\tau^2-3r^2)^9}
(B_{33}B_{11}B_{22}+B_{12}B_{23}B_{31}\\
{}-B_{13}B_{22}B_{31}-B_{12}B_{21}B_{33}
-B_{11}B_{23}B_{32}-B_{13}B_{21}B_{22})\biggr)^{1/2}.
\e
It is interesting to notice that the \cf\ $A_{dc}$ from Eq.~\er{a245} does not
enter $D\dr{reg}$.

Thus we have finally the \e
\beq aa89
\pp{}r\biggl(\wt D(r,\tau)\cdot \pp\Psi r(r,t)\biggr) - \pp{}\tau\biggl(\wt D(r,\tau)
\pp\Psi\tau (r,\tau)\biggr) =0
\e
where
\beq aa90
\wt D(r,\tau) = \frac{D\dr{reg}(r,\tau)}{\tau^4}.
\e

Let us consider Eq.~\er{aa89}. One gets
\bg aa91
\pp{^2\Psi(r,\tau)}{r^2} - \pp{^2\Psi(r,\tau)}{\tau^2} - \pp dr\,
\pp{\Psi(r,\tau)}r - \pp d\tau\,\pp{\Psi(r,\tau)}\tau= 0\\
d = \log\wt D \lb aa92
\e
Let us change the independent variables in \er{aa91}--\er{aa92}:
\bg aa93
x=r+\tau, \q y=r-\tau,\\
r=\tfrac12(x+y), \q \tau=\tfrac12(x-y). \lb aa94
\e
One gets
\bg aa95
\pp{^2\Phi}{x\pa y} + \frac12 \,\pp{\wt d}y\, \pp \Phi x +\frac12\,\pp{\wt d}
x\,\pp\Phi y=0 \\
\Phi = \Phi(x,y), \q \Psi(r,\tau)= \Phi(r+\tau,r-\tau). \lb aa96 \\
\wt d = \wt d(x,y) = \log \wt D (\tfrac12(x+y),\tfrac12(x-y)) \lb aa97
\e
and eventually
\bml aa98
\wt d = \log\Biggl[\frac{16}{(x^2-y^2)^4} \biggl(
1-\frac{\sqrt2\,H(x^2+y^2+xy)(x-y)^3}{2\pi\,x^3y^3}(B_{11}+B_{22}+B_{33})\\
{}+ \frac{H^2(x^2+y^2+xy)^2(x-y)^6}{2\pi x^6y^6}
(B_{11}B_{22}+B_{33}B_{22}+B_{33}B_{11}-B_{13}B_{31}-B_{12}B_{21}-B_{23}B_{32})\\
{}- \frac{\sqrt2\,H^3(x^2+y^2+xy)^3(x-y)^9}{4\pi^{3/2}x^9y^9}
(B_{33}B_{11}B_{22}+B_{12}B_{23}B_{31}\\
{}-B_{13}B_{22}B_{31}-B_{12}B_{21}B_{33}-B_{11}B_{23}B_{32}-B_{13}B_{21}B_{22})
\biggr)^{1/2}\Biggr]
\e

Let us consider Eq.~\er{aa95} in the case $x\to \iy$. One gets
\beq aa99
\pp{^2\Phi}{x\pa y} - \frac3{2y}\,\pp \Phi x =0.
\e
For $y\to\iy$
\beq aa100
\pp{^2\Phi}{x\pa y} + \frac3{2x}\,\pp \Phi y=0.
\e
In the first case it means $r\to\iy$, in the second case $r\to\iy$, $\tau\to
-\iy$. If $x\to\iy$, $y\to\iy$, one gets simply
\beq aa101
\pp{^2\Phi}{x\pa y} = 0,
\e
which is a string \e. A general solution of \er{aa101} reads
\beq aa102
\Phi(x,y) = g(x)+f(y)
\e
where $g$ and $f$ are arbitrary $C^1$ \f s of one real variable. Thus
\beq aa103
\Psi(r,\tau) = g(\tau+r)+ f(r-\tau) = g\biggl(r - \frac{a_0}H \exp(-Ht)\biggr)
+ f\biggl(r + \frac{a_0}H \exp(-Ht)\biggr).
\e
In this way we get for $r\to\iy$, $t\to\iy$
\beq aa104
\Psi(r,t) = g\biggl(r - \frac{a_0}H \exp(-Ht)\biggr)
+ f\biggl(r + \frac{a_0}H \exp(-Ht)\biggr).
\e

If we write $\Psi(r,t)=\Psi_0+\vf(r,t)$ we can get some kinds of \qe\ oscillations
around a $\Psi_0$ caused by primordial \gr al waves. We can use for convenience
also $Q$ or~$q$.

\let\th\theta
\def\<{\left\langle}
\def\>{\right\rangle}

Let us give simple applications of our \il\ theory to CMB anisotropy problem.
In order to do this we remind to the reader some notions of CMB anisotropy
theory:
\begin{equation}
\frac{\D T(\vec e)}T = \sum_{l,m}a_{lm}Y_{l,m}(\vec e) \eq342
\end{equation}
where $Y_{lm}(\vec e)=Y_{lm}(\th,\vf)$ are spherical harmonics and $a_{lm}$
are multipole coefficients of~CMB. $\frac{\D T(\vec e)}T$ means fluctuations
of a temperature of CMB measured in a direction of~$\vec e$ parametrized by
two angles $\th$ and~$\vf$. A~direction of a photon is $\vec n=-\vec e$.

After averaging one gets
\begin{equation}
\<a_{lm}a^*_{l'm'}\>=\d_{ll'}\d_{mm'}C_l. \eq343
\end{equation}
For a two point correlation function one gets
\begin{equation}
\<\frac{\D T}T(\vec e_1)\cdot \frac{\D T}T(\vec e_2)\>
_{\vec e_1\cdot\vec e_2=\cos\th}=
\frac1{4\pi}\sum_l(2l+1)C_lP_l(\cos\th) \eq344
\end{equation}
(see Refs \cite{alfa} for more details) where $P_l(\cos\th)$ is a Legendre
polynomial. Thus in order to compare a theory with observations it is
necessary to calculate $C_l$ in some region of~$l$ (i.e., $2<l<l_{\max}$).
Usually we subtract effects of relative movement and $C_0=C_1=0$.

$C_l$ can have different origins coming from several types of perturbations
(i.e.\ scalar, vector, tensor) (see Refs \cite{alfa} for details). In our theory we
have scalar perturbations due to \il\ and tensor perturbations due to \gr\
waves. The tensor fluctuations are scale invariant
\begin{equation}
n_T=n\dr{gr}-1=0.
\end{equation}
In the case of scalar fluctuations we calculate $n_s$ and $\frac{dn_s}{d\ln
k}$ (see Eqs \er{a55a}, \er{a62b}, \er{a56a}) and we
find conditions for $n_s=1$. Thus we get (under some conditions) a
Harrison--Zel$'$dovich spectrum. Moreover $\frac{dn_s}{d\ln k}$ can be in
general nonzero and it is possible to use it to compare with observations in
order to falsificate some models. In general we get in three cases
\begin{equation}
P^s_R(k)=P^s_0k^{n_s-1}. \eq345
\end{equation}
Thus via a standard procedure (see Ref.\ \cite{alfa}) one gets
\begin{equation}
C^s_l=\frac{\(P_0^s\)^2}9 \cdot \frac
{\G(3-n_s)\G(l-\frac12+\frac{n_s}2)}
{2^{3-n_s}\G^2(2-\frac{n_s}2)\G(l+\frac52-\frac{n_s}2)} \eq346
\end{equation}
where $n_s\simeq1$.

If we take $n_s=1$ we get
\begin{equation}
l(l+1)C^s_l=\text{const.} \eq347
\end{equation}
which is in an agreement with observational data. In the case of tensorial
(\gr\ waves) perturbations one gets
\begin{equation}
P^T(k)= P\gi{gr}_0 k^{n\dr{gr}-1}=P\gi{gr}_0k^{n_T} \eq348
\end{equation}
and by a standard procedure
\begin{equation}
C^T_l=C\gi{gr}_l= \(P\gi{gr}_0\)^2\cdot \frac{(l+2)!}{(l-2)!}
\cdot \frac{\G(6-n_T)\G(l-2+\frac{n_T}2)}
{2^{6-n_T}\G^2(\frac72-n_T)\G(l+4-\frac{n_T}2)}\,. \eq349
\end{equation}
For $n_T=0$ or $n\dr{gr}=1$ one gets
\begin{equation}
C\gi{gr}_l=\(P\gi{gr}_0\)^2\cdot\frac8{15\pi}\cdot
\frac{l^2(l+1)^2}{(l+3)(l-2)} \eq350
\end{equation}
(see Refs \cite{alfa}). The singularity for $l=2$ is not realistic, there is only
an amplification for $l=2$.

Let us notice that our power flat spectrum (scale-invariant) is a good
starting point for every structure formation theory in the Universe. Using our full
$P^s_R(k)$ spectra we can do more precise calculations for~$C^s_l$.
Especially in order to compare with very precise \co ical observations
(Planck satellite) (see Refs \cite{yy,yy1}) we should take under consideration $\frac{dn_s}{d\ln k}$
and deviations from scale invariant spectrum, which has been reported from
WMAP data. Some further developments of our $P_R(k)$ spectra are beyond the
scope of this Appendix for they strongly depend on details of \co ical models
involved (see Refs~\cite{alfa}, also Planck's 2018 \cite{yy1}). In particular we can consider $C^s_l$ and
$C\gi{gr}_l$ only for $l\lesssim100$. For $l\gtrsim100$, $\frac{\D T}T$ is
dominated by acoustic oscillation (see Refs~\cite{alfa}). $C\gi{gr}_l$
are very small for $l\ge60$ (they decay on sub-horizon
scales). Finally we write
\begin{gather}
C^s_l l(l+1) = \text{const.} \cong
\<\(\frac{\D T}T(\th_l)\)^2\>, \eq351
\\
\th_l=\frac\pi l \eq352
\end{gather}
and
\begin{equation}
C\gi{gr}_l l(l+1)=\(P\gi{gr}_0\)^2 \cdot \frac8{15\pi}\cdot\frac{l(l+1)}
{(l+3)(l-2)}\,. \eq353
\end{equation}
The behaviour of $C_l$ for $l>100$ is beyond the scope of this Appendix.

Let us consider an \e\ of state for a \qe. Suppose that a matter of a \qe\ is
only in a part (a~fraction) $\eta$ ($0<\eta<1$) stored as \pc s of a \qe. Let
us suppose also that a gas of \qe\ \pc\ is a perfect gas governed by the
Clapeyron \e
\beq a177
\frac{p_1}{\rho_1} = \frac{K_B T}{m_0} = \frac T{T_0}, \q
T_0 = \frac{m_0}{K_B}\simeq 0.11^\circ {\rm K}
\e
(if $m_0=10^{-5}\,{\rm eV}$, $K_B$ is the Boltzmann \ct).

One gets
\begin{gather}
\rho_1=\eta \rho_Q=\eta \rho \q (\rho_Q \hbox{ --- density of a \qe}) \eq 178 \\
p=p_Q+p_1=p_Q+\eta \frac T{T_0}\rho = -\rho(1-\eta)+\rho\frac T{T_0}\eta.
\eq179
\end{gather}

Eventually one gets
\begin{align}
p&=\rho\(\eta\(1+\frac T{T_0}\)-1\) \eq199 \\
p&=w\rho \eq200 \\
w&=\eta\(1+\frac T{T_0}\)-1. \eq201
\end{align}

Now we can calculate an isothermic speed of sound in a \qe.
\begin{equation}
c_1^2=\frac p\rho=w=\eta\(1+\frac T{T_0}\)-1. \eq202
\end{equation}
For
\begin{equation}
0<c_1^2<1 \eq203
\end{equation}
we get
\begin{equation}
\frac{T_0}{T+T_0}<\eta<\min\biggl[\frac{2T_0}{T_0+T},1\biggr]. \eq204
\end{equation}
Let us remind to the reader that an isothermic sound is appropriate for low
frequency of acoustic waves (we have to do with this sound in astrophysics).
In general we have to do with so called adiabatic sound. In order to
calculate a speed of an adiabatic sound in a \qe\ we should find an analogue
for a Poisson adiabate for our equation of state. One gets supposing that an
internal energy of a \qe\ is an energy of one-atomic gas of \qe\ \pc s
\begin{equation}
dU+p\,dV=0 \eq205
\end{equation}
or
\begin{equation}
\frac32\,dT=\frac{d\rho}{\rho}\(\eta-1+\frac T{T_0}\,\eta\) \eq206
\end{equation}
and finally
\begin{align}
&\frac p{\rho^\ka}=\text{const.} \eq207 \\
&\ka=\frac{2\eta+3}3 \eq208
\end{align}
and a speed of an adiabatic sound simply reads
\begin{equation}
c_2=\sqrt{\frac{\ka p}{\rho}}= \sqrt{\ka\(\eta\(1+\frac T{T_0}\)-1\)}\,. \eq209
\end{equation}
It is interesting to ask what kind of a polythrope $\ka$ represents. Let us
remind to the reader that in general
\begin{equation}
\ka=\frac{c_p-\ov c}{c_v-\ov c} \eq210
\end{equation}
where $\ov c$ is a specific heat of a polythrope in mind.

One gets in our case
\begin{equation}
\ov c=\frac{3(\eta-1)}{2\eta}\,\ov R \eq211
\end{equation}
where $\ov R$ is a universal gas \ct.

Thus we found both speeds of sound in a \qe\ for low frequency and for
high frequency of sound.
The measurement of both speeds can help us to find $\eta$ and~$T$. One
gets
\begin{equation}
\eta=\frac32\(\(\frac{c_2}{c_1}\)^2-1\) \eq212
\end{equation}
and
\begin{equation}
T=\frac{T_0}{3(c_2^2-c_1^2)}\(c_1^2(2c_1^2+5)-3c_2^2\). \eq213
\end{equation}
Let us notice that a gas of \qe\ \pc s is quite cold.
Moreover these \pc s are highly relativistic. For the temperature
$T\simeq T_0=0.11\,{}^\circ$K one sees that a speed of a \qe\ \pc \ is about
0.91 of the speed of light. Moreover we can consider lower temperatures.
It is interesting to notice that the speed of sound is of the same order.

Thus if we want to have both speeds smaller than a speed of light,
\begin{equation}
0<c_i^2<1, \quad i=1,2, \eq214
\end{equation}
we get
\begin{equation}
\eta>\frac{1+\sqrt{13+12\ov t}}{1+\ov t}=f(\ov t) \eq215
\end{equation}
where
\begin{equation}
\ov t=\frac T{T_0}\,. \eq216
\end{equation}
The condition
\begin{equation}
\ov t=\frac T{T_0}>13 \eq217
\end{equation}
guarantees that $f(\ov t)<1$. Moreover for $t=14$ one gets
\begin{equation}
\eta\ge 0.96357. \eq218
\end{equation}
This seems quite interesting, however maybe too much. Let us calculate a mean
scattering length of \qe\ \pc s for such a big $\eta=0.96357$
\begin{equation}
l\dr{scattering}=\frac1{\eta\sigma n}=1.0378\cdot l. \eq219
\end{equation}
We have still to do with a very dense gas of \qe\ \pc s (see Eqs
\er{4.162}--\er{4.164}).

It seems that it would be reasonable to repeat these calculations using a
different equation of state for \qe\ \pc s gas. In particular we can consider
a gas of \qe\ \pc s as a boson gas (with spin zero). For a mass of a \qe\
\pc\ is very small we can neglect this fact in an \e\ of state.
In this case the
equation of state looks:
\begin{equation}
p=p_Q+p_1=-(1-\eta)\rho+\frac{4\eta\sigma\rho}3\,T^4 \eq220
\end{equation}
and an adiabate equation
\begin{equation}
dU=\ov c_v\,dT+p\,dV=0 \eq221
\end{equation}
where
\begin{equation}
\ov c_v=16\sigma T^3 , \eq222
\end{equation}
$\sigma$ is the Stefan--Boltzmann \ct.

One easily integrates
\begin{equation}
\frac p{\rho^{\bar\ka}}=\text{const.} \eq223
\end{equation}
where
\begin{equation}
\ov\ka=\frac{3+\eta}3\,. \eq224
\end{equation}
We have as before two kinds of sound: an isothermic sound
\begin{equation}
\ov c_1=\sqrt{\frac p\rho}=\sqrt{\eta-1+\frac{4\eta\sigma}3\,T^4} \eq225
\end{equation}
and an adiabatic sound
\begin{equation}
\ov c_2=\sqrt{\ov\ka\(\eta-1+\frac{4\eta\sigma}3\,T^4\)}=\sqrt{\frac{\ov\ka p}\rho}\,.
\eq226
\end{equation}

From $0<\ov c_i^2<1$, $i=1,2$, one gets
\begin{equation}
\frac1{1+\(T/{\,\ov T_0\,}\)^4} < \eta < \min\biggl[1,\frac2{1+\(T/{\,\ov
T_0\,}\)^4} \biggr] \eq227
\end{equation}
and
\begin{equation}
\eta>\frac{-3\ov t{}^4-1+\sqrt{9\ov t{}^8+18\ov t{}^4+13}}{2(1+\ov t{}^4)}=\ov f(\ov t) \eq228
\end{equation}
with the condition
\begin{equation}
\ov t=\frac T{\,\ov T_0\,}>\frac1{\sqrt2}\simeq 0.707. \eq229
\end{equation}
This condition guarantees that
\begin{equation}
\ov f(\ov t)<1 \eq230
\end{equation}
where $T_0^4=\frac3{4\sigma}$ or
\begin{equation}
\ov T_0=\frac1{K_B}\root4\of{\frac{360}\pi}\simeq\frac{3.27}{K_B}\,. \eq231
\end{equation}

A \f\ $\ov f(\ov t)$ is going very quickly to zero if $\ov t\to\infty$.

From the measurements of $\ov c_1$ and $\ov c_2$ we can obtain as before $\eta$ and
$T$. One gets:
\begin{equation}
\eta=3\(\(\frac{\ov c_2}{\,\ov c_1\,}\)^2-1\) \eq232
\end{equation}
and
\begin{equation}
T=\frac{\ov T_0}{\root4\of3}\(\frac{4\ov c_1^2+\ov c_1^4-3\ov c_2^2}{\ov c_2^2-\ov
c_1^2}\)^{1/4}. \eq233
\end{equation}
Recently many papers have appeared concerning the speed of sound in a \qe\
(see Refs \cite{19c,19d,19e,19f,19g}). In some of them the authors propose to measure this speed.

Conditions \eqref{a217} and \eqref{a230}
give us a fraction of a Dark Matter with an accordance of Planck's measurement.
However, we cannot settle what is a fraction of \qe\ \pc s and skewon \pc s in a
full Dark Matter density. We need a value of~$T$ to settle this.

To be in accordance with Planck's Data we need $\eta<0.42105$.

The fraction $\eta$ can be connected with the part of an energy density of
\qe\ field $Q$ in such a way that it is a fraction of fluctuation energy density
around an equilibrium~$Q_0=\frac{\Ps_0}{\ov\b}$. In this way the $q_0$ field which can evolve due
to an evolution of~$Q$ and due to fluctuations of primordial \gr\ waves will
be a source of a gas of thermalized \pc s. It seems that an approach with a
boson equation of state is more appropriate to consider.

It is interesting to consider a fraction of a Dark Energy stored as boson
particles as a part of Dark Matter in the Universe.
One can consider an Einstein--Bose condensation of boson \pc s.

\section*{Appendix B}
\def\theequation{B.\arabic{equation}}
\setcounter{equation}0
\def\Beqno(#1){\label{#1}}
\def\Ker{\operatorname{Ker}\nolimits}
\def\esssup{\operatorname{ess~sup}\limits}
\def\Var{\operatorname{Var}\nolimits}
\def\N{\mathbb N}
\def\Tan{\operatorname{Tan}\nolimits}
\def\SL{\text{SL}}
\def\gote{\mathfrak e}

In the appendix we give a theory of an additional Dark Matter, appearing in
the \E\nos\ \KK (Jordan--Thiry) Theory. We use here the Friedrichs' theory in
order to get a mass spectrum for an infinite tower of scalar fields. We use
also a group representation theory to examine a spectrum of mass of scalar
\pc s.

Moreover it is interesting to consider a more general case with $\rho $
depending on $x\in E$ and $y\in M$ i.e.
\begin{equation}
\rho = \rho (x,y).\Beqno(5.39)
\end{equation}
In this case we expand $\rho $ into a complete set of real functions defined
on $M$.
\begin{equation}
\rho = \sum^{\infty }_{k=0}\rho _{k}(x) \chi _{k}(y)\Beqno(5.40)
\end{equation}
and consider such a $\rho $ that:
\begin{equation}
\|\widetilde{\rho }(x)\|^{2} = \sum^{\infty }_{k=0}\rho ^{2}_{k}(x) <
\infty.\Beqno(5.41)
\end{equation}
The condition \er{5.41} means that $\widetilde{\rho }(x) \in \ell ^{2}$
(a Hilbert space) $x\in E$ i.e.
$\ell ^{2}$--valued function defined on
$E$. If functions $\chi _{k}(y)$ defined on $M$ form
an orthogonal basis in $L^{2}(M, dm, \sqrt{|\widetilde{g}|})$ i.e.
\begin{equation}
\|\chi _{k}\|^{2} = \frac1{V_{2}}\int_{\rM}\sqrt{|\widetilde{g}|}
\chi ^{2}_{k}(y)\, dm(y) < \infty\Beqno(5.42)
\end{equation}
and
\begin{equation}
(\chi _{k}, \chi _{l}) = \frac{1}
{V_{2}}\int_{\rM}\sqrt{|\widetilde{g}|}\chi _{k}(y) \chi _{l}(y)\, dm(y)
= \|\chi _{k}\|^{2} \delta _{kl}\quad\ k,l = 0,1,2\ldots \Beqno(5.43)
\end{equation}
and for every real $f(y)$ such that
\bg5.44
\|f\|^{2} = \frac{1}{V_{2}}\int_{\rM}\sqrt{|\widetilde{g}|}
f^{2}(y)\, dm(y) < \infty,\\
f(y) = \sum^{\infty }_{k=0}f_{k} \chi_{k}(y),\label{5.45}\\
\sum^{\infty }_{k=0}f^{2}_{k} < \infty,\label{5.46}\\
f_{k} = \frac{1}{V_{2}}\int_{\rM}\sqrt{|\widetilde{g}|}
f(y) \chi _{k}(y)\, dm(y).\label{5.47}
\e
The convergence of \er{5.40}, \er{5.45} is understood in the sense of $L^{2}$
norm. Thus in this more general case we have to do with a tower of
scalar fields $\rho _{k}(x)$, $k=0,1,2,3\ldots $

We can arrange a basis $\chi _{k}(y)$ in such way that:
\begin{equation}
\chi _{0}(y) = 1.\Beqno(5.48)
\end{equation}
It is easy to see that this function is normalized to one:
\begin{equation}
\|\chi _{0}\|= 1.\Beqno(5.49)
\end{equation}
Using a Schmidt orthogonalization procedure one gets remaining $\chi
_{k}$, $k=1,2,\dots $ from generalized harmonics on $M$ such that:
\begin{equation}
(\chi _{k}, \chi _{l}) = \delta _{kl}.\Beqno(5.50)
\end{equation}
In this way the condition \er{4.38} means a truncation condition for $\rho $
i.e.
\begin{equation}
\rho (x) = \rho _{0}(x) + \sum^{\infty }_{k=1}\rho _{k}(x) \chi _{k}(y)
\Beqno(5.51)
\end{equation}
and we consider only the first term $\rho _{0}$
\begin{equation}
\rho (x) = \rho _{0}(x).\Beqno(5.52)
\end{equation}
The general treatment of $\rho $ is given below. 

The generalized harmonics on $(M,h^{0})$ are eigenfunctions of the usual
Beltrami--Laplace operator on $(M,h^{0})$ (which is an elliptic operator for
$M$ is compact and without boundaries).
\begin{equation}
{\varDelta} \eta _{k} = a_{k} \eta _{k},\Beqno(5.53)
\end{equation}
where $\eta _{k}$ is an eigenfunction of ${\varDelta} $ and $a_{k}$ is an eigenvalue, $k=0,1,2\ldots $
\begin{equation}
u{\varDelta} = \frac{1}{2} (d*d*+*d*d)\Beqno(5.54)
\end{equation}
is the Beltrami--Laplace operator on $M$, $*$---means a Hodge's star and $d$
is an exterior derivative on $M$, $u$ means a volume form on $M$ ($n_{1}$-form).
Sometimes we define $\delta = *d*$. One can write ${\varDelta} $ in a more convenient form:
\begin{equation}
{\varDelta} f = \ddiv(\grad(f)) = - \frac{1}{\sqrt{|h^0|}}
(h^{0\tilde a\tilde b}\sqrt{|h^0|}
f_{,\tilde b})_{,\tilde a},\Beqno(5.54a)
\end{equation}
where $h^{0} = \det (h^{0}_{\tilde a\tilde b})$.

It is easy to see that $a_{k} = 0$ corresponds to a constant function (a
manifold $M$ is compact).

Functions $\eta _{k}$ form a basis in a different Hilbert space of $L^{2}$-type i.e.
$\eta _{k} \in L^{2}(M,dm)$. Moreover $L^{2}(M,dm)$ and
$L^2(M,dm,\sqrt{|\widetilde{g}|})$ are unitary
equivalent since measures $d\mu _{1}=dm$ and $d\mu
_{2}=\sqrt{|\widetilde{g}|}\,dm$ are such that $\mu _{2}\ll\mu _{1}$
and $\mu _{1}\ll\mu _{2}$. The isomorphism
$$\displaylines{
{\mathbf U} : L^{2}(M,dm,\sqrt{|\widetilde{g}|}) \rightarrow
L^{2}(M,dm),\cr
f_{2} = {\mathbf U}(f_{1}) =|\widetilde{g}|^{1/4} f_{1}\cr}
$$
establishes this equivalence.

The truncation procedure can be obtained in a different way taking
\begin{equation}
\rho (x,y) = \rho _{0}(x) \rho _{1}(y) ,\quad\ x\in E ,\ y\in M.
\end{equation}
In this way after averaging over the manifold $M$ we get basically the
same shape of the lagrangian with exactly the same factors depending on
the scalar field $\rho _{0}(x)$.

Moreover it is more convenient to consider the
field ${\varPsi} $ in place of $\rho $.
\begin{equation}
\rho = e^{-{\varPsi} }.\Beqno(16.1)
\end{equation}
In this case one finds:
\begin{equation}
{\varPsi} (x,y) = \sum^{\infty }_{k=0}{\varPsi} _{k}(x) \chi
_{k}(y).\Beqno(16.2)
\end{equation}
Let us find a condition for the function $\rho (x,y)$ such that
${\varPsi} (x,y)$ is an
$L^{2}( M,dm,\sqrt{|\widetilde{g}|} )$-valued function of $x\in E$.
From \er{16.1} one finds that
$\ln (\rho )$ is an $L^{2}(M,dm,\sqrt{|\widetilde{g}|})$-valued function i.e.:
$$
\int_{M} (\ln (\rho ))^{2} \sqrt{|\widetilde{g}|} dm(y) <
\infty.\eqno(\text{\rm\ref{16.2}a})
$$
Thus the sufficient condition for $\rho $ is that $\rho $ and $\rho ^{-1}$ are bounded on
$M-A$, where $m(A)=0$. It means that $\widetilde{\rho }(x),
\widetilde{\rho }^{-1}(x) \in L^{\infty }(M,dm,\sqrt{|\widetilde{g}|})$
and $\|\widetilde{\rho }(x)\|_{\infty } =
\esssup_{y\in M}|\rho
(x,y)|< \infty $. The convergence $\widetilde{\rho }_{n}(x,\cdot )
\rightarrow \widetilde{\rho }(x,\cdot )$
with respect to the norm $\|\cdot \|_{\infty }$ means a uniform limit. We have the same
for $\widetilde{\rho }^{-1}(x)$. We can also consider a decomposition of $\rho (x,y)$ into a
series of generalized harmonics on $M$ i.e.
$$
\rho (x,y)\sim \rho _{0}(x) + \sum^{\infty }_{k=1}\rho _{k}(x) \chi _{k}(y)
\eqno(\text{\rm\ref{16.2}b})
$$
such that
$$
\begin{aligned}
\rho _{k}(x) &= \frac{1}{V_{2}}\int_{M}\sqrt{|\widetilde{g}|}\rho (x,y)
\chi _{k}(y) \,dm(y),\\
\rho _{0}(x) &= \frac{1}{V_{2}}\int_{M}\sqrt{|\widetilde{g}|}\rho (x,y)\,dm(y).
\end{aligned}\eqno(\text{\rm\ref{16.2}c})
$$
If ${\displaystyle\sum^{\infty }_{k=0}}\rho ^{2}_{k}(x) < \infty $, $x\in E$ the series
on the right-hand side of (\ref{16.2}b)
converges to the function $\rho (x,y)$ in
$L^{2}(M,dm,\sqrt{|\widetilde{g}|})$ sense. The limit
can be understood in a uniform sense if $\rho (x,y)$ is continuous and
$V(x)=\Var (\widetilde{\rho }(x)) < \infty $ in $M$ for every $x\in E$.

Moreover it is interesting to know properties of $\widetilde{{\varPsi}
}({x})$ if $\widetilde{\rho}({x})$ is an 
$L^{2}({M}{,}{dm}{,}\sqrt{|\widetilde{g}|})$-valued function. One gets:
$$
\|\widetilde{\rho }(x)\|^{2} \le V_{2} + \sum^{\infty }_{n=1}\frac{2^{n}}{n!}
(\|\widetilde{{\varPsi} }(x)\|_{n})^{n} < \infty,\eqno(\text{\ref{16.2}\rm d})
$$
where
\begin{equation}
\|{\varPsi} (x)\|_{n} = \Big(\int_{M}\sqrt{|\widetilde{g}|}
|{\varPsi} (x,y)|^{n} dm(y)\Big)^{1/n}
\end{equation}
is an $L^{n}(M,dm,\sqrt{|\widetilde{g}|})$ norm. Let us suppose that for
every $n\in {\N}^{\infty }_{1}$, $\widetilde{{\varPsi} }(x)$
is an $L^{n}(M,dm,\sqrt{|\widetilde{g}|})$-valued function. For
$$
V_{2} = \int_{M} \sqrt{|\widetilde{g}|} dm(y) < \infty
$$
one gets
$$
\|\widetilde{{\varPsi} }(x)\|_{n} \le \|\widetilde{{\varPsi} }(x)\|_{n' }
V^{1/n-1/n' }_{2},
$$
if $1 \le n \le n' \le \infty $ and
\begin{align*}
L^{\infty }(M,dm,\sqrt{|\widetilde{g}|}) &\subset
L^{n'}(M,dm,\sqrt{|\widetilde{g}|}) \subset
L^{n}(M,dm,\sqrt{|\widetilde{g}|})\subset\\
&\subset L^{1}(M,dm,\sqrt{|\widetilde{g}|}) =
L(M,dm,\sqrt{|\widetilde{g}|}).
\end{align*}
Thus if $n' \rightarrow \infty $ one obtains
$$
\|\widetilde{{\varPsi} }(x)\|_{n} \le \|\widetilde{{\varPsi} }(x)\|_{\infty }
V^{1/n}_{2} = (\esssup_{y\in M} |{\varPsi} (x,y)|)
V^{1/n}_{2}.\eqno(\text{\rm \ref{16.2}e})
$$
From (\ref{16.2}d) and (\ref{16.2}e) we have
$$
\|\widetilde{\rho }(x)\|^{2} \le V_{2} e^{2\|{\varPsi}
(\tilde{x})\|_{\infty}}.\eqno (\text{\rm \ref{16.2}f})
$$
Thus if $\widetilde{{\varPsi} }(x) \in L^{\infty
}(M,dm,\sqrt{|\widetilde{g}|}) \widetilde{\rho }(x)$ is an
$L^{2}(M,dm,\sqrt{|\widetilde{g}|})$-valued
function . This means that ${\varPsi} (x,y)$ is bounded on $M-A$ where
$m(A)=0$. In
this way we get an interesting duality. The sufficient condition for
$\widetilde{{\varPsi} }(x)$ to be an
$L^{2}(M,dm,\sqrt{|\widetilde{g}|})$-valued function is $\widetilde{\rho }(x),
\widetilde{\rho }^{-1}(x) \in
L^{\infty}(M,dm,\sqrt{|\widetilde{g}|})$ and the sufficient condition for $\widetilde{\rho }(x)$ to be an
$L^{2}(M,dm,\sqrt{|\widetilde{g}|})$ is $\widetilde{{\varPsi} }(x) \in
L^{\infty}(M,dm,\sqrt{|\widetilde{g}|})$. In the second case we can try
an expansion of ${\varPsi} (x,y)$ into a series of generalized harmonics similarly
as for $\rho$ (i.e.\ Eqs.~(\ref{16.2}b--c).

The $(n_{1}+4)$-dimensional lagrangian for the scalar field ${\varPsi} $ looks like:
\begin{equation}
{\cal L}_{\scal}({\varPsi} ) = (\ov{M}\widetilde{\gamma }^{(CM)}
{\varPsi} _{,C} {\varPsi} _{,M} + n^{2} \gamma ^{[MN]} \gamma _{DM}
\widetilde{\gamma }^{(DC)} {\varPsi} _{,N} {\varPsi} _{,C})\Beqno(16.3)
\end{equation}
or 
\begin{equation}
\aligned
{\cal L}_{\scal}({\varPsi} )
&= (\ov{M}\widetilde{g}^{(\gamma \mu )}
{\varPsi} _{,\gamma } {\varPsi} _{,\mu } + n^{2} g^{[\mu \nu ]}
g_{\delta \mu } \widetilde{g}^{(\delta \gamma )} {\varPsi} _{,\nu }
{\varPsi} _{,\gamma }) \\
&+ \frac{1}{r^{2}} (\ov{M}\widetilde{g}^{(\tilde c \tilde m )}
{\varPsi} _{,\tilde c }{\varPsi} _{,\tilde m }+n^{2}g^{[\tilde m
\tilde n ]}g_{\tilde d \tilde m }\widetilde{g}^{(\tilde d \tilde c )}
{\varPsi} _{,\tilde n }{\varPsi} _{,\tilde c })
=\widetilde{{\cal L}}_{\scal}({\varPsi} ) + Q({\varPsi}
)\endaligned
\Beqno(16.4)
\end{equation}
Moreover we should average over $M$ and $H$.

Thus one gets using (\ref{16.2}), \er{16.4}:
\beq 16.5
\frac{1}{V_{1}V_{2}}\int_{M} ({\cal L}_{\scal}({\varPsi} )+Q({\varPsi} )
)\sqrt{|\widetilde{g}|} \sqrt{|\ell|} dm(y) d\mu _{H}(h) =
 \sum^{\infty }_{k=0} {\cal L}_{\scal}({\varPsi} _{k}) + \frac{1}{2}
\sum^{\infty }_{k,l=1} \widehat{M}_{kl} {\varPsi} _{k} {\varPsi}_{l},
\e
where ${\cal L}_{\scal}({\varPsi} _{k})$ is a usual
lagrangian for the scalar field ${\varPsi} $ in our theory and
\beq 16.6
\frac{1}{2} \widehat{M}_{kl} = \frac{(-1)}{V_{2}r^{2}}
\int_{M}(\ov{M}\widetilde{g}^{(\tilde c \tilde m )} +
n^{2}g^{[\tilde m \tilde n ]} g_{\tilde d \tilde m }
\widetilde{g}^{(\tilde d \tilde c )}) \chi _{k,\tilde c }
\chi _{l,\tilde n }\sqrt{|\widetilde{g}|}\,dm(y) < \infty
\e
and $\widehat{M}_{kl} = \widehat{M}_{lk} $, $k,l=1,2,\dots $

It is easy to see that we get a tower of fields ${\varPsi} _{k}$ such that according
to \er{16.6} the field ${\varPsi} _{0}$ is massless and remaining fields are
massive with a scale of a mass ${\displaystyle\frac{1}{r}}$.

The infinite dimensional quadratic form
$$
\frac{1}{2} \sum^{\infty }_{k,l=1}\widehat{M}_{kl} {\varPsi} _{k}(x)
{\varPsi} _{l}(x)
$$
is convergent in $\ell ^{2}$ for every $x$. This is easily satisfied if the
derivatives of $\chi _{k}$ belong to\break
$L^{2}(M,dm,\sqrt{|\widetilde{g}|})$ and $g_{\tilde{a} \tilde{b} }$ are continuous
functions on $M$, which is always satisfied for our case $(M$ is compact
without boundaries and $g_{\tilde{a} \tilde{b} }$ defined in Section~4).
For $L^{2}(M,dm,\sqrt{|\widetilde{g}|})$
is unitary equivalent to $L^{2}(M,dm)$ we can consider $\chi _{k}$ eigenfunctions
of Beltrami--Laplace operator on $M$.

In some cases we can diagonalize the infinite dimensional matrix $\widehat{M}_{kl}$
getting some new fields 
\begin{equation} 
{\varPsi}'_{k'}(x) = \sum^{\infty }_{k=1}A_{k'k} {\varPsi} _{k}(x),
\Beqno(16.7)
\end{equation}
such that 
\begin{equation}
b_{k} \delta _{kl} = \sum^{\infty }_{k',l'=1}A_{kk'} A_{ll'}
\widehat{M}_{k'l'},\Beqno(16.8)
\end{equation}
where $A = (A_{k'k})$ is a unitary operator in $\ell ^{2}$ space i.e.
\begin{equation} 
\|Aa\|= \|a\|\Beqno(16.9)
\end{equation}
and $A(\ell ^{2}) = \ell ^{2}$.

The diagonalization procedure can be achieved if $\widehat{M}_{kl}$ is a symmetric
unbounded (Hermitian) operator in $\ell ^{2}$. This is equivalent to
\begin{equation}
\widehat{M}_{kl} = \widehat{M}_{lk}\Beqno(16.10)
\end{equation}
and $M(\ell ^{2}) = \ell ^{2}$.
Thus we get for the lagrangian:
\begin{equation}
{\cal L}_{\scal}({\varPsi} ) = \sum^{\infty }_{k=0}{\cal L}_{\scal}
({\varPsi}'_k) - \sum^{\infty }_{k=1} \frac{b_{k}}{2}
({\varPsi}'_{k})^{2},\Beqno(16.11)
\end{equation}
$\bigg({\displaystyle\frac{b_{k}}{\ov{M}}}\bigg)$ have an interpretation as
$m^{2}_{k}$ for ${\varPsi}'_k $. If $\widehat{M}_{kl}$ is a negatively
defined operator in $\ell ^{2}$ we have for every $k\ b_{k} \ge 0$ if $M_{kl}$ is invertible
$b_{k} > 0$ for every $k$. The latest condition means that every
${\varPsi}'_k $ is
massive with ${\displaystyle\frac{1}{r}}$ as a scale of the mass. Let us write down a
self-interaction term for ${\varPsi}'_k $ coming from cosmological terms. One gets:
\begin{equation}
V(\{{\varPsi}'_k\}) =
\frac{1}{V_{2}} \int_{M}\sqrt{|\widetilde{g}|}
dm(y) \bigg(\frac{\alpha ^{2}_{s}}{\ell ^{2}\pl} \widetilde{R}
(\widetilde{{\varGamma} }) e^{(n+2){\varPsi} (x,y)} +
\frac{e^{n{\varPsi} (x,y)}}{r^{2}} \widehat{\ov{R}}
(\widehat{\overline{\varGamma}})\bigg),\Beqno(16.12)
\end{equation}
where
\begin{equation}
{\varPsi} (x,y) = {\varPsi} _{0}(x) + \sum^{\infty }_{k=1}
{\varPsi} _{k}(x) \chi _{k}(y)\Beqno(16.13)
\end{equation}
and
\begin{equation}
{\varPsi}'_k = \sum^{\infty }_{k=1}A_{kk'} {\varPsi} _{k'}.\Beqno(16.14)
\end{equation}
The field ${\varPsi} _{0}(x)$ can get a mass from a cosmological background
(see Refs \cite{5, xx}) (or in a way described in Section~4). If we
suppose ${\varPsi} = {\varPsi} (x)$ it means really that ${\varPsi} (x) = {\varPsi} _{0}(x)$ i.e. a truncation.
In this more general case for the field $\rho$ (or ${\varPsi} )$ one can easily get
similar formulae for the remaining part of the lagrangian involving
this field i.e. averaging over the manifold $M$.

All the factors in front of lagrangian terms are exponential functions
of ${\varPsi} $. The sufficient condition to make all the integration over $M$
convergent is to suppose that $\widetilde{{\varPsi} }(x) \in
L^{2}(M,dm,\sqrt{|\widetilde{g}|}) \cap L^{\infty
}(M,dm,\sqrt{|\widetilde{g}|})$ i.e. $\widetilde{{\varPsi} }(x)$ is $L^{\infty }$-valued function.

In the case of a completely broken group $G$ we should put $G$ for
$M$.

Let us give a more rigorous justification for an intuitive procedure
concerning diagonalization of the infinite matrix $\widehat{M}$.

Let us consider a bilinear form in
$L^{2}(M,dm,\sqrt{|\widetilde{g}|}) = L^{2} \simeq L^{2}(M,dm)$
\begin{equation}
\widehat{M}(f,g) = \frac{1}{V_{2}r^{2}} \int_M (\ov{M}
\widetilde{g}^{(\tilde c \tilde m )}+n^{2}g^{[\tilde m (\tilde n ]}
g_{\tilde d \tilde n }\widetilde{g}^{(|\tilde d |\tilde c ))})
f_{,\tilde c } g_{,\tilde m }\sqrt{|\widetilde{g}|}
dm(y),\Beqno(16.15)
\end{equation}
for $f,g \in L^{2}(M,dm,\sqrt{|\widetilde{g}|}) \cap {\C}^{1}(M)$.

This form is always defined for $f$ and $g$, because $g_{\tilde{a}
\tilde{b} }$ are smooth
functions on a compact manifold and $f_{,\tilde c } , g_{,\tilde m }
\in L^{2}(M,dm,\sqrt{|\widetilde{g}|})$ (if
they exist). ${\C}^{1}(M)$ is linearly dense in
$L^{2}(M,dm,\sqrt{|\widetilde{g}|})$. Let us
suppose that the tensor $P^{(\tilde c \tilde m )}$, where
$$
P^{\tilde c \tilde m } = (\ov{M}\widetilde{g}^{\tilde c \tilde m } +
n^{2} g^{[\tilde m \tilde n ]} g_{\tilde d \tilde n }
\widetilde{g}^{(\tilde d \tilde c )})
$$
is invertible and negatively defined.

One notices that $P^{\tilde c \tilde m }$ is $G$-invariant on $M$.
The same is of course true
for $P_{(\tilde c \tilde m )}$ being an inverse tensor of
$\widetilde{P}^{(\tilde c \tilde m )}$. Moreover every
$G$-invariant metric is induced by a scalar product $\<.,.\>$ on
$\fg /\fg _{0} = \fm$
which is invariant under the action of $\Ad_{G_0}$ on $\fm$. Thus $P$ is induced by
such a scalar product. The tensor $P_{\tilde c \tilde m }$ induces on $M$ an additional
Einstein--Kaufmann geometry compatible with it. We can repeat some
consideration concerning symmetric and skewsymmetric forms on $M$ from
section~4. If $P_{[\tilde d \tilde{b} ]}$ is not degenerate and
$\widehat{\overline{\nabla}}_{\tilde{a} }P_{[\tilde d \tilde{b} ]} = 0$ we can get an
almost complex structure on $M$ induced by $P_{\tilde c \tilde m } =
P_{(\tilde c \tilde m )}+ P_{[\tilde c \tilde m ]}$ (if $\dim M$ is
even).

Let us consider a linear functional on $L^{2} \cap {\C}^{1}(M) = D$
\begin{equation}
F_{f}(g) = \widehat{M}(f,g) ,\quad\ g,f \in D \subset L^{2},\Beqno(16.16)
\end{equation}
$F_{f} \in (L^{2})^{*} = L^{2}$ (Riesz theorem).
Thus for every $f$ we have a functional $F$.

Thus we can define a linear operator on $L^{2}$ such that
\begin{equation}
L(f) = F_{f} ,\hbox{ or }(Lf,g) = \widehat{M}(f,g).\Beqno(16.17)
\end{equation}
It is easy to see that if $f_{1} = f_{2}$ then $F_{f_{1}} = F_{f_{2}}$. Thus $L$ is well
defined. Simultaneously we find that $\Ker L = \text{\rm
 constant}$ functions on $M \cong
{\R}^{1}$. This is a subspace of $L^{2}$. Thus we can consider $L$ a quotient space
of ${\cal H}$ and denote it $\widehat{L}$, where
\begin{equation}
{\cal H} = L^2/(\Ker L).\Beqno(16.18)
\end{equation}
Let $F_{f_{1}} = F_{f_{2}}$ then one gets
\begin{equation}
\bigwedge _{g\in L^2-(\Ker L)} \widehat{M}(f_{1},g) =
\widehat{M}(f_{2},g).\Beqno(16.19)
\end{equation}
Using \er{16.15} one gets that $f_{1,\tilde c } = f_{2,\tilde c }$ {\it
modulo}\/ a set of a zero
measure. Thus $[f_{1}] = [f_{2}] \in {\cal H}$. One can easily check that $\widehat{L}$ is
symmetric because $\widehat{M}(f,g) = \widehat{M}(g,f)$.

The operator $\widehat{L}$ is unbounded on $D$ and it is negatively strictly defined
i.e.
\begin{equation}
(-\widehat{L}(f),f) \ge C\|f\|^{2},\Beqno(16.20)
\end{equation}
$C$ is a positive constant.

For this we can apply Friedrichs' theory (Kurt O. Friedrichs (1901--1982)) for $\widehat{L}$. We denote the
Friedrichs' space for $\widehat{L}$, $L^{2}_{0}(M)$ and the scalar product in $L^{2}_{0}(M)$ is
given by :
\begin{equation}
(f,g)_{\rF} = (-\widehat{L}f,g),\Beqno(16.21)
\end{equation}
($D/(\Ker L)$ is dense in $L^{2}_{0}(M)$ in a sense of the norm $\|\cdot
\|_{F}$). Let $\widetilde{L}=\overline{\widehat{L}}$ (a
closure of $\widetilde{L})$ denotes a Hermitian extension of $\widehat{L}$
in $L^{2}_{0}(M)$, which
exists due to Friedrichs' theorem and it is invertible. Moreover $\widetilde{L}^{-1}$
is compact. $\wh L$ is called also a Friedrichs' extension (see Ref.~\cite{aaal, m7b}) in
distinction to a different extension called Krein--von Neumann extension, which we
do not consider here.

For this we can proceed a spectral decomposition of
$\widetilde{L}$. Such an
operator has a pure point unbounded spectrum on a negative part of a
real axis and eigenvectors (eigenfunctions) span $L^{2}_{0}(M)$.
One gets
\begin{equation}
\widetilde{L} = - \sum^{\infty }_{k=1}\mu ^{2}_{k} P_{k} ,\quad\
\mu _{k}\neq 0 ,\ \mu _{k}\rightarrow \infty ,\Beqno(16.22)
\end{equation}
where $P^{2}_{k} = 1$, $P_{k} P_{l} = \delta _{kl}I$ are projective operators in $L^{2}_{0}(M)$. The sum
is understood in a weak topology at a point.

The above considerations justify an intuition of a diagonalization
procedure for an infinite matrix $\widehat{M}_{kl}$.

The existence of the tower of scalar fields in our theory can help
in renormalization problems. The fields can work as regulator fields
if we change some parameters in our theory.

The Friedrichs' theory gives us the following properties of
eigenvectors for~$\widetilde{L}$,
\begin{equation}
\widetilde{L} \zeta _{k} = -\mu ^{2}_{k} \zeta _{k}\quad\ k=1,2,\dots
\Beqno(16.23)
\end{equation}
They are normalized and orthogonal:
\bea 16.24
\int_{M} \zeta ^{2}_{k}(y) \sqrt{|\widetilde{g}|} dm(y) &=1, \\
\int_{M} \zeta _{k}(y) \zeta _{l}(y) \sqrt{|\widetilde{g}|} dm(y) &=
0\quad \hbox{if}\ k\neq l. \label{16.25}
\e
If the spectrum is degenerated i.e. $\mu ^{2}_{k}$ is repeated $q$ times $(q$ is a
natural number)
\begin{equation}
\mu ^{2}_{k} = \mu ^{2}_{k+1} = \ldots = \mu
^{2}_{k+q-1},\Beqno(16.26)
\end{equation}
then $\zeta _{k}$, $\zeta _{k+1}$, \dots , $\zeta _{k+q-1}$ form an algebraic basis of a finite
dimensional linear space of functions $f \in L^{2}_{0}(M) \cap {\C}^{2}(M)$ satisfying
Eq.~\er{16.26}.  After the  orthogonalization  procedure  of
$\{\zeta _{k},\zeta _{k+1},\dots ,\zeta _{k+q-1}\}$ (for example Schmidt procedure) we can
define projectors:
$$
P_{k} = P_{\zeta _{k}}\hbox{ for every }k=1,2,\dots
$$
The orthonormal set of eigenfunctions $\zeta _{k}$, $k=1,2,\dots $ is complete in
$L^{2}_{0}(M)$. Thus
\begin{equation}
I = \sum^{\infty }_{k=1}P_{k} = \sum^{\infty }_{k=1}P_{\zeta _{k}},
\Beqno(16.27)
\end{equation}
where $I$ is an identity operator in $L^{2}_{0}(M)$. Thus we can expand the field
${\varPsi} $ into a complete set of functions $\zeta _{k}$
\beq 16.28
{\varPsi} (x,y) = {\varPsi} _{0}(x) + \sum^{\infty }_{k=1}{\varPsi} _{k}
(x) \zeta _{k}(y),
\e
such that
\beq16.29
\sum^{\infty }_{k=1}{\varPsi} ^{2}_{k}(x) < \infty ,
\e
for $x\in E$.

Let us find conditions of a diagonalization of the infinite matrix $\widehat{M}_{kl}$.
It means we are looking for conditions of the existence of the unitary
transformation from the basis $\chi _{k}$, $k=1,2,\dots $ to the basis $\zeta
_{k}$, $k=1,2,\dots $ in
${\cal H}$ (or in $L^{2}_{0}(M))$. Later we give those conditions.

For $\{\zeta _{k}\}$ is complete one always gets
\begin{equation}
\chi _{k} = \sum^{\infty }_{l=1}A_{kl} \zeta _{l}\quad\ \hbox{for every }k
\Beqno(16.30)
\end{equation}
and
\begin{equation}
\sum^{\infty }_{l=1}A^{2}_{kl} < \infty.\Beqno(16.31)
\end{equation}
Moreover one has
\begin{equation}
A_{kl} = (\chi _{k}, \zeta _{l}).\Beqno(16.32)
\end{equation}
For the same reasons we have
\bg16.33
\zeta _{k} = \sum^{\infty }_{l=1}B_{kl} \chi
_{l},\\
\sum^{\infty }_{l=1}B^{2}_{kl} < \infty \label{16.33'}
\e
and
\begin{equation}
B_{kl} = (\zeta _{k}, \chi _{l}).\Beqno(16.34)
\end{equation}
One gets
\begin{equation}
A_{kl} = B_{lk}\Beqno(16.35)
\end{equation}
and
\begin{equation}
\chi _{k} = \sum^{\infty }_{l=1}A_{kl} \sum^{\infty }_{p=1}B_{lp} \chi _{p} =
\sum^{\infty }_{p=1}\Big(\sum^{\infty }_{l=1} A_{kl} B_{lp}\Big)\chi
_{p}.
\Beqno(16.36)
\end{equation}
This is possible only if
\begin{equation}
\sum^{\infty }_{l=1}A_{kl} B_{lp} = \sum^{\infty }_{l=1}A_{kl} A_{pl} = \delta
_{kp}.\Beqno(16.37)
\end{equation}
Thus $A_{kl}$ is an isometry in the $\ell ^{2}$-space and $B_{kl}$ either. Moreover $A_{kl}$
and $B_{kl}$ are invertible and A is an inverse operation of $B$. This means
A is an unitary transformation. The transformation A diagonalizes the
infinite dimensional matrix $\widehat{M}_{kl}$. The only one condition for the
existence of $A$ is the following. The set $\{\chi _{k}\}$ or $\{\eta _{k}\}$ are complete
bases in $L^{2}_{0}(M)$.

Let us consider the operator $\widetilde{L}$ in more details and find its shape for
$f\in {\C}^{2}(M)/\break \Ker L)$. The operator $\widetilde{L}$ can be considered a Gateaux derivative of
the quadratic form $\widehat{M}$ in $L^{2}_{0}(M)$. One finds:
\beq16.38
\widetilde{L}f = {1\over \sqrt{|\widetilde{g}|}} (\sqrt{|\widetilde{g}|}
\ov{p}^{\tilde{a} \tilde{b} }f_{,\tilde{a}})_{,\tilde{a} }
= (\ln \sqrt{|\widetilde{g}|}_{,\tilde{b} }\ov{p}^{\tilde{a}
\tilde{b} } f_{,\tilde{a} } + \ov{p}^{\tilde{a} \tilde{b}
}{}_{,\tilde{b} } f_{,\tilde{a} } + \ov{p}^{\tilde{a} \tilde{b} }
f_{,\tilde{a} ,\tilde{b} }),
\e
where 
\begin{equation}
\ov{p}^{\tilde c \tilde m } =\ov{p}^{\tilde m \tilde c }
=\bigg[
\widetilde{g}^{(\tilde c \tilde m )} + {n^{2}\over 2} ( g^{[\tilde m
\tilde n ]} g_{\tilde d \tilde n } \widetilde{g}^{(\tilde d \tilde c )}
+ g^{[\tilde c \tilde n ]} g_{\tilde d \tilde n }
\widetilde{g}^{(\tilde d \tilde m )})\bigg]\Beqno(16.39)
\end{equation}
The operator $\widetilde{L}$ slightly differs from the Beltrami--Laplace operator
defined on $(M,\ov{p})$.
$$
\ov{p}= \ov{p}_{\tilde{a} \tilde{b} } \theta ^{\tilde{a} } \otimes
\theta ^{\tilde{b} },\quad\ \ov{p}_{\tilde{a} \tilde{b} }
\ov{p}^{\tilde{a} \tilde c } = \delta ^{\tilde c }_{\tilde{b} }
$$
because $\det (\ov{p}_{\tilde{a} \tilde{b} } ) \neq \det
(g_{\tilde{a} \tilde{b} }) = \widetilde{g}$.
It differs also from the Beltrami--Laplace operator on $(M,h^{0})$.

Moreover $\widetilde{L}$ is a left-invariant operator on $C^{2}(M)$ of the second order.
Supposing the reductive decomposition of $\fg=\fg_{0} \dot{+}\fm$ we have a
complete description of an algebra of $G$-invariant operators on $M$,
$D(G/G_{0})$ in terms of Lie algebra $\fg$ and $\fg_{0}$. The algebra is
commutative if $M$ is a
symmetric space. Let $M$ be a symmetric space (of compact type of course)
and let rank $(M) = K$. The algebra $D(G/G_{0})$ has finite numbers of
generators i.e.\ $D_{1}, D_{2},\dots, D_{K}$. In this case every
$D \in D(G/G_{0})$ is a
symmetric polynomial of $D_{i}$, $i = 1,2,\dots$ $K, D = W(D_{1}, D_{2},\dots
D_{K})$. Notice that degrees of $D_{i}$, $i = 1,2,\dots K$ $d_{i} = 1,2,\dots
K$ are canonically
established by $G$.

Thus $\Delta= W_{1}(D_{1},\dots
D_{K})$ and $\widetilde{L} = W_{2}(D_{1}\ldots D_{K},\zeta )$ and $W_{1} \neq W_{2}$, $W_{1}(\ldots
) = W_{2}(\ldots ,0)$.

The existence of the unitary operator $A$ in $\ell ^{2}$-space means that $\widehat{L}$ and
$\widehat{\Delta }=\Delta |_{{\cal H}}$ have the same Friedrichs' theory i.e. the same $L^{2}_{0}(M)$. For $M$ is
compact without boundaries and $h^{0}{}_{\tilde{a} \tilde{b} }
\theta ^{\tilde{a} } \otimes \theta ^{\tilde{b} }$,
$\widetilde{g}_{\tilde{a} \tilde{b} } \theta ^{\tilde{a} } \otimes
\theta ^{\tilde{b} }$, $p_{\tilde{a} \tilde{b} } \theta ^{\tilde{a} }
\otimes \theta ^{\tilde{b} }$ are smooth functions on $M$ one gets:
\begin{equation}
\ov{m}_{2}(-\widehat{\Delta }f,f) \le (-\widehat{L}f,f) \le \ov{m}_{1}
(-\widehat{\Delta }f,f),\Beqno(16.40)
\end{equation}
$\ov{m}_{1}$, $\ov{m}_{2}$ are positive constants.

This means that $L^{2}_{0}(M)$ for $\widehat{\Delta }$ and $\widehat{L}$ are equivalent and the conditions
we need to imposed for $g_{\tilde{a} \tilde{b} }$ and $p_{\tilde{a}
\tilde{b} }$ are
$$
\det ( g_{\tilde{a} \tilde{b} } ) \neq 0\quad \hbox{and}\quad \det
(\ov{p}_{\tilde{a} \tilde{b} }) \neq 0.
$$
These are the only conditions for the existence of the unitary operator
A in $\ell ^{2}$-space.

Let us consider $M = S^{2}$ with the nonsymmetric tensor (see Ref.~\cite5). One gets
\begin{equation}
\widetilde{L} f = \bigg(\ov{M} + {n^2\zeta ^2\over (\zeta^2+1)}\bigg)
\Delta f,\Beqno(16.41)
\end{equation}
where $\Delta$ is an ordinary Beltrami--Laplace operator on $S^{2}$
in spherical
coordinates. In this case we do not need any procedure described above,
because eigenfunction for $\widetilde{L}$ are simply spherical functions $Y_{\ell m}(\theta ,\varphi )$ and
\begin{equation}
\widetilde{L} Y_{\ell m} = - \bigg(\ov{M} + {n^2\zeta ^2\over
(\zeta^2+1)}\bigg)\ell (\ell + 1)Y_{\ell m}.\Beqno(16.42)
\end{equation}
The mass spectrum is degenerated for $m = -\ell $, $-(\ell - 1)\ldots
 0\ldots \ell - 1,\ell $, $\ell = 1,2,3,\dots $ and
\begin{equation}
\ov{m}(\ell ,m) = {1\over r}\sqrt{\bigg(1+{\displaystyle{n^2\zeta ^2\over
\ov{M}(\zeta^2+1)}}\bigg)\ell(\ell+1)},\quad\ \ell = 1,2,3.\Beqno(16.43)
\end{equation}
Let us notice the following fact: if the homogeneous space $M = G/G_{0}$ is a
two-point homogeneous space (see Ref.~\cite{24})
the operator $\widetilde{L}$ is proportional to the
Beltrami--Laplace operator on $M$. This fact is coming from
left-invariancy of $\widetilde{L}$ and from this fact that
$\widetilde{L}$ is of the second order
differential operator for $f \in C^{(2)}(M)$. In general this is not true.
Moreover for $\zeta = 0$
\begin{equation}
\widetilde{L} = \ov{M}\Delta.\Beqno(16.44)
\end{equation}
Thus $\ov{m}(\ell ,m) \simeq {1\over
r}\sqrt{(\ell +1)\ell}$. We have the following
two-point homogeneous spaces $\SU(p+1)/S(U(1)\times U(p))$, $\SO(n +
1)/\SO(n)$, $(f_{u(-52)}$, so(9)) of compact type, which can serve as the manifold $M =
G/G_{0}$. Only the first one is Hermitian (K\"ahlerian) and in this case one
gets
\begin{equation}
\widetilde{L}= \bigg(\ov{M} + {n^2\zeta ^2\over
(\zeta^2+1)}\bigg)\Delta\Beqno(16.45)
\end{equation}
the same formula as for $S^{2}$. Moreover for every two-point symmetric
space $\widetilde{L}$ is proportional to $\Delta$ (Beltrami--Laplace operator)
\begin{align}
\widetilde{L} &= \fg (\zeta )\Delta, \tag{\ref{16.45}\text{\rm a}}\\
\fg(0) &= \ov{M}. \tag{\ref{16.45}\text{\rm b}}
\end{align}
It is interesting to find $L^{2}_{0}(M)$ in terms of irreducible spaces of
representations of the group $G$. In the case of $S^{2} =
\SO(3)/\SO(2)$ one simply gets
\begin{equation}
L^{2}{}_{0}(M) =\sum^{\infty }_{i=1}\oplus H^{\ell },\Beqno(16.46)
\end{equation}
where $H^{\ell }$ is space of an irreducible representation of the group SO(3)
$$
\dim H^{\ell } = 2\ell + 1.
$$
In this case a dynamical group of $\widetilde{L}$ is SO(3,1) (the so called spectrum
generation group)\break SO(3)$\subset $SO(3,1).

${\displaystyle\sum^{\infty }_{i=0}}\oplus H^{\ell } = L^{2}(M)
=L^{2}{}_{0}(M) \oplus R$ is a representation space of a unitary
representation of SO(3,1).

In general the situation is more complex.

Let $G$ be a simple compact Lie group and let $G_{0}$ be its closed subgroup
such that it is a semisimple or $G_{0} = U(1)\otimes G_{0}$, where $G_{0}$ is semisimple.

Let $C = h_{ab}Y^{a}Y^{b}$ be a Casimir operator of $G$ and let $C_{0}$ be a Casimir
operator of $G_{0}$ (or $G_{0}'$). Let us suppose a reductive decomposition of
the Lie algebra $\fg$, $\fg = \fg_{0} \dot{+} \fm$ (or $\fg_{0}{}'$) and consider an operator
\begin{equation}
- D = C - C_{0} = h_{\tilde{a} \tilde{b} }Y^{\tilde{a} }Y^{\tilde{b}
}.\Beqno(16.47)
\end{equation}
This operator acts in the complement $\fm$, which is diffeomorphic to
$\Tan_{0}(G/G_{0}) $, $0 = \varphi (\varepsilon )$.

Let us define a left-invariant differential operator $\Delta$ on $M$
corresponding to $D$, via a pull-back of the left action of the group $G$
on $M$. Let us find eigenfunctions of this operator and its eigenvalues.
This can be done as follows. Let $H_{\bar{\lambda}_k}$ be invariant spaces of
irreducible representations of $G$ corresponding to the value of the
Casimir operator, $C$, $\overline{\lambda }_{k}$ and $H_{\bar{\mu }_{\ell }}$ be
invariant spaces of irreducible
representations of $G_{0}$ corresponding to the value of the Casimir
operator $C_{0}$, $\overline{\mu}_{\ell }$. Both groups are compact and those representations
are finite-dimensional and unitary. All invariant spaces $H_{\bar{\lambda}_k}$
and $H_{\bar{\lambda}_{\ell }}$ are Hilbert spaces. One gets
\begin{equation}
L^{2}(M) = \sum_{\bar{\lambda}_k} \oplus H_{\bar{\lambda}_k} =
\sum_{\bar{\lambda}_k} \oplus \sum_{\bar{\mu}_{\ell } \in \bar{\lambda}_k}
\oplus b(\bar{\lambda}_k,\bar{\mu}_{\ell })H_{\bar{\mu}_{\ell }},\Beqno(16.48)
\end{equation}
where $\overline{\mu}_{\ell } \in \overline{\lambda}_k$, means that $H_{\bar{\lambda}_k}$
is decomposed into some irreducible representation spaces $\overline{\mu}_{\ell }$
of the subgroup of $G$, $G_{0}$. $b(\overline{\lambda}_k,\overline{\mu}_{\ell })$
means a multiplicity of $H_{\bar{\mu}_{\ell }}$ in $H_{\bar{\lambda}_k}$.
Thus one gets
\begin{equation}
\Delta f_{k,\ell ,\ell '} =\overline{\eta} (k,\ell )f_{k,\ell ,\ell '},
\Beqno(16.49)
\end{equation}
where
\begin{equation}
\overline{\eta}(k,\ell ) = \overline{\lambda}_k - \overline{\mu}_{\ell } < 0,\Beqno(16.50)
\end{equation}
$f_{k,\ell ,\ell '} \in L^{2}(M)$ and $f_{k,\ell ,\ell '} \in H^{(\ell ')}
_{\bar{\mu}_{\ell }}$, $\overline{\mu}_{\ell } \in \overline{\lambda}_k$.
$$
H^{1}_{\bar{\mu}_{\ell }} \oplus \ldots \oplus H^{\ell '}\oplus \ldots
\oplus H^{b(\bar{\lambda}_k,\bar{\mu}_{\ell })}_{\bar{\mu}_{\ell }}
\subset H_{\bar{\lambda}_k}.
$$
The dimension of the space corresponding to $\eta _{(k,\ell )}$ can be easily
calculated.
\begin{equation}
\dim H_{\bar{\eta} (k,\ell )} = b(\overline{\lambda}_k,\overline{\mu}_{\ell })\dim
H_{\bar{\mu}_{\ell }}.\Beqno(16.51)
\end{equation}
Thus $\overline{\eta}(k,\ell )$ is in general degenerated. The most important result is
this that
\begin{equation}
L^{2}(M) =\sum^{}_{\bar{\lambda }_{k}}\oplus H_{\bar{\lambda
}_{k}},\quad\ \overline{\lambda }_k\ne 0\Beqno(16.52)
\end{equation}
or
$$
L^{2}(M) = L^{2}{}_{0}(M)\oplus H_{0},\quad\ \dim H_{0} = 1,\ H_{0} \simeq
R',\eqno(\text{\rm\ref{16.52}a})
$$
or
$$
L^{2}(M) = L^{2}{}_{0}(M)/R'.\eqno(\text{\rm\ref{16.52}b})
$$
In the case of a $U(1)$ factor we get similarly
\beq16.53
\Delta f_{(k,\ell ,\ell ',m)} =\overline{\eta}(k,\ell )f_{(k,\ell ,\ell',m)},
\e
where
\begin{equation}
f_{(k,\ell ,\ell ',m)} = f_{k,\ell ,\ell '}\fg_{m},\quad\ m = 0,\pm 1,\pm 2,\dots,
\Beqno(16.54)
\end{equation}
$\fg_{m}$ is a function of a one-dimensional irreducible representation of
$U(1)$, $m = 0$, $\pm 1$, $\pm 2,\dots $. Thus the construction of eigenfunctions
of $\Delta$ is known
from representation theory of $G$ and $G_{0}$ ($G'_0$) and the spectrum is known
as well. The case with $U(1)$ factor is important in our treatment of GSW model (see Ref.~\cite{11a}).
The interesting problem which arises here is as follows. What
is the group (noncompact in general) for which $L^{2}(M)$ $(L^{2}_{0}(M))$ is the
invariant space of a unitary, irreducible representation. In other
words what is an analogue of SO(3,1) for $\Delta$ on $S^{2}$ in a general case.
This group (if exists) we call dynamical group of $\Delta$, $\widehat{G}$ or a spectrum
generating group. We suppose that such a group is minimal for the above
requirement. In this way $G$ must be maximally compact subgroup of this
group, $G \subset \widehat{G}$. Thus $L^{2}(M)$ is a space of a unitary representation of
$\widehat{G}$ (which is up to now unknown) $T: \widehat{G} \to L(L^{2}(M),L^{2}(M))$ such that
\begin{equation}
T_{|G} = \sum^{}_{\lambda _{k}} \oplus H_{\bar{\lambda }_{k}},\Beqno(16.55)
\end{equation}
where the sum is over all the irreducible representation of $G$ with
multiplicity equal to one. Such an representation of $\widehat{G}$, $T$ is called
maximally degenerated (or most degenerated). Such a situation is
possible only if $G$ is a maximal compact subgroup of $\widehat{G}$. 

Thus the dynamical group $\widehat{G}$ for $\Delta$ is defined as follows:

\begin{itemize}
\item[1.] $G \subset \widehat{G}$ and is a maximal compact subgroup
$\widehat{G}$ ($\widehat{G}$ is noncompact, of course)

\item[2.] $\widehat{G}/G$ is a symmetric irreducible space of noncompact type, such that
for a given $G$ its rank is minimal.

\item[3.] The maximally degenerated, unitary, irreducible, infinite
dimensional representation of $\widehat{G}$, $T$ restricted to $G$ is equivalent to
the simple sum of all irreducible representation of the group $G$
(which are finite dimensional).
\end{itemize}

\noindent In our case $G$ is simple and compact. Thus we have the following
possibilities:
$$\vbox{
\halign{\strut\tabskip10pt#\hfil&\hfil#\hfil&#\hfil\tabskip0pt\cr
\omit\hfill {\bf Space}\hfill &&\omit\hfill {\bf Rank}\hfill\cr
$\SL(n,R)/\SO(n)$&,&$n-1$\cr
($\gote_{\text{\bf 6}(\text{\bf 6})},sp(4))$&,&6\cr
($\gote_{\text{\bf 6}(-\text{\bf 26})},\mathfrak f_{\text{\bf 4}})$&,&2\cr
($\gote_{\text{\bf 7}(\text{\bf 7})},su(8))$&,&7\cr
($\gote_{\text{\bf 8}(\text{\bf 8})},so(16))$&,&8\cr
($f_{\text{\bf 4}(-\text{\bf 20})},so(9))$&,&1\cr}}
$$
Except the listed above there are also some additional.
$$\vbox{\offinterlineskip\tabskip10pt plus5pt minus5pt
\halign {#\hfil&\hfil#\hfil&#\hfil\tabskip0pt\cr
SO$_{0}(p,1)$/SO$(p)$&,&1\cr}}
$$
SO$_{0}(3,1)$ is a Lorentz group and this is our example with $S^{2}$.
$$\vbox{\offinterlineskip\tabskip10pt plus5pt minus5pt
\halign {#\hfil&\hfil#\hfil&#\hfil\tabskip0pt\cr
SU$(p,1)$/U$(p)$&,&1\cr}}
$$
These are the only possibilities for which we have pairs $(\widehat{G},G)$
suspected to be a dynamical for $\Delta$ on $G/G_{0}$. Thus we need maximally
degenerated representations of SO$_{0}(p,1)$, SU(p,1), SL$(n,R)$,
$\gote_{\text{\bf 6}(\text{\bf 6})}$, $\gote_{\text{\bf 6}(-\text{\bf 26})}$
$\gote_{\text{\bf 7}(\text{\bf 7})}$, $\gote_{\text{\bf 8}(\text{\bf 8})}$, $\gote_{\text{\bf
4}(-\text{\bf 20})}$ and their decompositions, after a restrictions to
the maximally compact subgroup, to the irreducible representations of
those subgroups.

Let us consider SO$_{\text{\bf 0}}(p,1)$. The most degenerate discrete series
consists of
$$\widehat{T}(L),\ L = -\bigg\{{1\over 2}(p + 1) -4\bigg\}, -
\bigg\{{1\over 2}(p + 1) - 4\bigg\} + 1,
- \bigg\{{1\over 2}(p + 1) - 4\bigg\}+ 2, \dots
$$
in $L^{2}(H^{p,1},\mu )$.

$H^{p,1}$---means a hyperboloid in $R^{p+1}$,
${\displaystyle\sum^{p}_{i=1}}(x^{i})^{2} - (x^{p+1})^{2} = 1$ and $\mu $ is a
measure on $H^{p,1}$, quasiinvariant with respect to the action of the group
SO$_{\text{\bf 0}}(p)$ on $H^{p,1}$. In this case after a restriction of $\widehat{T}(L)$ to the
subgroup SO(p) every representation (finite dimensional, unitary,
irreducible) enters with a multiplicity equal to one (see 
Ref.~\cite{bet,gam}).
\begin{equation}
\widehat{T}_{\fg}(L)_{|G} = \sum_{\ell [p/2]}\oplus \widehat{T}^{\ell
[p/2]},\Beqno(16.56)
\end{equation}
where $\widehat{T}^{\ell [p/2]}$ are symmetric finite-dimensional representations of SO(p)
determined by the highest weight
$$\displaylines{
m= [\ell _{[p/2]},0,0,\dots ],\cr
\ell _{[p/2]} = L + 2n, n = 0,1,2,\dots \cr}
$$
Thus in the case of $G = \SO(p)$, $\widehat{G} = \SO(p,1)$.

In the case of $SU(p,1)$ the situation is even easier. Let us come back
to the operator $\widetilde{L}$. It is a $G$-invariant operator on $M$ of the second
order.

Thus it is a linear combination of $\ell = \rank M$, independent operators of
on $M$, left-invariant. If the rank of $M$ is 1 it is proportional to the
Beltrami--Laplace operator on $M$. Thus we have
\begin{equation}
\widetilde{L} = \ov{M} \sum^{\ell -1}_{i=0} \fg_{i}(\zeta
)\Delta_{i},\Beqno(16.57)
\end{equation}
such that $\Delta_{0} = \Delta$ (Beltrami--Laplace operator on $M)$ and
\bg 16.58
\fg_{0}(0) = 1\\
\fg_{i}(0) = 0,\quad\ i = 1,2,\dots \ell -1,\nonumber
\e
$\Delta_{i}$, $i =1,2,\dots,\ell -1$, are remaining $G$-invariant differential
operators on $M$.
Moreover if they commute we can find the spectrum of $\widetilde{L}$ using
eigenfunctions of $\Delta$ finding spectrum of masses for the tower of scalar
fields ${\varPsi} _{k}$. The matrix $M_{ke}$ can be diagonalized in $L^{2}(M,dm)$ which is
equivalent to $L^{2}(\sqrt{\widetilde{g}}, M, dm)$ and to Friedrichs' Hilbert space for $\Delta$.
Thus the operator has the same spectrum in $L_0{}^{2}(M)$, moreover the
eigenfunctions differ in their form from eigenfunctions of
Beltrami--Laplace operator due to different scalar product in $L_{0}{}^2(M)$.
\begin{equation}
m_{k} = {1\over 2} \sqrt{-2{\displaystyle\sum^{\ell-1}_{i=0}}
\mu_{i}\fg_{i}(\zeta)}.\Beqno(16.59)
\end{equation}
The spectrum generating group $\widehat{G}$ is the same for $\widetilde{L}$ as for the
Beltrami--Laplace operator on $M$.
$$
\widehat{G} \supset G \supset G_{0}.
$$

Let us remind to the reader that the maximally degenerated
representation means that all the Casimir operators are polynomials of
the Casimir operator of the lowest order. In our case of the Casimir of
the $2^{\text{\rm nd}}$ order. Our group $\widehat{G}$ is a spectrum generated group for the tower
of scalar field ${\varPsi} _{k}(x)$, $({\varPsi} (x,y)$ on $M\times E)$.

The Dark Matter in Appendix B is of course a cold Dark Matter.
This Dark Matter interacts only \gr ly. It is really a part of gravity
(multidimensional). The \ia s with ordinary matter are very weak, a coupling
\ct\ is the same as in the case of a skewon and a scalaron. All \pc s are
massive scalars, except~$\Ps_0$. Moreover, $\Ps_0$ corresponds to our $\Ps$
(or~$\vf$) massive scalaron. Masses and cross section of \ia s in a tree
\ap ion can be easily calculated. In some concrete cases this will be done.

Moreover, even if we get an infinite tower of scalar fields, we should
remember that our \pc s cannot be heavier than the Planck's mass, i.e.\
$\mu^2_k \le \mpl^2$. The scale of masses in our theory is~$\frac 1r$ or
$$
m_{\tA}=\frac{\hbar c}{r}, \q \wt\mu{}^2_k = \frac{\mu_k^2}{m_{\tA}^2}
$$
and
$$
\wt\mu{}_k^2 < \frac{\mpl^2}{m_{\tA}^2}\,.
$$
In this way we have a finite number of scalar \pc s such that their masses
are smaller than Planck's mass,
\beq B.78
\wt\mu{}_p^2 \le \frac{\mpl^2}{m_{\tA}^2}, \q\ \wt\mu{}_{p+1}^2 >
\frac{\mpl^2}{m_{\tA}^2}\,,
\e
$\wt\mu{}_k^2$, $k=1,2,\dots,p$.

An interesting problem is as follows. What will happen if a \co ical \ct\
obtained here is zero? Is it a danger for the theory? Certainly no! First of
all we do not have massive skewon and massive scalaron.
They are massless and are useless as Dark Matter \pc s. Moreover, we
still have a tower of massive scalar \pc s described in this appendix.
There is a \co ical problem with an accelerated expansion due to a
\co ical \ct. In the Robertson--Walker--Friedman Universe this is impossible.
We can use nonhomogeneous \co y described by Lema\^\i tre--Tolman
models \cite{xxy,xxz,xxr}. In the place of an accelerating expansion we have
inhomogeneous \co y. We do not need a \co ical \ct\ and we still have a Dark
Matter. From our point of view this would be a good solution for our Universe
is really inhomogeneous.

\section*{Appendix C}
\def\theequation{C.\arabic{equation}}
\setcounter{equation}0

Let us consider a bosonic part of QCD+GSW in the \E\nos\ \KK (Jordan--Thiry)
Theory, i.e.\ we give a geometrization and \un\ of gravity described by NGT with strong \ia s
described by $\SU(3)$-gauge field and with Glashow--Salam--Weinberg
model of electro-weak-\ia s.
This is a partial \un---a~path to a unified field theory. Let $P''$ be
a fibre bundle with a \sc al group $\SU(3)\ot G2$ over a base manifold
$V=E\tm S^2$, $E$---a \spt, and let $\o$~be a \cn\ defined on the bundle.
We have
\beq c1
\o=\o_{\SU(3)}+\o_{G2}.
\e
Let us suppose that $\o_{\SU(3)}$ has been obtained from a \cn\ $\o_{E\,\SU(3)}$
defined on a bundle~$P_1$ over a \spt~$E$ only with a structural group $\SU(3)$.
A~mentioned \cn\ has been moved from a bundle $P_1$ to~$P''$. For a~\cn\ $\o_{G2}$ we suppose
as usual \s y \wrt $\SO(3)$ rotations of~$S^2$. In this way we get $\o_E$ \cn\
with a \sc al group $\SU(2)\ot \U(1)$ and scalar fields---Higgs' fields from
GSW model (see Ref.~\cite{11a}). Let us metrize in a natural way the bundle manifold~$P''$
(see Sections 2 and~4).

One gets as usual (calculating a Moffat--Ricci scalar of \cvt\ $R(W)$ of a \cn\
$\gd\o,\tA,\tB,$) and changing variables, i.e.\ $\rho=e^{-\Ps}$ and
$g_{\mu\nu}\mapsto e^{n\Ps}\cdot g_{\mu\nu}$.
\bml c2
B\bigl(\ov W,g,A_{\SU(3)},A_{\SU(2)\ot \U(1)},\Ps,\F\bigr)\\
{}=\sqrt{-g} \biggl\{\ov R(\ov W)
+\frac{{\la'}^2}{4}\biggl(8\pi e^{-(n+2)\Ps}
\bigl(\cLY(\a_{\rm QCD}A_{\SU(3)})+\cLY(\a_sA_{\SU(2)\ot \U(1)})\bigr)\\
{}+\frac{2e^{-2\Ps}}{r_0^2}\cL_{\rm kin}(\gv\n \F) - \frac{e^{(n-2)\Ps}}{r_0^4}
\,\wh V(\F) - \frac{4e^{(n-2)\Ps}}{r_0^2}\cL\dr{int}(\F,\a_s A_{\SU(2)\ot \U(1)})\biggr)\\
{}+\frac{{\la'}^2}4 \,\cL\dr{scal}(\Ps)+ \frac1{{\la'}^2}\,e^{(n+2)\Ps}\,\wt R(\wt \G)
+\frac{e^{n\Ps}}{r_0^2}\,\wh{\ov P}_M\biggr\},
\e
$\a_{\rm QCD}$ is a \di less coupling \ct\ of $\SU(3)$ gauge fields (QCD) and
$\a_s$ is a \di less coupling \ct\ of $G2$-gauge fields-geometrization of electro-weak
\ia\ (GSW model).
\beq c3
\wt R(\wt\G)=\a^2\dr{QCD}\wt R_{\SU(3)}+\a_s^2 R_{G2},
\e
where we change as usual $\o_{\SU(3)}\to \frac{\a_{\rm QCD}}{\sqrt{\hbar c}}\o_{\SU(3)}$,
$\o_{G2}\to\frac{\a_s}{\sqrt{\hbar c}}\o_{G2}$, $\la'=2\ell\pl$,
$\ell\pl$ is a Planck's length, compare Eqs \er{4.80}--\er{4.81}.

\advance\abovedisplayskip by1pt
\advance\belowdisplayskip by1pt
The most interesting point in our calculations is a \co ical term. In order to
calculate them we should take under consideration $\wh V(0)$---the value of a
self-\ia\ \pt\ of zero Higgs' field. This value is not zero in our GSW model.
In our case $n=\dim\SU(3)+\dim G2= 8+14=22$,
\bml c4
\la_c=\frac{\a^2\dr{QCD}}{\ell\pl^2}\,e^{24\Ps}\,\wt R_{\SU(3)}
+\frac{\a_s^2}{\ell\pl^2}\,e^{24\Ps}\,\wt R_{G2}
+\frac{e^{22\Ps}}{r_0^2}\,\wh{\ov P}_{M} - \frac{\ell\pl^2}{r_0^4}\,\wh V(0)e^{20\Ps}\\
{}=\frac{e^{24\Ps}}{\ell\pl^2}\bigl(\a^2\dr{QCD}\wt R_{\SU(3)}
+\a_s^2 R_{G2}\bigr)+\frac{e^{22\Ps}}{r_0^2}\,\wh{\ov P}_{M} - \frac{\ell\pl^2}{r_0^4}\,\wh V(0)
e^{20\Ps}\\
{}=\frac{e^{22\Ps}}{r_0^2}\Biggl(\frac{16|\z|^3}{3(2\z^2+1)(1+\z^2)^{5/2}}\,
\z^2E\biggl(\frac{|\z|}{\sqrt{\z^2+1}}\biggr) - 2(\z^2+1)K\biggl(\frac{|\z|}
{\sqrt{1+\z^2}}\biggr)\\
{}+8\ln\biggl(|\z|\sqrt{\z^2+1} + \frac{4(1+9\z^2-8\z^4)|\z|^3}
{3(1+\z^2)^{3/2}}\biggr)\frac1{2|\z|\sqrt{1+\z^2}}\Biggr)\\
{}+\frac{\ell\pl^2}{r_0^4}\,e^{20\Ps}\,\frac\pi{\sqrt{1+\z^2}}\biggl(
4(1-2\z^2)+\frac{\z^2}{2\sqrt{1+\z^2}}\,\ov K(h_\a+h_\b,h_\a+h_\b)\biggr)\\
{}+\frac{e^{24\Ps}}{\ell\pl^2}\bigl(\a^2\dr{QCD}\wt R_{\SU(3)}+\a^2_s\wt R_{G2}\bigr),
\e
$M=\SO(3)/\SO(2)$, where $\wt R_{\SU(3)}$ is a Moffat--Ricci scalar for a group $\SU(3)$ for a
\nos\ tensor
\begin{gather*}
l_{ab}^{\SU(3)} = h_{ab}^{\SU(3)}+\mu k_{ab}^{\SU(3)}.\\
k_{ab}^{\SU(3)}=-k_{ba}^{\SU(3)},
\end{gather*}
a skewsymmetric right invariant tensor on $\SU(3)$.
$\wt R_{G2}$ is a Moffat--Ricci scalar for a group $G2$ for a \nos\ tensor
$$
l_{ab}^{G2} = h_{ab}^{G2}+i\xi k_{ab}^{G2},
$$
$\a\dr{QCD}$ is a coupling \ct\ for QCD at Electro-Weak (EW=$M_{Z^0}$) energy scale.
$\a_s$ is a coupling \ct\ of GSW model for the same scale, $\ell\pl$ is the
Planck's length, $r_0$~is a radius of a sphere~$S^2$.
$$
\bga
K(k) = \int_0^{\pi/2} \frac{d\theta}{\sqrt{1-k^2\sin^2\theta}}\\
E(k) = \int_0^{\pi/2} \sqrt{1-k^2\sin^2\theta}\,d\theta\\
0\le k^2\le 1
\ega
$$
are elliptic integrals of the first and the second order.
$$
\bga
\ov K(x,y) = \ov K_{de}x^dy^e = h^{G2\,cb}k^{G2}_{bd}k^{G2}_{ce}x^dy^e\\
k^{G2}_{ab}=-k^{G2}_{ba}
\ega
$$
(a skew-\s ic right invariant tensor on $G2$, see Ref.~\cite{11a} for details)
$$
\bal
h_\a &= \frac{2\a_i}{\a\cdot\a}\,H_i = [x_\a,x_{-\a}]\\
h_\b &= \frac{2\b_i}{\b\cdot\b}\,H_i = [x_\b,x_{-\b}],
\eal
$$
$\a,\b$---roots of a Lie algebra $G2$, $i=1,2$.

$\wt R_{\SU(3)}$ and $\wt R_{G2}$ can be calculated according to Section~2.
$\wh{\ov P}_M$~has been calculated in Ref.~\cite{1} and $\wh V(0)$ in Ref.~\cite{11a}.
In this way a \co ical term or a \pt\ for a field~$\Ps$ reads
\beq c5
U(\Ps) = \frac{e^{24\Ps}}{\ell\pl^2}\,\g + \frac{e^{22\Ps}}{r_0^2}\,\b+
\d\,\frac{\ell\pl^2}{r_0^4}\,e^{20\Ps}
\e
Let us look for a minimum (extremum) of $U(\Ps)$. One gets
\beq c6
\frac{dU}{d\Ps} = e^{20\Ps}\biggl(\frac{24\g}{\ell\pl^2}\,e^{4\Ps}
+\frac{22\b}{r_0^2}\,e^{2\Ps} + \frac{20\d \ell\pl^2}{r_0^4}\biggr)=0.
\e
Let $x=e^{2\Ps}$, then one gets $(\frac{dU}{dx}(x_0)=0)$
\beq c7
x_0=e^{2\Ps_0}=\frac{\ell\pl^2}{r_0^2}\biggl(\frac{-11\b+\sqrt{121\b^2-480\d\g}}
{24\g}\biggr)>0,
\e
where $121\b^2-480\d\g\ge0$, $\b>0$, $\g>0$, $\d<0$. In this way one gets
\beq c8
U(\Ps_0) = e^{20\Ps_0}\biggl(\frac{e^{4\Ps_0}}{\ell\pl^2}\,\g +
\frac{e^{2\Ps_0}}{r_0^2}\,\b + \frac{\d \ell\pl^2}{r_0^4}\biggr)
\e
and
\beq c9
\wt\la_c = -\frac12 \,U(\Ps_0).
\e
A mass of a \qe\ (scalar) \pc\ is equal to
\beq c10
m_s^2 = \frac{d^2 U}{d\Ps^2}(e^{2\Ps_0}) = 4e^{20\Ps_0}\biggl(144\g\frac
{e^{4\Ps_0}}{\ell\pl^2} + 121\b\,\frac{e^{2\Ps_0}}{r_0^2}+100\d\,\frac{\ell\pl^2}{r_0^4}
\biggr).
\e
We have as usual $\Ps=\Ps_0+\vf$.

Eventually
\beq c11
\wt\la_c = -\frac12\,U(\Ps_0) = -\frac1{576\g}\biggl(\frac{\ell\pl^2}{r_0^4}\biggr)
e^{20\Ps_0} \bigl(23\b\sqrt{121\b^2-480\d\g}-11\b^2+480\d\g\bigr)=-\wt\La
\e
and
\beq c12
m_{sk}^2 = \frac1{6\g}\biggl(\frac{\ell\pl^2}{r_0^4}\biggr)e^{20\Ps_0}\bigl[121\b^2
-480\d\g + 121\b\sqrt{121\b^2-480\d\g}\bigr]=2\wt\La,
\e
$\wt\La\ge0$ is a \co ical \ct\ and $m_{sk}$ is a mass of a skewon pseudovector \pc.
\bg c13
\g=\a\dr{QCD}^2\wt R_{\SU(3)}+\a_s^2\wt R_{G2}=\wt R(\wt \G)\\
\aligned
\b&=\wh{\ov P}_{M}=\Biggl(\frac{16|\z|^3}{3(2\z^2+1)(1+\z^2)^{5/2}}\cdot \z^2E
\biggl(\frac{|\z|}{\sqrt{\z^2+1}}\biggr)\\ & - 2(\z^2+1)K\biggl(\frac{|\z|}
{\sqrt{1+|\z|^2}}\biggr)+ 8\ln\biggl(|\z|\sqrt{\z^2+1}+ \frac
{4(1+9\z^2-8\z^4)|\z|^3}{3(1+\z^2)^{3/2}}\biggr)\frac1{2|\z|\sqrt{1+\z^2}}
\Biggr)\endaligned  \lb{c14} \\
\d=\frac\pi{\sqrt{1+\z^2}}\biggl(2(1-2\z^2)+\frac{\xi^2}{2\sqrt{1+\z^2}}
\,\ov K(h_\a+h_\b,h_\a+h_\b)\biggr). \lb{c15}
\e

Let us notice the following fact. Our Higgs' field is a singlet of $\SU(3)_c$
colour. The simplified \un\ model gives us a dielectric model of a \cfn\ of colour
and GSW model (a~bosonic part) with Higgs' field and a \ssb\ and \Hm. In this
way we geometrize bosonic part of a standard model with \un\ with NGT gravity,
obtaining a correct pattern of masses of $W^\pm$, $Z^0$ and Higgs' boson and correct
value of a Weinberg angle. However a \co ical model should be based on more
complicated model with a chain of groups (see Section~3). In this way we get
a geometrical interpretation of strong \ia s and electroweak \ia s
as \cvt s and torsions in higher
\di s. Our theory is 28-\di al and we mean \cvt s and torsions in 24~\di s.
Our partial \un\ from this Appendix is a \un\ of a bosonic part of Standard Model with NGT.

Geometrical \un\ means here a description of all physical \ia s (\gr al,
strong and electroweak) by an only one \cn\ $\gd W,\tA,\tB,$ defined
on a fibre bundle of frames over a \nos ally metrized fibre bundle.
A Dark Matter and a Dark Energy are ``interference effects'' in our \un.

Let us give an example of a different \un\ of gravity and gauge theory
(see Ref.~\cite{mbp}), i.e.\ metric-affine \un.

Moreover, we can get a different geometrical structure on $S^2$ coming from
a K\"ahlerian structure on~$S^2$ and a Hermitian version of our theory
(see Ref.~\cite{11a}). In particular
\beq c17
\wh{\ov P}_{S^2} = \frac1{V_2}\int_{S^2}\wh{\ov R}(\wh{\ov\G})\sqrt{|\wt g|}\,dm(\wt g)
=\frac1{V_2}\int_{S^2} \falg{}^{\ta\tb}\wt{\ov R}_{\ta\tb}(\wh{\ov \o})\,dm(\wt g),
\e
$V_2$ --- a volume of $S^2$.

$S^2$ is equipped with the Hermitian \nos\ tensor
\bg c18
g_{\ta\tb}=h_{\ta\tb} + \z k^0_{\ta\tb},\\
\ov \z=-\z \qh{(is pure imaginary),} \nonumber\\
\wt\na_{\td c}k^0_{\ta\tb}=0, \lb c19
\e
$\wt\na_{\td c}$ --- Riemannian covariant derivative on $S^2$ induced by a
metric
\beq c20
ds^2 = d\th^2 + \sin^2\th \,d\psi^2
\e
and $k^0_{\ta\tb}\th^{\ta}\land \th^{\tb}$ is a symplectic form
\bg c21
-\frac12\sin\th\,d\th\land d\psi \\
dm(g)=\sin\th\,d\th\,d\psi. \lb c22
\e
In this way one easily gets
\bg c23
\wh{\ov P}_{S^2}= \frac4{(1+\z^2)}\\
\aligned
\la_c &= \frac{e^{24\Ps}}{\ell\pl^2} \bigl(\a\dr{QCD}^2 \wt R_{\SU(3)}
+\a_s^2 R_{G2}\bigr) + \frac{4e^{22\Ps}}{r_0^2(1+\z^2)} \\
&+\frac{\ell\pl^2}{r_0^4}\,e^{20\Ps}\,\frac\pi{\sqrt{1+\z^2}}
\biggl(4(1-2\z^2)+\frac{\xi^2}{2\sqrt{1+\z^2}}\,\ov K(h_\a+h_\b,h_\a+h_\b)
\biggr).
\endaligned
\lb c24
\e
Moreover, according to Ref.~\cite{11a} we get
\beq c25
\z = \pm 0.911622 i.
\e

In the further development of the theory we consider $\z$ as a free parameter,
$r_0$~is a free parameter as usual.

On $\SU(3)$ we consider \nos\ real tensor, on $G2$ we consider \nos, Hermitian
tensor which is
right invariant \wrt $\SU(3)$ group and $G2$ group, respectively.
We remind to the reader that a metric tensor $g_{\mu\nu}$ is real \nos:
$g_{\mu\nu}=g_{(\mu\nu)}+ g_{[\mu\nu]}$.
\beq c26
\b=\wh{\ov P}_{S^2}=\frac4{(1+\z^2)} = 23.6763
\e
(if we use the value of $\z$ from \er{c25}, Ref. \cite{11a}),
\bg c27
\aligned
\d &= \frac\pi{\sqrt{1+\z^2}}(2(1-2\z^2)+\frac{\xi^2}{2\sqrt{1+\z^2}}\cdot
\ov K(h_\a+h_\b,h_\a+h_\b)\\ &= -0.47761 + 9.29766\xi^2 \cdot\ov
K(h_\a+h_\b,h_\a+h_\b)
\endaligned \\
\a^2\dr{QCD}=\a(M_{Z^0})=0.1179 \lb c28 \\
\a_s^2 = \a\dr{em}(M_{Z^0})=\frac1{127.955}=0.007815 \lb c29 \\
\bga
\g = 0.1185\cdot \wt R_{\SU(3)}+ 0.007815\cdot \wt R_{G2} \lb c30 \\
M_{Z^0}=91.18 \,{\rm GeV} \q\ \hbox{(see Ref.~\cite{m3k})}.
\ega
\e

It is interesting to find a self-\ia\ \pt\ for a field $\vf$. One gets for
the \pt\ of~$\vf$:
\bml c31
V(\vf) = -\frac12 U(\Ps_0+\vf) = -\frac12 e^{20\vf}\biggl(\frac{\ell\pl^2}{r_0^4}\biggr)
\frac1{24^{10}\g^{10}}\Bigl(-11\b+\sqrt{121\b^2-480\d\g}\Bigr)^{10}\\
{}\times\biggl(\d+e^{2\vf}\,\frac{(-11\b+\sqrt{121\b^2-480\d\g})}
{24\g}
+e^{4\vf}\frac{(121\b^2-240\d\g-11\b\sqrt{121\b^2-480\d\g})}
{288\g}\biggr).
\e

Let us notice a partial \un\ and geometrization of gravity (\nos---NGT), strong
(QCD) and electroweak \ia s (GSW) gives us also a \qe\ \pc\ and a skewon \pc.

Let us give a following comment. In Nature we have four fundamental \ia s:
\gr al, \elm c, weak and strong. \E\gr al \ia s are universal. They govern the
\Un\ on the \co ical level, movements of galaxies and also stars and planets in
galaxies, neutron stars and \gr al waves. \E\elm c \ia s are also widely present, e.g.\ in plasma physics,
\elm c spectra of atoms and molecules. They are source of chemical bonds and
a propagation of light, optics, \elm c waves phenomena.
In our contemporary every day live they are
fundaments of electric generators and engines and any phenomena of an electric
current. Weak \ia s are fundaments of $\beta$-decay of atomic nuclei. Strong
\ia s are responsible for the fact that nucleons are keeping together in atomic
nucleus. They are source of nuclear energy in atomic bombs, nuclear reactors and
stars. If we get some kind of a new \ia\ coming from higher dimensions of an
\un, this kind of \ia\ can be called the fifth \ia\ or even the fifth force.
In the \E\nos\ \KK (Jordan--Thiry) Theory this is the case. We get from higher dimension
$G_{\rm eff}$ (depending on a \qe\ field---a~scalar field~$\vf$, $\Psi$, $Q$
or~$q$). Thus we get the fifth force as an ``interference effect'' in our \un\
of fundamental \ia s. Simultaneously we get Dark Matter and Dark Energy with
a~\cn\ to the fifth force. According  to Ref.~\cite{Novd} we have several
possibilities to get $G_{\rm eff}$ depending on space and time. Simultaneously
we can measure an influence of the fifth force in geophysics. We are getting
$G_{\rm eff}$ from non-Newtonian \pt\ for an alternative theory of gravity.
In our approach (\E\nos\ \KK (Jordan--Thiry) Theory) the situation is similar.

There are many measurements of the Newton \gr al \ct\ ($G_{\eff}$), i.e.\ in
laboratory, in mines, in astronomy. Summing up we have
\beq c32
G_{\rm lab}< G_{\rm astr}\simeq G_{\rm geophys},
\e
where $G_{\rm lab}$ is a laboratory measurement (e.g.\ a torsion balance),
$G_{\rm astr}$ astronomical,
$G_{\rm geophys}$ a measurement in Australian mines (Hilton, Australia), in
submarines and bathyspheres. $G_{\rm astr}$ can depend on time (see Refs \cite{Novd,Novaa}).

According to Ref.~\cite{e} the authors use computer simulations of planets and
planetoids in the Solar System in the \gr al field of the Sun. The only
parameter of the Sun is $M_\odot G_{\rm astr}$. Moreover, $M_\odot$ is changing in time
due to the Solar wind $\dot M_\odot <0$. It seems that it is not enough to explain
a change of $M_\odot G_{\rm astr}$. A~measurement from Ref.~\cite{m8a} has
a different meaning.

Geophysical measurement of $G_N$ can be done by measurement of a changing of the Earth
acceleration using gravitmeters La~Costa Romberg G608 and~G20. It is necessary
to take under consideration a centrifugal force and a realistic model of the
Earth as consisting of ellipsoidal shells of the same density (see Ref.~\cite{Novaa}).

In the \E\nos\ \KK (Jordan--Thiry) Theory we have to do with the following effective
\gr al \pt
\beq c33
\wt V(R) = -\frac{Mm}R\, G_\infty \biggl(1 - \frac12 \exp\biggl(-\frac R{\la_s}
\biggr) + \frac12 \exp\biggl(-\frac R{\la_{sk}}\biggr)\biggr)
\e
where $\la_s$ is the range of a scalar (\qe) \pc\ and $\la_{sk}$ is the
range of a skewon.
\bea c34
\la_s &= \frac h{m_sc} = \frac {\frac hc}{2e^{10\Ps_0}\sqrt{144\g\, \dfrac
{e^{4\Ps_0}}{\ell\pl^2} + 121\b\,\dfrac{e^{2\Ps_0}}{r_0^2}+100\d\,\dfrac{\ell\pl^2}{r_0^4}}}\\
\la_{sk} &= \frac h{m_{sk}c} = \frac{\frac hc r_0^2\sqrt{6\g}}{
\ell\pl e^{10\Ps_0}\sqrt{121\b^2
-480\d\g + 121\b\sqrt{121\b^2-480\d\g}}} \label{c35}
\e
(It is easy to see that $G_\infty=G_{\rm astr}$, $R\to\infty$.)
Moreover our theory uses NGT as pure gravity theory. Thus we should add an
additional term to \er{c33} resulting in the \e\ for a force
\bg c36
F= -\frac{d\wt V}{dR} + \D F_{\rm NGT} \\
\D F_{\rm NGT} = \frac{2M^2G_\infty^2}{R^3}\,m + \frac{2l_M^4c^2}{R^5}\,m
- \frac{2l_m^2 l_M^2 Mc^2}{R^5} \label{c37a}
\e
where we take $G_N=G_\infty$, $l_M^2$ and $l_m^2$ are fermion charges
from NGT for a mass~$M$ and a mass~$m$ (see Refs \cite{6,Novy}).

An effective \gr al \ct\ is equal to
$$
G\dr{eff} = K(\Psi_0) = G_N \exp(-24\Psi_0),
$$
where $\Psi_0$ is an argument of minimum (extremum) of a \co ical term.

If we consider our world (our \Un) we can rescale $G\dr{eff}$ in such a way
that $G_N = G\dr{eff}(\Psi_0)$ and $G_N = G_0\exp(-24\Psi_0)$, $G_0=
G\dr{eff}(0)$. Thus an effective \gr al \ct\ is $G\dr{eff} = G_N\exp(-24
\vf)$. Moreover after an inflation in our partial approximation $\vf$ is equal
to zero. We have only small oscillation around an equilibrium. For many
possibilities of $G\dr{eff}$ see Section~5. Here we suppose that $\vf=0$.
We consider in this Appendix also $\wt G\dr{eff}$, which is not $G\dr{eff}$ and
has been obtained from Eq.~\er{c33}.

\def\(#1){{(#1)}}
\let\t\times
Let us consider full field equations in our partial \un\ theory.
This partial \un\ is really a \un\ of a bosonic part of a Standard Model with
NGT via \E\nos\ \KK (Jordan--Thiry) Theory (NKK(JT)T).
One gets field \e s from Palatini
\v al principle for \er{c2} \wrt $g_{\mu\nu},\Psi,\gd\ov W,\mu,\nu,,
A_{\SU(3)},A_{\SU(2)\ot \U(1)},\Phi$. We get
\beq c37
\ov R_{\mu\nu}(\ov W) - \frac12\,\ov R(\ov W)g_{\mu\nu}=8\pi G_N \nad{full}T_{\mu\nu}
\e
(\v\ \wrt $g_{\mu\nu}$) where $\nad{full}T_{\mu\nu}$ is a full energy momentum
tensor for all fields $A_{\SU(3)},A_{\SU(2)\ot \U(1)},\Phi,\Psi$ (look at
slightly different notation in Eq.~\er{4.109}).
\E\v s \wrt $\gd \ov W,\mu,\nu,$ are the same as in Section~4
and \e s are the same as \er{4.110}--\er{4.111}.
In general we consider for any energy momentum tensor in the place of~$T_{\a\b}$
the tensor $\nad{eff}T_{\a\b} = T_{\a\b}-\frac12 g_{\a\b}T_\m g^\m$.

The full set of remaining \e s is derived later in Appendix~C.
We are interested here in linearized \e s \wrt $h_{\mu\nu}$. One gets
\bg c38
g_{\mu\nu} = \eta_{\mu\nu}+h_{\mu\nu} =\eta_{\mu\nu}+h_{(\mu\nu)}+h_{[\mu\nu]},\q |h_{\mu\nu}|\ll1,\\
\ov h_\(\mu\nu) = h_\(\mu\nu) - \frac12 \eta^{\a\b} h_\(\a\b) \eta_{\mu\nu} \label{c39} \\
\gd h,(\mu\nu),{,\nu},=0 \label{c40}\\
\Box \ov h_\(\mu\nu)=-\frac{16\pi G_N}{c^4} (\nad{eff.full.lin}T_\(\mu\nu)), \label{c41}
\e
where $\nad{eff.full.lin}T_\(\mu\nu)$ is a linearization of $(\nad{full}T_{\mu\nu}
- \frac12 g_{\mu\nu} g^{\a\b}\nad{full}T_{\a\b})$ in the first order.

The field $h_\[{\mu\nu}]$ has a spin one (see Ref.~\cite{l}). In some sense $h_\[{\mu\nu}]$
is a different form of a vector field (Proca field with nonzero mass). In the case
of zero mass ($\wt\La=0$) it has spin zero. (It is a \gz\ (gauge abelian) Maxwell field.)
\beq c42a
\Box F_{\mu\nu\la} = 2\wt\La F_{\mu\nu\la} + \frac{16\pi G_N}{c^4}\, \nad{eff.full.lin}T
_{\tl[\mu\nu],\la\tp}, \q F_{\mu\nu\la}=h_{\tl[\mu\nu],\la\tp},
\e
$\nad{eff.full.lin}T_\[\mu\nu]$ is an anti\s ic part of the tensor mentioned above.

In this way a skewon field $h_\[\mu\nu]$ is massive in a linear \ap ion due to
\co ical term.

Let us consider a new field $a^\rho$ such that
\beq c42
\gd\ve,\mu\nu\la,\rho, F_{\mu\nu\la} = a_\rho, \quad a^\rho=\eta^{\a\rho}a_\a.
\e
One gets
\beq c43
\pp{a_\rho}{x^\rho} = \eta^{\a\b} a_{\a,\b}=0
\e
and
\beq c44
\gathered
\Box a_\rho=m^2_{sk} a_\rho + 16\pi \gd\ve,\mu\nu\la,\rho, \nad{eff.full.lin}T
_{\tl[\mu\nu],\la\tp}\\
\pp{a_\rho}{x^\rho}=0, \quad m^2_{sk}=2\wt\La.
\endgathered
\e
Thus $a_\rho$ is a Proca field with the same mass as a skewon field.
$\wt\La$~is a measured \co ical \ct.

Let us introduce a tensor of strength of $a_\rho$
\bg c45
f_{\mu\nu} = \partial_\mu a_\nu - \partial_\nu a_\mu\\
h_{\tl[\mu\nu],\la\tp} = F_{\mu\nu\la} = \frac16\,\ve_{\mu\nu\la\rho}a^\rho
\label{c46}
\e
Using $f_{\mu\nu}$ we get
\beq c47
\gd f,\mu,{\nu,\mu}, = \Box a_\nu
\e
and eventually
\beq c48
\gd f,\mu,{\nu,\mu}, = m^2_{sk}a_\rho + \frac{16\pi G_N}{c^4}\, \gd\ve,\mu\nu\la,\rho, \cdot
\nad{eff.full.lin}T_{\tl[\mu\nu],\la\tp}
\e
$\ve_{\mu\nu\la\rho}$ is an anti\s ic symbol \st $\ve_{1234}=1$.

Let us consider Eq.~\er{c48} with the first order of \ap ion. One gets
\beq c49
\Box a_\nu = m^2_{sk}a_\nu + \frac{16\pi G_N}{c^4}\, \nad{eff.full.lin}T _{\tl[\mu\a],\la\tp}
\gd\ve,\mu\a\la,\nu,.
\e
Let us notice the following fact: an \ia\ term
\beq c51
\wt V_{\rm int}= \frac{16\pi G_N}{c^4} \nad{eff.full.lin}T_{\tl[\mu\a],\la\tp} \gd\ve,\mu\a\la,\nu,a^\nu
\e
is very weak because of the \ct\ in front of $\frac{G_N}{c^4}$.

One can derive Eq.~\er{c49} from the following \lg\ in a Minkowski space
(remember we are in the first order of \ap ion of a weak field):
\beq c52
\cL\dr{skewon} = -\frac14\, f^{\mu\nu}f_{\mu\nu} - \frac{m_{sk}^2}2 \,a_\nu a^\nu
-\frac{8\pi G_N}{c^4} \nad{eff.full.lin}T_{\tl[\mu\a],\la\tp} \gd\ve,\mu\a\la,\nu, a^\nu.
\e
Thus the \ia\ of our ``skewon'' field $a_\nu$ is very weak with any field.
It means that ``skewon'' Dark Matter is extremely hard to be detected. It
interacts really only \gr ally.
$a_\mu$ is really a pseudovector because
of $\ve_{\mu\nu\a\la}$ anti\s ic symbol in the definition. We have
$\gd P,\la,\mu, a_\la(Px)=-a_\mu(x)$, where $\gd P,\la,\mu,$ is a~space inverse
operation
$$
\gd P,\la,\mu,=\left(\begin{matrix}
-1 &\ &0 &\ & 0 &\ & 0\\
0 && -1 && 0 && 0\\
0 && 0 && -1 && 0\\
0 && 0 && 0 && 1 \\
\end{matrix}\right), \quad Px= P(\bs x,\bs y, \bs z, t) =
(-\bs x,-\bs y,-\bs z,t)=(-\bs{\vec r},t).
$$

Similar situation we have to do with the field~$\vf$ (small oscillation
around the minimum of a self-\ia\ \pt\ of the field~$\Ps$).
The field~$\vf$ interacts only \gr ally. An \e\ and an effective \lg\ in a
weak field \ap ion look as follows (in Minkowski \spt). One gets
\begin{gather}
(2\ov M\Box - m_s^2)\vf - 192\pi e^{-24\Psi_0}\cLY(\a\dr{QCD}A_{\SU(3)})e^{-24\vf}
-4e^{-2\Psi_0}\cL\dr{kin}(\Phi,\a_s A_{\SU(2)\ot\U(1)})e^{-2\vf}\hskip30pt \nonumber\\
\hskip30pt {}+\frac{20}{r_0^4}\,e^{20\Psi_0}\wh V(\Phi)e^{20\vf}
+\frac{8}{r_0^2}\,e^{20\Psi_0}\cL\dr{int}(\Phi,\a_s A_{\SU(2)\ot\U(1)})e^{20\vf}=0 \label{c53} \\
\cL_\vf = \ov M \pa^\mu\vf \pa_{\mu}\vf -\frac12\,m_s^2 \vf^2
+8\pi e^{-24\Psi_0}\cLY e^{-24\vf}+2\,\frac{e^{2\Psi_0}}{r_0^2}
\cL\dr{kin}(\Phi,\a_s A_{\SU(2)\ot\U(1)})e^{-2\vf}\hskip30pt \nonumber \\
\hskip60pt {}-\frac{e^{20\Psi_0}}{r_0^4}\,\wh V(\Phi)e^{20\vf}
-\frac{4e^{20\Psi_0}}{r_0^2} \,\cL\dr{int}(\Phi,\a_s A_{\SU(2)\ot\U(1)})e^{20\vf}
\label{c54}
\end{gather}
where $m_s$ is the mass of a scalaron and
\bg c55
\aligned
\ov M &= \bigl((\mu \ell^{\SU(3)[ab]}\gd k,\SU(3),ab,+i\xi \ell^{G2[cd]}
\gd k,G2,cd,) -2\cdot 3^2\cdot 7\cdot11\bigr)\\
&=\bigl((\mu \ell^{\SU(3)[ab]}\gd k,\SU(3),ab,+i\xi \ell^{G2[cd]}
\gd k,G2,cd,) -1386\bigr) <0,
\endaligned\\
\cLY=\cLY(\a\dr{QCD}A_{\SU(3)})+\cLY(\a_sA_{\SU(2)\t\U(1)})\label{c56}\\
\ell^{\SU(3)ab}\gd \ell,\SU(3),ae, =\ell^{\SU(3)ba}\gd \ell,\SU(3),ea,=
\gd \d,b,e, \qquad a,b,e=7,8,9,\dots,14 \label{c57}\\
\ell^{G2ab}\gd \ell,G2,ae, = \ell^{G2ba}\gd \ell,G2,ea,=\gd \d,b,e,\qquad
 a,b,e=15,16,\dots,28 \label{c58}\\
\det(\gd\ell,\SU(3),ab,)\ne0,\qquad \det(\gd \ell,G2,cd,)\ne0. \label{c59}
\e

In a matrix form we have
\beq c60
\ell^{SU(3)\otimes G2}= \bma \ell^{\SU(3)}&0 \\\noalign{\hrule} 0&\ell^{G2}\ema
\e
where
\begin{align*}
\ell^{\SU(3)} &= (\gd\ell,\SU(3),ab,, \ a,b=7,8,\dots,14)\\
\ell^{G2} &= (\gd \ell,G2,cd,,\ c,d=15,\dots,28)
\end{align*}

Moreover now we have the following definition of \nos\ tensors defined on $\SU(3)$
and $G2$:
\begin{align*}
\gd \ell,\SU(3),ab, &= \gd h,\SU(3),ab, + \mu \gd k,\SU(3),ab,\q \hbox{(real \nos)}\\
\gd \ell,G2,cd, &= \gd h,G2,cd, + i\xi \gd k,G2,cd,\q \hbox{(complex Hermitian)}
\end{align*}
where $\mu$ and $\xi$ are real numbers, $\gd k,\SU(3),ab,$ and $\gd k,G2,cd,$
are as usual right-invariant skew\s ic tensors on $\SU(3)$ and~$G2$.
$\gd\ell,G2,cd,$ is a Hermitian tensor on~$G2$, $\gd\ell,\SU(3),ab,$ is a \nos\
and real tensor on SU(3). It is
easy to see that $\ell^{G2[cd]}$ is pure imaginary.
$\gd h,\SU(3),ab,$ and $\gd h,G2,cd,$ are Killing--Cartan tensors on Lie
algebras of $\SU(3)$ and $G2$, respectively.

We can tune $2\ov M$ to $-1$, i.e.\ $2\ov M=-1$. One gets for $n=14$
$$
m_{sk}= \sqrt{2\wt\La} \simeq 10^{-5}{\rm eV} \qh{and }
m_s = \sqrt{n(n+2)}\,\sqrt{2\wt\La} \simeq 15\cdot 10^{-5}{\rm eV}.
$$

In this way we have a bosonic part of the Standard Model in this partial \un.
Using $\gd \ell,\SU(3),ab,$, $\gd \ell,G2,ab,$, $\mu$, $\xi$ we can tune (in
principle) a \co ical \ct\ to observational value and from this we get $m_s$
and $m_{sk}$. It is easy to see that $m_s$ and $m_{sk}$ are induced by a \co
ical \ct. We can also redefine a scalar field~$\vf$ to
\beq c61
\ov\vf=\sqrt{2\ov M}\,\vf, \q \vf= \frac1{\sqrt{2\ov M}}\,\ov\vf.
\e
In this way one gets
\bml c62
\cL_{\bar\vf} = \frac12 \pa_\mu \ov\vf \,\pa^\mu \ov\vf - \frac12\,m_{\bar s}^2
\ov\vf{}^2 + 8\pi e^{-24\Psi_0}\cLY \exp\Bigl(-\frac{23\sqrt2}{\sqrt{\ov M}}\ov\vf\Bigr)\\
{}+2\,\frac{e^{2\Psi_0}}{r_0^2}\,\cL\dr{kin}(\Phi,\a_s A_{\SU(2)\t\U(1)})
\cdot \exp\Bigl(-\frac{\sqrt2\,\ov\vf}{\sqrt{\ov M}}\Bigr)
-\frac{e^{20\Psi_0}}{r_0^2}\,\wh V(\Phi)\exp\Bigl(\frac{10\sqrt2\,\ov\vf}{\sqrt{\ov M}}\Bigr)\\
{}-4e^{2\Psi_0}\cdot\cL\dr{int}(\Phi, \a_s A_{\SU(2)\t \U(1)})\cdot\exp
\Bigl(\frac{10\sqrt2}{\sqrt{\ov M}}\,\ov\vf\Bigr),
\e
where
\beq c63
m_{\bar s}=\frac{m_s}{\sqrt{2\ov M}} = \frac
{\sqrt2\,e^{10\Psi_0} \sqrt{144\g \frac{e^{4\Psi_0}}{\ell^2\pl} +
121\b \frac{e^{2\Psi_0}}{r_0^2} + 100\d \frac{\ell^2\pl}{r_0^4}}}
{\sqrt{i\bigl(\mu \ell^{\SU(3)[ab]} \gd k,\SU(3),ab, + \xi
\ell^{G2[cd]}\gd k,G2,cd,\bigr) -1386}}\,.
\e

Fermion fields can be added as external sources. In Kaluza--Klein Theory we
can have external sources as in Ref.~\cite{Deca}. Moreover fermion fields are
coupled only to the Levi-Civita part of the \cn\ (see Ref.~\cite{x}). In this
way fermions \ti{do not\/} interact classically with skewons.
After \ssb\ and \Hm\ we get an \ia\ of skewons and scalarons with Higgs' field
and $W^\pm,Z^0$ (massive) bosons and with photons and gluons (see Ref.~\cite{11a}).
For $\SU(3)$-gauge field we have a dielectric model of a \cfn\ (see Ref.~\cite{11a}).

In the case with zero \co ical \ct\ skewons and scalarons are radiation and
Dark Matter consists of scalar fields from Appendix~B.

In Ref.~\cite{xx} we consider a different
scheme of linearization of field \e s getting similar results.
J.~W.~Moffat in Ref.~\cite{Febb} considers New \E\nos\ \E\gr\ Theory. This
theory does not predict black holes (it seems they exist). In our \un\ we use
NGT (see Ref.~\cite6) and we can have black holes. We can have also nonsigular
particle-like \so s. Our theory is a classical field theory. Quantization can be
developed in nonlocal quantization approach (Ref.~\cite{11a}). In this way an
argument from Ref.~\cite{Feba} is not applied.

Let us come back to Eq.~\er{c33}. This is a \gr al \pt\ with an \ef\ \gr al
``\ct''. One easily gets an acceleration for a point mass $m$
\beq c64
\dot{\vec R} = -\frac{\wt G_{\eff} M}{R^3}\,\vec R
\e
where
\beq c65
\wt G_{\eff} = G_\iy \Bigl(1+\frac12\,R^2\Bigl(\frac1{\la^2_{sk}} - \frac1{\la_s^2}
\Bigr)\Bigr)
\e
for $R\ll \la_s, \la_{sk}$ ($\la_{sk}>\la_s$); for $R\to\iy$
\beq c66
\wt G_{\eff}=G_\iy.
\e
Formula \er{c65} can be used to fit a distance dependent \gr al ``\ct'' on
small distances.
In general
\beq c67
\wt G_{\eff} = G_\iy \biggl(1-\frac12\Bigl(\exp\Bigl(-\frac R{\la_{sk}}\Bigr)
-\exp\Bigl(-\frac R{\la_s}\Bigr)\Bigr) - \frac12\,\frac R{\la_s} \exp\Bigl(-\frac
R{\la_{s}}\Bigr) + \frac12\,\frac R{\la_{sk}} \exp\Bigl(-\frac R{\la_{sk}}\Bigr)\biggr).
\e
For $R\simeq0$ one gets $\wt G_{\eff} = G_\iy$ and $\wt G_{\eff} = G_\iy$ for $R\to\iy$.

Let us come back to \er{c32} and Eqs \er{c36}--\er{c37a}. For small distances
a $\wt G_{\eff}$ in Eq.~\er{c64} is influenced not only by \er{c65} but also \ef ly
by the term \er{c37a} coming from NGT. Thus in a laboratory experiments (very
small distances) it can lower the value $\wt G_{\eff}$ and $\wt G_{\eff}<G_\iy$,
$G_{\rm lab} <G_\iy$.

Moreover in \co ical applications we use the above formula for arbitrary $\la_s$
and $\la_{sk}$ putting $\ov\la_s$, $\ov\la_{sk}$.
$$
\wt G_{\eff} = G_\iy \biggl(1-\frac12\Bigl(\exp\Bigl(-\frac R{\ov\la_{sk}}\Bigr)
-\exp\Bigl(-\frac R{\ov\la_s}\Bigr)\Bigr) - \frac12\,\frac R{\ov\la_s} \exp\Bigl(-\frac
R{\ov\la_{s}}\Bigr) + \frac12\,\frac R{\ov\la_{sk}} \exp\Bigl(-\frac R{\ov\la_{sk}}\Bigr)\biggr).
\eqno(\rm\ref{c67}^*)
$$
Let us apply \er{c67} for Robertson--Walker pressureless \co ical model, i.e.\
for $\La$CDM model. One gets Friedman's \e s (see Ref.~\cite{21ga}):
\bg c68
\Bigl(\frac{\dot a}a\Bigr)^2 + \frac k{a^2} = \frac{8\pi \wt G_{\eff}}3 \,\rho
+\frac{\wt\La} 3 \\
\frac{\ddot a}a = -\frac{4\pi\wt G_{\eff}}3\,\rho +\frac{\wt\La} 3 \lb{c69}
\e
where $a=a(t)$ is the radius of the Universe, $\rho=\rho(t)=\frac{\rho_0}{a^3}$
is the density of barionic and a Cold Dark Matter density
\beq c70
\rho = \rho_{\rm B} + \rho_{\rm DM}.
\e
$\wt\La$ is a \co ical \ct\ and $c=1$ (velocity of light). We work with flat
spatial geometry. Thus $k=0$.

We take
\beq c71
\wt G_{\eff} = G_\iy \biggl(1-\frac12\Bigl(\exp\Bigl(-\frac a{\ov\la_{sk}}\Bigr)
-\exp\Bigl(-\frac a{\ov\la_s}\Bigr)\Bigr) - \frac12\,\frac a{\ov\la_s} \exp\Bigl(-\frac
a{\ov\la_{s}}\Bigr) + \frac12\,\frac a{\ov\la_{sk}} \exp\Bigl(-\frac a{\ov\la_{sk}}\Bigr)\biggr).
\e
In this way we consider the behaviour of the \Un\ on smaller distances
corresponding to $\wt G_{\eff}$ from Eq.~\er{c67}. This gives us a local value of the
Hubble parameter. $\rho_{\rm B}$~is
a barionic matter density and $\rho_{\rm DM}$ is a Cold Dark Matter density.

\E\e s \er{c68} and \er{c69} have also interpretation in Newtonian fluid
dynamics without pressure (see Ref.~\cite{21be}). However, $k$~has not an
interpretation as a spatial curvature. It is an integration \ct.

One easily gets
\beq c73
\rho(t) = \frac{\rho_0}{a^3}\,,\q \rho_0={\rm const}
\e
and
\beq c74
\rho_0 = \rho_{\rm B}^0 + \rho^0_{\rm DM}
\e
where $\rho_{\rm B}^0={\rm const}$, $\rho_{\rm DM}^0 = {\rm const}$.

This means that barionic matter and Cold Dark Matter do not interact during
an evolution of the \Un. Eventually we get
\bml c75
\Bigl(\frac{\dot a}a\Bigr)^2 = \frac{8\pi\rho_0G_\iy}3 \biggl(1-\frac12
\Bigl(\exp\Bigl(-\frac a{\ov\la_{sk}}\Bigr)-\exp\Bigl(-\frac a{\ov\la_s}\Bigr)\Bigr) \\
- \frac12\,\frac a{\ov\la_s} \exp\Bigl(-\frac a{\ov\la_{s}}\Bigr)
+ \frac12\,\frac a{\ov\la_{sk}} \exp\Bigl(-\frac a{\ov\la_{sk}}\Bigr)\biggr) +\frac{\wt\La}3
\e
or
\beq c76
\Bigl(\frac{\dot a}a\Bigr)^2 = \frac p{a^3} - \frac p{2a^3}\exp\biggl(
-\frac a{\ov\la_{sk}}\biggr) + \frac p{2a^3}\exp\biggl(-\frac a{\ov\la_s}\biggr)
-\frac{q_1}{a^2}\exp\biggl(-\frac a{\ov\la_s}\biggr) + \frac{q_2}{a^2}\exp
\biggl(-\frac a{\ov\la_{sk}}\biggr)+r
\e
where
\beq c77
p = \frac{8\pi}3\,G_\iy \rho_0, \q q_1= \frac{4\pi G_\iy\rho_0}{3\ov\la_s},
\q q_2= \frac{4\pi G_\iy\rho_0}{3\ov\la_{sk}}, \q r=\frac{\wt\La} 3.
\e
The definition of a Hubble parameter is $H=\frac{\dot a}a$.

In this way we can estimate the value of a Hubble parameter of our epoch for
small distances using $\ov\la_{sk}$ and $\ov\la_s$,
\beq c78
H_{01}^2 = \frac p{a^3} - \frac p{2a^3}\exp\biggl(
-\frac a{\ov\la_{sk}}\biggr) + \frac p{2a^3}\exp\biggl(-\frac a{\ov\la_s}\biggr)
-\frac{q_1}{a^2}\exp\biggl(-\frac a{\ov\la_s}\biggr) + \frac{q_2}{a^2}\exp
\biggl(-\frac a{\ov\la_{sk}}\biggr)+ \ov r
\e
For large distances one gets
\bg c79
\Bigl(\frac{\dot a}a\Bigr)^2 = \frac{8\pi G_\iy}3 \,\frac{\rho_0}{a^3}
+\frac {\wt\La}3\\
H_{02}^2 = \frac{p'}{a^3} +r \lb{c80} \\
p' = \frac{8\pi G_\iy\rho_0}3\,. \lb{c81}
\e

Using the approximation method by Mukhanov (see Ref.~\cite{21al}) we can find
the general expression for cosmic mean of the CMB temperature autocorrelation
\f\ expressed in terms of multiple $C_l$ not only for $l\ll200$ but for
large~$l$. It means we get a correct value of a Hubble parameter $H_{02}$
for contemporary epoch to compare it to~$H_{01}$ (for small distances) (see
Appendix~A for definitions of~$C_l$).

Let us make some very rough estimation of $H_{01}$. In order to do this, let
us take $a(t_0)=\a(t_0)\sqrt{\ov\la_s \ov\la_{sk}}$ for our contemporary epoch
(Present Day \Un). One gets
\begin{gather}
H_{01}^2 = \frac{8\pi G_\iy \rho_0}{3}\Biggl[\frac1{\a^3(t_0)(\ov\la_s \ov\la_{sk})^{3/2}}
\Biggl(1 - \frac12\exp\biggl(-\a(t_0)\sqrt{\frac{\ov\la_s}{\ov\la_{sk}}}\biggr)
+\frac12\exp\biggl(-\a(t_0)\sqrt{\frac{\ov\la_{sk}}{\ov\la_{s}}}\biggr)\Biggr)\nonumber\\
-\frac1{2\a^2(t_0)\ov\la_s\ov\la_{sk}}\Biggl(\frac1{\ov\la_s}\exp\biggl(-\a(t_0)\sqrt{\frac{\ov\la_{sk}}{\ov\la_{s}}}\biggr)
-\frac1{\ov\la_{sk}}\exp\biggl(-\a(t_0)\sqrt{\frac{\ov\la_s}{\ov\la_{sk}}}\biggr)\Biggr)\Biggr]
+\frac{\wt\La}3 \lb c82 \\
a(t_0)\simeq 10M_{pc}.\lb c83
\end{gather}
Using $\a,\ov\la_s$ and $\ov\la_{sk}$ we could (in principle) estimate $H_{01}$.

According to recent measurement of BAO and eBOSS, adding also microwave
background radiation measurements one gets for Hubble parameter
$67\frac{\rm km/s}{M_{pc}}$ (BAO\,=\,Baryon Acoustic Oscillations,
eBOSS\,=\,extended Baryon Oscillation Spectroscopy Survey). In our approach
it is $H_{02}$. Moreover, local measurements (local supernova distances)
show a value of~10\% bigger. In our approach it is $H_{01}$.

One can try to do more precise estimation of $H_{01}$ solving the differential
\e
\beq c84
\frac{da}{dt} = \Biggl(\frac p{a^3} - \frac p{2a^3}\exp\biggl(
-\frac a{\ov\la_{sk}}\biggr) + \frac p{2a^3}\exp\biggl(-\frac a{\ov\la_s}\biggr)
+\frac{q_1}{a^2}\exp\biggl(-\frac a{\ov\la_s}\biggr) + \frac{q_2}{a^2}\exp
\biggl(-\frac a{\ov\la_{sk}}\biggr)+ \ov r\Biggr)^{1/2}
\e

In particular (see Ref.~\cite{m2d}) we have the \fw\ measurement of the Hubble
\ct\ (Hubble parameter):
\bit
\item[$1^\circ$] Present Day \Un\\
$73.7\,\frac{{\rm km/s}}{M_{pc}}$ Quasars\\
$74.0\,\frac{{\rm km/s}}{M_{pc}}$ Cepheids stars\\
$73.9\,\frac{{\rm km/s}}{M_{pc}}$ Cosmic masers
\item[$2^\circ$] Early \Un\\
$67.4\,\frac{{\rm km/s}}{M_{pc}}$ Cosmic microwave background\\
$67.6\,\frac{{\rm km/s}}{M_{pc}}$ BAO
\eit
The difference between $1^\circ$ and $2^\circ$ is about 10\%. In our notation
it is $H_{01}$ and $H_{02}$.

Recently (see Ref.~\cite{m2d}) we have a new measurement (Present Day \Un):

$69.6\,\frac{{\rm km/s}}{M_{pc}}$ --- Tip of the red grant branch (TRGB).

Thus we should apply Eq.~\er{c84} to explain the data. This approach to cure
a Hubble \ct\ crisis is inspired by $\wt G_{\eff}$ from \E\nos\ \KK (Jordan--Thiry)
Theory. Parameters $\ov\la_s$ and $\ov\la_{sk}$ have nothing to do with $\la_s$
and $\la_{sk}$ from real $\wt G_{\eff}$ of NKK(J-T)T. We also use classical
(nonrelativistic, Newtonian) \co y (see Ref.~\cite{21be}).

According to recent observations (KiDS\,=\,Kilo Degree Survey) the \Un\ is
more homogeneous than it is coming from microwave background observations
(about 10\% more). It supports also SDSS (Sloan Digital Sky Survey). eBOSS was
a part of SDSS project (see Refs \cite{21k}, \cite{21l}). More data would be
obtained from the project DESI (Dark Energy Spectroscopic Instrument) in
Arizona and in a satellite mission Euclid which was launched by ESA
(European Space Agency) in~2022. All data can influence $H_{01}$ value.

A very important problem in our theory is a value of Newtonian \gr al \ct, in
particular some measurements of this \ct. In Ref.~\cite{m2a} this \ct\ has been
measured by two independent methods:
\bit
\item[1)] torsion pendulum experiments with the time-of-swing method,
\item[2)] torsion pendulum experiments with the angular acceleration-feedback
method.
\eit
Using the first method one has obtained
\beq c85
6.674184 \tm 10^{-11} \,\frac{\rm m^3}{\rm kg\,s^2}
\e
with standard uncertainties $11.64\tm 10^{-6}$.
The second method gives
\beq c86
6.674484 \tm 10^{-11} \,\frac{\rm m^3}{\rm kg\,s^2}
\e
with standard uncertainties $11.61\tm 10^{-6}$.

Both values are different. Simultaneously CODATA recommended value for $G_N$
is
\beq c87
(6.67408\pm 0.00031)\tm 10^{-11} \frac{\rm m^3}{\rm kg\,s^2}
\e
(CODATA 2014, see Ref.~\cite{m2b}) with uncertainty 47 per~$10^6$. Those
differences could be explained by the ``fifth force'' for $G_N$ enters formulas
of this force. In Ref.~\cite{m2c} there is a review of many measurements of~$G_N$
on small distances. There is also a discussion why two values of~$G_N$ from
Ref.~\cite{m2a} are different. The conclusion is that on small distances there
are many problems of measurement using small bodies. The value of the \ct\ is
the same independent on the measurement method. We have only one~$G_N$.

Moreover in our approach one can expect some dependence of $\wt G_{\eff}$ on
a distance between bodies. We have two Dark Matter \pc s of \gr al origin:
a~scalaron and a skewon with small masses and the \fw\ ranges:
\beq c88
\la_{sk}\approx 28.58\,{\rm m}, \q\ \la_s\approx0.95\,{\rm m}.
\e
It would be very interesting to try to measure their influence on~$G_N$ as
the fifth force. One gets
\beq c89
\wt G_{\eff} = G_\iy (1-0.5004R^2), \q R\ll0.95\,{\rm m}
\e
where $R$ is measured in meters. In all mentioned experiments $\wt G_{\eff}$ enters
as~$G_N$ and differences in some results could be explained by \er{c89}. Let us
notice a difference between $G_{\eff}$ and $\wt G_{\eff}$. The first one depends only
on scalar field (see Section~5), the second on scalar \pc\ and skewon \pc\ (see
Eq.~\er{c33}).

Let us come back to the \e\ of motion for test \pc s in our theory. This is
a bosonic part of QCD+GSW in the \E\nos\ \KK (Jordan--Thiry) Theory, i.e.\
geometrization and \un\ of gravity described by NGT with strong \ia s described
by $\SU(3)$-gauge field with GSW model. This is a partial \un. A~bundle~$P''$
has been metrized in a natural way introducing a scalar field~$\rho$. According
to the usual interpretation we write down a geodetic \e\ on~$P''$ \wrt a \LC
\cn\ induced by a \s ic part of $\ka_\(\tA\tB)$.

One writes
\beq3.169
u^{\tA}\wt\n_{\tA} u^\tB=0
\e
where $\wt\n_\tA$ means a \ci\ \dv\ \wrt a \LC \cn\ induced by $\ka_\(\tA\tB)$
on~$P''$.

One finds
\bg3.170
\aligned
\frac{\wt{\ov D}u^\a}{d\tau} + \Bigl(\frac{q^c}{m_0}\Bigr)u^\b h_{cd}\wt
g{}^\(\a\d) \gd H,d,\b\d, &+ \frac{c^i}{m_0}\,u^\b h_{ij}\wt g{}^{(\a\d)}
\gd H,j,\b\d, \\
& + \Bigl(\frac{q^c}{m_0}\Bigr)u^{\td b}h_{cd}
\wt g{}^\(\a\d)\gv{\n_\d}\Ft db
-\frac{\|\wh q\|^2}{8m_0^2}\wt g{}^\(\a\d)\Bigl(\frac1{\rho^2}\Bigr)_{,\d}=0
\endaligned \\
\frac{\wt Du^{\td a}}{d\tau} + \frac1{r^2}\Bigl(\frac{q^c}{m_0}\Bigr)
u^\b h_{cd}h^{0\td a\td d}\gv{\n_\b}\Ft dd + \frac1{r^2}
\Bigl(\frac{q^c}{m_0}\Bigr)u^{\td b}h_{cd}h^{0\td a\td d}\gd H,d,\td d\td b,
=0 \lb3.171 \\
\frac d{d\tau}\Bigl(\frac{q^b}{m_0}\Bigr)=0, \q\
\frac{d}{d\tau}\Bigl(\frac{c^i}{m_0}\Bigr)=0 \lb3.172 \\
2\rho^2u^a = \frac{q^a}{m_0}, \q\ 2\rho^2u^i = \frac{c^i}{m_0} \lb c94 \\
\|\wh q\|^2 = -h_{ab}q^aq^b - h_{ij}c^ic^j= {\rm const} \lb c95
\e
where $\wt{\ov D}$ means a \ci\ \dv\ along a line \wrt the \cn\
$\gd\wt{\ov\o},\a,\b,$ on~$E$. $\wt D$~means a \ci\ \dv\ along a line \wrt the
\cn\ $\gd\wt\o,a,b,$ on~$G/G_0$ ($r=\rm const$),
\beq3.173
u^\tA = (u^\a,u^{\td a},u^i,u^a),
\e
$q^a$ is a \YM' charge known from the \E\nA\ Kaluza--Klein Theory (colour
(isotopic) charge), $u^\a$ is a four-velocity of a test \pc, $c^i$ is a colour
charge for QCD.

Let us give some explanations for \er{3.173}. $u^a$ usually is a ``velocity''
of a test \pc\ in higher \di s connected to gauge \di s. They are not directly
observed. In our case of a partial \un\ those \di s are connected to groups
$\SU(3)_c$ and~$G2$. Thus we can write $u^a$ as a pair $(u^i,u^a)$, $i=1,2,
\dots,8$, $a=1,2,\dots,14$. This is dicted by the fact that we have a \cfn\
condition and $u^i$~should be put to zero as proportional to colour charges of
a test \pc. In a different way we can write $(u^{\ba},u^a)$, $\ba=7,8,\dots,14$,
$a=15,\dots,28$. We can use both notations.

$u^{\td a}$ is a charge associated with a Higgs' field.
Eq.~\er{3.171} describes a movement of a test \pc\ in a \gr al,
gauge and Higgs' field. Eq.~\er{3.172} is an \e\ for a charge associated with
Higgs' field. This charge describes a coupling between a test \pc\ and a
Higgs' field.
Eq.~\er{3.173} has a usual meaning (a constancy of colour isotopic charges).
In this way we get a \gn\ of \KWK \e\ to the presence of a Higgs' field.

\let\t\theta
Let us project the \e\ on a \spt~$E$, i.e.\ we take a section $e:E\to P''$. One
gets
\bg3.175
\aligned
\frac{\wt{\ov D}u^\a}{d\tau} + \Bigl(\frac{Q^c}{m_0}\Bigr)u^\b \wt g{}^\(\a\d)
\gd F,d,\b\d, &+\frac{C^i}{m_0}\,u^\b h_{ij}\wt g^{(\a\d)} \gd F,i,\b\d, \\
& + \Bigl(\frac{Q^c}{m_0}\Bigr)u^{\td b}h_{cd}\wt g{}^\(\a\d)
e^*(\gv{\n_\d}\Ft db) -\frac{\|\wh Q\|^2}{8m_0^2}\,\wt g{}^\(\a\d)
\Bigl(\frac1{\rho^2}\Bigr)_{,\d} =0
\endaligned\\
\frac{\wt Du^{\td a}}{d\tau} + \frac1{r^2}\Bigl(\frac{Q^c}{m_0}\Bigr)u^\b
h_{cd}h^{0\td a\td d}e^* (\gv{\n_\b}\Ft dd)
+\frac1{r^2}\Bigl(\frac{Q^c}{m_0}\Bigr)u^{\td b}h_{cd}h^{0\td a\td d}
e^*(\gd H,d,\td d\td b,)=0 \lb3.176 \\
e^*\o= \gd A,a,\mu,\ov\t{}^\mu X_a + \Ft ab\wt\t{}^b X_a + \gd A,i,\mu,
Y_i\ov\t{}^\mu \lb3.177 \\
e^* (q^cX_c + c^iY_i) = Q^cX_c + C^iY_i \lb3.178 \\
\|\wh Q\|^2 = -h_{ab}Q^aQ^b- h_{ij}C^iC^j= {\rm const.} \nonumber
\e
Equation \er{3.172} takes the form
\bg3.185
\bga
\frac{dQ^a}{d\tau} - \gd C,a,cb,Q^c \gd A,b,N, u^N=0,\\
\hbox{or}\q \frac{dQ^a}{d\tau} - \gd C,a,cb,Q^c\gd A,b,\nu, u^\nu
-\gd C,a,cb,Q^c\gd\F,b,\td n, u^{\td n}=0,
\ega \\
\hbox{and}\q \frac{dC^i}{d\tau} - \gd C,i,jk, C^k \gd A,j,\nu, u^\nu=0 \lb c102
\e
$\wt g^\(\a\b)$ is defined by Eq.~\er{2.24}.

Let us explain some notations. $h_{ij}$ is a Killing--Cartan tensor for $\SU(3)$
group, $h_{ij}=h_{ij}^{\SU(3)}$.
$[Y_i,Y_j] = \gd C,k,ij,Y_k$, where $Y_i$, $i=1,2,\dots,8$, are generators of
a Lie algebra of $\SU(3)_c$ group (a~QCD gauge colour group). $\gd A,i,\mu,$ are
gauge \pt\ of QCD and $\gd F,i,\mu\nu,$ its strength. $\gd H,i,\mu\nu,$~is a \cvt\
of a \cn~$\o_{\SU(3)}$ (see Eq.~\er{c1}). Here we are using $i,j,k$ indices
for~$h_{ij}$, for simplicity. We should write $\gd h,\SU(3),ab,$, $a,b=7,8,
\dots,14$, and for $\gd h,G2,ab,$, $a,b=15,16,\dots,28$.

Let us derive a \KWK \e\ in our partial \un. One gets (see Eqs \er{3.170}--\er{3.172})
\bg C.1
\aligned
\frac{\wt{\ov D}u^\a}{d\tau} &+ \frac{u^\b}{m_0} h(q,H_{\b\d})\wt g^{(\a\d)}
+\frac{u^\b}{m_0} h_{ij}c^i \gd H,j,\b\d, \wt g^{(\a\d)}\\
&+\frac1{m_0}\,\wt g{}^\(\a\d)\Bigl(u^5\cdot h(q,\gv{\n_\d}\F_5)
+u^6h(q,\gv{\n_\d}\F_6)\Bigr) - \frac{\|\wh q\|^2}{8m_0^2}
\gtn{\a\d}\biggl(\frac1{\rho^2}\biggr)_{,\d}=0 \endaligned \\
\frac{\wt Du^5}{d\tau} + \frac1{r^2}\,\frac{u^\b}{m_0}\,h(q,\gv{\n_\b}\F_5)
+\frac1{r^2}\,u^6h(q,H_{56})=0 \lb C.2 \\
\frac{\wt Du^6}{d\tau} +\frac1{r^2}\,\frac{u^\b}{m_0\sin^2\psi}\,h(q,\gv{\n_\b}\F_6)
+\frac1{r^2}\,\frac{u^5}{m_0\sin^2\t}\,h(q,H_{65})=0 \lb C.3 \\
\frac d{d\tau}\Bigl(\frac q{m_0}\Bigr)=0. \lb C.4
\e
$q$ is an isotopic charge belonging to a \Li\ of~$H$ ($\fh$), $u^{\td a}=
(u^5,u^6)$ is a charge which couples a test \pc\ to Higgs' field,
$C^i$ is a colour charge for QCD,
$H_{\b\d}$~is a strength of $\SU(2)\tm U(1)$ \YM' field, $\F_5,\F_6$ are
scalar fields before a \sn\ \s y breaking (see Ref.~\cite{11a}). $\frac{\wt{\ov
D}}{d\tau}$ is a \ci\ \dv\ along a line \wrt a \cn~$\gd\wt{\ov\o},\a,\b,$
on~$E$, $\frac{\wt D}{d\tau}$ is a \ci\ \dv\ \wrt a \LC \cn\ on~$S^2$.

Using some
additional fields $\F_1,\F_2,\F_3$ and also $\F$ and~$\wt\F$, we can write
$\gv{\n_\mu}\F_5$ and $\gv{\n_\mu}\F_6$ in terms of Higgs' fields $\vf_1$
and~$\vf_2$ (see Appendix~B of Ref.~\cite{11a}), $m_0$ is the mass of a test \pc.
For the convenience of the reader, i.e.\ to make our considerations friendly
for the reader, we repeat some results from Ref.~\cite{11a}. Moreover, our
case is more general for $\rho\ne1$ ($\Psi\ne0$).
\begin{align}
\gv{\n_\mu}\F_5&=\frac12\gv{\n_\mu}(\F+\wt\F)=\frac12\Bigl[\Bigl(\pa_\mu\vf_1
-\frac12\,iA_\mu^- \vf_2 - \frac12\,iA_\mu^3 \vf_1 -\frac12\,i\tan\ov\t B_\mu
\vf_1\Bigr)x_\a \nn \\
&+\Bigl(\pa_\mu \vf_2 - \frac12\,iA_\mu^+\vf_1 + \frac12\,iA_\mu^+\vf_2
-\frac12\,iB_\mu\vf_2\tan\ov\t\Bigr)x_\b \nn \\
&-\Bigl(\pa_\mu\vf_1^* + \frac12\,iA_\mu^+ \vf_2^* + \frac12\,iA_\mu^3\vf_1^*
+\frac12\,iB_\mu\vf_1^*\tan\ov\t\Bigr)x_{-\a} \nn \\
&-\Bigl(\pa_\mu\vf_2^* + \frac12 \,iA_\mu^- \vf_1^*
-\frac12\,i A_\mu^3 \vf_2^* +\frac12\,i\tan \ov\t B_\mu\vf_2^*\Bigr)x_{-\b}\Bigr]
\lb C.4a
\end{align}
\begin{align}
\gv{\n_\mu}\F_6 &= \frac{\sin\t}{2i} \gv{\n_\mu}(\F-\wt\F)=
\frac{\sin\t}{2i}
\Bigl[\Bigl(\pa_\mu\vf_1
-\frac12\,iA_\mu^- \vf_2 - \frac12\,iA_\mu^3 \vf_1 -\frac12\,i\tan\ov\t B_\mu
\vf_1\Bigr)x_\a \nn \\
&+\Bigl(\pa_\mu \vf_2 - \frac12\,iA_\mu^+\vf_1 + \frac12\,iA_\mu^+\vf_2
-\frac12\,iB_\mu\vf_2\tan\ov\t\Bigr)x_\b \nn \\
&-\Bigl(\pa_\mu\vf_1^* + \frac12\,iA_\mu^+ \vf_2^* + \frac12\,iA_\mu^3\vf_1^*
+\frac12\,iB_\mu\vf_1^*\tan\ov\t\Bigr)x_{-\a} \nn \\
&-\Bigl(\pa_\mu\vf_2^* + \frac12 \,iA_\mu^- \vf_1^*
-\frac12\,i A_\mu^3 \vf_2^* +\frac12\,i\tan \ov\t B_\mu\vf_2^*\Bigr)x_{-\b}\Bigr]
\lb C.5
\end{align}

Let
\beq C.6
q=q_\g x_\g + q_{-\g}x_{-\g}+qh+\wt qh_\g+q_\a x_\a+q_{-\a}x_{-\a}+q_\b x_\b
+q_{-\b}x_{-\b}.
\e
It is easy to see that the first part of $q$,
\bg C.7
q=q_1+q_2, \\
q_1=q_\g x_\g +q_{-\g}x_{-\g}+qh + \wt qh_\g \lb C.8
\e
couples to \YM' field and the second part
\bg C.9
q_2=q_\a x_\a+q_{-\a}x_{-\a}+q_\b x_\b +q_{-\b}x_{-\b}
\e
to scalar fields $\F_5$ and $\F_6$.

In this way in our partial \un\ a test \pc\ has a weak isotopic charge, weak
hypercharge which are equivalent to weak charge and an electric charge and
colour QCD charge. It
has also an additional weak charge which couples it to Higgs' field,
i.e.~$q_2$. Moreover, we have also $u^{\td a}=(u^5,u^6)$ charge. It would be
very interesting to observe this additional charges in an experiment.
All details of the formalism described here can be found in Ref.~\cite{11a}.
\bea C.10
H_{56}&= \pa_5\F_6 - \pa_6\F_5 + [\F_5,\F_6] \\
H_{65}&= \pa_6\F_5 - \pa_5\F_6 + [\F_6,\F_5] \lb C.11
\e

We get
\begin{gather}
\frac{\wt{\ov D}u^\a}{d\tau} + \frac{u^\b}{m_0} \,\wt g{}^\(\a\d)
\wt Q{}^i \d_{ij}\gd F,i,\b\d, + \frac{u^\b}{m_0}\,\wt g{}^\(\a\d)\cdot \wt Q
\cdot B_{\b\d} + \frac{u^\b}{m_0}h_{ij}C^i \gd H,j,\b\d,\wt g{}^{(\a\d)}
- \frac{\|\wh Q\|^2}{8m_0^2}\wt g{}^{(\a\d)}\Bigl(\frac1{\rho^2}\Bigr)_{,\d}
\nn \\
\hskip60pt {}+\frac1{m_0}\,\wt g{}^\(\a\d)\Bigl(u^5\cdot h\bigl(q,e^*(\gv{\n_\d}\F_5)\bigr)
+u^6 h\bigl(q,e^*(\gv{\n_\d}\F_6)\bigr)\Bigr)=0 \lb C.14  \\
\frac{\wt Du^5}{d\tau} + \frac1{r^2}\,\frac{u^\b}{m_0}\,h\bigl(Q,e^*(\gv{\n_\b})
\F_5\bigr) -\frac1{r^2}\,u^6h(Q,e^*(H_{56}))=0 \lb C.15 \\
\frac{\wt Du^6}{d\tau}+ \frac1{r^2}\,\frac{u^\b}{m_0\sin^2\t}\,h\bigl(Q,
e^*(\gv{\n_\b}\F_6)\bigr)
+\frac1{r^2}\,\frac{u^\b}{m_0\sin^2\t}\,h(Q,e^*(H_{65}))=0 \lb C.16 \\
\frac{dQ^a}{d\tau} - \gd C,c,cb,Q^cA^b_M u^M=0 \lb C.17
\end{gather}
where
\bg C.18
e^*\o_E = A^i_\mu \ov \t{}^\mu t_i + B_\mu \ov \t{}^\mu y \\
e^*(q^cX_c + c^jY_j) = Q^c X_c + C^jY_j \lb C.19 \\
e^*\o = \a^c_i A^i_\mu \ov \t{}^\mu \wt t_i + \F^a_{\td a}\t^{\td a}X_a
+\gd A,j,\mu, Y_j \ov\t{}^\mu,
\lb C.20 \\
\|\wh Q\|^2 = \|\wh q\|^2, \nn
\e
$\wt Q{}^i=\frac{Q^i}{\g\cdot \g}$ is an isotopic charge, $\wt
Q=\frac{Q}{\g\cdot \g}$ is a weak hypercharge,
\bg C.21
\wt t_i=t_i,\ i=1,2,3, \q \wt t_4=y, \\
h(x,y)=h_{ab}x^a y^b. \lb C.22
\e

We consider \tf s
\begin{gather}\lb{C.23b}
\vf(x) \mapsto \wt\vf(x)=U(x)\vf(x)=\frac1{\sqrt2}\left(
\begin{matrix} 0\\v+H^0(x)\end{matrix}\right),\q\ v=\frac{2\sqrt2}{rg}\cos\ov\t \\
\wt\vf(x) = \exp\biggl(i\frac1{2v} \si^a t^a(x)\biggr)\left(
\begin{matrix} 0\\\frac{v+H^0(x)}{\sqrt2}\end{matrix}\right).\lb{C.24b}
\end{gather}
$v$ is vacuum value of a Higgs field, $g$ is a coupling \ct\ such that
$\frac{g'}g=\tan\ov\t$, $q_0= g\sin\ov\t$, $q_0$~is an elementary charge,
for intermediate bosons one gets
\begin{gather}
M^2_{W^\pm}=\frac{g^2v^2}{4}, \q M^2_{Z^0}=\frac{({g'}^2+g^2)v^2}4 \q
\hbox{or}\q M_{Z^0}=\frac{M_{W^\pm}}{\cos\ov\theta},\lb{C.25b}  \\
M_{Z^0}=\frac{\hbar c}{r_0\sqrt{2\pi}\sqrt{1+\z^2}}, \q
M_{H^0}=\frac{\sqrt{1-2\z^2}}{\cos\ov\th}\,M_{W^\pm}.\lb{C.26b}
\end{gather}
\beq C.27b
U(x)=\exp\Bigl(-\frac1{2v}\,t^a(x)\si^a\Bigr),
\e
$\si^a$ are Pauli matrices,
$H^0(x)$ is the remaining neutral scalar field after a \s y breaking and a \Hm.
It is easy to see that a Higgs' field $H^0(x)$ is colourless. One gets
\beq C.28b
\gathered
A_\mu \mapsto A_\mu^u = {\rm ad}'_{U^{-1}(x)}A_\mu + U^{-1}(x)\pa_\mu U(x)\\
F_\m \mapsto F_\m^u = {\rm ad}'_{U^{-1}(x)}F_\m.
\endgathered
\e

Let us consider the following \tf s:
\beq C.23
\left(\begin{matrix} Z^0_\mu \\ A_\mu \end{matrix}\right) =
\left(\begin{matrix} \cos\ov\t &\ & -\sin\ov\t \\ \sin\ov\t && \cos\ov\t \end{matrix}\right)
\left(\begin{matrix} A_\mu \\ B_\mu \end{matrix}\right),\q\
\left(\begin{matrix} Q^0 \\ \ov q \end{matrix}\right) =
\left(\begin{matrix} \cos\ov\t &\ & \sin\ov\t \\ -\sin\ov\t && \cos\ov\t \end{matrix}\right)
\left(\begin{matrix} \wt Q{}^3 \\ Q \end{matrix}\right).
\e
$\ov\t$ plays of course a role of the Weinberg angle $\t_W$. $Q^0$~is a neutral
weak charge and $\ov q$~an electric charge. In this way we get in Eq.~\er{C.17} a
very familiar term
\beq C.23a
-\frac{u^\b}{m_0}\,\wt g{}^\(\a\d) \ov qF_{\b\d}
\e
where
\beq C.24
F_{\b\d}=\pa_\b A_\d - \pa_\d A_\b
\e
is a strength of an \elm c field and $\ov q$ an electric charge, i.e.\ a Lorentz
force term.

One gets for $H_{56}$
\bml C.25
H_{56}=-H_{65}=-i\sin\t\Bigl(\vf_1\vf_1^* \,\frac{\a_i}{\a\cdot \a}
+\vf_2\vf_2^*\,\frac{\b_i}{\b\cdot\b} \Bigr) H_i
-i\cos\t \vf_1x_\a - i\cos\t\vf_2 x_\b\\
{}-i\cos\t\vf_1^* x_{-\a} - i\cos\t\vf_2^* x_{-\b}
+\frac i2 \sin\t \vf_1\vf_2^* \gd C,\g,{\a,-\b},x_\g
+\frac i2 \sin\t \vf_2\vf_1^* \gd C,-\g,{\b,-\a},x_{-\g}.
\e

Let us proceed a \sn\ \s y breaking and \Hm\ in our \KWK\ \e. In this way we
transform
\beq C.26
\gv{\n_\mu}\F_{\td a} \mapsto {\rm ad}'_{U^{-1}(x)}\gv{\n_\mu}\F_{\td a} = \gv{\n_\mu}\F_{\td a}^u\,,
\q \wt a=5,6,
\e
where
\begin{gather}
\gv{\n_\mu}\F_5^u = \frac1{2\sqrt2}\Bigl[\pa_\mu H^0(x)(x_\b-x_{-\b})\hskip200pt
\nn\\
\hskip40pt{}+\frac i2(v+H^0(x))\bigl(A_\mu^{3u}(x_\b+x_{-\b})+B_\mu
\tan\ov\t(x_{-\b}-x_\b)
-A_\mu^{+u}x_{-\a}+A_\mu^{-u}x_\a\bigr)\Bigr] \lb C.27 \\
\gv{\n_\mu}\F_6^u = \frac{\sin\t}{2i}\Bigl[\pa_\mu H^0(x)(x_\b+x_{-\b})
\hskip200pt \nn \\
\hskip40pt{}+\frac i2(v+H^0(x))\bigl(A_\mu^{+u}x_{-\a}-A_\mu^{-u}x_\a
+A_\mu^{3u}(x_\b-x_{-\b})+B_\mu \tan\ov\t(x_{-\b}-x_\b)\bigr)\Bigr] \lb C.28 \\
H_{56}\mapsto {\rm ad}'_{U^{-1}(x)}H_{56}^u, \lb C.29
\end{gather}
where
\bg C.30
H_{56}^u =-\frac{\sin\t(v+H^0(x))}2 \Bigl((v+H^0(x))\frac{\b_i}{\b\cdot \b}
\,H_i +\sqrt2 \cos\t(x_\b+x_{-\b})\Bigr) \\
H_{56}^u = - H_{65}^u \lb C.31
\e
where
\bea C.32
A_\mu^+ \mapsto A_\mu^{+u} &= \bigl({\rm ad}'_{U^{-1}(x)}A_\mu\bigr)^+
+\frac i{2v}\,\pa_\mu t^+(x) \\
A_\mu^- \mapsto A_\mu^{-u} &= \bigl({\rm ad}'_{U^{-1}(x)}A_\mu\bigr)^-
+\frac i{2v}\,\pa_\mu t^-(x) \lb C.33 \\
A_\mu^3 \mapsto A_\mu^{3u} &= \bigl({\rm ad}'_{U^{-1}(x)}A_\mu\bigr)^3
+\frac i{2v}\,\pa_\mu t^3(x). \lb C.33b
\e
Simultaneously we proceed a \tf\ on charges
\beq C.33a
Q \mapsto Q^u ={\rm ad}'_{U^{-1}(x)}Q.
\e
In this way we have from Eqs \er{C.14}--\er{C.17}
\begin{gather}
\frac{\wt{\ov D}u^\a}{d\tau} + \frac{u^\b}{m_0}\,\wt g{}^\(\a\d) \wt Q{}^{iu}
\d_{ij}\gd W,i,\b\d, + \frac{C^i}{m_0}u^\b h_{ij}\wt g{}^{(\a\d)}\gd F,j,\b\d,
+ \frac{u^\b}{m_0}\,\wt g{}^\(\a\d)Q^{0u}\gd Z ,0,\b\d,
-\frac{u^\b}{m_0}\,\wt g{}^\(\a\d)\ov q F_{\b\d} \nn \\
{}+\frac{1}{m_0}\,\wt g{}^\(\a\d)\Bigl(u^5h(Q^u,\gv{\n_\d}\F_5^u)
+u^6 h(Q^u,\gv{\n_\d}\F_6^u)\Bigr) - \frac{\|\wh Q\|^2}{8m_0^2}\wt g{}^{(\a\d)}\Bigl(\frac1{\rho^2}\Bigr)_{,\d}
=0 \lb C.34 \\
\frac{\wt Du^5}{d\tau} + \frac1{r^2}\,\frac{u^\b}{m_0}\,
h(Q^u,\gv{\n_\b}\F_5^u) + \frac 1{r^2}\,u^6h(Q^u,H_{56}^u)=0 \lb C.35 \\
\frac{\wt Du^6}{d\tau} + \frac1{r^2}\,\frac{u^\b}{m_0\sin^2\t}\,
h(Q^u,\gv{\n_\b}\F_6^u) + \frac 1{r^2}\,\frac{u^5}{m_0\sin^2\t}\,
h(Q^u,H_{65}^u)=0. \lb C.36
\end{gather}
In Eqs \er{C.34}--\er{C.36} a test \pc\ is coupled to physical fields only,
i.e.\ $W_\m^i$, $F_\m$ and~$H^0$.

One derives a final form of $h(Q^u,e^*(\gv{\n_\mu}\F_5))$,
$h(Q^u,e^*(\gv{\n_\mu}\F_6))$, $h(Q^u,e^*(H_{56}))$, getting
\bml C.39
h(Q^u,e^*(\gv{\n_\mu}\F_5)) = \frac1{2\sqrt2}\Bigl(
\frac2{\a\cdot\a}\bigl(q_{-\a} W_\mu^{-u} - q_\a W_\mu^{+u}\bigr)
+ \frac2{\b\cdot\b}\Bigl({\pa_\mu H^0 (q_{-\b} -q_\b)}\\
{}+\frac i2\,(v+H^0(x))(Z_\mu^{0u}\cos\ov\t + A_\mu \sin\ov\t)(q_{-\b}+q_\b)
+(-Z_\mu^{0u}\sin\ov\t \tan\ov\t + A_\mu \sin\ov\t)(q_{-\b} - q_\b)\Bigr)\\
{}+ h(x_\a,x_\a)q_\a W_\mu^{-u}
+ h(x_\b,x_\b)q_\b \Bigl(\pa_\mu H^0 +\frac i2 (v+H^0(x))
(Z_\mu^{0u}\cos\ov\t + A_\mu\sin\ov\t)\\
{}\hskip200pt +(Z_\mu^{0u}\sin\ov\t \tan\ov\t -A_\mu\sin\ov\t)\Bigr)\\
{}+ h(x_{-\a},x_{-a}) q_{-\a} W_\mu^{+u}
+ h(x_{-\b,-\b})q_{-\b} \Bigl(-\pa_\mu H^0 + \frac i2(v+H^0(x))
(Z_\mu^{0u}\cos\ov\t + A_\mu \sin\ov\t)\\
{} +(-Z_\mu^{0u}\sin\ov\t \tan\ov\t +A_\mu\sin\ov\t)\Bigr)\Bigr)
\e
\bml C.40
h(Q^u,e^*(\gv{\n_\mu}\F_6)) = \frac{\sin\t}{2\sqrt2 \,i}\biggl(
\frac{2i(v+H^0(x))}{\a\cdot\a}\,(q_\a W_\mu^{+u} - q_{-\a} W_\mu^{-u})\\
{}+ \frac2{\b\cdot\b}\Bigl(\pa_\mu H^0(q_\b+q_{-\b})
+(q_{-\b} - q_\b)\,\frac{Z_\mu^{0u}}{\cos\ov\t}\Bigr)\\
{}-\frac i2\,h(x_\a,x_\a)q_\a W_\mu^{-u} + \frac i2\,h(x_{-\a},x_{-\a})
q_{-\a}(v+H^0(x))W_\mu^{+u} \\
{}+h(x_\b,x_\b)q_\b \Bigl(\pa_\mu H^0 + \frac{Z_\mu^{0u}}{\cos\ov\t}\Bigr)
+h(x_{-\b},x_{-\b})q_\b \Bigl(\pa_\mu H^0-\frac{Z_\mu^{0u}}{\cos\ov\t}
\Bigr)\biggr)
\e
\bml C.41
h(Q^u,e^*(H_{56})) = -h(Q^u,e^*(H_{65}))\\ {}=
-\frac{\sin2\t\sqrt2}4 \biggl((q_\b + q_{-\b})
\Bigl(\frac2{\b\cdot\b}+ h(x_\b,x_\b)+h(x_{-\b},x_{-\b})\Bigr)\biggr)
\e
Here the superscript $u$ means that all quantities are in a gauge $U$. In this
way all couplings of a test \pc\ are expressed by physical fields after a
\sn\ \s y breaking and \Hm.

Let us consider Eq.~\er{C.17} in more details and let us
change a gauge using a gauge changing \f\ $U(x)$. One finds
\begin{gather}
\frac{dQ_\g^u}{d\tau} - i\bigl((Z_\mu^{0u} \cos\ov\t - \sin\ov\t A_\mu)Q_\g^u
- W_\mu^{-u}\wt q\bigr)u^\mu + \frac{v+H^0(x)}{\sqrt2}\langle \g,\a\rangle
\Bigl(u^5 + i\,\frac{\sin\t}2\,u^6\Bigr)q_\a\nn\\
\hskip150pt {}-\frac i2\,u^6\cos\t
\bigl(\langle\g,\b\rangle+\langle\g,\a\rangle\bigr)
Q_\g^u =0 \lb C.42 \\
\frac{dQ_{-\g}^u}{d\tau} - i\bigl(W_\mu^{+u}\wt Q{}^u -(Z_\mu^{0u}\cos\ov\t
-\sin\ov\t A_\mu)Q_{-\g}^u\bigr) u^\mu +
\frac1{\sqrt2}(v+H^0(x))\langle\g,\a\rangle q_{-\a}
\Bigl(u^5-i\,\frac{\sin\t}2\,u^6\Bigr)\nn\\
\hskip150pt {}+
\frac i2\,u^6\cos\t(\langle\g,\a\rangle+\langle\g,\b\rangle)Q_{\g}^u=0
\lb C.43 \\
\frac{d\ov q}{d\tau}=0 \lb C.44 \\
\frac{d\wt Q{}^u}{d\tau}-\frac12(W_\mu^{+u}Q_\g^u - W_\mu^{-u}Q_{-\g}^u)=0
\lb C.45 \\
\frac{dq_\a}{d\tau}+\frac12\,i\Bigl((\cos\ov\t A_\mu - Z_\mu^{0u}\sin\ov\t)
\langle\a,\g\rangle q_\a - 2(Z_\mu^{0u}\cos\ov\t - \sin\ov\t A_\mu)q_\a
\,\frac{\g_1\a_2-\g_2\a_1}{\g\cdot\g}\Bigr)u^\mu \nn \\
\hskip80pt {}+\frac{u^5(v+H^0(x))}{\sqrt2}\,Q_\g^u \Bigl(u^5+\frac{i\sin\t}
{2\sqrt2}\,u^6\Bigr) - \frac i2\,u^6\cos\t q_\a(2+\langle\a,\b\rangle)=0
\lb C.46 \\
\frac{dq_{-\a}}{d\tau}+\frac12\,i\Bigl(2(\cos\ov\t A_\mu - Z_\mu^{0u}\sin\ov\t)
q_{-\a} \,\frac{\g_1\a_2-\g_2\a_1}{\g\cdot\g}
+ (Z_\mu^{0u}\cos\ov\t - \sin\ov\t A_\mu)q_{-\a}\langle\a,\g\rangle \Bigr)u^\mu \nn\\
\hskip100pt{}+\frac{v+H^0(x)}{\sqrt2}\,Q_{-\g}^u \Bigl(u^5 - \frac{i\sin\t u^6}2 \Bigr)
+\frac i2\,\cos\t q_{-\a}(2+\langle\a,\b\rangle )=0 \lb C.47 \\
\frac{dC^i}{d\tau} - \gd C,i,kj, C^k\gd A,j,\nu, u^\nu=0 \nn
\end{gather}
Simultaneously we get
\beq C.48
q_\b=q_{-\b}=0.
\e

In this way our \e s are simpler, e.g. $h(Q^u,e^*(H_{56}))=h(Q^u,e^*(H_{65}))
=0$. Let us notice that $q_\a$ and $q_{-\a}$ charges are not influenced by
the gauge \tf\ $U(x)$. The electric charge~$\ov q$ does not feel any movement of
additional charges.

Thus one gets eventually
\begin{gather}
\frac{\wt{\ov D}u^\mu}{d\tau} + \frac{\wt Q{}^{iu}}{m_0}\wt g{}^\(\mu\d)
\d_{ij}u^\b W_{\b\d}^{iu} + \frac{Q^{0u}}{m_0}\,\wt g{}^\(\mu\d)u^\b Z_{\b\d}^{0u}
+\frac q{m_0}\,\wt g{}^\(\mu\d)u^\b F_{\b\d} + \frac{c^i}{m_0}u^\b
h_{ij}\wt g{}^{(\a\d)}\gd F,i,\b\d, \nn \\
{}+\frac1{\sqrt2\,m_0}\,\wt g{}^\(\mu\d)\biggl(u^5\Bigl(\frac1{\a\cdot\a}
(q_{-\a}W_\d^{-u} {-} q_\a W_\d^{+u}) + \frac1{\b\cdot\b}\bigl(
h(x_\a,x_\a)q_\a W_\d^{-u}
+h(x_{-\a},x_{-\a})q_{-\a}W_\d^{+u}\bigr) \Bigr) \nn \\
\kern-19pt{}+u^6 \sin\t\Bigl(\frac{v{+}H^0(x)}{\a\cdot\a}(q_\a W_\d^{+u}{-}q_{-\a}W_\d^{-u}
)-\frac1{2(\b\cdot\b)}\bigl(h(x_\a,x_\a)q_{-\a}(v{+}H^0(x))\bigr)W_\d^{+u}\Bigr)
\biggr)\nn \\
\hskip200pt - \frac{\|\wh Q\|^2}{8m_0^2} \wt g{}^{(\a\d)} \Bigl(\frac1{\rho^2}\Bigr)_{,\d}=0\kern-3pt
\lb C.49 \\
\frac{\wt Du^5}{d\tau} + \frac1{r^2}\,\frac{u^\b}{\sqrt2\,m_0} \Bigl(
\frac1{\a\cdot\a}(q_{-\a}W_\b^{-u} - q_\a W_\b^{+u})\hskip80pt \nn \\
\hskip80pt {}+\frac1{\b\cdot\b}\bigl(h(x_\a,x_\a)q_\a W_\b^{-u} + h(x_{-\a},x_{-\a})
q_{-\a}W_\b^{+u}\bigr)\Bigr)=0 \lb C.50 \\
\frac{\wt Du^6}{d\tau} + \frac 1{r^2}\,\frac{u^\b}{\sqrt2\,m_0\sin\t}\Bigl(
\frac{v+H^0(x)}{\a\cdot\a}(q_\a W_\b^{+u} - q_{-\a}W_\b^{-u})\hskip80pt \nn \\
\hskip80pt {}+\frac1{2(\b\cdot\b)}\bigl(-h(x_\a,x_\a)q_\a W_\b^{-u} + h(x_{-\a},x_{-\a})
(v+H^0(x))W_\b^{+u}\bigr)\Bigr) =0. \lb C.51
\end{gather}

Let us suppose that $H=G2$. In this case one gets
\beq C.52
\bga
|\b|=|\a|=\sqrt2, \q |\g|=\sqrt6,\\
\a\cdot\a=\b\cdot\b=2, \q \g\cdot\g=6,\\
\langle \g,\a \rangle=3, \q \langle \g,\b \rangle=\langle \a,\b \rangle=-1,\\
\frac{\g_1\a_2-\g_2\a_1}{\g\cdot\g}=\frac{\sqrt3}6,\\
\ov\t=30^\circ, \q \cos\ov\t=\frac{\sqrt3}2\,, \q \sin\ov\t=\frac12\,.
\ega
\e
Thus one gets
\begin{gather}
\frac{dQ_\g^u}{d\tau} - i\Bigl(\frac12(Z_\g^{0u}\sqrt3 - A_\mu)Q_\g^u
-W_\g^{-u}\wt q\Bigr)u^\mu \hskip100pt \nn \\
\hskip100pt {}- \frac{3(v+H^0(x))}{\sqrt2}\,q_\a
\Bigl(u^5+i\,\frac{\sin\t}2\,u^6\Bigr) - i\cos\t Q_\g^u=0 \lb C.53 \\
\frac{dQ_{-\g}^u}{d\tau} - iu^\mu \Bigl(W_\mu^{+u}\wt Q{}^u
-\frac12(\sqrt3\,Z_\mu^{0u}-A_\mu)Q_{-\g}^u\Bigr)\hskip100pt \nn \\
\hskip100pt {}+\frac{3u^5}{\sqrt2}(v+H^0(x))q_{-\a}\Bigl(u^5-i\,\frac{\sin\t}
2\, u^6\Bigr) + iu^6\cos\t Q_{-\g}^u=0 \lb C.54 \\
\frac{d\ov q}{d\tau}=0 \lb C.55 \\
\frac{d\wt Q{}^u}{d\tau} - \frac12 (W_\mu^{+u}Q_\g^u - W_\mu^{-u}Q_{-\g}^u)=0
\lb C.56 \\
\frac{dq_\a}{d\tau} + \frac12\,i\Bigl(\frac32(\sqrt3\,A_\mu - Z_\mu^{0u})
-\frac{\sqrt3}6 (\sqrt3\,Z_\mu^{0u}-A_\mu)\Bigr)q_\a u^\mu \hskip100pt \nn \\
\hskip100pt {}+\frac{u^5(v+H^0(x))}{\sqrt2}\,Q_\g^u
\Bigl( u^5 + \frac{i\sin\t}{2\sqrt2}\,u^6\Bigr)
-\frac i2\,u^6\cos\t q_\a=0 \lb C.57 \\
\frac{dq_{-\a}}{d\tau} + \frac i2\Bigl(\frac{\sqrt3}6(\sqrt3\,A_\mu -
Z_\mu^{0u}) + \frac32(\sqrt3\,Z_\mu^{0u} - A_\mu)\Bigr) u^\mu q_{-\a}
\hskip100pt \nn \\
\hskip100pt {}+\frac{v+H^0(x)}{\sqrt2}\,Q_{-\g}^u \Bigl(u^5 - \frac{i\sin\t}2
\,u^6\Bigr)+\frac i2 \cos\t q_{-\a}=0. \lb C.58 \\
\frac{dC^i}{d\tau} - \gd C,i,kj, C^k \gd A,j,\nu, u^\nu=0 \lb c161 \\
\frac{dC^{\ba}}{d\tau} - \gd C,\ba,\bar b\bar c, C^{\bar b}A^{\bar c}u^\nu
=0. \lb c161b
\end{gather}

Eqs \er{C.49}--\er{C.51} and \er{C.53}--\er{c161} are generalized \KWK\ \e s
in our partial \un.

Moreover if we suppose a colour \cfn\ from QCD we should put $C^i=0$ or $C^{\bar a}=0$, and some
formulas are simpler. In this way our test \pc\ is colourless.

According to Section 6 we can consider the energy of a test \pc\
$$
E_p= m_0g_\(\a\b) u^\a u^\b
$$
getting the \fw\ \e s:
\bg c162
E_p+\frac{\|\wh Q\|^2}{4\rho^2 m_0} = {\rm const}' \\
\frac{dE_p}{d\tau} = -\frac{\|\wh Q\|^2}{4m_0}\,\frac d{d\tau}\Bigl(\frac1{\rho^2}\Bigr) \lb c163
\e
where
\bml c164
\|\wh Q\|^2 = -h_{ab}Q^aQ^b - h_{ij}C^iC^j = \ov q{}^2 - \wt Q{}^{iu}\wt Q{}^{ju}\d_{ij}
-(Q^{0u})^2 - \gd h,\SU(3),\ba \bar b,C^\ba C^{\bar b} = {\rm const}'', \\ \ba,\bar b=7,8,\dots,14
\e
(both notations for a Killing--Cartan tensor on $\SU(3)_c$).

\setbox9=\hbox{$\displaystyle{\frac{C^\ba}{m_0}\,\frac{dx^\b}{d\tau}\gd
h,\SU(3),\ba \bar b,\gtn{\a\d}\gd F,\bar b,\b\d,}$}
\setbox8=\hbox{\begin{picture}(119,17)(0,0)
\unitlength1pt
\put(0,0){\copy9}
\put(0,17){\line(5,-1){119}}
\end{picture}}

Let us write our generalized \KWK \e\ in the \fw\ form:
\begin{gather}
\frac{d^2x^\a}{d\tau^2} + {\a\brace \b\g}\frac{dx^\b}{d\tau}\,\frac{dx^\g}{d\tau}
+ \frac1{m_0}\,\frac{dx^\b}{d\tau}\gtn{\a\d}\wt Q{}^{iu}\d_{ij}W^{iu}_{\b\d}
+\frac{Q^{0u}}{m_0}\gtn{\mu\d}\frac{dx^\b}{d\tau}Z^{0u}{\b\d}
+\frac q{m_0}\gtn{\mu\d}\frac{dx^\b}{d\tau}u^\b F_{\b\d} \nn \\
{}+ \copy8 + \frac1{\sqrt2\,m_0}\gtn{\mu\d}\Biggl(\frac{d\t}{d\tau}
\biggl(\frac1{\a\cdot\a}(q_{-\a}W^{-u}_{\d} - q_\a W^{+u}_\d) \nn\\
{}+\frac1{\b\cdot\b}\bigl(h(x_\a,x_\a)q_\a W^{-u}_\d + h(x_{-\a},x_{-\a})
q_{-\a}W^{+u}_\d\bigr)\biggr) \nn\\
{}+ \frac{d\psi}{d\tau} \sin\t \biggl(
\frac{\g+H^0(x)}{\a\cdot\a}(q_\a W^{+u}_\d - q_{-\a} W^{-u}_\d)
- \frac1{2(\b\cdot\b)}\bigl(h(x_a,x_\a)q_{-\a}\cdot(v+H^0(x))\bigr)W^{+u}_\d\biggr)
\Biggr)\nn\\
\hskip200pt {}- \frac{\|\wh Q\|^2}{8m_0^2}\gtn{\a\d}\Bigl(\frac1{\rho^2}\Bigr)_{,\d}=0 \lb c165 \\
\frac{d^2\t}{d\tau^2} + \frac12\sin2\t\Bigl(\frac{d\psi}{d\tau}\Bigr)^2
-\frac1{r^2}\Bigl(\frac{dx^\b}{d\tau}\Bigr)\frac1{\sqrt2\,m_0}\biggl(
\frac1{\a\cdot\a}(q_{-\a}W^{-u}_\b - q_{-\a}W^{+u}_\b) \nn\\
{}+\frac1{\b\cdot\b}(h(x_\a,x_\a)q_\a W^{-u}_\b + h(x_{-\a},x_{-\a})q_{-\a}
W^{+u}_\b)\biggr) = 0 \lb c166 \\
{}+ \frac{d^2\psi}{d\tau^2} + 2\cot\t\Bigl(\frac{d\t}{d\tau}\Bigr)
\Bigl(\frac{d\psi}{d\tau}\Bigr) + \frac1{r^2}\Bigl(\frac{dx^\b}{d\tau}\Bigr)
\frac1{\sqrt2 \,m_0\sin\t}\biggl(\frac{v+H^0(x)}{\a\cdot\a}(q_\a W^{+u}_\b
-q_{-\a} W^{-u}_\b) \nn\\
+ \frac1{2(\b\cdot\b)}(-h(x_\a,x_\a)q_\a W^{-u}_\b
+ h(x_{-\a},x_{-\a})(v+H^0(x)) W^{+u}_\b)\biggr)=0 \lb c167
\end{gather}
where
\setbox9=\hbox{$\gd h,\SU(3),\ba \bar b, C^\ba C^{\bar b}$}
\setbox7=\hbox{\begin{picture}(62,11)(0,0)
\unitlength1pt
\put(0,0){\copy9}
\put(0,12){\line(5,-1){62}}
\end{picture}}
\beq c168
\|\wh Q\|^2 = \ov q{}^2 - \wt Q{}^{iu} \d_{ij} \wt Q{}^{iu} - (Q^{0u})^2
- \copy7 = {\rm const}
\e
\copy8\ in \er{c165} and \copy7\ in \er{c168} \vadjust{\vskip6pt}
mean that both terms should be
removed supposing colourless test \pc.
${\a\brace \b\g}$ are Christoffel symbols for $g_\(\a\b)$ and
$$
u^\mu = \frac{dx^\mu}{d\tau}, \q u^5=\frac{d\t}{d\tau}, \q u^6=\frac{d\psi}{d\tau},
\q {6\brace 56}=\cot\t, \q {5\brace66}=\frac12 \sin2\t.
$$
All remaining Christoffel symbols on $S^2$ for the metric \er{c20} are zero,
$\t\in\langle -\frac\pi2,\frac\pi2\rangle$, $\psi\in(0,2\pi\rangle$,
$\t$~is the fifth and $\psi$ is the sixth coordinate.

\advance\abovedisplayskip by-1pt
\advance\belowdisplayskip by-1pt
Eqs \er{c165}--\er{c168} give us \KWK \e s in our partial \un\ theory
of \gr al, electro-weak and strong \ia s in \E\nos\ \KK (Jordan--Thiry) Theory.
They are supplemented by Eqs \er{C.54}--\er{c161} or
\beq c161a
\frac{dC^\ba}{d\tau} - \gd C,\ba,\bar b\bar c, C^{\bar b}\gd A,\bar c,\nu,
\frac{dx^\nu}{d\tau} = 0, \q\ \ba,\bar b,\bar c = 7,8,\dots,14,
\e
in place of Eq.~\er{c161}.

Moreover in our theory we suppose a colour \cfn\ for strong \ia s described
by QCD ($\SU(3)_c$-gauge group). This means that colour charges of a test \pc\
$C^\ba$, $\ba=7,8,\dots,14$, are zero,
$$
C^\ba = 0.
$$
This results in removing a Lorentz-like term for colour-like charges
of Eqs \er{c161}, \er{c161a}.

Let us come back to Eqs \er{c161}, \er{c161b} in the case of $C^i=C^\ba=0$.
One gets
\setbox9=\hbox{$\displaystyle \frac{dC^i}{d\tau} - \gd C,i,kj, C^k \gd A,j,\nu,
\frac{dx^\nu}{d\tau}=0$}
\setbox7=\hbox{\begin{picture}(123,17)(0,0)
\unitlength1pt
\put(0,0){\copy9}
\put(0,15){\line(6,-1){123}}
\end{picture}}
$$
\copy7
$$
and we have the trivial \e
$$
0=0
$$
\setbox9=\hbox{$\displaystyle \frac{dC^\ba}{d\tau} - \gd C,\ba,\bar b\bar c, C^{\bar b} \gd A,\bar c,\nu,
\frac{dx^\nu}{d\tau}=0$}
or
$$
\begin{picture}(123,17)(0,0)
\unitlength1pt
\put(0,0){\copy9}
\put(0,15){\line(6,-1){123}}
\end{picture}
$$
with the same result
$$
0=0
$$
which is due to the supposed \cfn\ condition (colourless test \pc).

Let us consider the term
\beq c169
-\frac{\|\wh Q\|^2}{8m_0^2}\gtn{\a\d} \biggl(\frac1{\rho^2}\biggr)_{,\d}
\e
where $\|\wh Q\|^2$ is given by \er{c164}. One gets according to \er{6.13}
\beq c170
-\frac{\|\wh Q\|^2}{4m_0^2} e^{2(\Psi_0+\vf)}\gtn{\a\d}\vf_{,\d}.
\e
This term can be considered as a fluctuating force after an inflation and
\ssb. In this way we can consider a chaotic behaviour of a test \pc\ motion.
On the \co ical scale a field~$\vf$ is equal to zero ($\Psi=\Psi_0$). If $\vf$
fluctuates strongly (even $\vf$~is small) the term \er{c170} can be quite
large. This problem demands more investigations.

Simultaneously we shall examine the problem of travelling in higher dimensions,
i.e.\ in the fifth and sixth dimensions on~$S^2$ (see Eqs \er{c166}
and~\er{c167}). Both problems can be combined.

Let us apply some results from Appendix~B to our partial \un, i.e.\ a tower
of scalar fields as additional Dark Matter. In this case one has $M=S^2$,
$\rho=\rho(x,y)$, $x\in E$, $y\in S^2$. It is easy to see that it is better to
use $\Psi=\Psi(x,y)$. We have of course $y=(\t,\psi)$, as usual. An application
of Friedrichs theory is very easy for the operator $\wt L=-\bigl(\ov M
+\frac{n^2\z^2}{(\z^2+1)}\bigr)\Delta$, where $\Delta$ is a usual Laplace operator
on~$S^2$, i.e.\ Legendre operator in $\t$ and~$\psi$ coordinates,
$\D=-\bigl[\frac1{\sin\t}\,\pp{}\t(\sin\t\pp{}\t)+\frac1{\sin^2\t}\pp{^2}{\psi^2}\bigr]$.
Now $n=8+14 =22$, $n^2=484$. In particular we put $\z=\pm 0.91622i$,
moreover we will still consider $\z$ as a free parameter in further considerations.
\bg c172
\ov M+ \frac{n^2\z^2}{(\z^2+1)} \approx \bigl(((\mu\cdot \ell^{\SU(3)[ab]}
\gd k,\SU(3),ab, +i\xi \ell^{G2[cd]}\cdot \gd k,G2,cd, )-1386)-2398\bigr)\\
\ov M + \frac{n^2\z^2}{\z^2+1} = (\mu \ell^{\SU(3)[ab]}\gd k,\SU(3),ab, - 3784) \qh{for $\xi=0$}\nn
\e
where we use \er{16.41}--\er{16.43} and \er{c55}. In this way one gets a mass
spectrum for scalar  \pc s
\bg c173
\wt m(l,m) = \frac{\hbar c}r \sqrt{\biggl(1+\frac{484\z^2}{\ov M(1+\z^2)}
\biggr) l(l+1)},\\
m=-l,-l+1,-l+2,\dots,0,1,\dots,l-1,l, \q\ l=0,1,2,3,\dots. \lb c174
\e
Thus one gets eventually
\beq c175
\wt m(l,m) \approx \frac{\hbar c}r \biggl(\biggl( 1 - \frac{2398}{(\mu \ell^{\SU(3)[ab]}
\gd k,\SU(3),ab, +i\xi \ell^{G2[cd]}\gd k,G2,cd,)-1386}\biggr)l(l+1)\biggr)^{1/2}
\e
with the degeneracy \er{c174}.

The field $\Psi$ has the following shape:
\beq c176
\Psi(x,\t,\psi) =\sum_{\dwa{l=0\\|m|\le l}} \Psi_{lm}(x)Y_{lm}(\t,\psi),
\e
$Y_{lm}(\t,\psi)$ are usual harmonic \f s on~$S^2$.
Moreover, $\Psi(x,\t,\psi)$ is a real \f. Thus
\beq c177
\Psi(x,\t,\psi)=\frac1{\sqrt{2\pi}}\biggl[\Psi_{00}(x) + \sum_{\dwa{l=1\\0<m\le l}}^\iy
(\Psi_{lm}(x)+\Psi_{l-m})\Theta_{lm}(\t)\cos m\psi\biggr],
\e
where for $m\ge0$,
\begin{gather*}
\Theta_{lm}(\t) = (-1)^m \biggl[\frac{(2l+1)(l-m)!}{2(l+m)!}\biggr]^{1/2}
\sin^m\t \frac{d^m}{dt^m}P_l(t)|_{t=\cos\t},\\
\hbox{and }P_l(t)= \frac1{l!\,2^l}\,\frac{d^l}{dt^l}[(t^2-1)^l]
\end{gather*}
are Legendre polynomials.
The \lg\ for $\Psi(x,\t,\psi)$ in terms of $\Psi_{lm}(x)$ is as follows (after
averaging on~$S^2$):
\beq c178
\cL\dr{scal}(\Ps) = \sum_{\dwa{l=0\\|m|\le l}}(\cL\dr{scal}(\Ps_{lm})+
V(\{\Psi_{lm}\}))
\e
where
\bg c179
\cL\dr{scal}(\Ps_{lm}) = (\ov M\wt g{}^\(\g\nu) + 484 g^\[\mu\nu]
g_{\d\mu}\wt g^\(\d\g) )\cdot\Psi_{lm,\nu}\cdot\Psi_{lm,\g}
-\tfrac12 \wt m{}^2(l,m)\Psi^2_{lm}\\
\ov M = \bigl(\mu \ell^{\SU(3)[ab]}\gd k,\SU(3),ab, + i\xi \ell^{G2[cd]}\gd
k,G2,cd,\bigr) - 1386 \lb c180 \\
\hbox{for $\xi=0$,}\q \ov M = (\mu\ell^{\SU(3)[ab]}\gd k,\SU(3),ab,) - 1386 \nn\\
\hbox{and} \ V(\{\Psi_{lm}\}) = \frac1{4\pi}\intop_{-\pi/2}^{\pi/2} \,\intop_0^{2\pi}\sin\t
\,d\t\, d\psi\,\biggl(\frac{\a^2_s}{\ell^2\pl}\cdot \wt R(\wt \G)\cdot
e^{24\Psi(x,\t,\psi)}+ \frac1{r^2}\,e^{22\Psi(x,\t,\psi)}\wh{\ov R}(\wh{\ov\G})
\biggr). \lb c181
\e
This is Self-Interacting Dark Matter (SIDM).

\advance\abovedisplayskip by1pt
\advance\belowdisplayskip by1pt
In Appendix B we give a condition on a mass spectrum of scalar \pc s
(a special condition of a truncation)
$$
m \le m\pl.
$$
In our case it means
\beq c182
\wt m(l,m) \le m\pl.
\e
Using Eq.~\er{c175} one gets for large $l$
\beq c183
l \le \a\, 10^{17}
\e
where $\a$ is of order one. In the case of the zero \co ical \ct\ (it could
happen in the case of special arrangement of parameters) our Dark Matter
(scalarons and skewons)
becomes a radiation. Moreover, some inhomogeneous \co ical models can work
with our additional scalar \pc s as is described in Appendix~B (without an
accelerated expansion in Robertson--Walker--Friedman \Un). We can use \co
ical models of Lema\^\i tre--Tolman \cite{xxy,xxz,xxr} and be in agreement
with observation data. Moreover, it seems according to recent data that the
\Un\ is more homogeneous than it was supposed before.

It is interesting to consider some terms of the lowest order in Eq.~\er{c177},
\st
\beq c184
\wt m(\ell,m) \approx \frac{c\hbar}r = EW,
\e
i.e. their masses are about electroweak energy scale. One gets
\beq c185
\Psi(x,\t,\psi)\approx \frac1{\sqrt{2\pi}}\biggl(\Psi_{00}(x) - \frac{\sqrt3}2
\bigl((\Psi_{1,-1}(x) + \Psi_{1,1}(x))\cos\psi + \Psi_{10}(x)\bigr)\sin\t
\biggr).
\e
The field $\Psi_{00}(x)$ is a massless scalar field and corresponds to our
field $\Psi(x)$. Thus we can consider a mass generation for $\Psi_{00}(x)$
due to a \co ical \ct\  in such a way that
\beq c186
\Psi_{00}(x)=\Psi(x)=\Psi_0+\vf(x).
\e
The remaining fields in Eq.~\er{c177} are massive scalar fields with the mass
\beq c187
\wt m(1,-1) = \wt m(1,0) = \wt m(1,-1) = \wt m.
\e
If the \co ical \ct\ is zero, $\Psi_{00}$ is a radiation and we have as a low
energy Dark Matter $\Psi_{1,-1}(x)$, $\Psi_{1,0}(x)$, $\Psi_{1,1}(x)$ with
the common mass
\beq c188
\wt m \approx \frac{\sqrt2\,\hbar c}r \biggl( 1 - \frac{2398}{(\mu \ell^{\SU(3)[ab]}
\gd k,\SU(3),ab, +i\xi \ell^{G2[cd]}\gd k,G2,cd,)-1386}\biggr)^{1/2}.
\e
If
\beq c189
\wt m(\ell,m) \le M_G \le 2\cdot 10^{16}\,{\rm GeV}.
\e
($M_G$ is a Grand Unification Energy scale) then one finds
\beq c190
\ell \le \a_0\cdot 10^{14}
\e
where $\a_0$ is of order 1.
\beq c191
\Psi_{00}=\Psi=\Psi_0+\vf
\e
can fluctuate according to Eq.~\er{aa89}.

Let us come back to a \lg\ of a partial \un, i.e.\ Eq.~\er{c2}, and to a problem
of field \e\ in that theory. We get Eq.~\er{c37}. Moreover in this case we use
the notation from Eq.~\er{4.109}. Using results from Ref.~\cite{11a} and from
Sections~2 and~4 one gets
\beq c102b
\cLY(\a\dr{QCD}A_{\SU(3)}) = -\frac{\a^2\dr{QCD}}{8\pi}\,\ell^{\SU(3)}_{ab}
\bigl(2H^aH^b - L^{a\mu\nu}\gd H,b,\mu\nu,\bigr)
\e
where $\ell^{\SU(3)}_{ab}= h^{\SU(3)}_{ab}+ \mu k^{\SU(3)}_{ab}$
is a \nos\ right-invariant tensor on the group $\SU(3)$, $h^{\SU(3)}_{ab}$
is a Killing--Cartan tensor on SU(3),
\bea c103b
H^a &= \gtk{\m}\gd H,a,\m,\\
L^{a\m} &= g^{\a\mu}g^{\b\nu} \gd L,a,\a\b, \lb c104b
\e
and $L^{a\m}$ is given by Eq.~\er{2.125} for the group SU(3). This is a \lg\
for~QCD.

We have for the SU(3) group
\bml c105b
\gd L,n,\o\mu, = \gd H,n,\o\mu, + \mu h^{\SU(3)na}\gd k,\SU(3),[ad],\gd H,d,\o\mu,
+\bigl(\gd H,n,\a\o,\wt g{}^{(\a\d)}g\ink{\d\mu} - \gd H,n,\a\mu,\gtn{\a\d}g\ink{\d\o}\bigr)\\
-2\mu h^{\SU(3)na}\gd k,\SU(3),[ad],\gtn{\d\tau}\gtn{\a\b}\cdot \gd H,d,\d\a,
g\ink{\tau\o}g\ink{\b\mu}
-2\mu h^{\SU(3)}\gd k,\SU(3),ad, \gtn{\d\b}\gtn{\a\tau}\gd H,d,\b[\o,g_{\mu]\tau}
g\ink{\d\a}\\ + 2\mu^2h^{\SU(3)na}h^{\SU(3)bc}\gd k,\SU(3),[ac],\gd k,\SU(3),[bd],
\gd H,d,\a[\o,g_{|\mu|\b]}.
\e

Finally $\cLY(\a_{QCD}A_{\SU(3)})$ reads
\bml c106b
\cLY(\a_{QCD}A_{\SU(3)})\\
{} = -\frac{\a^2_{QCD}}{8\pi}\Bigl(\gd h,\SU(3),nk,
H^{k\o \mu}\gd H,n,\o\mu, - 2\gd h,\SU(3),cd, H^cH^d + 2\gd h,\SU(3),nk,
H^{k\o\mu}\gd H,n,\d\o, g\ink{\a\mu}\gtn{\a\d}\Bigr) \\
{}+ \mu\Bigl[2\gd k,\SU(3),[nk],
H^{k\o\mu}\gd H,n,\d\o,\gtn{\d\a}g\ink{\a\mu} - 2\gd k,\SU(3),[kd],H^{k\o\mu}
\gd H,d,\d\a,\gtn{\d\b}\gtn{\a\rho}g\ink{\b\o}g\ink{\rho\mu}\\
{} - \gd k,\SU(3),[kd],H^{k\o\mu}\gd H,d,\eta\mu,\gtn{\a\rho}g\ink{\mu\a}g\ink{\b\rho}
+ \gd k,\SU(3),[kd], H^{k\o\mu}\gd H,d,\eta\mu, \gtn{\eta\d}\gtn{\a\rho}g\ink{\d\b}g\ink{\o\d}\Bigr]\\{}
+ \mu^2\Bigl[ \gd k,\SU(3),[nk], \gd k,\SU(3)[n\cdot,d], H^{k\o\mu}\gd H,d,\eta\mu,
\gtn{\rho\b}\gtn{\eta\a}g\ink{\o\b}g\ink{\a\rho}\\
{} - 2\gd k,\SU(3),[nk],\gd k,\SU(3)[n\cdot,d],
H^{k\o\mu}\gd H,d,\d\a,\gtn{\d\eta}\gtn{\a\rho}g\ink{\eta\o}g\ink{\rho\mu}\\
{} - \gd k,\SU(3),[nk], \gd k,\SU(3)[n\cdot,d],
H^{k\o\mu}\gd H,d,\eta\o, \gtn{\rho\a}\gtn{\eta\b}g\ink{\mu\a}g\ink{\b\rho}
+ \gdg k,\SU(3),[k\cdot,b], \gd k,\SU(3),[nb], H^{k\o\mu} \gd H,d,\a\o,
\gtn{\a\b}g\ink{\o\b}\\{} + \gd k,\SU(3)[p\cdot,n], \gd k,\SU(3),[pk],H^{k\o\mu}
\gd H,n,\o\mu,\Bigr] + \mu^3 \Bigl[\gd k,\SU(3),[nk], k^{\SU(3)[nb]}\gd k,\SU(3),[bd],
\gd H,k,\o\mu, \gd H,d,\a\o, \gtn{\a\b}g\ink{\mu\b}\\{} - \gd k,\SU(3),[nk],
k^{\SU(3)[nb]}\gd k,\SU(3),[bd], H^{k\o\mu}\gd H,d,\a\mu,\gtn{\a\b}g\ink{\o\b}\Bigr]
\e
\vskip-18pt
\bml c107b
\cLY(\a_sA_{\SU(2)\ot \U(1)}) = -\frac{\a_s^2}{8\pi}\Bigl(\ell^{\SU(2)}_{ab}
(2\ov H{}^a\ov H{}^b - \ov L{}^{a\m}\gd \ov H{},b,\m,)\Bigr)\\{} + \frac{\a_s^2}{8\pi}\Bigl(2(\gtk{\m}
 G_{\m})^2 - (g^{\mu\a}g^{\nu\b} - g^{\nu\b}\gtn{\mu\a} +
g^{\nu\b}g^{\mu\o} - \gtn{\tau\a}g_{\o\tau})G_{\a\b} G_{\m}\Bigr)
\e
$\ell^{\SU(2)}_{ab}= h^{\SU(2)}_{ab}+i\xi k^{\SU(2)}_{ab}$ is a \nos\ hermitian
right-invariant tensor on the group SU(2), $h^{\SU(2)}_{ij}$ is a Killing--Cartan
tensor on SU(2):
\bea c108b
\ov H^a &= \gtk{\m} \gd \ov H,a,\m,\\
\ov L{}^{a\m} &= g^{a\mu}g^{\b\nu}\gd \ov L,a,\a\b, \lb c109b
\e
$G_{\a\b}=\pa_\a B_\b - \pa_\b B_\a$ is a strength of a gauge field for the
$\U(1)\dr Y$ group.

$\gd L,a,\a\b,$ is given by Eq.~\er{2.125} for the group SU(2).
\bml c107c
\gd \ov L{},n,\o\mu, = \gd \ov H{},n,\o\mu, + i\xi h^{\SU(2)na}\gd k,\SU(2),[ad],\gd \ov H{},d,\o\mu,
+\bigl(\gd \ov H{},n,\a\o,\wt g{}^{(\a\d)}g\ink{\d\mu} - \gd \ov H{},n,\a\mu,\gtn{\a\d}g\ink{\d\o}\bigr)\\
-2i\xi h^{\SU(2)na}\gd k,\SU(2),[ad],\gtn{\d\tau}\gtn{\a\b}\cdot \gd \ov H{},d,\d\a,
g\ink{\tau\o}g\ink{\b\mu}
-2i\xi h^{\SU(2)na}\gd k,\SU(2),ad, \gtn{\d\b}\gtn{\a\tau}\gd \ov H{},d,\b[\o,g_{\mu]\tau}
g\ink{\d\a}\\ - 2\xi^2h^{\SU(2)na}h^{\SU(2)bc}\gd k,\SU(2),[ac],\gd k,\SU(2),[bd],
\gd \ov H{},d,\a[\o,g_{|\mu|\b]}.
\e
One gets:
\bml c108c
\cLY(\a_{s}A_{\SU(2)})\\
{} = -\frac{\a^2_{s}}{8\pi}\Bigl(\gd h,\SU(2),nk,
\ov H{}^{k\o \mu}\gd \ov H{},n,\o\mu, - 2\gd h,\SU(2),cd, \ov H{}^c\ov H{}^d + 2\gd h,\SU(2),nk,
\ov H{}^{k\o\mu}\gd \ov H{},n,\d\o, g\ink{\a\mu}\gtn{\a\d}\Bigr) \\
{}+ i\xi\Bigl[2\gd k,\SU(2),[nk],
\ov H{}^{k\o\mu}\gd \ov H{},n,\d\o,\gtn{\d\a}g\ink{\a\mu} - 2\gd k,\SU(2),[kd],\ov H{}^{k\o\mu}
\gd \ov H{},d,\d\a,\gtn{\d\b}\gtn{\a\rho}g\ink{\b\o}g\ink{\rho\mu}\\
{} - \gd k,\SU(2),[kd],\ov H{}^{k\o\mu}\gd \ov H{},d,\eta\mu,\gtn{\a\rho}g\ink{\mu\a}g\ink{\b\rho}
+ \gd k,\SU(2),[kd], \ov H{}^{k\o\mu}\gd \ov H{},d,\eta\mu, \gtn{\eta\d}\gtn{\a\rho}g\ink{\d\b}g\ink{\o\d}\Bigr]\\{}
- \xi^2\Bigl[ \gd k,\SU(2),[nk], \gd k,\SU(2)[n\cdot,d], \ov H{}^{k\o\mu}\gd \ov H{},d,\eta\mu,
\gtn{\rho\b}\gtn{\eta\a}g\ink{\o\b}g\ink{\a\rho}\\
{} - 2\gd k,\SU(2),[nk],\gd k,\SU(2)[n\cdot,d],
\ov H{}^{k\o\mu}\gd \ov H{},d,\d\a,\gtn{\d\eta}\gtn{\a\rho}g\ink{\eta\o}g\ink{\rho\mu}\\
{} - \gd k,\SU(2),[nk], \gd k,\SU(2)[n\cdot,d],
\ov H{}^{k\o\mu}\gd \ov H{},d,\eta\o, \gtn{\rho\a}\gtn{\eta\b}g\ink{\mu\a}g\ink{\b\rho}
+ \gdg k,\SU(2),[k\cdot,b], \gd k,\SU(2),[nb], \ov H{}^{k\o\mu} \gd \ov H{},d,\a\o,
\gtn{\a\b}g\ink{\o\b}\\{} + \gd k,\SU(2)[p\cdot,n], \gd k,\SU(2),[pk],\ov H{}^{k\o\mu}
\gd \ov H{},n,\o\mu,\Bigr] - i\xi^3 \Bigl[\gd k,\SU(2),[nk], k^{\SU(2)[nb]}\gd k,\SU(2),[bd],
\gd \ov H{},k,\o\mu, \gd \ov H{},d,\a\o, \gtn{\a\b}g\ink{\mu\b}\\{} - \gd k,\SU(2),[nk],
k^{\SU(2)[nb]}\gd k,\SU(2),[bd], \ov H{}^{k\o\mu}\gd \ov H{},d,\a\mu,\gtn{\a\b}g\ink{\o\b}\Bigr]
\e
\vskip-18pt

\bml c110b
\cL\dr{kin}(\n\Phi) = \frac1{4\pi} \int_{S^2} \sin\th \,d\th\,d\psi\biggl[
\ell^{G2}_{nk} g^{\o\mu}g^{\td m\td p}\brgg\mu \gd\Phi,k,\td p,
\Bigl\{\brgg\o \gd\Phi,n,\td m,\\ {}+ i\xi\gd k,G2[n,d], \brgg\o \gd\Phi,d,\td m,
-\z \brgg\o\gd\Phi,d,\td a,\cdot h^{S^2\td a\td q}\gd k,S^2,\td q\td m,
-\brgg\a \gd\Phi,a,\td m,\gtn{\a\eta}g\ink{\eta\o}\\
{}-2i\xi\z\brgg\d \gd\Phi,d,\td a, \gd k,G2[n\cdot,d],\gtn{\d\a}g\ink{\a\o}
h^{S^2\td d\td q}\gd k,S^2,[\td q\td m],
- i\xi\bigl(\z^2 \gd k,G2[n\cdot,d],\brgg\o \gd\Phi,d,\td a,h^{S^2\td b\td q}
h^{S^2\td a\td \o}\cdot k^{S^2\td a\td q}\gd k,S^2,\td w\td b,\\ {}+ \gd k,G2[n\cdot,d],\cdot
\brgg\b \gd \Phi,d,\td m, \gtn{\a\nu}\gtn{\b\rho}g\ink{\nu\o}g\ink{\rho\a}\bigr)\\
{}-\xi^2\bigl(\z k^{G2[nb]}\gd k,G2,[bd],\brgg{}\gd\Phi,d,\td a,h^{S^2\td a\td w}
\gd k,S^2,\td q\td m, + \brgg\a \gd \Phi,d,\td m,\gtn{\a\b}g\ink{\b\o}\bigr)\Bigr\}\biggr]
\e
where $\ell^{G2}_{ab} = h^{G2}_{ab}+i\xi k^{G2}_{ab}$ is a \nos, right-\iv t, hermitian
tensor on the G2 group, $h^{G2}_{ab}$ is a Killing--Cartan tensor on~G2,
$\wt g_{\td a\td b}= h^{S^2}_{\td a\td b}+\z k^{S^2}_{\td a\td b}$ is a hermitian
tensor on~$S^2$. $h^{S^2}_{\td a\td b}$ is a Riemannian tensor on~$S^2$ and
$\gd k,S^2,[\td a\td b],$ a skew\s ic tensor on~$S^2$, $\z$~is a pure imaginary
\ct, $\z=-\ov \z$, $\wt a,\wt b=5,6$ or $\th,\psi$ (see Eqs \er{c21}--\er{c22}).

Let us consider \er{2.128} in our case. One gets
\bml c111b
\gd L,n,\td w\td m, = \gd H,n,\td w\td m, + i\xi \gd k,G2[n\cdot,d],\gd H,d,\td w\td m,
+ \z(h^{S^2\td a\td d}\gd H,n,\td a\td w,\gd k,S^2,\td d\td m, - h^{S^2\td a\td d}
\gd H,n,\td a\td d,\gd H,n,\td a\td m,\gd k,S^2,[\td d\td w],)\\
{}- 2i\xi \z^2 h^{S^2\td d\td c}h^{S^2\td a\td b}\gd H,d,\td d\td a,\gd k,S^2,\td c\td w,
\gd k,S^2,\td b\td m, - 2i\xi\z \gd k,G2[n\cdot,d], h^{S^2\td a\td p}h^{S^2\td d\td b}
\gd H,d,\td b[\td w, \gd k,S^2,\td m]\td p, \gd k,S^2,[\td d\td a],\\
{}- 2\xi^2\z k^{G2[nb]} \gd k,G2,[bd], \gd H,d,\td a[\td w, \gd k,S^2,\td m]\td p,
h^{S^2\td p\td a}
\e
\beq c112b
\gd H,b,\td a\td b, = \biggl(\a_s\,\frac1{\sqrt{\hbar c}}\,\gd C,G2d,cd,
\gd \Phi,c,\td a, \gd \Phi,d,\td b, - \frac1{\a_s}\sqrt{\hbar c}\,\gd\mu,b,\hat l,
\gd f,\SU(2)\hat l,\td a\td b, - \gd\F,b,\td c, \gd f,\SU(2)\td c,\td d\td b,
\biggr).
\e
$\gd C,G2d,cd,$ are structure \ct s of Lie algebra of $G2$, $\gd f,\SU(2)\hat \imath,
jk,$ are structure \ct s of Lie algebra of SU(2).

Thus we get for $\wh V(\Phi)$:
\bml c113b
\wh V(\F) = \frac1{4\pi}\int_{S^2} \sin\th \,d\th\,d\psi\biggl\{
g^{S^2\td w\td p}g^{S^2\td m\td q}\Bigl[\gd h,G2,nk, \gd H,n,\td w\td m,
+2\z \gd h,G2,nk, \gd H,n,\td d\td w, \gd k,S^2[\td d\cdot,\td m],\\
{}+ i\xi\z \bigl(2\gd k,G2,[nk],\gd H,n,\td d\td w,\gd k,S^2[\td d,\td m],
+\z(-2\gd k,G2,[kd],\gd H,d,\td d\td a, \gd k,S^2[\td d,\td w], \gd k,S^2[\td a,\td m],\\
{}- \gd k,G2,[kd], \gd H,d,\td l\td w, \gd k,S^2,[\td m\td a], k^{S^2[\td l\td a]}
+\gd k,G2,[kd], \gd H,d,\td l\td m, k^{S^2[\td l\td a]} \gd k,S^2,\td w\td a,)\bigr)
- \xi^2\z\bigl(\gdg k,G2,[n\cdot,b], \gd k,G2,[bd], \gd H,d,\td a\td w, \gdg k,S^2,[\td m\cdot,\td a],\\
{} - \gdg k,G2,[k\cdot,b], \gd k,G2,[bd], \gd H,d,\td a\td m, \gdg k,S^2,[\td w\cdot,\td a],
+ \z(\gd k,G2,[nk], \gd k,G2[n\cdot,d], \gd H,d,\td l\td m, \gdg k,S^2,[\td w\cdot,\td r], \gd k,S^2[l\cdot,\td r],\\
{} - 2\gd k,G2,[nk], \gd k,G2[n\cdot,d], \gd H,d,\td d\td a, \gd k,S^2[\td d\cdot,\td w], \gd k,S^2[\td a,\td m],
- \gd k,G2,[nk], \gd k,G2[n\cdot,d], \gd H,d,\td l\td w, \gdg k,S^2,[\td m\cdot,\td r], \gd k,S^2[\td d\cdot,\td r],)\bigr)\\
- i\xi^3\z\bigl(\gd k,G2,[nk], k^{G2[nb]} \gd k,G2,[kb], \gd H,d,\td a\td w, \gdg k,S^2,[\td m\cdot,\td a],
- \gd k,G2,[nk], k^{G2[nb]} \gd k,G2,[bd], \gd H,d,\td a\td m, \gdg k,S^2,[\td w\cdot,\td a], \bigr)\Bigr]
\cdot \gd H,k,\td p\td q, \\ {}- 2\z^2 \gd h,G2,cd, (\gd H,c,\td p\td q, \wt g{}^{S^2[\td p\td q]}
- \gd H,d,\td a\td b, \wt g{}^{S^2[\td a\td b]})\biggr\}
\e
\beq c114b
\cL\dr{int} (\F, A) = \a_s \gd h,G2,ab, \mu^a_i \wt H{}^i \wt g{}^{S^2[\td a\td b]}
\biggl(\frac1{\a_s} \sqrt{\hbar c}\cdot \gd C,G2b,cd, \gd \F,c,\td a, \gd \F,d,\td b,
- \a_s\,\frac{\hbar}c\, \mu^b_{\hat\imath} \gd f,\SU(2)\hat \imath,\td a\td b,
- \a_s\,\frac1{\sqrt{\hbar c}}\,\gd \F, b,\td d, \gd f,\SU(2)\td d,\td a\td b,\biggr)
\e
where $\wt H{}^i = \gd \wt H{},i,\m, g^{\m}$, $\gd \wt H,i,\m,= \gd\a,i,c,\gd\ov H,c,\m,$.

Let us remind to the reader that we are working in the \nos, real version of
the theory, i.e.\ $g_{\m}= g_{(\m)}+g\ink{\m}$ with hermitian additional
tensors $\gd\ell,G2,ab,$, $\gd\ell,\SU(2),ab,$, $\gd g,S^2,\ta\tb,$,
$\gd\ell,\SU(3),ab,$ remains also real \nos. $\gd\wt g{},S^2\td a\td b,,$
is an inverse tensor for $\gd h,S^2,\td a\td b,+\z \gd k,S^2,\td a\td b,$, i.e.
$$
\gd\wt g{},S^2\td a\td b,, = \frac1{\sin^2\th(1+\z^2)}\left(
\begin{array}{c|c}
-\sin^2\th & -\z\sin\th \\\hline \z\sin\th & -1
\end{array}\right).
$$
\E\co ical terms are given by Eqs \er{c4} and \er{c24}. Field \e s are given
by \er{4.109}--\er{4.111} and are supplied with a \so\ of Eq.~\er{4.111}
together with \er{D.3}--\er{D.6}.

For a field $\Psi$ (Eq.\ \er{4.112}) we have a small change. In the place of
$\cLY(\wt A)$ we have $\cLY(\a_{QCD}A_{\SU(3)}) + \cLY(\a_sA_{\SU(2)\ot\U(1)})$
given by Eqs \er{c106b} and \er{c107b}. In this way one gets
\bml c115b
2\bigl((242+\ov M)\gtn{\a\mu} - 242 g^{\nu\mu}g_{\d\nu}\gtn{\a\d}\bigr) \pap{^2\Psi}{x^\a \pa x^\mu} \\*
{}+ \frac2{\sqrt{-g}}\,\pa_\mu\biggl\{\sqrt{-g}\Bigl[242\gtn{\a\mu}
-121g_{\d\nu}(g^{\nu\a}\gtn{\mu\d}+ g^{\nu\d}\gtn{\mu\a}) - \ov M\gtn{\mu\a}
\Bigr]\biggr\} \pap\Psi{x^\a} \\
{}-96\pi e^{24\Psi}\bigl(\cLY(\a_{QCD}A_{\SU(3)}) + \cLY(\a_sA_{\SU(2)\ot\U(1)})\bigr)
-2e^{-2\Psi}\cL\dr{kin}(\br{gauge(SU(2))}{\n\F})\\ {}+ \frac{10}{r^4}\,e^{20\Psi}
\wh V(\F) + \frac{40}{r^2}\,e^{20\Psi}\cL\dr{int}(\F,A) - \frac{11}{r^2}e^{22\Psi}\wh{\ov P}_M
- \frac{12}{\ell\pl^2}\,e^{24\Psi}(\a_s^2\wt R_{G2} + \a_{QCD}^2\wt R_{\SU(3)})\\
{}-10\,\frac{e^{20\Psi}\ell^2\pl}{r^4}\,\wh V(0)=0.
\e

The \e\ for \YM' fields with $G=\SU(3)$ reads:
\bml c116b
\br{gauge(SU(3))}{\n_\mu} (\gd\ell,\SU(3),ab,\fal L^{a\a\mu}) = 2\falg^{[\a\b]}
\br{gauge(SU(3))}{\n_\b} (\gd h,\SU(3),ab, \gtk{\m} \gd H,a,\m,) \\
{}+24\pa_\b \Psi\bigl[\gd \ell,\SU(3),ab,\cdot \fal L^{a\b\a} - 2\falg^{[\b\a]}
(\gd h,\SU(3),ab, \gtk{\m} \gd H,a,\m,)\bigr].
\e
For SU(2) one gets
\bml c117b
\br{gauge(SU(2))}{\n_\mu}(\gd \ell,\SU(2),ij, \fal{\ov L}{}^{i\a\mu})
= 2\falg^{[\a\b]}\br{gauge(SU(2))}{\n_\b}(\gd h,\SU(2),ij, \gtk{\m} \gd \ov H{},i,\m,)\\
{}+ 2\sqrt{-g}\,\a_s \,\frac1{\sqrt{\hbar c}}\,\frac{e^{22\Psi}}{r^2}
\bigl(\gd\ell,G2,ab, g^{S^2\td b\td n} \gd L,a,\mu\td b, (\gd\F,d,\td c,
\gd C,G2,dc, \a^c_j + \gd\F,b,\td a, \gd f,\td a,\td nj,]_{av}\\
{}+ \left(\frac{\d \gd L,a,\b\td b,}{\d\br{gauge(SU(2))}{\n_\a}\gd\F,w,\td v,}\right)
\gd\ell,G2,ab, g^{\b\mu}(\br{gauge(SU(2))}{\n_\mu}\gd \F,b,\td n,)
(\gd\F,d,\td w, \gd C,G2w,dc,\a^c_j + \gd\F,w,\td a, \gd f,\SU(2)\td a,\td nj,)]_{av}\\
{}+4\sqrt{-g}\,\frac{e^{44\Psi}}{r^2}\, \gd h,G2,ab, \mu^a_k \gd\ell,\SU(2),ij,
\falg^{[\td a\td b]} \br{gauge(SU(2))}{\n_\mu}\biggl\{\gtk{\mu\a}\biggl[\frac1{\a_s}
\sqrt{\hbar c}\,\gd C,G2b,cd, \gd \F,c,\td a, \gd \F,d,\td b,\\ {}-\a_s\biggl(
\frac{\hbar}{c}\, \mu^b_{\hat l} \cdot \gd f,\SU(2)\hat l,\td a\td b,
- \a_s\,\frac1{\sqrt{\hbar c}}\, \gd\F,b,\td d, \gd f,\SU(2)\td d,\td a\td b,
\biggr)\biggr]\biggr\}\\
{}+ 24\pa_\b \Psi\bigl[\gd\ell,\SU(2),ij, \fal {\ov L}{}^{i\b\a} - 2\falg^{[\b\a]}
(\gd h,\SU(2),ij, \gtk{\m}\cdot \gd\ov H{},i,\m,)\bigr]=0
\e
\bml c118b
\br{gauge(SU(2))}{\n_\mu}(\gd \ell,G2,ab, \gd\falkaL,a\mu,\td b,)
= -\sqrt{-g}\,\frac{e^{22\Psi}}{2r^2} \biggl\{\frac{\d V'}{\d \gd\F,b,\td a,}
\cdot \gd g,S^2,\td b\td n,\\ {}- 2\sqrt{-g}\,e^{22\Psi}\mu^c_i \biggl(\gd \ov H{},i,\m,
\gtk{\m} \gd h,G2,cd,\biggl(\frac2{\a_s}\sqrt{\hbar c}\,\ul g^{S^2[\td a\td n]}
\gd C,G2d,cb, \gd\F,c,\td a, \gd g,S^2,\td b\td n,\\ {}- \a_s\,\frac1{\sqrt{\hbar c}}
\,g^{S^2[\td c\td d]} \gd f,\SU(2)\td n,\td c\td d, \gd g,S^2,\td b\td n, \biggr)
+2\pa_\mu \Psi \gd\ell,G2,ab, \gd \ov L{},\a\mu,\td b,\biggr\}_{av}
\e
where
$$
(\cdots)_{av} = \frac1{4\pi}\int_{S^2}\sin\th\,d\th\,d\psi\,(\cdots),\q
\ul g^{S^2[\td n\td m]}= \frac1{4\pi}\int_{S^2}\sin\th \,d\th\,d\psi\,
g^{S^2[\td n\td m]}.
$$

In field \e\ \er{4.109} we have
\begin{gather}
8\pi \mathop{T_{\mu\nu }}\limits^{\rm eff.full}
= \frac{8\pi K}{ c^{4}} \Bigl(\mathop{T_{\mu\nu }}\limits^{\gauge(\SU(3))} +
\mathop{T_{\mu\nu }}\limits^{\gauge(\SU(2))} +\mathop{T_{\mu\nu }}\limits^{\gauge(\U(1))}
+\bigl(T_{\m}(\Phi)- \frac12 \,g_{\m}T_{\a\b}(\Phi)g^{\a\b}\bigr)\hskip60pt\nonumber\\
\hskip100pt +
\bigl(\mathop{T_{\mu \nu }}\limits^{\scal}(\Ps) - \frac12 \,g_{\m}\mathop{T_{\a\b }}\limits^{\scal}(\Ps)g^{\a\b}\bigr)
+\mathop{T_{\mu \nu }}\limits^{\Int} +g_{\mu \nu }\wt{\varLambda }\Bigr)\lb c119b \\
K= G_N e^{-24\Psi} = G\dr{eff}(\Psi). \lb c120b
\end{gather}
In order to get \e s in the form $R_{\b\a} = 8\pi \nad{eff.full}{T_{\b\a}}$ we
should transform every energy momentum tensor or a sum of them $\nad{eff.full}{T_{\b\a}}
\mapsto \nad{eff.full}{T_{\b\a}} - \frac12g_{\b\a}(\nad{eff.full}{T_\m}g^\m)$.
\bml c121b
\mathop{T_{\a\b }}\limits^{\gauge(\SU(3))} = -\a_{QCD}^2\, \frac{\gd\ell,\SU(3),ij,}{4\pi}
\biggl\{ g_{\g\b}g^{\tau\rho} g^{\ve\g} \gd L,i,\rho\a, \gd L,j,\tau\ve,
-2\gtk{\m} \gd H,(i,\m, \gd H,j),\a\b, \\{}- \frac14 \,g_{\a\b}\Bigl[L^{i\m}
\gd H,j,\m, -2(\gtk{\m}\gd H,i,\m,)(\gtk{\g\d}\gd H,j,\g\si,)\Bigr]\biggr\}
\e
\vskip-18pt
\bml c122b
\mathop{T_{\a\b }}\limits^{\gauge(\SU(2))} = -\a_s^2\, \frac{\gd\ell,\SU(2),ij,}{4\pi}
\biggl\{g_{\g\b}g^{\tau\rho}g^{\ve\g} \gd \ov L{},i,\rho\a, \gd \ov L{},j,\tau\ve,
- 2\gtk{\m} \gd \ov H{},(i,\m, \gd \ov H{},j),\a\b, \\{}- \frac14\,g_{\a\b}
\Bigl[\ov L{}^{i\m} \gd\ov H{},j,\m, -2 (\gtk{\m}\gd\ov H{},i,\m,)
(\gtk{\g\si} \gd\ov H{},j,\g\si,)\Bigr]
\biggr\}
\e
\vskip-18pt
\bml c123b
\mathop{T_{\a\b}}\limits^{\scal} ({\varPsi} ) - \frac12\,g_{\a\b}
\mathop{T_{\m}}\limits^{\scal} (\Psi)g^{\m} = -\frac{e^{24\Psi}}{16\pi}
\Bigl\{(g_{\ka\a}g_{\o\b} + g_{\o\a}g_{\ka\b}) \cdot \gtn{\g\ka}\gtn{\nu\o}
[242(g^{\xi\mu}g_{\nu\xi} - \d^\mu_\nu)\varPsi_{,\mu} \\{}+ \d^\mu_\nu \ov M\varPsi_{,\nu}]
\Ps_{,\g} - g_{\a\b}[\ov M \gtn{\nu\mu}\Ps_{,\mu}\Ps_{,\nu} + 484\gtk{\m}g_{\d\mu}\gtn{\g\d}
\Ps_{,\nu}\Ps_{,\g}]\\ + g_{\a\b}\Ps_{,\mu}\Ps_{,\g}\bigl\{(242+\ov M)\gtn{\g\mu}
+242(\gtn{\g\tau}g^{\xi\mu}g_{\nu\xi} - 4\gtk{\a\mu}g_{[\d\a]}\gtn{\g\d})\bigr\}\Bigr\}
\e
\vskip-18pt
\bml c124b
T_{\m}(\F) = \frac{e^{22\Psi}}{4\pi r^2}\,\gd\ell,G2,ab, \bigl(g^{S^2\td b
\td n}\gd L,a,\mu \td b,\cdot \br{gauge(G2)}{\n_\nu} \gd\F,b,\td n,\bigr)_{av}\\
{}- \frac12\,g_{\m}\biggl(-\frac{e^{40\Psi}}{8\pi r^4}\,\wh V(\F) + \frac{e^{18\Psi}}{4\pi r^2}
\,\gd\ell,G2,ab, (g^{S^2\td b\td n}g^{\a\b} \gd L,a,\a\td b, \br{gauge(G2)}{\n_\b}
\gd \F,b,\td n,)_{av}\biggr)
\e
\vskip-18pt
\bml c123c
\mathop{T_{\mu \nu }}\limits^{\Int} = -\frac{e^{40\Psi}\a_s}{2\pi r^2}\,\gd h,G2,ab,
\mu^a_i \gd \wt H{},i,\m, \fal{\wt g}{}^{S^2[\td a\td b]}\biggl(\frac1{\a_s}\sqrt{\hbar c}
\,\gd C,G2b,cd, \gd\F,c,\td a, \gd\F,d,\td b, - \a_s\,\frac1{\sqrt{\hbar c}}\,
\gd\mu,b,\hat\imath, \gd f,\SU(2)\hat\imath,\td a\td b,\\
{}-\a_s\,\frac1{\sqrt{\hbar c}}\,\gd\F,b,\td d, \gd f,\SU(2)\td d,\td a\td b,\biggr)
+ \frac{e^{40\Psi}}{4\pi r^2}\,g_{\m} \gd h,G2,ab, \mu^a_i (\gd \wt H{},i,\a\b,\gtk{\a\b})
\fal{\wt g}{}^{S^2[\td a\td b]}\biggl(\frac1{\a_s}\sqrt{\hbar c}\,\gd C,G2b,cd,
\gd\F,c,\td a, \gd\F,d,\td b, \\{}- \a_s\,\frac{\hbar}c\, \mu^b_{\hat \imath} \,
\gd f,\SU(2)\hat\imath,\td a\td b,
-\a_s\,\frac1{\sqrt{\hbar c}}\,\gd\F, b,\td d, \gd f,\SU(2)\td d,\td a\td b,\biggr)
\e
\beq c124c
\wt{\varLambda } = \frac1{16\pi G_N} \biggl(\frac{e^{48\Psi}}{\ell\pl^2}(\a_{QCD}^2
\wt R_{\SU(3)} + \a_s^2\wt R_{G2}) + \frac{e^{24\Psi}}{r^2}\cdot\wh{\ov P}_M - \frac{e^{20\Psi}\ell^2\pl}{r^4}
\wh V(0)\biggr).
\e

From Eq.~\er{c37} we simply get Eqs \er{D.15} and \er{D.17}.

$\dfrac{\d \gd L,\a,\b\td b,}{\d \br{gauge(SU(2))}{\n_\a} \gd\F,w,\td a,}$ satisfies the following \e:
\bml c125b
\gd\ell,G2,dc, g_{\mu\b}g^{\g\mu} \left(\frac{\d \gd L,d,\g\td a,}{\d \br{gauge(SU(2))}{\n_\a}\gd\F,w,\td v,}\right)
+ \gd\ell,G2,cd, \gd g,S^2,\td a\td m, g^{S^2\td m\td c}
\left(\frac{\d \gd L,d,\b\td c,}{\d \br{gauge(SU(2))}{\n_\a}\gd\F,m,\td v,}\right)\\*
= 2\gd\ell,G2,cd, \gd g,S^2,\td a\td m, g^{S^2\td m\td c}\d^\a_\b \d^{\td v}_{\td c}\d^d_w.
\e

We have the following notations
\bg c126b
\wh V{}' = \wh V + \gd\psi,\hat\imath\td d,c, \bigl(\gd\F,c,\td b,
\gd f,\SU(2)\td b,\hat\imath\td d, - \mu^a_{\hat\imath}\gd \F,b,\td d, \cdot \gd C,G2c,ab,\bigr)\\
\mathop{T_{\a\b}}\limits^{\rm gauge(U(1)\dr Y)} = \frac{\a_s^2}{4\pi}\Bigl\{
g_{\g\b}g^{\tau\rho}g^{\ve\g} \wh L_{\rho\a}\wh L_{\tau\ve}
-2\gtk{\m} G_{\m}G_{\a\b}\hskip100pt\nonumber \\ \hskip100pt {}- \frac14\,g_{\a\b}\bigl[\wh L{}^{\m}G_{\m}
-2(\gtk{\m}G_{\m})(\gtk{\g\si}G_{\g\si})\bigr]\Bigr\} \lb c127b
\e
where $\wh L_{\a\b}$ satisfies the \e
\beq c128b
g_{\mu\b}g^{\g\mu} \wh L_{\g\a} + g_{\a\mu}g^{\mu\g}\wh L_{\b\g} = 2g_{\a\mu}g^{\mu\g}G_{\b\g},
\e
that is,
\beq c229c
\wh L_{\nu\mu} = G_{\nu\mu} - \gtn{\tau\a} G_{\a\nu}g_{[\mu\tau]}+ \gtn{\tau\a}G_{\a\mu}g_{[\nu\tau]},
\e
and in a place of \er{c127b}
\bml c230c
\mathop{T_{\a\b}}\limits^{\rm gauge(U(1)\dr Y)} = \frac{\a_s^2}{4\pi}\Bigl(\gd G,\tau,\a,G_{\tau\b}
-\frac14\,g_{\a\b}G^{\m}G_{\m} + g_{\g\b}\gd G,\tau,\nu, G_{\o\tau}g^{\ve\g}\gtn{\rho\nu}
\gtn{\d\o}g_{[\a\rho]}g_{[\ve\d]} \\
{}- g_{\g\b}\gtn{\rho\nu}(G^{\mu\g}G_{\nu\mu}g_{[\a\rho]} + G_{\mu\ve}\dg G,\nu,\mu,g^{\ve\g}g_{[\a\rho]})
-2\gtk{\m}G_{\m}G_{\a\b}\\ {}+ \frac14\,g_{\a\b}\bigl(2(\gtk{\m}G_{\m})^2
-(g^{\nu\d}g^{\mu\o}\gtn{\tau\ve}g_{\o\tau} - g^{\nu\d}\gtn{\mu\ve}+g^{\nu\d}g^{\mu\ve})G_{\ve\d}G_{\m}\bigr)\Bigr)
\e
$$
\gd G,\tau,\nu, = g^{\tau\g}G_{\g\nu}.
$$
One easily gets
\beq c129b
\mathop{T_{\a\b}}\limits^{\rm gauge(\SU(3))}\!\!\!g^{\a\b} = \mathop{T_{\a\b}}\limits^{\rm gauge(\SU(2))}\!\!\!g^{\a\b}
= \mathop{T_{\a\b}}\limits^{\rm gauge(U(1))}\!\!\!g^{\a\b}= \mathop{T_{\a\b}}\limits^{\rm int} g^{\a\b} =0.
\e
$L^{ad\mu}$ and $\ov L{}^{i\a\mu}$ satisfy the following \e s:
\bg c130b
\gd\ell,\SU(3),dc, g_{\mu\b}g^{\g\mu} \gd L,d,\g\a, + \gd \ell,\SU(3),cd, g_{\a\mu}g^{\mu\g}\gd L,d,\b\g,
= 2\gd\ell,\SU(3),cd, g_{\a\mu}g^{\mu\g} \gd H,d,\b\g, \\
\gd\ell,\SU(2),ij, g_{\mu\b}g^{\g\mu} \gd\ov L{},i,\g\a, + \gd\ell,\SU(2),ij, g_{\a\mu}g^{\mu\g} \gd\ov L{},i,\b\g,
= 2\gd\ell,\SU(2),ji, g_{\a\mu}g^{\mu\g}\gd\ov H{},i,\b\g,. \lb c131b
\e
Both \e s can be exactly solved.

$\gd L,a\mu,\td b,$ satisfies the \e
\beq c132b
\gd\ell,G2,dc, g_{\mu\b} g^{\g\mu} \gd L,d,\g\td a, + \gd\ell,G2,cd, \gd g,S^2,\td a\td m, \gd g,S^2\td m\td c,, \gd L,d,\b\td c,
= 2\gd\ell,G2,cd, g_{\td a\td m}g^{\td m\td c}\mathop{\n_{\b}}\limits^{\rm gauge(G2)}\gd\F,d,\td c,.
\e
This \e\ can also be explicitly solved.

From Eq.~\er{4.109} via Eqs \er{D.15} and \er{D.17} we get eventually
Eqs \er{4.172} and \er{4.173}. They are final \e s for \gr al field with
$T^{\rm eff.full}_{\m}$ as a source. The remaining \e s are \e s for non\gr al
fields: $\SU(3)_c$ gauge field, $\SU(2)_L\ot \U(1)\dr{Y}$ gauge field, Higgs' field~$\F$
and scalar field~$\Ps$. For $\U(1)$ gauge field the \e\ reads
\beq c133b
\ov\n_\mu \wh L^{\a\mu} = 2\falg^{[\a\b]} \ov\n_\b (\gtk{\m}G_{\m}) + 24 \pa_\b\Ps\bigl[
\fal {\wh L}{}^{\b\a} - 2\falg^{[\b\a]}(\gtk{\m}G_{\m})\bigr]
\e
where $\wh L^{\a\mu}$ is given by Eq.~\er{c128b}.

Let us recapitulate field \e s of a partial \un\ of \fn\ \ia s (i.e.\ \un\ of
a bosonic part of a Standard Model with NGT).
\bml c134b
\frac14\tv\n _{\tl\o} \wt g{}^\(\nu\d) \biggl\{\Bigl(-\tv\n_\d g^{\hp{[(\b\tp}\a}
_{[(\b\tp\cdt]} - \tv\n_{(\b\tp} g^{\hp{[}\a}_\[\cdt \d] + \wt g{}^\(\a\rho)
\tv\n_\rho g_{[\d(\b]\tp}\Bigr) g_\[|\nu|)\a] \\
{}+g^{\hp{[\rho}\b}_\[\rho\cdt] \Bigl[ g^{\hp{(|\nu|}\rho}_{(|\nu|\cdt]} \Bigl(
-\tv\n_{\d)}g_\[\a\b\tp] - \tv\n_\a g_\[\b\tp\d)] +\tv\n_{\b\tp}
g_\[|\d|)\a] \Bigr) g^{\hp{[\d}\a} _\[\d\cdt] \Bigr]\biggr\}_{,\mu\tp}\\
{}-\frac14 \tv\n_\nu\biggl(-\tv\n_{\tl\b} g^{\hp{[\o}\nu}_\[\o\cdt]
-\tv\n_{\tl \o} g^{\hp{[}\nu}_\[\cdt\tl\b] + \wt g{}^\(\nu\rho)
\tv\n_\rho g_\[\tl \b\o]\\
{}+ 2g^{\hp{\tl[\o}\a}_{\tl[\o\cdt]} \Bigl(-\tv\n_{[\b\tp}g_\[\a\rho]
-\tv\n_\a g_\[\rho\tp\b] +\tv\n_\rho g_\[[\b\tp\a]\Bigr) g^\[\nu\rho]\biggr)
_{,\mu\tp} = 8\pi \nad{eff.full}T_{\tl[\b\o],\mu\tp}
\e
$\tv \n$ is a covariant \dv\ \wrt \LC \cn\ generated by $g_{(\a\b)}$ on~$E$.
\bml c135b
\tv R_{\b\g} = 8\pi \mathop{T_\(\b\g)}\limits^{\rm eff.full} + \frac34 \,\tv\n_\nu \biggl(\wt g{}^\(\nu\d)
\Bigl\{\Bigl(-\tv\n_\d g^{\hp{[(\b}\a}_\[(\b\cdt]
-\tv\n_{(\b} g^{\hp{[}\a}_\[\cdt\d] + \wt g{}^\(\a\rho)\tv\n_\rho
g_\[\d(\b]\Bigr) \cdot g_\[\g)\a] \\
{}+ g^{\hp{[\rho}\si}_\[\rho\cdt]\Bigl[ g^{\hp{([\g}\rho}_{([\g\cdt]}
\Bigl(-\tv\n_{(\b} g_\[\a\si] - \tv\n_\a g_\[\si(\b] + \tv\n_\si g_{(\b\a]}\Bigr)
g^{\hp{[\d}\a}_\[\d\cdt] \\
{}-\Bigl( -\tv\n_\d g_\[\a\si] - \tv\n_\a g_\[\si\d] + \tv\n_\si g_\[\d\a]\Bigr)
\cdot g^{\hp{[\g}\a}_\[\g\cdt] g^{\hp{[\g)}\rho}_\[\g)\cdt] \Bigr]\Bigr\}\biggr)\\
{}-\frac14\, \tv\n_{(\g} \biggl\{\wt g{}^\(\nu\d) \biggl\{\Bigl(-\tv\n_\d g^{\hp{[(\b}\a}
_\[(\b\cdt] -\tv\n_\b g^{\hp{[}\a}_\[\cdt\d] + \wt g{}^\(\a\rho) \tv\n_\rho
g_\[\d(\b] \Bigr)\cdot g_\[\nu)\a] \\
{}+ g^{\hp{[\rho}\si}_\[\rho\cdt] \Bigl( g^{\hp{([\nu}\rho} _{([\nu\cdt]}
\Bigl(-\tv\n_{\b)}g_\[\a\si] - \tv\n_\a g_\[\si\b)] + \tv\n_\si g_\[\b)\a]
\Bigr) g^{\hp{[\d} \a}_\[\d\cdt]\\
{}- \Bigl(-\tv\n_{\b)} g_\[\a\si] -\tv\n_\a g_\[\si\b)]
+\tv\n_\si g_\[\b)\a] \Bigr) g^{\hp{[\d}\a}_\[\d\cdt] \\
{}-\Bigl( -\tv\n_{\d} g_\[\a\si] - \tv\n_\a g_\[\si\d] + \tv\n_\si g_\[\d\a]
\Bigr)\cdot g^{\hp{[\g}\a}_\[\g\cdt] g^{\hp{[\nu)}\rho}_\[\nu)\cdt]
\Bigr]\biggr\}\biggr\}
\e
$\tv R_{\b\g}$ is a Ricci tensor for a \LC \cn~$E$ generated by $g_{(\a\b)}$,
where
\bml c136b
\nad{eff.full}T_{\m} = \frac{8\pi G_N}{c^4} \,e^{-24\Psi}\biggl( -\frac{\a^2_{QCD}}{4\pi}
\gd\ell,\SU(3),ij, \Bigl(g_{\g\mu}g^{\tau\rho}g^{\ve\g}\gd L,i,\rho\nu, \gd L,j,\tau\ve,
- 2\gtk{\a\b} \gd H,(i,\a\b, \gd H,j),\m,\\ {}- \frac14\,g_{\m}
\bigl[L^{i\a\b} \gd H,j,\a\b, - 2(\gtk{\a\b} \gd H,i,\a\b,)(\gtk{\g\si} \gd H,j,\g\si,)\bigr]\Bigr)\\
{}-\frac{\a^2_s}{4\pi}\,\gd\ell,\SU(2),ij, \Bigl\{g_{\g\mu}g^{\tau\rho}g^{\ve\g} \gd\wh L{},i,\rho\nu, \gd\wh L{},j,\tau\ve,
-2\gtk{\a\b} \gd \ov H,(i,\a\b, \gd\ov H,j),\m,\\ {}-\frac14\,g_{\m}\Bigl[\ov L{}^{i\m} \gd\ov H,j,\m,
-2(\gtk{\a\b} \gd\ov H{},i,\a\b, )(\gtk{\g\si} \gd \ov H{},j,\g\si,)\Bigr]\Bigr\}\\
{}+\frac{\a_s^2}{4\pi} \Bigl\{g_{\g\nu}g^{\tau\rho}g^{\ve\g} \wh L_{\rho\nu}\wh L_{\tau\ve}
-2 \gtk{\a\b}G_{\a\b}G_{\m} - \frac14\,g_{\m}\Bigl[\wh L{}^{\a\b}G_{\a\b}
-2 (\gtk{\a\b}G_{\a\b})(\gtk{\g\si}G_{\g\si})\Bigr]\Bigr\} \\{}- \frac{e^{24\Psi}}{16\pi}
\Bigl\{(g_{\g\mu}g_{\o\nu}+g_{\o\mu}g_{\ka\nu}) \gtn{\g\ka}\gtn{\nu\o}\Bigl[242(g^{\xi\a}g_{\b\rho}- \d^\a_b)\Ps_{,\a}
+\ov M\Ps_{,\b}\Bigr]\Ps_{,\g} \\{} - g_{\m}\bigl(\ov M\gtn{\b\a}+ 484\cdot \gtk{\a\b}g_{\d\a} \gtn{\g\d}\cdot\Ps_{,\a}\Ps_{,\b}\bigr)\Bigr\}\\
{}+\frac{e^{22\Psi}}{4\pi r^2}\,\gd\ell,G2,ab, (g^{S^2\wt b\wt n} \gd L,a,\mu\wt b, \brr{{\rm gauge}(G2)}{\n_\nu}\gd\F,b,\td n,)_{av}
-\frac12\,g_{\m} \biggl( -\frac{e^{40\Psi}}{8\pi r^4}\,\wh V(\F)\biggr)\\
{}+ \frac{e^{18\Psi}}{4\pi r^2}\,\gd\ell,G2,ab, \bigl(g^{S^2\td b\td n}g^{\a\b} \gd L,a,\a\td b, \cdot \brr{{\rm gauge}(G2)}{\n_\b}
\gd\F,b,\td n,\bigr)_{av}\\ {}- \frac{e^{40\Psi}\a_s}{2\pi r^2}\,\gd h,G2,ab, \mu^a_i \gd\wt H{},i,\m, g^{S^2[\td a\td b]}
\biggl(\frac1{\a_s}\sqrt{\hbar c}\,\gd C,G2b,cd, \gd\F,c,\td a, \gd\F,d,\td b, - \frac{\a_s}{\sqrt{\hbar c}}\,\mu^b_{\hat\imath}
\gd f,\SU(2)\hat\imath,\td d\td b, - \frac{\a_s}{\sqrt{\hbar c}}\,\gd\F,b,\td d, \gd f,\SU(2)\td d,\td a\td b,\biggr)\\
{}+\frac{e^{40\Psi}\a_s}{4\pi r^2}\,g_{\m} \gd h,G2,ab, \mu^a_i \biggl(\gd\wt H{},i,\a\b, g^{[\a\b]}\biggl[\fal{\ov g}{}^{S^2[\td a\td b]}
\biggl(\frac{\sqrt{\hbar c}}{\a_s}\cdot \gd C,G2b,cd, \gd\F,c,\td a, \gd\F,d,\td b, - \frac{\a_s \hbar}c \,\mu^b_{\hat\imath}
\gd f,\SU(2)\hat\imath,\td a\td b, - \frac{\a_s}{\sqrt{\hbar c}}\cdot \gd\F,b,\td d, \gd f,\SU(2)\td d,\td a\td b,\biggr)\biggr]\\
{}+\frac{g_{\m}}{16\pi G_N}\biggl(\frac{e^{48\Psi}}{\ell\pl} (\a_{QCD}^2 \wt R_{\SU(3)} + \a_s^2 \wt R_{G2}) +\frac{e^{24\Psi}}{r^2}\,\wh{\ov P}_{S^2}
-\frac{e^{20\Psi}}{r^4}\,\wh V(0)\biggr)\biggr).
\e

For $\Ps$ field one gets
\bml c137b
\bigl((242+\ov M)\gtn{\a\mu} - 242\gtn{\a\mu} - 121 g^{\m}g_{\d\nu}\gtn{\a\d}\bigr)\pap{^2\Ps}{x^\a \pa x^\mu}\\
{}+\frac1{\sqrt{-g}}\,\pa_\mu\Bigl\{\sqrt{-g}\bigl[242\gtn{\a\mu} - 121g_{\d\nu}\cdot
(g^{\nu\a}\gtn{\mu\d}+\g^{\nu\mu}\gtn{\mu\a}) - \ov M \gtn{\mu \a}\bigr]\Bigr\}\pap\Ps{x^\a}\\
{}-96\pi e^{-24\Psi}\bigl(\cLY(\a_{QCD}A_{\SU(3)})+\cLY(\a_sA_{\SU(2)_L\ot \U(1)\dr{Y}})\bigr) \\
{}-2e^{-2\Psi} \cL\dr{kin}(\br{gauge(SU(2)_L\ot \U(1)\dr{Y})}{\n\F})+\frac{10}{r_0^4}\,e^{20\Psi}\wh V(\F) +
\frac{40}{r_0^2}\,e^{20\Psi}\cL\dr{int}-\frac{11}{r_0^2}\,e^{22\Psi}\cdot P_{S^2} \\
{}- \frac{12}{\ell\pl^2}\,e^{24\Psi}(\a_s^2 \wt R_{G2} + \a_{QCD}^2\wt R_{\SU(3)})
-10\,\frac{e^{20\Psi}}{r_0^4}\,\ell\pl^2 \wh V(0) = 0
\e
(compare \er{c115b}).

The field $\Ps$ is in some sense a part of an extended \gr.
\bml c138b
\br{gauge(SU(3))}{\n_\mu} (\gd\ell,\SU(3),ab, \fal L^{a\a\mu}) = 2\falg^{[\a\b]}
\br{gauge(SU(3))}{\n_\b} (\gd h,\SU(3),ab, \gtk{\m} \gd H,a,\m,) \\
{}+24\pa_\b \Psi\bigl[\gd \ell,\SU(3),ab,\cdot \fal L^{a\b\a} - 2\falg^{[\b\a]}
(\gd h,\SU(3),ab, \gtk{\m} \gd H,a,\m,)\bigr].
\e
\vskip-18pt
\bml c139b
\br{gauge(SU(2))}{\n_\mu}(\gd \ell,\SU(2),ij, \fal{\ov L}{}^{i\a\mu})
= 2\falg^{[\a\b]}\br{gauge(SU(2))}{\n_\b}(\gd h,\SU(2),ij, \gtk{\m} \gd \ov H{},i,\m,)\\
{}+ 2\sqrt{-g}\,\a_s \,\frac1{\sqrt{\hbar c}}\,\frac{e^{22\Psi}}{r^2}
\Bigl[\gd\ell,G2,ab, g^{S^2\td b\td n} \gd L,a,\mu\td b, (\gd\F,d,\td c,
\gd C,G2\,b,dc, \a^c_j + \gd\F,b,\td a, \gd f,\SU(2)\td a,\td nj,)_{av}\\
{}+ \left(\frac{\d \gd L,a,\b\td b,}{\d\br{gauge(SU(2))}{\n_\a}\gd\F,w,\td v,}\right)
(\gd\F,d,\td w, \gd C,G2w,dc,\a^c_j + \gd\F,w,\td a, \gd f,\SU(2)\td a,\td nj,)\Bigr]_{av}\\
{}+4\sqrt{-g}\,\frac{e^{44\Psi}}{r^2}\, \gd h,G2,ab, \mu^a_k \gd\ell,\SU(2),ij,
\gtn{\td a\td b} \br{gauge(SU(2))}{\n_\mu}\biggl\{\gtk{\mu\a}\biggl[\frac1{\a_s}
\sqrt{\hbar c}\,\gd C,G2b,cd, \gd \F,c,\td a, \gd \F,d,\td b,\\ {}-\a_s\biggl(
\frac{\hbar}{c}\, \mu^b_{\hat l} \cdot \gd f,\SU(2)\hat l,\td a\td b,
- \a_s\,\frac1{\sqrt{\hbar c}}\, \gd\F,b,\td d, \gd f,\SU(2)\td d,\td a\td b,
\biggr)\biggr]\biggr\}\\
{}+ 24\pa_\b \Psi\bigl[\gd\ell,\SU(2),ij, \fal {\ov L}{}^{i\b\a} - 2\falg^{[\b\a]}
(\gd h,\SU(2),ij, \gtk{\m}\cdot \gd\ov H{},i,\m,)\bigr]
\e
$\gd C,G2\,w,dc,$ are
structure \ct s for $G2$, $\gd f,\SU(2)\,i,jk,$ are structure \ct s for SU(2), i.e.\ $\gd f,\SU(2)i,jk, = \ve^i_{jk}$.
\er{c138b} is an \e\ for SU(3) gauge field and \er{c139b} for SU(2) gauge field.
\bml c141b
\br{gauge(SU(2))}{\n_\mu}(\gd \ell,G2,ab, \gd\falkaL,a\mu,\td b,)
= -\sqrt{-g}\,\frac{e^{22\Psi}}{2r^2} \biggl\{\frac{\d V'}{\d \gd\F,b,\td a,}
\cdot \gd g,S^2,\td b\td n,\\ {}- 2\sqrt{-g}\,e^{22\Psi}\mu^c_i (\gd H,i,\m,
\gtk{\m}) \gd h,G2,cd,\biggl(\frac2{\a_s}\sqrt{\hbar c}\,\ul g^{S^2[\td a\td n]}
\gd C,G2d,cb, \gd\F,c,\td a, \gd g,S^2,\td b\td n,\\ {}- \a_s\,\frac1{\sqrt{\hbar c}}
\,g^{S^2[\td c\td d]} \gd f,\SU(2)\td n,\td c\td d, \gd g,S^2,\td b\td n, \biggr)
+2\pa_\mu \Psi \gd\ell,G2,ab, \gd \ov L,\a\mu,\td b,\biggr\}_{av}
\e
is an \e\ for a Higgs' field $\F$.

\def\ggg#1{\!\!\mathop{\n_{#1}}\limits^{\rm gauge(G2)}\!\!}
Let us do some specification of our \e s. First of all we have
\bg c235
\cL(\F) = \cL\dr{kin}(\ggg\mu \!\F) -\wh V(\F)\\
\wh V(\F) = -\frac{\pi(1+\z)}{(1+\z^2)} \bigl((1-\z)\gd h,G2,dk, - \z^2
\gd k,G2\,b,k, \gd k,G2,bd,\bigr)\cdot H^k_{56}H^d_{56} \lb c237 \\
H^d_{56} = -H^d_{65} = \frac{\a_s}{\sqrt{\hbar c}} \gd C,G2\,d,ab,
\F^a_5 \F^b_6 - \frac{\sqrt{\hbar c}}{\a_s}\mu^d_{\hi}\gd f,\SU(2)\,\hi,56,\hskip40pt \nonumber \\
\hskip120pt {}= \pa_5\F_6 - \pa_6\F_5 + [\F_5,\F_6] = \pa_\th\F_\psi - \pa_\psi\F_\th
+[\F_\th,\F_\psi] \lb c238 \\
\cL\dr{int}(\F,A_{\SU(2)}) = -\frac{\a_s}{(1+\z^2)}\, \gd h,G2,ab, \mu^a_{i}\wt H{}^{i}\cdot \gd H,b,56, \lb c239 \\
\wt H{}^i = \gtk{\m} \gd\wt H,i,\m, , \q
\gd\wt H,i,\m, =\gd\a,i,c, \gd\ov H,c,\m, \lb c240 \\
\cL\dr{kin}\bigl(\ggg\mu \F\bigr) = \frac{\pi}{(1+\z^2)}\biggl\{
\gd\ell,G2,nk, g^{\o\mu}\biggl[\ggg\mu \F^k_5
\Bigl\{\ggg\o \F^n_5 \nonumber \\
{}+\gd k,G2\,n,d, \ggg\a\F^d_5 \bigl((\xi^2-1)\gtn{\mu\eta}g_{[\eta\o]} - i\xi \gtn{\a\nu}\gtn{\b\rho}
g\ink{\nu\o}g\ink{\rho\a}\bigr) +i\xi (1-\z^2)\gd k,G2\,n,d, \ggg\o \F^d_5\Bigr\} \nonumber \\
{}- \ggg\mu \F^k_6 \Bigl\{\ggg\o \F^n_6
+ \gd k,G2\,n,d,\bigl((\xi^2-1)\ggg\a\F^d_5 \gtn{\a\eta}g\ink{\eta\o} \nonumber \\
{}- i\xi\ggg\b \F^d_6 \gtn{\a\nu}\gtn{\b\rho}g\ink{\nu\o}g\ink{\rho\a}\bigr)\Bigr\}\biggr\} \lb c236
\e

Let us calculate an energy momentum tensor for $\cL(\F)$ (i.e.\ for scalar field~$\F$). One gets
\bg c241
T_{\psi\vf}(\F) = T_{\psi\vf}(\cL(\F)) = T_{\psi\vf}\bigl(\cL\dr{kin}(\ggg\mu
\F - \wh V(\F)\bigr) \cr
{}= \frac\pi{(1+\z^2)}\biggl\{\ell_{nk}\biggl[\ggg\psi\F^k_5\Bigl\{\ggg\vf \F^n_5
+\tfrac12\,\gd k,n,d, \ggg\a \F^d_5\bigl((\xi^2-1)\cr
{}\cdot [g\ink{\eta\o}(\gtn{\a\b}\gtn{\o\eta}g_{\b\vf}g_{\o\psi} + \gtn{\o\eta}\gtn{\a\b}
g_{\o\psi}g_{\b\psi}) - \gtn{\a\eta}g_{\o\vf}g_{\d\psi} + \gtn{\a\eta}g_{\eta\vf}g_{\o\psi}]\cr
{}+i\xi \bigl[(\gtn{\a\b}\gtn{\o\nu}g_{\b\vf}g_{\o\psi} + \gtn{\o\nu}\gtn{\a\b}g_{\o\vf}g_{\b\psi})
\gtn{\b\rho}g\ink{\nu\o}g\ink{\rho\a} \cr {}+ \gtn{\a\nu}(\gtn{\o\b}\gtn{\xi\nu}g_{\b\vf}g_{\xi\psi}
+ \gtn{\xi\nu}\gtn{\o\b}g_{\xi\vf}g_{\b\psi})g\ink{\nu\o}g\ink{\rho\a}\cr
{}+\tfrac12\, \gtn{\a\nu}\gtn{\b\g}(g_{\nu\vf}g_{\o\psi} - g_{\o\vf}g_{\nu\psi})g\ink{\rho\a}
+\tfrac12\,\gtn{\a\nu}\gtn{\b\g}g\ink{\nu\o}(g_{\rho\vf}g_{\a\psi} - g_{\a\vf}g_{\rho\psi})\bigr]\bigr)\cr
{}+i\xi(1-\z^2)\gd k,G2\,n,d, \ggg\psi \F^d_5 \Bigr\}\cr
{}-\ggg\psi \F^k_6 \Bigl\{ \ggg\vf \F^n_6 + \tfrac12\,\gd k,G2\,n,d, \bigl((\xi^2-1)\ggg\a\F^d_5
\bigl[g\ink{\eta\o}\gtn{\a\b}\gtn{\o\eta}g_{\b\vf}g_{\o\nu}\cr {}+ \gtn{\o\eta}\gtn{\a\b}
g_{\o\vf}g_{\b\nu} - \gtn{\a\eta}g_{\o\vf}g_{\d\psi} + \gtn{\a\eta}g_{\eta\vf}g_{\o\psi}\bigr]\cr
{}-i\xi\cdot \ggg\b \F^d_6 \bigl[(\gtn{\a\b}\gtn{\o\nu} g_{\b\vf}g_{\o\psi} +
\gtn{\o\nu}\gtn{\a\b}g_{\o\vf}g_{\b\psi}) \gtn{\b\rho}g\ink{\nu\o}g\ink{\rho\a}\cr
{}+\gtn{\a\nu}(\gtn{\o\b}\gtn{\xi\nu}g_{\b\vf}g_{\xi\psi} + \gtn{\xi\nu}\gtn{\o\b}g_{\ve\vf}g_{\b\psi})g\ink{\nu\o}g\ink{\rho\a}
+ \tfrac12\,\gtn{\a\nu}\gtn{\b\g}(g_{\nu\vf}g_{\o\psi} - g_{\o\vf}g_{\nu\psi})g\ink{\rho\a}\cr
{}+ \tfrac12\,\gtn{\a\nu}\gtn{\b\g}g\ink{\nu\o} (g_{\rho\vf}g_{\a\psi} - g_{\a\vf}g_{\rho\psi})\bigr]\bigr\}\cr
{}-\tfrac14\, g_{\psi\vf} \biggl\{\gd\ell,G2,nk, g^{\o\mu}\biggl[\ggg\mu \F^k_5\bigl\{\ggg\o\F^n_5
+ \gd k,G2\,n,d, \ggg\a \F^d_5 \bigl((\xi^2-1)\gtn{\a\eta}g\ink{\eta\o} \cr
{}- i\xi \gtn{\a\nu}\gtn{\b\rho} g\ink{\nu\o}g\ink{\rho\a}\bigr) + i\xi(1-\z^2)\gd k,G2\,n,d, \ggg\o \F^d_5\bigr\}\cr
{} - \ggg\mu \F^k_6 \Bigl\{\ggg\o \F^n_6 + \gd k,G2\,n,d, \bigl((\xi^2-1)\ggg\a\F^d_5
\gtn{\a\eta}g\ink{\eta\o}\cr {}- i\xi \ggg\b \F^d_6 \gtn{\a\nu}\gtn{\b\rho}g\ink{\nu\o}g\ink{\rho\a}\bigr)\Bigr\}\biggr]\cr
{}+\frac1{(1+\z^2)} \biggl(\gd h,G2,kd, \,\frac{(1-\z^4+4\z^2)}{(1+\z^2)} - \z(1+\z)\xi^2
\gd k,G2\,b,k, \gd k,G2,bd, \biggr)\gd H,d,56, \gd H,k,56,\biggr\}.
\e
\beq c242
{\mathop{T}\limits^{\rm int}}_{\m} = T_{\m}(\cL\dr{int}(\F,A_{\SU(2)})) =
-\frac{\a_s}{(1+\z^2)} \bigl[\gd h,G2,ab, \mu^a_i \gd H,b,56, \gd\wt H,i,\m,
-\frac14\,g_{\m}(\gd h,G2,ab, \mu^a_i \wt H{}^i \gd H,b,56,)\bigr]
\e
During calculations of energy-momentum tensor we are using the following formulae:
\bea c243
\d g_{\d\nu} &= -g_{\nu\g} g_{\d\o}\d g^{\o\g} \\
\d \gtn{\a\d} &= \tfrac12\bigl[\gtn{\a\b}\gtn{\o\d}g_{\b\g}g_{\o\psi} + \gtn{\o\d}\gtn{\a\b}g_{\o\g}g_{\b\psi}\bigr]\d g^{\psi\g} \lb c244 \\
\d g\ink{\m} &= \tfrac12\,(g_{\mu\g}g_{\nu\o} - g_{\nu\g}g_{\mu\o})\d g^{\o\g} \lb c245 \\
\d \gtk{\m} &= \tfrac12(\d g^{\m}- \d g^{\nu\mu}) \lb c246
\e

In Ref.~\cite{11a} some constraints on fields $\F_5$ ($=\F_\t$) and $\F_6$ ($=\F_\psi$)
have been solved (the situation in the case of 6-\di al GSW model is easier than
in a general case). One gets
\bea c247
\F_5 &= \tfrac12(\vf_1x_\a + \vf_2x_\b - \vf_1^\ast x_{-\a} - \vf_2^\ast x_{-\b}) \\
\F_6 &= -\tfrac12\,i\bigl[(\vf_1x_\a + \vf_2x_\b + \vf_1^\ast x_{-\a} + \vf_2^\ast x_{-\b})\sin\th
+2(h_\a+h_\b)\cos\th\bigr]. \lb c248
\e
$\vf_1$ and $\vf_2$ are usual Higgs' fields in GSW model
\beq c249
\wt\vf = \biggl(\begin{matrix}\vf_1\\\vf_2\end{matrix}\biggr).
\e

Formulae \er{C.4a}--\er{C.5} give us gauge \dv s in terms of $\gd A,i,\mu,$ ($\gd A,-,\mu,$,
$\gd A,+,\mu,$, $\gd A,3,\mu,$) of SU(2) gauge fields and $B_\mu$ of $\U(1)\dr{Y}$.
In the formulae we get also a Weinberg angle~$\ov\th$, which is equal to $\frac\pi6$
in the case of $G2$.

As usual we consider a gauge \tf\ $U(x)$ which gives us $Z^0_\mu$ and $A_\mu$ (\elm c
field). After \tf s \er{C.23b}--\er{C.28b} we get for $\ggg\mu \F^u_5$, $\ggg\mu
\F^u_6$ and $H^u_{56}$ Eqs \er{C.27}, \er{C.28} and \er{C.30} only $H^0$ scalar
Higgs' field after \ssb. In this way one gets
\bml c250
\cL_{\rm kin}(\gw\n \F)=\frac1{2\pi r^2(1+\z^2)}\Biggl\{
\gw{\n_\mu}\F_5^k \gw{\n_\o}\F_5^d\Bigl(2(i\xi\z^2(\z-1)-1-i\xi)\gd k,G2,kd,\\
{}+\bigl(i\xi\z(-2\xi^2-\z-2i\xi\z-2)
+i(\pi-2)\xi\bigr)\gd k,G2\,b,d, \gd k,G2,bk,
-2h_{kd}-i\xi^3\z(-2\xi^2-\pi)k^{G2\,nb}\gd k,G2,nk, \gd k,G2,bd,\Bigr)g^{\o\mu}\\
{}+\gw{\n_\mu}\F_5^d \gw{\n_\g}\F_5^n \Bigl(2\xi^2\gd h,G2,nd,\wt g{}^\(\g\b)
g_\[\b\o] +2i\xi^3\wt g{}^\(\g\b)g_\[\b\o]\gd k,G2,nd,
+ 2\wt g^\(\a\nu)\wt g^\(\g\rho)g_\[\nu\o]g_\[\rho\a]\gd k,G2,nd,\\
\hskip60pt {}-2\xi^2 \gd k,G2,nd, \gd k,G2\,n,d,\wt g{}^\(\a\nu)\wt g{}^\(\g\rho)
g_\[\nu\o]g_\[\rho\a]\Bigr)g^{\o\mu}\\
+\Bigl(-2\z \gw{\n_\mu}\F_5^k \gw{\n_\o}\wh \F{}_6^n
+2\z\gw{\n_\mu}\wh\F{}_6^k \gw{\n_\o}\F_5^n
+2\z(\z+1)\gd k,G2\,n,d,\gw{\n_\mu}\F_5^k \gw{\n_\o}\wh\F{}_6^d\\
{}-2\xi^2(i\z\xi+1)\gd k,G2\,n,d,\gw{\n_\mu}\F_5^k \gw{\n_\o}\wh\F{}_6^d
-2\xi^2\z k^{G2\,nb}\gd k,G2,bd, \gw{\n_\mu}\wh\F{}_6^k\gw{\n_\o}\F_5^d\Bigr)g^{\o\mu}
\gd\ell,G2,nk,\\
{}+\gd\ell,G2,nk, g^{\o\mu}\gd k,n,d,\bigl(\pi+4\z-4\z\xi^2+2\pi i\xi\z^2+4i\z\xi\bigr)
g^{\a\eta}g_\[\eta\o]\gw{\n_\mu}\F_5^n \gw{\n_\a}\wh\F{}_6^d\\
{}+2i\xi\z\gd \ell,G2,nk, g^{\o\mu}\wt g{}^\(\a\nu)\wt g{}^\(\g\rho)g_\[\nu\o]
g_\[\rho\a]\bigl(\gw{\n_\mu}\F_5^k \gw{\n_\g}\wh\F{}_6^d
+\z^2\gw{\n_\mu}\wh\F{}_6^k\gw{\n_\g}\F_5^d\\
+2\gw{\n_\mu}\wh\F{}_6^k\gw{\n_\g}\F_5^d \bigr) \\
{}+\gd\ell,G2,nk, g^{\o\mu}\Bigl[\Bigl(-2\gw{\n_\mu}\wh\F{}_6^k\gw{\n_\o}\wh\F{}_6^n
-2i\xi(\z-1)\gd k,G2\,n,d,\gw{\n_\mu}\wh\F{}_6^k\gw{\n_\o}\wh\F_6^d\\
{}-2\xi^2\z k^{G2\,nb} \gd k,G2,bd, \gw{\n_\mu}\wh\F{}_6^k\gw{\n_\o}\wh\F{}_6^d\Bigr)
+\wt g^\(\a\eta)g_\[\eta\o]\bigl(2\gw{\n_\a}\wh\F{}_6^d\gw{\n_\mu}\wh\F{}_6^k\\
{}+2\xi^2\gw{\n_\mu}\wh\F{}_6^k\gw{\n_\a}\wh\F{}_6^d
-4i\z^2\xi \gd k,G2\,n,d, \gw{\n_\mu}\wh\F{}_6^k\gw{\n_\a}\wh\F{}_6^d\bigr)
\Bigr]\Biggr\}
\e
where $\gw{\n_\mu}\F_5$ is given by
\begin{gather}
\gw{\n_\mu}\F_5 = \frac1{2\sqrt2}\Bigl[
-\frac12(v+H^0(x))(W_\mu^{-u}x_\a + W_\mu^{+u}x_{-\a})\hskip100pt \nn \\
{}+\Bigl(\pa_\mu H^0+\frac12\,i(v+H^0(x))\Bigl(W_\mu^{+u}+\frac12(\sqrt3\,Z_\mu^{0u}
-A_\mu)\Bigr)x_\b\Bigr) \nn \\
\hskip100pt{}-\Bigl(\pa_\mu H^0+\frac i{\sqrt3}(v+H^0(x))Z_\mu^{0u}\Bigr)
x_{-\b}\Bigr]  \lb c251
\end{gather}
We have
\beq c252
\gw{\n_\mu}\wh\F{}_6= \frac1{\sin\th}\gw{\n_\mu}\F_6
\e
and eventually
\begin{gather}
\gw{\n_\mu}\wh\F_6= \frac1{2\sqrt2}\Bigl(\frac12\,i(v+H^0(x))(W_\mu^{-u}x_\a
+W_\mu^{+u}x_{-\a})\Bigr) \nn \hskip100pt \\
{}-\Bigl(\pa_\mu H^0+ \frac12\,i(v+H^0(x))\Bigl(W_\mu^{+u}+\frac12(\sqrt3\,Z_\mu^{0u}
-A_\mu)\Bigr)x_\b\Bigr) \nn \\
\hskip100pt{}+\Bigl(\pa_\mu H^0+\frac i{\sqrt3}(v+H^0(x))Z_\mu^{0u}\Bigr)
x_{-\b}. \lb c253
\end{gather}
\beq c254
\cL_{\rm int}=\frac{i\z\sqrt6}{16\pi r^2(1+\z^2)}\bigl(3F_\m^3 g^\m
(\vf_1\vf_1^* - \vf_2\vf_2^*) - \vf_1\vf_2^*(F_\m^-g^\m+F_\m^+g^\m)\bigr)
\e
or
\beq c255
\cL_{\rm int}=-\frac{i\z 3\sqrt6\,(v+H^0(x))^2}{64\pi r^2(1+\z^2)}
(Z_\m^{0u}+\sqrt3\,F_\m)g^\m.
\e
In the case of a Higgs' \pt\ one gets
\bg c256
V(\vf_1,\vf_2)=V_1(\vf_1,\vf_2)+V_2(\vf_1,\vf_2)\\
V_1(\vf_1,\vf_2)=-\frac{\pi(1-2\z^2)}{2(1+\z^2)r_0^2}
\Bigl[\frac9{\g\cdot\g}\, |\vf_1|^2|\vf_2|^2
-\frac1{\a\cdot\a}\,(2-|\vf_1|^2)^2\hskip100pt \nn \\
\hskip100pt {}-\frac1{\b\cdot\b}\,(2-|\vf_2|^2)^2
-\frac{2(\a\cdot\b)}{(\a\cdot\a)(\b\cdot\b)}\,(2-|\vf_1|^2)(2-|\vf_2|^2)
\Bigr] \lb c257 \\
V_2(\vf_1,\vf_2)=\frac{-\pi\xi^2}{2(1+\z^2)r_0^2}\biggl[
-\frac14(2-|\vf_1|^2)^2 \ov K(h_\a,h_\a)
-\frac12(2-|\vf_1|^2)(2-|\vf_2|^2)\ov K(h_\a,h_\b)\hskip30pt\nn \\
+\frac32(2-|\vf_1|^2)\vf_1\vf_2^* \ov K(h_\a,x_\g)
+\frac32(2-|\vf_1|^2)\vf_1\vf_2^* \ov K(h_\a,x_{-\g}) \nn \\
{}-\frac14(2-|\vf_2|^2)^2 \ov K(h_\b,h_\b)
+\frac32(2-|\vf_2|^2)\vf_1\vf_2^* \ov K(h_\b,x_\g)
+\frac32(2-|\vf_2|^2)\vf_1\vf_2^* \ov K(h_\b,x_{-\g}) \nn \\
\hskip50pt{}-\frac94(\vf_1\vf_2^*)^2 \ov K(x_\g,x_\g)
-\frac94(\vf_1^*\vf_2)^2 \ov K(x_{-\g},x_{-\g})
-\frac94|\vf_1|^2|\vf_2|^2\ov K(x_{-\g},x_{-\g})\biggr] \lb c258
\e
where
\beq c259
\bga
\ov K(x,y)=\ov k_{de}x^dy^e = \gd k,c,d, k_{ce}x^dy^e,\\
\gd k,c,d,=h^{cb}k_{bd}, \q \ov k_{dc}=\ov k_{cd}, \q k_{ab}=-k_{ba}
\ega
\e
We have of course
\bg c260
\ka(x,y)=\ka_{ad}x^ay^d\\
\ka_{ad}=(1-2\z^2)h_{ad}-\xi^2\gd k,c,d,k_{ce} \lb c261 \\
\ggg\mu\F^u_5 = \frac1{2\sqrt2}\biggl[\pa_\mu H^0(x_\b-x_{-\b}) + \frac i4(v+H^0(x))
\biggl((\sqrt3\,\gd Z,u0,\mu, + \gd A,u,\mu,)(x_\b+x_{-\b}) \hskip50pt \nn \\
\hskip70pt {} +(-\gd Z,u0,\mu,+\sqrt3\,\gd A,u,\nu,)\frac{\sqrt3}3(x_{-\b}+x_\b)+ \gd W,u+,\mu,x_{-\a} + \gd W,u-,\mu,x_\a\biggr)\biggr] \lb c262 \\
\gd W,u\mp,\mu, = \gd A,u\pm,\mu, = \frac1{\sqrt2}(\gd A,u1,\mu, \pm i\gd A,u2,\mu,) \lb c263 \\
\ggg\mu \F^u_6 = \frac{\sin\th}{2i} \biggl[\pa_\mu H^0(x) (x_\b+x_{-\b})
+ \frac i2(v+H^0(x))(\gd W,u+,\mu, x_{-\a} - \gd W,u-,\mu, x_\a)\hskip80pt \nn \\
\hskip10pt {}+ \biggl(\frac{\sqrt3}2\,\gd Z,0u,\mu, + \frac12\,A_\mu\biggr) (x_\b-x_{-\b})
+ \biggl(-\frac12\,\gd Z,0u,\mu, + \frac{\sqrt3}2 \,A_\mu\biggr)\cdot \frac{\sqrt3}3(x_{-\b}+x_\b)\biggr] \lb c264 \\
G_{\m} = \frac12\bigl(\sqrt3\,\gd Z,0u,\m, - F_{\m}\bigr).\lb c265
\e
One also gets
\bml c266
\gd L,n,\o\td m,=\gw{\n_\o}\Ft nm + i\xi\gd k,G2\,n,d, \gw{\n_\o}\Ft dm
-\wt g{}^\(\a\mu)\gw{\n_\a}\Ft nm g_\[\mu\o]\\
{}+i\xi \gd k,G2\,n,d,\gw{\n_\b}\Ft dm\wt g{}^\(\d\b)g_\[\d\a]g_\[\o\mu]
\wt g{}^\(\a\mu) + \xi^2 k^{G2\,nb}\gd k,G2,bd,\wt g{}^\(\a\b)\gw\n \Ft dmg_\[\o \b].
\e
or
\bml c267
\gd L,n,\o\td m,=\gw{\n_\o}\F^n_{\td m}+ i\xi \gd k,G2\,n,d,\gw{\n_\o}
\F^d_{\td m} - \bigl(\z\gw{\n_\o}\F^n_{\td a}h^{S^2\td a\td d}k_{S^2\td d\td m}
+\wt g{}^\(\a\mu)\gw{\n_\a}\F^n_{\td m}g_\[\mu\o]\bigr)\\
{}-2i\xi\z\gd k,\SU(2)n,d,\gw{\n_\o}\F^d_{\td d}\wt g{}^\(\d\a)g_\[\a\o]
h^{S^2\td d\td a}\gd k,S^2,\td a\td m,+i\xi\gd k,\SU(2)n,d,\bigl(\z^2h^{S^2\td d\td a}
\gw{\n_\o}\F^d_{\td a}\gd k,S^2,\td d\td b,\gd k,S^2,\td m\td c,h^{S^2\td c\td b}\\
{}+\gw{\n_\b}\F^d_{\td m}\wt g{}^\(\d\b)g_\[\d\a]g_\[\o\mu]\wt g{}^\(\a\mu)
\bigr)
+\xi^2k^{\SU(2)nb}\gd k,\SU(2),bd, \bigl(\z\gw{\n_\o}\F^d_{\td a}h^{S^2\td a\td b}\gd k,S^2,\td m\td b, \\
{}+ \wt g{}^\(\a\b)\gw{\n_a} \F^d_{\td m}g_\[\o\b]\bigr).
\e
One also gets
\bml c268
\gd L,n,\o\mu,=\gd H,n,\o\mu, +\mu h^{\SU(3)na}\gd k,\SU(3),ad, \gd H,d,\o\mu,
+\bigl(\gd H,n,\a\o, \wt g{}^\(\a\d) g_\[\d\mu] - \gd H,n,\a\mu,
\wt g{}^\(\a\d)g_\[\d\o]\bigr)\\
{}-2\mu h^{\SU(3)na}\gd k,\SU(3),ad,\wt g{}^\(\d\tau)\wt g{}^\(\a\b)\gd H,d,\d\a, g_\[\tau\o]
g_\[\b\mu]
-2\mu h^{\SU(3)na}\gd k,\SU(3),ad, \wt g{}^\(\d\b)\wt g{}^\(\a\tau)
\gd H,d,\b\tl \o,g_{\mu\tp\tau}g_\[\d\a]\\
{}+2\mu^2 h^{\SU(3)na}h^{\SU(3)bc}\gd k,\SU(3),ac, \gd k,\SU(3),bd, \wt g{}^\(\a\b)\gd H,d,\a\tl\o,
g_{[\mu\tp\b]}
\e
and of course
\bml c269
\cLY=\frac{\a\dr{QCD}^2}{8\pi}\biggl(\gd h,\SU(3),nk, H^{k\o\mu}\gd H,n,\o\mu,
-2\gd h,\SU(3),cd, H^cH^d+ 2\gd h,\SU(3),nk, H^{k\o\mu}\gd H,n,\d\o,g_\[\a\mu]\wt g{}^\(\a\d)\\
{}+\mu\Bigl[2\gd k,\SU(3),nk, H^{k\o\mu}\gd H,n,\d\o,\wt g{}^\(\d\a)g_\[\a\mu]
-2\gd k,\SU(3),kd, H^{k\o\mu}\gd H,d,\d\a,\wt g{}^\(\d\b)\wt g{}^\(\a\rho)
g_\[\b\o]g_\[\rho\mu]\\
{}-\gd k,\SU(3),kd, H^{k\o\mu}\gd H,d,\eta\o,\wt g{}^\(\eta\b)\wt g{}^\(\a\rho)
g_\[\mu\a]g_\[\b\rho]
+\gd k,\SU(3),kd, H^{k\o\mu}\gd H,d,\eta\o,\wt g{}^\(\eta\d)\wt g{}^\(\a\rho)
g_\[\d\b]g_\[\o\d]\Bigr]\\
{}+\mu^2\Bigl[\gd k,\SU(3),nk, \gd k,\SU(3)n,d,H^{k\o\mu}\gd H,d,\eta\mu,\wt g{}^\(\rho\b)
\wt g{}^\(\eta\a)g_\[\o\b]g_\[\a\rho]
-2\gd k,\SU(3),nk, \gd k,\SU(3)n,d,H^{k\o\mu}\gd H,d,\d\a,\wt g{}^\(\d\eta)\wt g{}^\(\a\rho)
g_\[\eta\o]g_\[\rho\mu]\\
{}-\gd k,\SU(3),nk, \gd k,\SU(3)n,d,H^{k\o\mu}\gd H,d,\eta\o,\wt g{}^\(\rho\a)
\wt g{}^\(\eta\b)g_\[\mu\a]g_\[\b\rho]
+\gdg k,\SU(3),k,b, \gd k,\SU(3),bd, H^{k\o\mu}\gd H,d,\a\o,\wt g{}^\(\a\b)g_\[\mu\a]\\
{}-\gdg k,\SU(3),k,b,k_{bd}H^{k\o\mu}\gd H,d,\a\mu,\wt g{}^\(\a\b)g_\[\o\b]
+\gd k,p,n,k_{pk}H^{k\o\mu}\gd H,n,\o\mu,\Bigr]\\
{}+\mu^3\Bigl[\gd k,\SU(3),nk, k^{\SU(3)nb} \gd k,\SU(3),bd, H^{k\o\mu}\gd H,d,\a\o,\wt g{}^\(\a\b)g_\[\mu\b]\\
{}-\gd k,\SU(3),nk, k^{\SU(3)nb} \gd k,\SU(3),bd, H^{k\o\mu}\gd H,d,\a\mu,\wt g{}^\(\a\b)g_\[\o\b]\Bigr]
\biggr).
\e

In this way we have completed field \e s for a partial \un\ (i.e.\ a \un\ of
a bosonic part of a Standard Model with NGT).

Eqs \er{c134b}--\er{c141b} give us the full set of field \e s in the partial \un\
of a bosonic part of \fn\ \ia s.

\def\zz{(1+\z^2)}
\def\zy{(1-2\z^2)}
\def\zf#1{(\sqrt3\,Z^{u0}_{#1} - F_{#1})}
\def\ggs#1{\!\!\!\mathop{\n_{#1}}\limits^{\rm gauge(\SU(2)_L\ot \U(1)\dr{Y})}\!\!\!}
\def\ww#1{W^{u+}_{#1}{+}W^{u-}_{#1}}
\def\wy#1{W^{u+}_{#1}{-}W^{u-}_{#1}}
\def\az#1{(A_{#1}^u+\sqrt3\,Z^{u0}_{#1})}
\def\bgg#1{\biggl(#1\biggr)}
\def\rc{Ref.~\cite}
\def\NJ{\E\nos\ \KK (Jordan--Thiry) Theory}
\def\h{\hskip2pt minus0.5pt}
Let us consider the problem of a \co ical \ct\ in our partial \un\ of \fn\ \ia s (i.e.\ a \un\ of NGT with a bosonic part of a Standard Model).
For up to now  we have not obtained in experiment any new Higgs' boson phenomena (see Ref.~\cite{m3k}) we put in all formulae $\xi=0$ ($\xi\ne0$
is going to additional Higgs' phenomena). One gets a \co ical term
\beq c276
\la_c = \frac{e^{24\Psi}}{\ell\pl^2}(\a^2\dr{QCD} \wt R{\SU(3)} + \a^2_s R{G2}) + \frac{4e^{22\Psi}}{r_0^2(1+\z^2)} + \frac{\ell\pl^2}{r_0^4}
e^{20\Psi}\,\frac{4\pi}{\sqrt{1+\z^2}}(1-2\z^2).
\e
One gets
\beq c277
\frac{d\la_c}{d\Psi} = 8e^{20\Psi}\biggl(\frac{3e^{4\Psi}}{\ell\pl^2}(\a^2\dr{QCD} \wt R{\SU(3)} + \a^2_s R{G2})
+ \frac{11e^{2\Psi}}{r_0^2(1+\z^2)} + \frac{10\ell^2\pl}{r_0^4}\cdot \frac\pi{\sqrt{1+\z^2}}(1-2\z^2)\biggr)
\e
In our meaning a \co ical \ct\ is the minimum value of a \co ical term \wrt a scalar field~$\Psi$.

From $\frac{d\la_c}{d\Psi}=0$ one gets
\beq c278
\frac{3e^{4\Psi}}{\ell^2\pl}(\a^2\dr{QCD} \wt R{\SU(3)} + \a^2_s R{G2}) + \frac{11e^{2\Psi}}{r_0^2(1+\z^2)} + \frac{10\ell^2\pl}{r_0^4}
\cdot \frac{\pi(1-2\z^2)}{\sqrt{1+\z^2}} = 0.
\e
Taking $x=e^{2\Psi}$ we get the quadratic \e
\bg c278n
\frac{3(\a^2\dr{QCD} \wt R{\SU(3)} + \a^2_s R{G2})}{\ell^2\pl} x^2 + \frac{11}{r_0^2(1+\z^2)} x + \frac{10\ell^2\pl}{r_0^4}\cdot
\frac{\pi(1-2\z^2)}{\sqrt{1+\z^2}} = 0 \\
\D = \frac1{r_0^4(1+\z^2)^{1/2}} \biggl(\frac{121}{(1+\z^2)^{3/2}} - 120 \pi (\a^2\dr{QCD} \wt R{\SU(3)} + \a^2_s R{G2})(1-2\z^2)\biggr)>0. \lb c279
\e
Eq.\ \er{c279} gives
\beq c280
\frac{121}{120\pi} > (1+\z^2)^{1/2}(1-\z^2-2\z^4)(\a^2\dr{QCD} \wt R{\SU(3)} + \a^2_s R{G2}).
\e

One gets
\beq c281
x_{01,2} = \biggl(\frac{\ell^2\pl}{r_0^2}\biggr) \frac{\bigl(-11 \pm \bigl(121{-}120\pi(1+\z^2)^{1/2}(1{-}\z^2{-}2\z^4)
(\a^2\dr{QCD} \wt R{\SU(3)} + \a^2_s R{G2})\bigr)^{1/2}\bigr)}{6(1+\z^2)(\a^2\dr{QCD} \wt R{\SU(3)} + \a^2_s R{G2})}.
\e
We impose the condition
\beq c282
\a^2\dr{QCD} \wt R{\SU(3)} + \a^2_s R{G2} < 0.
\e
Due to this we have $x_0>0$ and
\beq c283
x_0 = \biggl(\frac{\ell\pl}{r_0}\biggr)^2 \frac{\bigl(11+\bigl(121-120\pi(1+\z^2)^{1/2}(1-\z^2-2\z^4)(\a^2\dr{QCD} \wt R{\SU(3)} + \a^2_s R{G2})
\bigr)^{1/2}\bigr)}{6(1+\z^2)(\a^2\dr{QCD} \wt R{\SU(3)} + \a^2_s R{G2})}
\e
and
\bml c284
\Psi_0 = \frac12\log x_0 \\
{}= \frac{\ell\pl}{r_0} + \frac12 \log\biggl[
\frac{\bigl(11+\bigl(121-120\pi(1+\z^2)^{1/2}(1-\z^2-2\z^4)(\a^2\dr{QCD} \wt R{\SU(3)} + \a^2_s R{G2})
\bigr)^{1/2}\bigr)}{6(1+\z^2)}\biggr]\\ {}+ \frac12 \log(-(\a^2\dr{QCD} \wt R{\SU(3)}+ \a^2_s R{G2})).
\e

For $\la_c(\Psi_0) = \la_c(x_0)>0$ we get
\bml c285
\la_c(\Psi_0) = x_0^{10}\, \frac{\ell\pl^2}{r_0^4}\biggl[
\frac{\bigl(11+\bigl(121-120\pi(1+\z^2)^{1/2}(1-\z^2-2\z^4)(\a^2\dr{QCD} \wt R{\SU(3)} + \a^2_s R{G2})
\bigr)^{1/2}\bigr)^2}{36(1+\z^2)(\a^2\dr{QCD} \wt R{\SU(3)}+ \a^2_s R{G2})^2}\\
{}- \frac{2\bigl(11+\bigl(121-120\pi(1+\z^2)^{1/2}(1-\z^2-2\z^4)(\a^2\dr{QCD} \wt R{\SU(3)} + \a^2_s R{G2})
\bigr)^{1/2}\bigr)}{3(1+\z^2)^2(\a^2\dr{QCD} \wt R{\SU(3)}+ \a^2_s R{G2})}
+ \frac{\pi(1-2\z^2)}{\sqrt{1+\z^2}}\biggr]
\e
Eventually
\bml c286
\la_c(\Psi_0) = \biggl(\frac{\ell\pl}{r_0}\biggr)^{22} \frac
{\bigl(11+\bigl(121-120\pi(1+\z^2)^{1/2}(1-\z^2-2\z^4)(\a^2\dr{QCD} \wt R{\SU(3)} + \a^2_s R{G2})
\bigr)^{1/2}\bigr)^{10}}{6982656r_0^2 (1+\z^2)^{10}(\a^2\dr{QCD} \wt R{\SU(3)} + \a^2_s R{G2})^{10}}\\
\biggl[\frac{\bigl(11+\bigl(121-120\pi(1+\z^2)^{1/2}(1-\z^2-2\z^4)(\a^2\dr{QCD} \wt R{\SU(3)} + \a^2_s R{G2})
\bigr)^{1/2}\bigr)^{2}}{36(1+\z^2)(\a^2\dr{QCD} \wt R{\SU(3)} + \a^2_s R{G2})^2}\\
{}- \frac{4\bigl(11+\bigl(121-120\pi(1+\z^2)^{1/2}(1-\z^2-2\z^4)(\a^2\dr{QCD} \wt R{\SU(3)} + \a^2_s R{G2})\bigr)^{1/2}\bigr)}
{3(1+\z^2)(\a^2\dr{QCD} \wt R{\SU(3)} + \a^2_s R{G2})} + \frac{4\pi(1-2\z^2)}{\sqrt{1+\z^2}}\biggr]\\
{}= \frac12 m_{sk}^2 = \wt\La.
\e

Let us consider $\pd{^2\la_c}{\Psi^2}$. One gets
\begin{gather}
\pd{^2\la_c}{\Psi^2}(\Psi_0)\hskip400pt  \nn \\ {}= \frac5{7746r_0^2} \biggl(\frac{\ell\pl}{r_0}\biggr)^{22} \frac
{\bigl(11+\bigl(121-120\pi(1+\z^2)^{1/2}(1-\z^2-2\z^4)(\a^2\dr{QCD} \wt R{\SU(3)} + \a^2_s R{G2})
\bigr)^{1/2}\bigr)^{10}}{(1+\z^2)^{11}(\a^2\dr{QCD} \wt R{\SU(3)} + \a^2_s R{G2})^{10}} \nn \\
{}\cdot \bigl[55\bigl(121-120\pi(1+\z^2)^{1/2}(1-\z^2-2\z^4)(\a^2\dr{QCD} \wt R{\SU(3)} + \a^2_s R{G2})\bigr)^{1/2} \nn \\
\hskip100pt {}+ 12\pi(5-6(1+\z^2)^{1/2})(\a^2\dr{QCD} \wt R{\SU(3)} + \a^2_s R{G2})-121\bigr] \lb c287 \\
\pd{^2\la_c}{\Psi^2}(\Psi_0)>0. \lb c288
\end{gather}
Thus $\la_c(\Psi_0)$ is the minimum of the \co ical term.

Moreover we have
\bg c289
m_s^2 = \pd{^2\la_c}{\Psi^2}(\Psi_0) \\
m_{\bar s}^2 = \frac{m_s^2}{2\ov M} = \frac{-\pd{^2\la_c}{\Psi^2}(\Psi_0)}
{2(\mu\ell^{\SU(3)[ab]}\gd k,\SU(3),[ab], - 1386)}
= \pd{^2\la_c}{\Psi^2}(\Psi_0)\,\frac{\mu^2+4}{38703\mu^2 + 1544816}.
\e
The \co ical \ct\ can be estimated using the value of $y=\a^2\dr{QCD} \wt R{\SU(3)} + \a^2_s R{G2}$ to tune it to the experimental value~$\wt\La$.

Finally, let us notice that $RG2$ can easily be calculated ($\xi=0$):
\beq c291
RG2 = -\frac14 h^{G2\,ab}\gd h,G2,ab, = -\frac{14}4 = -\frac72
\e
and
\beq c292
y = \a^2\dr{QCD}\wt R\SU(3) - \frac72\a_s^2.
\e

Moreover this problem can be reversed. It means, to find such $y$ that we have the desired~$\wt\La$. In order to do it we should find an \e\
for~$y$ explicitly
\beq c293
y = \frac{22t-1}{t^2} \cdot \frac1{120\pi\zz^{1/2}(1-\z^2-2\z^4)}
\e
where $t$ \sf ies the \e
\bg c294
at^{12} + bt^{11} + ct^{10} = k \\
\aligned
a_{12} &= a = \biggl(\frac{\ell\pl}{r_0}\biggr)^{22} \frac{(120\pi)^{11}(1-\z^2-2\z^4)^{10}}{6982656\cdot36}\\
a_{11} &= b = - \biggl(\frac{\ell\pl}{r_0}\biggr)^{22} \frac{(120\pi)^{11}(1-2\z^2)^{11}\zz^5\zz^{1/2}}{3\cdot 1745664} \\
a_{10} &= c  = \biggl(\frac{\ell\pl}{r_0}\biggr)^{22}\cdot 4\pi(120\pi)^{10}\zz^{14}\zz^{1/2}(1-2\z^2)^{11} \\
-a_0 &= k = \wt\La r_0^2
\endaligned \lb c295
\e

Let us prove the \fw\ fact: $\wt\La$ cannot be zero ($\wt\La\ne0$ or $k\ne0$). We prove it by \ti{reductio ad absurdum}. Thus we suppose that
$\wt\La=k=0$. We get from Eq.~\er{c294}
\beq c296d
t^{10}(at^2+bt+c)=0.
\e
Thus $t=0$ or $at^2+bt+c=0$. If $t=0$ then $y\to-\iy$ which is impossible. Thus $at^2+bt+c=0$. But this quadratic \e\  has no real roots because
$\D<0$. It means, we obtain a contradiction and $\wt\La\ne0$. Let us calculate $y$ in terms of our geometrical quantities. We get
\beq c297d
y = \a\dr{QCD}^2 \wt R\SU(3) - \tfrac72 \a_s^2.
\e
In this way a value of a \co ical \ct\ has been connected to the geometrical parameter $\wt R\SU(3)$ (Eq.~\er{c283}) connected to $\SU(3)$-gauge
field (QCD). This is a perfect holistic physics (see Ref.~\cite{a}). In this way because $\wt\La\ne0$ we have really massive Dark Matter
\pc s (scalarons and skewons), not radiation.

Let us consider Eq.~\er{c294}. This is an algebraic \e\ of twelfth order and it cannot be solved by radicals. Moreover, it can be solved by
$\mathcal A$-hypergeometric series (see Refs \cite{mbk,mbl,mbm,mbn}. Let us define
\beq c298d
f(t) = a_0 + a_{10}t^{10} + a_{11}t^{11} + a_{12}t^{12}.
\e
Every root of $f(t)$ is an algebraic \f\ of \cf s $a_0,a_1,a_2,\dots,a_9,a_{10},a_{11},a_{12}$, i.e.
\beq c299d
T(a_0,a_1,\dots,a_9,a_{10},a_{11},a_{12})
\e
where $a_1=a_2=a_3=\ldots=a_9=0$. According to Ref.~\cite{mbk} the roots of Eq.~\er{c298d} \sf y the \fw\ system of linear partial differential \e s:
\bg c300d
\pp{^2T}{a_i\pa a_j} = \pp{^2T}{a_k\pa a_l}, \q \hbox{whenever }i+j=k+l,\ i,j,k,l=0,1,\dots,12, \\
\aligned
\sum_{i=0}^{12} i a_i\pp T{a_i} &= -T, \\
\sum_{i=0}^{12} a_i\pp T{a_i} &= -T,
\endaligned \lb c301d
\e
We can derive the \fw\ \e\ using a logarithmic \dv\ of $f(t)$:
\bg c302d
\pd {}t(\log f(t)) = \frac{f'(t)}{f(t)} \\
\pp{^2}{a_i\pa a_j}(\log f(t))' = -\frac{t^{i+j}}{f^2(t)}\,. \lb c303d
\e
In this way $\pd{}t[\log f(t)]$ \sf ies $\mathcal A$-hypergeometric \e s \er{c300d}--\er{c301d}. One gets
\beq c304d
T = \frac1{2\pi i} \int_\G \frac{zf'(z)}{f(z)}\,dz, \q z\in\C,
\e
where $\G$ is a sufficiently small loop on the complex plane. In this way we get a \so\ of equation \er{c294} and also $y$ from Eq.~\er{c293}.
In particular one gets
\bg c305d
f(T) = aT^{12} + bT^{11} + cT^{10} - k=0 \\
T = T(-k,c,b,a) \lb c306d \\
12a\,\pp Ta + 11b\,\pp Tb + 10c\,\pp Tc = -T \lb c307d \\
a\,\pp Ta + b\,\pp Tb + c\,\pp Tc = 0 \lb c308d
\e

According to Ref.~\cite{mbk} we write a \so\ of our twelfth order algebraic \e\ \er{c305d} as~$T$ (\er{c306d} with \er{c307d}--\er{c308d}). Let us
consider the matrix
\beq c309d
\mathcal A = \left(\begin{matrix}
0\ &\ 10\ &\ 11\ &\ 12 \\ 1\ & 1 & 1 &\ 1
\end{matrix}\right)
\e
and a subset $J \subset \{0,10,11,12\}$ doing a triangulation of $\mathcal A$ indices by $-\{0,12\}$, i.e.\ $0=i_0<i_1<\ldots<i_r=12$,
$$
[i_0,i_1], [i_1,i_2], \dots, [i_{r-1},i_r]
$$
defining $d_j=i_j-i_{j-1}$, i.e.\ the length of $j$-th segment of the triangulation. $d_1+d_2+\ldots+d_r=12$ (see Ref.~\cite{mbk} for details).
In our case $r=1$ and $d_1=12$. In this way we can write a \so\ of our \e\ according to Ref.~\cite{mbk}:
$$\dsl{
\refstepcounter{equation}\lb c310d
\indent T_{1,\o} = \o [a_0^{1/12}\cdot a_{12}^{-1/12}] + \frac1{12}\biggl(\frac1{a_0}[a_{10}\cdot a_0^{-1/12}a_{12}^{-11/12}]
+ [a_{11}\cdot a_0^0 a_{10}^{-12/12}]\biggr)\hfill \cr
\hfill{}=\o [(-k)^{1/12}\cdot a^{-1/12}] + \frac1{12} \biggl(\frac1{c_0}[(-k)^{-1/12}c^{1/12}] + [b\cdot c^{-1}]\biggr).\indent\rm(\theequation)\cr
\refstepcounter{equation}\lb c311d
\hfill \o = (-1)^{1/12} = \cos\biggl(\frac\pi{12}+\frac{k\pi}6\biggr) + i \sin\biggl(\frac1{12}+\frac{k\pi}6\biggr),\q k=0,1,\dots,11\hfill
\indent\rm(\theequation)\cr
\hfill \o^{12} = -1. \hfill}
$$
We choose any twelfth root of $(-1)$ (see Ref.~\cite{mbk}).

A condition for the convergence of the series $T_{1,\o}$ is \sf ied if we choose
\beq c312d
M = \max(A+1,B+1)
\e
where
\beq c313d
A= \frac{k^2 |a|^{10}}{|c|}\,, \q\ B= \frac{|k|a^{11}}{|b|}
\e
(see Ref.~\cite{mbk}). Finally we get:
\begin{align}
T_{1,\o} &= \o \sum_{v\in\cL}\frac{\G(\frac{13}{12})\G(\frac{11}{12})}{\G(\frac{13}{12}+v_0)\G(1+v_0)\G(1+v_{11})\G(\frac{11}{12}+v_{12})}\nn\\
&\qquad\ {}\cdot (-k)^{1/12+v_0} c^{v_{10}} b^{v_{11}} a^{11/12+v_{12}} \nn\\
&+ \frac1{12\o} \sum_{v\in\cL} \frac{\G(\frac{11}{12})\G(\frac{13}{12})\G(\frac{17}{6})\G(-\frac{5}{6})}
{\G(\frac{11}{12}+v_0)\G(\frac{13}{12}+v_{10})\G(\frac{17}{6}+v_{11})\G(-\frac{5}{6}+v_{12})}\nn \\
&\qquad\ {}\cdot (-k)^{-1/12+v_0}c^{1/12+v_0}b^{11/6+v_{11}}a^{-11/6+v_{12}} \lb c314d
\end{align}
$T_{1,\o}$ \sf ies Eqs \er{c307d} and \er{c308d} and trivially \er{c300d}. In order to get Eq.~\er{c307d} we solved the \e
\beq c315d
\bal
&\mathcal A(u_0,u_{10},u_{11},u_{12}) = \binom{\g_1}{\g_2} = \binom{-1}0, \\
&v = (v_0,v_{10},v_{11},v_{12})
\eal
\e
and $\cL$ is a 3-\di al sublattice of $Z^4$ (spanned by $(e_0,e_{10},e_{11},e_{12})$) spanned by $e_0$, $-2e_{10}+e_{11}$, $e_{10}-2e_{11}+e_{12}$
(see Ref.~\cite{mbk}), $\cL\subset Z^4$,
\beq c316d
f(T_{1,\o}) = 0.
\e

Let us consider a full \lg\ of our \un\ (i.e.\ \un\ of a \nos\ \gr al theory with bosonic part of a Standard Model in the \KK (Jordan--Thiry)
Theory framework. In the \lg\ we multiplied several parts by factors depending on a scalar field~$\Psi$ (which is a part of an extended gravity).
In our contemporary approach we consider a scalar field as a \ct\ $\Psi=\Psi_0$, because we use $\Psi$ as a source of a \co ical \ct\ and an
inflaton field. A~\co ical \ct\ is the minimum value of a self\ia\ \pt\ for~$\Psi$ at $\Psi=\Psi_0$. In this way some factors depending on~$\Psi_0$
can influence a value of $M_H$ and~$M_{Z^0}$.

In particular one gets $e^{-2\Psi_0}\cL\dr{kin}$ and $e^{(n-2)\Psi_0}V(\F)$. In our case we have $n=22$ and one gets
$e^{-2\Psi_0}\cL\dr{kin}$ and $e^{20\Psi_0}V(\F)$. The relations between $M_W$ and $M_{Z^0}$ are the same as for $\Psi=0$ (an absence of a scalar
field~$\Psi$). Moreover,
\bmlg
M_{Z^0} = \ov M_{Z^0}e^{-2\Psi_0} = \ov M_{Z^0}\,\frac1{x_0}\\{} = -\ov M_{Z^0}\biggl(\frac{r_0^2}{\ell^2\pl}\biggr)\frac{6\zz(\a^2\dr{QCD}\wt R\SU(3)
-\frac72\a_s^2)}{11+(121 - 120\pi\zz(1-\z^2-2\z^4)(\a^2\dr{QCD}\wt R\SU(3)-\frac72\a_s^2))^{1/2}}
\e
where $\ov M_{Z^0}$ is the mass of a $Z^0$ boson without a scalar field $\Psi$ correction.

Let us consider an important parameter of the theory for $\z=0$:
\beq c317d
\frac{M_H}{M_{Z^0}} = e^{11\Psi_0} = x_0^{11} = \biggl(\frac{\ell\pl}{r_0}\biggr)^{11}
\frac{(11+(121-120\pi y)^{1/2})^{11/2}}{6^{11/2}(-y)^{1/2}}
\e
where $y=\a^2\dr{QCD} \wt R\SU(3) - \frac72\a_s^2<0$.

One can find the \fw\ fact:
\bg c318d
y = \frac{-(132\a^{2/11}\b^2 + 120\pi)}{36\a^{11/2}\b^4},\\
\a=\frac{M_H}{M_{Z^0}}, \q \b=\biggl(\frac{r_0}{\ell\pl}\biggr). \lb{c322e} \\
y = \a^2\dr{QCD}R\SU(3) + \a^2_sRG2 = \frac{2\a^2\dr{QCD}}{(\mu^2+4)^2} (2\mu^3+7\mu^2+25\mu+20) - \frac12\biggl(\frac52\,\a^2\dr{QCD}+7\a^2_s
\biggr) \lb{c323e}
\e
One gets
\bml c319d
\wt\La = \frac12\,m_{sk}^2 = \la_c(\Psi_0) = \biggl(\frac{\ell\pl}{r_0}\biggr)^2
\frac{(11+(121-120\pi y)^{1/2})^{10}}{6982656r_0^2y^{10}}\\
{}\cdot\biggl[\frac{(11+(121-120\pi y)^{1/2})^2}{36y^2} - \frac{4(11+(121-120\pi y)^{1/2})}{3y} + 4\pi\biggr]
\e
\beq c320d
\pd{^2\la_0}{\Psi^2} (\Psi_0) = \frac5{774r_0^2}\biggl(\frac{\ell\pl}{r_0}\biggr)^{22}
\frac{(11+(121-120\pi y)^{1/2})^{11}}{y^{10}}\bigl[55(121-120\pi y)^{1/2}-12y -121\bigr].
\e
It means, we can find $y$ for any $\frac{M_H}{M_{Z^0}}$, i.e.\ for an experimental value of $\frac{M_H}{M_{Z^0}}$.

For $\z\ne0$ one gets
\beq c321d
\frac{M_H}{M_{Z^0}} = \sqrt{\zy} e^{11\Psi_0}.
\e
Eventually one gets
\beq c322d
\frac{M_H}{M_{Z^0}} = \zy^{1/2} \biggl(\frac{\ell\pl}{\sqrt{6\zz}\,r_0}\biggr)^{11}
\biggl(\frac{11+(121-120\pi \zz^{1/2}(1-\z^2-2\z^4)y)^{1/2}}{-y}\biggr)^{11/2}.
\e

We can parametrize $\frac{M_H}{M_{Z^0}}$ in several ways taking
\bg c325d
\sqrt{1-\rho_0^2} = e^{11\Psi_0} \\
\rho_0= \sqrt{\frac{1-e^{22\Psi_0}}2} \lb c326d \\
\rho_0 = \sqrt{\frac12\biggl(1 - \biggl(\frac{\ell\pl}{r_0}\biggr)^{22}\,\frac{11+(121-120\pi y)^{1/2}}{6y}\biggr)^{11}} \nn
\e
and $\rho_0=\z\dr{eff}$ for $\z=0$ ($\z\dr{eff}$ is defined below).

For $\z\ne0$ one gets
\bg c327d
\z\dr{eff} = \sqrt{\frac12\biggl(1 - \biggl(\frac{\ell\pl}{r_0}\biggr)^{22}\biggl(\frac{11+(121-120\pi \zz^{1/2}(1-\z^2-2\z^4)y)^{1/2}}
{6y\zz}\biggr)^{11} \biggr)}\\
\z\dr{eff} = \z\dr{eff}(\z,y), \q\ y = \a^2\dr{QCD}\wt R\SU(3) - \frac72\,\a_s^2. \nn
\e
In this way one gets
\beq c328d
\frac{M_H}{M_{Z^0}} = \sqrt{\zy} \qh{in NKKT}
\e
(see Ref.~\cite{11a}),
\bg c329d
\frac{M_H}{M_{Z^0}} = \sqrt{1-2\rho_0^2} \qh{in NKKT(JT)T for $\z=0$}\\
\frac{M_H}{M_{Z^0}} = \sqrt{1-2\z^2\dr{eff}} \qh{in NKKT(JT)T for $\z\ne0$.}\lb c330d
\e

In Ref.~\cite{11a} we are able to get $\frac{M_H}{M_{Z^0}}$ in an agreement with the experiment and a Weinberg angle $\t_W$ in an agreement with
the experiment using $\z$ as a free parameter and $\t_W=\frac\pi6$ (a~value without correction) to get a full agreement with the experiment using
a finite renormalization scheme. We can repeat all considerations from Appendix~E of Ref.~\cite{11a} to get the same results with redefinition of
$r_0^2\zz$ as an electroweak energy scale (i.e.\ $\d$~deviation from $\frac\pi6$ for $\t_W$ to a correct value equal to $\frac\pi6+\d$).

Moreover in Ref.~\cite{11a} we get those results in the \E\nos\ \KK Theory ($\rho=1$) which is different in general from the \E\nos\ \KK
(Jordan--Thiry) Theory ($\rho\ne1$). Thus for a convenience for the reader we repeat those considerations. Supposing as in Ref.~\cite{11a} that
$H=G2$ we get
\beq c335e
\frac{M_H}{M_W} = \frac1{\cos\th_W}\sqrt{1-2\z^2\dr{eff}} = \frac{2\sqrt{1-2\z^2\dr{eff}}}{\sqrt3}\,.
\e
$\z\dr{eff}$ is considered as an arbitrary \ct. If we take $M_H\simeq 125{\rm GeV}$ and $M_W=80{\rm GeV}$ we get
\beq c336e
\z\dr{eff} = \pm 0.911622i
\e
($\z\dr{eff}$ is not $\z$). $\z\dr{eff}$ is pure imaginary and is given by Eq.~\er{c327d} where $\z$~is now a free parameter and can be calculated.
One gets
\bml c337e
\z = \pm \frac i{2\ell\pl} \biggl(\biggl(-3(\z^2\dr{eff}-\tfrac12)^{1/11}r_0^2 + 10\pi \\{}+ \sqrt{3(\z^2\dr{eff}-\tfrac12)^{2/11}r_0^4(3-80\pi y)
+9\o \pi^2\ell\pl^4 - 94(\z^2\dr{eff}-\tfrac12)^{1/11}r_0^2\ell\pl^2}\biggr)/(10\pi)\biggr)^{1/2}.
\e
Now $y$ is given by the \e
$$
y = \frac{2\a^2\dr{QCD}}{(\mu^2+4)^2}(2\mu^3+7\mu^2+25\mu+20) - \frac12\biggl(\frac52\,\a^2\dr{QCD}+7\a_s^2\biggr)
$$
($\z$ is pure imaginary as before).

$r_0$ is a free parameter ($\ov M_{Z^0}$ is without a scalar field correction). We can define
\beq c338e
r_{0\rm eff} = \frac{\hbar c}{M_{Z^0}\sqrt{2\pi}\sqrt{1+\z^2\dr{eff}}}
\e
where
\beq c339e
r\dr{0eff} = \frac{\hbar c}{\sqrt{2\pi}\,M_{Z^0}\sqrt{1+\z\dr{eff}^2}} \simeq 273126 \tm 10^{-18}\,{\rm m}.
\e
$r_{0{\rm eff}}$ has a geometrical interpretation as a ``radius'' of a sphere $S^2$ in 5th and 6th \di s.

$r_0$ can be expressed in terms of $r\dr{0eff}$, $\z\dr{eff}$, $\a\dr{QCD}$, $\a_s$ and $\mu$
\bml c358n
r_0 = \bgg{\frac{\ell^2\pl}{r\dr{0eff}}} \\{}\cdot \frac{\Bigl(11{+}\Bigl(121{-}120\pi(1{+}\z^2\dr{eff})^{1/2}(1{-}\z^2\dr{eff}{-}2\z^4\dr{eff})
\Bigl(\frac{2\a^2\dr{QCD}}{(\mu^2+4)^2}(2\mu^3{+}7\mu^2{+}25\mu{+}20){-}\frac12\bigl(\frac52\a^2\dr{QCD}{+}7\a_s^2\bigr)\Bigr)\Bigr)^{1/2}\Bigr)}
{3\a_s^2(1+\z^2\dr{eff})\Bigl(\frac12\bigl(\frac52\a^2\dr{QCD} + 7\a_s^2\bigr) - \frac{2\a^2\dr{QCD}}{(\mu^2+4)^2}
(2\mu^3+7\mu^2+25\mu+20)\Bigr)}
\e

Now we are working as in Ref.~\cite{11a}. We see from the theory that $\th_W=\frac\pi6$. Let us remind to the reader that previously we used
$\ov\th$ for a Weinberg angle. We switch off a gravity and we are working in Minkowski \spt,
$g_\m=\eta_\m$,
\beq c340e
M_{Z^0} = \frac{M_W}{\cos \th_W} = \frac2{\sqrt3}\,M_W \simeq 92.4\,{\rm GeV}
\e
and $\sin^2\th_W = 0.25$. However from the experiment we get
\beq c341e
\sin^2 \th_W = 0.2397\pm0.0013,
\e
which is not $0.25$.

The value $0.25$ is without radiation correction and it is possible to turn it out at $Q=91.2\frac{\rm GeV}c$ in the MS scheme to get a desired value.
Let us notice the \fw\ fact. In electroweak theory we have a \lg\ for neutral current \ia
\beq c342e
\cL_N = q J^{\rm em}_\mu A^\mu + \sum_f \ov\psi_f\g_\mu(g^f_V - g^f_A\g^5)\psi_fZ^{0\mu} = qJ^{\rm em}_\mu A^\mu + \frac g{\cos\th_W}
(J^3_\mu - \sin^2\th_W J^{\rm em}_\mu)Z^{0\mu}
\e
where $g^f_V$ and $g^f_A$ are coupling \ct s for vector and axial \ia s for a fermion~$f$. One gets
\bea c342f
g^f_V &= \frac{2q}{\sin^2\th_W} (T^3_f - 2q_f \sin^2\th_W) \\
g^f_A &= \frac{2q}{\sin^2\th_W} \lb{c343e} \\
q_f &= T^3_f + \frac{Y_f}2 \lb{c344e}
\e
where $T^3_f$ is the third component of a weak isospin of a fermion~$f$, $q_f$~is its electric charge measured in \el ary charge~$q$, $Y_f$~is a weak
hypercharge of~$f$.

It is easy to see that for an electron we get $g^f_V=0$ if $\th_W=\frac\pi6$. Moreover, we know from an experiment that
\beq c345e
g^e_V \ne 0.
\e
Now we refer the reader to Ref.~\cite{11a} for further consideration on a correction~$\d$ to~$\frac\pi6$, $\th_W=\frac\pi6+\d$. One gets
\beq c346e
\d = -25'44'' \q\hbox{and}\q \th_W= 29^\circ 34' 16''
\e
getting
\beq c347e
M_W = M_{Z^0}\cos\th_W, \q\ \frac{M_H}{M_{Z^0}} = \sqrt{1-2\z^2\dr{eff}}\,.
\e
Now we have in the place of $\z$
\beq c348e
\z\dr{eff} = \pm 0.948735 i
\e
and
\beq c349e
M_W = M_{Z^0} \cos\bigl(\tfrac\pi6+\d\bigr).
\e
The value $\d$ can be improved getting
\bg c350e
\d = -0.0074855 \\
\hbox{and }\ \sin^2\th_W = 0.249162 \lb{c351e} \\
M_W = 79.3321\,{\rm GeV}. \lb{c352e}
\e
Moreover, in Appendix E of Ref.~\cite{11a} we give an improvement of those results using $\D r$ theory up to the second order. We get really perfect
agreement with an experiment:
\beq c353e
\th_W = 29^\circ 3' 11.898''.
\e
This value is consistent with values of masses
\beq c354e
\bal
M_W &= 79.7067\,{\rm GeV} \\
M_H &= 125.7\,{\rm GeV} \\
M_{Z^0} &= 91.18\,{\rm GeV}
\eal
\e
and with values of a running coupling \ct\ $\a\dr{em}$
\bea c355e
\a\dr{em}(0) &= \a\dr{em}(m_e^2) = \frac1{137.035} \\
\a\dr{em}(M^2_{Z^0}) &= \a\dr{em}(M^2_H) = \frac1{128}\,. \lb{c356e}
\e
This means we get a selfconsistency also in the case of $\z\dr{eff}$ in the place of~$\z$, i.e.\ in the \E\nos\ \KK (Jordan--Thiry) Theory which is
a partial \un\ of NGT and bosonic part of Standard Model. This result has been obtained in the case of switching off gravity, i.e.\ in Minkowski
\spt. In this way it belongs to ``interference effects'' in our \un.

For the convenience of the reader we give some details of our calculations. Let us define a differential cross-section $f^+f^- \to f^+f^-$ scattering
\beq c357e
\pd \si t\bigl(f^-(P)f^+ \to {f'}^-(P)'f^+\bigr) = \frac{4\pi \a\dr{em}}s \,\ka^2_{pp'}\cdot |M_{pp'}(-s)|^2
\e
where $\ka^2_{pp'}$ is a kinematic factor from the Dirac algebra equal to $\bigl(\frac us\bigr)$ for ${\rm L(left)} \to {\rm L(left)}$,
${\rm R(right)} \to {\rm R(right)}$ and $\bigl(\frac ts\bigr)^2$ for ${\rm L(left)}\to {\rm R(right)}$ and vice versa. At~$Z^0$ mass energies we can
ignore mass of fermion $f$ and~$f'$ ($M_{Z^0}>2m_f$ and $M_{Z^0}>2m_{f'}$). In this way the helicity is conserved. $M_{pp'}$~is an invariant amplitude
which contains all nontrivial information about a coupling. It is derived in such a way that $M_{pp'}$~is equal to~$1$ independently of~$p,p'$ for
a simple $s$-channel photon exchange diagram of the lowest order QED for electrons. In GSW theory one gets
\beq c358e
M_{pp'}(Q^2) = q_f\biggl(\frac{-s}{Q^2}\biggr)q_{f'} + \frac{T_f^3 - q_f\sin^2\th_W}{\cos\th_W \sin\th_W} \,
\frac{-s}{Q^2+M^2_{Z^0} - {\rm Im}(\Pi^{\text{1-loop}}_{Z^0Z^0}(Q^2))} \,\frac{T^3_{f'}-q_{f'}\sin^2\th_W}{\cos\th_W\sin\th_W}
\e
where
\bml c359e
{\rm Im}(\Pi^{\text{1-loop}}_{Z^0Z^0}) = \G^0_{Z^0}M_{Z^0} = \frac{\a\dr{em}}{3\sin^2\th_W \cos^2\th_W}\cdot
Z\Biggl[\biggl[\biggr(\frac{T^3_{fL}}2 - q_f\sin^2\th_W\biggr) \biggl(1+\frac{2m_f^2}{M^2_{Z^0}}\biggr)\\
{}+\biggl(\frac{T^3_{fL}}2\biggr)^2\biggl(1-4\frac{m_f^2}{M^2_{Z^0}}\biggr)\biggr]\biggl(1-\frac{4m_f^2}{M^2_{Z^0}}\biggr)C\dr{QCD}(f)\Biggr].
\e
where $T^3_{fL}$ is a left-handed isospin component for a fermion~$f$. The factor $C\dr{QCD}$ is\break $3(1+\a\dr{QCD}(M^2_{Z^0})/\pi)$ for quarks
and~$1$ for leptons. These formulae are very well known in all textbooks and as we mention above $\th_W$ is an arbitrary parameter. If we evaluate the
formulae for $\th_W=\frac\pi6$ we get
\beq c360e
M_{pp'}(Q^2) = q_f\biggl(-\frac s{Q^2}\biggr)q_{f'}
\e
which simply means $g^f_V=0$ for $\sin^2\th_W=0.25$.

Moreover we can introduce an effective Weinberg angle $\th_W = \frac\pi6+\d$ in such a way that all formulae are \sf ied. In this way radiative
corrections can be considered as corrections to $\frac\pi6$ Weinberg angle. The formula \er{c358e} can be evaluated in the \fw\ way
\beq c361e
M_{pp'}(Q^2) = q_f\biggl(-\frac s{Q^2} + 4\d^2\biggl(-\frac1{(Q^2+M^2_{Z_0}-{\rm Im}(\Pi^{\text{1-loop}}_{Z^0Z^0}(Q^2)))}\biggr)\biggr)q_{f'}
\e
($\d$ is a small correction to $\frac\pi6$).

One can use also some achievements from GSW model. Let us notice that
\beq c362e
M^2_W = \frac{\pi \a\dr{em}(m_e^2)G_F}{\sqrt2 \sin^2\th_W (1-\D r)}
\e
where $\D r$ is a 1-loop correction and its dominant contribution is
\bea c363e
\D r&= \D r_0 - \frac{1-\sin^2\th_W}{\sin^2\th_W}\,\D\rho + \D r\dr{rem} \\
\D r_0 &= 1-\frac{\a\dr{em}(m_e^2)}{\a\dr{em}(M^2_{Z^0})} \lb{c364e} \\
\D \rho &= \frac{3G_F\Si_f(m^2_{f_1}-m^2_{f_2})}{8\pi^2\sqrt2} \simeq \frac{3G_F(m_t^2-m_b^2)}{8\sqrt2\,\pi^2} \lb{c365e} \\
\D r\dr{rem} &= \frac{\sqrt2\,G_F M^2_W}{16\pi^2} \cdot \frac{11}3 \biggl(2\ln\biggl(\frac{M_H}{M_W}\biggr)-\frac56\biggr). \lb{c366e}
\e
$G_F$ is the Fermi \ct, $m_t$ and $m_b$ are top and bottom quark masses. The term $\D r_0$ corresponds to the running of $\a\dr{em}$ from $Q^2=m_e^2$
to the electroweak scale $Q^2=M^2_{Z^0}$.

$\D\rho$ depends quadratically on the mass difference between the members  of the same fermion doublet $\D r\dr{rem}$ (the remainder), is dominated
by Higgs' boson effects and depends logarithmically on~$M_H$. We evaluate a formula for~$M_W$ for $\th_W=\frac\pi6$ getting
\beq c367e
M^2_W = \frac{4\pi \a\dr{em}G_F}{\sqrt2(1-\D r)}
\e
(the so called $\D r$ theory, see Ref.~\cite{mcb}).

Now we proceed as before writing
\beq c368e
M_W^2 = \frac{4\a\dr{em}G_F}{\sqrt2 \sin^2\th_W}
\e
where $\th_W=\frac\pi6+\d$. $\d$ is not a new phenomenological parameter. It is an effect of 1-loop corrections and a running of~$\a\dr{em}$. In
Eq.~\er{c363e} we write $\th_W=\frac\pi6$ getting
\beq c369e
\D r = \D r_0 - 3\D\rho+ \D r\dr{rem}.
\e
In formula \er{c366e} we put value of Higgs' boson mass and a bare value of~$M_W$ (Eq.~\er{c368e}). In this way we get the desired value of
$\sin^2\th_W$:
\beq c370e
\sin^2\th_W = 4(1 - \D r).
\e
In terms of $\d$ one gets
\bea c371e
\d &= -\frac{\sqrt3\,\D r}6 \\
\hbox{or} \q \d &= -\frac{\sqrt3}6 \biggl(1-\frac{\a\dr{em}(m_e^2)}{\a\dr{em}(M^2_{Z^0})} - 3\D\rho + \D r\dr{rem}\biggr). \lb{c372e}
\e
We get exactly the same results if we use the $\ov{\text{MS}}$ definition $\sin^2\th_W = 1- \frac{M^2_W}{M^2_{Z^0}}$ which is also an effective value
of $\sin^2\th_W$. Using results from \E\pc\ Data Group we can evaluate~$\d$ from Eq.~\er{c372e} getting
\beq c373e
\d = -0.00748550
\e
or $\d = -25'44''$ and $\th_W= 29^\circ 34' 16''$. This gives
\beq c374e
\sin^2\th_W = 0.249162.
\e
The formula for $\d$ can be improved
\beq c375e
\d = -\frac{\sqrt3}6\bgg{1 - \frac{\a\dr{em}(m_e^2)}{\a\dr{em}(M^2_{Z^0})} - 3\D\rho + \D r\dr{rem}}\cdot\frac1{1-4\D\rho}\,.
\e
We get
\beq c376e
\d = -0.0074855, \q \sin^2\th_W = 0.249162.
\e

More precise quantum calculations can improve the result. The conclusion is as follows. The Weinberg angle is coming from the \un\ with $G2$ group.
The value of this parameter is equal to~$\frac\pi6$. The $\d$~correction is coming from radiative corrections. Now we have a physical relevance and
correct description of Nature (a~\un\ of NGT and a Standard Model-\E\nos\ \KK (Jordan--Thiry) Theory). The result can be improved starting from the
formula
\beq c377e
M^2_W \bgg{1-\frac{M_W^2}{M^2_{Z^0}}} = \frac{\pi \a\dr{em}(m_e^2)}{\sqrt2\,G_F}(1+\D r).
\e
We give some remarks.

We have here to do with a finite renormalization of a parameter in the theory, i.e.\ with a finite renormalization of a Weinberg angle. According to
the idea of a renormalization of any parameter due to quantum \ia s this is correct. We should renormalize not only masses and charges (as in~QED,
an electron charge and its mass, which is an infinite renormalization), but really any physical quantity, as in solid state physics (an effective mass
of an electron). An infinite renormalization in QED showed us an impossibility to avoid a renormalization in general.

The second remark is as follows. In classical field theory as in our model for $g_\m=\eta_\m$ we have to do with parameters which have an
interpretation as tree values. They should be renormalized. Only in a superrenormalizable theory they can remain the same in any order of perturbation
calculus. Our theory is not superrenormalizable.

In our approach we have on a classical level the \fw\ parameters $\z,r_0,\mu$ ($\sin^2\th_W$ is known from the theory). In order to get
precise predictions we should translate them into~$G_F$ (in particular $G_F=G_\mu$), $\a\dr{em}(m_e^2)$ (i.e.~$\a_s^2$) and~$\wt\La$ (a~\co ical \ct).
We take from \E\pc\ Data (see Ref.~\cite{m3k}) all the \ia\ \ct s (coupling parameters) which can change running parameters. Let us use the above
results to recalculate $\z\dr{eff}$ and~$r_0$ in terms of $M_H$, $M_{Z^0}$. One gets
\bea c378e
\frac{M_H}{M_{Z^0}} &= \sqrt{1-2\z^2\dr{eff}} \\
M_W &= M_{Z^0} \cos\th_W. \lb{c379e}
\e
$\z\dr{eff}$ is a \f\ of more \fn\ parameters, i.e.\ $\z$ and $y=y(\mu)$.

One gets
\bg c380e
\z\dr{eff} = \pm 0.948735 i \\
r\dr{0eff} = \frac{\hbar c}{M_{Z^0}\sqrt{2\pi}\,\sqrt{1+\z^2\dr{eff}}} = 273126 \tm 10^{-18}\,{\rm m}. \lb{c381e}
\e
In this way
\beq c382e
M_W = M_{Z^0} \cos\bigl(\tfrac\pi6+\d\bigr)
\e
gives the value of $\sin^2\th_W$ in Eq.~\er{c376e}.

We have of course
\beq c383e
M_H =125.7\,{\rm GeV}, \q M_{Z^0}=91.19\,{\rm GeV}.
\e
This means that if $\z\dr{eff}$ is given by \er{c380e} and $r_0$~given in Eq.~\er{c381e} we get $M_W = 79.3321\,$GeV which is a little smaller than
the experimental value $80.385\,$GeV. Moreover, considerations of higher order corrections of perturbation calculus (2-loop corrections) improve the
result to tune it to experimental value.

Let us notice the \fw\ fact. In further improvement using QCD corrections we neglect corrections coming from an extended \lg\ with $\mu\ne0$ (\nos\
QCD) as of higher order corrections.

Let us proceed the improvement. In order to do it we consider $\D r$ theory up to the second order known in the literature (see Ref.~\cite{mcb})
and references therein, see also Refs~\cite{mcc,mcd}. We have
\beq c384e
\D r = \bgg{1-\frac{\a\dr{em}(m_e^2)}{\a\dr{em}(M^2_{Z^0})}}\bgg{1-\frac{C_W^2}{S^2_W}\D\rho} + \D r\dr{rem}
\e
where
\bg c385e
\D\rho = 3x_t(1+x_t\rho^{(2)}(z) + i\rho\dr{QCD}) \\
x_t = \frac{G_F m_t^2}{8\pi^2\sqrt2} \lb{c386e} \\
\d\rho\dr{QCD} = -\frac{\a^2\dr{QCD}(M^2_{Z^0})}\pi \,C_1 +\bgg{\frac{\a^2\dr{QCD}(M^2_{Z^0})}\pi}^2 C_2(M^2_{Z^0}) \lb{c387e} \\
C_1= \frac23\bgg{\frac{\pi^2}3 +1 } \lb{c388e} \\
C_2 = -14.59 \lb{c389e} \\
\kern-30pt \rho^{(2)}(z)= -\frac{49}4 + \pi^2 + \frac{27}2 \log z + \frac32 \log^2z + \frac z3 (2-12\pi^2 +12\log z - 27\log^2 z) \nn \\
\hskip160pt {} +\frac{z^2}{48} (1613 - 240\pi^2 -1500 \log z - 720\log^2 z) \lb{c390e} \\
z = \frac{m_t^2}{M^2_H} \lb{c391e} \\
\D r\dr{rem} = \frac{\sqrt2\,G_FM^2_W}{16\pi^2} \cdot \frac{11}3\bgg{2\log\biggl(\frac{M_H}{M_W}\biggr) - \frac56}. \lb{c392e}
\e
From the definition
\beq c393e
\sin^2\th_W = \sin^2\bigl(\tfrac\pi6+ \d\bigr) = 4(1-\D r).
\e

One gets
\beq c394e
\d = -\frac{\sqrt3\Bigl(1-\frac{\a\dr{em}(m_w^2)}{\a\dr{em}(M^2_{Z^0})} - 3\D\rho\Bigl(1+\frac{\a\dr{em}(m_w^2)}{\a\dr{em}(M^2_{Z^0})}\Bigr)
+\D r\dr{rem}\Bigr)}
{6\Bigl(1-4\D\rho\Bigl(1-\frac{\a\dr{em}(m_w^2)}{\a\dr{em}(M^2_{Z^0})}\Bigr)\Bigr)}.
\e
Taking
\bea c395e
\a^2\dr{QCD}(M^2_{Z^0}) &= 0.1179 \\
G_F&= G_\mu = 1.66378 \tm 10^{-5} \,({\rm GeV})^{-2} \lb{c396e} \\
m_t&= 173.21\,{\rm GeV} \lb{c397e} \\
M_H&= 125.71\,{\rm GeV} \lb{c398e} \\
M_W&= 80.385\,{\rm GeV} \lb{c399e} \\
M_{Z^0} &= 91.18\,{\rm GeV} \lb{c400e} \\
\a\dr{em}(m_e^2) &= \frac1{137.035} = \a_s^2(m_e^2) \lb{c401e} \\
\a\dr{em}(M^2_{Z^0}) &= \frac1{128} = \a^2_s(M^2_{Z^0}) \lb{c402e} \\
\D\rho &= -0.000044702563 \lb{c403e}
\e
we eventually obtain
\bg c404e
\d = -0.01652297 \\
\d = -56'48.108'' \lb{c405e} \\
\sin^2\th_W = \sin^2\bigl(\tfrac\pi6+\d) = 0.23583 \lb{c406e} \\
M_W = 79.7067 \,{\rm GeV} \lb{c407e}
\e
which is almost correct value of a mass $W^\pm$ bosons
\beq c408e
\th_W = 29^\circ 3' 11.898''.
\e
It seems that this is selfconsistent.

Let us notice the \fw\ fact. Our partial \un\ (i.e. an \un\ of bosonic part of a Standard Model and NGT in the \KK (Jordan--Thiry) Theory) neglects
a possibility of \exi\ of additional bosons which can influence fermion physics. We incorporate fermions in the theory (see \rc{x}). Moreover, some
additional gauge bosons are not considered.

It seems that they could exist. Some experimental results coming from BaBar, LHCb, Belle, BelleII (a~``beauty physics'') in particular decays:
$b\mapsto sl^+l^-$, where $b$~is a beauty (bottom) quark, $s$~is a strange quark and $l^+,l^-$ are leptons (i.e.\ $e^+,e^-,\mu^+,\mu^-$)
indicate a possibility of an \exi\ of an additional $Z^0$~boson (i.e.\ $Z^{0\prime}$).

This can be also evident in $b\mapsto cl\nu$ decays ($l=e,\mu$). Such a boson can be incorporated in our scheme. Similarly there is also the so-called
Cabbibo angle anomaly for $qq\mapsto l^+l^-$ decays, $l=e,\mu,\tau$, which also breaks a university of leptons couplings. This can be explained by
an \exi\ of additional gauge bosons, e.g.\ leptoquarks. It could be incorporated in our \un\ by extending a gauge sector, i.e.\ taking a bigger gauge
group (Grand \E\un). Our approach is a~path to \un\ of \fn\ physical \ia s and can be extended by a larger group as in Section~3.

Let us notice that
\beq c330e
11t-1 < 0 \qh{and} 22t-1 = (11t-1) + 11t <0
\e
This can be easily \sf ied if $T_{1,\o}<0$ (see next formulae). We have
\beq c331e
\o(-1)^{1/12} =1
\e
which can be \sf ied if
\bg c332d
\o = \o_{k_1} = \exp\biggl(\frac{i\pi}{12} + \frac{2k_1\pi i}{12}\biggr),\q k_1=0,1,\dots,11,\\
(-1)^{1/12} = \o_{k_2} = \exp\biggl(\frac{i\pi}{12} + \frac{2k_2\pi i}{12}\biggr),\q k_2=0,1,\dots,11, \lb c333d \\
1 = \o_{k_1}\o_{k_2} = \exp\biggl(\frac{\pi i}6 (k_1+k_2+1)\biggr), \lb c334d \\
k_1+k_2+1=12, \q k_1+k_2=11. \nn
\e
In this way
\bml c335d
t = T(-k,c,b,a) = \sum_{v\in\cL} \frac{\G(\frac{13}{12})\G(\frac{11}{12})(-1)^{r_0}k^{1/12+v_0}c^{v_{10}}b^{v_{11}}a^{11/12+v_{12}}}
{\G(\frac{11}{12}+v_0)\G(1+v_{10})\G(1+v_{11})\G(\frac{11}{12}+v_{12})}\\
{}+ \frac1{12} \sum_{v\in\cL}\frac{\G(\frac{11}{12})\G(\frac{13}{12})\G(\frac{17}{6})\G(-\frac{5}{6})(-1)^{v_0}}
{\G(\frac{11}{12}+v_0)\G(\frac{13}{12}+v_{10})\G(\frac{17}6+v_{11})\G(-\frac56+v_{12})}\\
{}\cdot k^{-1/12+v_0}c^{1/12+v_{10}}b^{11/6+v_{11}}a^{-11/6+v_{12}} < 0
\e
$$
f(t)=0.
$$
\bml c336d
t = -12\pi(\sqrt6+\sqrt2)\sum_{v\in\cL}(-1)^{v_0}k^{v_0}c^{v_{10}}b^{v_{11}}a^{v_{12}}\\
\Biggl((\wt\La r_0)^{1/12}\biggl(\frac{\ell\pl}{r_0}\biggr)^{121/6}
\frac{(120\pi)^{121/12}(1-\z^2-2\z^4)^{55/2}}{6982656^{11/2}\cdot 6^{11/6}} \\ {}\cdot
\frac1{(v_0+\frac1{12})v_0v_{10}v_{11}(v_{12}-\frac1{12})\G(v_0+\frac1{12})\G(v_0)\G(v_{10})\G(v_{11})\G(v_{12}-\frac1{12})}\\
{}+ \frac{11\pi(4\pi)^{1/12}}{36\wt\La^{1/12}r_0^{1/12}\zz^{157/24}}\biggl(\frac{\ell\pl}{r_0}\,\pi^2\zy 691200\biggr)^{11/6} \\ {}\cdot
\frac1{(v_0-\frac1{12})(v_{10}+\frac1{12})(v_{11}+\frac{11}6)\G(v_0-\frac1{12})\G(v_{10}+\frac1{12})\G(v_{11}+\frac{11}6)\G(v_{12}-\frac{11}6)}\Biggr)
\e
We also get
\bml c337d
t= -12\pi(\sqrt6+\sqrt2)\sum_{v\in\cL} \frac{(-1)^{v_0+v_{11}}(r_0\wt\La)^{v_0}\bigl(120\pi\bigl(\frac{\ell\pl}{r_0}\bigr)^2 \zy\bigr)
^{11(v_{10}+v_{11}+v_{12})}}{36^{v_{12}}\bigl(48(1-\z^2-2\z^4)\bigr)^{v_{10}}\cdot 52336992^{v_{10}+v_{11}}\zz^{(11v_{11}-3v_{12})/2}}\\
\Biggl(\frac{(\wt \La r_0)^{1/12}(\frac{\ell\pl}{r_0})^{121/6}(120\pi)^{121/12}(1-\z^2-2\z^4)^{55/3}6982656^{-11/12}6^{-11/6}}
{(v_0+\frac1{12})v_0v_{10}v_{11} \G(v_0+\frac1{12})\G(v_0)\G(v_{10})\G(v_{11})\G(v_{12}+\frac1{12})(v_{12}-\frac1{12})}\\
{}+ \frac{11\pi(4\pi)^{1/12}\bigl(\frac{\ell\pl}{r_0}\pi^2\zy 691200\bigr)^{11/6}36^{-1}(\wt\La r_0)^{-1/12}\zz^{-157/24}}
{(v_0-\frac1{12})(v_{10}+\frac1{12})(v_{11}+\frac{11}6)\G(v_0-\frac1{12})\G(v_{10}+\frac1{12})\G(v_{11}+\frac{11}6)\G(v_{12}-\frac{11}6)}\Biggr)
\e

Using results from Section 2 one gets
\beq c299e
y = \a^2\dr{QCD} \wt R\SU(3) + \a^2_sRG2 = \frac{2\a^2\dr{QCD}}{(\mu^2+4)^2} (2\mu^3 +7 \mu^2 +25\mu+20) - \frac12\biggl(\frac52
\a^2\dr{QCD} + 7\a_s^2\biggr).
\e
Let us define
\beq c300e
s= \biggl(y+\frac72\a_s^2\biggr)\frac1{\a^2\dr{QCD}}+\frac54\,, \q
s = \frac{2(2\mu^3+7\mu^2+25\mu+20)}{(\mu^2+4)^2}\,.
\e
One gets the \e
\beq c301e
s\mu^4 - 4\mu^3 + 2(4s-7)\mu^2 - 50\mu + 8(2s-5)=0.
\e
This is a quartic \e\ for $\mu$, a parameter coming from strong \ia s ($s$~is a parameter involving~$y$ connected to a \co ical \ct). In this way
we have a holistic picture of physical \fn\ \ia. Let us solve Eq.~\er{c301e}. In order to do it we \tr\ \er{c300e} into a canonical form
\bg c302e
\mu = \eta+ \frac1s \\
\eta^4 + \frac2{s^2}\,\eta^2(4s^2\h{-}\h7s \h{-}\h 3) - \frac2{s^3}\,\eta (4\h{+}\h 17s^2\h{+}\h 14s) + \frac1{s^4}(14s^4 \h{-}\h 40s^3
\h{-}\h 42s^2 \h{-}\h 14s \h{+}\h1)=0 \lb{c303e} \\
\eta^4 + p\eta^2 + q\eta + r =0 \nn \\
p = \frac2{s^2}(4s^2-7s-3), \q q=-\frac2{s^3}(17s^2+14s+4), \q r=\frac1{s^4}(14s^4-40s^3-14s+1). \lb{c304e}
\e
In order to solve \er{c304e} we define a resolvent \e
\beq c305e
-\D(\la) = \la^3 - p\la^2 -4r \la + (4pr-q^2) =0
\e
and \tr\ it to a canonical form
\beq c305ae
\la = y+\frac p3\,.
\e
One gets
\beq c306e
\bga
y^3 + \ov p y+\ov q =0\\
\ov p = -\frac4{3s^4} (2s^4 - 16 s^3 + 67s^2 + 56s + 8) \\
\ov q = \frac2{9s^6} (1616s^6 + 737s^5 - 19436 s^4 - 1727s^3 + 1221s^2 - 233s - 369).
\ega
\e
To use Cardano method one should calculate a discriminant of our \e
\bml c307e
\ov D = \frac{\ov q{}^2}4 + \frac{\ov p{}^3}{27}  = \frac1{81s^{12}} \bigl[ 9(1616s^6 + 737s^5 - 19436 s^4 - 1727s^3 + 1221s^2 - 233s - 369)^2\\
{}- 64(2s^4 - 16s^3 + 67s^2 - 233s - 269)^3\bigr]
\e
or
\bml c307ae
\ov D = \frac1{81s^{12}}\bigl[18357292s^{12} + 2087856s^{11} - 564970768s^{10} - 372105128s^9 + 4051578 s^8\\ {}+ 73738438 s^7 - 487513380s^6
+20041305s^5 + 477179055s^4 - 289157780s^3\\ {}+ 3412942547s^2 - 738794224s + 1258057772 \bigr].
\e

We have two possibilities:

$1^\circ$: $\ov D>0$
\beq c308e
\la_1 = \frac1{6s^2}(4s^2-7s^3)+ \root3 \of{-\frac{\ov q}2 + \sqrt{\ov D}} - \root3\of {\frac{\ov q}2 + \sqrt{\ov D}}
\e
(the only real root of the resolvent \e) or
\bml c308ae
\la_1 = \frac1{6s^2}(4s^2-7s^3) + \frac{\root3\of3}{3s^2} \Bigl\{\Bigl[-(1616s^6 + 737s^5 - 19436 s^4 - 1727s^3 + 1221s^2 - 233s - 369)\\
{}+ \bigl(9(1616s^6 + 737s^5 - 19436 s^4 - 1727s^3 + 1221s^2 - 233s - 369)^2 \\{}- 64(2s^4-16s^3+67s^2-233s-269)^3\bigr)^{1/2}\Bigl]^{1/3} \\
{}-\Bigl[(1616s^6 + 737s^5 - 19436 s^4 - 1727s^3 + 1221s^2 - 233s - 369) \\
{}+ \bigl(9(1616s^6 + 737s^5 - 19436 s^4 - 1727s^3 + 1221s^2 - 233s - 369)^2 \\
{}-64(2s^4 - 16s^3 + 67s^2 - 233s - 269)^3\bigr)^{1/2}\Bigr]^{1/3}\Bigr\}.
\e

$2^\circ$: $\ov D<0$, $\ov p<0$ (\ti{casus irreducibilis})
\bg c309e
\la_k = \frac1{6s^2}(4s^2-7s-3) + \frac43\sqrt{2s^4 - 16s^3 + 67s^2 + 56s +8}\cdot \cos\biggl(\frac \vf3+\frac{2k\pi}3\biggr), \q
k=0,1,2, \\
\cos\vf = -\frac{3(1616s^6 + 737s^5 - 19436 s^4 - 1727s^3 + 1221s^2 - 233s - 369)}
{16s^2(2s^4 - 16s^3 + 67s^2 - 233s - 269)^{3/2}}\,. \lb{c310e}
\e
A quartic \e\ written in a canonical form
$$
\eta^4 + p\eta^2 + q\eta + r =0
$$
can be \tr ed into a product of two quadratic \e s
\beq c311e
\bca
\eta^2 + \eta\sqrt{\la-p} + \biggl(\dfrac\la2 - \dfrac q{2\sqrt{\la - p}}\biggr) = 0 \\
\eta^2 - \eta\sqrt{\la-p} + \biggl(\dfrac\la2 + \dfrac q{2\sqrt{\la - p}}\biggr) = 0
\eca
\e
where $\la$ is one of the roots of the resolvent \e\ $-\D(\la)=0$. In this way we get
\beq c312e
\bal
\mu &= \eta+\frac1s,\\
\mu_{1,2} &= \frac{(\la-p)^{3/4} \pm \sqrt{2q - (p+\la)\sqrt{\la-p}}}{2(\la-p)^{1/4}} + \frac1s \\
\mu_{3,4} & = \frac{(\la-p)^{3/4} \pm \sqrt{-(2q+(p+\la)\sqrt{\la -p })}}{2(\la-p)^{1/4}} + \frac1s
\eal
\e
or
\beq c313e
\bal
&\mu_{1,2} = \frac1s + {}\\
& \frac{\bigl(\la-\frac2{s^2}(4s^2{-}7s{-}3)\bigr)^{3/4} \pm \sqrt{-\frac4{s^3}(17s^2+14s+4)-\bigl(\bigl(\frac2{s^2}{-}7s{-}3\bigr)+\la\bigr)
\sqrt{\la-\frac2{s^2}(4s^2{-}7s{-}3)}}}{2\bigl(\la - \frac2{s^2}(4s^2{-}7s{-}3)\bigr)^{1/4}}  \\
&\mu_{3,4} = \frac1s + {}\\
& \frac{\bigl(\la-\frac2{s^2}(4s^2{-}7s{-}3)\bigr)^{3/4} \pm \sqrt{-\bigl(-\frac4{s^3}(17s^2+14s+4)+\bigl(\bigl(\frac2{s^2}{-}7s{-}3\bigr)+\la\bigr)
\sqrt{\la-\frac2{s^2}(4s^2{-}7s{-}3)}\bigr)}}{2\bigl(\la - \frac2{s^2}(4s^2{-}7s{-}3)\bigr)^{1/4}}
\eal
\e
One remembers that
$$
s=\frac{\bigl(y+\tfrac72\a_s^2\bigr)}{\a^2\dr{QCD}} + \frac54
$$
where
$$
y = \frac{22t-1}{t^2} \cdot \frac1{(120\pi \zz^{1/2}(1-\z^2-2\z^4))}
$$
and
\bml c314e
t = -12\pi \sqrt2\bigl(1+\sqrt3\bigr)\sum_{v\in\cL}
\frac{(-1)^{v_0+v_{11}}(r_0\wt\La)^{v_0}\Bigl(120\pi\bigl(\frac{\ell\pl}{r_0}\bigr)^2 \zy\Bigr)^{11(v_{10}+v_{11}+v_{12})}}
{36^{v_{12}}\bigl(48(1-\z^2-2\z^4)\bigr)^{v_{10}} (52336992)^{v_{10}+v_{11}}\zz^{(11v_{11}-3v_{12})/2}}\\
\cdot
\Biggl(\frac{(\wt \La r_0)^{1/12}(\frac{\ell\pl}{r_0})^{121/6}(120\pi)^{121/12}(1-\z^2-2\z^4)^{55/3}6982656^{-11/12}6^{-11/6}}
{(v_0+\frac1{12})v_0v_{10}v_{11} \G(v_0+\frac1{12})\G(v_0)\G(v_{10})\G(v_{11})\G(v_{12}-\frac1{12})(v_{12}-\frac1{12})}\\
{}+ \frac{11\pi(4\pi)^{1/12}\bigl(\frac{\ell\pl}{r_0}\pi^2\zy 691200\bigr)^{11/6}36^{-1}(\wt\La r_0)^{-1/12}\zz^{-157/24}}
{(v_0-\frac1{12})(v_{10}+\frac1{12})(v_{11}+\frac{11}6)\G(v_0-\frac1{12})\G(v_{10}+\frac1{12})\G(v_{11}+\frac{11}6)\G(v_{12}-\frac{11}6)}\Biggr)
\e
In this way we connect a \ct\ $\mu$ from extended QCD (\nos) with a \co ical \ct. We can tune $\mu$ to the desired value of~$\wt\La$.

Let us consider field \e s for our partial \un\ (a~\un\ of NGT and a bosonic part of a Standard Model) once again. For after ten years from the
discovery of a Higgs' boson we have not obtained any new Higgs' phenomena (see Ref.~\cite{m3k}) we put $\xi=0$. In this way we get the \fw\
formulae:
\bml c296
\cL\dr{kin}\bigl(\ggs{}\F\bigr) \\ {}= \frac{\a_s^2}{2\pi \hbar cr_0^2\zz} \Bigl(\bigl((2g^{\o\mu}-\gtn{\a\mu})\bigl(\ka
\bigl(\ggs\o\F_5,\ggs\mu\wh\F_6\bigr) \\ {}- \ka\bigl(\ggs\o \wh\F_6, \ggs\mu \F_6\bigr)\bigr)\bigr) \\
{} - 2\z(g^{\o\mu}-\gtn{\o\mu})\ka\bigl(\ggs\o \F_5,\ggs\mu\wh\F_6\bigr)\Bigr)
\e
where
\beq c297
\ka(x,y) = h_{ab}x^ay^b
\e
and
\bml c298
\ggs\mu \F_5 = \frac12\biggl(\pa_\mu\vf_1 - \frac{\a_s i}{2\sqrt{\hbar c}}A^-_\mu\vf_2 - \frac{\a_s i}{2\sqrt{\hbar c}}A^3_\mu\vf_1
-\frac{\a_s i}{6\sqrt{\hbar c}}\sqrt3\,B_\mu\vf_1\biggr)x_\a \\
{}+\biggl(\pa_\mu\vf_2 - \frac{\a_s i}{2\sqrt{\hbar c}}A^+_\mu \vf_1 + \frac{\a_s i}{2\sqrt{\hbar c}}A^3_\mu\vf_2
- \frac{\a_s i}{6\sqrt{\hbar c}}B_\mu\vf_2\sqrt3\biggr)x_\b \\
{}- \biggl(\pa_\mu\vf_1^* + \frac{\a_s i}{2\sqrt{\hbar c}}A^+_\mu\vf_2^* + \frac{\a_s i}{2\sqrt{\hbar c}}A^3_\mu\vf_1^*
+ \frac{\a_s i}{6\sqrt{\hbar c}}B_\mu\vf_1^* \sqrt3\biggr)x_{-\a} \\
{}+\biggl(\pa_\mu\vf_2^* + \frac{\a_s i}{2\sqrt{\hbar c}}A^-_\mu\vf_1^* - \frac{\a_s i}{2\sqrt{\hbar c}}A^3_\mu\vf_2^*
+ \frac{\a_s i}{6\sqrt{\hbar c}}B_\mu\vf_1^*\sqrt3\biggr)x_{-\b}
\e
\vskip-18pt
\bml c299
\ggs\mu \wh\F_6 = \frac i2\biggl(\biggl(\pa_\mu\vf_1 - \frac{\a_s i}{2\sqrt{\hbar c}}A^-_\mu\vf_2 - \frac{\a_s i}{2\sqrt{\hbar c}}A^3_\mu\vf_1
- \frac{\a_s i}{6\sqrt{\hbar c}}\sqrt3\,B_\mu\vf_1\biggr)x_\a \\
+ \biggl(\pa_\mu\vf_2 - \frac{\a_s i}{2\sqrt{\hbar c}}A^+_\mu\vf_1 + \frac{\a_s i}{2\sqrt{\hbar c}}A^3_\mu\vf_2
- \frac{\a_s i}{6\sqrt{\hbar c}}B_\mu\vf_2\sqrt3\biggr)x_\b \\
- \biggl(\pa_\mu\vf_1^* + \frac{\a_s i}{2\sqrt{\hbar c}}A^+_\mu\vf_2^* + \frac{\a_s i}{2\sqrt{\hbar c}}A^3_\mu\vf_1^*
+ \frac{\a_s i}{6\sqrt{\hbar c}}B_\mu\sqrt3\biggr)x_{-\a} \\
- \biggl(\pa_\mu\vf_2^* + \frac{\a_s i}{2\sqrt{\hbar c}}A^-_\mu\vf_1^* - \frac{\a_s i}{2\sqrt{\hbar c}}A^3_\mu\vf_2^*
+ \frac{\a_s i}{6\sqrt{\hbar c}}B_\mu\sqrt3\biggr)x_{-\b}\biggr)
\e
Finally we get a very useful formula for $\cL\dr{kin}(\n\F)$:
\refstepcounter{equation} \lb c300
$$
\displaylines{
\indent \cL\dr{kin}(\n\F) =\bigl(2g^{\mu\b} -\gtn{\mu\b} +2\z(g^{\mu\b}-\gtn{\mu\b})\bigr)\biggl(\pa_\mu\wt\vf -\frac12\,\frac{\a_s}{\sqrt{\hbar c}}\,
A^a_\mu \si^a\wt\vf - \frac12\,\frac{\a_s}{\sqrt{3\hbar c}}\,B_\mu\wt\vf\biggr)\hfill\cr
\hfill \biggl(\pa_\b\wt\vf - \frac12\,\frac{\a_s}{\sqrt{\hbar c}}\,A^a_\b \si^a\wt\vf - \frac12\,\frac{\a_s}{\sqrt{\hbar
c}}\,B_\b\wt\vf\biggr)\indent\cr
\hfill \wh V(\F) = \wh V(\vf_1,\vf_2) = \frac{\pi(1-2\z^2)}{2r_0^2\zz}(4-2(\wt\vf{}^+\wt\vf)+(\wt\vf^+\wt\vf)^2). \hfill\rm(\theequation)
}
$$
\bml c301
T_{\o\mu}(\cL\dr{kin}(\ggs{}\F)) = \frac1{2\pi r_0^2\zz}\Bigl[2\bigl(\ka(\ggs\o\F_5, \ggs\mu\F_5)\\{} - \ka(\ggs\o\wh\F_6,\ggs\mu\wh\F_6)
- \z\ka(\ggs\o\F_5, \ggs\mu\wh\F_6)\bigr)\\ {}+ \tfrac12 (\gtn{\a\b}\gtn{\xi\d}g_{\b\mu}g_{\xi\o} + \gtn{\xi\d}\gtn{\a\b}g_{\xi\mu}g_{\b\o})
\Bigl(-\ka(\ggs\a\F_5,\ggs\d\F_5) \\ {}+ \ka(\ggs\a\wh\F_6,\ggs\d\wh\F_6) + 2\z\ka(\ggs\a\F_5,\ggs\d\wh\F_6)\Bigr)\\
{}-\tfrac14 g_{\o\mu}\Bigl((2g^{\a\d}-\gtn{\a\d})\bigl(\ka(\ggs\a\F_5,\ggs\d\F_5) \\{}- \ka(\ggs\a\wh\F_6,\ggs\d\wh\F_6)\bigr)\\
{}-2\z(g^{\a\d} - \gtn{\a\d}) \ka(\ggs\a\F_5,\ggs\d\wh\F_6)\Bigr)\Bigr]
\e
\beq c302
T_{\o\mu}(\wh V(\F)) = -\frac{\pi (1-2\z^2)}{8\zz r_0^2} (4-2(\wt\vf{}^+\wt\vf) + (\wt\vf{}^+\wt\vf)^2)g_{\o\mu}.
\e

$\cL\dr{int}$ is given by the formula \er{c254}.
\bml c303
T_{\o\mu}(\cL\dr{int}) = \frac{i\z\sqrt6}{16\pi r_0^2\zz}\bigl[(3F^3_{\o\mu}(|\vf_1|^2-|\vf_2|^2) - \vf_1\vf_2^*(F^-_{\o\mu}+F^+_{\o\mu}))\\
{}-\tfrac14g_{\o\mu}(3F^3_{\a\b}g^{\a\b}(|\vf_1|^2-|\vf_2|^2) - \vf_1\vf_2^*(F^-_{\a\b}g^{\a\b}+F^+_{\a\b}g^{\a\b}))\bigr].
\e
Now the field \e\ looks as in \er{4.109}. In \er{c119b} we put in the place of $T_{\m}(\F)$
$$
T_{\m}(\cL\dr{kin}(\ggs{}\F) + T_{\m}(\wh V(\F))
$$
according to formulae \er{c301} and \er{c302}. For $\br{int}T_\m$ we put $T_{\m}(\cL\dr{int})$ according to Eq.~\er{c303}.

For further investigations we need a new form of field \e s \er{4.109}--\er{4.111}:
\bg c304
\ov R_{(\a\b)}(\ov\G) = 8\pi\nad{eff.full}{T_{(\a\b)}} \\
\ov R_{[[\a\b],\g]} = 8\pi \nad{eff.full}{T_{[[\a\b],\g]}} \lb c305 \\
\ov \G_\mu = 0 \lb c306 \\
g_{\m,\si} - g_{\xi\nu}\gd \ov\G,\xi,\mu\si, - g_{\mu\xi}\gd\ov\G,\xi,\nu\si, = 0 \lb c307 \\
\ov\G_\nu = \gd\ov\G,\si,[\mu\si], \lb c308
\e
$\ov R_{\a\b}(\ov \G)$ is a Moffat--Ricci tensor for the \cn
\beq c309
\gd \ov\o,\a,\b, = \gd\ov\G,\a,\b\g, \ov\th^\g,
\e
in the place of \er{c306} we also use
\beq c310
\gd \falg,[\m],{,\nu}, = 0
\e
or
\bg c311
\ov\n_\mu \gtk{\m} =0 \\
8\pi \nad{eff.full}{T_{\a\b}} = \frac{8\pi K}{c^4} \bigl(\nad{gauge(\SU(3))}{T_{\a\b}} + \nad{gauge(\SU(2)_L\ot \U(1)\dr{Y})}{T_{\a\b}}
+ (\nad{scal}T_{\a\b}(\Psi) - \tfrac12 g_{\a\b}\nad{scal}T_{\m}(\Psi)g^{\m}) \nn \\
{}+ T_{\a\b}(\cL\dr{int}) + (T_{\a\b}(\cL\dr{kin}(\ggs{}\F) + T_{\a\b}(\wh V(\F)) \nn \\
{}- \tfrac12 g_{\a\b} (g^{\m}(T_{\m}(\cL\dr{kin}(\ggs{}\F)) + T_{\m}(\wh V(\F)))) \lb c312
\e
We have
\bml c313
\cLY(\a_sA_{\SU(2)_L\ot\U(1)\dr{Y}}) = \frac{\a_s^2}{24\pi}\bigl[\d_{ij}(\ov L{}^{i\,\m}\gd \ov H,j,\m, - 2(\gtk{\m}\gd \ov H,i,\m,)(\gtk{\g\si}
\gd\ov H,j,\g\nu, ))\\ {}+ (\wh L{}^{\m}G_{\m} - 2(\gtk{\m}G_{\m})(\gtk{\g\d}G_{\g\d}))\bigr].
\e
For $\br{gauge(\SU(3))}{T_{\a\b}}$ we have
\bml c314
\mathop{T_{\a\b }}\limits^{\gauge(\SU(3))} = -\a_{QCD}^2\, \frac{\gd\ell,\SU(3),ij,}{4\pi}
\biggl\{ g_{\g\b}g^{\tau\rho} g^{\ve\g} \gd L,i,\rho\a, \gd L,j,\tau\ve,
-2\gtk{\m} \gd H,(i,\m, \gd H,j),\a\b, \\{}- \frac14 \,g_{\a\b}\Bigl[L^{i\m}
\gd H,j,\m, -2(\gtk{\m}\gd H,i,\m,)(\gtk{\g\d}\gd H,j,\g\si,)\Bigr]\biggr\}
\e
For $\br{scal}T_{\a\b}(\Psi) - \frac12 g_{\a\b}\br{scal}T_{\m}(\Psi)g^{\m}$ we have
\bml c315
\mathop{T_{\a\b}}\limits^{\scal} ({\varPsi} ) - \frac12\,g_{\a\b}
\mathop{T_{\m}}\limits^{\scal} (\Psi)g^{\m} = -\frac{e^{24\Psi}}{16\pi}
\Bigl\{(g_{\ka\a}g_{\o\b} + g_{\o\a}g_{\ka\b}) \cdot \gtn{\g\ka}\gtn{\nu\o}
[242(g^{\xi\mu}g_{\nu\xi} - \d^\mu_\nu)\varPsi_{,\mu} \\{}+ \d^\mu_\nu \ov M\varPsi_{,\nu}]
\Ps_{,\g} - g_{\a\b}[\ov M \gtn{\nu\mu}\Ps_{,\mu}\Ps_{,\nu} + 484\gtk{\m}g_{\d\mu}\gtn{\g\d}
\Ps_{,\nu}\Ps_{,\g}]\\ + g_{\a\b}\Ps_{,\mu}\Ps_{,\g}\bigl\{(242+\ov M)\gtn{\g\mu}
+242(\gtn{\g\tau}g^{\xi\mu}g_{\nu\xi} - 4\gtk{\a\mu}g_{[\d\a]}\gtn{\g\d})\bigr\}\Bigr\}
\e
\bml c316
\br{gauge(\SU(2)_L\ot\U(1)\dr{Y})}{T_{\a\b}} = \frac{\a_s^2}{24\pi}
\Bigl\{\d_{ij}\Bigl\{ g_{\g\b}g^{\tau\rho}\gd\ov L,i,\rho\a, \gd\ov L,j,\tau\ve, - 2\gtk{\m}\gd\ov H,(i,\m, \gd\ov H,j),\a\b,\\
{}- \tfrac14 g_{\a\b}[\ov L{}^{i\,\m}\gd\ov H,j,\m, - 2(\gtk{\m}\gd\ov H,i,\m,)(\gtk{\g\d}\gd\ov H,j,\g\d,)]\Bigr\}
+ \Bigl\{g_{\nu\b}g^{\ve\g} \wh L_{\rho\a}\wh L_{\tau\ve} - 2\gtk{\m}G_{\m}G_{\a\b} \\ {}- \tfrac14 g_{\a\b}[\wh L{}^{\m}G_\m
-2(g^\m G_\m)(\gtk{\g\d}G_{\g\d})\bigr]\Bigr\}\Bigr\}.
\e
For $T_{\a\b}(\cL\dr{int})$ we get Eq.~\er{c303} with some rearrangement of indices.
\bml c317
T_{\a\b}(\cL\dr{int}) = \frac{i\z\sqrt6}{16\pi r_0^2\zz} \bigl[(3F^3_{\a\b}(|\vf_1|^2 -|\vf_2|^2) - \vf_1\vf_2^*(F^-_{\a\b}+F^+_{\a\b}))\\
{}- \tfrac14g_{\a\b}(g^\m(3F^3_\m(|\vf_1|^2-|\vf_2|^2) - \vf_1\vf_2^*(F^-_\m +F^+_\m)))\bigr].
\e
The last term in Eq.~\er{c312} is given via formulas \er{c301}--\er{c302} with \er{c296}--\er{c297} and \er{c298}--\er{c299}.

Let us write down \e s for the remaining fields. For $\SU(3)$ gauge field we get \er{c138b} which we repeat here:
\bml c318
\br{gauge(SU(3))}{\n_\mu} (\gd\ell,\SU(3),ab, \fal L^{a\a\mu}) = 2\falg^{[\a\b]}
\br{gauge(SU(3))}{\n_\b} (\gd h,\SU(3),ab, \gtk{\m} \gd H,a,\m,) \\
{}+24\pa_\b \Psi\bigl[\gd \ell,\SU(3),ab,\cdot \fal L^{a\b\a} - 2\falg^{[\b\a]}
(\gd h,\SU(3),ab, \gtk{\m} \gd H,a,\m,)\bigr].
\e
For scalar field we have Eq.~\er{c137b} which we repeat here with some improvements:
\bml c319
\bigl((242+\ov M)\gtn{\a\mu} - 242\gtn{\a\mu} - 121 g^{\m}g_{\d\nu}\gtn{\a\d}\bigr)\pap{^2\Ps}{x^\a \pa x^\mu}\\
{}+\frac1{\sqrt{-g}}\,\pa_\mu\Bigl\{\sqrt{-g}\bigl[242\gtn{\a\mu} - 121g_{\d\nu}\cdot
(g^{\nu\a}\gtn{\mu\d}+\g^{\nu\mu}\gtn{\mu\a}) - \ov M \gtn{\mu \a}\bigr]\Bigr\}\pap\Ps{x^\a}\\
{}-96\pi e^{-24\Psi}\bigl(\cLY(\a_{QCD}A_{\SU(3)})+\cLY(\a_sA_{\SU(2)_L\ot \U(1)\dr{Y}})\bigr) \\
{}-2e^{-2\Psi} \cL\dr{kin}(\br{gauge(SU(2)_L\ot \U(1)\dr{Y})}{\n\F})+\frac{10}{r_0^4}\,e^{20\Psi}\wh V(\F) +
\frac{40}{r_0^2}\,e^{20\Psi}\cL\dr{int}-\frac{11}{r_0^2}\,e^{22\Psi}\cdot \wh{\ov P}_{S^2} \\
{}- \frac{12}{\ell\pl^2}\,e^{24\Psi}(\a_s^2 \wt R_{G2} + \a_{QCD}^2\wt R_{\SU(3)})
-10\,\frac{e^{20\Psi}}{r_0^4}\,\ell\pl^2 \wh V(0) = 0.
\e
Let us consider a \lg\ for $\SU(2)_L \ot \U(1)\dr{Y}$ gauge fields and scalar fields, i.e.
\beq c320
\cL(\a_s A_{\SU(2)_L\ot\U(1)\dr{Y}}) + \cL\dr{kin}(\br{gauge(SU(2)_L\ot \U(1)\dr{Y})}{\n\F}) - V(\F) = \cL(A_{\SU(2)_L\ot\U(1)\dr{Y}},\F).
\e
In this way one gets
\bml c321
\cLY(\a_s A_{\SU(2)_L\ot\U(1)\dr{Y}}) = \frac{\a_s^2}{48\pi\hbar c}\bigl[\d_{ij}[\ov L{}^{i\,\m}\gd\ov H,j,\m, - 2(\gtk\m \gd\ov H,i,\m, )
(\gtk{\a\b}\gd\ov H,j,\a\b,)] \\ {}+ [\ov L{}^\m G_\m - 2(\gtk\m G_\m)^2]\bigr].
\e
For the \pt\ $\wh V(\F) = \wh V(\wt\vf) = \wh V(\vf_1,\vf_2)$ one gets
\beq c322
\wh V(\F) = \wh V(\vf_1,\vf_2) = \frac{(1-2\z^2)\pi}{2\zz^{1/2} r_0^2} (4-\wt\mu{}^2 (\wt\vf{}^+ \wt\vf) + \la'(\wt\vf{}^+ \wt\vf)^2), \q\wt\mu{}^2=2,
\ \la'=1.
\e

Let us come back to a \ssb\ and \Hm\ in our \un. In order to do it we switch off a gravity. It means, we are working in a Minkowski space. The \pt\
for a scalar field $\wt\vf=\binom{\vf_1}{\vf_2}$ is as follows (see Eq.~\er{c320}) or in a classical form
\bg c323
U(\vf_1,\vf_2) = -\mu^2(\wt\vf{}^+ \wt\vf)^2 + \la(\wt\vf{}^+\wt\vf)^2 + {\rm const.}\\
\mu^2 = \frac{2(1-2\z^2)\a_s^4}{64r_0^4 (\hbar c)^2 \zz^2} \lb c324 \\
\la = \frac{(1-2\z^2)\a_s^2}{64r_0^4 (\hbar c)^2 \zz^2} \nn
\e

Let us parametrize a doublet $\wt\vf$ of scalar fields in the \fw\ way (we are using here a formalism more connecting to \el ary \pc\ physics)
(see Ref.~\cite{maa}:
\bg c325
\wt\vf(x) = U^{-1}\biggl(\frac12 t(x)\biggr)\left(\begin{matrix}
0 \\
\dfrac{v+H^0(x)}{\sqrt2}
\end{matrix}\right)\\
U\biggr(\frac12 t(x)\biggr) = \exp\biggl(-\frac i2\,t^a(x)\frac{\si^a}v\biggr), \q v=\biggl(\frac{\mu^2}\la\biggr)^{1/2}=\sqrt2, \nn
\e
where $\si^a$ are Pauli matrices.

In this way one gets in a gauge $U$
\bg c326
\wt\vf(x) \to \wt\vf{}^u(x) = U\biggl(\frac12t(x)\biggr)\wt\vf(x) = \left(\begin{matrix}
0 \\
\dfrac{v+H^0(x)}{\sqrt2}
\end{matrix}\right) \\
\hbox{or}\q \wt\vf{}^u(x) = \frac{v+H^0(x)}{\sqrt2}\d, \q \d=\binom01. \lb c327
\e
One gets in a gauge $U$
\bg c328
\frac{\si^a A^{ua}_\mu}2 = U\biggl(\frac12t(x)\biggr)\biggl(\frac{\si^a A^a_\mu}2\biggr)U^{-1}\biggl(\frac12t(x)\biggr)
- \frac ig\biggl[\pa_\mu\biggl(U\biggl(\frac12t(x)\biggr)\biggr)\biggr] U^{-1}\biggl(\frac12t(x)\biggr) \\
B^u_\mu = B_\mu. \nn
\e
One gets
\bml c329
\frac{v^2}{2\pi r_0^2\hbar c\zz}\,\d^+ \biggl(\frac g2 \si^a A^{ua}_\mu + \frac{g'}2B^u_\mu\biggr)\biggl(\frac g2 \si^a A^{ua}_\mu +
\frac{g'}2B^{u\mu}\biggr)\d \\
{}= \frac{v^2g^2}{16\pi r_0^2\hbar c\zz}\biggl([(A^{u1}_\mu)^2+(A^{u2}_\mu)^2]+\biggl(A^{u3}_\mu- \frac1{\sqrt3}B^u_\mu\biggr)^2\biggr) =
M^2_W W^{u+}_\mu W^{u{-}\mu} + \frac12M_{Z^0}^2Z^{u0}_\mu Z^{u0\mu}
\e
where
\begin{gather}
M_W^2 W^{u+}_\mu W^{u{-}\mu} = \frac{g^2v^2}{16\pi r_0^2\hbar c\zz}[(A^{u1}_\mu)^2+(A^{u2}_\mu)^2] \nn \\
W^{u\pm}_\mu = \frac1{\sqrt2}(A^{u1}_\mu \mp iA^{u2}_\mu) \lb c330 \\
M^2_W = \frac{g^2v^2}{8\pi r_0^2\hbar c\zz} = \frac{\a_s^2}{4\pi r_0^2\zz} \nn \\
\frac12 M_{Z^0}^2 Z^{u0}_\mu Z^{u0\mu} = \frac{\a_s^2}{16\pi r_0^2\zz}\biggl(A^{u3}_\mu - \frac1{\sqrt3}B^u_\mu\biggr)^2. \lb c331
\end{gather}
We have of course
\beq c332
\bal
Z^{u0}_\mu &= \frac{\sqrt3}2\,A^{u3}_\mu - \frac12B^u_\mu\\
A^u_\mu &= \frac12 A^{u3}_\mu + \frac{\sqrt3}2 B^u_\mu \\
M_{Z^0}^2 &= \frac{\a_s^2}{8\pi r_0^2\zz \hbar c}, \q \frac{g'}g = \frac1{\sqrt3}\,,
\eal
\e
see Eqs \er{C.23b}--\er{C.26b} and \er{C.27b}--\er{C.23} in a different formalism.

\beq c333
\frac{\a_s^2 v^2}{16\pi r_0^2\zz}(A^{u3}_\mu,B^u_\mu)
\left(\begin{matrix} 1 & -\frac1{\sqrt3} \\[5pt] -\frac1{\sqrt3} & \frac13 \end{matrix}\right)
\left(\begin{matrix} A^{u3\mu} \\ B^{u\mu} \end{matrix}\right) =
\frac12(Z^{u0}_\mu, A^u_\mu)
\left(\begin{matrix} M_{Z^0}^2 & 0 \\ 0 & 0 \end{matrix}\right)
\left(\begin{matrix} Z^{u0\mu} \\ A^{u\mu} \end{matrix}\right)
\e
For $\wh V(\F)$ we get
\bg c335
M_{H^0} = \sqrt2\, \mu \frac{\zy^{1/2} \a_s^2}{8r_0^2\zz \hbar c} \\
\wh V(\F) = \wh V(H^0) = \frac{\zy\a_s^2v^2}{r_0^2\zz 16\pi}(4-\mu^2(H^0)^2 + \la(H^0)^4)
\lb{c336}
\e
for a Fermi \ct
\beq c337g
G_F = \frac{\sqrt2\,g^2}{8M_W^2} = \frac{\sqrt2 \not{\!\!\a_s^2} \hbar c \not8 \pi r_0^2\zz}{\not8 \not{\!\!\a_s^2} v^2}
= \frac{\sqrt2 \hbar c \pi r_0^2\zz}{v^2} = \frac{\sqrt2 \hbar c \pi \zz r_0^2}{2}.
\e

Coming back to our \nos, curved and twisted \spt\ one gets
\bml c340d
\frac{v^2}{2\pi r_0^2 \sqrt{\hbar c}\zz}\,\d^+\biggl(\frac{\a_s}{2\sqrt{\hbar c}}\,\si^a A^{ua}_\mu + \frac{\a_s}{2\sqrt{3\hbar c}}
\,B^u_\mu\biggr)\\ {}\cdot \bigl(2g^{\mu\b}-\gtn{\mu\b}+2\z (g^{\mu\b}-\gtn{\mu\b})\bigr)
\biggl(\frac{\a_s}{2\sqrt{\hbar c}}\,\si^a A^{ua}_\b+\frac{\a_s}{2\sqrt{3\hbar c}}\,B^u_\b\biggr)\d \\
{}= \frac{v^2\a_s^2}{16\pi \hbar c\zz} \bigl((A^{u1}_\mu + A^{u2}_\mu) + (A^{u3}_\mu - \tfrac1{\sqrt3}\,B^u_\mu)\bigr)
(2g^{\mu\b}-\gtn{\mu\b}-2\z(g^{\mu\b}-\gtn{\mu\b}))\\ {}\cdot \bigl((A^{u1}_\b+A^{u2}_\b)+(A^{u3}_\b-\tfrac1{\sqrt3}\,B^u_\b)\bigr) \\
{}= M^2_W \bigl(2g^{\mu\b}-\gtn{\mu\b}+2\z(g^{\mu\b}-\gtn{\mu\b})(W^{u+}_\mu W^{u-}_\b + W^{u+}_\b W^{u-}_\mu)\\
{}+\tfrac12\,M^2_{Z^0}(2g^{\mu\b}-\gtn{\mu\b}+2\z(g^{\mu\b}-\gtn{\mu\b}))Z^{u0}_\mu Z^{u0}_\b.
\e
\bml c341d
\frac{\a_s^2v^2}{16\pi r_0^2\zz} (A^{u3}_\mu, B^u_\mu) \left(\begin{matrix}
1 & -\frac1{\sqrt3} \\[2pt] -\frac1{\sqrt3} & \frac13
\end{matrix}\right)
(2g^{\mu\b}-\gtn{\mu\b}+2\z(g^{\mu\b}-\gtn{\mu\b}))
\left(\begin{matrix}
A^{u3} \\ B^u_b
\end{matrix}\right)\\
{}= \frac12(2g^{\mu\b}-\gtn{\mu\b}+2\z(g^{\mu\b}-\gtn{\mu\b}))(Z^{u0},A^u_\mu)
\left(\begin{matrix}
M^2_{Z^0} & 0 \\ 0 & 0
\end{matrix}\right)
\left(\begin{matrix}
Z^{u0} \\ A^u_\b
\end{matrix}\right)
\e

Let us recapitulate the situation after \ssb\ and \Hm\ in our partial \un. In an electroweak \ia\ sector we have only physical fields:
$H^0(x)$---Higgs' field, $W^\pm_\mu$---charged intermediate boson fields, $Z^0_\mu$---neutral intermediate boson field, $A_\mu$---\elm c field.
For Higgs' field one gets a \lg\
$$
\cL\dr{kin}(\br{gauge(SU(2)_L\ot \U(1)\dr{Y})}{\n\F})-\wh V(\F)
$$
in the \fw\ form: $\wh V(\F)$ is given by Eq.~\er{c336} and $\cL\dr{kin}(\ggs\F)$ by Eq.~\er{c337}.
\bml c337
\cL\dr{kin}(\br{gauge(SU(2)_L\ot \U(1)\dr{Y})}{\n\F}) = \frac1{\zz}\cdot \frac1{2\pi r_0^2} \biggl\{\\
-\frac18 (2g^{\o\mu}-\gtn{\o\mu})\biggl(
\frac{\a_s}{2\sqrt{\hbar c}}(v+H^0)^2(W^{u-}_\o W^{u+}_\mu + W^{u+}_\o W^{u-}_\mu)\\
{}+2\biggl(\pa_\o H^0 + \frac{\a_si}{2\sqrt{\hbar c}}(v+H^0)\bigl(W^{u+}_\o + \tfrac12(\sqrt3\,Z^{ua}_\o -A^u_\o)\bigr)\biggr)
\biggl(\pa_\mu H^0+ \frac{\a_si}{2\sqrt{\hbar c}}(v+H^0)Z^{u0}_\mu\biggr)\\
{}+\biggl(\pa_\o H^0 + \frac{\a_si}{\sqrt{3\hbar c}}(v+H^0)Z^{u0}_\o\biggr)\biggl(\pa_\mu H^0 +
\frac{\a_si}{2\sqrt{\hbar c}}(v+H^0)\bigl(W^{u+}_\mu + \tfrac12(\sqrt3\,Z^{u0}_\mu - A^u_\mu)\bigr)\biggr)\\
{}+\frac14\z(g^{\o\mu}-\gtn{\o\mu})\biggl(-\frac{\a_si}{4\sqrt{\hbar c}}(v+H^0)^2(W^{u-}_\mu W^{u+}_\o + W^{u+}_\mu W^{u-}_\o)\\
{}+\biggl(\pa_\mu H^0 + \frac{\a_si}{2\sqrt{\hbar c}}(v+H^0)\bigl(W^{u+}_\mu + \tfrac12(\sqrt3\,Z^{u0}_\mu - A^u_\mu)\bigl)\biggr)
\biggl(\pa_\o H^0+ \frac{\a_si}{\sqrt{3\hbar c}}(v+H^0)Z^{u0}_\o\biggr)\\
\hskip-20pt {}+\biggl(\pa_\mu H^0 + \frac{\a_si}{2\sqrt{\hbar c}}(v+H^0)Z^{u0}_\mu\biggr)
\biggl(\pa_\o H^0 + \frac{\a_si}{2\sqrt{\hbar c}}(v+H^0)\bigl(W^{u+}_\o + \tfrac12(\sqrt3\,Z^{u0}_\mu - A^u_\mu)\bigr)\biggr)\biggr)\biggr)\biggr\}
\kern-5pt
\e
and $\wh V(\F)$ is given by Eq.~\er{c336}.

One can derive an energy-momentum tensor for a Higgs' field
\bml c338
T_{\o\mu}(\F) = -\frac18 \biggl(2t_{\o\mu} - \frac{\d \gtn{\a\b}}{\d g^{\o\mu}}t_{\a\b}\biggr)\frac1{\zz 2\pi r_0^2}
+ \frac14\z\biggl(\ov t_{\o\mu} - \frac{\d \gtn{\a\b}}{\d g^{\o\mu}}\ov t_{\a\b}\biggr)\frac1{\zz 2\pi r_0^2}\\
{}- \frac14g_{\o\mu}\bigl(\cL\dr{kin}(\br{gauge(SU(2)_L\ot \U(1)\dr{Y})}{\n\F})- \wh V(\F)\bigr)\\
{} =\frac1{\zz 2\pi r_0^2} \biggl[\biggl(-\frac14 t_{\o\mu} + \frac1{16}[\gtn{\a\xi}\gtn{\nu\b}g_{\xi\mu}g_{\nu\o}
+ \gtn{\nu\b}\gtn{\a\xi}g_{\nu\mu}g_{\xi\o}]t_{\a\b}\biggr)\\
{}+\frac14\z\biggl(\ov t_{\o\mu} - \frac12 [\gtn{\a\xi}\gtn{\nu\b}g_{\xi\mu}g_{\nu\o} + \gtn{\nu\b}\gtn{\a\xi}g_{\nu\mu}g_{\xi\o}]\ov t_{\a\b}\biggr)
\biggr] - \frac14g_{\o\mu}(\cL\dr{kin}(\br{gauge(SU(2)_L\ot \U(1)\dr{Y})}{\n\F}) - \wh V(\F))\\
{}= \frac1{\zz 2\pi r_0^2} \Biggl[-\frac14\biggl(\frac{\a_si}{2\sqrt{\hbar c}}(v+H^0)^2(W^{u-}_\o W^{u+}_\mu + W^{u+}_\o W^{u-}_\mu)\\
{}+2\biggl(\pa_\o H^0 + \frac{\a_si}{2\sqrt{\hbar c}}(v+H^0)(W^{u+}_\o + \tfrac12(\sqrt3\,Z^{u0}_\o - A^u_\o))\biggr)
\biggl(\pa_\mu H^0 + \frac{\a_si}{\sqrt{3\hbar c}}(v+H^0)Z^{u0}_\mu\biggr)\\
{}+\biggl(\pa_\o H^0 + \frac{\a_si}{\sqrt{3\hbar c}}(v+H^0)Z^{u0}_\o\biggr)\biggl(\pa_\mu H^0 +
\frac{\a_si}{2\sqrt{\hbar c}}(v+H^0)\bigl(W^{u+}_\mu + \tfrac12(\sqrt3\,Z^{u0}_\mu-A^u_\mu)\bigr)\biggr)\biggr)\\
{}+\frac1{16}\bigl[\gtn{\a\xi}\gtn{\nu\b}g_{\xi\mu}g_{\nu\o} + \gtn{\nu\b}\gtn{\a\xi}g_{\nu\mu}g_{\xi\o}\bigr]
\biggl(\frac{\a_si}{2\sqrt{\hbar c}}(v+H^0)^2 (W^{u-}_\a W^{u+}_\b + W^{+u}_\a W^{-u}_\b)\\
{}+ 2\biggl(\pa_\a H^0 + \frac{\a_si}{2\sqrt{\hbar c}}(v+H^0)
(W^{u+}_\a + \tfrac12(\sqrt3\,Z^{u0}_\a - A^u_\a))\biggr)\biggl(\pa_\b H^0 + \frac{\a_si}{\sqrt{3\hbar c}}(v+H^0)Z^{u0}_\b\biggr)\\
{}+ \biggl(\pa_\a H^0 + \frac{\a_si}{\sqrt{3\hbar c}}(v+H^0)Z^{u0}_\a\biggr) \biggl(\pa_\b H^0 + \frac{\a_si}{2\sqrt{\hbar c}}(v+H^0)
\bigl(W^{u+}_\b + \tfrac12(\sqrt3\,Z^{u0}_\b - A^u_\b)\bigr)\biggr)\\
{}+\frac14\z \biggl(-\frac{\a_si}{4\sqrt{\hbar c}}(v+H^0)^2 (W^{u-}_\mu W^{u+}_\o + W^{u+}_\mu W^{u-}_\o)\\
{}+\biggl(\pa_\mu H^0 + \frac{\a_si}{2\sqrt{\hbar c}}(v+H^0)\bigl(W^{u+}_\mu + \tfrac12(\sqrt3\,Z^{u0}_\mu - A^u_\mu)\bigr)\biggr)
\biggl(\pa_\o H^0 + \frac{\a_si}{\sqrt{3\hbar c}}(v+H^0)Z^{u0}_\o\biggr)\\
+ \biggl(\pa_\mu H^0+ \frac{\a_si}{2\sqrt{\hbar c}}(v+H^0)Z^{u0}_\mu\biggr)
\biggl(\pa_\o H^0 + \frac{\a_si}{2\sqrt{\hbar c}}(v+H^0)\bigl(W^{u+}_\o + \tfrac12(\sqrt3\, Z^{u0}_\o -
A^u_\o\bigr)\biggr)\biggr)\biggr)\\
{}-\frac12 \bigl[ \gtn{\a\xi}\gtn{\nu\b}g_{\z\mu}g_{\nu\o} + \gtn{\nu\b}\gtn{\a\xi}g_{\nu\mu}g_{\xi\o}\bigr]
\biggl(\frac{\a_si}{4\sqrt{\hbar c}}(v+H^0)^2(W^{u-}_\b W^{u+}_\a + W^{u+}_\b W^{u-}_\a)\\
{}+ \biggl(\pa_\b H^0 + \frac{\a_si}{2\sqrt{\hbar c}}(v+H^0)\bigl(W^{u+}_\b + \tfrac12(\sqrt3\,Z^{u0}_\b - A^u_\b)\bigr)\biggr)
\biggl(\pa_\a H^0 + \frac{\a_si}{\sqrt{3\hbar c}}(v+H^0)Z^{u0}_\a\biggr)\\
{}+ \biggl(\pa_\b H^0 + \frac{\a_si}{2\sqrt{\hbar c}}(v+H^0)Z^{u0}_\b\biggr)\biggl(\pa_\a H^0 + \frac{\a_si}{2\sqrt{\hbar c}}(v+H^0)
\bigl(W^{u+}_\a + \tfrac12(\sqrt3\,Z^{u0}_\a - A^u_\a)\bigr)\biggr)\biggr)\Biggr]\\
{}-\frac14 g_{\o\mu}\Biggl(\frac1{\zz 2\pi r_0^2}\biggl\{-\frac18(2g^{\a\b} - \gtn{\a\b})\biggl(\frac{\a_si}{2\sqrt{\hbar c}}(v+H^0)^2
(W^{u+}_\a + W^{u+}_\b + W^{u-}_\a W^{u-}_\b)\\ {}+ 2\biggl(\pa_\a H^0 + \frac{\a_si}{\sqrt{3\hbar c}}(v+H^0)
\bigl(W^{u+}_\a + \tfrac12(\sqrt3\,Z^{u0}_\a - A^u_\a)\bigr)\biggr)\biggl(\pa_\b H^0 + \frac{\a_si}{\sqrt{3\hbar c}}(v+H^0)
Z^{u0}_\b\biggr)\\ + \biggl(\pa_\a H^0 + \frac{\a_si}{2\sqrt{\hbar c}}(v+H^0)Z^{u0}_\a\biggr)
\biggl(\pa_\b H^0 + \frac{\a_si}{2\sqrt{\hbar c}}(v+H^0)\bigl(W^{u+}_\b + \tfrac12(\sqrt3\,Z^{u0}_\b - A^u_\b)\bigr)\biggr)\biggr)\\
{} + \frac14\z(g^{\a\b} - \gtn{\a\b})\biggl(-\frac{\a_si}{4\sqrt{\hbar c}}(v+H^0)^2 (W^{u-}_\b W^{u+}_\a + W^{u+}_\b W^{u-}_\a)\\
{}+ \biggl(\pa_\b H^0 + \frac{\a_si}{2\sqrt{\hbar c}}(v+H^0)\bigl(W^{u+}_\b + \tfrac12(\sqrt3\,Z^{u0}_\b - A^u_\b)\bigr)\biggr)
\biggl(\pa_\a H^0 + \frac{\a_si}{\sqrt{3\hbar c}}(v+H^0)Z^{u0}_\a\biggr)\\
{} + \biggl(\pa_\b H^0 + \frac{\a_si}{2\sqrt{\hbar c}}(v+H^0)Z^{u0}_\b\biggr)\biggl(\pa_\a H^0 + \frac{\a_si}{2\sqrt{\hbar c}}(v+H^0)
\bigl(W^{u+}_\a + \tfrac12(\sqrt3\,Z^{u0}_\a - A^u_\a)\bigr)\biggr)\biggr)\biggr\}\\
{}-\frac{\zy \pi}{2\zz r_0^2} \biggl[4 - \frac6{g^2}(H_0)^2 + \frac{3\sqrt6}{g^3}(H_0)^3 + \frac9{4g^4}(H^0)^4\biggr]\Biggr)
\e

In new physical variables, i.e.\ $W^{u\pm}_\mu$, the \lg\ for $\SU(2)_L\ot\U(1)\dr{Y}$ gauge fields looks as follows:
\bml c339
\cL(\a_sA_{\SU(2)_L\ot\U(1)\dr{Y}}) = \frac{\a_s^2}{192\pi}\Biggl(2\gtk\m\biggl(W^{u+}_\m + W^{u-}_\m - \frac{i\a_s}{\sqrt{\hbar c}}
\bigl(2(W^{u+}_{[\mu} - W^{u-}_{[\mu})(A_{\nu]}+\sqrt3\,Z^{u0}_{\nu]})\\
{}- \ww\m\bigr) + F_\m + \sqrt3\,Z^{u0}_\m + 2W^{u+}_{[\mu}W^{u-}_{\nu]}\biggr)^2
- (g^{\nu\b}(g^{\mu\a}-\gtn{\mu\a}+g^{\mu\o}\gtn{\tau\a}g_{\o\tau}))\biggl\{(\ww\m \\
{}- \frac{i\a_s}{\sqrt{\hbar c}}(W^{u+}_\mu-W^{u-}_\mu)
(A_\nu+\sqrt3\,Z^{u0}_\nu)) \biggl(\ww{\a\b}- \frac{i\a_s}{\sqrt{\hbar c}}(W^{u+}_\a - W^{u-}_\a)(A_\b+\sqrt3\,Z^{u0}_\b) \\
{}- \biggl(W^{u+}_\m -W^{u-}_\m + \frac{i\a_s}{\sqrt{\hbar c}}(W^{u+}_\mu-W^{u-}_\mu)(A_\nu+\sqrt3\,Z^{u0}_\nu)\biggr)
\biggl(W^{u+}_{\a\b}-W^{u-}_{\a\b} + \frac{i\a_s}{\sqrt{\hbar c}}(W^{u+}_\a-W^{u-}_\a)(A_\b+\sqrt3\,Z^{u0}_\b)\biggr)\\
{}+\frac{i\a_s}{\sqrt{\hbar c}}\bigl(F_\m + \sqrt3\,Z^{u0}_\m + W^{u+}_\mu W^{u+}_\nu + W^{u+}_\mu W^{u-}_\nu - W^{u-}_\mu W^{u+}_\nu -
W^{u-}_\mu W^{u-}_\nu \bigr)\\
\bigl(F_{\a\b} + \sqrt3\,Z^{u0}_{\a\b} + W^{u+}_\a W^{u+}_\b + W^{u+}_\a W^{u-}_\b - W^{u-}_\a W^{u+}_\b -
W^{u-}_\a W^{u-}_\b\bigr) + \frac{3\a_s}{\sqrt{\hbar c}}(F_\m - Z^{u0}_\m)(F_{\a\b}-Z^{u0}_\a\b)\biggr)\biggr\}
\e

One can derive an energy-momentum tensor for $\cL(\a_s A_{\SU(2)_L\ot\U(1)\dr{Y}})$ as follows:
\bml c339n
\br{gauge(\SU(2)_L \ot\U(1)\dr{Y})}{T_{\rho\g}} = \frac{\d\cL(\a_s A_{\SU(2)_L \ot\U(1)\dr{Y}})}{\d g^{\rho\g}}
- \frac14g_{\rho\g}(\cL(\SU(2)_L \ot\U(1)\dr{Y})) \\
{} = \frac{\a_s^2}{48\pi} \Biggl\{\biggl(\ww{\rho\g} + F_{\rho\g} +\sqrt3\,Z^{u0}_{\rho\g} + i\biggl(\wy{\rho\g} + \frac{2\a_s}{\sqrt{\hbar c}}
(\wy{[\rho})\az{\g]}\biggr)\biggr)\\
{}+\frac14 \bigl(g^{\nu\b}\d^\mu_\rho \d^\a_\g + (g^{\mu\a}-\gtn{\mu\a})\d^\nu_\rho \d^\b_\g + \tfrac12g^{\nu\b}\gtn{\o\a}
((g_{\xi\g}g_{\o\rho} + g_{\o\g}g_{\xi\rho})(\gtn{\mu\xi}-g^{\mu\a}\gtn{\tau\xi}g_{\o\tau})) + 2g^{\mu\d}g_{\o\g}g_{\rho\rho}\bigr)\\
\biggl\{\biggl[\ww\m - \frac{i\a_s}{\sqrt{\hbar c}}\az\nu(\wy\mu)\biggr]\biggl[\ww{\a\b} - \frac{i\a_s}{\sqrt{\hbar c}}\az\b(\wy\a)\biggr]\\
{}- \biggl[\ww\m + \frac{\a_s}{\sqrt{\hbar c}}(\wy\mu)\az\nu\biggr]\biggl[\wy{\a\b}+ \frac{\a_s}{\sqrt{\hbar c}}(\wy\a)\az\b\biggr]\\
{}+ \biggl(F_\m + \sqrt3\,Z^{u0}_\m + \frac{\a_s}{\sqrt{\hbar c}}(W^{u+}_\mu W^{u+}_\nu + W^{u+}_\mu W^{u-}_\nu - W^{u-}_\mu W^{u+}_\nu
- W^{u-}_\mu W^{u-}_\nu)\biggr) \\
\biggl(F_{\a\b} + \sqrt3\,Z^{u0}_{\a\b} + \frac{\a_s}{\sqrt{\hbar c}}(W^{u+}_\mu W^{u+}_\nu + W^{u+}_\mu W^{u-}_\nu - W^{u-}_\mu W^{u+}_\nu
- W^{u-}_\mu W^{u-}_\nu)\biggr)\\
{}+ \frac{3\a_s}{\sqrt{\hbar c}}(F_\m-Z^{u0}_\m)(F_{\a\b}-Z^{u0}_{\a\b})\biggr\}\Biggr\}\\
{}- \frac1{16}g_{\rho\g}\Biggl(2\biggl(\gtk\m \biggl(\ww\m - \frac{i\a_s}{\sqrt{\hbar c}}(2(\wy{[\mu})\az{\nu]}) - \ww\m\biggr)\\
{}+F_\m +\sqrt3\,Z^{u0}_\m + \frac{2\a_s}{\sqrt{\hbar c}}W^{u+}_{[\mu}W^{u-}_{\nu]}\biggr)\Biggr)^2
- (g^{\nu\b}(g^{\mu\a}-\gtn{\mu\a} + g^{\mu\o}\gtn{\tau\a}g_{\o\tau}))\\
\Biggl\{\biggl(\!\ww\m - \frac{i\a_s}{\sqrt{\hbar c}}(\wy\mu)\az\nu \!\biggr)\biggl(\!\ww{\a\b} - \frac{i\a_s}{\sqrt{\hbar c}}(\wy\a)\az\b\!\biggr)\\
{}- \biggl(\!\wy\m - \frac{\a_s}{\sqrt{\hbar c}}(\wy\mu)\az\nu\!\biggr)\biggl(\!\wy{\a\b} + \frac{\a_s}{\sqrt{\hbar c}}(\wy\a)\az\b\!\biggr)\\
{}+ \biggl(F_\m + \sqrt3\,Z^{u0}_\m + \frac{\a_s}{\sqrt{\hbar c}}(W^{u+}_\mu W^{u+}_\nu + W^{u+}_\mu W^{u-}_\nu - W^{u-}_\mu W^{u+}_\nu
- W^{u-}_\mu W^{u-}_\nu)\biggr)\\ \biggl(F_{\a\b} + \sqrt3\,Z^{u0}_{\a\b} + \frac{\a_s}{\sqrt{\hbar c}}(W^{u+}_\a W^{u+}_\b + W^{u+}_\a W^{u-}_\b
- W^{u-}_\a W^{u+}_\b - W^{u-}_\a W^{u-}_\b )\biggr)\\ {}+ \frac{3\a_s}{\sqrt{\hbar c}}(F_\m - Z^{u0}_\m)(F_{\a\b}-Z^{u0}_{\a\b})\biggr)\Biggr\}
\e
Let us remind to the reader that the trace of this tensor is zero.

After a \ssb\ and \Hm\ $\cL\dr{int}$ changes into
\beq c340
\cL\dr{int} = -\frac{i\z 3\sqrt6(v+H^0(x))^2}{64\pi r_0^2\zz} (Z^{u0}_{\m} + \sqrt3\,F_\m)g^\m.
\e
One can derive an energy-momentum tensor for $\cL\dr{int}$ and get
\beq c341
T_\m(\cL\dr{int}) = \frac{i\z 3\sqrt6(v+H^0(x))^2}{256\pi r_0^2\zz}(g_\m(Z^{u0}_{\a\b}+\sqrt3\,F_{\a\b})g^{\a\b} - 4(Z^{u0}_\m+F_\m)).
\e
It is easy to see that the trace of this energy-momentum tensor is zero.

Let us consider an ingredient of mass terms to a \lg
\bml c342
\D\cL = ((2g^{\o\mu}-\gtn{\o\mu}) + 2\z(g^{\o\mu}-\gtn{\o\mu}))(M^2_W W^{u+}_\o W^{u-}_\mu + \tfrac12 M^2_{Z^0} Z^{u0}_\o Z^{u0}_\mu)
+\tfrac12 M^2_{H^0}(H^0)^2\\
{}= (2(\z+1)g^{\o\mu}-(2\z+1)\gtn{\o\mu})(M^2_W W^{u+}_\o W^{u-}_\mu + \tfrac12 M^2_{Z^0} Z^{u0}_\o Z^{u0}_\mu)+\tfrac12 M^2_{H^0}(H^0)^2.
\e
One can write an energy-momentum tensor for $\D\cL$. $\D\cL$ and $T_\m(\D\cL)$ complete all sources of \gz\ Einstein \e s in our partial \un\ (\un\
of NGT and bosonic part of a Standard Model)
\bml c343
T_{\psi\g}(\D\cL) = 2(\z+1)(M^2_W W^{u+}_\psi W^{u-}_\g + \tfrac12 M^2_{Z^0} Z^{u0}_\psi Z^{u0}_\g) + \tfrac12M^2_{H^0}(H^0)^2g_{\psi\g}\\
-\tfrac12 (2\z+1)\gtn{\o\b}\gtn{\rho\mu}(g_{\b\g} g_{\rho\psi} + g_{\rho\g} g_{\b\psi})(M^2_W W^{u+}_\o W^{u-}_\mu + \tfrac12 M^2_{Z^0}
Z^{u0}_\o Z^{u0}_\mu) \\ {}- \tfrac14\bigl[((M^2_W W^{u+}_\o W^{u-}_\mu + \tfrac12 M^2_{Z^0} Z^{u0}_\o Z^{u0}_\mu))
(2(\z+1)g^{\o\mu} - (2\z+1)\gtn{\o\mu})\bigr] g_{\psi\g}.
\e
Let us calculate an \ef\ energy-momentum tensor for $\D\cL$. One gets
\beq c344
\br{eff}T_\m(\D\cL) = T_\m(\D\cL) - \frac12g_\m(T_{\g\psi}(\D\cL)\cdot g^{\psi\g}) = T_\m(\D\cL)
\e
for the trace of $T_\m(\D\cL)$ is zero.

Let us write field \e s for $W^{u\pm}_\mu$, $Z^{u0}_\mu$, $A^u_\mu$, $H^0$, $\Psi$. One gets for $W^{u\pm}_\mu$:
\bml c345
\br{gauge(\SU(2)_L)}{\n_\mu} \bigl(\falg^{\a\b}g^{\mu\rho}(\ov H{}^1_{\b\rho}+ \gtn{\tau\d}(\ov H{}^1_{\d\g}g\ink{\b \tau}
- \ov H{}^1_{\g\b}g\ink{\rho \tau}))\bigr) =
\falg^{\a\b}\br{gauge(\SU(2)_L)}{\n_\b} (\gtk\m \ov H{}^1_\m)\\ {}+ 24\pa_b \Psi \bigl[\falg^{\b\nu}g^{\a\mu}(\ov H{}^1_\m + \gtn{\tau\a}
(\ov H{}^1_{\a\mu}g\ink{\nu\tau} - \ov H{}^1_{\a\nu}g\ink{\tau\nu})) - \falg^{[\b\a]}(\gtk\m \ov H{}^1_\m)\bigr]\\
{}+\frac1{\sqrt2} M^2_W(2(\z+1)\falg^{\a\mu} - (2\z+1)\gtn{\a\mu})(\ww\mu)
\e
where
\bml c346
\ov H{}^1_\m = \frac1{\sqrt2}\biggl[(\pa_\mu W^{u+}_\nu - \pa_\nu W^{u+}_\mu) + (\pa_\mu W^{u-}_\nu - \pa_\nu W^{u-}_\mu) +
\frac{\a_si}{2\sqrt{\hbar c}}\bigl((A^u_{[\nu}(\wy{\mu]})\\ {}+ \sqrt3\,Z^{u0}_{[\nu}(\wy{\mu]})\bigr)
=\frac1{\sqrt2} (\ww\m)\\ {}+ \frac{\a_si}{2\sqrt{\hbar c}}\bigl((A^u_{[\nu}(\wy{\mu]}) + \sqrt3(Z^{u0}_{[\nu}(\wy{\mu]}))
\e
\bml c347
\br{gauge(\SU(2)_L)}{\n_\mu} \biggl(\falg^{\a\b}g^{\mu\rho}\biggl(\frac1{\sqrt2}(\ww{\b\rho}) + \frac{\a_si}{2\sqrt{\hbar c}}
(A^u_{[\b}(\wy{\rho]}) + \sqrt3(Z^{u0}_{[\b} (\wy{\rho]}))\\ {}+ \gtn{\tau\d}\biggl(\biggl(\frac1{\sqrt2}(\ww{\d\rho}) +
\frac{\a_si}{2\sqrt{\hbar c}}\bigl(A^u_{[\d}(\wy{\rho]}) + \sqrt3(Z^{u0}_{[\d}(\wy{\rho]}))\bigr)g\ink{\b\tau}\\
{}+\biggl(\frac1{\sqrt2}(\ww{\d\b})-\frac{\a_si}{\sqrt{\hbar c}}\bigl(\bigl(A^u_{[\d}(\wy{\b]})+ \sqrt3(Z^{u0}_{[\d}(\wy{\b]}))\bigr)g\ink{\rho\tau}\\
{}= \falg^{\a\b} \br{gauge(\SU(2)_L)}{\n_\b} \biggl(\gtk\m \biggl(\frac1{\sqrt2}(\ww\m) -
\frac{\a_si}{2\sqrt{\hbar c}}\bigl(A^u_{[\mu}(\wy{\nu]}) + \sqrt3(Z^{u0}_{[\mu}(\wy{\nu]}))\bigr)^2\\
+24\pa_\b \Psi \biggl[\falg^{\b\nu}g^{\a\mu}\biggl(\frac1{\sqrt2}(\ww{\nu\mu}) - \frac{\a_si}{2\sqrt{\hbar c}}\bigl(A^u_{[\nu}(\wy{\mu]})
+\sqrt3\,Z^{u0}_{[\nu}(\wy{\mu]})\bigr)\biggr)\\
{}+\wt g{}^{(\tau\a)}\biggl(\frac1{\sqrt2}(\ww{\a\mu}) - \frac{\a_si}{2\sqrt{\hbar c}}\bigl(A^u_{[\a}(\wy{\mu]})
+ \sqrt3\,Z^{u0}_{[\a}(\wy{\mu]})\bigr)\biggr)g\ink{\nu\tau}\\
{}- \biggl(\frac1{\sqrt2}(\ww{\a\nu}) - \frac{\a_si}{2\sqrt{\hbar c}}\bigl(A^u_{[\a}(\wy{\nu]}) + \sqrt3\,Z^{u0}_{[a}(\wy{\nu]})\bigr)\biggr)
g\ink{\tau\mu}\\ {}- \falg^{[\b\a]}\biggl(\gtk\m\biggl(\frac1{\sqrt2}(\ww\m) - \frac{\a_si}{2\sqrt{\hbar c}} \bigl(A^u_{[\mu}(\wy{\nu]})
+\sqrt3\,Z^{u0}_{[\mu}(\wy{\nu]})\bigr)\biggr)\\
\hfill {}+ \frac1{\sqrt2} M^2_W (2(\z+1)\falg^{\a\mu} - (2\z+1)\wt{\falg}^{(\a\mu)})(\ww\mu))\indent \\
\hfill W^\pm_\m = \pa_\mu W^\pm_\nu - \pa_\nu W^\pm_\mu \hfill
\e

We continue \e s for $W^\pm_\mu$
\bml c348
\br{gauge(\SU(2)_L)}{\n_\mu}\bigl(\falg^{\a\b}g^{\mu\rho}(\ov H{}^2_{\b\rho} + \gtn{\tau\d}(\ov H{}^2_{\d\rho}g\ink{\b\tau} - \ov H{}^2_{\d\b}
g\ink{\rho\tau}))\bigr)\\
{}= \falg^{[\a\b]}(\br{gauge(\SU(2)_L)}{\n_\b}(\gtk\m \ov H{}^2_\m)) + 24\pa_\b \Psi \bigl[\falg^{\b\nu}g^{\a\mu}(\ov H{}^2_{\nu\mu}
+\gtn{\tau\a}(\ov H{}^2_{\a\mu} g\ink{\nu\tau}
- {\ov H}^2_{\a\nu} g\ink\m) \\ {}- \falg^{[\b\a]}(\gtk\m \ov H{}^2_\m)\bigr]
- i\sqrt2(2(\z+1)\falg^{\a\mu} - (2\z+1)\fal{\ov g}^{(\a\mu)}(W^{u-}_\mu {-} W^{u+}_\mu)M^2_W
\e
where
\bml c349
\ov H{}^2_\m = \sqrt2\,i((\pa_\mu W^{u+}_\nu - \pa_\nu W^{u-}_\mu) - (\pa_\mu W^{u-}_\nu - \pa_\nu W^{u-}_\mu))
- \frac{i\a_s}{\sqrt{2\hbar c}}\bigl[(\ww{[\mu}) A^u_{\nu]} \\ {}+ \sqrt3 (\ww{[\mu})Z^{ua}_{\nu]}\bigr]
 = \sqrt2\,i (\wy{[\mu}) \\ {}- \frac{\a_s}{\sqrt{2\hbar c}}\bigl[(\ww{[\mu})A^u_{\nu]} + \sqrt3 (\ww{[\mu})Z^{u0}_{\nu]}\bigr]
\e
or
\bml c349n
\br{gauge(\SU(2)_L)}{\n_\mu} \biggl(\falg^{\a\b}g^{\mu\rho}\biggl(\sqrt2\,i(\wy{\b\rho}) -
\frac{\a_s}{\sqrt{2\hbar c}}\bigl[(\ww{[\b})A^u_{\rho]}+ \sqrt3(\ww{[\b})Z^{u0}_{\rho]}\bigr]\\ {} + \gtn{\tau\d}\biggl(\sqrt2\,i(\ww{\a\rho})
-\frac{\a_s}{\sqrt{2\hbar c}}\bigl[(\ww{[\d})A^u_{\rho]}+ \sqrt3(\ww{[\d})Z^{u0}_{\rho]}\bigr]g\ink{\b\tau}\\ {} - \biggl(\sqrt2\,i(\ww{\d\b})
-\frac{\a_s}{\sqrt{2\hbar c}}(\ww{[\d})A^u_{\b]} + \sqrt3(\ww{[\d})Z^{u0}_{\b]}\biggr)\biggr)g\ink{\rho\tau}\biggr)\biggr)\\
{}= \falg^{[\a\b]}\biggl(\br{gauge(\SU(2)_L)}{\n_\b} \biggl(\gtk\m\biggl(\sqrt2\,i(\wy\m)-\frac{\a_s}{\sqrt{2\hbar c}}
\bigl[(\ww{[\mu})A^u_{\nu]}\\ {}+ \sqrt3(\ww{[\mu})Z^{u0}_{\nu]}\bigr]\biggr)\biggr)\\
{}+ 24\pa_\b \Psi\biggl[ \falg^{\b\nu}g^{\a\mu}\biggl(\sqrt2\,i(\wy{\nu\mu}) - \frac{\a_s}{\sqrt{2\hbar c}}\bigl[(\ww{[\mu})A^u_{\mu]}
+ \sqrt3(\ww{[\nu})Z^{u0}_{\mu]}\bigr]\biggr)\\
{}+ \gtn{\tau\a}\biggl(\biggl(\sqrt2\,i\biggl((\ww{\a\mu}) - \frac{\a_s}{\sqrt{2\hbar c}}\bigl[(\ww{[\a})A^u_{\mu]}
+ \sqrt3 (\ww{[\a})Z^{u0}_{\mu]}\bigr]\biggr) \\ {}- \biggl(\sqrt2\,i(\ww{\a\nu}) - \frac{\a_s}{\sqrt{2\hbar c}}\bigl[(\ww{[\a})A^u_{\nu]}
+ \sqrt3 (\ww{[\a})Z^{u0}_{\nu]}\bigr]\biggr) g\ink{\mu\tau}\biggr)\\
{}- \gtk{\b\a} \biggl(\gtk\m\biggl(\sqrt2\,i(\ww\m) - \frac{\a_s}{\sqrt{2\hbar c}}\bigl[(\ww{[\mu})A^u_{\nu]} + \sqrt3 (\ww{[\mu})Z^{u0}_{\nu]}\bigr]
\biggr)\biggr)\\ {}+ i\sqrt2(2(\z+1)\falg^{\a\mu} - (2\z+1)\gtn{\a\mu})M^2_W (W^{u-}_\mu {-} W^{u+}_\mu).
\e

For $Z^{u0}_\mu$ one gets
\bml c350
\br{gauge(\SU(2)_L)}{\n_\mu}\bigl(\falg^{\a\b}g^{\mu\rho}(\ov H{}^3_{\b\rho} + \gtn{\tau\d}(\ov H{}^3_{\d\rho}g\ink{\b\tau} - \ov H{}^3_{\d\b}
g\ink{\rho\tau}))\bigr) \\
{} = 2\falg^{[\a\b]}(\br{gauge(\SU(2)_L)}{\n_\b} (\gtk\m \ov H{}^3_\m)) + 24\pa_\b\Psi\bigl[\falg^{\b\nu}g^{\a\mu}(\ov H{}^3_{\nu\mu}
+ \gtn{\tau\a}(\ov H{}^3_{\a\mu}g\ink{\nu\tau} - \ov H{}^3_{\a\nu}g\ink{\mu\tau}) \\ {}- \falg^{[\b\a]}(\ov H{}^3_{\m}\gtk\m))\bigr]
+ \frac{\sqrt3}2 M^2_{Z^0}(2(\z+1)\falg^{\a\mu} - (2\z+1)\fal{\wt g}{}^{(\a\mu)})Z^{0u}_\mu
\e
where
\beq c351
\ov H{}^3_\m = \frac12(F_\m + \sqrt3\,Z^0_\m) + \frac{i\a_s}{\sqrt{\hbar c}} (\ww{[\nu})(\wy{\mu]})
\e
or
\bml c352
\br{gauge(\SU(2)_L)}{\n_\mu} \biggl(\falg^{\a\b}g^{\mu\rho}\biggl(\frac12(F_{\b\rho} + \sqrt3\,Z^{u0}_{\b\rho}) +
\frac{i\a_s}{\sqrt{\hbar c}}(\ww{[\b})(\wy{\rho]}) \\ {}+ \gtn{\tau\d}\biggl(\biggl(\frac12(F_{\d\rho}+ \sqrt3\,Z^{u0}_{\d\rho})
+ \frac{i\a_s}{\sqrt{\hbar c}} (\ww{[\d})(\wy{\rho]})\biggr)g\ink{\b\tau}\\
- \biggl(\biggl(\frac12(F_{\d\b}+ \sqrt3\,Z^{u0}_{\d\b}) + \frac{i\a_s}{\sqrt{\hbar c}}(\ww{[\d})(\wy{\b]})\biggr)g\ink{\d\tau}\biggr)\biggr)\\
{}= 2\falg^{[\a\b]}\biggl(\br{gauge(\SU(2)_L)}{\n_\mu} \biggl(\gtk\m \biggl(\frac12(F_\m + \sqrt3\,Z^{0u}_\m) + \frac{\a_s}{\sqrt{\hbar c}}
(\ww{[\mu})(\wy{\nu]})\biggr)\biggr)\\
{}+24 \pa_\b \Psi \biggl[\falg^{\b\nu}g^{\a\mu}\biggl(\biggl(\frac12(F_\m + \sqrt3\,Z^{u0}_m) + \frac{i\a_s}{\sqrt{\hbar c}}
(\ww{[\nu})(\wy{\mu]})\biggr) \\
{}+ \gtn{\tau\a}\biggl(\biggl(\frac12(F_{\a\mu}+\sqrt3\,Z^{u0}_{\a\mu}) + \frac{i\a_s}{\sqrt{\hbar c}}(\ww{[\a})(\wy{\mu]}) g\ink{\nu\tau}\\
{} - \biggl(\frac12(F_{\a\nu}+ \sqrt3\,Z^{u0}_{\a\nu}) + \frac{i\a_s}{\sqrt{\hbar c}}(\ww{[\a})(\wy{\nu]})\biggr)g\ink{\mu\tau}\biggr)\\
{} - \falg^{[\b\a]}\biggl(\biggl(\frac12(F_\m + \sqrt3\,Z^{u0}_\m) + \frac{\a_s}{\sqrt{\hbar c}}(\ww{[\mu})(\wy{\nu]})\biggr)\gtk\m\biggr)\biggr]\\
{}+ \frac{\sqrt3}2 M^2_{Z^0}(2(\z+1)\falg^{(\a\mu)} - (2\z+1)\fal{\ov g}{}^{(\a\mu)}) Z^{u0}_\mu.
\e

For a $B_\mu$ field and an \elm c field $F_\m$ one gets
\bml c353
\pa_\mu \bigl(\falg^{\a\b}g^{\mu\rho}(G_{\b\rho} + \gtn{\tau\d}(G_{\d\rho}g\ink{\b\tau}-G_{\d\b}g\ink{\rho\tau}))\bigr)\\
{}= 2\falg^{[\a\b]}(\pa_\b (\gtk\m G_\m)) + 24\pa_b\Psi \bigl[\falg^{\b\nu}g^{\a\mu}(G_{\nu\mu} + \gtn{\tau\a}(G_{\a\mu}g\ink{\nu\tau} -
G_{\a\nu}g\ink{\mu\tau})\\ {}+ 2\falg^{[\b\a]}(\gtk{\mu\d}G_\m))\bigr]
- \frac12M^2_{Z^0}(2(\z+1)\falg^{\a\mu} - (2\z+1) \fal{\wt g}{}^{(\o\mu)})Z^{u0}_\mu + \frac{i\z3\sqrt2(v+H^0(x))}{16\pi r_0^2\zz}
\e
or
\bml c354
\pa_\mu\bigl(\falg^{\a\b}g^{\mu\rho}(\zf{\b\rho} + \gtn{\tau\d}(\zf{\d\rho}g\ink{\b\tau}\zf{\d\b}g\ink{\rho\tau}))\bigr) \\
{}= 2\falg^{[\a\b]}(\pa_\b (\gtk\m\zf\m)) + 24\pa_\b \Psi\bigl[\falg^{\b\nu}g^{\a\mu}\zf{\nu\mu}\\{} + \gtn{\tau\a}(\zf{\a\mu}g\ink{\nu\tau}
- \zf{\a\nu}g\ink{\mu\tau}) + 2\falg^{[\b\a]}(\gtk\m\zf\m) \\
{}- M^2_{Z^0}(2(\z+1)\falg^{\a\mu} - (2\z+1)\fal{\wt g}{}^{(\o\mu)}) Z^{u0}_\mu + \frac{i\z 3\sqrt2(v+H^0(x))}{8\pi r_0^2\zz}\,\falg^{\a\mu}
\pa_\mu H^0(x).
\e
For Higgs' field $H^0$ one gets
\bg c355
(2(\z+1)g^{(\o\mu)} - (2\z+1)\gtn{\o\mu})\ov \n_\mu \ov \n_\o H^0(x) - M^2_{H^0}H^0 - 3\la' v(H^0)^2 - \la'(H^0)^3 = 0 \\
2\mu^2 = M^2_{H^0} = \frac{\pi\zy}{2\zz r_0^2} \nn \\
\la' = \frac{\pi\zy}{2\zz} \nn
\e
For scalar field $\Psi$ one gets
\bml c356
2((242+\ov M)\gtn{\a\mu} - 242g^{\nu\mu}g_{\d\nu}\gtn{\a\d}) \pp{^2\Psi}{x^\a \pa x^\mu} + \frac2{\sqrt{-g}} \pa_\mu
\bigl\{\sqrt{-g}\,[242\gtn{\a\mu}\\ {}- 121 g_{\d\nu}(g^{\nu\a}\gtn{\mu\d} + g^{\nu\d}\gtn{\mu\a}) - \ov M\gtn{\mu\a}]\bigr\}\pp\Psi{x^\a}\\
{}- 96\pi e^{24\Psi}\bigl(\cLY(\a\dr{QCD}A_{\SU(3)}) + \cLY(\a_sA_{\SU(2)_L \ot\U(1)\dr{Y}})\bigr)\\ {}- 2e^{-2\Psi}\cL\dr{kin}(H^0)
+ \frac{10}{r_0^4}\,\wh V(H^0)e^{20\Psi} + \frac{40}{r_0^2}e^{20\Psi}\cL\dr{int} - \frac{11}{r_0^2} \,\wh{\ov P}_{S^2}e^{22\Psi}\\
{}- \frac{12}{\ell\pl^2}\bigl(-\tfrac72 \a_s^2 + \a^2\dr{QCD}\cdot \wt R\SU(3)\bigr)e^{24\Psi} - 40\cdot\frac{e^{20\Psi}}{r_0^4}=0
\e
where
\bg c357
\cL\dr{kin}(H^0) = (2(\z+1)g^{(\o\mu)} - (2\z+1)\gtn{\o\mu})\cdot \ov \n_\mu H^0\ov \n_\o H^0(x) \\
\wh V(H^0) = \frac{\pi\zy}{2\zz r_0^2} \biggl(2(H^0)^2 + \la' v(H^0)^3 + \frac{\la'} 4(H^0)^4\biggr) \lb c358 \\
\cL\dr{int} = -\frac{i\z 3\sqrt6(v+H^0(x))^2}{64\pi r_0^2\zz} (Z^{u0}_\m + \sqrt3\,F_\m)g^\m \lb c359 \\
G_\m = \frac12\zf\m \lb c360 \\
\cLY(\a\dr{QCD}A_{\SU(3)}) = -\frac{\a\dr{QCD}^2}{8\pi} \ell^{\SU(3)}_{cd} (2H^c H^d - L^{c\m}H^d_\m) \hskip60pt \nn \\
\hskip60pt {}= -\frac{\a^2\dr{QCD}}{8\pi} \bigl(2h^{\SU(3)}_{cd}(\gtk\m H^c_\m)(\gtk{\a\b}H^a_{\a\b}) - \ell^{\SU(3)}_{cd} L^c_{\a\b}H^d_\m g^{\a\mu}
g^{\b\nu}\bigr) \lb c361
\e
and
\bml c362
\cLY(\a_s A_{\SU(2)_L}) = -\frac{\a_s^2}{8\pi}\bigl(2h^{\SU(2)}_{cd}(\gtk\m \ov H{}^c_\m)(\gtk{\a\b}\ov H{}^d_{\a\b})\\ {}- h^{\SU(2)}_{cd}
(\ov H{}^c_{\a\b}- \gtn{\tau\g}(\ov H{}^c_{\g\a}g\ink{\b\tau} - \ov H{}^c_{\g\b}g\ink{\a\tau})) \ov H{}^d_\m g^{\a\mu}g^{\b\nu}\bigr)
= -\frac{\a_s^2}{8\pi} \bigl(2h^{\SU(2)}_{cd} (\gtk\m \ov H{}^c_\m)(\gtk{\a\b}\ov H{}^d_{\a\b}) \\{}- h^{\SU(2)}_{cd} g^{\a\mu}g^{\b\nu}
(\ov H{}^c_{\a\b} - \gtn{\tau\g}(\ov H{}^c_{\g\a}g\ink{\b\tau} - \ov H{}^c_{\g\b}g\ink{\a\tau})) \ov H{}^d_\m\bigr)
\e

\vskip-18pt
\bml c363
\cLY(\a_s A_{\U(1)\dr{Y}}) = \frac{\a_s^2}{8\pi}\bigl(2(\gtk\m G_\m)^2 - g^{\a\mu}g^{\b\nu}G_\m (G_{\a\b} - \gtn{\tau\g}(G_{\g\a}g\ink{\b\tau}
- G_{\g\b}g\ink{\a\tau}))\bigr)\\
= \frac{\a_s^2}{32\pi}\bigl(2(\gtk\m\zf\m)^2 - g^{\a\mu}g^{\b\nu}\bigl[\zf{\a\b}\\{}-\gtn{\tau\g}(\zf{\g\a}g\ink{\b\tau} -
\zf{\g\b}g\ink{\a\tau})\bigr] \zf\m\bigr)
\e

For the $A_{\SU(3)}$ gauge field one gets
\bml c364
\br{gauge(\SU(3))}{\n_\mu} \gd\ell,\SU(3),ab,  L{}^{a\a\mu} = 2\falg^{\a\b}\br{gauge(\SU(3))}{\n_\b}(\gd h,\SU(3),ab, (\gtk\m H^a_\m))\\
{}+ 24\pa_\b \Psi\bigl[\gd\ell,\SU(3),ab,\cdot \fal L^{a\b\a} - 2\falg^{[\b\a]}(\gd h,\SU(3),ab, \gtk\m H^a_\m)\bigr]
\e
or
\bml c365
\br{gauge(\SU(3))}{\n_\g}\bigl(\gd\ell,\SU(3),nf, \falg^{(\d\o)}g^{\g\mu} \bigl(H^n_{\o\mu} + \mu h^{\SU(3)\,na}\gd k,\SU(3),[ad], H^d_{\o\mu}\\
{}+ (H^n_{\a\o} \gtn{\a\d}g\ink{\d\mu} - H^n_{\a\mu}\gtn{\a\d}g\ink{\d\o}) - 2\mu h^{\SU(3)\,na}\gd k,\SU(3),[ad],
\gtn{\d\tau}\gtn{\a\b} H^d_{\d\a}g\ink{\tau\o}g\ink{\b\mu}\\ {}- 2\mu h^{\SU(3)\,na}\gd k,\SU(3),[ad], \gtn{\d\b}\gtn{\a\tau} H^d_{\b[\o}g_{\mu]\tau}
g\ink{\d\a}\\ {}+ 2\mu^2 h^{\SU(3)\,na}h^{\SU(3)\,bc}\gd k,\SU(3),[ac], \gd k,\SU(3),[bd], H^d_{\a[\o}g_{\tl\mu]\b\tp}\bigr)\\
{}= 2\falg^{\a\g}\br{gauge(\SU(3))}{\n_\g}(\gd h,\SU(3),mf, (\gtk\m H^m_\m))
+ 24\pa_\g \Psi \Bigl[\gd\ell,\SU(3),nf, \falg^{\g\o}g^{\chi\mu}\bigl(H^n_{\o\mu} + \mu h^{\SU(3)\,na}\gd k,\SU(3),[cd], H^d_{\o\mu}\\
{}+ (H^n_{\a\o}\gtn{\a\d}g\ink{\d\mu} - H^n_{\a\mu}\gtn{\a\d}g\ink{\d\o})
- 2\mu h^{\SU(3)\,na}\gd k,\SU(3),[ad], \gtn{\d\tau}\gtn{\a\b} H^d_{\d\a} g\ink{\tau\o}g\ink{\b\mu}\\
{}- 2\mu h^{\SU(3)\,na}\gd k,\SU(3),[ad], \gtn{\d\b}\gtn{\d\tau} H^d_{\b[\o} g_{\mu]\tau} g\ink{\d\a}
+ 2\mu^2 h^{\SU(3)\,na} h^{\SU(3)\,bc}\gd k,\SU(3),[ac], \gd k,\SU(3),[bd], H^d_{\a[\o}g\ink{\mu]\b} \bigr)\\
{}- 2\falg^{[\g\a\chi]}(\gd h,\SU(3),mf, (\gtk\m H^m_\m))\Bigr]
\e

It is interesting to find some \so s of those \e s
in the case of stationary and spherically \s ic (or axially \s ic). In the \elm c
case we get finite mass \so\ (a~model of an electron, see Refs~\cite{4,8}). In this
extended case we can get some models of elementary particles.

Let us consider spherically \s ic, static field configuration of our \un\ of NGT and a bosonic part of a Standard Model. One gets in spherical \cd s
(see Ref.~\cite{mab}):
\beq c366
g_\m = \left(\begin{matrix}
-\a\ &\ 0\ &\ 0\ &\ \o\\
0\ &\ -\b\ &\ f\sin\t\ &\ 0\\
0\ &\ -f\sin\t\ &\ -\b\sin^2\t\ &\ 0\\
-\o\ &\ 0\ &\ 0\ &\ \g
\end{matrix}\right)
\e
where $\a,\b,f,\o,\g$ are \f s of $r$ only, $\a>0$, $\g>0$. For $g^\m$ one easily gets
\beq c367
\aligned
g^{11} &= \frac\g{\o^2-\a\g}\\
g^{22} &= g^{33}\sin^2\t = -\frac \b{\b^2+f^2} \\
\gtk{14} &= \frac \o{\o^2-\a\g} \\
g^{44} &= -\frac\a{\o^2-\a\g} \\
\gtk{23}\sin\t &= \frac f{\b^2+f^2}, \q g=\det(g_\m) = -(f^2+\b^2)(\o^2-\a\g)\sin^2\t.
\endaligned
\e
We suppose
\beq c368
\o^2-\a\g \ne 0 \qh{and} \b^2+f^2\ne0.
\e

We use for calculations field \e s \er{4.109}--\er{4.111}. Using Eq.~\er{4.110} one easily gets from \er{c264}
\beq c369
\frac{\o^2}{\a\g-\o^2} = \frac{\ell^2}{\b^2+f^2},
\e
$\ell$ is an integration \ct. Moreover, we really use, in the place of Eqs~\er{4.109} or \er{c37}, Eqs \er{c304}--\er{c311} with $\br{eff.full}
{T_\m}$ where $\br{eff.full}{T_\m}$ contains all fields from the bosonic part of a Standard Model (see Eq.~\er{c230c} or Eqs \er{c345}--\er{c365}).
We suppose spherically \s ic configurations for all fields, i.e.\ for $\SU(3)$-gauge field, $\SU(2)_L$-gauge field, $\U(1)\dr{Y}$-gauge field
and for $\Psi$ and~$\F$ scalar fields. In this way we get for $\U(1)\dr Y$ field
\beq c370
\wh L_{\nu\mu} = G_{\nu\mu}, \q G_{14} = \wt E(r), \q G_{23} = \wt B \sin\t.
\e
The remaining components of $G_\m$ vanish. For an \elm c field we suppose:
\beq c371
H_{14} = E_r \ne 0, \q H_{23} = B \sin\t\ne0.
\e
We have:
\bea c372
L^n_{14} &= H^n_{14} + \mu h^{\SU(3)\,na}\gd k,\SU(3),ad, H^d_{14} = D^n_r \\
L^n_{23} &= H^n_{23} + \mu h^{\SU(3)\,na}\gd k,\SU(3),ad, H^d_{23} = H^n_r \sin\t \lb c373
\e
$H^n_{14} = E^n_r$, $H^n_{23} = B^n_r\sin\t$ for $\SU(3)$-gauge field and
\bea c374
\ov H{}^n_{14} &= \ov E{}^n_r \ne 0, \q \ov H{}^n_{23} = \ov B{}^n_r \sin\t \ne 0 \\
\ov L{}^n_{14} &= \ov H{}^n_{14} + i\xi h^{\SU(2)\,na}\gd k,\SU(2),nd, \ov H{}^d_{14} = \ov D{}^a_r \lb c375 \\
\ov L{}^n_{23} &= \ov H{}^n_{23} + i\xi h^{\SU(2)\,na}\gd k,\SU(2),nd, \ov H{}^d_{23} = \ov H{}^n_r \sin\t \lb c376
\e
for $\SU(2)_L$-gauge field.

The remaining components of $H,\ov H,L,\wh L$ are zero.
Scalar fields $\F=\F(r)$, $\Psi=\Psi(r)$ are \f s of~$r$ only.

For we have spherically \s ic configuration we have also special dependence of \cd s of \pt\ for $\SU(3)$-gauge field, $\SU(2)_L$-gauge field and
$\U(1)\dr{Y}$-gauge field.

Simultaneously we take $\wh B{}^n_r = \ov B{}^n_r = 0$. In this way $\wh L_{23}=\ov L_{23}=0$ and $L_{23}\ne 0$. We suppose $\xi=0$.

Let us write Eq.~\er{c345} in a spherical and static configuration for $\a=4$. One gets
\bml c377
-\pd{\ov E{}^1_r}r \sqrt{\frac{f^2+\b^2}{\o^2-\a\g}}\,\o(\a-\o) - \ov E{}^1_r \biggl(\pd{}r\biggl(\sqrt{\frac{f^2+\b^2}{\o^2-\a\g}}\,
\o^2\biggr) - \sqrt{\frac{f^2+\b^2}{\o^2-\a\g}}\,\a \pd{}r\biggl(\frac \o{\o^2-\a\g}\biggr)\biggr)\\
{} = 24\pd\Psi r \sqrt{\frac{f^2+\b^2}{\o^2-\a\g}}\,\ov E{}^1_r\,\frac{\o(\a-\o)}{\o^2-\a\g}
- \sqrt2\,M^2_W \sqrt{\frac{f^2+\b^2}{\o^2-\a\g}}\,(\z+1)\o-(2\z+1)\o(\ww r)
\e
Let us write Eq.~\er{c348} in the spherical and static case. One gets
\bml c378
-\pd{\ov E{}^2_r}r \sqrt{\frac{f^2+\b^2}{\o^2-\a\g}}\,(\a-\o)\o - \ov E{}^2_r\cdot \pd{}r\biggl(\sqrt{\frac{f^2+\b^2}{\o^2-\a\g}}\,\o^2\biggr)
- \sqrt{-g}\,\frac\a{\o^2-\a\g}\,\pd{}r\biggl(\frac\o{\o^2-\a\g}\biggr)\\
{}= 24\pd\Psi r \sqrt{\frac{f^2+\b^2}{\o^2-\a\g}}\,\ov E{}^2_r \o(\a-\o) - 2\sqrt2\,M^2_W \sqrt{\frac{f^2+\b^2}{\o^2-\a\g}}\,(\z+1)\o
(W^{u-}_r - W^{u+}_r).
\e
Let us write Eq.~\er{c350} in the spherical and static case for $\a=4$. One gets
\bml c379
-\pd{\ov E{}^3}r \sqrt{\frac{f^2+\b^2}{\o^2-\a\g}}\,(\a-\o)\o - \ov E{}^3_r \biggl(\pd{}r \sqrt{\frac{f^2+\b^2}{\o^2-\a\g}}\,\o^2\biggr)
- \sqrt{\frac{f^2+\b^2}{\o^2-\a\g}}\,\a\cdot\pd{}r\biggl(\frac\o{\o^2-\a\g}\biggr)\\
{} = 24\pd\Psi r \sqrt{\frac{f^2+\b^2}{\o^2-\a\g}}\,\ov E{}^3_r \o(\a-\o) + \sqrt3 \sqrt{\frac{f^2+\b^2}{\o^2-\a\g}}\,M^2_{Z^0}(\z+1)\o
Z^{u0}_r.
\e

Let us write Eq.~\er{c353} in a spherical and static configuration. One gets (for $\a=4$)
\bml c380
-\pd{}r(\sqrt 3\, \wh E_r - E_r)\sqrt{\frac{f^2+\b^2}{\o^2-\a\g}}\,\o(\a-\o) - (\sqrt3\,\wh E_r-E_r)\biggl(\pd {}r
\sqrt{\frac{f^2+\b^2}{\o^2-\a\g}}\,\o^2\biggr)\\ {}- \sqrt{\frac{f^2+\b^2}{\o^2-\a\g}}\,\a\,\pd {}r\biggl(\frac\o{\o^2-\a\g}\biggr)
= 24\pd\Psi r \sqrt{\frac{f^2+\b^2}{\o^2-\a\g}}\,(\sqrt3\,\wh E_r - E_r)\o(\a-\o)\\ {}+ M^2_{Z^0}(\z+1)\sqrt{\frac{f^2+\b^2}{\o^2-\a\g}}\,
Z^{u0}_r - \frac{i\z 3\sqrt2(v+H^0(r))}{16\pi^2 r_0^2\zz}\, \pd{H^0}r \sqrt{\frac{f^2+\b^2}{\o^2-\a\g}}\,\o.
\e

Let us write Eq.~\er{c355} in a spherical and static configuration. One gets
\beq c381
\biggl(2(\z+1)\frac\g{\o^2-\a\g} + \frac{2\z+1}\a\biggr) \pd{^2H^0}{r^2} - M^2_{H^0}H^0 - 3\la v(H^0)^2 - \la(H^0)^3 = 0.
\e

Let us consider Eq.~\er{c364} in the spherical and static configuration for $H^a_{14}=0$ and
\beq c382
H^a_{23} = B^a_r \sin\t \ne 0.
\e
One gets
\beq c383
\pd{}r \biggl(\frac f{f^2+\b^2}\,B^a_r\biggr) = 24\pd\Psi r\biggl(\frac{\o fB^a_r}{(\o^2-\a\g)(f^2+\b^2)}\biggr).
\e

Let us consider Eq.~\er{c356} in a spherical and static configuration. One gets
\bml c384
-\frac2\a\biggl[(242+\ov M)+ 242\frac{\o^2+\a\g}{\o^2-\a\g}\biggr]\pd{^2\Psi}{r^2}
- \frac2{\sqrt{(f^2+\b^2)(\o^2-\a\g)}} \,\pd{}r\biggl\{\frac{\sqrt{(f^2+\b^2)(\o^2-\a\g)}}\a\\ \biggl[242+ 121\cdot\biggl(
\frac{3(\o^2+\a\g)}{\o^2-\a\g} + \frac{2(\b^2-f^2)}{\b^2+f^2}\biggr) - \ov M\biggr]\biggr\}\pd \Psi r\\
{}- 40\frac{e^{20\Psi}}{r_0^4} - \frac{12}{\ell\pl^2} \bigl(\a^2\dr{QCD}\wt R\SU(3) - \tfrac72\a^2_s\bigr)e^{24\Psi}
- \frac{11}{r_0^2}\,\wh{\ov P}_{S^2} e^{22\Psi}\\ {}+ \frac{10}{r_0^4}e^{20\Psi}\,\frac{\pi\zy}{2\zz r_0}\biggl(2(H^0)^2+\la v(H^0)^3 + \frac\la4
(H^0)^4\biggr)\\ {}- e^{-2\Psi}\,\frac2{\a(\o^2-\a\g)}\cdot(2(\z+1)\a\g + (2\z+1)(\o^2-\a\g))\cdot\pd{^2H^0}{r^2}\\
{}- \frac{10i}{r_0^2}e^{20\Psi}\cdot \frac{\z 3\sqrt6(v+H^0(r))^2\o}{16\pi r_0^2\zz (\o^2-\a\g)}(\sqrt3\,E^0_r + \wh E)\\
{}+12 e^{24\Psi}\biggl(\a^2\dr{QCD}\gd\ell,\SU(3),cd, \biggl(\frac{\b^2}{\b^2+f^2}\biggr)(2B^c_rB^d_r - (B^c_r + \mu h^{\SU(3)\,cd}
\gd k,\SU(3),de, B^e_r)B^d) \\{}+ \a_s^2\,\frac{2\o^2+\a\g}{(\o^2-\a\g)^2}(\wh E{}^2_r + \gd h,\SU(2),cd, \ov E{}^c_r E^d_r)\biggr)=0.
\e

Let us consider tensors of energy-momentum in our partial \un\ in the case of spherical \s y and static. One gets for $\SU(2)_L$-gauge field
\bea c385
\br{gauge(\SU(2)_L)}{T_{11}} &= -\frac{\a_s^2}{16\pi}\,\gd h,\SU(2),ab, \ov E{}^a_r \ov E{}^b_r \,\frac{12\o^2 - 4\a\g+\a\g^2+\g\o}{(\o^2-\a\g)^2}\\
\br{gauge(\SU(2)_L)}{T_{22}} &= \frac{\a_s^2}{8\pi}\,\frac{(4\o^2+\a\g)\b}{(\o^2-\a\g)^2}\,\gd h,\SU(2),ab, \ov E{}^a_r \ov E{}^b_r \lb c386 \\
\br{gauge(\SU(2)_L)}{T_{33}} &= -\frac{\a_s^2}{8\pi}\,\gd h,\SU(2),ab,\ov E{}^a_r \ov E{}^b_r\,\frac{\sin^2\t(8\o^2+\a\g)}{(\o^2-\a\g)^2} \lb c387 \\
\br{gauge(\SU(2)_L)}{T_{44}} &= -\frac{\a_s^2}{8\pi} \,\gd h,\SU(2),ab, \ov E{}^a_r \ov E{}^b_r \,\frac{\a\g-2\o^2}{(\o^2-\a\g)^2} \lb c388 \\
\br{gauge(\SU(2)_L)}{T_{23}} &= -\br{gauge(\SU(2)_L)}{T_{32}} = \frac{\a_s^2 f\sin\t}{8\pi}\,\gd h,\SU(2),ab, \ov E{}^a_r \ov E{}^b_r
\frac{\a\g+4\o^2}{(\o^2-\a\g)^2}. \lb c389
\e

For $\cL\dr{int}$ one gets
\begin{align}
\br{int}{T_{11}} & = -\frac{\a\o}{\o^2-\a\g}\,\frac{i\z\sqrt6\,(v+H^0(r))^2}{256\pi r_0^2\zz}(Z^{u0}_{14} + \sqrt3\,E_r) \nn \\
\br{int}{T_{22}} & = -\frac{\b\o}{\o^2-\a\g}\,\frac{i\z\sqrt6\,(v+H^0(r))^2}{256\pi r_0^2\zz}(Z^{u0}_{14} + \sqrt3\,E_r) \nn \\
\br{int}{T_{33}} & = -\frac{\b\sin^2\t \o}{\o^2-\a\g}\,\frac{i\z\sqrt6\,(v+H^0(r))^2}{256\pi r_0^2\zz}(Z^{u0}_{14} + \sqrt3\,E_r) \nn \\
\br{int}{T_{44}} & = \frac{\g\o}{\o^2-\a\g}\,\frac{i\z\sqrt6\,(v+H^0(r))^2}{256\pi r_0^2\zz}(Z^{u0}_{14} + \sqrt3\,E_r) \nn \\
\br{int}{T_{[23]}} & = \frac{f\o \sin\t}{\o^2-\a\g}\,\frac{i\z\sqrt6\,(v+H^0(r))^2}{256\pi r_0^2\zz}(Z^{u0}_{14} + \sqrt3\,E_r) \nn \\
\br{int}{T_{[14]}} & = \frac{i\z 3\sqrt6(v+H^0(r))^2}{64\pi r_0^2\zz}(Z^{u0}_{14}+\sqrt3\,E_r)\biggl(\frac{\o^2}{4(\o^2-\a\g)}-1\biggr) \lb c390
\end{align}
For $\cL\dr{kin}$ one gets
\beq c391
\cL\dr{kin}(\Psi) = \frac1\a\biggl(-484\frac{\o^2}{\o^2-\a\g} + \ov M\biggr)\biggl(\pd\Psi r\biggr)^2.
\e
For $\cL\dr{int}$ one gets
\beq c392
\cL\dr{int} = -\frac\o{\o^2-\a\g}\cdot\frac{15\sqrt6}{8\pi r_0^2}e^{20\Psi} i \z(v+H^0(r))^2(Z^{u0}_{14} + \sqrt3\,E_r).
\e

For $\br{scal}{T_{\a\b}} - \frac12 g_{\a\b}\br{scal}T g^\m = \wt T_{\a\b} = \br{eff.scal}{T_{\a\b}}$ one gets
\bea c393
\wt T_{11} &= -\frac{e^{24\Psi}}{16\pi} \biggl\{\ov M+484 - 968\frac{\o^2}{\o^2-\a\g}\biggr\}\biggl(\pd\Psi r\biggr)^2 \\
\wt T_{22} &= -\frac{e^{24\Psi}}{16\pi} \biggl(\frac{\b}\a\biggr) \biggl\{\ov M+484 - 968\frac{\o^2}{\o^2-\a\g}\biggr\}\biggl(\pd\Psi r\biggr)^2
\lb c394 \\
\wt T_{33} &= -\frac{e^{24\Psi}}{16\pi} \biggl(\frac{\b\sin^2\t}\a\biggr) \biggl\{\ov M+484 - 968\frac{\o^2}{\o^2-\a\g}\biggr\}
\biggl(\pd\Psi r\biggr)^2 \lb c395 \\
\wt T_{44} &= \frac{e^{24\Psi}}{16\pi} \biggl(\frac{\g}\a\biggr) \biggl\{\ov M+484 - 968\frac{\o^2}{\o^2-\a\g}\biggr\}\biggl(\pd\Psi r\biggr)^2
\lb c396 \\
\wt T_{32} &= -\wt T_{23} = -\frac{e^{24\Psi}}{16\pi} \biggl(\frac{f\sin\t}\a\biggr) \biggl\{\ov M+484 - 968\frac{\o^2}{\o^2-\a\g}\biggr\}
\biggl(\pd\Psi r\biggr)^2 \lb c397
\e

For a gauge $\U(1)\dr Y$ field one gets
\bea c398
\br{gauge(\U(1)\dr Y)}{T_{11}} &= -\frac{G^2_{14}\a_s^2}{8\pi(\o^2-\a\g)^2}(\a^2\g +2\a\g\o+2\o^3-5\a\o^2) \\
\br{gauge(\U(1)\dr Y)}{T_{22}} &= -\frac{\b G_{14}^2\a_s^2}{8\pi(\o^2-\a\g)^2}(\a\g+\a^2-4\o^2) \lb c399 \\
\br{gauge(\U(1)\dr Y)}{T_{33}} &= -\frac{\b\sin^2\t \a_s^2}{8\pi(\o^2-\a\g)^2}(\a\g+\a^2-4\o^2) \lb c400 \\
\br{gauge(\U(1)\dr Y)}{T_{44}} &= -\frac{\a_s^2 G^2_{14}\g}{8\pi(\o^2-\a\g)^2}(\a^2-3\a\g-4\o^2) \lb c401
\e
For $\SU(3)$ gauge field one gets
\bml c402
\kern-8pt
\br{gauge(\SU(3))}{T_{11}}\!\!\! = \frac{\a^2\dr{QCD}}{8\pi}\,\frac\a {\sin^2\t(\b^2{+}f^2)}\bigl(\b^2(\gd h,\SU(3),ab, +\mu^2 \gd k,\SU(3),ea,
h^{\SU(3)\,ep}\gd k,\SU(3),eb,) - 4f^2\gd h,\SU(3),ab,\bigr) H^a_{23}H^b_{23} \\
{}= \frac{\a^2\dr{QCD}}{8\pi}\,\frac\a{\b^2+f^2}\bigl(\b^2(\gd h,\SU(3),ab, +\mu^2 \gd k,\SU(3),ea, h^{\SU(3)\,ep}\gd k,\SU(3),eb,)
- 4f^2\gd h,\SU(3),ab,\bigr) B^a B^b
\e
\bml c403
\br{gauge(\SU(3))}{T_{22}} = -\frac{\a^2\dr{QCD}}{4\pi} \, \frac{\b^2}{\b^2+f^2} \, \frac1{\sin^2\t}\bigl[\tfrac12\b(\gd h,\SU(3),ab,
+ \mu^2\gd k,\SU(3),ed,\cdot h^{\SU(3)\,ef}\cdot \gd k,\SU(3),fb,) H^a_{23}H^b_{23}\\ {}+ 2f^2\gd h,\SU(3),ab, H^a_{23}H^b_{23}\bigr]\\
{}= -\frac{\a^2\dr{QCD}}{4\pi}\,\frac{\b^2}{\b^2+f^2}\bigl[ \tfrac12 \b(\gd h,\SU(3),ab, + \mu^2 \gd k,\SU(3),ed, \cdot h^{\SU(3)\,ef}
\gd k,\SU(3),fb,) + 2f^2\gd h,\SU(3),ab,) B^a_rB^b_r\bigr]
\e
\bml c404
\br{gauge(\SU(3))}{T_{33}} = -\frac{\a^2\dr{QCD}\b^2}{4\pi \sin^2\t(\b^2+f^2)}\bigl(\tfrac12\b (\gd h,\SU(3),ab, + \mu^2 \gd k,\SU(3),ea,
\cdot h^{\SU(3)\,ep}\gd k,\SU(3),pb,)H^a_{23}H^b_{23}\\ {}+ 2f^2 \gd h,\SU(3),ab,H^a_{23}H^b_{23}\bigr)\\
{}= -\frac{\a^2\dr{QCD}}{4\pi(\b^2+f^2)} \bigl(\tfrac12\b(\gd h,\SU(3),ab, + \mu^2\gd k,\SU(3),ea, \cdot h^{\SU(3)\,ep}\cdot
\gd k,\SU(3),pb, )B^a_rB^b_r\bigr)
\e
\bml c405
\br{gauge(\SU(3))}{T_{44}} = -\frac{\a^2\dr{QCD}\g}{4\pi \sin^2\t (\b^2+f^2)}\bigl(\tfrac12 \b^2(\gd h,\SU(3),ab, + \mu^2\gd k,\SU(3),ea,
h^{\SU(3)\,ep}\gd k,\SU(3),pb,)H^a_{23}H^b_{23}\\ {}+ 4f^2\gd h,\SU(3),ab, H^a_{23}H^b_{23}\bigr) \\
{} = -\frac{\a^2\dr{QCD}\g}{4\pi(\b^2+f^2)}\bigl(\tfrac12 \b^2(\gd h,\SU(3),ab, + \mu^2\gd k,\SU(3),ea,h^{\SU(3)\,ep}\cdot \gd k,\SU(3),pb,)
+ \gd h,\SU(3),ab,\bigr) B^a_r B^b_r
\e
\beq c405n
\br{gauge(\SU(3))}{T_{23}} =0
\e
and
\bml c405a
\br{gauge(\SU(3))}{T_{14}} = \frac{\a^2\dr{QCD}\o}{16\pi\sin^2\t}\,\gd\ell,\SU(3),ab, (\b^2(H^a_{23} + \mu h^{\SU(3)\,ae,}\gd k,\SU(3),ed,
H^d_{23}H^b_{23} - f^2 H^{(a}_{23}H^{b)}_{23}\\ {}= \frac{\a^2\dr{QCD}\o}{16\pi} \bigl(\b^2(\gd h,\SU(3),ab, +\mu^2 \dg k,ed,\SU(3), h^{\SU(3)\,ef}
\gd k,\SU(3),fb,) - 2f^2\gd h,\SU(3),ab,\bigr)B^a_r B^b_r.
\e
For a group $\SU(3)$ we have of course
\beq c407
\gd \ell,\SU(3),ab, = \gd h,\SU(3),ab, + \mu \gd k,\SU(3),ab,
\e
where $\gd h,\SU(3),ab,$ is a \KC tensor for Lie algebra of $\SU(3)$ group and $\gd k,\SU(3),ab,$ is a skew-\s ic real tensor on $\SU(3)$
right-invariant \wrt an action of a group $\SU(3)$. We have here
\bg c408
H^a_{14} = L^a_{14} =0 \\
H^a_{23} = \sin\t B^a \lb c409 \\
L^a_{23} = H^a_{23} + \mu h^{\SU(3)\,ae}\gd k,\SU(3),ed, H^d_{23} = \sin\t(B^a_r + \mu h^{\SU(3)\,ae}\gd k,\SU(3),ed, B^d_r) = \sin\t H^a_r \lb c410
\e

For $\SU(2)_L$ gauge field we have
\beq c411
\gd\ell,\SU(2),ab, = \gd h,\SU(2),ab, + i\xi \gd k,\SU(2),ab,.
\e
Moreover, we suppose now that $\xi=0$, $\ov H{}^a_{14} = \ov L{}^a_{14}$. In this case for spherically \s ic and static case we have
\beq c412
\cL(\SU(2)_L) = \frac{\a^2_s}{4\pi} \gd h,\SU(2)_L,cd, \ov H{}^c_{14}\ov H{}^d_{14}\cdot \frac{\o^2\g - \a^2\g - 2\o^2\a}{\a(\o^2-\a\g)^2}
= \frac{\a_s^2}{4\pi}\,\gd h,\SU(2),cd, \ov E{}^c_r \ov E{}^d_r \frac{\o^2\g - \a^2\g - 2\o^2\a}{\a(\o^2-\a\g)^2}.
\e
For $\U(1)\dr Y$ gauge field one gets in the case of spherical and static \s y
\beq c413
\cL(\U(1)\dr Y) = - \frac{\a_s^2E_r^2}{8\pi(\o^2-\a\g)} \biggl(\frac{\g(\o-\a)-6\o^2}{\o^2-\a\g} + (2\o^2+1)\biggr), \q G_{14}=E_r.
\e

Let us consider a \lg\ $\D\cL$ (see Eq.~\er{c342}) and an energy-momentum tensor for $\D\cL$ (see Eq.~\er{c343}). One gets
\bea c414
\D\cL &= \frac{(2\z+1)\o^2+\a\g}{\a(\o^2-\a\g)}\bigl(M^2_W W^{u+}_r W^{u-}_r + \tfrac12\,M^2_{Z^0}(Z^{u0}_r)^2\bigr) + \tfrac12 M^2_{H^0}(H^0)^2 \\
T_{11}(\D\cL) &= \frac{(2\z+5)\o^2 - 3\a\g}{4(\o^2-\a\g)} \bigl(M^2_W W^{u+}_r W^{u-}_r + \tfrac12M^2_{Z^0}(Z^{u0}_r)^2\bigr)
+ \tfrac12 M^2_{H^0}(H^0)^2 \lb c415 \\
T_{22}(\D\cL) &= \frac \b2 \biggl[\frac{(2\z+1)\o^2+\a\g}{\a(\o^2-\a\g)}\bigl(M^2_W W^{u+}_r W^{u-}_r + \tfrac12M^2_{Z^0}(Z^{u0}_r)^2\bigr)
- M^2_{H^0}(H^0)^2\biggr] \lb c416 \\
T_{33}(\D\cL) &= -\frac{\b\sin^2\t}2 \biggl[\frac{(2\z+1)\o^2+\a\g}{\a(\o^2-\a\g)}\bigl(M^2_W W^{u+}_r W^{u-}_r + \tfrac12 M^2_{Z^0}(Z^{u0}_r)^2\bigr)
- M^2_{H^0}(H^0)^2\biggr] \lb c417 \\
T_{32}(\D\cL) &= \frac{f\sin\t}2 \biggl[\frac{(2\z+1)\o^2+\a\g}{\a(\o^2-\a\g)}\bigl(M^2_W W^{u+}_r W^{u-}_r + \tfrac12M^2_{H^0}(Z^{u0}_r)^2\bigr)
- M^2_{H^0}(H^0)^2 \biggr] \nn \\
&= -T_{23}(\D\cL) \lb c418 \\
T_{21}(\D\cL) &= T_{12}(\D\cL) = 0 \lb c419 \\
T_{31}(\D\cL) &= T_{13}(\D\cL) = 0 \lb c420 \\
T_{44}(\D\cL) &= \frac\g2 M^2_{H^0}(H^0)^2 - \bigl(M^2_W W^{u+}_r W^{u-}_r \nn \\
& + \tfrac12M_{Z^0}(Z^{u0}_r)^2\bigr)
\biggl(\frac{(2\z+1)(4\o^4 - 3\o^2\a\g + \a^2\g^2)}{4\a^2(\o^2-\a\g)}\biggr) \lb c421 \\
T_{24}(\D\cL) &= T_{42}(\D\cL) = 0 \lb c422 \\
T_{14}(\D\cL) &= -T_{41}(\D\cL) = \o\biggl(\tfrac12M^2_{H^0}(H^0)^2 + (M^2_W W^{u+}_r W^{u-}_r \nn \\
& + \tfrac12 M^2_{Z^0}(Z^{u0}_r))
\frac{3(2\z+1)\o^2 - (8\z+5)\a\g}{4\a(\o^2-\a\g)}\biggr) \lb c423 \\
T_{43}(\D\cL) &= T_{34}(\D\cL) = 0 \lb c424
\e

Let us consider a spherical and static configuration taking under consideration the case with $\Psi=\Psi_0$ (see Eq.~\er{c284}). This means, we have
to do with a \co ical \ct~$\wt\La$. We suppose also that $f=0$ and $\b=r^2$ (this is a simple \cd\ choice). Simultaneously in order to simplify
our \so\ we consider an \elm c field and $\SU(3)$-gauge field only. In this way we contemporary neglect weak \ia s and Higgs' phenomena. Our \so\
of field \e s will be made of \gr al, \elm c and strong \ia\ (a~\co ical \ct\ is included to an extended gravity via the field $\Psi$).

One gets:
\bea c425
\br{gauge(\SU(3))}{T_{11}} &= \frac{\a^2\dr{QCD}}{8\pi}\a(\gd h,\SU(3),ab, +\mu^2\gd k,\SU(3),ea, h^{\SU(3)\,ep} \gd k,\SU(3),eb,)B^a_rB^b_r \\
\br{gauge(\SU(3))}{T_{22}} &= -\frac{\a^2\dr{QCD}}{8\pi}\b(\gd h,\SU(3),ab, +\mu^2\gd k,\SU(3),ea, h^{\SU(3)\,ep} \gd k,\SU(3),eb,)B^a_rB^b_r
= \br{gauge(\SU(3))}{T_{33}} \lb c426 \\
\br{gauge(\SU(3))}{T_{44}} &=-\frac{\a^2\dr{QCD}}{8\pi}\g(\gd h,\SU(3),ab,+\mu^2\gd k,\SU(3),ea,h^{\SU(3)\,ep} \gd k,\SU(3),eb,)B^a_rB^b_r \lb c427 \\
\br{gauge(\SU(3))}{T_{23}} &= 0. \lb c428
\e
Moreover we can also use a notion of $H^a_r = B^a_r + \mu h^{\SU(3)\,ae}\gd k,\SU(3),ed, B^d_r$.

Bianchi identities give us
\beq c429
B^a_r = B^a_{0r} = {\rm const} \qh{and} H^a_r = H^a_{0r} = {\rm const}.
\e
We will use the notation
\beq c430
(\gd h,\SU(3),ab, +\mu^2 \gd k,\SU(3),ed, h^{\SU(3)\,ep}\gd k,\SU(3),eb,)B^a_{0r}B^b_{0r} = M_0 = {\rm const.}
\e
Let us consider a full field \e\ with $\SU(3)$-gauge and $\U(1)\dr{em}$ fields
\beq c431
\br{eff.full}{T_\m} = \br{gauge(\SU(3))}{T_\m} + \br{gauge(\U(1)\dr{em})}{T_\m}.
\e
We add later a term with a \co ical \ct\ $g_\m \wt \La$.

Using results from Refs~\cite{4} and \cite{mab} one gets
\bea c432
E(r) &= -\o\,\frac Q{\ell^2}\cdot \frac{r^4}{r^4+4\ell^4} \\
\intertext{$Q$ is an integration \ct}
E(r) &= F_{14}. \lb c433
\e
We have $f=B_0=0$ and thus one gets
\bea c433n
8\pi \br{eff.full}{T_{11}} &= \a^2\dr{QCD}\a\cdot M_0 + \a\cdot\a_s^2\,\frac{Q^2\b^4 - 4Q^2\ell^4\b^2}{\b^2(\b^2+4\ell^4)} \\
\frac{8\pi}{\sin\t}\br{eff.full}{T_{23}} &= \frac{8\pi}{\sin\t}\br{eff.full}{T_{[23]}} = 0 \lb c434 \\
8\pi \br{eff.full}{T_{44}} &= -\frac\g\a\,8\pi \br{eff.full}{T_{11}} \lb c435 \\
8\pi \br{eff.full}{T_{22}} &= 8\pi \br{eff.full}{T_{33}} = \frac \b\a 8\pi \br{eff.full}{T_{11}} \lb c436 \\
8\pi \br{eff.full}{T_{14}} &= -8\pi \br{eff.full}{T_{41}} \lb c437
\e

Eventually we get the \fw\ \e s
\bea c438
A_{11}(\ov \G) &= 8\pi \br{eff.full}{T_{11}} \\
A_{44}(\ov \G) &= 8\pi \br{eff.full}{T_{44}}. \lb c439
\e
$A_\m(\ov\G)$ is an ordinary Ricci tensor used by Pant (see Ref.~\cite{mab}) for the second contraction of a \cvt\ tensor is zero. In this case
the Moffat--Ricci tensor is equal to the ordinary Ricci tensor
\bea c440
A_{22}(\ov\G) &= 8\pi \br{eff.full}{T_{22}} \\
A_{33}(\ov\G) &= 8\pi \br{eff.full}{T_{33}} \lb c441 \\
A_{[23]}(\ov\G) &- 8\pi \br{eff.full}{T_{[23]}} = C_1\sin\t \lb c442 \\
A_{22}(\ov\G) &- 8\pi \br{eff.full}{T_{22}} = \frac1{\sin^2\t} \bigl[A_{33}(\ov\G)-8\pi\br{eff.full}{T_{33}}\bigr] \lb c443
\e
(see Refs \cite{4} and \cite{mab}).

We have also
\beq c444
\o = \frac{\ell^2}{r^2}.
\e
Substituting $\b=r^2$ one gets
\bea c445
E(r) &= -Q\,\frac{\a_s r^2}{r^4+4\ell^4} \\
A_{11}(\ov\G) &= \a_s^2 \,\frac{\a Q^2(r^4-4\ell^4)}{(r^4+4\ell^4)^2} - \a^2\dr{QCD}\a M_0 \lb c446 \\
A_{44}(\ov\G) &= \a_s^2 \,\frac{\g Q^2(r^4-4\ell^4)}{(r^4+4\ell^4)^2} - \a^2\dr{QCD}\g M_0 \lb c447 \\
A_{22}(\ov\G) &= -\a_s^2 \,\frac{r^2 Q^2(r^4-4\ell^4)}{(r^4+4\ell^4)^2} - \a^2\dr{QCD}r^2 M_0 \lb c448 \\
A_{[14]} &= 0 \lb c449 \\
A_{[23]}(\ov\G) &- 8\pi \br{eff.full}{T_{[23]}} = C_1\sin\t \lb c450 \\
8\pi \br{eff.full}{T_{[14]}} &= 8\pi \br{eff.full}{T_{14}} = \a_s \,\frac{3\ell^2}{r^2}\,Q^2\,\frac{7r^4-16\ell^4}{(r^4+4\ell^4)^2}
+ \frac12\a^2\dr{QCD}\,\frac{\ell^2}{r^2}M_0 \lb c451 \\
\frac1\a A_{11} &+ \frac1\g A_{44} =0 \lb c452 \\
\pd{}r (\log \a\g) &= -\frac4r\,\frac{\ell^4}{\ell^4+r^4} \lb c453 \\
\a\g &= B\biggl(1+\frac{\ell^4}{r^4}\biggr), \q B=1 \lb c454 \\
A_{11}(\ov\G) &= 8\pi T_{11} \lb c455 \\
A_{11}(\ov\G) &= \a\biggl(\frac{\a_s^2Q^2(r^4-4\ell^4)}{(r^4+4\ell^4)^2} - \a^2\dr{QCD}M_0\biggr)
= \a\biggl(\frac{\ov Q{}^2(r^4-4\ell^4)}{(r^4+4\ell^4)^2} - \ov M_0\biggr) \lb c456
\e
Finally one gets (if $\ov M_0=0$)
\beq c457
\hbox to0.3\textwidth{\hss$\displaystyle \a^{-1}(r) = 1+ \frac Cr + \frac{\ov Q{}^2}{br} g\biggl(\frac rb\biggr)\q (\ov Q{}^2 = \a_s^2Q^2)$,\hss}
\e
where
\beq c458
g(x) = \frac1{4\sqrt2}\log\biggl(\frac{x^2+\sqrt2\,x+1}{x^2-\sqrt2\,x+1}\biggr) - \frac1{2\sqrt2}\bigl(\arctan(\sqrt2\,x+1)
+ \arctan(\sqrt2\,x-1)\bigr), \q b^2=4\ell^2
\e
and we put $C=0$.

For $\g$ one gets
\beq c459
\g = \biggl(1+\frac{\ov Q^2}{br} g\biggl(\frac rb\biggr)\biggr)\biggl(1+\frac{\ell^4}{r^4}\biggr).
\e
If we add $\ov M_0$ into \er{c456} we get
\bg c457n
\a^{-1}(r) = \biggl(1 + \frac{\ov Q^2}{br} g\biggl(\frac rb\biggr) + \frac{\ov M_0r^2}3 \biggr) \\
\g = \biggl(1+\frac{\ov Q^2}{br}g\biggl(\frac rb\biggr) + \frac{\ov M_0r^2}3\biggr)\biggl(1+\frac{\ell^4}{r^4}\biggr) \lb c458n
\e

$\SU(3)$-gauge field renormalizes a \co ical \ct\ in such a way that
\beq c459n
\wt \La \mapsto \wt \La+ \ov M_0.
\e
One finds a \so\ with a \co ical \ct\ in Ref.~\cite8. The \so\ found here \sf ies a \cfn\ condition for $H^a_{14} = L^a_{14} = 0$ and we have
only a chromomagnetic field $H^a_{23} = \sin\t\, B^a_r$, $L^a_{23} = \sin\t\, H^a_r$. The \cfn\ condition is a dielectric (see Ref.~\cite{11a}) \cfn\
condition for we have no colour change distribution ($D^a_r=0$, $L^a_{14}= \sin\t\, D^a_r$, $H^a_{14}= \sin\t\, E^a_r$).

Let us recapitulate the properties of our \so. First of all we write down $\a$ and~$\g$:
\bea c460
\a^{-1} &= 1+ \frac{\ov Q^2}{br}g\biggl(\frac rb\biggr) + \frac{\wt\La r^2}{3} \\
\g &= \biggl(1+\frac{\ell^4}{r^4}\biggr)\biggl(1+ \frac{\ov Q^2}{br}g\biggl(\frac rb\biggr) + \frac{\wt\La r^2}{3}\biggr) \lb c461
\e
For electric field one gets
\beq c462
E = -\ov Q\frac{r^2}{r^4+4\ell^4}.
\e
For chromomagnetic field one finds
\beq c463
\aligned
H^a_{23} &= \sin \t\, B^a_{0r}, \q B^a_{0r} = {\rm const} \\
L^a_{23} &= \sin \t\, H^a_{0r}, \q H^a_{0r} = {\rm const}
\endaligned
\e
and zero colour change distribution
\beq c464
L^a_{14} = \sin\t\, D^a_r, \q D^a_r = 0.
\e
In the case of a \co ical \ct\ $\wt\La$ caused by the minimum of a self\ia\ \pt\ for $\Psi=\Psi_0$ one gets as in \er{c460} and \er{c461}.
Moreover, we get a term which behaves as a \co ical \ct. One gets:
\begin{gather}
\La\dr{eff} = \wt\La + \a^2\dr{QCD} M_0 = \wt\La + \a^2\dr{QCD}(\gd h,\SU(3),ab, + \mu^2\gd k,\SU(3),ea, h^{\SU(3)\,ep,}\gd k,\SU(3),pb,)
\cdot B^a_{0r}B^b_{0r}\hskip60pt \nn \\
\hskip80pt {} = -2\a^2\dr{QCD}\Bigl((1+\mu^2)((B^1_{0r})^2 + (B^2_{0r})^2) + \sum_{n=3}^8 (B^n_{0r})^2\Bigr) \lb c465 \\
E(r) \mathrel{\mathop{\kern0pt \approx}\limits_{r\to\iy}} -\frac {\ov Q}{r^2} \lb c466
\end{gather}
One gets $E(0)=0$ and
$$
|E(r)| \le E\dr{max} = |E(\sqrt2\,\ell)| = \frac{|\ov Q|}{2\ell^2}
$$
The electric field is nonsingular at $r=0$ and bounded.

Let us examine $\a$ and $\g$:
\bg c467
\a^{-1} \mathrel{\mathop{\kern0pt \approx}\limits_{r\to\iy}} 1 - \frac{2m_N}r + \frac{\ov Q{}^2}{r^2} + \frac{\La\dr{eff}r^2}3 \\
m_N c^2 = \frac{\pi \ov Q{}^2}{2\sqrt{2b}} \nn \\
\lim_{r\to0}\a^{-1} = 1. \nn
\e
It means, we get asymptotic behaviour of Reissner--Nordstr\"om metric.

The total charge is equal to
\beq c468
Q\dr{tot} = \int_{\R^3} \sqrt{-g}\,\rho \,d^3x = -16Q\ell^4 \int_0^\iy \frac1r\cdot \frac{r^4}{r^4+2\ell^4}\,dr = -Q,
\e
the last means that the total electric charge defined above is the same as the charge obtained from the asymptotic properties of the electric field.
We use the \fw\ formulae
\bg c469
\sqrt{-g}\,\rho = \gd \fal H,4i,{,i}, = \mathop{\rm div}\os D \\
\fal H^{4i} = \sqrt{-g}\,\frac E{\a\g-\o^2} \lb c470 \\
\sqrt{-g}\,\rho = -\frac1\pi\,\frac Qr\,\frac{4\ell^4r^4}{r^4+4\ell^4} \sin\t \nn
\e
where $\rho$ is the distribution of the charge. Thus the total electric charge defined above is the same as the charge obtained from asymptotic
properties of an electric field and \f s $\a$ and~$\g$.

It is interesting to calculate the energy of an \elm c field. One gets
\bg c471
\frac12\bigl(g^{4\mu}\br{gauge(\U(1)\dr{em})}{T_{4\mu}} + g^{\mu4}\br{gauge(\U(1)\dr{em})}{T_{\mu4}}\bigr) = \gd \br{gauge(\U(1)\dr{em})}T,4,4,
= \frac1{8\pi}\,Q^2\,\frac1{r^4+4\ell^4}\\
E\dr{tot} = 4\pi \int_0^\iy r^2 \gd\br{gauge(\U(1)\dr{em})}T,4,4,\,dr = \frac\pi{2\sqrt2}\,\frac{Q^2}b = m_N c^2 \lb c472
\e
(where $b^4=4\ell^4$). Thus the total energy of an \elm c field is equal to the Newtonian mass in Reissner--N\"ordstr\"om metric.

The determinant of the \s ic part of the metric $g_{(\a\b)}$ is equal to
\beq c473
\wt g = \det(g_{(\a\b)}) = -(r^4+\ell^4)\sin^2\t.
\e
The determinant of a full metric
\beq c474
g = \det(g_{\a\b)}) = r^4\sin^2\t.
\e
Thus there is a singularity at $r=0$. Moreover, there is no singularity in~$\a$ and only one singularity in~$\g$ due to $\bigl(1+\frac{\ell^4}
{r^4}\bigr)$ factor, $\o$~is also singular at $r=0$, which is a part of a skew-\s ic part of~$g_{\a\b}$.

In Ref.~\cite4 we have plots for \f s $g(x)$ (Fig.~2), $\wt E(R)$ (Fig.~3), where $R=\frac rb$, $E=\bigl(\frac Qb\bigr)^2\wt E$. In Ref.~\cite8
we plot also the \f
$$
f(x) = \a^{-1} = 1 + a\,\frac{g(x)}x + bx^2, \q x=\frac rb,
$$
for several values of parameters $a$~and~$b$ (Figs 1--5). For a zero \co ical \ct\ ($b=0$) we have a plot in Ref.~\cite4 (Fig.~6).

In Ref.~\cite{mac} R.~B.~Mann found \so s of the \NK\ in the spherically \s ic, static case. In his approach $f=u$ in Eq.~\er{c366}. The \so\ for
a \gr al field is in his paper in (4.14), which we quote here with our extension with a \co ical \ct:
\beq c475
g_\m = \left( \begin{matrix}
\a & 0 &\ 0 & \dfrac{\ell^2}{r^2\sqrt{1+u_0^2}} \\
0 & -r^2 &\ u_0 r^2 \sin\t & 0 \\
0 & -u_0r^2\sin\t &\ -r^2\sin^2\t & 0 \\
-\dfrac{\ell^2}{r^2\sqrt{1+u_0^2}} & 0 &\ 0 & \biggl(1+\dfrac{\ell^4}{(1+u_0^2)r^4}\biggr)\a^{-1}
\end{matrix} \right)
\e
where
\beq c476
\a = \biggl(1+\frac{\La\dr{eff} r^2}3 + \biggl[\frac{Q^2}{1+u_0^2}r^{-1}\biggr] K(r,L)\biggr)^{-1}
\e
and
\bg c477
K(r,L) = -\int r^2(r^4+4L^4)^{-2}\,dr \\
L^4 = \frac{\ell^4}{1+u_0^2} \nn \\
K(r,L) = \frac1{8L} \biggl[\log\biggl(\frac{x^2+\sqrt2\,x+1}{x^2 - \sqrt2\,x+1}\biggr) + 2\bigl[\arctan(\sqrt2\,x+1)
+ \arctan(\sqrt2\,x-1)\bigr]\biggr] \lb c478 \\
\ov b{}^4 = 4L^4, \q x= \frac r{\,\ov b\,}.
\e
We put $\a_0=0$. For our purposes the most interesting case is case~A from Table~1 from Ref.~\cite{mac}. This generalizes our \so. In this \so\
we have also nonzero \ct\ magnetic field~$B_0$ (if we want). We put $B_0=0$. The asymptotic behaviour of~$\a$ and~$\g$ is the same as before. Moreover
a Newtonian mass is a little different:
$$
m_N = \frac{Q^2\pi}{16\sqrt2\,L^4} = \frac{Q^2\pi}{16\sqrt2\,\ell^4}\cdot(1+u_0^2) = \frac{Q_0^2\pi}{16\sqrt2\,\ell^4}, \q
Q_0 = Q\sqrt{1+u_0^2}.
$$
The electric field behaves as
\beq c481
E(r) = -\frac{Qr^2}{(1+u_0^2)^{1/2}}\cdot\frac1{r^4+4L^4}
\e
and is nonsingular at $r=0$.

Using $Q_0 = Q\sqrt{1+u_0^2}$ one gets
\beq c482
E(r) = - \frac{Q_0r^2}{r^4+4L^4}.
\e

It is necessary to mention that the electric charge $Q$ (or~$Q_0$) is an integration \ct. Thus the Newtonian mass is the same as in our \so\ found
before. $u_0$ is also an integration \ct.

Now we can pose a \fn, philosophical question: \ti{What is a matter?}. The answer is: \ti{It is an energy of \fn\ \ia s} (see Ref.~\cite{a}).

We will try to extend this formalism to include $\SU(2)_L\ot \U(1)\dr Y$-gauge fields and Higgs' phenomena in spherical, static case and also to
include axially \s ic \so s.

According to current ideas (see Ref.~\cite{11a} and references therein) the \cfn\ of a colour could be connected to the dielectricity of the
vacuum (dielectric model of \cfn). Due to the so-called antiscreening mechanism, the effective dielectric \ct\ is equal to zero (see \rc{11a}).
This means that the energy of an isolated charge is going to infinity. Now there are also the so-called classical dielectric models of \cfn\
(see \rc{11a}). The \cfn\ is induced by a special type of dielectricity \st $\os E\ne0$ and $\os D=0$ ($\os E{}^a\ne0$, $\os D{}^0=0$).
In this case we have not a distribution of a charge. This is similar to electric-type of Meissner effect.

It is easy to see that in our case (the \E\nos\ \KK (Jordan--Thiry) Theory) the dielectricity is induced by \nos\ tensors $\ell_{ab}$
and~$g_{\a\b}$. If $g\ink{\a\b}=0$, $\os D=\os E$, $\os B=\os H$. The factor $G\dr{eff}$ depending on a scalar field~$\Psi$ also induced some kind
of dielectricity of vacuum. The \gr al field described by the \nos\ tensor $g_{\a\b}$ behaves as a medium for an \elm c field (gauge field). The
condition $\os E\ne0$, $\os D=0$ ($\os E{}^a=0$, $\os D{}^0=0$) can be \sf ied in the axial, stationary case for $F_\m$, $H_\m$ ($H^a_\m$,
$L^a_\m$), $g_\m$. Thus it is interesting to find \so s of field \e s for the \E\nos\ \KK (Jordan--Thiry) Theory (a~partial \un\ of \fn\ \ia s
--- \un\ of NGT with a bosonic part of a Standard Model, including a gauge field with $G=\SU(3)_c$ (QCD)). This could offer a model of \el ary \pc s
(e.g.\ hadrons). The axially \sy ic, stationary case seems to be very interesting from a more general point of view. We have in GR very peculiar
properties of stationary, axially \sy ic \so s of the Einstein--Maxwell \e s. Those \so s describe the \gr al and \elm c fields of the rotating
charged bodies. Thus we get a magnetic field component. Asymptotically (those \so s are asymptotically flat) the magnetic field behaves as a dipole
field. We can calculate a gyromagnetic ratio at infinity, i.e.\ a ratio of a magnetic dipole moment and an angular momentum moment. It is worth to
notice that we get an anomalous gyromagnetic ratio (see \rc{mcw}), i.e.\ a gyromagnetic ratio for a fermion (e.g.\ an electron) for a charged Dirac
\pc. We cannot interpret the Kerr--Newman \so\ as a model of a fermion, for we have a singularity. In the \E\nos\ \KK (Jordan--Thiry) Theory in the
case of a \un\ of~NGT with a bosonic part of a Standard Model (a~partial \un) we can expect a completely nonsingular \so. We can also expect an
asymptotic behaviour of Einstein--Maxwell theory. Thus it seems that we get probably \so s with an anomalous gyromagnetic ratio. Such a \so\ could be
treated as a model (classical) of a spin-$\frac12$ \pc. In the nonabelian case with $G=\SU(3)_c \tm \U(1)\dr{em}$ (our partial \un) this could offer
us a model of a charged baryon (e.g.\ proton), where a skewon $g\ink\m$ induces a colour \cfn. In this way the skewon field $g\ink\m$ plays a double
role (see \rc{11a}): $1^\circ$~additional \gr\ \ia\ from NGT, $2^\circ$~a~strong \ia\ field responsible for a \cfn. In our theory, where a \co ical
\ct\ \ti{must be nonzero}, $g\ink\m$ in a linear approximation is a spin~1 \gz\ Maxwell field with nonzero mass or a Proca pseudovector field.

In the theory there exists a field $\ov W_\mu$ which does not propagate.

What is a spin content in the theory (NGT)? It is $(2,0)$. Thus we get a graviton and a skewon. In extension to higher \di s (NKKT, NKK(JT)T) the
situation changes. We have $(2,1,0)$, i.e.\ graviton, skewon-pseudovector field and scalar field.

Thus it is natural to expect an exchange of spin one and spin zero \pc s in the nucleon-nucleon \ia\ at low and intermediate energy (in a \pt). We do
not observe such \pc s (in our theory they are Dark Matter \pc s). However we cannot fit experimental data for a nucleon-nucleon \pt\ without
a mysterious $\si$-\pc, see Refs \cite{mcr,mcs,mct}. It happens we need two such \pc s to fit data. In our proposal, they are connected to $g\ink\m$
and to a scalar field~$\Psi$. Simultaneously they are Dark Matter \pc s. Our \lg\ for strong \ia\ field and scalar field~$\vf$, $\Psi=\Psi_0+\vf$
resumbles the bosonic part of a soliton bag model \lg
\beq c538
\cL = -\frac14\,K(\vf) h_{ab}H^{a\m}\gd H,b,\m, - \frac12\,\pa_\mu\vf \pa^\mu\vf - U(\vf)
\e
in a Minkowski space limit.

In an approach from \rc{mct} $\vf$ is equal to $\si$.

In the \E\nos\ \KK (Jordan--Thiry) Theory we have an effective \lg\ for strong \ia\ only.
\beq c539
\cL\dr{strong} = \frac{\a^2\dr{QCD}}{8\pi} \,K(\vf) \gd\ell,\SU(3),ab, (2H^aH^b - \gd H,a,\m,L^{b\m}) + \cL\dr{kin}(\vf) - U(\vf)
\e
which is more general than the \lg\ from a soliton bag model. In general we take
\begin{gather*}
\cL(\vf) = \cL\dr{kin}(\vf) - U(\vf), \\
\cL(\vf) = \cL(\Psi_0+\vf), \q K(\vf) = \frac1{G_N}\,G\dr{eff}(\Psi_0+\vf).
\end{gather*}
In particular one obtains
\bml c540
\cL\dr{strong} = \frac{\a^2\dr{QCD}}{8\pi} e^{-24\vf}\gd\ell,\SU(3),ab,(2H^aH^b -L^{a\m}\gd H,b,\m,) \\
{}-\bgg{\bgg{1386 - \frac{\mu^2}{64(\mu^2+4)}}\gtn{\nu\g} - 484\gtk{\nu\mu} g_{\d\nu}\gtn{\d\g}}\vf_{,\nu}\cdot\vf_{,\g} \\
{}+ e^{20\Psi_0}
\bgg{\bgg{\frac{2\a^2\dr{QCD}}{(\mu^2+4)^2}(2\mu^3+7\mu^2+25\mu+20)
- \frac12\bgg{\frac52\,\a^2\dr{QCD}+7\a_s^2}}\frac{e^4\Psi_0}{\ell^2\pl}\,e^{4\vf} \\
{} + \frac{4e^{2\Psi_0}}{r_0^2\zz}\,e^{2\vf} + \frac{4\pi \zy\ell^2\pl}{r_0^4\sqrt{1+\z^2}}}e^{20\vf},
\e
where
\bg c541
H^a = \gtk\m\cdot \gd H,a,\m,, \q L^{a\m} = g^{\mu\a}g^{\nu\a}\gd L,a,\a\b,\\
\ov M = -\wt M = 1386 - \frac{\mu^2}{64(\mu^2+4)} >0 \lb c542
\e
for any real $\mu$.
\begin{gather}
e^{2\Psi_0} = x_0 = \bgg{\frac{\ell\pl}{r_0}}^2\frac{11+(121-120\pi\zz^{1/2}(1-\z^2-2\z^4)y)^{1/2}}{-6\zz y} \nn \\
y = \a^2\dr{QCD}\wt R\SU(3)+\a_s^2R G2 = \frac{2\a^2\dr{QCD}}{(\mu^2+4)^2}(2\mu^3+7\mu^2+25\mu+20) - \frac12\bgg{\frac52\a^2\dr{QCD}+7\a_s^2} = y(\mu)
\lb c543 \\
\lim_{\mu\to\iy} y(\mu) = -\frac12\bgg{\frac52\a^2\dr{QCD}+7\a_s^2} < 0, \lb c544 \\
y(0) = -\frac12\bgg{-\frac52\a^2\dr{QCD}+7\a_s^2}, \lb c545 \\
y(\mu) \searrow \hbox{in }(0,+\iy). \nn
\end{gather}
For sufficiently large $\mu$
\beq c546
y(\mu)<0.
\e

The above \lg\ has been considered by Friedberg and Lee (see \rc{mct}) in their soliton bag model, see also Refs \cite{mcp,mcq,mcz,mcs}. Our \lg\ is
more general because it contains a \gr al-\nos\ field ($g_\m$) and also $L^{a\m}$ which has been obtained in \E\nos\ \KK (Jordan--Thiry) Theory. They
considered~$\vf$ (in their approach it is called~$\si$) as an additional dynamic field with self\ia\ given $U(\vf)$. The quantity $K(\vf)$
is a dielectric \ct\ of a vacuum, also $G\dr{eff} = G\dr{eff}(\Psi_0+\vf)$. It is interesting to notice that $K$~plays a double role (effective
\gr al ``\ct'' and an effective dielectric ``\ct'').

Thus we propose the \lg\ of the \NJ\ (a~partial \un\ of NGT and a bosonic part of a Standard Model as a strong \ia\ \lg). Why? It seems that something
is missing in the QCD \lg. Exactly $\si$-\pc s are lacking (our $\vf$ and $g\ink\m$). Thus we propose to extend QCD to the \NJ. In this way we get
dielectric model of a colour \cfn\ and soliton bag-like model.

Thus we propose the \fw\ program of investigations:

$1^\circ$: Find exact \so s of the \NJ\ (a partial \un, \un\ of NGT with a bosonic part of a Standard Model) in the case of spherical and axial \s y
using inverse scattering method, B\"acklund \tf, Hirota method, AKNS (Albowitz, Kaup, Newell, Segur) method etc.

$2^\circ$: Find an effective \ia\ of two axially \s ic \so s exactly or using some numerical methods. This could offer a nucleon-nucleon \pt\ \ia\
similar to nucleon-nucleon \ia\ in a Skyrme model. The \so s should be treated as \pc s using a collective \cd\ method.

\smallskip
Let us consider field \e s with sources given by $\cL\dr{strong}$. In this way one gets
$$
R_{\a\b} = \frac{2\pi G_N}{c^4}\cdot \nad{strong.eff}{T_{\a\b}}
$$
where:
\bml c547
\nad{strong.eff}{T_{\a\b}} = e^{-24\vf} \nad{gauge\,SU(3)}{T_{\a\b}} + \nad{scal.eff}{T_{\a\b}}
= -\frac{\a^2\dr{QCD}}{4\pi}\, e^{-24\vf}\gd\ell,\SU(3),ij, \\
\cdot\biggl\{g_{\g\b}g^{\tau\rho}g^{\ve\g}\gd L,i,\rho\a, \gd L,j,\tau\ve, -2 \gtk\m \gd H,(i,\m, \gd H,j),\a\b,
- \frac14\,g_{\a\b}\bigl[L^{i\m}\gd H,j,\m, - 2(\gtk\m \gd H,i,\m,)(\gtk{\g\d}\gd H,j,\g\d,)\bigr]\biggr\}\\
{}+ \frac{e^{24\Psi_0}e^{24\vf}}{16\pi} \biggl\{(g_{\ka\a}g_{\o\b} + g_{\o\a}g_{\ka\b})\gtn{\g\ka}\gtn{\nu\o}
\bigl[242(g^{\xi\mu}g_{\nu\xi}-\d^\mu_\nu)\vf_{,\mu} - \d^\mu_\nu \wt M\vf_{,\nu}\bigr]\vf_{,\g}\\
- g_{\a\b}\bigl[ -\wt M\gtn{\nu\mu}\vf_{,\mu}\vf_{,\nu} + 484\gtk\m g_{\d\mu}\gtn{\g\mu}\vf_{,\nu}\vf_{,\g}\bigr]\\
{}+ g_{\a\b}\vf_{,\mu}\vf_{,\g}\bigl[(242-\wt M)\gtn{\g\mu} + 242(\gtn{\g\tau}g^{\xi\mu}g_{\nu\xi} -4 \gtk\mu g\ink{\d\a}\gtn{\g\d})\bigl]\biggr\}\\
{}+\frac14\,g_{\a\b}e^{20\Psi_0}\bgg{N\,\frac{e^{4\Psi_0}}{\ell^2\pl} \,e^{4\vf} + \frac{4e^{2\Psi_0}}{r_0^2\zz}\,e^{2\vf}
+ \frac{4\pi\zy \ell^2\pl}{r_0^4\zz^{1/2}}}e^{20\vf},
\e
\beq c548
\bga
\wt M = 1386 - \frac{\mu^2}{64(\mu^2+4)} \\
\nad{strong.eff}{T_{\a\b}} = \nad{gauge(SU(3))}{T_{\a\b}} + \nad{scal.eff}{T_{\a\b}}, \q
\nad{scal.eff}{T_{\a\b}} = \nad{scal}{T_{\a\b}} - \tfrac12 g_{\a\b}(\nad{scal}{T_\m}g^\m) \\
N = \frac{2\a^2\dr{QCD}}{(\mu^2+4)^2}(2\mu^3+7\mu^2+25\mu+20) - \frac12\bgg{\frac52\,\a^2\dr{QCD}+7\a_s^2}.
\ega
\e

One gets as field \e s:
$$\displaylines{\indent
\refstepcounter{equation} \lb c549
\nad{gauge(SU(3))}{\n_\mu}\gd \ell,\SU(3),ab, L^{a\a\mu} = 2\falg^{\a\b}\nad{gauge(SU(3))}{\nabla_\b}\bigl(\gd h,\SU(3),ab,(\gtk\m\gd H,a,\m,)
\bigr)\hfill\cr
\hfill {}+24\pa_b \vf\bigl[\gd\ell,\SU(3),ab, \fal L^{a\b\a} - 2\falg^{[\b\a]}(\gd h,\SU(3),ab,\gtk\m \gd H,a,\m,)\bigr] \indent \rm(\theequation)\cr
\refstepcounter{equation} \lb c550
\indent 2\bigl((242 - \wt M)\gtn{\a\mu} - 242 g^{\nu\mu}g_{\d\nu}g_{\d\nu}\gtn{\a\d}\bigr)\pp{^2\vf}{x^\a \pa x^\mu} \hfill\cr
+ \frac2{\sqrt{-g}}\,\pa_\mu\bigl(\sqrt{-g}\bigr[242\gtn{\a\mu} - 121g_{\d\nu}(g^{\nu\a}\gtn{\mu\d} + g^{\nu\d}\gtn{\mu\a})
{}+ \wt M \gtn{\mu\a}\bigr]\bigr)\pp\vf{x^\a} \cr
\hfill {}- 8 e^{20\Psi_0}\bgg{3N\frac{e^{4\Psi_0}}{\ell^2\pl}\,e^{4\vf}
+ \frac{11e^{2\Psi_0}}{r_0^2\zz}\,e^{2\vf} + \frac{10\pi\zy \ell^2\pl}{\zz^{1/2}r_0^4}} e^{20\vf}
+24 e^{-24\vf}\cLY^{\SU(3)} =0. \indent \rm(\theequation) }
$$
where
\beq c551
\cLY^{\SU(3)} = \frac{\a^2\dr{QCD}}{4\pi} \bigl(\d_{ab}H^aH^b + 2\d_{ab} L^{a\m}\gd H,b,\m, - \mu(L^{1\m}\gd H,2,\m, - L^{2\m}\gd H,1,\m,)\bigr)
\e
where for $n\ne 1,2$
\begin{align}
\gd L,n,\o\mu, &= \gd H,n,\o\mu, + (\gd H,n,\a\o, \gtn{\a\d}g\ink{\d\mu} - \gd H,n,\a\mu, g^{(\a\d)}g\ink{\d\a}) \\
\gd L,1,\o\mu, &= \gd H,1,\o\mu, + (\gd H,1,\a\o, \gtn{\a\d}g\ink{\d\mu} - \gd H,1,\a\mu, \gtn{\a\d}g\ink{\d\o})
- 2\mu \gd H,2,\a\mu, + 4\mu \gtn{\d\tau}\gtn{\a\b} \gd H,2,\d\a, g\ink{\tau\o}g\ink{\b\mu}\nn \\
&+ 4\mu \gtn{\d\b}\gtn{\a\tau} \gd H,2,\b[\o, g_{\mu]\tau} g\ink{\d\a} - 8\mu^2\gtn{\a\b}\gd H,1,\a\tl\o, g_{[\mu\tp\b]} \lb c552 \\
\gd L,2,\o\mu, &= \gd H,2,\o\mu, + (\gd H,2,\a\o, \gtn{\a\d}g\ink{\d\mu} - \gd H,2,\a\mu, \gtn{\a\d}g\ink{\d\o})
+ 2\mu \gd H,1,\o\mu, - 4\mu \gtn{\a\d}\gtn{\a\b} \gd H,1,\d\a, g\ink{\tau\a}g\ink{\b\mu} \nn \\
&- 4\mu \gtn{\d\b}\gtn{\a\tau} \gd H,1,\b[\o, g_{\mu]\tau} g\ink{\d\a} - 8\mu^2 \gd H,2,\a\tl\o,g_{[\mu\tp\b]}. \lb c553
\end{align}

Let us consider a case with spherical \s y and static with a \cfn\ of colour condition. One gets
\begin{align}
\gd L,a,23, &= \gd H,a,23, \qh{for} a\ne1,2 \\
\gd L,1,23, &= \gd H,1,23, - \frac \mu2\,\gd H,2,23,,\q \gd H,1,r, = \gd B,1,r, - \frac\mu2\, \gd B,2,r, \\
\gd L,2,23, &= \gd H,2,23, + \frac \mu2\,\gd H,1,23,,\q \gd H,2,r, = \gd B,2,r, + \frac\mu2\, \gd B,1,r, \lb c554 \\
\gd H,a,14, &= \gd L,a,14, = 0 \lb c555 \\
\cLY &= \frac{\a^2\dr{QCD}}{4\pi \b^2} \bgg{\sum_{n=1}^8 (\gd B,n,r,)^2 + \bgg{1+\frac{\mu^2}2}((\gd B,1,r,)^2 + (\gd B,2,r,)^2)} \lb c556 \\
\gd B,n,r, &= \gd B,n,r0, = {\rm const.} \lb c557
\end{align}

For the field $\vf$ one gets the \fw\ \e
\bml c558
\bgg{\frac{484\a\g}{\o^2-\a\g} + 1386 - \frac{\mu^2}{64(\mu^2+4)}}\pd {^2\vf}{r^2}\\
{} - \frac2{\b\sqrt{\o^2-\a\g}}\,\pd{}r\bgg{\b\sqrt{\o^2-\a\g}\bgg{968\,\frac{\o^2+\a\g}{\o^2-\a\g} - \frac{242}\b - 1386 + \frac{\mu^2}{64(\mu^2+4)}}
\pd \vf r \\
{}+ \frac{6\a}{\pi\b^2}\,e^{-24\vf} \bgg{\sum_{n=3}^8 (\gd B,n,r0,)^2 + \bgg{1+\frac{\mu^2}2}\bigl((\gd B,1,r0,)^2+(\gd B,2,r0,)^2\bigr)}}\\
{} - 8a e^{20\Psi_0}\bgg{3\bgg{\frac{2\a^2\dr{QCD}}{(\mu^2+4)^2}(2\mu^3+7\mu^2+25\mu+20) - \frac12\bgg{\frac52\,\a^2\dr{QCD}+7\a_s^2}}
\frac{e^{4\Psi_0}} {\ell^2\pl}\,e^{4\vf} \\
{}+ \frac{11e^{2\Psi_0}}{r_o^2\zz}\,e^{2\vf} + \frac{10\pi\zy\ell\pl^2}{\zz^{1/2}r_0^4}}e^{20\vf} =0.
\e

Our theory (i.e.\ a partial \un\ of NGT and a Standard Model) is a classical field theory. The problem which arises now is: How to quantize the
theory? We can consider our theory as GR (General Relativity) with additional geometrical sources and develop it into nonlocal theory.

There are several approaches of quantization of nonlocal theories (see Refs \cite{na,nb}). In this case we can avoid infinities appearing in
perturbation calculus, getting a theory which is renormalizable, superrenormalizable or even finite. Nonlocal theories, roughly speaking, are
equivalent to theories with higher \dv s up to an infinite order. An integral \tf\ is equivalent to a differential operator of an infinite
order.

Moreover, introducing nonlocality or a differential operator of an infinite order can be considered. A divergence of a one loop in the case of gravity
can be removed using \di al regularization-renormalization procedure. In order to avoid massive ghosts we should carefully design higher order \dv\
corrections to gravity, \YM' fields, Higgs' field, scalar field using differential operator of infinite order. According to Refs \cite{ng,nd} we
should add to the \lg\ \er{c2} ($B(\ov W,g,A_{\SU(3)},A_{\SU(2)\dr L\ot \U(1)\dr Y},\Psi,\Phi)$) the \fw\ terms
\beq c559
\ov R(\ov W)h_1\bgg{-\frac{\n^2}{\La^2}}\ov R(\ov W) + \ov R_\m h_2\bgg{-\frac{\n^2}{\La^2}}\ov R{}^\m.
\e
$\ov R(\ov W)$ and $\ov R{}^\m$ should be expressed by $\wt{\ov R}$, $\wt{\ov R}_\m$ and additional fields, i.e.\ $g\ink\m$ and $\ov W_\mu$
($\La$~is a~scale),
\beq c560
\n^2 = \gtn{\a\b}\wt{\ov \n}_\a \wt{\ov \n}_\b.
\e
Taking under consideration $A_{\SU(3)}$ gauge field we should add
\bg c561
e^{-24\vf}\gd\ell,\SU(3),ab, L^{a\m} h_3\bgg{-\frac{\nad{SU(3)}{\D}}{\La^2}}\gd H,b,\m, \\
\hbox{and }\ e^{24\vf}\gd h,\SU(3),ab, H^a h_4\bgg{-\frac{\nad{SU(3)}\D}{\La^2}}H^b \lb c562
\e
where
\beq c563
\nad{SU(3)}\D = \gtn{\a\b} \nad{gauge(SU(3))}{\wt{\ov\n}_\a} \nad{gauge(SU(3))}{\wt{\ov\n}_\b}.
\e
$\nad{SU(3)}\D$ is a Laplace operator gauge(SU(3)) covariant and covariant \wrt a \LC \cn\ on~$E$.

In the case of $A_{\SU(2)\dr L\ot \U(1)\dr Y}$ gauge field we have similarly (remember now $\xi=0$)
\beq c564
e^{-24\vf} \gd h,\SU(2),ab, \ov L{}^{a\m}h_5\bgg{-\frac{\nad{SU(2)}\D}{\La^2}}\gd \ov H,a,\m,
\e
where
\beq c565
\nad{SU(2)}\D = \gtn{\a\b}\nad{gauge(SU(2)\dr L)}{\wt{\ov\n}_\a} \nad{gauge(SU(2)\dr L)}{\wt{\ov\n}_\b}
\e
is a Laplace operator gauge($\SU(2)\dr L$) covariant and \ci\ \wrt \LC \cn\ on~$E$. We add also
\beq c566
e^{-24\vf} \gd h,\SU(2),ab, \ov H{}^a h_6 \bgg{-\frac{\nad{SU(2)}\D}{\La^2}}\ov H{}^b.
\e
In the case of $\U(1)\dr Y$ gauge field we have simply
\bg c567
\ov L{}^\m h_7\bgg{-\frac{\n^2}{\La^2}}G_\m \\
e^{24\vf}\gtk\m G_\m h_8 \bgg{-\frac{\n^2}{\La^2}} \gtk{\a\b}G_{\a\b}
\e
where $\n^2$ is given above. We add also
\bg c569
e^{-24\vf}\cdot \wt\vf{}^+ h_9\bgg{-\frac{\n^2}{\La^2}}\wt \vf \\
e^{-24\vf}\bigl(2g^{\mu\b}- \gtn{\mu\b}+ 2\z(g^{\mu\b}-\gtn{\mu\b})\bigr)\bgg{\pa_\mu \wt\vf - \frac12\,\frac{\a_s}{\sqrt{\hbar c}} \gd A,a,\mu,
\si^a \wt\vf - \frac12\,\frac{\a_s}{\sqrt{3\hbar c}}\,B_\mu \wt\vf}\hskip30pt \nn \\
\hskip60pt {}\cdot h_{10} \bgg{-\frac{\n^2+\nad{SU(2)}\D}{\La^2}}
\bgg{\pa_\b\wt\vf - \frac12\,\frac{\a_s}{\sqrt{\hbar c}}\,\gd A,a,\b, \si^a\wt\vf - \frac12\,\frac{\a_s}{\sqrt{3\hbar c}}\,B_\b \wt \vf} \lb c570 \\
\bgg{\bgg{1386 - \frac{\mu^2}{64(\mu^2+4)}}\gtn{\nu\g} - 484\gtk{\nu\mu}g_{\d\nu}\gtn{\d\g}}\vf_{,\nu}h_{11}\bgg{-\frac{\n^2}{\La^2}}\vf_{,\g}
\lb c571 \\
e^{12\vf} h_{12} \bgg{-\frac{\n^2}{\La^2}} e^{12\vf} \lb c572 \\
e^{11\vf} h_{13} \bgg{-\frac{\n^2}{\La^2}} e^{11\vf} \lb c573 \\
e^{10\vf} h_{14} \bgg{-\frac{\n^2}{\La^2}} e^{10\vf}, \lb c574
\e
All geometrical quantities must be expressed by $g_{(\a\b)}$, $g_{[\a\b]}$, $\ov W_\mu$.
$h_i$, $i=1,2,\dots,14$, are entire transcendental \f s of complex variable.

The problem which arises now is as follows: Is it possible to choose $h_i$, $i=1,2,\dots,15$, in such a way that no physical poles are introduced
while the theory will be (super-)renormalizable and unitary. It seems that such entire \f s can be defined (see also Refs \cite{na,nb,ng,nd}).
This will be examined elsewhere ($h_{15}$ is connected to $\cL\dr{int}$).

We hope to find them to get (super-)renormalizable or even finite theory unifying NGT and a bosonic part of a Standard Model (a~partial \un) with
``interference effects'' obtained on a level of the classical field theory. This will be in some sense geometrical (geometrization of physical
\ia s) and also nonlocal. This nonlocality should be of course causal and this depends on \f s $h_i$, $i=1,2,\dots,15$.

Such \f s are entire transcendental \f s. They are not polynomial, i.e.
\beq c575
h(z) = \sum_{n=0}^\iy a_nz^n \qh{and } \lim_{n\to\iy}\root n \of{|a_n|} =0.
\e
It means they are defined on the whole complex plane and according to the Liouville theorem they have a pole or an essential singularity at infinity.
The construction of such \f s can be done according to Refs \cite{ng,nd,mcpi,no}. In any case we can write
\beq c576
h(z) = 1 + \exp\bgg{\int_0^{p_\g(z)} \frac{1-\z(w)}w\, dw - 1},
\e
where $p_\g(z)$ is a real polynomial of degree~$\g$ and $p_\g(0)=0$, $\z(w)$~is an entire \f\ and real on the real axis
and $\z(0)=1$
\beq c577
|\z(w)|\to\iy \qh{for } |z|\to\iy,\ z\in\C.
\e
There are several propositions for such \f s in some applications of GR and \YM' fields.

We can perform a perturbation calculus using Feynman diagrams for S-matrix which is unitary. The full program will be developed below. Let us give
the \fw\ remarks. The Nonlocal Fields can be described by some integral kernels in the place of differential operators of infinite order. Some ideas
of nonlocal quantum fields have been introduced by H.~Yukawa (see Refs \cite{mcr1,mct1}). After a completion of nonlocal quantization of our \un\ one
can pose a question: Are our previous results on renormalization on Weinberg angle $\th_W$ still valid? It seems: Yes!

Let us consider a full \lg\ in the theory
\bml c578
L = \ov R(\ov W)h_1\bgg{-\frac{\n^2}{\La^2}}\ov R(\ov W) + \ov R_\m h_2\bgg{-\frac{\n^2}{\La^2}} \ov R{}^\m
-\frac{\a^2\dr{QCD}}{8\pi}\,e^{-24\vf}\gd\ell,\SU(3),ab, (H^aH^b - L^{a\m}\gd H,b,\m,)\\
{}+\frac{\a^2\dr{QCD}}{8\pi}\,e^{-24\vf}\gd\ell,\SU(3),ab, L^{a\m} h_3\bgg{-\frac{\nad{SU(3)}\D}{\La^2}}\gd H,b,\m,
-\frac{\a^2\dr{QCD}}{8\pi}\,\gd h,SU(3),ab, H^a h_4\bgg{-\frac{\nad{SU(3)}\D}{\La^2}}H^b\\
{}-\frac{\a_s^2}{8\pi}\,e^{-24\vf}\gd h,\SU(2),ij, (\ov H{}^i \ov H{}^j - \ov L{}^{i\m}\gd \ov H,j,\m,)
+ \frac{\a_s^2}{8\pi}\,e^{-24\vf}\gd h,\SU(2),ij, \ov L{}^{i\m}h_5\bgg{-\frac{\nad{SU(2)}\D}{\La^2}}\gd \ov H,j,\m, \\
{}-\frac{\a_s^2}{8\pi}\,e^{-24\vf}\gd h,\SU(2),ij, \ov H{}^i h_6\bgg{-\frac{\nad{SU(2)}\D}{\La^2}}\ov H{}^j
+ \frac{\a_s^2}{8\pi}\,e^{-24\vf}\bigl(2\gtk\m G_\m \gtk{\a\b}G_{\a\b}-L^\m G_\m\bigr)\\
{}+\frac{\a_s^2}{8\pi}\,e^{-24\vf}L^\m h_7\bgg{-\frac{\n^2}{\La^2}}G_\m
-\frac{\a_s^2}{4\pi} \,e^{-24\vf}\gtk\m G_\m h_8\bgg{-\frac{\n^2}{\La^2}}\gtk{\a\b}G_{\a\b} e^{-24\vf}V(\wt \vf)\\
{}+e^{-24\vf}\wt\vf{}^+ h_9\bgg{-\frac{\n^2}{\La^2}}\wt\vf
+ e^{-24\vf}(2g^{\mu\b}-\gtn{\mu\b} + 2\z(g^{\mu\b}-\gtn{\mu\b}))\\
{}\cdot\bgg{\pa_\mu\wt\vf - \frac{\a_s}{2\sqrt{\hbar c}}\gd A,a,\mu, \si^a\wt\vf - \frac{\a_s}{2\sqrt{3\hbar c}}B_\mu\wt\vf}
\bgg{\pa_\b\wt\vf - \frac{\a_s}{2\sqrt{\hbar c}}\gd A,a,\b, \si^a\wt\vf - \frac{\a_s}{2\sqrt{3\hbar c}}B_\b\wt\vf}\\
{}+ e^{-24\vf}(2g^{\mu\b}-\gtn{\mu\b}+2\z(g^{\mu\b}-\gtn{\mu\b}))
h_{10}\bgg{-\frac{\n^2+\nad{SU(2)}\D}{\La^2}}\bgg{\pa_\b \wt\vf - \frac{\a_s}{2\sqrt{\hbar c}}\gd A,a,\b, \si^a \wt\vf
- \frac{\a_s}{2\sqrt{3\hbar c}}\,B_\b \wt\vf}\\
{}+\bgg{\bgg{1386 - \frac{\mu^2}{64(\mu^2+4)}}\gtn{\nu\g} - 4849\gtk\m g_{\g\nu}\gtn{\d\g}}\vf_{,\nu}\vf_{,\g}
\bgg{\bgg{1386 - \frac{\mu^2}{64(\mu^2+4)}}\gtn{\nu\g}\\ {}- 4849\gtk{\nu\mu} g_{\d\nu}\gtn{\d\g}}\vf_{,\nu} h_{11}\bgg{-\frac{\n^2}{\La^2}}
\vf_{,\g} + e^{20\Psi_0}\bgg{\frac{2\a^2\dr{QCD}}{(\mu^2+4)^2}(2\mu^3+7\mu^2+25\mu+20) \\
{}-\frac12\bgg{\frac52\,\a^2\dr{QCD}+7\a_s^2}}\frac{e^{4\Psi_0}}{\ell^2\pl}\,e^{4\vf}
+\frac{4e^{2\Psi_0}}{r_0\zz}\,e^{2\vf} + \frac{4\pi\zy \ell\pl}{r_0^4\zz^{1/2}}\,e^{20\vf}\\
{}+ e^{20\Psi_0}\bgg{\frac{2\a\dr{QCD}^2}{(\mu^2+4)^2}(2\mu^3+7\mu^2+25\mu+20) - \frac12\bgg{\frac52\,\a\dr{QCD}^2 + 7\a_s^2}}
e^{12\vf}h_{12}\bgg{-\frac{\n^2}{\La^2}}e^{12\vf}\\ {}+ \frac{4e^{2\Psi_0}}{r_0\zz}\,e^{11\vf} h_{13}\bgg{-\frac{\n^2}{\La^2}}e^{11\vf}
+\frac{4\pi\zy \ell\pl}{r_0^4\zz^{1/2}} \,e^{10\vf}h_{14}\bgg{-\frac{\n^2}{\La^2}}e^{10\vf}\\
{}-i\z\sqrt6\cdot\frac{e^{-24\vf}\bigl(3F^3_\m g^\m(\wt\vf_1\wt\vf_1{}^\ast - \wt\vf_2\vf_2{}^\ast)
-\vf_1\vf_2^\ast (F_\m^- g^\m+ F^+_\m g^\m)\bigr)}{16\pi r_0^2\zz}\\
{}-i\z\sqrt6\cdot\frac{e^{-24\vf}}{16\pi r_0^2\zz}\,h_{15}\bgg{-\frac{\n^2+\nad{SU(2)}\D}{\La^2}}\bigl(3F^3_\m g^\m(\wt\vf_1\wt\vf_1{}^\ast
-\wt\vf_2\wt\vf_2{}^\ast) - \vf_1\wt\vf_2{}^\ast (F^-_\m g^\m + F^+_\m g^\m)\bigr)
\e

One gets
\bg c579
\ov R_{\b\g}(\ov W) = \wt{\ov R}_{\b\g}(\ov\G) - \frac34\,\wt{\ov\n}_\d \gd\D,\d,\b\g, - \frac12\,\wt{\ov\n}_\d \gd\ov Q,\d,\b\g,
+\frac14\,\wt{\ov\n}_\g \gd \D,\a,\b\a, + \frac23\,\ov W\ink{\b,\g} \\
R^\m(\ov W) = g^{\mu\b}g^{\nu\g}R_{\b\g}(\ov W) \lb c580 \\
\ov R(\ov W) = \wt{\ov R}(\wt{\ov \G}) + \wt{\ov R}_{\b\g}(g^{\b\g}-\gtn{\b\g}) - \frac34\,g^{\b\g}\wt{\ov\n}_\d \gd\D,\d,\b\g,\hskip80pt \nn \\
\hskip80pt {}- \frac12\,
\gtk{\b\g}\wt{\ov \n}_\d \gd \ov Q,\d,\b\g, + \frac14\,g^{\b\g}\wt{\ov\n}_\g \gd\D,\a,\b\a, + \frac23\,\gtk{\b\g}\ov W\ink{\b,\g} \lb c581
\e
where
\bg c582
\gd\ov Q,\nu,\g\mu, = \frac12(k_{\g\mu\rho}g^{\rho\nu} - 2g_{\tl[\mu\rho]}k_{\g\tp \a\b} \gtk{\nu\b}g^{\rho\a}) \\
\gd\D,\nu,\g\mu, = \bigl\{k_{\d(\g\rho}g\ink{\mu)\a}g^{\rho\a} + g_{([\mu\rho]}k_{\g)\a\b} g\ink{\d\z}g^{\rho\b}g^{\z\a}
- k_{\d\a\b}g_{([\g\rho]}g\ink{\mu)\z} g^{\rho\a}g^{\z\b}\bigr\} \lb c583 \\
k_{\a\b\g} = -\wt{\ov{\n}}_\a g\ink{\b\g} - \wt{\ov\n}_\b g\ink{\g\a} + \wt{\ov\n}_\g g\ink{\a\b} \lb c584
\e

Let us apply a path-integral method to quantize fields in our \un. Write rather formally (see \rc{mcx})
\beq c585
Z= \int e^{iS(A_{\SU(3)},A_{\SU(2)\ot\U(1)},g,W,\wt\vf,\vf)}\cdot DA_{\SU(3)}\,DA_{\SU(2)}\,DA_{\U(1)}\,Dg\,DW\,D\wt\vf\,D\vf
\e
where $S$ is a classical action
\beq c586
\bal
DA_{\SU(3)} &= \prod_\mu DA_{\SU(3)\mu} = \prod_{x,\mu} dA_{\SU(3)\mu}(x)\\
DA_{\SU(2)} &= \prod_\mu DA_{\SU(2)\mu} = \prod_{x,\mu} dA_{\SU(2)\mu}(x)\\
DA_{\U(1)} &= \prod_\mu DA_{\U(1)\mu} = \prod_{x,\mu} dA_{\U(1)\mu}(x)\\
Dg &= \prod_{\dwa{\a,\b\\\a\le\b}} Dg_{(\a\b)} \prod_{\dwa{\a,\b\\\a<\b}} g\ink{\a\b} = \prod_{\dwa{x,\a,\b\\\a\le\b}}dg_{(\a\b)}(x)
\prod_{\dwa{x,\a,\b\\\a<\b}} dg\ink{\a\b}(x) \\
D\ov W&= \prod_\mu D\ov W_\mu = \prod_{\mu,x} d\ov W_\mu(x)
\eal
\e
and
\beq c587
\bal
D\wt \vf &= D \vf_1 D\vf_2 = \prod_x d\vf_1(x) \prod_x d\vf_2(x) \\
D\vf &= \prod_x d\vf(x)
\eal
\e
mean \f al (nonexisting) measures for our fields. According to \rc{mcx} we add gauge fixing terms
\bml c588
L = -\frac{\eta^\m}{2\b_1}\,f_\nu[g] W_g\bgg{-\frac{\Box}{\La^2}}f_\mu[g] - \frac1{2\b_2} f_a[A_{\SU(3)}]W\dr{YMSU(3)}
\bgg{-\frac{\Box}{\La^2}} h^{ab\SU(3)}f_b[A_{\SU(3)}]\\
{}-\frac1{2\b'_2}\,f_a[A_{\SU(2)}]W\dr{YMSU(2)}\bgg{-\frac{\Box}{\La^2}}h^{ab\SU(2)}f_b[A_{\SU(2)}]\\
{}-\frac1{2\b_2''}\,f_a[A_{\U(1)\dr{Y}}]W\dr{YMU(1)\dr Y}\bgg{-\frac{\Box}{\La^2}}f_b[A_{\U(1)\dr Y}]
- \frac1{2\b_3}\, h[\ov W]\ov W \bgg{-\frac{\Box}{\La^2}}h(\ov W)
\e
where $\b_1,\b_2,\b_2',\b_2'',\b_3$ are \ct s and $\Box$ is an ordinary d'Alembert operator in Minkowski \spt, $f^\mu[g]=f[g_{\a\b}]$ in a
gauge-fixing \f\ for \gr al field, $f_a[A]$ for gauge fields ($A_{\SU(3)}$, $A_{\SU(2)}$, $A_{\U(1)}$). $W_g$~is a gravity gauge-fixing weight \f,
$W\dr{YM}$~is a gauge field gauge-fixing weight \f\ ($A_{\SU(3)}$, $A_{\SU(2)}$, $A_{\U(1)}$) (for \YM\ field). We add also a gauge-fixing \f\
for $\ov W_\mu$ field with its weight~$\ov W$. We can also add gauge-fixing terms for Higgs' fields $\wt\vf=(\wt\vf_1,\wt\vf_2)$.

Sometimes it is possible to consider a gauge condition which involves gauge and Higgs' fields together. In this way we can get also additional
Faddeev--Popov ghosts. But this does not threaten us. These ghosts are easily exorcized. The most important problems in this theory are possible
massive ghosts which could appear if \f s $h_i$ are not properly chosen. The FP (Faddeev--Popov) ghosts are not dangerous as we mention above from
quantum field theory point of view. They are also exorcized from geometrical point of view, i.e. they can be geometrized (see \rc{mcz}). According to
Refs \cite{mcz,mcla} a gauge field (in a specific fixed gauge, i.e. in a section of a principal bundle) plus a ghost field is a globally defined \cn\
on the principal bundle. The anticommuting property of a ghost field can be easily derived and a nilpotent BRST charge obtained as a differential
operator.

In order to proceed a \f al integration we apply a well-known Faddeev--Popov trick in order to do an integration over those configurations which
\sf y a gauge-fixing conditions. In this approach ghost fields appear with \lg. We have two kinds of ghosts --- gauge fields ghosts and gravity field
ghosts. Thus we have ghost field \lg
\bml c582n
L\dr{gh} = \ov c_a(\SU(3)) M_{ab}(\SU(3))c_b(\SU(3)) + \ov c_a(\SU(2))M_{ab}(\SU(2))c_b(\SU(2)) \\
{}+ \ov c_a(\U(1))M_{ab}(\U(1))c_b(\U(1)) + \ov c{}^\mu N_\m c^\nu + \ov c Mc
\e
coming from exponentiation of a Faddeev--Popov determinant (an infinite analogue of a ``Jacobian'') \st
\begin{gather}
\left. \bal
M_{ab}(\SU(3))c_b(\SU(3)) &= \d_c f_a[A_{\SU(3)},x]\\
M_{ab}(\SU(2))c_b(\SU(2)) &= \d_c f_a[A_{\SU(2)},x]\\
M_{ab}(\U(1))c_b(\U(1)) &= \d_c f_a[A_{\U(1)},x]
\eal \right\} \q\ \hbox{(scalars)} \nn \\
\bal
N_\m c_\mu &= \d_c f_\nu[g,x] \q\ \hbox{(vector)}\\
M_c &= \d h[\ov W,x] \q\ \hbox{(scalar)} \lb c583n
\eal
\end{gather}
where $\d_cf_a$ is an infinitesimal \tf \ of~$f_a$ with gauge parameters $c_b$ (for all gauge fields $A_{\SU(3)}$, $A_{\SU(2)}$, $A_{\U(1)}$),
$\d_cf_\nu$~is the infinitesimal \tf \ of~$f_\nu$ with a changing of a frame with parameter $c_\mu$ and $\d h$ is an analogue of a gauge changing
of~$\ov W_\nu$. They do not depend on weighting \f s. The Faddeev--Popov ghosts are ghost fields in the sense that they do not have a right
statistics. In order to get ghost \lg\ we should integrate using anticommuting fields. In this way they are anticommuting bosons. Thus one gets
\bml c589
Z = \int e^{iS[A_{\SU(3)},A_{\SU(2)},A_{\U(1)},g,W,c_a(\SU(3)),c_a(\SU(2)),c_a(\U(1)),c_\mu,c,\td\vf,\vf]}\,DF\,Dh\\
{}=V'\int e^{iS[A_{\SU(3)},A_{\SU(2)},A_{\U(1)},g,W,c_a(\SU(3)),c_a(\SU(2)),c_a(\U(1)),c_\mu,c,\td\vf,\vf]}\,DF
\e
where
\bml c590
DF = \prod_x dA\dq{fix}_{\SU(3)}(x) \prod_x dA\dq{fix}_{\SU(2)}(x) \prod_x dA\dq{fix}_{\U(1)}(x) \prod_x d\ov W{}\dq{fix}(x)\\
\prod_x \bigl(dc_a(\SU(3))(x)\, d\bar c_a(\SU(3))(x)\bigr) \prod_x \bigl(dc_a(\SU(2))(x)\, d\bar c_a(\SU(2))(x)\bigr)
\prod_x \bigl(dc_a(\U(1))\, d\bar c_a(\U(1))\bigr) \\ \prod_x (dc_\mu(x)\,d\bar c_\mu(x)\bigr) \prod_x (dc(x)\,d\bar c(x))
\prod_x d\wt\vf_1(x) \prod_x d\wt\vf_2(x) \prod_x d\vf(x),
\e
$Dh = \prod_x dh(x)$ (integration over gauge groups), $\bar c_a,\bar c_\mu,\bar c$ mean antighost fields. $V'$~is an ``infinite volume'' of a local
gauge group ($V_{\SU(3)}\tm V_{\SU(2)}\tm V_{\U(1)}\tm V\dr{GL(4,R)}$). $A\dq{fix}(x)$ means that the gauge for $A_{\SU(3)}$, $A_{\SU(2)}$, or
$A_{\U(1)}$ has been fixed.

The same for $g\dq{fix}(x)$ and for $\ov W{}\dq{fix}(x)$. (In a more geometrical language we say that an integration is over an orbit space of a
local gauge group.)
\bml c591
S[A_{\SU(3)},A_{\SU(2)},A_{\U(1)},g,\ov W,c_a(\SU(3)),c_a(\SU(2)),c_a(\U(1)),c_\mu,c,\td\vf,\vf] \\
= \int d^4x \sqrt{-g}\bigl(L_{\SU(3)}+L_{\SU(2)}+L_{\U(1)} + Lg + Lgh + L(\wt \vf) + L(\vf)\bigr)
\e

The above formulae are starting points for path-integral quantization of our \un\ of NGT and a Standard Model via \E\nos\ \KK (Jordan--Thiry) Theory
with ``interference effects''. After inclusions of fermion fields this program accomplishes the Einstein idea of Unified Field Theory of all \ia s,
which is geometrical (geometrization of physical \ia s), nonlinear (nonlinear field \e s) and also nonlocal.

We can develop a path integral formalism in Euclidean \spt. In this case $\Box=-\D_4$ (where $\D_4$ is a four-\di al Laplacian) and we have $e^{-S}$
in the place of~$e^{iS}$. This formalism after an analytical continuation $t\mapsto it$ can give some sound results. We change \gr al field variables
in hyperbolic case in the \fw\ way. Let
\beq c691
\ov h{}^\m = \sqrt{-g}\,g^\m, \q g=\det g_\m, \q \ov h=\det \ov h{}^\m.
\e
Let us define the \fw\ measure of path integration
\beq c692
\prod_x g^{5/2}(x)\prod_{\mu\le\nu} dg^{(\m)} (x)\prod_{\mu<\nu} d\gtk\m(x)
= \prod_x \ov h{}^{-5/2}(x) \prod_{\mu\le\nu} d\ov h{}^{(\m)}(x) \prod_{\mu<\nu}d\ov h{}^{[\m]}(x).
\e
This measure is ${\rm GL(4,R)}$-gauge invariant.

In order to do perturbation calculus for \gr al field we define
\beq c693
\ov h{}^\m = \eta^\m+\ka u^\m,
\e
where
\beq c694
\ka =\frac{G_N}{c^4} \qh{and } \eta^\m \eta_{\mu\a} = \d^\nu_\a,
\e
$\eta_{\mu\a}$ is a Minkowski tensor.

\section*{Acknowledgement}
I would like to thank Professor B. Lesyng for the opportunity to carry out
computations using Mathematica\TM~9\footnote{Mathematica\TM\ is the
registered mark of Wolfram Co.} in the Centre of Excellence
BioExploratorium, Faculty of Physics, University of Warsaw, Poland.
I~would like to thank the referees for critical remarks to improve my paper.

\end{document}